%% file: hatp58-64.tex
\documentclass[apjl,ip,twocolumn]{aastex63}
\usepackage{comment}
\usepackage{ifthen}
\usepackage[switch]{lineno}
\modulolinenumbers[5]

\input{newcommand_gen.tex}

\newcommand{\loopand}{\ifnum\value{planetcounter}=2 and \else\fi}
\newcommand{\loopcomma}{\ifnum\value{planetcounter}<2 ,\else. \fi}
\newcommand{\loopcommanoperiod}{\ifnum\value{planetcounter}<2 ,\else \space\fi}
\newcommand{\loopcommanospace}{\ifnum\value{planetcounter}<2 ,\else \fi}

\input{newcommand_spec_multi.tex}

\newcounter{planetcounter}

\newboolean{emulateapj}
\setboolean{emulateapj}{true}

\newboolean{rvtablelong}
\setboolean{rvtablelong}{true}

\newboolean{astroph}
\setboolean{astroph}{true}

\newboolean{longabstract}
\setboolean{longabstract}{false}

\shortauthors{Bakos et al.}
\shorttitle{\hatcur{58}\lowercase{b}--\hatcur{64}\lowercase{b}}
\ifthenelse{\boolean{emulateapj}}{
    
}{
    
}

\begin{document}

\title{
\hatcur{58}\lowercase{b}--\hatcur{64}\lowercase{b}: Seven Planets
 Transiting Bright Stars\footnote{Based on observations of
 the Hungarian-made Automated Telescope Network and observations obtained
 at the following observatories: W.~M.~Keck Observatory, the 1.5\,m and
 the 1.2\,m telescopes at the Fred Lawrence Whipple Observatory of the
 Smithsonian Astrophysical Observatory, the Kitt Peak National
 Observatory, the 1.93\,m telescope at Observatoire de Haute-Provence,
 the Subaru Telescope of the National Astronomical Observatory of
 Japan, the Nordic Optical Telescope in the Spanish Observatorio del
 Roque de los Muchachos of the Intituto de Astrof\'isica de Canarias,
 and the Apache Point Observatory 3.5\,m telescope.}
}

\correspondingauthor{G\'asp\'ar Bakos}
\email{gbakos@astro.princeton.edu}

\author[0000-0001-7204-6727]{G.~\'A.~Bakos}
\altaffiliation{Packard Fellow}
\affil{Department of Astrophysical Sciences, Princeton University, NJ 08544, USA}
\affil{MTA Distinguished Guest Fellow, Konkoly Observatory, Research Centre for Astronomy and Earth Sciences, H-1121 Budapest, Konkoly Thege Miklós út 15-17, Hungary}

\author[0000-0001-8732-6166]{J.~D.~Hartman}
\affil{Department of Astrophysical Sciences, Princeton University, NJ 08544, USA}

\author[0000-0002-0628-0088]{W.~Bhatti}
\affil{Department of Astrophysical Sciences, Princeton University, NJ 08544, USA}

\author[0000-0002-8423-0510]{Z.~Csubry}
\affil{Department of Astrophysical Sciences, Princeton University, NJ 08544, USA}

\author[0000-0003-4464-1371]{K.~Penev}
\affil{Department of Physics, University of Texas at Dallas, Richardson, TX 75080, USA}


\author[0000-0001-6637-5401]{A.~Bieryla}
\affiliation{Center for Astrophysics $\vert$ Harvard \& Smithsonian, 60 Garden St, Cambridge, MA 02138, USA}

\author[0000-0001-9911-7388]{D.~W.~Latham}
\affiliation{Center for Astrophysics $\vert$ Harvard \& Smithsonian, 60 Garden St, Cambridge, MA 02138, USA}

\author[0000-0002-8964-8377]{S.~Quinn}
\affiliation{Center for Astrophysics $\vert$ Harvard \& Smithsonian, 60 Garden St, Cambridge, MA 02138, USA}

\author[0000-0003-1605-5666]{L.~A.~Buchhave}
\affiliation{DTU Space, National Space Institute, Technical University of Denmark, Elektrovej 328, DK-2800 Kgs.~Lyngby, Denmark}

\author{G.~Kov\'acs}
\affiliation{Konkoly Observatory, Research Centre for Astronomy and Earth Sciences, H-1121 Budapest, Konkoly Thege Miklós út 15-17, Hungary}

\author[0000-0002-5286-0251]{Guillermo Torres}
\affiliation{Center for Astrophysics $\vert$ Harvard \& Smithsonian, 60 Garden St, Cambridge, MA 02138, USA}

\author{R.~W.~Noyes}
\affiliation{Center for Astrophysics $\vert$ Harvard \& Smithsonian, 60 Garden St, Cambridge, MA 02138, USA}

\author[0000-0002-7061-6519]{E.~Falco}
\affiliation{Center for Astrophysics $\vert$ Harvard \& Smithsonian, 60 Garden St, Cambridge, MA 02138, USA}

\author{Bence B\'eky}
\affiliation{Google, Googleplex, 1600 Amphitheatre Parkway, Mountain View, CA 94043, USA}

\author[0000-0002-5610-7697]{T.~Szklen\'ar}
\affil{Konkoly Observatory, Research Centre for Astronomy and Earth Sciences, H-1121 Budapest, Konkoly Thege Miklós út 15-17, Hungary}

\author[0000-0002-9789-5474]{G.~A.~Esquerdo}
\affiliation{Center for Astrophysics $\vert$ Harvard \& Smithsonian, 60 Garden St, Cambridge, MA 02138, USA}

\author[0000-0001-8638-0320]{A.~W.~Howard}
\affil{Department of Astronomy, California Institute of Technology, Pasadena, CA, USA}

\author[0000-0002-0531-1073]{H.~Isaacson}
\affil{Department of Astronomy, University of California, Berkeley, CA, USA}

\author[0000-0002-2909-0113]{G.~Marcy}
\affiliation{Department of Astronomy, University of California, Berkeley, CA, USA}

\author[0000-0001-8033-5633]{B.~Sato}
\affiliation{Department of Earth and Planetary Sciences, Tokyo Institute of Technology, 2-12-1 Ookayama, Meguro-ku, Tokyo 152-8551, Japan}

\author[0000-0001-8388-8399]{I.~Boisse}
\affil{Aix Marseille Universit\'e, CNRS, LAM (Laboratoire d'Astrophysique de Marseille) UMR 7326, F-13388, Marseille, France}

\author[0000-0002-3586-1316]{A.~Santerne}
\affil{Aix Marseille Universit\'e, CNRS, LAM (Laboratoire d'Astrophysique de Marseille) UMR 7326, F-13388, Marseille, France}

\author{G.~H\'ebrard}
\affil{Institut d'Astrophysique de Paris, UMR7095 CNRS, Universit\'e Pierre \& Marie Curie, 98bis boulevard Arago, 75014 Paris, France}

\author[0000-0003-2935-7196]{M.~Rabus}
\affil{Las Cumbres Observatory Global Telescope Network, 6740 Cortona Dr.~Suite 102, Goleta, CA 93117}
\affil{Department of Physics, University of California, Santa Barbara, CA 93106-9530, USA}

\author[0000-0002-8590-007X]{D.~Harbeck}
\affil{Las Cumbres Observatory Global Telescope Network, 6740 Cortona Dr.~Suite 102, Goleta, CA 93117}

\author[0000-0001-5807-7893]{C.~McCully}
\affil{Las Cumbres Observatory Global Telescope Network, 6740 Cortona Dr.~Suite 102, Goleta, CA 93117}

\author[0000-0002-0885-7215]{M.~E.~Everett}
\affiliation{NSF's Optical-Infrared Astronomy Research Lab, Tucson, AZ 85719 USA}

\author[0000-0003-2159-1463]{E.~P.~Horch}
\affiliation{Department of Physics, Southern Connecticut State University, 501 Crescent Street, New Haven, CT 06515, USA}

\author[0000-0001-8058-7443]{L.~Hirsch}
\affiliation{Kavli Institute for Particle Astrophysics and Cosmology, Stanford University, Stanford, CA, USA}

\author[0000-0002-2532-2853]{S.~B.~Howell}
\affiliation{NASA Ames Research Center, Moffett Field, CA, 94035, USA}

\author[0000-0003-0918-7484]{C.~X.~Huang}
\affiliation{Department of Physics, and Kavli Institute for Astrophysics and Space Research, Massachusetts Institute of Technology, Cambridge, MA 02139, USA}

\author{J.~L\'az\'ar}
\affil{Hungarian Astronomical Association, 1451 Budapest, Hungary}

\author{I.~Papp}
\affil{Hungarian Astronomical Association, 1451 Budapest, Hungary}

\author{P.~S\'ari}
\affil{Hungarian Astronomical Association, 1451 Budapest, Hungary}


\begin{abstract}

\setcounter{footnote}{1}
\ifthenelse{\boolean{longabstract}}{
	We report the discovery and characterization of 7 transiting
	extrasolar planets from the HATNet survey. The planets, which are hot
	Jupiters and Saturns transiting bright sun-like stars, include: \hatcurb{58} (with
	mass $\mpl = \hatcurPPm{58}$\,\mjup, radius $\rpl =
	\hatcurPPr{58}$\,\rjup, and orbital period $P =
	\hatcurLCPshort{58}$\,days), \hatcurb{59} ($\mpl =
	\hatcurPPm{59}$\,\mjup, $\rpl = \hatcurPPr{59}$\,\rjup, $P =
	\hatcurLCPshort{59}$\,days), \hatcurb{60} ($\mpl =
	\hatcurPPm{60}$\,\mjup, $\rpl = \hatcurPPr{60}$\,\rjup, $P =
	\hatcurLCPshort{60}$\,days), \hatcurb{61} ($\mpl =
	\hatcurPPm{61}$\,\mjup, $\rpl = \hatcurPPr{61}$\,\rjup, $P =
	\hatcurLCPshort{61}$\,days), \hatcurb{62} ($\mpl =
	\hatcurPPm{62}$\,\mjup, $\rpl = \hatcurPPr{62}$\,\rjup, $P =
	\hatcurLCPshort{62}$\,days), \hatcurb{63} ($\mpl =
	\hatcurPPm{63}$\,\mjup, $\rpl = \hatcurPPr{63}$\,\rjup, $P =
	\hatcurLCPshort{63}$\,days), and \hatcurb{64} ($\mpl =
	\hatcurPPm{64}$\,\mjup, $\rpl = \hatcurPPr{64}$\,\rjup, $P =
	\hatcurLCPshort{64}$\,days). These planets orbit the F and G stars:
	\hatcur{58} (with mass $\mstar = \hatcurISOm{58}$\,\msun, radius
	$\rstar = \hatcurISOr{58}$\,\rsun, and apparent optical magnitude $V =
	\hatcurCCtassmv{58}$\,mag), \hatcur{59} ($\mstar =
	\hatcurISOm{59}$\,\msun, $\rstar = \hatcurISOr{59}$\,\rsun, and $V =
	\hatcurCCtassmv{59}$\,mag), \hatcur{60} ($\mstar =
	\hatcurISOm{60}$\,\msun, $\rstar = \hatcurISOr{60}$\,\rsun, and $V =
	\hatcurCCtassmv{60}$\,mag), \hatcur{61} ($\mstar =
	\hatcurISOm{61}$\,\msun, $\rstar = \hatcurISOr{61}$\,\rsun, and $V =
	\hatcurCCtassmv{61}$\,mag), \hatcur{62} ($\mstar =
	\hatcurISOm{62}$\,\msun, $\rstar = \hatcurISOr{62}$\,\rsun, and $G =
	\hatcurCCgaiamG{62}$\,mag), \hatcur{63} ($\mstar =
	\hatcurISOm{63}$\,\msun, $\rstar = \hatcurISOr{63}$\,\rsun, and $V =
	\hatcurCCtassmv{63}$\,mag), and \hatcur{64} ($\mstar =
	\hatcurISOm{64}$\,\msun, $\rstar = \hatcurISOr{64}$\,\rsun, and $V =
	\hatcurCCtassmv{64}$\,mag). With $V = \hatcurCCtassmv{60}$\,mag,
	\hatcur{60} is an especially bright transiting planet host, and an
	excellent target for additional follow-up observations. With $\rpl =
	\hatcurPPr{64}$\,\rjup, \hatcurb{64} is a highly inflated hot Jupiter
	around a star nearing the end of its main-sequence lifetime, and is
	among the largest known planets. Five of the seven systems have
	long-cadence observations by {\em TESS} which are included in the
	analysis. Of particular note is \hatcur{59} (TOI-1826.01) which is within the
	northern continuous viewing zone of the {\em TESS} mission, and for
	which seven sectors of observations are included in the analysis, and
	\hatcur{60}, which is a candidate identified by the MIT
	quick-look pipeline as TOI-1580.01.
}{
	We report the discovery and characterization of 7 transiting
	exoplanets from the HATNet survey. The planets, which are hot
	Jupiters and Saturns transiting bright sun-like stars, include: 
	\hatcurb{58} (with mass $\mpl = \hatcurPPmshort{58}$\mjup, radius $\rpl = \hatcurPPrshort{58}$\rjup, and orbital period $P = \hatcurLCPshort{58}$days), 
	\hatcurb{59} ($\mpl = \hatcurPPmshort{59}$\mjup, $\rpl = \hatcurPPrshort{59}$\rjup, $P = \hatcurLCPshort{59}$days), 
	\hatcurb{60} ($\mpl = \hatcurPPmshort{60}$\mjup, $\rpl = \hatcurPPrshort{60}$\rjup, $P = \hatcurLCPshort{60}$days), 
	\hatcurb{61} ($\mpl = \hatcurPPmshort{61}$\mjup, $\rpl = \hatcurPPrshort{61}$\rjup, $P = \hatcurLCPshort{61}$days), 
	\hatcurb{62} ($\mpl = \hatcurPPmshort{62}$\mjup, $\rpl = \hatcurPPrshort{62}$\rjup, $P = \hatcurLCPshort{62}$days), 
	\hatcurb{63} ($\mpl = \hatcurPPmshort{63}$\mjup, $\rpl = \hatcurPPrshort{63}$\rjup, $P = \hatcurLCPshort{63}$days), and 
	\hatcurb{64} ($\mpl = \hatcurPPmshort{64}$\mjup, $\rpl = \hatcurPPrshort{64}$\rjup, $P = \hatcurLCPshort{64}$days). 
	The typical errors on these quantities are $0.06\mjup$,
	$0.03\rjup$, and 0.2seconds, respectively. We also provide accurate stellar
	parameters for each of the hosts stars. With $V =
	\hatcurCCtassmv{60}$mag, \hatcur{60} is an especially bright
	transiting planet host, and an excellent target for additional
	follow-up observations. With $\rpl = \hatcurPPr{64}$\rjup,
	\hatcurb{64} is a highly inflated hot Jupiter around a star nearing
	the end of its main-sequence lifetime, and is among the largest
	known planets. Five of the seven systems have long-cadence
	observations by {\em TESS} which are included in the analysis. Of
	particular note is \hatcur{59} (TOI-1826.01) which is within the
	Northern continuous viewing zone of the {\em TESS} mission, 
%
%
	and \hatcur{60}, which is the TESS candidate 
%
%
TOI-1580.01.
}
\setcounter{footnote}{0}
\end{abstract}

\keywords{
    planetary systems ---
    stars: individual (
\setcounter{planetcounter}{1}
\hatcur{58},
\hatcurCCgsc{58}\loopcommanoperiod
\setcounter{planetcounter}{2}
\hatcur{59},
\hatcurCCgsc{59}\loopcommanoperiod
\setcounter{planetcounter}{3}
\hatcur{60},
\hatcurCCgsc{60}\loopcommanoperiod
\setcounter{planetcounter}{4}
\hatcur{61},
\hatcurCCgsc{61}\loopcommanoperiod
\setcounter{planetcounter}{5}
\hatcur{62},
\hatcurCCgsc{62}\loopcommanoperiod
\setcounter{planetcounter}{6}
\hatcur{63},
\hatcurCCgsc{63}\loopcommanoperiod
\setcounter{planetcounter}{7}
\hatcur{64},
\hatcurCCgsc{64}\loopcommanoperiod
\setcounter{planetcounter}{8}
) 
    techniques: spectroscopic, photometric
}


\section{Introduction}
\label{sec:introduction}

The Hungarian-made Automated Telescope Network
\citep[HATNet;][]{bakos:2004:hatnet} began initial operations in 2003,
with the primary science goal of discovering and accurately
characterizing transiting extrasolar planets (TEPs) around bright
stars. It is one of four ongoing ground-based wide field transiting
planet surveys with more than ten planet discoveries, the others being
HATSouth \citep[][although led by the same PI, this project is
independent from the northern HATNet survey]{bakos:2013:hatsouth},
SuperWASP \citep{pollacco:2006} and KELT \citep{pepper:2007}.

HATNet consists of six 11\,cm diameter telephoto lenses coupled to
front-side-illuminated charged-coupled device (CCD) imagers, each in a
separate mount and enclosure. Four of the units (called HAT-5, -6, -7,
and -10) are located at Fred Lawrence Whipple Observatory (FLWO) in Arizona,
while the other two (called HAT-8 and -9) are located on the roof of
the Submillimeter Array service building at Mauna Kea Observatory (MKO) in
Hawaii. The system has been fully operational in an autonomous fashion
since 2004, and has remained nearly homogenous, with only a few
changes to the instrumentation and observing procedures since that
time. To date a total of 63 TEP discoveries have been published based
on HATNet observations (the most recent being
\citealp{zhou:2019:hat6970}). Here we present the discovery of 7 new TEP
systems identified using HATNet, together with an accurate determination
of the system parameters, including precise radial velocity (RV) observations
used to measure the planetary masses. Before delving into a detailed
discussion of these new discoveries, we first provide a brief update
on the status of HATNet.

Since 2004 there have been four different combinations of CCD cameras
and filters used by HATNet. The initial setup (until the summer of
2007) made use of Apogee AP10 $2K \times 2K$ CCDs and Cousins $I$-band
filters. This provided a $8\fdg 2 \times 8\fdg 2$ field of view
(FOV) and a plate scale of 14\arcsec\,pixel$^{-1}$. In September 2007
we replaced the CCDs to Apogee U16m $4K \times 4K$ imagers, providing a
larger field of view ($10\fdg 6 \times 10\fdg 6$) and higher spatial
resolution (9\arcsec\,pixel$^{-1}$). We also changed the filters to
Bessel $R$-band to better match the peak QE of the CCD, and a year later
(in September 2008), we changed the
filters to Sloan $r$ band to have better overall response, and sharp
wavelength boundaries. Majority of the HATNet survey was performed
with this setup, i.e.~the Apogee U16m $4K \times 4K$ imagers and the
Sloan $r$ band filters. The most recent modification was in October
2013, when the imager on HAT-7 at FLWO was changed to an FLI
back-side-illuminated $2K \times 2K$ CCD device. The other units
continue to use the Apogee U16m $4K \times 4K$ devices.

HATNet follows a point-and-stare mode of observations, where each unit
is assigned a primary field (one of 838 discrete pointings which tile
the full $4\pi$ steradian celestial sphere), which it observes
continuously over the night using 3\,min integrations, so long as the
field is above $30^{\circ}$ elevation, and not too close to the moon. A
secondary field is also assigned to each instrument, which is
observed when the primary field is not visible. In recent years we have
adopted a strategy where all of the units are assigned the same primary
and secondary fields, which we have found to significantly increase the
sensitivity to small radius planets. This is in contrast to our earlier
mode of observing where different units are assigned different fields
to maximize the sky coverage. The total time spent on a field varies
significantly, from a minimum of 2,000 observations, to as many as
40,000 observations collected (the median is 6000 observations). As of
May 2020, a total of 185 fields, corresponding to 148 unique pointing
positions\footnote{We have revisited some sky positions with a
different instrumental configuration leading to multiple ``fields'' for
these positions.}, and covering approximately 35\% of the Northern sky,
have been observed, reduced, and searched for transiting planets. Some
9.3 million light curves have been generated from these images for 5.9
million stars ranging in brightness from $r \approx 9.5$\,mag to $r =
14.5$. The trend-filtered light curves reach a precision of $\sim
3$\,mmag at cadence for the brightest sources. Based on these light
curves a total of 2460 candidate transiting planets have been selected.

The majority of the candidates (approximately 2200 to date) have
received at least some follow-up spectroscopic and/or photometric
observations using a variety of facilities (e.g.,
\citealp{latham:2009:hat8}). Based on these observations, some 1950 of
the candidates have been set aside as false positives or false alarms
(i.e., cases where we suspect that the candidate transit signal
detected in the HATNet light curve is spurious). 
In addition to those planets presented here, more than a dozen other
planets have been confirmed, but have not yet been published. Some 250
candidates have received some follow-up, but require additional
follow-up observations for confirmation and characterization.

The seven planets that are the focus of this paper are quite typical of
the population of transiting planets that have been discovered thus far
by HATNet. With planetary masses between $\hatcurPPm{58}$\,\mjup\
(\hatcurb{58}) and $\hatcurPPm{59}$\,\mjup\ (\hatcurb{59}), orbital
periods between $\hatcurLCPshort{61}$\,days (\hatcurb{61}) and
$\hatcurLCPshort{60}$\,days (\hatcurb{60}), and host star masses
between $\hatcurISOm{63}$\,\msun\ (\hatcur{63}) and
$\hatcurISOm{64}$\,\msun\ (\hatcur{64}), these are all hot Jupiters
transiting Sun-like stars. The host stars are all relatively bright,
particularly \hatcur{60} at $V = \hatcurCCtassmv{60}$\,mag, enabling
the accurate determination of the orbital parameters, and planetary and
stellar physical parameters, that we provide in this paper for each of
these systems. The targets are also amenable to additional follow-up
observations that may be carried out to characterize the orbital
geometries (e.g., spin--orbit alignment measurements via the
Rossiter-McLaughlin effect, \citealp{queloz:2000}) and planetary
atmospheres (e.g., transmission spectroscopy,
\citealp{charbonneau:2002}). The continued discovery and
characterization of TEPs such as these increases the sample that may be
used for statistical analysis of the population, which in turn provides
insights into the physical processes involved in their formation and
evolution. In fact, the planets reported here have already been
included in a statistical analysis carried out by
\citet{hartman:2016:hat6566}, which revealed observational evidence for
the re-inflation of close-in giant planets.

In the next section (\ref{sec:obs}) we describe the observations
collected to identify, confirm, and characterize the seven
transiting planet systems presented here. The analysis carried out to
measure the parameters of each system and to rule out blended stellar
eclipsing binary false positive scenarios is described in
\refsecl{analysis}. We discuss the results in \refsecl{discussion}.

\section{Observations}
\label{sec:obs}

\subsection{Photometric detection}
\label{sec:detection}

Periodic transit events were first identified for all seven systems
based on time series photometric observations obtained with the HATNet
wide-field photometric network \citep{bakos:2004:hatnet}. The
instruments and filters used, number of measurements collected and date
range over which they were collected, observational cadence, and
photometric precision achieved are all listed in
Tables~\ref{tab:photobs} and~\ref{tab:photobs2} for each of the seven
systems. The raw HATNet images were reduced to light curves following
\citet{bakos:2004:hatnet}, making use of aperture and image subtraction
photometry routines based on the FITSH software package
\citet{pal:2012}. Following \citet{bakos:2010:hat11} we filtered
variations from the light curves that are correlated with a variety of
auxiliary parameters, and we then applied the Trend-Filtering Algorithm
(TFA) of \citet{kovacs:2005:TFA}. The latter operates by fitting each
light curve as a linear combination of ``template'' light curves (in
our case these are light curves for a random sample of non-variable
stars distributed across the image plane and in magnitude) and then
subtracting the best fit model from the light curve being filtered. In
our initial pass we apply the filtering in signal recovery mode, where
we assume the light curve contains no astrophysical variations. We then
use the Box Least Squares \citep[BLS;][]{kovacs:2002:BLS} method to
search the filtered light curves for periodic transits. Once recovered,
we then re-apply the trend filtering, this time in signal
reconstruction mode, where we simultaneously fit to the light curve the
linear filter and a periodic box-shaped transit model. This produced a
filtered light curve without distorting the transit signal. The final
trend-filtered photometric data for each system are shown phase-folded
in Figure~\ref{fig:hatp58}, and Figures~\ref{fig:hatp59}--\ref{fig:hatp64}, 
while the measurements are available in \reftabl{phfu}.

We used the {\sc vartools} package \citep{hartman:2016:vartools} to
search the residual HATNet light curves of each target for additional
periodic transit signals using BLS, but do not find any significant
signals attributable to additional transiting planets around these
stars. For \hatcur{58} the highest peak in the BLS spectrum (in the
residual light curve) is at
$P=38.5$\,d with a signal-to-pink noise ratio (${\rm S}/{\rm N}_{\rm
  pink}$) of 5.5 (we require ${\rm S}/{\rm N}_{\rm pink} > 7.0$ for
detection) and a transit depth of 6.3\,mmag. For \hatcur{59} we detect
a signal at the sidereal frequency, which is presumably due to
systematic errors in the photometry that are not fully removed through
EPD and TFA.  The first harmonic of this same signal is also detected
with the Generalized Lomb-Scargle periodogram
\citep[GLS;][]{zechmeister:2009}, and when it is filtered from the
light curve using a Fourier series fit, we find no other significant
transit signals with BLS. Altogether, we find the following peaks,
significances and transit depths in the residual light curves:
\begin{itemize}
\item  \hatcur{58}, $P=38.5$\,d, ${\rm S}/{\rm N}_{\rm pink} = 5.5$, 6.3\,mmag;
\item  \hatcur{59}, $P=$1.59\,d, ${\rm S}/{\rm N}_{\rm pink} = 6.0$, 2.3\,mmag; 
\item  \hatcur{60}, $P=2.48$\,d, ${\rm S}/{\rm N}_{\rm pink} = 6.1$, 2.3\,mmag; 
\item  \hatcur{61}, $P=61.8$\,d,  ${\rm S}/{\rm N}_{\rm pink} = 5.2$, 2.7\,mmag;
\item  \hatcur{62}, $P=0.146$\,d,  ${\rm S}/{\rm N}_{\rm pink} = 6.1$, 2.9\,mmag;
\item  \hatcur{63}, $P=0.194$\,d, ${\rm S}/{\rm N}_{\rm pink} = 6.0$, 8.3\,mmag; 
\item  \hatcur{64}, $P=0.438$\,d, ${\rm S}/{\rm N}_{\rm pink} = 6.7$, 2.3\,mmag;
\end{itemize}

We also used {\sc vartools} to search the residual HATNet light curves
for continuous periodic variability with GLS. For \hatcur{58},
\hatcur{60}, and \hatcur{62}--\hatcur{64} we do not detect any periodic
signals, and place 95\% confidence upper limits on the peak-to-peak
amplitudes of such signals of 2.0\,mmag for \hatcur{58}, 0.96\,mmag for
\hatcur{60}, 1.2\,mmag for \hatcur{62}, 3.9\,mmag for \hatcur{63}, and
2.0\,mmag for \hatcur{64}. For \hatcur{59} a strong signal with a
period of $P = 0.49976 \pm 0.00086$\,days is detected with 
%
%
a peak-to-peak amplitude of
16.6\,mmag. Given the close proximity of the period to twice the
sidereal frequency, we suspect that this signal is most likely be due
to systematic errors in the photometry that are not fully corrected
through EPD and TFA. After subtracting a Fourier series model from the
light curve, GLS finds no additional signals present in the data, and
we place a 95\% confidence upper limit of 1.5\,mmag on the peak-to-peak
amplitude of any such signals. For \hatcur{61} we detect a possible
signal with a period of $10.6 \pm 0.5$\,days and with a formal false
alarm probability of 0.16\% and peak-to-peak amplitude of 2.6\,mmag.
The GLS periodogram is shown in Figure~\ref{fig:hat61gls}. This may
correspond to the photometric rotation period of the star, in which
case the equatorial rotation velocity of 4.7\,\kms\ is $2\sigma$ larger
than the spectroscopically measured projected rotation velocity of
$\vsini = \hatcurSMEiivsin{61}$\,\kms.

\ifthenelse{\boolean{emulateapj}}{
    \begin{figure*}[!p]
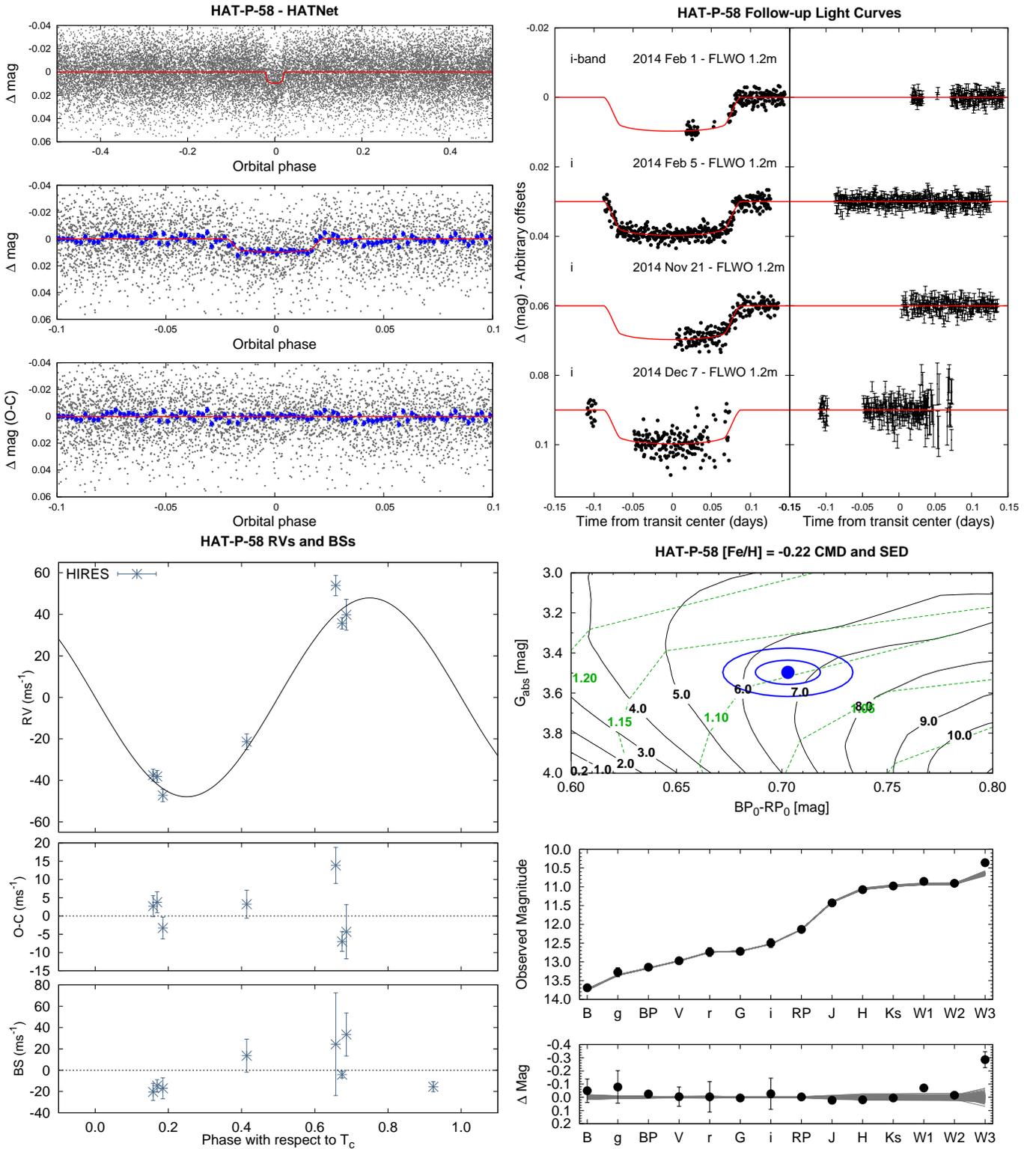

}{
    \begin{figure}[!p]
}
 {
 \centering
 \leavevmode
 \includegraphics[width={1.0\linewidth}]{\hatcurhtr{58}-banner}
}
 {
 \centering
 \leavevmode
 \includegraphics[width={0.5\linewidth}]{\hatcurhtr{58}-hatnet}%
 \hfil
 \includegraphics[width={0.5\linewidth}]{\hatcurhtr{58}-lc}%
 }
 {
 \centering
 \leavevmode
 \includegraphics[width={0.5\linewidth}]{\hatcurhtr{58}-rv}%
 \hfil
 \includegraphics[width={0.5\linewidth}]{\hatcurhtr{58}-iso-bprp-gabs-isofeh-SED}%
 }                        
\caption{
    Observations used to confirm the transiting planet system
    \hatcur{58}, excluding data from the NASA {\em TESS} mission which
    are shown in Figure~\ref{fig:hatp58tess}. {\em Top Left:}
    Phase-folded unbinned HATNet light curve. The top panel shows the
    full light curve, the middle panel shows the light curve zoomed-in
    on the transit, and the bottom panel shows the residuals from the
    best-fit model zoomed-in on the transit. The solid lines show the
    model fits to the light curves. The dark filled circles show the
    light curves binned in phase with a bin size of 0.002. (Caption
    continued on next page.)
\label{fig:hatp58}
}
\ifthenelse{\boolean{emulateapj}}{
    \end{figure*}
}{
    \end{figure}
}

\addtocounter{figure}{-1}
\ifthenelse{\boolean{emulateapj}}{
    \begin{figure*}[!p]
}{
    \begin{figure}[!p]
}
\caption{
    (Caption continued from previous page.)
{\em Top Right:} Unbinned follow-up transit light curves
    corrected for instrumental trends fitted
    simultaneously with the transit model, which is overplotted.
    The dates, filters and instruments used are indicated.  The
    residuals are shown on the right-hand-side in the same order as
    the original light curves.  The error bars represent the photon
    and background shot noise, plus the readout noise. Note that these
    uncertainties are scaled up in the fitting procedure to achieve a
    reduced $\chi^2$ of unity, but the uncertainties shown in the plot
    have not been scaled.  {\em Bottom Left:} High-precision RVs
    phased with respect to the mid-transit time. The instruments used
    are labelled in the plot.  The top panel shows the phased
    measurements together with the best-fit model.  The center-of-mass
    velocity has been subtracted. The second panel shows the velocity
    $O\!-\!C$ residuals.  The error bars include the estimated jitter.
    The third panel shows the bisector spans.  {\em Bottom Right:}
    Color-magnitude diagram (CMD) and spectral energy distribution
    (SED). The top panel shows the absolute $G$ magnitude vs.\ the
    de-reddened $BP - RP$ color compared to theoretical isochrones
    (black lines) and stellar evolution tracks (green lines) from the
    PARSEC models interpolated at the best-estimate value for the
    metallicity of the host. The age of each isochrone is listed in
    black in Gyr, while the mass of each evolution track is listed in
    green in solar masses. The filled blue circles show the measured
    reddening- and distance-corrected values from Gaia DR2, while the
    blue lines indicate the $1\sigma$ and $2\sigma$ confidence
    regions, including the estimated systematic errors in the
    photometry. Note that the determination of the final age of the
    system is informed by other input parameters, such as the
    spectroscopic effective temperature, the broad-band photometry in
    additional bandpasses and the stellar density from the light curves.
    The middle panel shows the SED as measured via
    broadband photometry through the listed filters. Here we plot the
    observed magnitudes without correcting for distance or
    extinction. Overplotted are 200 model SEDs randomly selected from
    the MCMC posterior distribution produced through the global
    analysis (gray lines).  The model makes use of the predicted
    absolute magnitudes in each bandpass from the PARSEC isochrones,
    the distance to the system (constrained largely via Gaia DR2) and
    extinction (constrained from the SED with a prior coming from the
    {\sc mwdust} 3D Galactic extinction model).  The bottom panel
    shows the $O\!-\!C$ residuals from the best-fit model SED.
\label{fig:hatp58:labcontinue}}
\ifthenelse{\boolean{emulateapj}}{
    \end{figure*}
}{
    \end{figure}
}


\ifthenelse{\boolean{emulateapj}}{
    \begin{figure*}[!ht]
}{
    \begin{figure}[!ht]
}
 {
 \centering
 \leavevmode
 \includegraphics[width={1.0\linewidth}]{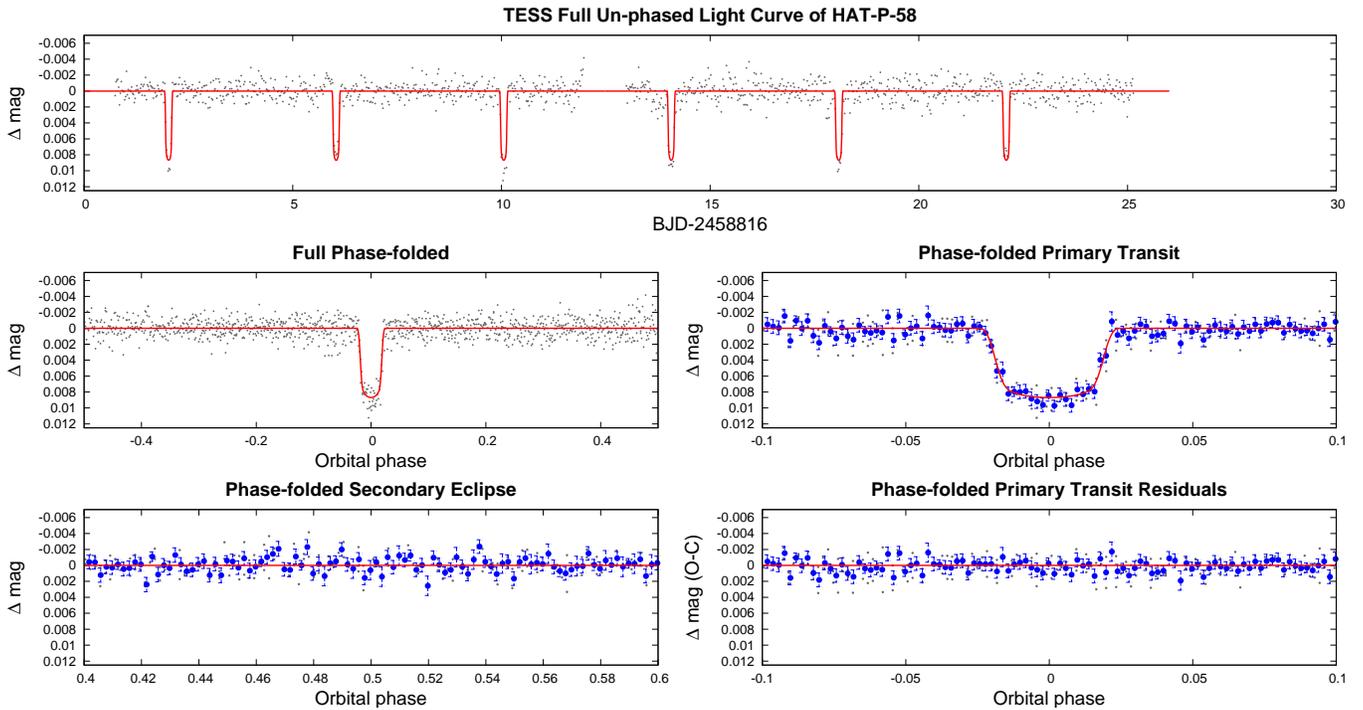}
}
\caption{
    {\em TESS} long-cadence light curve for \hatcur{58}. We show the
	full un-phased light curve as a function of time ({\em top}), the
	full phase-folded light curve ({\em middle left}), the phase-folded
	light curve zoomed-in on the planetary transit ({\em middle
	right}), the phase-folded light curve zoomed-in on the secondary
	eclipse ({\em bottom left}), and the residuals from the best-fit
	model, phase-folded and zoomed-in on the planetary transit ({\em
	bottom right}). The solid line in each panel shows the model fit to
	the light curve, account for the 30\,min integrations. The dark
	filled circles show the light curve binned in phase with a bin size
	of 0.002. Other observations included in our analysis of this
	system are shown in Figure~\ref{fig:hatp58}.
\label{fig:hatp58tess}
}
\ifthenelse{\boolean{emulateapj}}{
    \end{figure*}
}{
    \end{figure}
}


\begin{figure}[!ht]
\plotone{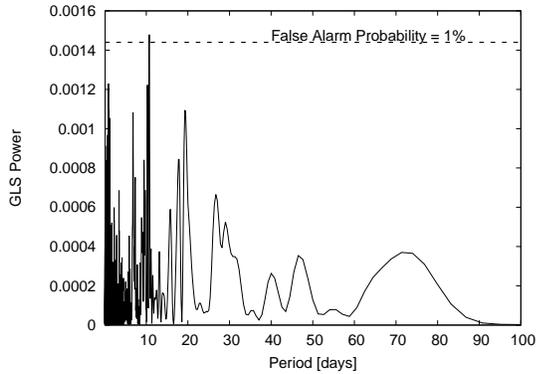}
\caption{
    Generalized Lomb-Scargle periodogram of the HATNet observations of
    \hatcur{61} showing the possible detection of a $P = 10.6$\,day
    periodic signal in the light curve of this star.
\label{fig:hat61gls}}
\end{figure}


\ifthenelse{\boolean{emulateapj}}{
    \begin{deluxetable*}{llrrrr}
}{
    \begin{deluxetable}{llrrrr}
}
\tablewidth{0pc}
\tabletypesize{\scriptsize}
\tablecaption{
    Summary of photometric observations \hatcur{58}--\hatcur{61}
    \label{tab:photobs}
}
\tablehead{
\noalign{\vskip 3pt}
    \multicolumn{1}{c}{Instrument/Field\tablenotemark{a}} &
    \multicolumn{1}{c}{Date(s)} &
    \multicolumn{1}{c}{\# Images\tablenotemark{b}} &
    \multicolumn{1}{c}{Cadence\tablenotemark{c}} &
    \multicolumn{1}{c}{Filter} &
    \multicolumn{1}{c}{Precision\tablenotemark{d}} \\
    \multicolumn{1}{c}{} &
    \multicolumn{1}{c}{} &
    \multicolumn{1}{c}{} &
    \multicolumn{1}{c}{(sec)} &
    \multicolumn{1}{c}{} &
    \multicolumn{1}{c}{(mmag)}
}
\startdata
\sidehead{\textbf{\hatcur{58}}}
~~~~HAT-5/G093 & 2012 Sep--2013 Apr & 9254 & 213 & $r$ & 21.3 \\
~~~~HAT-7/G093 & 2012 Sep & 238 & 213 & $r$ & 18.3 \\
~~~~HAT-8/G093 & 2012 Jul--2013 Apr & 11078 & 217 & $r$ & 14.8 \\
~~~~FLWO~1.2\,m/KeplerCam & 2014 Feb 01 & 157 & 48 & $i$ & 1.5 \\
~~~~FLWO~1.2\,m/KeplerCam & 2014 Feb 05 & 378 & 48 & $i$ & 1.5 \\
~~~~FLWO~1.2\,m/KeplerCam & 2014 Nov 21 & 207 & 51 & $i$ & 1.8 \\
~~~~FLWO~1.2\,m/KeplerCam & 2014 Dec 07 & 188 & 51 & $i$ & 3.1 \\
~~~~TESS/Sector~19 & 2019 Nov 29--2019 Dec 23 & 1117 & 1798 & $T$ & 1.1 \\
\sidehead{\textbf{\hatcur{59}}}
~~~~HAT-5/G081 & 2012 Oct--2012 Dec & 1963 & 213 & $r$ & 11.1 \\
~~~~HAT-6/G081 & 2012 Sep--2012 Dec & 2500 & 214 & $r$ & 9.1 \\
~~~~HAT-7/G081 & 2012 Jul--2012 Dec & 2340 & 213 & $r$ & 9.3 \\
~~~~HAT-8/G081 & 2012 Sep--2012 Dec & 2121 & 214 & $r$ & 9.1 \\
~~~~HAT-9/G081 & 2012 Sep--2012 Dec & 2158 & 213 & $r$ & 8.1 \\
~~~~FLWO~1.2\,m/KeplerCam & 2013 Nov 12 & 189 & 26 & $i$ & 2.9 \\
~~~~FLWO~1.2\,m/KeplerCam & 2014 Feb 19 & 177 & 26 & $i$ & 2.5 \\
~~~~FLWO~1.2\,m/KeplerCam & 2014 Mar 16 & 314 & 27 & $i$ & 2.1 \\
~~~~FLWO~1.2\,m/KeplerCam & 2014 May 13 & 642 & 26 & $i$ & 2.5 \\
~~~~TESS/Sector~14 & 2019 Jul 18--2019 Aug 14 & 1233 & 1799 & $T$ & 0.75 \\
~~~~TESS/Sector~15 & 2019 Aug 15--2019 Sep 8 & 821 & 1799 & $T$ & 0.72 \\
~~~~TESS/Sector~16 & 2019 Sep 12--2019 Oct 6 & 999 & 1799 & $T$ & 0.66 \\
~~~~TESS/Sector~17 & 2019 Oct 8--2019 Oct 31 & 938 & 1799 & $T$ & 0.64 \\
~~~~TESS/Sector~18 & 2019 Nov 4--2019 Nov 27 & 1036 & 1799 & $T$ & 0.63 \\
~~~~TESS/Sector~20 & 2019 Dec 25--2020 Jan 20 & 1175 & 1799 & $T$ & 0.66 \\
~~~~TESS/Sector~21 & 2020 Jan 23--2020 Feb 18 & 1189 & 1799 & $T$ & 0.72 \\
\sidehead{\textbf{\hatcur{60}}}
~~~~HAT-7/G089 & 2009 Sep--2010 Mar & 5577 & 225 & $r$ & 4.4 \\
~~~~FLWO~1.2\,m/KeplerCam & 2013 Oct 20 & 873 & 25 & $z$ & 3.6 \\
~~~~FLWO~1.2\,m/KeplerCam & 2014 Sep 11 & 840 & 22 & $i$ & 2.9 \\
~~~~FLWO~1.2\,m/KeplerCam & 2014 Oct 10 & 781 & 22 & $z$ & 2.9 \\
~~~~TESS/Sector~18 & 2019 Nov 4--2019 Nov 27 & 1031 & 1799 & $T$ & 0.38 \\
\sidehead{\textbf{\hatcur{61}}}
~~~~HAT-5/G094 & 2007 Oct--2008 Mar & 3526 & 384 & $R$ & 11.2 \\
~~~~HAT-5/G093 & 2012 Sep--2013 Apr & 9476 & 213 & $r$ & 18.6 \\
~~~~HAT-7/G093 & 2012 Sep & 240 & 213 & $r$ & 17.3 \\
~~~~HAT-8/G093 & 2012 Jul--2013 Apr & 11084 & 217 & $r$ & 15.6 \\
~~~~FLWO~1.2\,m/KeplerCam & 2014 Sep 21 & 165 & 58 & $i$ & 1.5 \\
~~~~FLWO~1.2\,m/KeplerCam & 2014 Oct 10 & 280 & 59 & $i$ & 2.4 \\
~~~~TESS/Sector~19 & 2019 Nov 28--2019 Dec 23 & 1145 & 1799 & $T$ & 1.1 \\
\enddata
\tablenotetext{a}{
    For HATNet data we list the HATNet unit and field name from which
    the observations are taken. HAT-5, -6, -7 and -10 are located at
    Fred Lawrence Whipple Observatory in Arizona. HAT-8 and -9 are
    located on the roof of the Smithsonian Astrophysical Observatory
    Submillimeter Array hangar building at Mauna Kea Observatory in
    Hawaii. Each field corresponds to one of 838 fixed pointings used
    to cover the full 4$\pi$ celestial sphere. All data from a given
    HATNet field are reduced together, while detrending through
    External Parameter Decorrelation (EPD) is done independently for
    each unique unit+field combination.
}
\tablenotetext{b}{
    Excluding outliers and other images that were not included when
	modelling the light curves.
}
\tablenotetext{c}{
    The median time between consecutive images rounded to the nearest
    second. Due to factors such as weather, the day--night cycle,
    guiding and focus corrections the cadence is only approximately
    uniform over short timescales.
}
\tablenotetext{d}{
    The RMS of the residuals from the best-fit model.
}
\ifthenelse{\boolean{emulateapj}}{
    \end{deluxetable*}
}{
    \end{deluxetable}
}

\ifthenelse{\boolean{emulateapj}}{
    \begin{deluxetable*}{llrrrr}
}{
    \begin{deluxetable}{llrrrr}
}
\tablewidth{0pc}
\tabletypesize{\scriptsize}
\tablecaption{
    Summary of photometric observations \hatcur{62}--\hatcur{64}
    \label{tab:photobs2}
}
\tablehead{
\noalign{\vskip 3pt}
    \multicolumn{1}{c}{Instrument/Field\tablenotemark{a}} &
    \multicolumn{1}{c}{Date(s)} &
    \multicolumn{1}{c}{\# Images\tablenotemark{b}} &
    \multicolumn{1}{c}{Cadence\tablenotemark{c}} &
    \multicolumn{1}{c}{Filter} &
    \multicolumn{1}{c}{Precision\tablenotemark{d}} \\
    \multicolumn{1}{c}{} &
    \multicolumn{1}{c}{} &
    \multicolumn{1}{c}{} &
    \multicolumn{1}{c}{(sec)} &
    \multicolumn{1}{c}{} &
    \multicolumn{1}{c}{(mmag)}
}
\startdata
\sidehead{\textbf{\hatcur{62}}}
~~~~HAT-5/G093 & 2012 Sep--2013 Apr & 9472 & 213 & $r$ & 15.1 \\
~~~~HAT-7/G093 & 2012 Sep & 240 & 213 & $r$ & 13.7 \\
~~~~HAT-8/G093 & 2012 Jul--2013 Apr & 11093 & 217 & $r$ & 12.9 \\
~~~~FLWO~1.2\,m/KeplerCam & 2014 Dec 01 & 192 & 41 & $z$ & 2.1 \\
~~~~FLWO~1.2\,m/KeplerCam & 2014 Dec 09 & 376 & 41 & $i$ & 1.7 \\
~~~~FLWO~1.2\,m/KeplerCam & 2015 Jan 10 & 136 & 40 & $i$ & 2.1 \\
~~~~FLWO~1.2\,m/KeplerCam & 2015 Mar 04 & 363 & 39 & $i$ & 3.5 \\
~~~~FLWO~1.2\,m/KeplerCam & 2015 Sep 26 & 335 & 41 & $i$ & 2.1 \\
\sidehead{\textbf{\hatcur{63}}}
~~~~HAT-5/G384 & 2009 May--2009 Jun & 389 & 416 & $r$ & 12.4 \\
~~~~HAT-9/G384 & 2009 May--2009 Sep & 2361 & 356 & $r$ & 9.6 \\
~~~~FLWO~1.2\,m/KeplerCam & 2013 Mar 13 & 111 & 86 & $i$ & 2.1 \\
~~~~FLWO~1.2\,m/KeplerCam & 2013 Mar 30 & 68 & 175 & $i$ & 1.5 \\
~~~~FLWO~1.2\,m/KeplerCam & 2013 Apr 16 & 157 & 86 & $i$ & 2.4 \\
\sidehead{\textbf{\hatcur{64}}}
~~~~HAT-6/G357 & 2009 Sep--2010 Mar & 3885 & 343 & $r$ & 14.1 \\
~~~~HAT-8/G357 & 2009 Sep--2010 Mar & 9097 & 224 & $r$ & 14.6 \\
~~~~FLWO~1.2\,m/KeplerCam & 2011 Feb 02 & 93 & 105 & $i$ & 1.5 \\
~~~~FLWO~1.2\,m/KeplerCam & 2011 Oct 12 & 182 & 73 & $i$ & 2.3 \\
~~~~TESS/Sector~5 & 2018 Nov 15--2018 Dec 11 & 1149 & 1799 & $T$ & 0.99 \\
\enddata
\tablenotetext{a}{
    For HATNet data we list the HATNet unit and field name from which
    the observations are taken. HAT-5, -6, -7 and -10 are located at
    Fred Lawrence Whipple Observatory in Arizona. HAT-8 and -9 are
    located on the roof of the Smithsonian Astrophysical Observatory
    Submillimeter Array hangar building at Mauna Kea Observatory in
    Hawaii. Each field corresponds to one of 838 fixed pointings used
    to cover the full 4$\pi$ celestial sphere. All data from a given
    HATNet field are reduced together, while detrending through
    External Parameter Decorrelation (EPD) is done independently for
    each unique unit+field combination.
}
\tablenotetext{b}{
    Excluding outliers and other images that were not included when
	modelling the light curves.
}
\tablenotetext{c}{
    The median time between consecutive images rounded to the nearest
    second. Due to factors such as weather, the day--night cycle,
    guiding and focus corrections the cadence is only approximately
    uniform over short timescales.
}
\tablenotetext{d}{
    The RMS of the residuals from the best-fit model.
}
\ifthenelse{\boolean{emulateapj}}{
    \end{deluxetable*}
}{
    \end{deluxetable}
}

\subsection{Spectroscopic Observations}
\label{sec:obsspec}

Spectroscopic observations of the TEP systems were carried out using
the Tillinghast Reflector Echelle Spectrograph
\citep[TRES;][]{furesz:2008} on the 1.5\,m Tillinghast Reflector at
FLWO, the SOPHIE spectrograph \citep{bouchy:2009} on the Observatoire
de Haute Provence (OHP)~1.93\,m in France, HIRES \citep{vogt:1994} on
the Keck-I~10\,m at MKO together with its I$_2$ absorption cell, the
High Dispersion Spectrograph \citep[HDS;][]{noguchi:2002} and its
I$_{2}$ cell \citep{kambe:2002} on the Subaru~8\,m at MKO, the
Astrophysical Research Consortium Echelle Spectrometer
\citep[ARCES;][]{wang:2003} on the ARC~3.5\,m telescope at Apache Point
Observatory (APO) in New Mexico, the FIbre-fed \'Echelle Spectrograph
(FIES) on the Nordic Optical Telescope (NOT)~2.5\,m
\citep{djupvik:2010} in La Palma, Spain, and the Network of Robotic
Echelle Spectrographs \citep[NRES;][]{siverd:2018:nres} on the
LCOGT~1\,m network. \reftabl{specobs} summarizes the spectroscopic
observations collected for each TEP system. Phased high-precision RV
and bisector (BS) measurements are shown for each system in
Figures~\ref{fig:hatp58}--\ref{fig:hatp64}. The data are listed in
\reftabl{rvs} at the end of the paper.

The TRES observations were reduced to spectra and cross-correlated
against synthetic stellar templates to measure the RVs and to estimate
\teffstar, \logg, and \vsini. Here we followed the procedure of
\cite{buchhave:2010:hat16}, initially making use of a single order
containing the gravity and temperature-sensitive Mg~b lines. Based on
these observations we quickly ruled out common false positive
scenarios, such as transiting M dwarf stars, or blends between giant
stars and pairs of eclipsing dwarf stars. For \hatcur{59} through
\hatcur{63} the initial TRES RVs exhibited low amplitude variations
consistent with planetary mass companions, so we continued to collect
spectroscopic observations with TRES for these objects with the aim of
confirming them as TEP systems, measuring the masses of the planets,
and providing high precision stellar atmospheric parameters, including
the stellar metallicities. For this work high precision RVs and
spectral line bisector spans (BSs) were determined based on a
multi-order analysis of the spectra
\citep[e.g.,][]{bieryla:2014:hat49}, while the atmospheric parameters
were determined using the Stellar Parameter Classification
\citep[SPC;][]{buchhave:2012:spc} method. For \hatcur{58} and
\hatcur{64} the TRES observations were used solely for reconnaissance
and were not included in the analysis described in
Section~\ref{sec:globmod}.

The SOPHIE observations of \hatcur{59}, \hatcur{60}, \hatcur{63} and
\hatcur{64} were reduced to RVs and BSs following
\citet{boisse:2013:hat42hat43}. In all cases the RVs show variations
consistent with planetary mass companions, and with the variations
seen using other spectrographs.

The HIRES observations of \hatcur{58}, \hatcur{60}, \hatcur{61}, and
\hatcur{64} were reduced to relative RVs following the method of
\citet{butler:1996}, and to BSs following \citet{torres:2007:hat3}. We
also measured Ca~II HK chromospheric emission indices (the so-called
$S$ and $\log_{10}R^{\prime}_{\rm HK}$ indices) following
\citet{isaacson:2010} and \citet{noyes:1984}. For \hatcur{64} we
measured stellar atmospheric parameters from the I$_{2}$-free template
spectra using SPC.

The HDS observations of \hatcur{63} were reduced to relative RVs
following \cite{sato:2002,sato:2012:hat38} and to BSs following
\citet{torres:2007:hat3}. The RVs are seen to vary in phase with the
photometric ephemeris of the TEP, and are consistent with the
variations seen with the TRES and SOPHIE spectrographs for this
system.

The ARCES spectrum of \hatcur{63} was reduced following
\citet{hartman:2015:hat50hat53} and \citet{buchhave:2012:spc} and was
used for reconnaissance. The RV and atmospheric parameters of
\hatcur{63} determined from this spectrum are consistent with the
results from TRES.

The FIES spectra of \hatcur{63} and \hatcur{64} were reduced following
\citet{buchhave:2010:hat16}. For \hatcur{63} the first spectrum was
obtained using the medium resolution fiber, while the other spectra
were obtained with the high resolution fiber. For \hatcur{64} all four
spectra were obtained with the high resolution fiber. While the
spectra were intended to be used for measuring the masses of the
planetary companions, the resulting RV precision was insufficient for
this purpose, given the small number of observations obtained. We
therefore do not include these measurements in our analyses of
\hatcur{63} or \hatcur{64}.

NRES spectra of \hatcur{60} were collected from the McDonalds
Observatory and Wise Observatory LCOGT~1\,m facilities. We
obtained 22 useful spectra with an SNR between 32 and 65, measured at
$\sim$5150 \AA. The exposure time for all spectra was 1800 sec. In
order to obtain the wavelength calibrated spectra, we adapted the CERES
pipeline \citep{brahm:2017:ceres}. We limited the order extraction to
the central 50 orders, covering the wavelength range from 4194\,\AA\ to
7445\,\AA.

\ifthenelse{\boolean{emulateapj}}{
    \begin{deluxetable*}{llrrrrr}
}{
    \begin{deluxetable}{llrrrrrrrr}
}
\tablewidth{0pc}
\tabletypesize{\scriptsize}
\tablecaption{
    Summary of spectroscopic observations
    \label{tab:specobs}
}
\tablehead{
\noalign{\vskip 3pt}
    \multicolumn{1}{c}{Instrument}          &
    \multicolumn{1}{c}{UT Date(s)}             &
    \multicolumn{1}{c}{\# Spec.}   &
    \multicolumn{1}{c}{Res.}          &
    \multicolumn{1}{c}{S/N Range\tablenotemark{a}}           &
    \multicolumn{1}{c}{$\gamma_{\rm RV}$\tablenotemark{b}} &
    \multicolumn{1}{c}{RV Precision\tablenotemark{c}} \\
    &
    &
    &
    \multicolumn{1}{c}{($\lambda$/$\Delta \lambda$)/1000} &
    &
    \multicolumn{1}{c}{(\kms)}              &
    \multicolumn{1}{c}{(\ms)}
}
\startdata
\sidehead{\textbf{\hatcur{58}}}
FLWO~1.5\,m/TRES & 2014 Jan 14--16 & 2 & 44 & 16--19 & $-35.96$ & 97 \\
Keck-I/HIRES+I$_{2}$ & 2014 Aug--Sep & 7 & 55 & 35--115 & $\cdots$ & 8.2 \\
Keck-I/HIRES & 2014 Aug 25 & 1 & 55 & 166 & $\cdots$ & $\cdots$ \\
\sidehead{\textbf{\hatcur{59}}}
FLWO~1.5\,m/TRES & 2013 Oct--Nov & 13 & 44 & 13--25 & $-20.35$ & 27 \\
OHP~1.93\,m/SOPHIE & 2013 Oct--Nov & 10 & 39 & $\cdots$ & $-21.16$ & 20 \\
\sidehead{\textbf{\hatcur{60}}}
FLWO~1.5\,m/TRES & 2013 Feb--Oct & 13 & 44 & 20--61 & $6.58$ & 17 \\
OHP~1.93\,m/SOPHIE & 2013 Oct--Nov & 8 & 39 & $\cdots$ & $6.03$ & 14 \\
Keck-I/HIRES+I$_{2}$ & 2013 Dec--2016 Jan & 8 & 55 & 140--196 & $\cdots$ & 12 \\
Keck-I/HIRES & 2015 Nov 29 & 1 & 55 & 306 & $\cdots$ & $\cdots$ \\
LCO~1m+ELP/NRES & 2019 Dec-2020 Jan & 12 & 53 & 32--65 & $5.92$ & 63 \\
LCO~1m+TLV/NRES & 2019 Dec-2020 Jan & 10 & 53 & 32--65 & $5.84$ & 57 \\
\sidehead{\textbf{\hatcur{61}}}
FLWO~1.5\,m/TRES & 2014 Sep--Nov & 18 & 44 & 12--22 & $4.81$ & 53 \\
Keck-I/HIRES+I$_{2}$ & 2015 Nov 27--29 & 3 & 55 & 63--95 & $\cdots$ & 9.3 \\
Keck-I/HIRES & 2015 Nov 29 & 1 & 55 & 119 & $\cdots$ & $\cdots$ \\
\sidehead{\textbf{\hatcur{62}}}
FLWO~1.5\,m/TRES & 2014 Jan--Nov & 15 & 44 & 15--25 & $50.42$ & 37 \\
\sidehead{\textbf{\hatcur{63}}}
FLWO~1.5\,m/TRES \tablenotemark{d} & 2012 Apr 6--28 & 3 & 44 & 13--15 & $-68.92$ & 33 \\
APO~3.5\,m/ARCES & 2012 Apr 30 & 1 & 31.5 & 18 & $-69.57$ & 500 \\
Subaru~8\,m/HDS & 2012 Sep 19 & 4 & 60 & 41--44 & $\cdots$ & $\cdots$ \\
Subaru~8\,m/HDS+I$_{2}$ & 2012 Sep 20--22 & 12 & 60 & 37--55 & $\cdots$ & 4.7 \\
NOT~2.5\,m/FIES & 2013 May 14 & 1 & 46 & 50 & $-69.11$ & 100 \\
NOT~2.5\,m/FIES & 2013 May 15--17 & 2 & 67 & 15--24 & $-69.045$ & 66 \\
OHP~1.93\,m/SOPHIE & 2013 Jun 3--13 & 7 & 39 & $\cdots$ & $-69.60$ & 23 \\
\sidehead{\textbf{\hatcur{64}}}
FLWO~1.5\,m/TRES & 2010 Oct--2011 Jan & 2 & 44 & 25--28 & $25.220$ & 58 \\
NOT~2.5\,m/FIES & 2011 Oct 9--25 & 4 & 67 & 44--54 & $25.142$ & 65 \\
Keck-I/HIRES & 2011 Jan--Sep & 2 & 55 & 96--138 & $\cdots$ & $\cdots$ \\
Keck-I/HIRES+I$_{2}$ & 2011 Jan--2012 Jan & 7 & 55 & 80--113 & $\cdots$ & 22 \\
OHP~1.93\,m/SOPHIE & 2011 Dec 5--12 & 6 & 39 & $\cdots$ & $24.49$ & 35 \\
\enddata 
\tablenotetext{a}{
    S/N per resolution element near 5180\,\AA. This was not reported for the OHP~1.93\,m/SOPHIE observations.
}
\tablenotetext{b}{
    For high-precision RV observations included in the orbit
    determination this is the zero-point RV from the best-fit
    orbit. For other instruments it is the mean value. We do not
    provide this quantity for the Keck-I/HIRES observations, from
    which we have only measured relative RVs.
}
\tablenotetext{c}{
    For high-precision RV observations included in the orbit
    determination this is the scatter in the RV residuals from the
    best-fit orbit (which may include astrophysical jitter), for other
    instruments this is either an estimate of the precision (not
    including jitter), or the measured standard deviation. We do not
    provide this quantity for the I$_{2}$-free templates obtained with
    Keck-I/HIRES or Subaru/HDS.
}
\tablenotetext{d}{
	One of the TRES spectra of \hatcur{63} was low S/N and did not
	permit high precision RVs, so only two of the TRES RVs of this
	object are included in the analysis.
}
\ifthenelse{\boolean{emulateapj}}{
    \end{deluxetable*}
}{
    \end{deluxetable}
}


\subsection{Ground-based photometric follow-up observations}
\label{sec:phot}

In order to determine the physical parameters of each TEP system, we
conducted follow-up photometric time-series observations of each object
using KeplerCam on the 1.2\,m telescope at FLWO. These observations are
summarized in Tables~\ref{tab:photobs} and~\ref{tab:photobs2}, where we
list the dates of the observed transit events, the number of images
collected for each event, the cadence of the observations, the filters
used, and the per-point photometric precision achieved. The images were
reduced to light curves following \citet{bakos:2010:hat11}, which are
plotted in Figures~\ref{fig:hatp58}--\ref{fig:hatp64}. The data are
provided in \reftabl{phfu}.

\subsection{TESS Space-Based Photometry}
\label{sec:tess}

Five of the seven planetary systems presented here were observed by
the NASA {\em TESS} mission \citep{ricker:2015}, as summarized in
Tables~\ref{tab:photobs} and~\ref{tab:photobs2}.  Of particular note
is \hatcur{59} which is located in the northern {\em TESS} continuous
viewing zone, and had data from Sectors 14, 15, 16, 17, 18, 20 and 21
that we included in the analysis. We were not able to extract useful
photometry for this system from the Sector 19 observations. The two
systems that did not have {\em TESS} observations were either too
close to the ecliptic plane (\hatcur{63}), or located only on the edge
of a CCD in Sector 19, with no useful data collected (\hatcur{62}).

We note that \hatcurb{59} and \hatcurb{60} have both been
independently identified as candidate transiting planets based on the
{\em TESS} observations. \hatcurb{59} (a.k.a.\ TOI-1826.01) is listed
as a community-identified candidate on ExoFOP-TESS, while \hatcurb{60}
(a.k.a.\ TOI-1580.01) is listed as a candidate identified by the MIT
quick-look pipeline. All of the systems presented here were detected
and confirmed as planets by the HATNet team prior to the launch of the
{\em TESS} mission.

The five systems with {\em TESS} observations were all observed in
long-cadence mode, and we extracted simple aperture photometry for them
from the {\em TESS} Full-Frame Image (FFI) data using the Lightkurve
tool \citep{lightkurve:2018}. Here we made use of the TESSCut API
\citep{brasseur:2019} to download $10 \times 10$ pixel FFI cutouts
around each source, and made use of the automated mask routine in
Lightkurve to generate the apertures using a threshold of 3.0, and to
generate the background regions using a threshold of 0.001. We then
used VARTOOLS \citep{hartman:2016:vartools} to apply a moving median
filter to remove large systematic variations from the light curves.
This was done by first manually removing regions from the light curves
with excessive systematic behavior, then masking the transits and
performing a median filter with a 0.5\,day window. We then performed a
monotonic spline interpolation over the masked regions of the light
curves to estimate the systematic corrections to apply to the
in-transit portions of the data. Note that the procedures above 
will likely erase the rotation induced and other long-term variation of the stars. 
The resulting light curves are shown, together with the best-fit models, in
Figures~\ref{fig:hatp58tess}--\ref{fig:hatp64tess}. These data are also
made available in \reftabl{phfu}.

As for the HATNet observations, we used the {\sc vartools} package to
search the residual {\em TESS} light curves of each target for
additional periodic transit signals using BLS, and for additional
sinusoidal periodic signals using GLS. Table~\ref{tab:tessbls} gives
the ephemeris information and significance for the top peak in the BLS
spectrum of the {\em TESS} residuals for each system. None of the
systems show strong evidence for additional periodic transit signals.
In a few cases (\hatcur{58} and \hatcur{59}) there is marginal evidence
for signals with signal-to-pink noise ratio S/N$> 7$ (see
\citealp{hartman:2016:vartools} for a definition of this measure as
used in {\sc vartools}), though these are likely false alarms, and
future observations by {\em TESS} in its extended mission should
confirm or refute these. None of the systems shows evidence for a
continuous periodic variation detected by GLS, though any such
variations would likely be removed by the median-filtering procedure
that we applied to the light curves.

\ifthenelse{\boolean{emulateapj}}{
    \begin{deluxetable*}{lrrrrrr}
}{
    \begin{deluxetable}{lrrrrrr}
}
\tablewidth{0pc}
\tabletypesize{\normalsize}
\tablecaption{
    BLS search for additional transits in the residual {\em TESS} light curves
    \label{tab:tessbls}
}
\tablehead{
    \multicolumn{1}{c}{System} &
    \multicolumn{1}{c}{Period} &
    \multicolumn{1}{c}{T$_{C}$} &
    \multicolumn{1}{c}{duration} &
    \multicolumn{1}{c}{depth} &
    \multicolumn{1}{c}{N$_{\rm transits}$} &
    \multicolumn{1}{c}{S/N\tablenotemark{a}} \\
    &
    \multicolumn{1}{c}{(d)} &
    \multicolumn{1}{c}{(BJD$_{TDB}-245000$)} &
    \multicolumn{1}{c}{(hr)} &
    \multicolumn{1}{c}{(mmag)} &
    &
}
\startdata
\hatcur{58} & $22.130$\tablenotemark{b} & $8829.949$ & $20.5$ & $1.4$ & $1$ & $7.75$ \\
\hatcur{59} & $19.956$ & $8702.753$ & $10.1$ & $0.48$ & $8$ & $7.66$ \\
\hatcur{60} & $6.7248$ & $8799.080$ & $6.9$ & $0.45$ & $4$ & $6.84$ \\
\hatcur{61} & $17.447$ & $8816.517$ & $8.3$ & $1.4$ & $2$ & $6.75$ \\
\hatcur{64} & $0.2151$ & $8438.105$ & $0.072$ & $1.4$ & $17$ & $5.80$ \\
\enddata
\tablenotetext{a}{
    The signal-to-pink-noise ratio as calculated by {\sc vartools}
	\citep{hartman:2016:vartools}.
}
\tablenotetext{b}{
    In this case only a single transit event is identified by BLS, and
	the period is not meaningful.
}
\ifthenelse{\boolean{emulateapj}}{
    \end{deluxetable*}
}{
    \end{deluxetable}
}

\subsection{Speckle imaging observations}
\label{sec:luckyimaging}

\begin{figure*}[!ht]
{
\plottwo{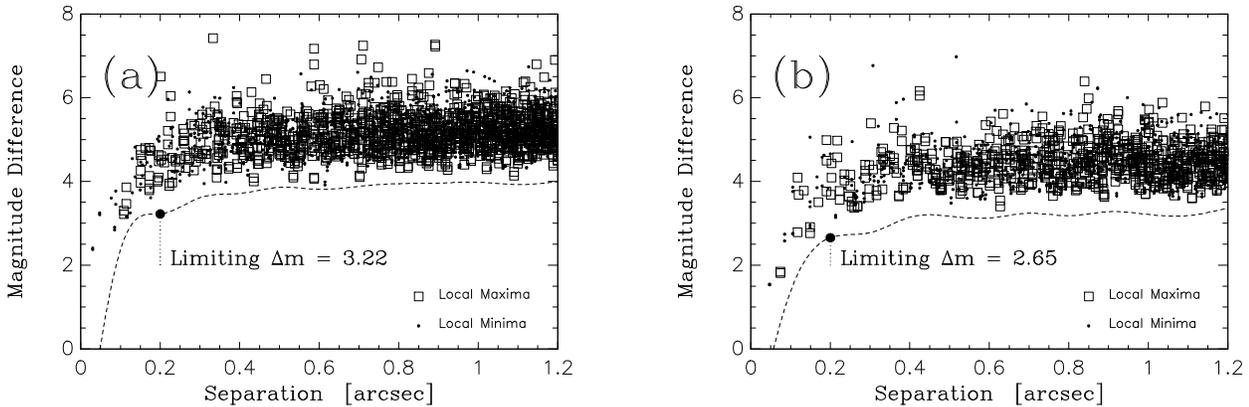}{\hatcurhtr{58}_f1b}
}
\caption{
	Limits on the relative magnitude of a resolved companion to
	\hatcur{58} as a function of angular separation based on speckle
	imaging observations from WIYN~3.5\,m/DSSI. The left panel shows
	the limits for the 692\,nm filter, the right shows limits for the
	880\,nm filter.
}
\label{fig:luckyimage58}
\end{figure*}


In order to detect nearby stellar companions which may be diluting the
transit signals, we obtained high spatial resolution speckle imaging
observations of all seven systems. For \hatcur{58}--\hatcur{62} and
\hatcur{64} we used the Differential Speckle Survey Instrument
\citep[DSSI;][]{horch:2009,howell:2011,horch:2011,horch:2012}, while for
\hatcur{63} we used the newer NN-explore Exoplanet Stellar Speckle
Imager \citep[NESSI;][]{scott:2018}. Both instruments were used with
the WIYN~3.5\,m telescope\footnote{The WIYN Observatory is a joint
facility of the University of Wisconsin-Madison, Indiana University,
the National Optical Astronomy Observatory and the University of
Missouri.} at Kitt Peak National Observatory in Arizona.

The DSSI observations were gathered between the nights of UT 26
September 2015 and UT 3 October 2015. A dichroic beamsplitter is used
to obtain simultaneous imaging through 692\,nm and 880\,nm filters.
Each observation consists of a sequence of 1000 40\,ms exposures
read-out on $128 \times 128$ pixel ($2\farcs8 \times 2\farcs8$)
subframes, which are reduced to reconstructed images following
\citet{horch:2011}. These images are searched for companions, and when
none are detected, $5\sigma$ lower limits on the differential magnitude
between a putative companion and the primary star are determined as a
function of angular separation as described in \citet{horch:2011}.

The NESSI observation was gathered on the night of UT 7 September 2017,
in this case using a dichroic beamsplitter to image at 562\,nm and
832\,nm. The observing mode and reduction method are similar to those
used for DSSI, and have been detailed in \citet{scott:2018}. In this
case the $256 \times 256$ pixel subframe has a field of view of
$4\farcs6 \times 4\farcs6$.

\begin{figure*}[!ht]
{
\plottwo{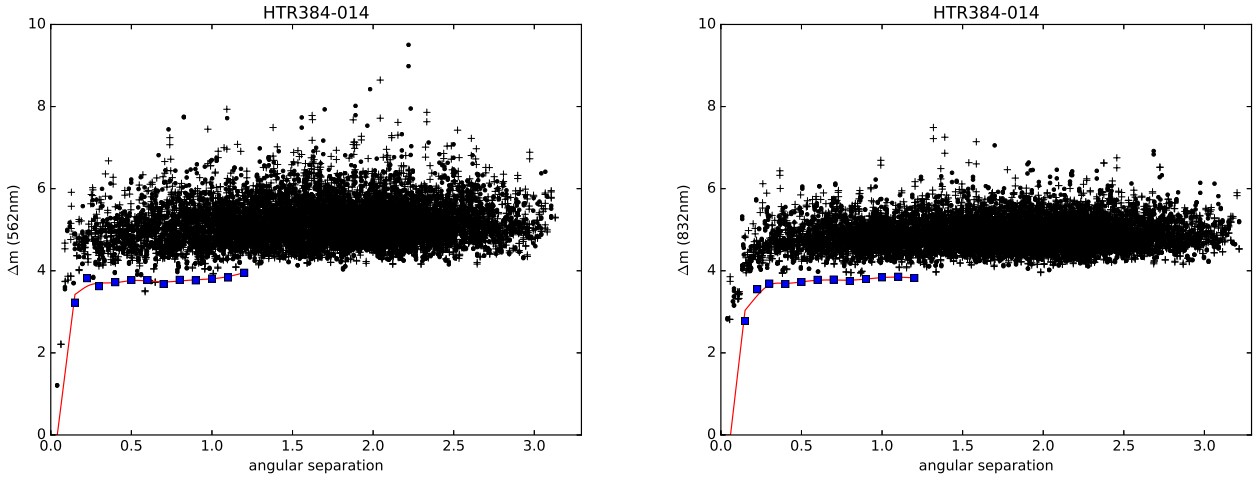}{\hatcurhtr{63}_20170907r}
}
\caption{
	Similar to Figure~\ref{fig:luckyimage58}, here we show the results
	for \hatcur{63} obtained with the NESSI instrument on the WIYN~3.5\,m.
	For this instrument the filters used have wavelengths of 562\,nm (left)
	and 832\,nm (right).
}
\label{fig:luckyimage63}
\end{figure*}

For \hatcur{60} we obtained a single observation, while for the other
six objects we obtained five observations apiece. In all cases no
companions are detected within $1\farcs2$, and we place limits on the
differential magnitudes in the blue and red filters as shown in
Figures~\ref{fig:luckyimage58}--\ref{fig:luckyimage64}. We find
limiting magnitude differences at $\sim 0\farcs2$ of
\begin{itemize}
\item {\bf \hatcur{58}} - $\Delta m_{692} > 3.22$ and $\Delta m_{880} > 2.65$
\item {\bf \hatcur{59}} - $\Delta m_{692} > 3.14$ and $\Delta m_{880} > 2.74$
\item {\bf \hatcur{60}} - $\Delta m_{692} > 4.04$ and $\Delta m_{880} > 3.41$
\item {\bf \hatcur{61}} - $\Delta m_{692} > 2.85$ and $\Delta m_{880} > 2.62$
\item {\bf \hatcur{62}} - $\Delta m_{692} > 3.16$ and $\Delta m_{880} > 2.81$
\item {\bf \hatcur{63}} - $\Delta m_{562} > 3.82$ and $\Delta m_{832} > 3.55$
\item {\bf \hatcur{64}} - $\Delta m_{692} > 2.60$ and $\Delta m_{880} > 2.80$
\end{itemize}

In addition to the companion limits based on the WIYN~3.5\,m/DSSI
observations, we also queried the UCAC~4 catalog
\citep{zacharias:2013:ucac4} for neighbors within 20\arcsec\, and the
Gaia~DR2 catalog \citep{gaiadr2} for neighbors within 10\arcsec\ that
may dilute either the HATNet or KeplerCam photometry. We find that
\hatcur{60}, \hatcur{62}, and \hatcur{64} have fainter neighbors in
Gaia~DR2, while only the neighbor for \hatcur{62} is also detected in
UCAC~4. The neighbors have separations and G-band magnitude differences
as follows:
\begin{itemize}
\item {\bf \hatcur{60}} - $9\farcs088$ southeast, $\Delta G = 10.79$\,mag
\item {\bf \hatcur{62}} - $5\farcs565$ northwest, $\Delta G = 2.10$\,mag, $\Delta V = 2.18$\,mag
\item {\bf \hatcur{64}} - $2\farcs510$ northwest, $\Delta G = 6.38$\,mag
\end{itemize}
Based on the Gaia~DR2 parallaxes the neighbors to \hatcur{60} and
\hatcur{62} are background objects that are not physically associated
with the planet hosts. No parallax, proper motion, or color information
is available for the neighbor to \hatcur{64}. This neighbor is at a
projected separation of 1667\,AU from the planet host, if it is
physically associated. The neighbors to \hatcur{60} and \hatcur{64} are
too faint to significantly affect the photometry and the resulting
planet and stellar parameters, and can be ruled out as the source of
the detected transit signals. We do account for the neighbor to
\hatcur{62} ($\Delta G = 2.10$\,mag) in the analysis of this system as described in
Section~\ref{sec:globmod}.



%
%
\ifthenelse{\boolean{emulateapj}}{
    \begin{deluxetable*}{llrrrrl}
}{
    \begin{deluxetable}{llrrrrl}
}
\tablewidth{0pc}
\tablecaption{
    Light curve data for \hatcur{58}--\hatcur{63}\label{tab:phfu}.
}
\tablehead{
    \colhead{Object\tablenotemark{a}} &
    \colhead{BJD$_{TDB}$\tablenotemark{b}} & 
    \colhead{Mag\tablenotemark{c}} & 
    \colhead{\ensuremath{\sigma_{\rm Mag}}} &
    \colhead{Mag(orig)\tablenotemark{d}} & 
    \colhead{Filter} &
    \colhead{Instrument} \\
    \colhead{} &
    \colhead{\hbox{~~~~(2,400,000$+$)~~~~}} & 
    \colhead{} & 
    \colhead{} &
    \colhead{} & 
    \colhead{} &
    \colhead{}
}
\startdata
\input{phfu_tab_combined_short.tex}
\enddata
\tablenotetext{a}{
    Either \hatcur{58}, \hatcur{59}, \hatcur{60}, \hatcur{61},
	\hatcur{62}, \hatcur{63} or \hatcur{64}.
}
\tablenotetext{b}{
    Barycentric Julian Date on the dynamical time system, including the
	correction for leap seconds.
}
\tablenotetext{c}{
    The out-of-transit level has been subtracted. For observations
    made with the HATNet instruments (identifed by ``HN'' in the
    ``Instrument'' column) these magnitudes have been corrected for
    trends using the EPD and TFA procedures applied either {\em prior}
    to fitting the transit model, or in conjunction with fitting a
    box-shaped transit. This procedure, together with blending for
    nearby stars, may lead to an artificial dilution in the transit
    depths. The blend factors for the HATNet light curves are listed
    in Tables~\ref{tab:planetparam}~and~\ref{tab:planetparamtwo}. For
    observations made with follow-up instruments (anything other than
    ``HN'' in the ``Instrument'' column), the magnitudes have been
    corrected for a quadratic trend in time, for variations correlated
    with three PSF shape parameters, and for trends correlated with
    variations seen in the light curves of other stars in the field
    (the TFA method) fit simultaneously with the transit.
}
\tablenotetext{d}{
    Raw magnitude values without correction for the quadratic trend in
    time, for trends correlated with the shape of the PSF, or
    application of TFA. These are only reported for the follow-up
    observations.
}
\tablecomments{
    This table is available in a machine-readable form in the online
    journal.  A portion is shown here for guidance regarding its form
    and content.
}
\ifthenelse{\boolean{emulateapj}}{
    \end{deluxetable*}
}{
    \end{deluxetable}
}

\section{Analysis}
\label{sec:analysis}

We analyzed the photometric and spectroscopic observations of
\hatcur{58}--\hatcur{64} to determine the parameters of each system.
The analysis followed the methods discussed in detail most recently by
\citet{hartman:2019:hats60hats69}. Here we give a brief summary of the
procedure.

\subsection{Properties of the parent star}
\label{sec:stelparam}

High-precision atmospheric parameters, including the effective surface
temperature \teffstar, the surface gravity \logg, the metallicity
\feh, and the projected rotational velocity \vsini, were determined by
applying SPC to our high resolution spectra. For \hatcur{58} through
\hatcur{63} this analysis was performed on the TRES spectra, while for
\hatcur{64} we made use of the Keck-I/HIRES I$_{2}$-free template
spectra. The analysis is performed seperately on each spectrum and we
take the weighted average of the results over all spectra obtained for
each target. Here we assume minimum uncertainties of 50\,K on
\teffstar, 0.10\,dex on \logg, 0.08\,dex on \feh, and 0.5\,\kms\ on
\vsini, which reflects the systematic uncertainty in the method, and
is based on applying the SPC analysis to observations of spectroscopic
standard stars. Following \citet{torres:2012}, we then revised the
atmospheric parameters of the stars in an iterative fashion. We
carried out a joint analysis of the light curves and RV curves to
determine the mean stellar density \rhostar\ for each host. We then
combined the \teffstar\ and \feh\ from the spectra with \rhostar\ to
determine the surface gravities via interpolation within the
Yonsei-Yale theoretical stellar isochrones \citep[Y2;][]{yi:2001}. The
surface gravities were then fixed to the values from this procedure in
a second iteration of SPC where only \teffstar, \feh\ and \vsini\ were
allowed to vary. Note that this procedure for determining the fixed
value of \loggstar\ was performed prior to the release of Gaia~DR2,
and we choose not to perform an additional iteration of SPC making
use of the Gaia~DR2 parallax. The expected change in the atmospheric
parameters are in all cases smaller than the systematic uncertainties.

The final spectroscopic parameters, together with catalog astrometry and
photometry are listed for the host stars in
Tables~\ref{tab:stellarobserved}~and~\ref{tab:stellarobservedtwo}.

The final atmospheric parameters are then treated as observations which
are simultaneously fitted, together with the light curves, RV curves,
parallaxes, and catalog broad-band photometry as described in
Section~\ref{sec:globmod}. Here the fitting procedure makes use of the
PARSEC stellar evolution models \citep{marigo:2017} to constrain the
physical properties of the stars. The final derived physical parameters
of the stars, based on this method, including \mstar, \rstar,
\loggstar, \rhostar, \lstar, \teffstar, \feh, the age of the system,
the $V$-band extinction A$_{V}$, and the distance to the system are
listed in
Tables~\ref{tab:stellarderived}~and~\ref{tab:stellarderivedtwo}. Note
that the values of \teffstar\ and \feh\ listed here are the optimized
values that are varied in the joint analysis, and may differ from the
values for these parameter determined from modeling the spectra listed
in Tables~\ref{tab:stellarobserved}~and~\ref{tab:stellarobservedtwo}.
Figures~\ref{fig:hatp58}--\ref{fig:hatp64} show the de-reddened Gaia
DR2 $BP-RP$ colors vs.\ absolute $G$ magnitudes for each star compared
to the PARSEC stellar evolution models, and also show the broad-band
spectral energy distribution of each star compared to the PARSEC
models.

%
%
\ifthenelse{\boolean{emulateapj}}{
    \begin{deluxetable*}{lccccl}
}{
    \begin{deluxetable}{lccccl}
}
\tablewidth{0pc}
\tabletypesize{\footnotesize}
\tablecaption{
    Astrometric, Spectroscopic and Photometric parameters for
    \hatcur{58}, \hatcur{59}, \hatcur{60} and \hatcur{61}
    \label{tab:stellarobserved}
}
\tablehead{
\noalign{\vskip 3pt}
    \multicolumn{1}{c}{} &
    \multicolumn{1}{c}{\bf HAT-P-58} &
    \multicolumn{1}{c}{\bf HAT-P-59} &
    \multicolumn{1}{c}{\bf HAT-P-60} &
    \multicolumn{1}{c}{\bf HAT-P-61} &
    \multicolumn{1}{c}{} \\
    \multicolumn{1}{c}{~~~~~~~~Parameter~~~~~~~~} &
    \multicolumn{1}{c}{Value}                     &
    \multicolumn{1}{c}{Value}                     &
    \multicolumn{1}{c}{Value}                     &
    \multicolumn{1}{c}{Value}                     &
    \multicolumn{1}{c}{Source}
}
\startdata
\noalign{\vskip -3pt}
\sidehead{Astrometric properties and cross-identifications}
~~~~TIC-ID\dotfill                 & 9443323 & 229400092 & 354469661 & 259506033 \\
~~~~TOI-ID\dotfill                 & $\cdots$ & 1826.01 & 1580.01 & $\cdots$ \\
~~~~2MASS-ID\dotfill               & \hatcurtwomassshort{58}  & \hatcurtwomassshort{59} & \hatcurtwomassshort{60} & \hatcurtwomassshort{61} & \\
~~~~GSC-ID\dotfill                 & \hatcurCCgsc{58}      & \hatcurCCgsc{59}     & \hatcurCCgsc{60}     & \hatcurCCgsc{61}     & \\
~~~~GAIA~DR2-ID\dotfill                 & \hatcurCCgaiadrtwo{58}      & \hatcurCCgaiadrtwo{59}     & \hatcurCCgaiadrtwo{60}     & \hatcurCCgaiadrtwo{61}     & \\
~~~~R.A. (J2000)\dotfill            & \hatcurCCra{58}       & \hatcurCCra{59}    & \hatcurCCra{60}    & \hatcurCCra{61}    & GAIA DR2\\
~~~~Dec. (J2000)\dotfill            & \hatcurCCdec{58}      & \hatcurCCdec{59}   & \hatcurCCdec{60}   & \hatcurCCdec{61}   & GAIA DR2\\
~~~~$\mu_{\rm R.A.}$ (\masy)              & \hatcurCCpmra{58}     & \hatcurCCpmra{59} & \hatcurCCpmra{60} & \hatcurCCpmra{61} & GAIA DR2\\
~~~~$\mu_{\rm Dec.}$ (\masy)              & \hatcurCCpmdec{58}    & \hatcurCCpmdec{59} & \hatcurCCpmdec{60} & \hatcurCCpmdec{61} & GAIA DR2\\
~~~~parallax (mas)              & \hatcurCCparallax{58}    & \hatcurCCparallax{59} & \hatcurCCparallax{60} & \hatcurCCparallax{61} & GAIA DR2\\
\sidehead{Spectroscopic properties}
~~~~$\teffstar$ (K)\dotfill         &  \hatcurSMEteff{58}   & \hatcurSMEteff{59} & \hatcurSMEteff{60} & \hatcurSMEteff{61} & SPC\tablenotemark{a}\\
~~~~$\feh$\dotfill                  &  \hatcurSMEzfeh{58}   & \hatcurSMEzfeh{59} & \hatcurSMEzfeh{60} & \hatcurSMEzfeh{61} & SPC               \\
~~~~$\vsini$ (\kms)\dotfill         &  \hatcurSMEvsin{58}   & \hatcurSMEvsin{59} & \hatcurSMEvsin{60} & \hatcurSMEvsin{61} & SPC                \\
~~~~$\vmac$ (\kms)\dotfill          &  $1.0$   & $1.0$ & $1.0$ & $1.0$ & Assumed              \\
~~~~$\vmic$ (\kms)\dotfill          &  $2.0$   & $2.0$ & $2.0$ & $2.0$ & Assumed              \\
~~~~$\gamma_{\rm RV}$ (\ms)\dotfill&  \hatcurRVgammaabs{58}  & \hatcurRVgammaabs{59} & \hatcurRVgammaabs{60} & \hatcurRVgammaabs{61} & TRES\tablenotemark{b}  \\
~~~~$S_{\rm HK}$\dotfill           & \hatcurSindex{58} & $\cdots$ & \hatcurSindex{60} & \hatcurSindex{61} & HIRES \\
~~~~$\log R^{\prime}_{\rm HK}$\dotfill           & \hatcurRHKindex{58} & $\cdots$ & \hatcurRHKindex{60} & \hatcurRHKindex{61} & HIRES \\
\sidehead{Photometric properties}
~~~~$G$ (mag)\tablenotemark{c}\dotfill               &  \hatcurCCgaiamG{58}  & \hatcurCCgaiamG{59} & \hatcurCCgaiamG{60} & \hatcurCCgaiamG{61} & GAIA DR2 \\
~~~~$BP$ (mag)\tablenotemark{c}\dotfill               &  \hatcurCCgaiamBP{58}  & \hatcurCCgaiamBP{59} & \hatcurCCgaiamBP{60} & \hatcurCCgaiamBP{61} & GAIA DR2 \\
~~~~$RP$ (mag)\tablenotemark{c}\dotfill               &  \hatcurCCgaiamRP{58}  & \hatcurCCgaiamRP{59} & \hatcurCCgaiamRP{60} & \hatcurCCgaiamRP{61} & GAIA DR2 \\
~~~~$B$ (mag)\dotfill               &  \hatcurCCtassmB{58}  & \hatcurCCtassmB{59} & \hatcurCCtassmB{60} & \hatcurCCtassmB{61} & APASS\tablenotemark{d} \\
~~~~$V$ (mag)\dotfill               &  \hatcurCCtassmv{58}  & \hatcurCCtassmv{59} & \hatcurCCtassmv{60} & \hatcurCCtassmv{61} & APASS\tablenotemark{d} \\
~~~~$I$ (mag)\dotfill               &  $\cdots$  & \hatcurCCtassmI{59} & \hatcurCCtassmI{60} & \hatcurCCtassmI{61} & TASS Mark IV\tablenotemark{e} \\
~~~~$g$ (mag)\dotfill               &  \hatcurCCtassmg{58}  & \hatcurCCtassmg{59} & $\cdots$ & \hatcurCCtassmg{61} & APASS\tablenotemark{d} \\
~~~~$r$ (mag)\dotfill               &  \hatcurCCtassmr{58}  & \hatcurCCtassmr{59} & $\cdots$ & \hatcurCCtassmr{61} & APASS\tablenotemark{d} \\
~~~~$i$ (mag)\dotfill               &  \hatcurCCtassmi{58}  & \hatcurCCtassmi{59} & \hatcurCCtassmi{60} & \hatcurCCtassmi{61} & APASS\tablenotemark{d} \\
~~~~$J$ (mag)\dotfill               &  \hatcurCCtwomassJmag{58} & \hatcurCCtwomassJmag{59} & \hatcurCCtwomassJmag{60} & \hatcurCCtwomassJmag{61} & 2MASS           \\
~~~~$H$ (mag)\dotfill               &  \hatcurCCtwomassHmag{58} & \hatcurCCtwomassHmag{59} & \hatcurCCtwomassHmag{60} & \hatcurCCtwomassHmag{61} & 2MASS           \\
~~~~$K_s$ (mag)\dotfill             &  \hatcurCCtwomassKmag{58} & \hatcurCCtwomassKmag{59} & \hatcurCCtwomassKmag{60} & \hatcurCCtwomassKmag{61} & 2MASS           \\
\enddata
\tablenotetext{a}{
    SPC = Stellar Parameter Classification procedure for the analysis
    of high-resolution spectra \citep{buchhave:2012:spc}, applied to
    the TRES spectra of \hatcur{58}--\hatcur{61}. These parameters
    rely primarily on SPC, but have a small dependence also on the
    iterative analysis incorporating the isochrone search and global
    modeling of the data.
}
\tablenotetext{b}{
    In addition to the uncertainty listed here, there is a $\sim
    0.1$\,\kms\ systematic uncertainty in transforming the velocities
    to the IAU standard system.
} 
\tablenotetext{c}{
    The listed uncertainties for the Gaia DR2 photometry are taken from
	the catalog. For the analysis we assume additional systematic
	uncertainties of 0.002\,mag, 0.005\,mag and 0.003\,mag for the G,
	BP and RP bands, respectively.
}
\tablenotetext{d}{
    From APASS DR6 for as
    listed in the UCAC 4 catalog \citep{zacharias:2013:ucac4}.  
}
\tablenotetext{e}{
    From the Amateur Sky Survey (TASS) catalog release IV \citep{droege:2006}.
}
\ifthenelse{\boolean{emulateapj}}{
    \end{deluxetable*}
}{
    \end{deluxetable}
}

%
%
\ifthenelse{\boolean{emulateapj}}{
    \begin{deluxetable*}{lcccl}
}{
    \begin{deluxetable}{lcccl}
}
\tablewidth{0pc}
\tabletypesize{\footnotesize}
\tablecaption{
    Astrometric, Spectroscopic and Photometric parameters for
    \hatcur{62}, \hatcur{63} and \hatcur{64}
    \label{tab:stellarobservedtwo}
}
\tablehead{
\noalign{\vskip 3pt}
    \multicolumn{1}{c}{} &
    \multicolumn{1}{c}{\bf HAT-P-62} &
    \multicolumn{1}{c}{\bf HAT-P-63} &
    \multicolumn{1}{c}{\bf HAT-P-64} &
    \multicolumn{1}{c}{} \\
    \multicolumn{1}{c}{~~~~~~~~Parameter~~~~~~~~} &
    \multicolumn{1}{c}{Value}                     &
    \multicolumn{1}{c}{Value}                     &
    \multicolumn{1}{c}{Value}                     &
    \multicolumn{1}{c}{Source}
}
\startdata
\noalign{\vskip -3pt}
\sidehead{Astrometric properties and cross-identifications}
~~~~TIC-ID\dotfill                 & 453064665 & 1635721458 & 455036659 \\
~~~~TOI-ID\dotfill                 & $\cdots$ & $\cdots$ & $\cdots$ \\
~~~~2MASS-ID\dotfill               & \hatcurtwomassshort{62}  & \hatcurtwomassshort{63} & \hatcurtwomassshort{64} & \\
~~~~GSC-ID\dotfill                 & \hatcurCCgsc{62}      & \hatcurCCgsc{63}     & \hatcurCCgsc{64}     & \\
~~~~GAIA~DR2-ID\dotfill                 & \hatcurCCgaiadrtwo{62}      & \hatcurCCgaiadrtwo{63}     & \hatcurCCgaiadrtwo{64}     & \\
~~~~R.A. (J2000)\dotfill            & \hatcurCCra{62}       & \hatcurCCra{63}    & \hatcurCCra{64}    & GAIA DR2\\
~~~~Dec. (J2000)\dotfill            & \hatcurCCdec{62}      & \hatcurCCdec{63}   & \hatcurCCdec{64}   & GAIA DR2\\
~~~~$\mu_{\rm R.A.}$ (\masy)              & \hatcurCCpmra{62}     & \hatcurCCpmra{63} & \hatcurCCpmra{64} & GAIA DR2\\
~~~~$\mu_{\rm Dec.}$ (\masy)              & \hatcurCCpmdec{62}    & \hatcurCCpmdec{63} & \hatcurCCpmdec{64} & GAIA DR2\\
~~~~parallax (mas)              & \hatcurCCparallax{62}    & \hatcurCCparallax{63} & \hatcurCCparallax{64} & GAIA DR2\\
\sidehead{Spectroscopic properties}
~~~~$\teffstar$ (K)\dotfill         &  \hatcurSMEteff{62}   & \hatcurSMEteff{63} & \hatcurSMEteff{64} & SPC\tablenotemark{a}\\
~~~~$\feh$\dotfill                  &  \hatcurSMEzfeh{62}   & \hatcurSMEzfeh{63} & \hatcurSMEzfeh{64} & SPC               \\
~~~~$\vsini$ (\kms)\dotfill         &  \hatcurSMEvsin{62}   & \hatcurSMEvsin{63} & \hatcurSMEvsin{64} & SPC                \\
~~~~$\vmac$ (\kms)\dotfill          &  $1.0$   & $1.0$ & $1.0$ & Assumed              \\
~~~~$\vmic$ (\kms)\dotfill          &  $2.0$   & $2.0$ & $2.0$ & Assumed              \\
~~~~$\gamma_{\rm RV}$ (\ms)\dotfill&  \hatcurRVgammaabs{62}  & \hatcurRVgammaabs{63} & \hatcurRVgammaabs{64} & FEROS or HARPS\tablenotemark{b}  \\
~~~~$S_{\rm HK}$\dotfill           & $\cdots$ & $\cdots$ & \hatcurSindex{64} & HIRES \\
~~~~$\log R^{\prime}_{\rm HK}$\dotfill           & $\cdots$ & $\cdots$ & \hatcurRHKindex{64} & HIRES \\
\sidehead{Photometric properties}
~~~~$G$ (mag)\tablenotemark{c}\dotfill               &  \hatcurCCgaiamG{62}  & \hatcurCCgaiamG{63} & \hatcurCCgaiamG{64} & Gaia DR2 \\
~~~~$BP$ (mag)\tablenotemark{c}\dotfill               &  \hatcurCCgaiamBP{62}  & \hatcurCCgaiamBP{63} & \hatcurCCgaiamBP{64} & Gaia DR2 \\
~~~~$RP$ (mag)\tablenotemark{c}\dotfill               &  \hatcurCCgaiamRP{62}  & \hatcurCCgaiamRP{63} & \hatcurCCgaiamRP{64} & Gaia DR2 \\
~~~~$B$ (mag)\dotfill               &  $\cdots$  & \hatcurCCtassmB{63} & \hatcurCCtassmB{64} & APASS\tablenotemark{d} \\
~~~~$V$ (mag)\dotfill               &  $\cdots$  & \hatcurCCtassmv{63} & \hatcurCCtassmv{64} & APASS\tablenotemark{d} \\
~~~~$I$ (mag)\dotfill               &  $\cdots$  & $\cdots$ & \hatcurCCtassmI{64} & TASS Mark IV\tablenotemark{e} \\
~~~~$g$ (mag)\dotfill               &  $\cdots$  & \hatcurCCtassmg{63} & \hatcurCCtassmg{64} & APASS\tablenotemark{d} \\
~~~~$r$ (mag)\dotfill               &  $\cdots$  & \hatcurCCtassmr{63} & \hatcurCCtassmr{64} & APASS\tablenotemark{d} \\
~~~~$i$ (mag)\dotfill               &  $\cdots$  & \hatcurCCtassmi{63} & \hatcurCCtassmi{64} & APASS\tablenotemark{d} \\
~~~~$J$ (mag)\dotfill               &  \hatcurCCtwomassJmag{62} & \hatcurCCtwomassJmag{63} & \hatcurCCtwomassJmag{64} & 2MASS           \\
~~~~$H$ (mag)\dotfill               &  \hatcurCCtwomassHmag{62} & \hatcurCCtwomassHmag{63} & \hatcurCCtwomassHmag{64} & 2MASS           \\
~~~~$K_s$ (mag)\dotfill             &  \hatcurCCtwomassKmag{62} & \hatcurCCtwomassKmag{63} & \hatcurCCtwomassKmag{64} & 2MASS           \\
\enddata
\tablenotetext{a}{
    SPC = Stellar Parameter Classification procedure for the analysis
    of high-resolution spectra \citep{buchhave:2012:spc}, applied to
    the TRES spectra of \hatcur{62} and \hatcur{63}, and to the HIRES
    I$_{2}$-free template spectra of \hatcur{64}. These parameters
    rely primarily on SPC, but have a small dependence also on the
    iterative analysis incorporating the isochrone search and global
    modeling of the data.
}
\tablenotetext{b}{
    In addition to the uncertainty listed here, there is a $\sim
    0.1$\,\kms\ systematic uncertainty in transforming the velocities
    to the IAU standard system.
} 
\tablenotetext{c}{
    The listed uncertainties for the Gaia DR2 photometry are taken from
the catalog. For the analysis we assume additional systematic
uncertainties of 0.002\,mag, 0.005\,mag and 0.003\,mag for the G, BP
and RP bands, respectively.
}
\tablenotetext{d}{
    From APASS DR6 for as
    listed in the UCAC 4 catalog \citep{zacharias:2013:ucac4}.  
}
\tablenotetext{e}{
    From the Amateur Sky Survey (TASS) catalog release IV \citep{droege:2006}.
}
\ifthenelse{\boolean{emulateapj}}{
    \end{deluxetable*}
}{
    \end{deluxetable}
}

%
%
\ifthenelse{\boolean{emulateapj}}{
    \begin{deluxetable*}{lcccc}
}{
    \begin{deluxetable}{lcccc}
}
\tablewidth{0pc}
\tabletypesize{\footnotesize}
\tablecaption{
    Derived stellar parameters for \hatcur{58}, \hatcur{59}, \hatcur{60} and \hatcur{61}
    \label{tab:stellarderived}
}
\tablehead{
\noalign{\vskip 2pt}
    \multicolumn{1}{c}{} &
    \multicolumn{1}{c}{\bf HAT-P-58} &
    \multicolumn{1}{c}{\bf HAT-P-59} &
    \multicolumn{1}{c}{\bf HAT-P-60} &
    \multicolumn{1}{c}{\bf HAT-P-61} \\
    \multicolumn{1}{c}{~~~~~~~~Parameter~~~~~~~~} &
    \multicolumn{1}{c}{Value}                     &
    \multicolumn{1}{c}{Value}                     &
    \multicolumn{1}{c}{Value}                     &
    \multicolumn{1}{c}{Value}                     
}
\startdata
~~~~$\mstar$ ($\msun$)\dotfill      &  \hatcurISOmlong{58}   & \hatcurISOmlong{59} & \hatcurISOmlong{60} & \hatcurISOmlong{61}\\
~~~~$\rstar$ ($\rsun$)\dotfill      &  \hatcurISOrlong{58}   & \hatcurISOrlong{59} & \hatcurISOrlong{60} & \hatcurISOrlong{61}\\
~~~~$\loggstar$ (cgs)\dotfill       &  \hatcurISOlogg{58}    & \hatcurISOlogg{59} & \hatcurISOlogg{60} & \hatcurISOlogg{61}\\
~~~~$\rhostar$ (\gcmc)\dotfill       &  \hatcurISOrho{58}    & \hatcurISOrho{59} & \hatcurISOrho{60} & \hatcurISOrho{61}\\
~~~~$\lstar$ ($\lsun$)\dotfill      &  \hatcurISOlum{58}     & \hatcurISOlum{59} & \hatcurISOlum{60} & \hatcurISOlum{61}\\
~~~~$\teffstar$ (K)\dotfill      &  \hatcurISOteff{58} &  \hatcurISOteff{59} & \hatcurISOteff{60} & \hatcurISOteff{61} \\
~~~~$\feh$\dotfill      &  \hatcurISOzfeh{58} &  \hatcurISOzfeh{59} & \hatcurISOzfeh{60}  & \hatcurISOzfeh{61} \\
~~~~Age (Gyr)\dotfill               &  \hatcurISOage{58}     & \hatcurISOage{59} & \hatcurISOage{60} & \hatcurISOage{61}\\
~~~~$A_{V}$ (mag)\dotfill               &  \hatcurXAv{58}     & \hatcurXAv{59} & \hatcurXAv{60} & \hatcurXAv{61}\\
~~~~Distance (pc)\dotfill           &  \hatcurXdistred{58}\phn  & \hatcurXdistred{59} & \hatcurXdistred{60} & \hatcurXdistred{61}\\
\enddata
\tablecomments{
	The listed parameters are those determined through the joint
	differential evolution Markov Chain analysis described in
	Section~\ref{sec:globmod}. For all four systems the
	fixed-circular-orbit model has a higher Bayesian evidence than the
	eccentric-orbit model. We therefore assume a fixed circular orbit
	in generating the parameters listed here.
}
\ifthenelse{\boolean{emulateapj}}{
    \end{deluxetable*}
}{
    \end{deluxetable}
}

%
%
\ifthenelse{\boolean{emulateapj}}{
    \begin{deluxetable*}{lccc}
}{
    \begin{deluxetable}{lccc}
}
\tablewidth{0pc}
\tabletypesize{\footnotesize}
\tablecaption{
    Derived stellar parameters for \hatcur{62}, \hatcur{63} and \hatcur{64}
    \label{tab:stellarderivedtwo}
}
\tablehead{
\noalign{\vskip 3pt}
    \multicolumn{1}{c}{} &
    \multicolumn{1}{c}{\bf HAT-P-62} &
    \multicolumn{1}{c}{\bf HAT-P-63} &
    \multicolumn{1}{c}{\bf HAT-P-64} \\
    \multicolumn{1}{c}{~~~~~~~~Parameter~~~~~~~~} &
    \multicolumn{1}{c}{Value}                     &
    \multicolumn{1}{c}{Value}                     &
    \multicolumn{1}{c}{Value}                     
}
\startdata
~~~~$\mstar$ ($\msun$)\dotfill      &  \hatcurISOmlong{62}   & \hatcurISOmlong{63} & \hatcurISOmlong{64}\\
~~~~$\rstar$ ($\rsun$)\dotfill      &  \hatcurISOrlong{62}   & \hatcurISOrlong{63} & \hatcurISOrlong{64}\\
~~~~$\loggstar$ (cgs)\dotfill       &  \hatcurISOlogg{62}    & \hatcurISOlogg{63} & \hatcurISOlogg{64}\\
~~~~$\rhostar$ (\gcmc)\dotfill       &  \hatcurISOrho{62}    & \hatcurISOrho{63} & \hatcurISOrho{64}\\
~~~~$\lstar$ ($\lsun$)\dotfill      &  \hatcurISOlum{62}     & \hatcurISOlum{63} & \hatcurISOlum{64}\\
~~~~$\teffstar$ (K)\dotfill      &  \hatcurISOteff{62} &  \hatcurISOteff{63} & \hatcurISOteff{64} \\
~~~~$\feh$\dotfill      &  \hatcurISOzfeh{62} &  \hatcurISOzfeh{63} & \hatcurISOzfeh{64} \\
~~~~Age (Gyr)\dotfill               &  \hatcurISOage{62}     & \hatcurISOage{63} & \hatcurISOage{64}\\
~~~~$A_{V}$ (mag)\dotfill               &  \hatcurXAv{62}     & \hatcurXAv{63} & \hatcurXAv{64}\\
~~~~Distance (pc)\dotfill           &  \hatcurXdistred{62}\phn  & \hatcurXdistred{63} & \hatcurXdistred{64}\\
\enddata
\tablecomments{
	The listed parameters are those determined through the joint
	differential evolution Markov Chain analysis described in
	Section~\ref{sec:globmod}. For all three systems the
	fixed-circular-orbit model has a higher Bayesian evidence than the
	eccentric-orbit model. We therefore assume a fixed circular orbit
	in generating the parameters listed here.
}
\ifthenelse{\boolean{emulateapj}}{
    \end{deluxetable*}
}{
    \end{deluxetable}
}

\subsection{Excluding blend scenarios}
\label{sec:blend}

In order to exclude blend scenarios we carried out an analysis
following \citet{hartman:2012:hat39hat41}, as updated in
\citet{hartman:2019:hats60hats69}. Here we attempt to model the
available photometric data (including light curves and catalog
broad-band photometric measurements) for each object as a blend between
an eclipsing binary star system and a third star along the line of
sight (either a physical association, or a chance alignment). The
physical properties of the stars are constrained using the Padova
isochrones \citep{girardi:2002}, while we also require that the
brightest of the three stars in the blend have atmospheric parameters
consistent with those measured with SPC. We also simulate composite
cross-correlation functions (CCFs) and use them to predict RVs and BSs
for each blend scenario considered.

Based on this analysis we rule out blended stellar eclipsing binary
scenarios for all seven systems. The results for each object are as
follows:
\begin{itemize}
\item {\em \hatcur{58}}: All blend models tested yield higher $\chi^2$
  fits to the photometry than the model of a single star with a
  transiting planet, and can be rejected with $\sim 1\sigma$
  confidence. Those models which cannot be rejected with at least
  $5\sigma$ confidence based solely on the photometry predict BS
  variations in excess of 1\,\kms\ (however, the measured BS r.m.s.\ scatter
  from HIRES is 21\,\ms).
\item {\em \hatcur{59}}: All blend models tested yield higher $\chi^2$
  fits to the photometry than the model of a single star with a
  transiting planet, and can be rejected with $3\sigma$
  confidence. Those models which cannot be rejected with at least
  $5\sigma$ confidence based solely on the photometry predict BS
  variations in excess of 100\,\ms\ (however, the measured BS r.m.s.\ scatter
  from TRES is 50\,\ms) and RV variations that do not reproduce the
  observed sinusoidal variation.
\item {\em \hatcur{60}}: All blend models tested can be rejected with
  at least $5\sigma$ confidence based solely on the photometry.
\item {\em \hatcur{61}}: Similar to \hatcur{59}, all blend models
  tested yield higher $\chi^2$ fits to the photometry than the model
  of a single star with a transiting planet, and can be rejected with
  $2\sigma$ confidence based on the photometry alone. Those models
  which cannot be rejected with at least $5\sigma$ confidence based
  solely on the photometry predict HIRES BS variations in excess of
  100\,\ms\ (the measured BS r.m.s.\ scatter from HIRES is 5\,\ms),
  TRES BS variations in excess of 200\,\ms\ (the measured BS
  r.m.s.\ scatter from TRES is 50\,\ms) and RV variations that do not
  reproduce the observed sinusoidal variation.
\item {\em \hatcur{62}}: All blend models tested have higher $\chi^2$
  fits to the photometry than the model of a single star with a
  transiting planet, and can be rejected with at least $1\sigma$
  confidence. Those models which cannot be rejected with at least
  $5\sigma$ confidence can be rejected based on the BS
  observations. These blend models yield an r.m.s.\ scatter for the
  BSs in excess of 390\,\ms, whereas the measured TRES BS
  r.m.s.\ scatter is 35\,\ms.
\item {\em \hatcur{63}}: Similar to \hatcur{59}, all blend models
  tested yield higher $\chi^2$ fits to the photometry than the model
  of a single star with a transiting planet, and can be rejected with
  $1.5\sigma$ confidence based on the photometry alone. Those models
  which cannot be rejected with at least $5\sigma$ confidence based
  solely on the photometry predict HDS BS variations in excess of
  60\,\ms\ (the measured BS r.m.s.\ scatter from HDS is 13\,\ms),
  SOPHIE BS variations in excess of 400\,\ms\ (the measured BS
  r.m.s.\ scatter from SOPHIE is 26\,\ms) and RV variations in excess
  of $\sim 200$\,\ms\ that do not reproduce the observed sinusoidal
  variation.
\item {\em \hatcur{64}}: All blend models tested have higher $\chi^2$
  fits to the photometry than the model of a single star with a
  transiting planet, and can be rejected with at least $1\sigma$
  confidence. Those models which cannot be rejected with at least
  $5\sigma$ confidence predict a BS r.m.s.\ scatter of at least
  160\,\ms, compared to the measured BS r.m.s.\ of 35\,\ms\ for the
  Keck/HIRES observations.
\end{itemize}

The analysis described above was carried out before the release of
Gaia~DR2 or {\em TESS} data. The consistency between the distance
inferred for each source by this method, assuming it is a single star
with a planet, and the Gaia~DR2 distance only bolsters the basic
conclusion that none of these systems is a blended stellar eclipsing
binary. Moreover, the {\em TESS} light curves showed no features (such
as secondary eclipses or large ellipsoidal variations) that would be
indicative of a blended eclipsing binary that might motivate a
re-analysis.

\subsection{Global modeling of the data}
\label{sec:globmod}

In order to determine the physical parameters of the TEP systems, we
carried out a global modeling of the HATNet, KeplerCam, and {\em TESS}
photometry, the high-precision RV measurements, the SPC \teffstar\ and
\feh\ measurements, the Gaia DR2 parallax, and the Gaia DR2, APASS,
TASS Mark IV, 2MASS and WISE broad-band photometry ($G$, $BP$, $RP$,
$B$, $V$, $g$, $r$, $i$, $R$, $I_{C}$, $J$, $H$, $K_{S}$, $W_1$, $W_2$,
$W_3$, $W_4$; where available).

We fit \citet{mandel:2002} transit models to the light curves assuming
quadratic limb darkening. The limb darkening coefficients are allowed
to vary in the fit, but we use the tabulations from
\citet{claret:2012,claret:2013} and \citet{claret:2018} to place
informative Gaussian prior constraints on their values, assuming a
prior uncertainty of $0.2$ for each coefficient.

We allow for a dilution of the HATNet transit depth in cases where
there are neighbors blended with the targets in the low spatial
resolution survey images (\hatcur{61}--\hatcur{64}). For {\em TESS} we
allowed for dilution for all five observed systems, and also binned
the model to account for the 30\,min exposure time \citep{kipping:2010}. For
the KeplerCam light curves we include a quadratic trend in time,
linear trends with up to three parameters describing the shape of the
PSF, and a simultaneous application of the Trend Filtering Algorithm
\citep{kovacs:2005:TFA} in our model for each event to correct for
systematic errors in the photometry. For \hatcur{62} we also include
dilution factors in the KeplerCam model to account for the blending
with the $5\farcs21$ neighbor. To do this we simulate KeplerCam images
of the primary target and its neighbor using the observed PSF and
drawing $i$-band magnitudes for each component from Normal
distributions with means and standard deviations based on the measured
$i$ magnitudes for each source from APASS. We also simulate images
without the neighbor. We then carry out aperture photometry on the
simulated images and compare the flux measured with and without the
neighbor to determine the expected dilution. The median and standard
deviation of the dilution are then calculated from all simulations for
a given night to establish Gaussian priors which are placed on the
dilution parameters which we vary in our modeling.

We fit Keplerian orbits to the RV curves allowing the zero-point for
each instrument to vary independently in the fit, and allowing for RV
jitter which we also vary as a free parameter for each
instrument. 

To model the additional stellar atmospheric, parallax and photometry
observations we introduce four new model parameters which are allowed
to vary in the fit: the distance modulus $(m-M)_{0}$, the $V$-band
extinction $A_{V}$, and the stellar atmospheric parameters \teffstar\
and \feh. Each link in the Markov Chain yields a combination of
(\teffstar, \rhostar, \feh) which we use to determine the stellar mass,
radius, $\logg$, luminosity, and absolute magnitude in various
bandpasses by comparison with the PARSEC stellar evolution models
(specifically PARSEC realease v1.2S + CLIBRI release PR16, as in
\citealp{marigo:2017}) which we generated using the CMD 3.0 web
interface by
L.~Girardi\footnote{\url{http://stev.oapd.inaf.it/cgi-bin/cmd}}. Note
that $\rhostar$ is not varied directly in the fit, but rather can be
computed from the other transit and orbital parameters which are
varied. These absolute magnitudes, together with the model distance
modulus and polynomial relations for the extinction in each bandpass as
a function of $A_{V}$ and \teffstar\ are used to compute model values
for the broad-band photometry measurements to be compared to the
observations. Here we assume systematic errors of $0.002$\,mag,
$0.005$\,mag and $0.003$\,mag on the $G$, $BP$ and $RP$ photometry,
respectively, following \citet{evans:2018}. These systematic
uncertainties are added in quadrature to the statistical uncertainties
on the measurements listed in the Gaia DR2 catalog.

For $A_{V}$ we made use of the MWDUST 3D Galactic extinction
model \citep{bovy:2016} to tabulate the extinction in 0.1\,kpc steps
in the direction of the source. For a given $(m-M)_{0}$ we then
perform linear interpolation among these values to estimate the
expected $A_{V}$ at that distance. We treat this expected value as a
Gaussian prior, with a $1\sigma$ uncertainty of 20\% the maximum
value.

We used a Differential Evolution Markov Chain Monte Carlo procedure to
explore the fitness landscape and to determine the posterior
distribution of the parameters. When a proposed link in the Markov
Chain falls outside of the parameter values spanned by the stellar
evolution models (e.g., if a star with a density greater than what is
allowed by the stellar evolution models at a given temperature and
metallicity is proposed) the link is rejected and the
previous link is retained. In this manner the fitting procedure used
here forces the solutions to match to the theoretical stellar
evolution models.  We tried fitting both fixed-circular-orbits and
free-eccentricity models to the data, and for all seven systems find
that the data are consistent with a circular orbit. 
We therefore adopt the parameters that come from the
fixed-circular-orbit models for all of the systems. The resulting
parameters for \hatcurb{58}, \hatcurb{59}, \hatcurb{60}, and
\hatcurb{61} are listed in \reftabl{planetparam}, while for
\hatcurb{62}, \hatcurb{63} and \hatcurb{64} they are listed in
\reftabl{planetparamtwo}.

\startlongtable
%
\ifthenelse{\boolean{emulateapj}}{
    \begin{deluxetable*}{lcccc}
}{
    \begin{deluxetable}{lcccc}
}
\tabletypesize{\footnotesize}
\tablecaption{Orbital and planetary parameters for \hatcurb{58}, \hatcurb{59}, \hatcurb{60} and \hatcurb{61}\label{tab:planetparam}}
\tablehead{
\noalign{\vskip 3pt}
    \multicolumn{1}{c}{} &
    \multicolumn{1}{c}{\bf HAT-P-58b} &
    \multicolumn{1}{c}{\bf HAT-P-59b} &
    \multicolumn{1}{c}{\bf HAT-P-60b} &
    \multicolumn{1}{c}{\bf HAT-P-61b} \\ 
    \multicolumn{1}{c}{~~~~~~~~~~~~~~~Parameter~~~~~~~~~~~~~~~} &
    \multicolumn{1}{c}{Value} &
    \multicolumn{1}{c}{Value} &
    \multicolumn{1}{c}{Value} &
    \multicolumn{1}{c}{Value}
}
\startdata
\noalign{\vskip -3pt}
\sidehead{\underline{\Lc{} parameters}}
~~~$P$ (days)             \dotfill    & $\hatcurLCP{58}$ & $\hatcurLCP{59}$ & $\hatcurLCP{60}$ & $\hatcurLCP{61}$ \\
~~~$T_c$ (${\rm BJD_{TDB}}-2450000$)    
      \tablenotemark{a}   \dotfill    & $\hatcurLCTshort{58}$ & $\hatcurLCTshort{59}$ & $\hatcurLCTshort{60}$ & $\hatcurLCTshort{61}$ \\
~~~$T_{14}$ (days)
      \tablenotemark{a}   \dotfill    & $\hatcurLCdur{58}$ & $\hatcurLCdur{59}$ & $\hatcurLCdur{60}$ & $\hatcurLCdur{61}$ \\
~~~$T_{12} = T_{34}$ (days)
      \tablenotemark{a}   \dotfill    & $\hatcurLCingdur{58}$ & $\hatcurLCingdur{59}$ & $\hatcurLCingdur{60}$ & $\hatcurLCingdur{61}$ \\
~~~$\arstar$              \dotfill    & $\hatcurPPar{58}$ & $\hatcurPPar{59}$ & $\hatcurPPar{60}$ & $\hatcurPPar{61}$ \\
~~~$\zrstar$ \tablenotemark{b}             \dotfill    & $\hatcurLCzeta{58}$\phn & $\hatcurLCzeta{59}$\phn & $\hatcurLCzeta{60}$ & $\hatcurLCzeta{61}$\phn \\
~~~$\rpl/\rstar$          \dotfill    & $\hatcurLCrprstar{58}$ & $\hatcurLCrprstar{59}$ & $\hatcurLCrprstar{60}$ & $\hatcurLCrprstar{61}$ \\
~~~$b^2$                  \dotfill    & $\hatcurLCbsq{58}$ & $\hatcurLCbsq{59}$ & $\hatcurLCbsq{60}$ & $\hatcurLCbsq{61}$ \\
~~~$b \equiv a \cos i/\rstar$
                          \dotfill    & $\hatcurLCimp{58}$ & $\hatcurLCimp{59}$ & $\hatcurLCimp{60}$ & $\hatcurLCimp{61}$ \\
~~~$i$ (deg)              \dotfill    & $\hatcurPPi{58}$\phn & $\hatcurPPi{59}$\phn & $\hatcurPPi{60}$ & $\hatcurPPi{61}$\phn \\
%
\sidehead{\underline{HATNet blend factors \tablenotemark{c}}}
~~~Blend factor 1 \dotfill & $\cdots$ & $\cdots$ & $\cdots$ & $\hatcurLCiblendA{61}$ \\
~~~Blend factor 2 \dotfill & $\cdots$ & $\cdots$ & $\cdots$ & $\hatcurLCiblendB{61}$ \\
\sidehead{\underline{TESS blend factors \tablenotemark{c}}}
~~~Blend factor 1 \dotfill & $\hatcurLCiblendB{58}$ & $\hatcurLCiblendB{59}$ & $\hatcurLCiblendB{60}$ & $\hatcurLCiblendC{61}$ \\
~~~Blend factor 2 \dotfill & $\cdots$ & $\hatcurLCiblendC{59}$ & $\cdots$ & $\cdots$ \\
~~~Blend factor 3 \dotfill & $\cdots$ & $\hatcurLCiblendD{59}$ & $\cdots$ & $\cdots$ \\
~~~Blend factor 4 \dotfill & $\cdots$ & $\hatcurLCiblendE{59}$ & $\cdots$ & $\cdots$ \\
~~~Blend factor 5 \dotfill & $\cdots$ & $\hatcurLCiblendF{59}$ & $\cdots$ & $\cdots$ \\
~~~Blend factor 6 \dotfill & $\cdots$ & $\hatcurLCiblendG{59}$ & $\cdots$ & $\cdots$ \\
~~~Blend factor 7 \dotfill & $\cdots$ & $\hatcurLCiblendH{59}$ & $\cdots$ & $\cdots$ \\
\sidehead{\underline{Limb-darkening coefficients \tablenotemark{d}}}
~~~$c_1,R$                  \dotfill    & $\cdots$ & $\cdots$ & $\cdots$ & $\hatcurLBiR{61}$ \\
~~~$c_2,R$                  \dotfill    & $\cdots$ & $\cdots$ & $\cdots$ & $\hatcurLBiiR{61}$ \\
~~~$c_1,r$                  \dotfill    & $\hatcurLBir{58}$ & $\hatcurLBir{59}$ & $\hatcurLBir{60}$ & $\hatcurLBir{61}$ \\
~~~$c_2,r$                  \dotfill    & $\hatcurLBiir{58}$ & $\hatcurLBiir{59}$ & $\hatcurLBiir{60}$ & $\hatcurLBiir{61}$ \\
~~~$c_1,i$                  \dotfill    & $\hatcurLBii{58}$ & $\hatcurLBii{59}$ & $\hatcurLBii{60}$ & $\hatcurLBii{61}$ \\
~~~$c_2,i$                  \dotfill    & $\hatcurLBiii{58}$ & $\hatcurLBiii{59}$ & $\hatcurLBiii{60}$ & $\hatcurLBiii{61}$ \\
~~~$c_1,z$                  \dotfill    & $\cdots$ & $\cdots$ & $\hatcurLBiz{60}$ & $\cdots$ \\
~~~$c_2,z$                  \dotfill    & $\cdots$ & $\cdots$ & $\hatcurLBiiz{60}$ & $\cdots$ \\
~~~$c_1,T$                  \dotfill    & $\hatcurLBiT{58}$ & $\hatcurLBiT{59}$ & $\hatcurLBiT{60}$ & $\hatcurLBiT{61}$ \\
~~~$c_2,T$                  \dotfill    & $\hatcurLBiiT{58}$ & $\hatcurLBiiT{59}$ & $\hatcurLBiiT{60}$ & $\hatcurLBiiT{61}$ \\
%
\sidehead{\underline{RV parameters}}
~~~$K$ (\ms)              \dotfill    & $\hatcurRVK{58}$\phn\phn & $\hatcurRVK{59}$\phn\phn & $\hatcurRVK{60}$\phn\phn & $\hatcurRVK{61}$\phn\phn \\
%
~~~$e$ \tablenotemark{e}               \dotfill    & $\hatcurRVeccentwosiglimeccen{58}$ & $\hatcurRVeccentwosiglimeccen{59}$ & $\hatcurRVeccentwosiglimeccen{60}$ & $\hatcurRVeccentwosiglimeccen{61}$ \\
~~~RV jitter HIRES (\ms) \tablenotemark{f}       \dotfill    & \hatcurRVjittertwosiglim{58} & $\cdots$ & \hatcurRVjitterA{60} & \hatcurRVjittertwosiglimA{61} \\
~~~RV jitter TRES (\ms)        \dotfill    & $\cdots$ & \hatcurRVjittertwosiglimA{59} & \hatcurRVjittertwosiglimB{60} & \hatcurRVjitterB{61} \\
~~~RV jitter SOPHIE (\ms)        \dotfill    & $\cdots$ & \hatcurRVjittertwosiglimB{59} & \hatcurRVjittertwosiglimC{60} & $\cdots$ \\
~~~RV jitter NRES/ELP (\ms)        \dotfill    & $\cdots$ & $\cdots$ & $\hatcurRVjitterD{60}$ & $\cdots$ \\
~~~RV jitter NRES/TLV (\ms)        \dotfill    & $\cdots$ & $\cdots$ & $\hatcurRVjitterE{60}$ & $\cdots$ \\
\sidehead{\underline{Planetary parameters}}
~~~$\mpl$ ($\mjup$)       \dotfill    & $\hatcurPPmlong{58}$ & $\hatcurPPmlong{59}$ & $\hatcurPPmlong{60}$ & $\hatcurPPmlong{61}$ \\
~~~$\rpl$ ($\rjup$)       \dotfill    & $\hatcurPPrlong{58}$ & $\hatcurPPrlong{59}$ & $\hatcurPPrlong{60}$ & $\hatcurPPrlong{61}$ \\
~~~$C(\mpl,\rpl)$
    \tablenotemark{g}     \dotfill    & $\hatcurPPmrcorr{58}$ & $\hatcurPPmrcorr{59}$ & $\hatcurPPmrcorr{60}$ & $\hatcurPPmrcorr{61}$ \\
~~~$\rhopl$ (\gcmc)       \dotfill    & $\hatcurPPrho{58}$ & $\hatcurPPrho{59}$ & $\hatcurPPrho{60}$ & $\hatcurPPrho{61}$ \\
~~~$\log g_p$ (cgs)       \dotfill    & $\hatcurPPlogg{58}$ & $\hatcurPPlogg{59}$ & $\hatcurPPlogg{60}$ & $\hatcurPPlogg{61}$ \\
~~~$a$ (AU)               \dotfill    & $\hatcurPParel{58}$ & $\hatcurPParel{59}$ & $\hatcurPParel{60}$ & $\hatcurPParel{61}$ \\
~~~$T_{\rm eq}$ (K)        \dotfill   & $\hatcurPPteff{58}$ & $\hatcurPPteff{59}$ & $\hatcurPPteff{60}$ & $\hatcurPPteff{61}$ \\
~~~$\Theta$ \tablenotemark{h} \dotfill & $\hatcurPPtheta{58}$ & $\hatcurPPtheta{59}$ & $\hatcurPPtheta{60}$ & $\hatcurPPtheta{61}$ \\
%
~~~$\log_{10}\langle F \rangle$ (cgs) \tablenotemark{i}
                          \dotfill    & $\hatcurPPfluxavglog{58}$ & $\hatcurPPfluxavglog{59}$ & $\hatcurPPfluxavglog{60}$ & $\hatcurPPfluxavglog{61}$ \\
\enddata
\tablenotetext{a}{
    Times are in Barycentric Julian Date on the dynamical time system, including the correction for leap seconds.
    \ensuremath{T_c}: Reference epoch of
    mid transit that minimizes the correlation with the orbital
    period.
    \ensuremath{T_{14}}: total transit duration, time
    between first to last contact;
    \ensuremath{T_{12}=T_{34}}: ingress/egress time, time between first
    and second, or third and fourth contact.
}
\tablecomments{
	For all four systems the fixed-circular-orbit model has a higher
	Bayesian evidence than the eccentric-orbit model. We therefore
	assume a fixed circular orbit in generating the parameters listed
	here.
}
\tablenotetext{b}{
	Reciprocal of the half duration of the transit used as a jump
	parameter in our MCMC analysis in place of $\arstar$. It is related
	to $\arstar$ by the expression $\zrstar =
	\arstar(2\pi(1+e\sin\omega))/(P\sqrt{1-b^2}\sqrt{1-e^2})$
	\citep{bakos:2010:hat11}.
}
\tablenotetext{c}{
    Scaling factor applied to the model transit that is fit to the
	HATNet and {\em TESS} light curves. This factor accounts for
	dilution of the transit due to blending from neighboring stars and
	over-filtering of the light curve (in cases where we do not apply
	signal-reconstruction TFA). These factors are varied in the fit,
	and we allow independent factors for observations obtained for
	different HATNet fields and different {\em TESS} sectors. For
	\hatcur{58}--\hatcur{60} we do not include these factors for HATNet
	because the stars are well isolated on the HATNet images, and we
	applied signal-reconstruction TFA to preserve the signal shape
	while filtering the light curves.
}
\tablenotetext{d}{
    Values for a quadratic law. These are allowed to vary in the fit,
	using the tabulations from \citet{claret:2012,claret:2013} and
	\citet{claret:2018} to place informative Gaussian prior constraints
	on their values.
}
\tablenotetext{e}{
    The 95\% confidence upper limit on the eccentricity determined when
    $\sqrt{e}\cos\omega$ and $\sqrt{e}\sin\omega$ are allowed to vary
    in the fit.
}
\tablenotetext{f}{
    Term added in quadrature to the formal RV uncertainties for each
    instrument. This is treated as a free parameter in the fitting
    routine. In cases where the jitter is consistent with zero we list
    the 95\% confidence upper limit.
}
\tablenotetext{g}{
    Correlation coefficient between the planetary mass \mpl\ and radius
    \rpl\ estimated from the posterior parameter distribution.
}
\tablenotetext{h}{
    The Safronov number is given by $\Theta = \frac{1}{2}(V_{\rm
    esc}/V_{\rm orb})^2 = (a/\rpl)(\mpl / \mstar )$
    \citep[see][]{hansen:2007}.
}
\tablenotetext{i}{
    Incoming flux per unit surface area, averaged over the orbit.
}
\ifthenelse{\boolean{emulateapj}}{
    \end{deluxetable*}
}{
    \end{deluxetable}
}
\clearpage
\newpage

\startlongtable
%
\ifthenelse{\boolean{emulateapj}}{
    \begin{deluxetable*}{lccc}
}{
    \begin{deluxetable}{lccc}
}
\tabletypesize{\footnotesize}
\tablecaption{Orbital and planetary parameters for \hatcurb{62}, \hatcurb{63} and \hatcurb{64}\label{tab:planetparamtwo}}
\tablehead{
\noalign{\vskip 3pt}
    \multicolumn{1}{c}{} &
    \multicolumn{1}{c}{\bf HAT-P-62b} &
    \multicolumn{1}{c}{\bf HAT-P-63b} &
    \multicolumn{1}{c}{\bf HAT-P-64b} \\ 
    \multicolumn{1}{c}{~~~~~~~~~~~~~~~Parameter~~~~~~~~~~~~~~~} &
    \multicolumn{1}{c}{Value} &
    \multicolumn{1}{c}{Value} &
    \multicolumn{1}{c}{Value}
}
\startdata
\noalign{\vskip -3pt}
\sidehead{\underline{\Lc{} parameters}}
~~~$P$ (days)             \dotfill    & $\hatcurLCP{62}$ & $\hatcurLCP{63}$ & $\hatcurLCP{64}$ \\
~~~$T_c$ (${\rm BJD_{TDB}}-2450000$)    
      \tablenotemark{a}   \dotfill    & $\hatcurLCTshort{62}$ & $\hatcurLCTshort{63}$ & $\hatcurLCTshort{64}$ \\
~~~$T_{14}$ (days)
      \tablenotemark{a}   \dotfill    & $\hatcurLCdur{62}$ & $\hatcurLCdur{63}$ & $\hatcurLCdur{64}$ \\
~~~$T_{12} = T_{34}$ (days)
      \tablenotemark{a}   \dotfill    & $\hatcurLCingdur{62}$ & $\hatcurLCingdur{63}$ & $\hatcurLCingdur{64}$ \\
~~~$\arstar$              \dotfill    & $\hatcurPPar{62}$ & $\hatcurPPar{63}$ & $\hatcurPPar{64}$ \\
~~~$\zrstar$ \tablenotemark{b}             \dotfill    & $\hatcurLCzeta{62}$\phn & $\hatcurLCzeta{63}$\phn & $\hatcurLCzeta{64}$\phn \\
~~~$\rpl/\rstar$          \dotfill    & $\hatcurLCrprstar{62}$ & $\hatcurLCrprstar{63}$ & $\hatcurLCrprstar{64}$ \\
~~~$b^2$                  \dotfill    & $\hatcurLCbsq{62}$ & $\hatcurLCbsq{63}$ & $\hatcurLCbsq{64}$ \\
~~~$b \equiv a \cos i/\rstar$
                          \dotfill    & $\hatcurLCimp{62}$ & $\hatcurLCimp{63}$ & $\hatcurLCimp{64}$ \\
~~~$i$ (deg)              \dotfill    & $\hatcurPPi{62}$\phn & $\hatcurPPi{63}$\phn & $\hatcurPPi{64}$\phn \\
%
\sidehead{\underline{HATNet blend factors \tablenotemark{c}}}
~~~Blend factor \dotfill & $\hatcurLCiblend{62}$ & $\cdots$ & $\hatcurLCiblendA{64}$ \\
\sidehead{\underline{TESS blend factors \tablenotemark{c}}}
~~~Blend factor \dotfill & $\cdots$ & $\cdots$ & $\hatcurLCiblendB{64}$ \\
\sidehead{\underline{Limb-darkening coefficients \tablenotemark{d}}}
~~~$c_1,r$                  \dotfill    & $\hatcurLBir{62}$ & $\hatcurLBir{63}$ & $\hatcurLBir{64}$ \\
~~~$c_2,r$                  \dotfill    & $\hatcurLBiir{62}$ & $\hatcurLBiir{63}$ & $\hatcurLBiir{64}$ \\
~~~$c_1,i$                  \dotfill    & $\hatcurLBii{62}$ & $\hatcurLBii{63}$ & $\hatcurLBii{64}$ \\
~~~$c_2,i$                  \dotfill    & $\hatcurLBiii{62}$ & $\hatcurLBiii{63}$ & $\hatcurLBiii{64}$ \\
~~~$c_1,T$                  \dotfill    & $\cdots$ & $\cdots$ & $\hatcurLBiT{64}$ \\
~~~$c_2,T$                  \dotfill    & $\cdots$ & $\cdots$ & $\hatcurLBiiT{64}$ \\
%
\sidehead{\underline{RV parameters}}
~~~$K$ (\ms)              \dotfill    & $\hatcurRVK{62}$\phn\phn & $\hatcurRVK{63}$\phn\phn & $\hatcurRVK{64}$\phn\phn \\
%
~~~$e$ \tablenotemark{e}               \dotfill    & $\hatcurRVeccentwosiglimeccen{62}$ & $\hatcurRVeccentwosiglimeccen{63}$ & $\hatcurRVeccentwosiglimeccen{64}$ \\
~~~RV jitter HIRES (\ms) \tablenotemark{f}       \dotfill    & $\cdots$ & $\cdots$ & \hatcurRVjitterA{64} \\
~~~RV jitter TRES (\ms)        \dotfill    & $\hatcurRVjitter{62}$ & \hatcurRVjittertwosiglimB{63} & $\cdots$ \\
~~~RV jitter SOPHIE (\ms)        \dotfill    & $\cdots$ & \hatcurRVjitterC{63} & \hatcurRVjittertwosiglimB{64} \\
~~~RV jitter HDS (\ms)        \dotfill    & $\cdots$ & \hatcurRVjittertwosiglimA{63} & $\cdots$ \\
\sidehead{\underline{Planetary parameters}}
~~~$\mpl$ ($\mjup$)       \dotfill    & $\hatcurPPmlong{62}$ & $\hatcurPPmlong{63}$ & $\hatcurPPmlong{64}$ \\
~~~$\rpl$ ($\rjup$)       \dotfill    & $\hatcurPPrlong{62}$ & $\hatcurPPrlong{63}$ & $\hatcurPPrlong{64}$ \\
~~~$C(\mpl,\rpl)$
    \tablenotemark{g}     \dotfill    & $\hatcurPPmrcorr{62}$ & $\hatcurPPmrcorr{63}$ & $\hatcurPPmrcorr{64}$ \\
~~~$\rhopl$ (\gcmc)       \dotfill    & $\hatcurPPrho{62}$ & $\hatcurPPrho{63}$ & $\hatcurPPrho{64}$ \\
~~~$\log g_p$ (cgs)       \dotfill    & $\hatcurPPlogg{62}$ & $\hatcurPPlogg{63}$ & $\hatcurPPlogg{64}$ \\
~~~$a$ (AU)               \dotfill    & $\hatcurPParel{62}$ & $\hatcurPParel{63}$ & $\hatcurPParel{64}$ \\
~~~$T_{\rm eq}$ (K)        \dotfill   & $\hatcurPPteff{62}$ & $\hatcurPPteff{63}$ & $\hatcurPPteff{64}$ \\
~~~$\Theta$ \tablenotemark{h} \dotfill & $\hatcurPPtheta{62}$ & $\hatcurPPtheta{63}$ & $\hatcurPPtheta{64}$ \\
%
~~~$\log_{10}\langle F \rangle$ (cgs) \tablenotemark{i}
                          \dotfill    & $\hatcurPPfluxavglog{62}$ & $\hatcurPPfluxavglog{63}$ & $\hatcurPPfluxavglog{64}$ \\
\enddata
\tablecomments{
	For all three systems the fixed-circular-orbit model has a higher
	Bayesian evidence than the eccentric-orbit model. We therefore
	assume a fixed circular orbit in generating the parameters listed
	here. For all further tablenotes please refer to
	Table.~\ref{tab:planetparam}. 
}
\ifthenelse{\boolean{emulateapj}}{
    \end{deluxetable*}
}{
    \end{deluxetable}
}

\section{Discussion}
\label{sec:discussion}

We presented the discovery of seven hot Jupiters transiting bright
stars. These planets were first identified as transiting planet
candidates by the HATNet survey from among some 6 million stars that
have been observed to date since 2004. They were subsequently confirmed
and accurately characterized using high-precision time-series
photometry from FLWO~1.2\,m/KeplerCam, and the NASA {\em TESS} mission,
and high-resolution spectroscopy, enabling high-precision radial
velocity measurements, carried out with the FLWO~1.5\,m/TRES, Keck-I/HIRES,
OHP~1.93\,m/SOPHIE, Subaru~8\,m/HDS, APO~3.5\,m/ARCES,
NOT~2.5\,m/FIES, and LCOGT~1\,m/NRES telescopes/instruments.

The planets discovered here contribute to the growing sample of
transiting planets with precisely measured masses and radii. All seven
planets have radii measured to better than $\sim$10\% precision, and
six of them have masses measured to this level of precision as
well. Such planets are valuable contributions to the growing sample of
well-characterized exoplanets which may be used in statistical studies
to test theories of planet formation and evolution. In fact, the
planets presented here have already been included in one such study
\citep{hartman:2016:hat6566}.

Close-in giant planets transiting bright stars, such as these, can
also be followed-up in a modest amount of time using current
facilities to measure their orbital (mis-)alignments and probe the
planetary atmospheres. We estimate that the amplitude of the
Rossiter-McLaughlin effect is: 35\,\ms, 18\,\ms, 36\,\ms, 23\,\ms,
30\,\ms, 44\,\ms, and 128\,\ms, for \hatcurb{58}--\hatcurb{64},
respectively. Given the host star brightnesses, measured RV jitter
values, and transit durations, the effect would be detectable using
facilities ranging from FLWO~1.5\,m/TRES (\hatcurb{60} which orbits a
$V = \hatcurCCtassmv{60}$\,mag host star, and \hatcurb{64} with its
large amplitude signal and long-lasting transits), to Keck-I/HIRES
(\hatcurb{59}). With $\arstar > 9$, and $\teffstar < 6000$\,K,
\hatcurb{59} and \hatcurb{63} may be particularly interesting objects
for which to observe this effect, in an effort to determine whether giant
planets transiting cool stars become less well aligned as the strength
of the tidal interaction with their host stars decreases
\citep[e.g.,][]{albrecht:2012}.

As regards atmospheric characterization, with its 1\% deep transits
lasting almost five hours, and large atmospheric scale height ($\log
g_p = \hatcurPPlogg{64}$), \hatcurb{64} is perhaps the most promising
of the planets discovered here for having readily detectable features
in its transmission spectrum. These may be atomic or molecular
absorption features as seen, for example, in the spectrum of the
inflated Neptune HAT-P-26b, \citep{wakeford:2017}, among many other
planets. Alternatively, this may be evidence of an atmospheric haze
revealed through Rayleigh-scattering, as seen, for example, in the
spectrum of the highly inflated hot Jupiter HAT-P-32b,
\citep{mallonn:2017}, again among many planets. With a planetary radius
of $\hatcurPPr{64}$\,\rjup, \hatcurb{64} is also one of the largest
known transiting exoplanets (as of 2018 July there are only 23
transiting planets listed in the NASA exoplanet archive with larger
radii). The planet follows the well-established trend between
high-equilibrium temperature and inflated radius
\citep[e.g.,][]{fortney:2007,enoch:2011,kovacs:2010:hat15,beky:2011:hat27,enoch:2012}.

Including the systems presented here, a total of 67 transiting planets
have now been discovered and published by HATNet. In addition to these,
some 17 planets discovered by other teams have been independently
detected in HATNet light curves (KELT-1, KELT-3, Kepler-6, Kepler-12,
KOI-13, Qatar-1, TrES-2, TrES-3, TrES-5, WASP-2, WASP-10, WASP-13,
WASP-24, WASP-33, WASP-48, XO-3, and XO-5), and more than a dozen
additional transiting planets have been detected by HATNet and
confirmed through follow-up observations, but have not yet been
published. Altogether at least $\sim 100$ transiting exoplanets have
been detected by HATNet, and certainly more planets remain to be
discovered among the 500 remaining candidates that have not yet been
confirmed or set aside as false positives or false alarms. The
Hungarian-made Automated Telescope Network (HATNet) continues to
operate in a fully autonomous manner, and will continue to produce
high-precision high-cadence time-series photometry for millions of
stars over a large swath of the Northern sky. Over the past 16 years it
has amassed a rich database of light curves for six million stars.

The NASA {\em TESS} mission \citep{ricker:2015} uses a set of four
lenses, very similar in diameter to those used by HATNet, to survey the
entire sky. Although the HATNet light curves are of lower photometric
precision than {\em TESS}, the observations are made at higher spatial
resolution than {\em TESS}, and are useful for identifying {\em TESS}
candidates that are actually blended stellar eclipsing binary objects.
The HATNet light curves may also be used in conjunction with the {\em
TESS} data to search for longer period planets than could be found in
the typical 27.4\,d {\em TESS} observing windows alone.

The planet \hatcurb{59} presented has made for a particularly fruitful
synergy between HATNet and {\em TESS}. This planet lies $10\fdg4$ from
the northern ecliptic pole, and is thus within the Northern continuous
viewing zone of {\em TESS}. It will be observed continuously for
approximately 1\,yr by {\em TESS}, and we have already included seven
sectors of data in our analysis of this system.

We plan to continue operating HATNet for the foreseeable future, and
anticipate widening the region of parameter space to which we are
sensitive to planets (i.e., toward finding sub-Neptune-size planets and
planets with periods of several tens of days), by combining HATNet and
{\em TESS} data, and by extending the time coverage of regions on the
sky previously observed by HATNet.


\acknowledgements 

HATNet operations have been funded by NASA grants NNG04GN74G,
NNX08AF23G, and NNX13AJ15G. Follow-up of HATNet targets has been
partially supported through NSF grant AST-1108686. G.\'A.B, J.H.,
Z.C.~and K.P.~acknowledge partial support from NASA grant NNX17AB61G.
G.B.~acknowledges support from the Hungarian Academy of Sciences, and
thanks for the warm hospitality of Konkoly Observatory in carrying out
some of his research. J.H.~acknowledges support from NASA grant
NNX14AF87G. K.P.~acknowledges support from NASA grant 80NSSC18K1009.
G.K.~thanks the support from the National Research, Development and
Innovation Office (grant K~129249). We acknowledge partial support also
from the Kepler Mission under NASA Cooperative Agreement NNX13AB58A
(D.W.L., PI). Data presented in this paper are based on observations
obtained at the HAT station at the Submillimeter Array of SAO, and the
HAT station at the Fred Lawrence Whipple Observatory of SAO. We
acknowledge J.A.~Johnson in supporting the Keck HIRES observations. The
authors wish to acknowledge the very significant cultural role and
reverence that the summit of Mauna Kea has always had within the
indigenous Hawaiian community. We are most fortunate to have the
opportunity to conduct observations from this mountain. This research
has made use of Keck telescope time granted through NOAO (programs:
A245Hr, A202Hr; PI: G.B) and NASA (programs: N154Hr, N133Hr,
N136Hr, N143Hr, N169Hr, N186Hr; PI: G.B).
  Based on observations at Kitt Peak National Observatory, National
  Optical Astronomy Observatory (NOAO Prop.~ID: 2015B-0156; PI:
  J.H.), which is operated by the Association of Universities for
  Research in Astronomy (AURA) under a cooperative agreement with the
  National Science Foundation. Based on radial velocities obtained with
  the SOPHIE spectrograph mounted on the 1.93\,m telescope at
  Observatoire de Haute-Provence. Based on data collected at Subaru
  Telescope, which is operated by the National Astronomical Observatory
  of Japan. Based on observations made with the Nordic Optical
  Telescope, operated on the island of La Palma jointly by Denmark,
  Finland, Norway, Sweden, in the Spanish Observatorio del Roque de los
  Muchachos of the Intituto de Astrof\'isica de Canarias. Based on
  observations obtained with the Apache Point Observatory 3.5\,m
  telescope, which is owned and operated by the Astrophysical Research
  Consortium.
This research was made possible through the use of the AAVSO
Photometric All-Sky Survey (APASS), funded by the Robert Martin Ayers
Sciences Fund.
This research has made use of the NASA Exoplanet Archive, which is
operated by the California Institute of Technology, under contract with
the National Aeronautics and Space Administration under the Exoplanet
Exploration Program.


\clearpage
\bibliographystyle{apj}
\bibliography{htrbib}

\clearpage

\ifthenelse{\boolean{emulateapj}}{
    \begin{figure*}[!p]
}{
    \begin{figure}[!p]
}
{
 \centering
 \leavevmode
 \includegraphics[width={1.0\linewidth}]{\hatcurhtr{59}-banner}
}
{
 \centering
 \leavevmode
 \includegraphics[width={0.5\linewidth}]{\hatcurhtr{59}-hatnet}%
 \hfil
 \includegraphics[width={0.5\linewidth}]{\hatcurhtr{59}-lc}%
}
{
 \centering
 \leavevmode
 \includegraphics[width={0.5\linewidth}]{\hatcurhtr{59}-rv}%
 \hfil
 \includegraphics[width={0.5\linewidth}]{\hatcurhtr{59}-iso-bprp-gabs-isofeh-SED}%
}                        
\caption{
    Observations of \hatcur{59} together with our best-fit model.
	Please refer to Figure~\ref{fig:hatp58} for a more detailed
	caption. The {\em TESS} light curve for this system is shown in
	Figure~\ref{fig:hatp59tess}.
\label{fig:hatp59}
}
\ifthenelse{\boolean{emulateapj}}{
    \end{figure*}
}{
    \end{figure}
}

\ifthenelse{\boolean{emulateapj}}{
    \begin{figure*}[!p]
}{
    \begin{figure}[!p]
}
 {
 \centering
 \leavevmode
 \includegraphics[width={1.0\linewidth}]{\hatcurhtr{60}-banner}
}
 {
 \centering
 \leavevmode
 \includegraphics[width={0.5\linewidth}]{\hatcurhtr{60}-hatnet}%
 \hfil
 \includegraphics[width={0.5\linewidth}]{\hatcurhtr{60}-lc}%
 }
 {
 \centering
 \leavevmode
 \includegraphics[width={0.5\linewidth}]{\hatcurhtr{60}-rv}%
 \hfil
 \includegraphics[width={0.5\linewidth}]{\hatcurhtr{60}-iso-bprp-gabs-isofeh-SED}%
 }                        
\caption{
	Observations of \hatcur{60} together with our best-fit model.
	Please refer to Figure~\ref{fig:hatp58} for a more detailed
	caption. The {\em TESS} light curve for this system is shown in
	Figure~\ref{fig:hatp60tess}.
\label{fig:hatp60}
}
\ifthenelse{\boolean{emulateapj}}{
    \end{figure*}
}{
    \end{figure}
}

\ifthenelse{\boolean{emulateapj}}{
    \begin{figure*}[!p]
}{
    \begin{figure}[!p]
}
 {
 \centering
 \leavevmode
 \includegraphics[width={1.0\linewidth}]{\hatcurhtr{61}-banner}
}
 {
 \centering
 \leavevmode
 \includegraphics[width={0.5\linewidth}]{\hatcurhtr{61}-hatnet}%
 \hfil
 \includegraphics[width={0.5\linewidth}]{\hatcurhtr{61}-lc}%
 }
 {
 \centering
 \leavevmode
 \includegraphics[width={0.5\linewidth}]{\hatcurhtr{61}-rv}%
 \hfil
 \includegraphics[width={0.5\linewidth}]{\hatcurhtr{61}-iso-bprp-gabs-isofeh-SED}%
 }                        
\caption{
	Observations of \hatcur{61} together with our best-fit model.
	Please refer to Figure~\ref{fig:hatp58} for a more detailed
	caption. The {\em TESS} light curve for this system is shown in
	Figure~\ref{fig:hatp61tess}.
\label{fig:hatp61}
}
\ifthenelse{\boolean{emulateapj}}{
    \end{figure*}
}{
    \end{figure}
}

\ifthenelse{\boolean{emulateapj}}{
    \begin{figure*}[!p]
}{
    \begin{figure}[!p]
}
 {
 \centering
 \leavevmode
 \includegraphics[width={1.0\linewidth}]{\hatcurhtr{62}-banner}
}
 {
 \centering
 \leavevmode
 \includegraphics[width={0.5\linewidth}]{\hatcurhtr{62}-hatnet}%
 \hfil
 \includegraphics[width={0.5\linewidth}]{\hatcurhtr{62}-lc}%
 }
 {
 \centering
 \leavevmode
 \includegraphics[width={0.5\linewidth}]{\hatcurhtr{62}-rv}%
 \hfil
 \includegraphics[width={0.5\linewidth}]{\hatcurhtr{62}-iso-bprp-gabs-isofeh-SED}%
 }                        
\caption{
	Observations of \hatcur{62} together with our best-fit model.
	Please refer to Figure~\ref{fig:hatp58} for a more detailed
	caption. \label{fig:hatp62}
}
\ifthenelse{\boolean{emulateapj}}{
    \end{figure*}
}{
    \end{figure}
}

\ifthenelse{\boolean{emulateapj}}{
    \begin{figure*}[!p]
}{
    \begin{figure}[!p]
}
 {
 \centering
 \leavevmode
 \includegraphics[width={1.0\linewidth}]{\hatcurhtr{63}-banner}
}
 {
 \centering
 \leavevmode
 \includegraphics[width={0.5\linewidth}]{\hatcurhtr{63}-hatnet}%
 \hfil
 \includegraphics[width={0.5\linewidth}]{\hatcurhtr{63}-lc}%
 }
 {
 \centering
 \leavevmode
 \includegraphics[width={0.5\linewidth}]{\hatcurhtr{63}-rv}%
 \hfil
 \includegraphics[width={0.5\linewidth}]{\hatcurhtr{63}-iso-bprp-gabs-isofeh-SED}%
 }                        
\caption{
	Observations of \hatcur{63} together with our best-fit model.
	Please refer to Figure~\ref{fig:hatp58} for a more detailed
	caption.
\label{fig:hatp63}
}
\ifthenelse{\boolean{emulateapj}}{
    \end{figure*}
}{
    \end{figure}
}
\clearpage

\ifthenelse{\boolean{emulateapj}}{
    \begin{figure*}[!ht]
}{
    \begin{figure}[!ht]
}
 {
 \centering
 \leavevmode
 \includegraphics[width={1.0\linewidth}]{\hatcurhtr{64}-banner}
}
 {
 \centering
 \leavevmode
 \includegraphics[width={0.5\linewidth}]{\hatcurhtr{64}-hatnet}%
 \hfil
 \includegraphics[width={0.5\linewidth}]{\hatcurhtr{64}-lc}%
 }
 {
 \centering
 \leavevmode
 \includegraphics[width={0.5\linewidth}]{\hatcurhtr{64}-rv}%
 \hfil
 \includegraphics[width={0.5\linewidth}]{\hatcurhtr{64}-iso-bprp-gabs-isofeh-SED}%
 }                        
\caption{
	Observations of \hatcur{64} together with our best-fit model.
	Please refer to Figure~\ref{fig:hatp58} for a more detailed
	caption. The {\em TESS} light curve for this system is shown in
	Figure~\ref{fig:hatp64tess}.
\label{fig:hatp64}
}
\ifthenelse{\boolean{emulateapj}}{
    \end{figure*}
}{
    \end{figure}
}

\ifthenelse{\boolean{emulateapj}}{
    \begin{figure*}[!ht]
}{
    \begin{figure}[!ht]
}
 {
 \centering
 \leavevmode
 \includegraphics[width={1.0\linewidth}]{\hatcurhtr{59}-TESS}
}
\caption{
    Similar to Figure~\ref{fig:hatp58tess}, here we show the {\em TESS}
	long-cadence light curve for \hatcur{59}. Other observations
	included in our analysis of this system are shown in
	Figure~\ref{fig:hatp59}.
\label{fig:hatp59tess}
}
\ifthenelse{\boolean{emulateapj}}{
    \end{figure*}
}{
    \end{figure}
}

\ifthenelse{\boolean{emulateapj}}{
    \begin{figure*}[!ht]
}{
    \begin{figure}[!ht]
}
 {
 \centering
 \leavevmode
 \includegraphics[width={1.0\linewidth}]{\hatcurhtr{60}-TESS}
}
\caption{
    Similar to Figure~\ref{fig:hatp58tess}, here we show the {\em TESS}
	long-cadence light curve for \hatcur{60}. Other observations
	included in our analysis of this system are shown in
	Figure~\ref{fig:hatp60}.
\label{fig:hatp60tess}
}
\ifthenelse{\boolean{emulateapj}}{
    \end{figure*}
}{
    \end{figure}
}

\ifthenelse{\boolean{emulateapj}}{
    \begin{figure*}[!ht]
}{
    \begin{figure}[!ht]
}
 {
 \centering
 \leavevmode
 \includegraphics[width={1.0\linewidth}]{\hatcurhtr{61}-TESS}
}
\caption{
    Similar to Figure~\ref{fig:hatp58tess}, here we show the {\em TESS}
	long-cadence light curve for \hatcur{61}. Other observations
	included in our analysis of this system are shown in
	Figure~\ref{fig:hatp61}.
\label{fig:hatp61tess}
}
\ifthenelse{\boolean{emulateapj}}{
    \end{figure*}
}{
    \end{figure}
}

\ifthenelse{\boolean{emulateapj}}{
    \begin{figure*}[!ht]
}{
    \begin{figure}[!ht]
}
 {
 \centering
 \leavevmode
 \includegraphics[width={1.0\linewidth}]{\hatcurhtr{64}-TESS}
}
\caption{
    Similar to Figure~\ref{fig:hatp58tess}, here we show the {\em TESS}
	long-cadence light curve for \hatcur{64}. Other observations
	included in our analysis of this system are shown in
	Figure~\ref{fig:hatp64}.
\label{fig:hatp64tess}
}
\ifthenelse{\boolean{emulateapj}}{
    \end{figure*}
}{
    \end{figure}
}


\begin{figure*}[!ht]
{
\plottwo{\hatcurhtr{59}_f1a}{\hatcurhtr{59}_f1b}
}
\caption{
Same as Figure~\ref{fig:luckyimage58}, here we show the results for \hatcur{59}.
}
\label{fig:luckyimage59}
\end{figure*}

\begin{figure*}[!ht]
{
\plottwo{\hatcurhtr{60}_1a}{\hatcurhtr{60}_1b}
}
\caption{
	Same as Figure~\ref{fig:luckyimage58}, here we show the results for \hatcur{60}.
}
\label{fig:luckyimage60}
\end{figure*}

\begin{figure*}[!ht]
{
\plottwo{\hatcurhtr{61}_f1a}{\hatcurhtr{61}_f1b}
}
\caption{
	Same as Figure~\ref{fig:luckyimage58}, here we show the results for \hatcur{61}.
}
\label{fig:luckyimage61}
\end{figure*}

\begin{figure*}[!ht]
{
\plottwo{\hatcurhtr{62}_f1a}{\hatcurhtr{62}_f1b}
}
\caption{
	Same as Figure~\ref{fig:luckyimage58}, here we show the results for \hatcur{62}.
}
\label{fig:luckyimage62}
\end{figure*}

\begin{figure*}[!ht]
{
\plottwo{\hatcurhtr{64}_f1a}{\hatcurhtr{64}_f1b}
}
\caption{
	Same as Figure~\ref{fig:luckyimage58}, here we show the results for \hatcur{64}.
}
\label{fig:luckyimage64}
\end{figure*}
\clearpage

%
%
\startlongtable
\tabletypesize{\scriptsize}
\ifthenelse{\boolean{emulateapj}}{
    \begin{deluxetable*}{llrrrrrrl}
}{
    \begin{deluxetable}{llrrrrrrl}
}
\tablewidth{0pc}
\tablecaption{
    Relative radial velocities and bisector spans for \hatcur{58}--\hatcur{64}.
    \label{tab:rvs}
}
\tablehead{
\noalign{\vskip 3pt}
    \colhead{Star} &
    \colhead{BJD} &
    \colhead{RV\tablenotemark{a}} &
    \colhead{\ensuremath{\sigma_{\rm RV}}\tablenotemark{b}} &
    \colhead{BS} &
    \colhead{\ensuremath{\sigma_{\rm BS}}} &
    \colhead{S$_{\rm HK}$ \tablenotemark{c}} &
    \colhead{Phase} &
    \colhead{Instrument}\\
    \colhead{} &
    \colhead{\hbox{(2,450,000$+$)}} &
    \colhead{(\ms)} &
    \colhead{(\ms)} &
    \colhead{(\ms)} &
    \colhead{(\ms)} &
    \colhead{} &
    \colhead{} &
    \colhead{}
}
\startdata
\multicolumn{9}{c}{\bf HAT-P-58} \\
\hline\\
    \input{\hatcurhtr{58}_rvtable.tex}
\cutinhead{\bf HAT-P-59}
    \input{\hatcurhtr{59}_rvtable.tex}
\cutinhead{\bf HAT-P-60}
    \input{\hatcurhtr{60}_rvtable.tex}
\cutinhead{\bf HAT-P-61}
    \input{\hatcurhtr{61}_rvtable.tex}
\cutinhead{\bf HAT-P-62}
    \input{\hatcurhtr{62}_rvtable.tex}
\cutinhead{\bf HAT-P-63}
    \input{\hatcurhtr{63}_rvtable.tex}
\cutinhead{\bf HAT-P-64}
    \input{\hatcurhtr{64}_rvtable.tex}
\enddata
\tablenotetext{a}{
    The zero-point of these velocities is arbitrary. An overall offset
    $\gamma_{\rm rel}$ fitted independently to the velocities from
    each instrument has been subtracted.
}
\tablenotetext{b}{
    Internal errors excluding the component of astrophysical jitter
    considered in \refsecl{globmod}.
}
\tablenotetext{c}{
    Ca II HK line core emission index measured from the Keck-I/HIRES spectra
    following \citet{isaacson:2010}.
}
\ifthenelse{\boolean{rvtablelong}}{
}{
} 
\ifthenelse{\boolean{emulateapj}}{
    \end{deluxetable*}
}{
    \end{deluxetable}
}

\end{document}

%% file: newcommand_gen.tex
\newcommand{\forloop}[5][1]%
{%
\setcounter{#2}{#3}%
\ifthenelse{#4}%
	{%
	#5%
	\addtocounter{#2}{#1}%
	\forloop[#1]{#2}{\value{#2}}{#4}{#5}%
	}%
	{%
	}%
}%

\newcommand{\ctbd}[1]{}

\newcommand{\Lc}{Light curve}

\newcommand{\masy}{\ensuremath{\rm mas\,yr^{-1}}}
\newcommand{\kms}{\ensuremath{\rm km\,s^{-1}}}
\newcommand{\ms}{\ensuremath{\rm m\,s^{-1}}}

\newcommand{\gcmc}{\ensuremath{\rm g\,cm^{-3}}}

\newcommand{\C}{\ensuremath{^{\circ}C\;}}

\newcommand{\logg}{\ensuremath{\log{g}}}
\newcommand{\vsini}{\ensuremath{v \sin{i}}}
\newcommand{\feh}{\ensuremath{\rm [Fe/H]}}

\newcommand{\vmac}{\ensuremath{v_{\rm mac}}}
\newcommand{\vmic}{\ensuremath{v_{\rm mic}}}

\newcommand{\rsun}{\ensuremath{R_\sun}}
\newcommand{\msun}{\ensuremath{M_\sun}}
\newcommand{\lsun}{\ensuremath{L_\sun}}

\newcommand{\rstar}{\ensuremath{R_\star}}
\newcommand{\mstar}{\ensuremath{M_\star}}
\newcommand{\lstar}{\ensuremath{L_\star}}

\newcommand{\teffstar}{\ensuremath{T_{\rm eff\star}}}
\newcommand{\rhostar}{\ensuremath{\rho_\star}}
\newcommand{\loggstar}{\ensuremath{\log{g_{\star}}}}

\newcommand{\rpl}{\ensuremath{R_{p}}}
\newcommand{\mpl}{\ensuremath{M_{p}}}

\newcommand{\rhopl}{\ensuremath{\rho_{p}}}

\newcommand{\arstar}{\ensuremath{a/\rstar}}
\newcommand{\zrstar}{\ensuremath{\zeta/\rstar}}

\newcommand{\rjup}{\ensuremath{R_{\rm J}}}
\newcommand{\mjup}{\ensuremath{M_{\rm J}}}

\newcommand{\refsecl}[1]{\mbox{Section \ref{sec:#1}}}

\newcommand{\reftabl}[1]{Table~\ref{tab:#1}}



%% file: newcommand_spec_multi.tex
\input{planetinfo_multi.tex}
\input{planetinfo_multi_eccen.tex}
\input{planetinfo_multi_orig.tex}
\input{planetinfo_multi_eccen_orig.tex}
\input{newcommand_spec_HTR062-011.tex}
\input{newcommand_spec_HTR080-003.tex}
\input{newcommand_spec_HTR089-018.tex}
\input{newcommand_spec_HTR094-024.tex}
\input{newcommand_spec_HTR130-001.tex}
\input{newcommand_spec_HTR384-014.tex}
\input{newcommand_spec_HTR405-001.tex}
\newcommand{\hatcur}[1]{\ifnum#1=58 %
\hatcurxxxxxA
\else
\ifnum#1=59 %
\hatcurxxxxxB
\else
\ifnum#1=60 %
\hatcurxxxxxC
\else
\ifnum#1=61 %
\hatcurxxxxxD
\else
\ifnum#1=62 %
\hatcurxxxxxE
\else
\ifnum#1=63 %
\hatcurxxxxxF
\else
\ifnum#1=64 %
\hatcurxxxxxG
\else
??????\fi
\fi
\fi
\fi
\fi
\fi
\fi
}
\newcommand{\hatcurb}[1]{\ifnum#1=58 %
\hatcurbxxxxxA
\else
\ifnum#1=59 %
\hatcurbxxxxxB
\else
\ifnum#1=60 %
\hatcurbxxxxxC
\else
\ifnum#1=61 %
\hatcurbxxxxxD
\else
\ifnum#1=62 %
\hatcurbxxxxxE
\else
\ifnum#1=63 %
\hatcurbxxxxxF
\else
\ifnum#1=64 %
\hatcurbxxxxxG
\else
??????\fi
\fi
\fi
\fi
\fi
\fi
\fi
}
\newcommand{\hatcurc}[1]{\ifnum#1=58 %
\hatcurcxxxxxA
\else
\ifnum#1=59 %
\hatcurcxxxxxB
\else
\ifnum#1=60 %
\hatcurcxxxxxC
\else
\ifnum#1=61 %
\hatcurcxxxxxD
\else
\ifnum#1=62 %
\hatcurcxxxxxE
\else
\ifnum#1=63 %
\hatcurcxxxxxF
\else
\ifnum#1=64 %
\hatcurcxxxxxG
\else
??????\fi
\fi
\fi
\fi
\fi
\fi
\fi
}
\newcommand{\hatcurCCtassvi}[1]{\ifnum#1=58 %
\hatcurCCtassvixxxxxA
\else
\ifnum#1=59 %
\hatcurCCtassvixxxxxB
\else
\ifnum#1=60 %
\hatcurCCtassvixxxxxC
\else
\ifnum#1=61 %
\hatcurCCtassvixxxxxD
\else
\ifnum#1=62 %
\hatcurCCtassvixxxxxE
\else
\ifnum#1=63 %
\hatcurCCtassvixxxxxF
\else
\ifnum#1=64 %
\hatcurCCtassvixxxxxG
\else
??????\fi
\fi
\fi
\fi
\fi
\fi
\fi
}
\newcommand{\hatcurisocite}[1]{\ifnum#1=58 %
\hatcurisocitexxxxxA
\else
\ifnum#1=59 %
\hatcurisocitexxxxxB
\else
\ifnum#1=60 %
\hatcurisocitexxxxxC
\else
\ifnum#1=61 %
\hatcurisocitexxxxxD
\else
\ifnum#1=62 %
\hatcurisocitexxxxxE
\else
\ifnum#1=63 %
\hatcurisocitexxxxxF
\else
\ifnum#1=64 %
\hatcurisocitexxxxxG
\else
??????\fi
\fi
\fi
\fi
\fi
\fi
\fi
}
\newcommand{\hatcurisofull}[1]{\ifnum#1=58 %
\hatcurisofullxxxxxA
\else
\ifnum#1=59 %
\hatcurisofullxxxxxB
\else
\ifnum#1=60 %
\hatcurisofullxxxxxC
\else
\ifnum#1=61 %
\hatcurisofullxxxxxD
\else
\ifnum#1=62 %
\hatcurisofullxxxxxE
\else
\ifnum#1=63 %
\hatcurisofullxxxxxF
\else
\ifnum#1=64 %
\hatcurisofullxxxxxG
\else
??????\fi
\fi
\fi
\fi
\fi
\fi
\fi
}
\newcommand{\hatcurisoshort}[1]{\ifnum#1=58 %
\hatcurisoshortxxxxxA
\else
\ifnum#1=59 %
\hatcurisoshortxxxxxB
\else
\ifnum#1=60 %
\hatcurisoshortxxxxxC
\else
\ifnum#1=61 %
\hatcurisoshortxxxxxD
\else
\ifnum#1=62 %
\hatcurisoshortxxxxxE
\else
\ifnum#1=63 %
\hatcurisoshortxxxxxF
\else
\ifnum#1=64 %
\hatcurisoshortxxxxxG
\else
??????\fi
\fi
\fi
\fi
\fi
\fi
\fi
}
\newcommand{\hatcurjhkfilset}[1]{\ifnum#1=58 %
\hatcurjhkfilsetxxxxxA
\else
\ifnum#1=59 %
\hatcurjhkfilsetxxxxxB
\else
\ifnum#1=60 %
\hatcurjhkfilsetxxxxxC
\else
\ifnum#1=61 %
\hatcurjhkfilsetxxxxxD
\else
\ifnum#1=62 %
\hatcurjhkfilsetxxxxxE
\else
\ifnum#1=63 %
\hatcurjhkfilsetxxxxxF
\else
\ifnum#1=64 %
\hatcurjhkfilsetxxxxxG
\else
??????\fi
\fi
\fi
\fi
\fi
\fi
\fi
}
\newcommand{\hatcurlumind}[1]{\ifnum#1=58 %
\hatcurlumindxxxxxA
\else
\ifnum#1=59 %
\hatcurlumindxxxxxB
\else
\ifnum#1=60 %
\hatcurlumindxxxxxC
\else
\ifnum#1=61 %
\hatcurlumindxxxxxD
\else
\ifnum#1=62 %
\hatcurlumindxxxxxE
\else
\ifnum#1=63 %
\hatcurlumindxxxxxF
\else
\ifnum#1=64 %
\hatcurlumindxxxxxG
\else
??????\fi
\fi
\fi
\fi
\fi
\fi
\fi
}
\newcommand{\hatcurplanetnum}[1]{\ifnum#1=58 %
\hatcurplanetnumxxxxxA
\else
\ifnum#1=59 %
\hatcurplanetnumxxxxxB
\else
\ifnum#1=60 %
\hatcurplanetnumxxxxxC
\else
\ifnum#1=61 %
\hatcurplanetnumxxxxxD
\else
\ifnum#1=62 %
\hatcurplanetnumxxxxxE
\else
\ifnum#1=63 %
\hatcurplanetnumxxxxxF
\else
\ifnum#1=64 %
\hatcurplanetnumxxxxxG
\else
??????\fi
\fi
\fi
\fi
\fi
\fi
\fi
}
\newcommand{\hatcurRHKindex}[1]{\ifnum#1=58 %
\hatcurRHKindexxxxxxA
\else
\ifnum#1=60 %
\hatcurRHKindexxxxxxC
\else
\ifnum#1=61 %
\hatcurRHKindexxxxxxD
\else
\ifnum#1=64 %
\hatcurRHKindexxxxxxG
\else
??????\fi
\fi
\fi
\fi
}
\newcommand{\hatcurRVgammaabs}[1]{\ifnum#1=58 %
\hatcurRVgammaabsxxxxxA
\else
\ifnum#1=59 %
\hatcurRVgammaabsxxxxxB
\else
\ifnum#1=60 %
\hatcurRVgammaabsxxxxxC
\else
\ifnum#1=61 %
\hatcurRVgammaabsxxxxxD
\else
\ifnum#1=62 %
\hatcurRVgammaabsxxxxxE
\else
\ifnum#1=63 %
\hatcurRVgammaabsxxxxxF
\else
\ifnum#1=64 %
\hatcurRVgammaabsxxxxxG
\else
??????\fi
\fi
\fi
\fi
\fi
\fi
\fi
}
\newcommand{\hatcurRVgammainst}[1]{\ifnum#1=58 %
\hatcurRVgammainstxxxxxA
\else
\ifnum#1=59 %
\hatcurRVgammainstxxxxxB
\else
\ifnum#1=60 %
\hatcurRVgammainstxxxxxC
\else
\ifnum#1=61 %
\hatcurRVgammainstxxxxxD
\else
\ifnum#1=62 %
\hatcurRVgammainstxxxxxE
\else
\ifnum#1=63 %
\hatcurRVgammainstxxxxxF
\else
\ifnum#1=64 %
\hatcurRVgammainstxxxxxG
\else
??????\fi
\fi
\fi
\fi
\fi
\fi
\fi
}
\newcommand{\hatcurSindex}[1]{\ifnum#1=58 %
\hatcurSindexxxxxxA
\else
\ifnum#1=60 %
\hatcurSindexxxxxxC
\else
\ifnum#1=61 %
\hatcurSindexxxxxxD
\else
\ifnum#1=64 %
\hatcurSindexxxxxxG
\else
??????\fi
\fi
\fi
\fi
}
\newcommand{\hatcurSMElogg}[1]{\ifnum#1=58 %
\hatcurSMEloggxxxxxA
\else
\ifnum#1=59 %
\hatcurSMEloggxxxxxB
\else
\ifnum#1=60 %
\hatcurSMEloggxxxxxC
\else
\ifnum#1=61 %
\hatcurSMEloggxxxxxD
\else
\ifnum#1=62 %
\hatcurSMEloggxxxxxE
\else
\ifnum#1=63 %
\hatcurSMEloggxxxxxF
\else
\ifnum#1=64 %
\hatcurSMEloggxxxxxG
\else
??????\fi
\fi
\fi
\fi
\fi
\fi
\fi
}
\newcommand{\hatcurSMEteff}[1]{\ifnum#1=58 %
\hatcurSMEteffxxxxxA
\else
\ifnum#1=59 %
\hatcurSMEteffxxxxxB
\else
\ifnum#1=60 %
\hatcurSMEteffxxxxxC
\else
\ifnum#1=61 %
\hatcurSMEteffxxxxxD
\else
\ifnum#1=62 %
\hatcurSMEteffxxxxxE
\else
\ifnum#1=63 %
\hatcurSMEteffxxxxxF
\else
\ifnum#1=64 %
\hatcurSMEteffxxxxxG
\else
??????\fi
\fi
\fi
\fi
\fi
\fi
\fi
}
\newcommand{\hatcurSMEversion}[1]{\ifnum#1=58 %
\hatcurSMEversionxxxxxA
\else
\ifnum#1=59 %
\hatcurSMEversionxxxxxB
\else
\ifnum#1=60 %
\hatcurSMEversionxxxxxC
\else
\ifnum#1=61 %
\hatcurSMEversionxxxxxD
\else
\ifnum#1=62 %
\hatcurSMEversionxxxxxE
\else
\ifnum#1=63 %
\hatcurSMEversionxxxxxF
\else
\ifnum#1=64 %
\hatcurSMEversionxxxxxG
\else
??????\fi
\fi
\fi
\fi
\fi
\fi
\fi
}
\newcommand{\hatcurSMEvmac}[1]{\ifnum#1=58 %
\hatcurSMEvmacxxxxxA
\else
\ifnum#1=59 %
\hatcurSMEvmacxxxxxB
\else
\ifnum#1=60 %
\hatcurSMEvmacxxxxxC
\else
\ifnum#1=61 %
\hatcurSMEvmacxxxxxD
\else
\ifnum#1=62 %
\hatcurSMEvmacxxxxxE
\else
\ifnum#1=63 %
\hatcurSMEvmacxxxxxF
\else
\ifnum#1=64 %
\hatcurSMEvmacxxxxxG
\else
??????\fi
\fi
\fi
\fi
\fi
\fi
\fi
}
\newcommand{\hatcurSMEvmic}[1]{\ifnum#1=58 %
\hatcurSMEvmicxxxxxA
\else
\ifnum#1=59 %
\hatcurSMEvmicxxxxxB
\else
\ifnum#1=60 %
\hatcurSMEvmicxxxxxC
\else
\ifnum#1=61 %
\hatcurSMEvmicxxxxxD
\else
\ifnum#1=62 %
\hatcurSMEvmicxxxxxE
\else
\ifnum#1=63 %
\hatcurSMEvmicxxxxxF
\else
\ifnum#1=64 %
\hatcurSMEvmicxxxxxG
\else
??????\fi
\fi
\fi
\fi
\fi
\fi
\fi
}
\newcommand{\hatcurSMEvsin}[1]{\ifnum#1=58 %
\hatcurSMEvsinxxxxxA
\else
\ifnum#1=59 %
\hatcurSMEvsinxxxxxB
\else
\ifnum#1=60 %
\hatcurSMEvsinxxxxxC
\else
\ifnum#1=61 %
\hatcurSMEvsinxxxxxD
\else
\ifnum#1=62 %
\hatcurSMEvsinxxxxxE
\else
\ifnum#1=63 %
\hatcurSMEvsinxxxxxF
\else
\ifnum#1=64 %
\hatcurSMEvsinxxxxxG
\else
??????\fi
\fi
\fi
\fi
\fi
\fi
\fi
}
\newcommand{\hatcurSMEzfeh}[1]{\ifnum#1=58 %
\hatcurSMEzfehxxxxxA
\else
\ifnum#1=59 %
\hatcurSMEzfehxxxxxB
\else
\ifnum#1=60 %
\hatcurSMEzfehxxxxxC
\else
\ifnum#1=61 %
\hatcurSMEzfehxxxxxD
\else
\ifnum#1=62 %
\hatcurSMEzfehxxxxxE
\else
\ifnum#1=63 %
\hatcurSMEzfehxxxxxF
\else
\ifnum#1=64 %
\hatcurSMEzfehxxxxxG
\else
??????\fi
\fi
\fi
\fi
\fi
\fi
\fi
}
\newcommand{\hatcurSMEzfehshort}[1]{\ifnum#1=58 %
\hatcurSMEzfehshortxxxxxA
\else
\ifnum#1=59 %
\hatcurSMEzfehshortxxxxxB
\else
\ifnum#1=60 %
\hatcurSMEzfehshortxxxxxC
\else
\ifnum#1=61 %
\hatcurSMEzfehshortxxxxxD
\else
\ifnum#1=62 %
\hatcurSMEzfehshortxxxxxE
\else
\ifnum#1=63 %
\hatcurSMEzfehshortxxxxxF
\else
\ifnum#1=64 %
\hatcurSMEzfehshortxxxxxG
\else
??????\fi
\fi
\fi
\fi
\fi
\fi
\fi
}
\newcommand{\hatcurtwomassshort}[1]{\ifnum#1=58 %
\hatcurtwomassshortxxxxxA
\else
\ifnum#1=59 %
\hatcurtwomassshortxxxxxB
\else
\ifnum#1=60 %
\hatcurtwomassshortxxxxxC
\else
\ifnum#1=61 %
\hatcurtwomassshortxxxxxD
\else
\ifnum#1=62 %
\hatcurtwomassshortxxxxxE
\else
\ifnum#1=63 %
\hatcurtwomassshortxxxxxF
\else
\ifnum#1=64 %
\hatcurtwomassshortxxxxxG
\else
??????\fi
\fi
\fi
\fi
\fi
\fi
\fi
}

%% file: planetinfo_multi.tex
\input{HTR062-011.tex}
\input{HTR080-003.tex}
\input{HTR089-018.tex}
\input{HTR094-024.tex}
\input{HTR130-001.tex}
\input{HTR384-014.tex}
\input{HTR405-001.tex}
\newcommand{\hatcurCCbbHmag}[1]{\ifnum#1=58 %
\hatcurCCbbHmagxxxxxA
\else
\ifnum#1=59 %
\hatcurCCbbHmagxxxxxB
\else
\ifnum#1=60 %
\hatcurCCbbHmagxxxxxC
\else
\ifnum#1=61 %
\hatcurCCbbHmagxxxxxD
\else
\ifnum#1=62 %
\hatcurCCbbHmagxxxxxE
\else
\ifnum#1=63 %
\hatcurCCbbHmagxxxxxF
\else
\ifnum#1=64 %
\hatcurCCbbHmagxxxxxG
\else
??????\fi
\fi
\fi
\fi
\fi
\fi
\fi
}
\newcommand{\hatcurCCbbJmag}[1]{\ifnum#1=58 %
\hatcurCCbbJmagxxxxxA
\else
\ifnum#1=59 %
\hatcurCCbbJmagxxxxxB
\else
\ifnum#1=60 %
\hatcurCCbbJmagxxxxxC
\else
\ifnum#1=61 %
\hatcurCCbbJmagxxxxxD
\else
\ifnum#1=62 %
\hatcurCCbbJmagxxxxxE
\else
\ifnum#1=63 %
\hatcurCCbbJmagxxxxxF
\else
\ifnum#1=64 %
\hatcurCCbbJmagxxxxxG
\else
??????\fi
\fi
\fi
\fi
\fi
\fi
\fi
}
\newcommand{\hatcurCCbbKmag}[1]{\ifnum#1=58 %
\hatcurCCbbKmagxxxxxA
\else
\ifnum#1=59 %
\hatcurCCbbKmagxxxxxB
\else
\ifnum#1=60 %
\hatcurCCbbKmagxxxxxC
\else
\ifnum#1=61 %
\hatcurCCbbKmagxxxxxD
\else
\ifnum#1=62 %
\hatcurCCbbKmagxxxxxE
\else
\ifnum#1=63 %
\hatcurCCbbKmagxxxxxF
\else
\ifnum#1=64 %
\hatcurCCbbKmagxxxxxG
\else
??????\fi
\fi
\fi
\fi
\fi
\fi
\fi
}
\newcommand{\hatcurCCcitHmag}[1]{\ifnum#1=58 %
\hatcurCCcitHmagxxxxxA
\else
\ifnum#1=59 %
\hatcurCCcitHmagxxxxxB
\else
\ifnum#1=60 %
\hatcurCCcitHmagxxxxxC
\else
\ifnum#1=61 %
\hatcurCCcitHmagxxxxxD
\else
\ifnum#1=62 %
\hatcurCCcitHmagxxxxxE
\else
\ifnum#1=63 %
\hatcurCCcitHmagxxxxxF
\else
\ifnum#1=64 %
\hatcurCCcitHmagxxxxxG
\else
??????\fi
\fi
\fi
\fi
\fi
\fi
\fi
}
\newcommand{\hatcurCCcitJmag}[1]{\ifnum#1=58 %
\hatcurCCcitJmagxxxxxA
\else
\ifnum#1=59 %
\hatcurCCcitJmagxxxxxB
\else
\ifnum#1=60 %
\hatcurCCcitJmagxxxxxC
\else
\ifnum#1=61 %
\hatcurCCcitJmagxxxxxD
\else
\ifnum#1=62 %
\hatcurCCcitJmagxxxxxE
\else
\ifnum#1=63 %
\hatcurCCcitJmagxxxxxF
\else
\ifnum#1=64 %
\hatcurCCcitJmagxxxxxG
\else
??????\fi
\fi
\fi
\fi
\fi
\fi
\fi
}
\newcommand{\hatcurCCcitKmag}[1]{\ifnum#1=58 %
\hatcurCCcitKmagxxxxxA
\else
\ifnum#1=59 %
\hatcurCCcitKmagxxxxxB
\else
\ifnum#1=60 %
\hatcurCCcitKmagxxxxxC
\else
\ifnum#1=61 %
\hatcurCCcitKmagxxxxxD
\else
\ifnum#1=62 %
\hatcurCCcitKmagxxxxxE
\else
\ifnum#1=63 %
\hatcurCCcitKmagxxxxxF
\else
\ifnum#1=64 %
\hatcurCCcitKmagxxxxxG
\else
??????\fi
\fi
\fi
\fi
\fi
\fi
\fi
}
\newcommand{\hatcurCCdec}[1]{\ifnum#1=58 %
\hatcurCCdecxxxxxA
\else
\ifnum#1=59 %
\hatcurCCdecxxxxxB
\else
\ifnum#1=60 %
\hatcurCCdecxxxxxC
\else
\ifnum#1=61 %
\hatcurCCdecxxxxxD
\else
\ifnum#1=62 %
\hatcurCCdecxxxxxE
\else
\ifnum#1=63 %
\hatcurCCdecxxxxxF
\else
\ifnum#1=64 %
\hatcurCCdecxxxxxG
\else
??????\fi
\fi
\fi
\fi
\fi
\fi
\fi
}
\newcommand{\hatcurCCesoHKmag}[1]{\ifnum#1=58 %
\hatcurCCesoHKmagxxxxxA
\else
\ifnum#1=59 %
\hatcurCCesoHKmagxxxxxB
\else
\ifnum#1=60 %
\hatcurCCesoHKmagxxxxxC
\else
\ifnum#1=61 %
\hatcurCCesoHKmagxxxxxD
\else
\ifnum#1=62 %
\hatcurCCesoHKmagxxxxxE
\else
\ifnum#1=63 %
\hatcurCCesoHKmagxxxxxF
\else
\ifnum#1=64 %
\hatcurCCesoHKmagxxxxxG
\else
??????\fi
\fi
\fi
\fi
\fi
\fi
\fi
}
\newcommand{\hatcurCCesoHmag}[1]{\ifnum#1=58 %
\hatcurCCesoHmagxxxxxA
\else
\ifnum#1=59 %
\hatcurCCesoHmagxxxxxB
\else
\ifnum#1=60 %
\hatcurCCesoHmagxxxxxC
\else
\ifnum#1=61 %
\hatcurCCesoHmagxxxxxD
\else
\ifnum#1=62 %
\hatcurCCesoHmagxxxxxE
\else
\ifnum#1=63 %
\hatcurCCesoHmagxxxxxF
\else
\ifnum#1=64 %
\hatcurCCesoHmagxxxxxG
\else
??????\fi
\fi
\fi
\fi
\fi
\fi
\fi
}
\newcommand{\hatcurCCesoJHmag}[1]{\ifnum#1=58 %
\hatcurCCesoJHmagxxxxxA
\else
\ifnum#1=59 %
\hatcurCCesoJHmagxxxxxB
\else
\ifnum#1=60 %
\hatcurCCesoJHmagxxxxxC
\else
\ifnum#1=61 %
\hatcurCCesoJHmagxxxxxD
\else
\ifnum#1=62 %
\hatcurCCesoJHmagxxxxxE
\else
\ifnum#1=63 %
\hatcurCCesoJHmagxxxxxF
\else
\ifnum#1=64 %
\hatcurCCesoJHmagxxxxxG
\else
??????\fi
\fi
\fi
\fi
\fi
\fi
\fi
}
\newcommand{\hatcurCCesoJKmag}[1]{\ifnum#1=58 %
\hatcurCCesoJKmagxxxxxA
\else
\ifnum#1=59 %
\hatcurCCesoJKmagxxxxxB
\else
\ifnum#1=60 %
\hatcurCCesoJKmagxxxxxC
\else
\ifnum#1=61 %
\hatcurCCesoJKmagxxxxxD
\else
\ifnum#1=62 %
\hatcurCCesoJKmagxxxxxE
\else
\ifnum#1=63 %
\hatcurCCesoJKmagxxxxxF
\else
\ifnum#1=64 %
\hatcurCCesoJKmagxxxxxG
\else
??????\fi
\fi
\fi
\fi
\fi
\fi
\fi
}
\newcommand{\hatcurCCesoJmag}[1]{\ifnum#1=58 %
\hatcurCCesoJmagxxxxxA
\else
\ifnum#1=59 %
\hatcurCCesoJmagxxxxxB
\else
\ifnum#1=60 %
\hatcurCCesoJmagxxxxxC
\else
\ifnum#1=61 %
\hatcurCCesoJmagxxxxxD
\else
\ifnum#1=62 %
\hatcurCCesoJmagxxxxxE
\else
\ifnum#1=63 %
\hatcurCCesoJmagxxxxxF
\else
\ifnum#1=64 %
\hatcurCCesoJmagxxxxxG
\else
??????\fi
\fi
\fi
\fi
\fi
\fi
\fi
}
\newcommand{\hatcurCCesoKmag}[1]{\ifnum#1=58 %
\hatcurCCesoKmagxxxxxA
\else
\ifnum#1=59 %
\hatcurCCesoKmagxxxxxB
\else
\ifnum#1=60 %
\hatcurCCesoKmagxxxxxC
\else
\ifnum#1=61 %
\hatcurCCesoKmagxxxxxD
\else
\ifnum#1=62 %
\hatcurCCesoKmagxxxxxE
\else
\ifnum#1=63 %
\hatcurCCesoKmagxxxxxF
\else
\ifnum#1=64 %
\hatcurCCesoKmagxxxxxG
\else
??????\fi
\fi
\fi
\fi
\fi
\fi
\fi
}
\newcommand{\hatcurCCgaia}[1]{\ifnum#1=58 %
\hatcurCCgaiaxxxxxA
\else
\ifnum#1=59 %
\hatcurCCgaiaxxxxxB
\else
\ifnum#1=60 %
\hatcurCCgaiaxxxxxC
\else
\ifnum#1=61 %
\hatcurCCgaiaxxxxxD
\else
\ifnum#1=62 %
\hatcurCCgaiaxxxxxE
\else
\ifnum#1=63 %
\hatcurCCgaiaxxxxxF
\else
\ifnum#1=64 %
\hatcurCCgaiaxxxxxG
\else
??????\fi
\fi
\fi
\fi
\fi
\fi
\fi
}
\newcommand{\hatcurCCgaiadrtwo}[1]{\ifnum#1=58 %
\hatcurCCgaiadrtwoxxxxxA
\else
\ifnum#1=59 %
\hatcurCCgaiadrtwoxxxxxB
\else
\ifnum#1=60 %
\hatcurCCgaiadrtwoxxxxxC
\else
\ifnum#1=61 %
\hatcurCCgaiadrtwoxxxxxD
\else
\ifnum#1=62 %
\hatcurCCgaiadrtwoxxxxxE
\else
\ifnum#1=63 %
\hatcurCCgaiadrtwoxxxxxF
\else
\ifnum#1=64 %
\hatcurCCgaiadrtwoxxxxxG
\else
??????\fi
\fi
\fi
\fi
\fi
\fi
\fi
}
\newcommand{\hatcurCCgaiamBP}[1]{\ifnum#1=58 %
\hatcurCCgaiamBPxxxxxA
\else
\ifnum#1=59 %
\hatcurCCgaiamBPxxxxxB
\else
\ifnum#1=60 %
\hatcurCCgaiamBPxxxxxC
\else
\ifnum#1=61 %
\hatcurCCgaiamBPxxxxxD
\else
\ifnum#1=62 %
\hatcurCCgaiamBPxxxxxE
\else
\ifnum#1=63 %
\hatcurCCgaiamBPxxxxxF
\else
\ifnum#1=64 %
\hatcurCCgaiamBPxxxxxG
\else
??????\fi
\fi
\fi
\fi
\fi
\fi
\fi
}
\newcommand{\hatcurCCgaiamG}[1]{\ifnum#1=58 %
\hatcurCCgaiamGxxxxxA
\else
\ifnum#1=59 %
\hatcurCCgaiamGxxxxxB
\else
\ifnum#1=60 %
\hatcurCCgaiamGxxxxxC
\else
\ifnum#1=61 %
\hatcurCCgaiamGxxxxxD
\else
\ifnum#1=62 %
\hatcurCCgaiamGxxxxxE
\else
\ifnum#1=63 %
\hatcurCCgaiamGxxxxxF
\else
\ifnum#1=64 %
\hatcurCCgaiamGxxxxxG
\else
??????\fi
\fi
\fi
\fi
\fi
\fi
\fi
}
\newcommand{\hatcurCCgaiamRP}[1]{\ifnum#1=58 %
\hatcurCCgaiamRPxxxxxA
\else
\ifnum#1=59 %
\hatcurCCgaiamRPxxxxxB
\else
\ifnum#1=60 %
\hatcurCCgaiamRPxxxxxC
\else
\ifnum#1=61 %
\hatcurCCgaiamRPxxxxxD
\else
\ifnum#1=62 %
\hatcurCCgaiamRPxxxxxE
\else
\ifnum#1=63 %
\hatcurCCgaiamRPxxxxxF
\else
\ifnum#1=64 %
\hatcurCCgaiamRPxxxxxG
\else
??????\fi
\fi
\fi
\fi
\fi
\fi
\fi
}
\newcommand{\hatcurCCgsc}[1]{\ifnum#1=58 %
\hatcurCCgscxxxxxA
\else
\ifnum#1=59 %
\hatcurCCgscxxxxxB
\else
\ifnum#1=60 %
\hatcurCCgscxxxxxC
\else
\ifnum#1=61 %
\hatcurCCgscxxxxxD
\else
\ifnum#1=62 %
\hatcurCCgscxxxxxE
\else
\ifnum#1=63 %
\hatcurCCgscxxxxxF
\else
\ifnum#1=64 %
\hatcurCCgscxxxxxG
\else
??????\fi
\fi
\fi
\fi
\fi
\fi
\fi
}
\newcommand{\hatcurCCmag}[1]{\ifnum#1=58 %
\hatcurCCmagxxxxxA
\else
\ifnum#1=59 %
\hatcurCCmagxxxxxB
\else
\ifnum#1=60 %
\hatcurCCmagxxxxxC
\else
\ifnum#1=61 %
\hatcurCCmagxxxxxD
\else
\ifnum#1=62 %
\hatcurCCmagxxxxxE
\else
\ifnum#1=63 %
\hatcurCCmagxxxxxF
\else
\ifnum#1=64 %
\hatcurCCmagxxxxxG
\else
??????\fi
\fi
\fi
\fi
\fi
\fi
\fi
}
\newcommand{\hatcurCCparallax}[1]{\ifnum#1=58 %
\hatcurCCparallaxxxxxxA
\else
\ifnum#1=59 %
\hatcurCCparallaxxxxxxB
\else
\ifnum#1=60 %
\hatcurCCparallaxxxxxxC
\else
\ifnum#1=61 %
\hatcurCCparallaxxxxxxD
\else
\ifnum#1=62 %
\hatcurCCparallaxxxxxxE
\else
\ifnum#1=63 %
\hatcurCCparallaxxxxxxF
\else
\ifnum#1=64 %
\hatcurCCparallaxxxxxxG
\else
??????\fi
\fi
\fi
\fi
\fi
\fi
\fi
}
\newcommand{\hatcurCCpm}[1]{\ifnum#1=58 %
\hatcurCCpmxxxxxA
\else
\ifnum#1=59 %
\hatcurCCpmxxxxxB
\else
\ifnum#1=60 %
\hatcurCCpmxxxxxC
\else
\ifnum#1=61 %
\hatcurCCpmxxxxxD
\else
\ifnum#1=62 %
\hatcurCCpmxxxxxE
\else
\ifnum#1=63 %
\hatcurCCpmxxxxxF
\else
\ifnum#1=64 %
\hatcurCCpmxxxxxG
\else
??????\fi
\fi
\fi
\fi
\fi
\fi
\fi
}
\newcommand{\hatcurCCpmdec}[1]{\ifnum#1=58 %
\hatcurCCpmdecxxxxxA
\else
\ifnum#1=59 %
\hatcurCCpmdecxxxxxB
\else
\ifnum#1=60 %
\hatcurCCpmdecxxxxxC
\else
\ifnum#1=61 %
\hatcurCCpmdecxxxxxD
\else
\ifnum#1=62 %
\hatcurCCpmdecxxxxxE
\else
\ifnum#1=63 %
\hatcurCCpmdecxxxxxF
\else
\ifnum#1=64 %
\hatcurCCpmdecxxxxxG
\else
??????\fi
\fi
\fi
\fi
\fi
\fi
\fi
}
\newcommand{\hatcurCCpmra}[1]{\ifnum#1=58 %
\hatcurCCpmraxxxxxA
\else
\ifnum#1=59 %
\hatcurCCpmraxxxxxB
\else
\ifnum#1=60 %
\hatcurCCpmraxxxxxC
\else
\ifnum#1=61 %
\hatcurCCpmraxxxxxD
\else
\ifnum#1=62 %
\hatcurCCpmraxxxxxE
\else
\ifnum#1=63 %
\hatcurCCpmraxxxxxF
\else
\ifnum#1=64 %
\hatcurCCpmraxxxxxG
\else
??????\fi
\fi
\fi
\fi
\fi
\fi
\fi
}
\newcommand{\hatcurCCra}[1]{\ifnum#1=58 %
\hatcurCCraxxxxxA
\else
\ifnum#1=59 %
\hatcurCCraxxxxxB
\else
\ifnum#1=60 %
\hatcurCCraxxxxxC
\else
\ifnum#1=61 %
\hatcurCCraxxxxxD
\else
\ifnum#1=62 %
\hatcurCCraxxxxxE
\else
\ifnum#1=63 %
\hatcurCCraxxxxxF
\else
\ifnum#1=64 %
\hatcurCCraxxxxxG
\else
??????\fi
\fi
\fi
\fi
\fi
\fi
\fi
}
\newcommand{\hatcurCCtassmB}[1]{\ifnum#1=58 %
\hatcurCCtassmBxxxxxA
\else
\ifnum#1=59 %
\hatcurCCtassmBxxxxxB
\else
\ifnum#1=60 %
\hatcurCCtassmBxxxxxC
\else
\ifnum#1=61 %
\hatcurCCtassmBxxxxxD
\else
\ifnum#1=62 %
\hatcurCCtassmBxxxxxE
\else
\ifnum#1=63 %
\hatcurCCtassmBxxxxxF
\else
\ifnum#1=64 %
\hatcurCCtassmBxxxxxG
\else
??????\fi
\fi
\fi
\fi
\fi
\fi
\fi
}
\newcommand{\hatcurCCtassmBshort}[1]{\ifnum#1=58 %
\hatcurCCtassmBshortxxxxxA
\else
\ifnum#1=59 %
\hatcurCCtassmBshortxxxxxB
\else
\ifnum#1=60 %
\hatcurCCtassmBshortxxxxxC
\else
\ifnum#1=61 %
\hatcurCCtassmBshortxxxxxD
\else
\ifnum#1=62 %
\hatcurCCtassmBshortxxxxxE
\else
\ifnum#1=63 %
\hatcurCCtassmBshortxxxxxF
\else
\ifnum#1=64 %
\hatcurCCtassmBshortxxxxxG
\else
??????\fi
\fi
\fi
\fi
\fi
\fi
\fi
}
\newcommand{\hatcurCCtassmg}[1]{\ifnum#1=58 %
\hatcurCCtassmgxxxxxA
\else
\ifnum#1=59 %
\hatcurCCtassmgxxxxxB
\else
\ifnum#1=60 %
\hatcurCCtassmgxxxxxC
\else
\ifnum#1=61 %
\hatcurCCtassmgxxxxxD
\else
\ifnum#1=62 %
\hatcurCCtassmgxxxxxE
\else
\ifnum#1=63 %
\hatcurCCtassmgxxxxxF
\else
\ifnum#1=64 %
\hatcurCCtassmgxxxxxG
\else
??????\fi
\fi
\fi
\fi
\fi
\fi
\fi
}
\newcommand{\hatcurCCtassmgshort}[1]{\ifnum#1=58 %
\hatcurCCtassmgshortxxxxxA
\else
\ifnum#1=59 %
\hatcurCCtassmgshortxxxxxB
\else
\ifnum#1=60 %
\hatcurCCtassmgshortxxxxxC
\else
\ifnum#1=61 %
\hatcurCCtassmgshortxxxxxD
\else
\ifnum#1=62 %
\hatcurCCtassmgshortxxxxxE
\else
\ifnum#1=63 %
\hatcurCCtassmgshortxxxxxF
\else
\ifnum#1=64 %
\hatcurCCtassmgshortxxxxxG
\else
??????\fi
\fi
\fi
\fi
\fi
\fi
\fi
}
\newcommand{\hatcurCCtassmi}[1]{\ifnum#1=58 %
\hatcurCCtassmixxxxxA
\else
\ifnum#1=59 %
\hatcurCCtassmixxxxxB
\else
\ifnum#1=60 %
\hatcurCCtassmixxxxxC
\else
\ifnum#1=61 %
\hatcurCCtassmixxxxxD
\else
\ifnum#1=62 %
\hatcurCCtassmixxxxxE
\else
\ifnum#1=63 %
\hatcurCCtassmixxxxxF
\else
\ifnum#1=64 %
\hatcurCCtassmixxxxxG
\else
??????\fi
\fi
\fi
\fi
\fi
\fi
\fi
}
\newcommand{\hatcurCCtassmI}[1]{\ifnum#1=58 %
\hatcurCCtassmIxxxxxA
\else
\ifnum#1=59 %
\hatcurCCtassmIxxxxxB
\else
\ifnum#1=60 %
\hatcurCCtassmIxxxxxC
\else
\ifnum#1=61 %
\hatcurCCtassmIxxxxxD
\else
\ifnum#1=62 %
\hatcurCCtassmIxxxxxE
\else
\ifnum#1=63 %
\hatcurCCtassmIxxxxxF
\else
\ifnum#1=64 %
\hatcurCCtassmIxxxxxG
\else
??????\fi
\fi
\fi
\fi
\fi
\fi
\fi
}
\newcommand{\hatcurCCtassmishort}[1]{\ifnum#1=58 %
\hatcurCCtassmishortxxxxxA
\else
\ifnum#1=59 %
\hatcurCCtassmishortxxxxxB
\else
\ifnum#1=60 %
\hatcurCCtassmishortxxxxxC
\else
\ifnum#1=61 %
\hatcurCCtassmishortxxxxxD
\else
\ifnum#1=62 %
\hatcurCCtassmishortxxxxxE
\else
\ifnum#1=63 %
\hatcurCCtassmishortxxxxxF
\else
\ifnum#1=64 %
\hatcurCCtassmishortxxxxxG
\else
??????\fi
\fi
\fi
\fi
\fi
\fi
\fi
}
\newcommand{\hatcurCCtassmIshort}[1]{\ifnum#1=58 %
\hatcurCCtassmIshortxxxxxA
\else
\ifnum#1=59 %
\hatcurCCtassmIshortxxxxxB
\else
\ifnum#1=60 %
\hatcurCCtassmIshortxxxxxC
\else
\ifnum#1=61 %
\hatcurCCtassmIshortxxxxxD
\else
\ifnum#1=62 %
\hatcurCCtassmIshortxxxxxE
\else
\ifnum#1=63 %
\hatcurCCtassmIshortxxxxxF
\else
\ifnum#1=64 %
\hatcurCCtassmIshortxxxxxG
\else
??????\fi
\fi
\fi
\fi
\fi
\fi
\fi
}
\newcommand{\hatcurCCtassmr}[1]{\ifnum#1=58 %
\hatcurCCtassmrxxxxxA
\else
\ifnum#1=59 %
\hatcurCCtassmrxxxxxB
\else
\ifnum#1=60 %
\hatcurCCtassmrxxxxxC
\else
\ifnum#1=61 %
\hatcurCCtassmrxxxxxD
\else
\ifnum#1=62 %
\hatcurCCtassmrxxxxxE
\else
\ifnum#1=63 %
\hatcurCCtassmrxxxxxF
\else
\ifnum#1=64 %
\hatcurCCtassmrxxxxxG
\else
??????\fi
\fi
\fi
\fi
\fi
\fi
\fi
}
\newcommand{\hatcurCCtassmrshort}[1]{\ifnum#1=58 %
\hatcurCCtassmrshortxxxxxA
\else
\ifnum#1=59 %
\hatcurCCtassmrshortxxxxxB
\else
\ifnum#1=60 %
\hatcurCCtassmrshortxxxxxC
\else
\ifnum#1=61 %
\hatcurCCtassmrshortxxxxxD
\else
\ifnum#1=62 %
\hatcurCCtassmrshortxxxxxE
\else
\ifnum#1=63 %
\hatcurCCtassmrshortxxxxxF
\else
\ifnum#1=64 %
\hatcurCCtassmrshortxxxxxG
\else
??????\fi
\fi
\fi
\fi
\fi
\fi
\fi
}
\newcommand{\hatcurCCtassmv}[1]{\ifnum#1=58 %
\hatcurCCtassmvxxxxxA
\else
\ifnum#1=59 %
\hatcurCCtassmvxxxxxB
\else
\ifnum#1=60 %
\hatcurCCtassmvxxxxxC
\else
\ifnum#1=61 %
\hatcurCCtassmvxxxxxD
\else
\ifnum#1=62 %
\hatcurCCtassmvxxxxxE
\else
\ifnum#1=63 %
\hatcurCCtassmvxxxxxF
\else
\ifnum#1=64 %
\hatcurCCtassmvxxxxxG
\else
??????\fi
\fi
\fi
\fi
\fi
\fi
\fi
}
\newcommand{\hatcurCCtassmvshort}[1]{\ifnum#1=58 %
\hatcurCCtassmvshortxxxxxA
\else
\ifnum#1=59 %
\hatcurCCtassmvshortxxxxxB
\else
\ifnum#1=60 %
\hatcurCCtassmvshortxxxxxC
\else
\ifnum#1=61 %
\hatcurCCtassmvshortxxxxxD
\else
\ifnum#1=62 %
\hatcurCCtassmvshortxxxxxE
\else
\ifnum#1=63 %
\hatcurCCtassmvshortxxxxxF
\else
\ifnum#1=64 %
\hatcurCCtassmvshortxxxxxG
\else
??????\fi
\fi
\fi
\fi
\fi
\fi
\fi
}
\newcommand{\hatcurCCtwomass}[1]{\ifnum#1=58 %
\hatcurCCtwomassxxxxxA
\else
\ifnum#1=59 %
\hatcurCCtwomassxxxxxB
\else
\ifnum#1=60 %
\hatcurCCtwomassxxxxxC
\else
\ifnum#1=61 %
\hatcurCCtwomassxxxxxD
\else
\ifnum#1=62 %
\hatcurCCtwomassxxxxxE
\else
\ifnum#1=63 %
\hatcurCCtwomassxxxxxF
\else
\ifnum#1=64 %
\hatcurCCtwomassxxxxxG
\else
??????\fi
\fi
\fi
\fi
\fi
\fi
\fi
}
\newcommand{\hatcurCCtwomassHmag}[1]{\ifnum#1=58 %
\hatcurCCtwomassHmagxxxxxA
\else
\ifnum#1=59 %
\hatcurCCtwomassHmagxxxxxB
\else
\ifnum#1=60 %
\hatcurCCtwomassHmagxxxxxC
\else
\ifnum#1=61 %
\hatcurCCtwomassHmagxxxxxD
\else
\ifnum#1=62 %
\hatcurCCtwomassHmagxxxxxE
\else
\ifnum#1=63 %
\hatcurCCtwomassHmagxxxxxF
\else
\ifnum#1=64 %
\hatcurCCtwomassHmagxxxxxG
\else
??????\fi
\fi
\fi
\fi
\fi
\fi
\fi
}
\newcommand{\hatcurCCtwomassJmag}[1]{\ifnum#1=58 %
\hatcurCCtwomassJmagxxxxxA
\else
\ifnum#1=59 %
\hatcurCCtwomassJmagxxxxxB
\else
\ifnum#1=60 %
\hatcurCCtwomassJmagxxxxxC
\else
\ifnum#1=61 %
\hatcurCCtwomassJmagxxxxxD
\else
\ifnum#1=62 %
\hatcurCCtwomassJmagxxxxxE
\else
\ifnum#1=63 %
\hatcurCCtwomassJmagxxxxxF
\else
\ifnum#1=64 %
\hatcurCCtwomassJmagxxxxxG
\else
??????\fi
\fi
\fi
\fi
\fi
\fi
\fi
}
\newcommand{\hatcurCCtwomassKmag}[1]{\ifnum#1=58 %
\hatcurCCtwomassKmagxxxxxA
\else
\ifnum#1=59 %
\hatcurCCtwomassKmagxxxxxB
\else
\ifnum#1=60 %
\hatcurCCtwomassKmagxxxxxC
\else
\ifnum#1=61 %
\hatcurCCtwomassKmagxxxxxD
\else
\ifnum#1=62 %
\hatcurCCtwomassKmagxxxxxE
\else
\ifnum#1=63 %
\hatcurCCtwomassKmagxxxxxF
\else
\ifnum#1=64 %
\hatcurCCtwomassKmagxxxxxG
\else
??????\fi
\fi
\fi
\fi
\fi
\fi
\fi
}
\newcommand{\hatcurCCWfourmag}[1]{\ifnum#1=58 %
\hatcurCCWfourmagxxxxxA
\else
\ifnum#1=59 %
\hatcurCCWfourmagxxxxxB
\else
\ifnum#1=60 %
\hatcurCCWfourmagxxxxxC
\else
\ifnum#1=61 %
\hatcurCCWfourmagxxxxxD
\else
\ifnum#1=62 %
\hatcurCCWfourmagxxxxxE
\else
\ifnum#1=63 %
\hatcurCCWfourmagxxxxxF
\else
\ifnum#1=64 %
\hatcurCCWfourmagxxxxxG
\else
??????\fi
\fi
\fi
\fi
\fi
\fi
\fi
}
\newcommand{\hatcurCCWonemag}[1]{\ifnum#1=58 %
\hatcurCCWonemagxxxxxA
\else
\ifnum#1=59 %
\hatcurCCWonemagxxxxxB
\else
\ifnum#1=60 %
\hatcurCCWonemagxxxxxC
\else
\ifnum#1=61 %
\hatcurCCWonemagxxxxxD
\else
\ifnum#1=62 %
\hatcurCCWonemagxxxxxE
\else
\ifnum#1=63 %
\hatcurCCWonemagxxxxxF
\else
\ifnum#1=64 %
\hatcurCCWonemagxxxxxG
\else
??????\fi
\fi
\fi
\fi
\fi
\fi
\fi
}
\newcommand{\hatcurCCWthreemag}[1]{\ifnum#1=58 %
\hatcurCCWthreemagxxxxxA
\else
\ifnum#1=59 %
\hatcurCCWthreemagxxxxxB
\else
\ifnum#1=60 %
\hatcurCCWthreemagxxxxxC
\else
\ifnum#1=61 %
\hatcurCCWthreemagxxxxxD
\else
\ifnum#1=62 %
\hatcurCCWthreemagxxxxxE
\else
\ifnum#1=63 %
\hatcurCCWthreemagxxxxxF
\else
\ifnum#1=64 %
\hatcurCCWthreemagxxxxxG
\else
??????\fi
\fi
\fi
\fi
\fi
\fi
\fi
}
\newcommand{\hatcurCCWtwomag}[1]{\ifnum#1=58 %
\hatcurCCWtwomagxxxxxA
\else
\ifnum#1=59 %
\hatcurCCWtwomagxxxxxB
\else
\ifnum#1=60 %
\hatcurCCWtwomagxxxxxC
\else
\ifnum#1=61 %
\hatcurCCWtwomagxxxxxD
\else
\ifnum#1=62 %
\hatcurCCWtwomagxxxxxE
\else
\ifnum#1=63 %
\hatcurCCWtwomagxxxxxF
\else
\ifnum#1=64 %
\hatcurCCWtwomagxxxxxG
\else
??????\fi
\fi
\fi
\fi
\fi
\fi
\fi
}
\newcommand{\hatcurfield}[1]{\ifnum#1=58 %
\hatcurfieldxxxxxA
\else
\ifnum#1=59 %
\hatcurfieldxxxxxB
\else
\ifnum#1=60 %
\hatcurfieldxxxxxC
\else
\ifnum#1=61 %
\hatcurfieldxxxxxD
\else
\ifnum#1=62 %
\hatcurfieldxxxxxE
\else
\ifnum#1=63 %
\hatcurfieldxxxxxF
\else
\ifnum#1=64 %
\hatcurfieldxxxxxG
\else
??????\fi
\fi
\fi
\fi
\fi
\fi
\fi
}
\newcommand{\hatcurhtr}[1]{\ifnum#1=58 %
\hatcurhtrxxxxxA
\else
\ifnum#1=59 %
\hatcurhtrxxxxxB
\else
\ifnum#1=60 %
\hatcurhtrxxxxxC
\else
\ifnum#1=61 %
\hatcurhtrxxxxxD
\else
\ifnum#1=62 %
\hatcurhtrxxxxxE
\else
\ifnum#1=63 %
\hatcurhtrxxxxxF
\else
\ifnum#1=64 %
\hatcurhtrxxxxxG
\else
??????\fi
\fi
\fi
\fi
\fi
\fi
\fi
}
\newcommand{\hatcurISOage}[1]{\ifnum#1=58 %
\hatcurISOagexxxxxA
\else
\ifnum#1=59 %
\hatcurISOagexxxxxB
\else
\ifnum#1=60 %
\hatcurISOagexxxxxC
\else
\ifnum#1=61 %
\hatcurISOagexxxxxD
\else
\ifnum#1=62 %
\hatcurISOagexxxxxE
\else
\ifnum#1=63 %
\hatcurISOagexxxxxF
\else
\ifnum#1=64 %
\hatcurISOagexxxxxG
\else
??????\fi
\fi
\fi
\fi
\fi
\fi
\fi
}
\newcommand{\hatcurISOlogg}[1]{\ifnum#1=58 %
\hatcurISOloggxxxxxA
\else
\ifnum#1=59 %
\hatcurISOloggxxxxxB
\else
\ifnum#1=60 %
\hatcurISOloggxxxxxC
\else
\ifnum#1=61 %
\hatcurISOloggxxxxxD
\else
\ifnum#1=62 %
\hatcurISOloggxxxxxE
\else
\ifnum#1=63 %
\hatcurISOloggxxxxxF
\else
\ifnum#1=64 %
\hatcurISOloggxxxxxG
\else
??????\fi
\fi
\fi
\fi
\fi
\fi
\fi
}
\newcommand{\hatcurISOlum}[1]{\ifnum#1=58 %
\hatcurISOlumxxxxxA
\else
\ifnum#1=59 %
\hatcurISOlumxxxxxB
\else
\ifnum#1=60 %
\hatcurISOlumxxxxxC
\else
\ifnum#1=61 %
\hatcurISOlumxxxxxD
\else
\ifnum#1=62 %
\hatcurISOlumxxxxxE
\else
\ifnum#1=63 %
\hatcurISOlumxxxxxF
\else
\ifnum#1=64 %
\hatcurISOlumxxxxxG
\else
??????\fi
\fi
\fi
\fi
\fi
\fi
\fi
}
\newcommand{\hatcurISOlumshort}[1]{\ifnum#1=58 %
\hatcurISOlumshortxxxxxA
\else
\ifnum#1=59 %
\hatcurISOlumshortxxxxxB
\else
\ifnum#1=60 %
\hatcurISOlumshortxxxxxC
\else
\ifnum#1=61 %
\hatcurISOlumshortxxxxxD
\else
\ifnum#1=62 %
\hatcurISOlumshortxxxxxE
\else
\ifnum#1=63 %
\hatcurISOlumshortxxxxxF
\else
\ifnum#1=64 %
\hatcurISOlumshortxxxxxG
\else
??????\fi
\fi
\fi
\fi
\fi
\fi
\fi
}
\newcommand{\hatcurISOm}[1]{\ifnum#1=58 %
\hatcurISOmxxxxxA
\else
\ifnum#1=59 %
\hatcurISOmxxxxxB
\else
\ifnum#1=60 %
\hatcurISOmxxxxxC
\else
\ifnum#1=61 %
\hatcurISOmxxxxxD
\else
\ifnum#1=62 %
\hatcurISOmxxxxxE
\else
\ifnum#1=63 %
\hatcurISOmxxxxxF
\else
\ifnum#1=64 %
\hatcurISOmxxxxxG
\else
??????\fi
\fi
\fi
\fi
\fi
\fi
\fi
}
\newcommand{\hatcurISOmlong}[1]{\ifnum#1=58 %
\hatcurISOmlongxxxxxA
\else
\ifnum#1=59 %
\hatcurISOmlongxxxxxB
\else
\ifnum#1=60 %
\hatcurISOmlongxxxxxC
\else
\ifnum#1=61 %
\hatcurISOmlongxxxxxD
\else
\ifnum#1=62 %
\hatcurISOmlongxxxxxE
\else
\ifnum#1=63 %
\hatcurISOmlongxxxxxF
\else
\ifnum#1=64 %
\hatcurISOmlongxxxxxG
\else
??????\fi
\fi
\fi
\fi
\fi
\fi
\fi
}
\newcommand{\hatcurISOmshort}[1]{\ifnum#1=58 %
\hatcurISOmshortxxxxxA
\else
\ifnum#1=59 %
\hatcurISOmshortxxxxxB
\else
\ifnum#1=60 %
\hatcurISOmshortxxxxxC
\else
\ifnum#1=61 %
\hatcurISOmshortxxxxxD
\else
\ifnum#1=62 %
\hatcurISOmshortxxxxxE
\else
\ifnum#1=63 %
\hatcurISOmshortxxxxxF
\else
\ifnum#1=64 %
\hatcurISOmshortxxxxxG
\else
??????\fi
\fi
\fi
\fi
\fi
\fi
\fi
}
\newcommand{\hatcurISOr}[1]{\ifnum#1=58 %
\hatcurISOrxxxxxA
\else
\ifnum#1=59 %
\hatcurISOrxxxxxB
\else
\ifnum#1=60 %
\hatcurISOrxxxxxC
\else
\ifnum#1=61 %
\hatcurISOrxxxxxD
\else
\ifnum#1=62 %
\hatcurISOrxxxxxE
\else
\ifnum#1=63 %
\hatcurISOrxxxxxF
\else
\ifnum#1=64 %
\hatcurISOrxxxxxG
\else
??????\fi
\fi
\fi
\fi
\fi
\fi
\fi
}
\newcommand{\hatcurISOrho}[1]{\ifnum#1=58 %
\hatcurISOrhoxxxxxA
\else
\ifnum#1=59 %
\hatcurISOrhoxxxxxB
\else
\ifnum#1=60 %
\hatcurISOrhoxxxxxC
\else
\ifnum#1=61 %
\hatcurISOrhoxxxxxD
\else
\ifnum#1=62 %
\hatcurISOrhoxxxxxE
\else
\ifnum#1=63 %
\hatcurISOrhoxxxxxF
\else
\ifnum#1=64 %
\hatcurISOrhoxxxxxG
\else
??????\fi
\fi
\fi
\fi
\fi
\fi
\fi
}
\newcommand{\hatcurISOrholong}[1]{\ifnum#1=58 %
\hatcurISOrholongxxxxxA
\else
\ifnum#1=59 %
\hatcurISOrholongxxxxxB
\else
\ifnum#1=60 %
\hatcurISOrholongxxxxxC
\else
\ifnum#1=61 %
\hatcurISOrholongxxxxxD
\else
\ifnum#1=62 %
\hatcurISOrholongxxxxxE
\else
\ifnum#1=63 %
\hatcurISOrholongxxxxxF
\else
\ifnum#1=64 %
\hatcurISOrholongxxxxxG
\else
??????\fi
\fi
\fi
\fi
\fi
\fi
\fi
}
\newcommand{\hatcurISOrlong}[1]{\ifnum#1=58 %
\hatcurISOrlongxxxxxA
\else
\ifnum#1=59 %
\hatcurISOrlongxxxxxB
\else
\ifnum#1=60 %
\hatcurISOrlongxxxxxC
\else
\ifnum#1=61 %
\hatcurISOrlongxxxxxD
\else
\ifnum#1=62 %
\hatcurISOrlongxxxxxE
\else
\ifnum#1=63 %
\hatcurISOrlongxxxxxF
\else
\ifnum#1=64 %
\hatcurISOrlongxxxxxG
\else
??????\fi
\fi
\fi
\fi
\fi
\fi
\fi
}
\newcommand{\hatcurISOrshort}[1]{\ifnum#1=58 %
\hatcurISOrshortxxxxxA
\else
\ifnum#1=59 %
\hatcurISOrshortxxxxxB
\else
\ifnum#1=60 %
\hatcurISOrshortxxxxxC
\else
\ifnum#1=61 %
\hatcurISOrshortxxxxxD
\else
\ifnum#1=62 %
\hatcurISOrshortxxxxxE
\else
\ifnum#1=63 %
\hatcurISOrshortxxxxxF
\else
\ifnum#1=64 %
\hatcurISOrshortxxxxxG
\else
??????\fi
\fi
\fi
\fi
\fi
\fi
\fi
}
\newcommand{\hatcurISOspec}[1]{\ifnum#1=58 %
\hatcurISOspecxxxxxA
\else
\ifnum#1=59 %
\hatcurISOspecxxxxxB
\else
\ifnum#1=60 %
\hatcurISOspecxxxxxC
\else
\ifnum#1=61 %
\hatcurISOspecxxxxxD
\else
\ifnum#1=62 %
\hatcurISOspecxxxxxE
\else
\ifnum#1=63 %
\hatcurISOspecxxxxxF
\else
\ifnum#1=64 %
\hatcurISOspecxxxxxG
\else
??????\fi
\fi
\fi
\fi
\fi
\fi
\fi
}
\newcommand{\hatcurISOteff}[1]{\ifnum#1=58 %
\hatcurISOteffxxxxxA
\else
\ifnum#1=59 %
\hatcurISOteffxxxxxB
\else
\ifnum#1=60 %
\hatcurISOteffxxxxxC
\else
\ifnum#1=61 %
\hatcurISOteffxxxxxD
\else
\ifnum#1=62 %
\hatcurISOteffxxxxxE
\else
\ifnum#1=63 %
\hatcurISOteffxxxxxF
\else
\ifnum#1=64 %
\hatcurISOteffxxxxxG
\else
??????\fi
\fi
\fi
\fi
\fi
\fi
\fi
}
\newcommand{\hatcurISOzfeh}[1]{\ifnum#1=58 %
\hatcurISOzfehxxxxxA
\else
\ifnum#1=59 %
\hatcurISOzfehxxxxxB
\else
\ifnum#1=60 %
\hatcurISOzfehxxxxxC
\else
\ifnum#1=61 %
\hatcurISOzfehxxxxxD
\else
\ifnum#1=62 %
\hatcurISOzfehxxxxxE
\else
\ifnum#1=63 %
\hatcurISOzfehxxxxxF
\else
\ifnum#1=64 %
\hatcurISOzfehxxxxxG
\else
??????\fi
\fi
\fi
\fi
\fi
\fi
\fi
}
\newcommand{\hatcurLBiB}[1]{\ifnum#1=58 %
\hatcurLBiBxxxxxA
\else
\ifnum#1=59 %
\hatcurLBiBxxxxxB
\else
\ifnum#1=60 %
\hatcurLBiBxxxxxC
\else
\ifnum#1=61 %
\hatcurLBiBxxxxxD
\else
\ifnum#1=62 %
\hatcurLBiBxxxxxE
\else
\ifnum#1=63 %
\hatcurLBiBxxxxxF
\else
\ifnum#1=64 %
\hatcurLBiBxxxxxG
\else
??????\fi
\fi
\fi
\fi
\fi
\fi
\fi
}
\newcommand{\hatcurLBiC}[1]{\ifnum#1=58 %
\hatcurLBiCxxxxxA
\else
\ifnum#1=59 %
\hatcurLBiCxxxxxB
\else
\ifnum#1=60 %
\hatcurLBiCxxxxxC
\else
\ifnum#1=61 %
\hatcurLBiCxxxxxD
\else
\ifnum#1=62 %
\hatcurLBiCxxxxxE
\else
\ifnum#1=63 %
\hatcurLBiCxxxxxF
\else
\ifnum#1=64 %
\hatcurLBiCxxxxxG
\else
??????\fi
\fi
\fi
\fi
\fi
\fi
\fi
}
\newcommand{\hatcurLBig}[1]{\ifnum#1=58 %
\hatcurLBigxxxxxA
\else
\ifnum#1=59 %
\hatcurLBigxxxxxB
\else
\ifnum#1=60 %
\hatcurLBigxxxxxC
\else
\ifnum#1=61 %
\hatcurLBigxxxxxD
\else
\ifnum#1=62 %
\hatcurLBigxxxxxE
\else
\ifnum#1=63 %
\hatcurLBigxxxxxF
\else
\ifnum#1=64 %
\hatcurLBigxxxxxG
\else
??????\fi
\fi
\fi
\fi
\fi
\fi
\fi
}
\newcommand{\hatcurLBiH}[1]{\ifnum#1=58 %
\hatcurLBiHxxxxxA
\else
\ifnum#1=59 %
\hatcurLBiHxxxxxB
\else
\ifnum#1=60 %
\hatcurLBiHxxxxxC
\else
\ifnum#1=61 %
\hatcurLBiHxxxxxD
\else
\ifnum#1=62 %
\hatcurLBiHxxxxxE
\else
\ifnum#1=63 %
\hatcurLBiHxxxxxF
\else
\ifnum#1=64 %
\hatcurLBiHxxxxxG
\else
??????\fi
\fi
\fi
\fi
\fi
\fi
\fi
}
\newcommand{\hatcurLBii}[1]{\ifnum#1=58 %
\hatcurLBiixxxxxA
\else
\ifnum#1=59 %
\hatcurLBiixxxxxB
\else
\ifnum#1=60 %
\hatcurLBiixxxxxC
\else
\ifnum#1=61 %
\hatcurLBiixxxxxD
\else
\ifnum#1=62 %
\hatcurLBiixxxxxE
\else
\ifnum#1=63 %
\hatcurLBiixxxxxF
\else
\ifnum#1=64 %
\hatcurLBiixxxxxG
\else
??????\fi
\fi
\fi
\fi
\fi
\fi
\fi
}
\newcommand{\hatcurLBiI}[1]{\ifnum#1=58 %
\hatcurLBiIxxxxxA
\else
\ifnum#1=59 %
\hatcurLBiIxxxxxB
\else
\ifnum#1=60 %
\hatcurLBiIxxxxxC
\else
\ifnum#1=61 %
\hatcurLBiIxxxxxD
\else
\ifnum#1=62 %
\hatcurLBiIxxxxxE
\else
\ifnum#1=63 %
\hatcurLBiIxxxxxF
\else
\ifnum#1=64 %
\hatcurLBiIxxxxxG
\else
??????\fi
\fi
\fi
\fi
\fi
\fi
\fi
}
\newcommand{\hatcurLBiiB}[1]{\ifnum#1=58 %
\hatcurLBiiBxxxxxA
\else
\ifnum#1=59 %
\hatcurLBiiBxxxxxB
\else
\ifnum#1=60 %
\hatcurLBiiBxxxxxC
\else
\ifnum#1=61 %
\hatcurLBiiBxxxxxD
\else
\ifnum#1=62 %
\hatcurLBiiBxxxxxE
\else
\ifnum#1=63 %
\hatcurLBiiBxxxxxF
\else
\ifnum#1=64 %
\hatcurLBiiBxxxxxG
\else
??????\fi
\fi
\fi
\fi
\fi
\fi
\fi
}
\newcommand{\hatcurLBiiC}[1]{\ifnum#1=58 %
\hatcurLBiiCxxxxxA
\else
\ifnum#1=59 %
\hatcurLBiiCxxxxxB
\else
\ifnum#1=60 %
\hatcurLBiiCxxxxxC
\else
\ifnum#1=61 %
\hatcurLBiiCxxxxxD
\else
\ifnum#1=62 %
\hatcurLBiiCxxxxxE
\else
\ifnum#1=63 %
\hatcurLBiiCxxxxxF
\else
\ifnum#1=64 %
\hatcurLBiiCxxxxxG
\else
??????\fi
\fi
\fi
\fi
\fi
\fi
\fi
}
\newcommand{\hatcurLBiig}[1]{\ifnum#1=58 %
\hatcurLBiigxxxxxA
\else
\ifnum#1=59 %
\hatcurLBiigxxxxxB
\else
\ifnum#1=60 %
\hatcurLBiigxxxxxC
\else
\ifnum#1=61 %
\hatcurLBiigxxxxxD
\else
\ifnum#1=62 %
\hatcurLBiigxxxxxE
\else
\ifnum#1=63 %
\hatcurLBiigxxxxxF
\else
\ifnum#1=64 %
\hatcurLBiigxxxxxG
\else
??????\fi
\fi
\fi
\fi
\fi
\fi
\fi
}
\newcommand{\hatcurLBiiH}[1]{\ifnum#1=58 %
\hatcurLBiiHxxxxxA
\else
\ifnum#1=59 %
\hatcurLBiiHxxxxxB
\else
\ifnum#1=60 %
\hatcurLBiiHxxxxxC
\else
\ifnum#1=61 %
\hatcurLBiiHxxxxxD
\else
\ifnum#1=62 %
\hatcurLBiiHxxxxxE
\else
\ifnum#1=63 %
\hatcurLBiiHxxxxxF
\else
\ifnum#1=64 %
\hatcurLBiiHxxxxxG
\else
??????\fi
\fi
\fi
\fi
\fi
\fi
\fi
}
\newcommand{\hatcurLBiii}[1]{\ifnum#1=58 %
\hatcurLBiiixxxxxA
\else
\ifnum#1=59 %
\hatcurLBiiixxxxxB
\else
\ifnum#1=60 %
\hatcurLBiiixxxxxC
\else
\ifnum#1=61 %
\hatcurLBiiixxxxxD
\else
\ifnum#1=62 %
\hatcurLBiiixxxxxE
\else
\ifnum#1=63 %
\hatcurLBiiixxxxxF
\else
\ifnum#1=64 %
\hatcurLBiiixxxxxG
\else
??????\fi
\fi
\fi
\fi
\fi
\fi
\fi
}
\newcommand{\hatcurLBiiI}[1]{\ifnum#1=58 %
\hatcurLBiiIxxxxxA
\else
\ifnum#1=59 %
\hatcurLBiiIxxxxxB
\else
\ifnum#1=60 %
\hatcurLBiiIxxxxxC
\else
\ifnum#1=61 %
\hatcurLBiiIxxxxxD
\else
\ifnum#1=62 %
\hatcurLBiiIxxxxxE
\else
\ifnum#1=63 %
\hatcurLBiiIxxxxxF
\else
\ifnum#1=64 %
\hatcurLBiiIxxxxxG
\else
??????\fi
\fi
\fi
\fi
\fi
\fi
\fi
}
\newcommand{\hatcurLBiiJ}[1]{\ifnum#1=58 %
\hatcurLBiiJxxxxxA
\else
\ifnum#1=59 %
\hatcurLBiiJxxxxxB
\else
\ifnum#1=60 %
\hatcurLBiiJxxxxxC
\else
\ifnum#1=61 %
\hatcurLBiiJxxxxxD
\else
\ifnum#1=62 %
\hatcurLBiiJxxxxxE
\else
\ifnum#1=63 %
\hatcurLBiiJxxxxxF
\else
\ifnum#1=64 %
\hatcurLBiiJxxxxxG
\else
??????\fi
\fi
\fi
\fi
\fi
\fi
\fi
}
\newcommand{\hatcurLBiiK}[1]{\ifnum#1=58 %
\hatcurLBiiKxxxxxA
\else
\ifnum#1=59 %
\hatcurLBiiKxxxxxB
\else
\ifnum#1=60 %
\hatcurLBiiKxxxxxC
\else
\ifnum#1=61 %
\hatcurLBiiKxxxxxD
\else
\ifnum#1=62 %
\hatcurLBiiKxxxxxE
\else
\ifnum#1=63 %
\hatcurLBiiKxxxxxF
\else
\ifnum#1=64 %
\hatcurLBiiKxxxxxG
\else
??????\fi
\fi
\fi
\fi
\fi
\fi
\fi
}
\newcommand{\hatcurLBiikep}[1]{\ifnum#1=58 %
\hatcurLBiikepxxxxxA
\else
\ifnum#1=59 %
\hatcurLBiikepxxxxxB
\else
\ifnum#1=60 %
\hatcurLBiikepxxxxxC
\else
\ifnum#1=61 %
\hatcurLBiikepxxxxxD
\else
\ifnum#1=62 %
\hatcurLBiikepxxxxxE
\else
\ifnum#1=63 %
\hatcurLBiikepxxxxxF
\else
\ifnum#1=64 %
\hatcurLBiikepxxxxxG
\else
??????\fi
\fi
\fi
\fi
\fi
\fi
\fi
}
\newcommand{\hatcurLBiiM}[1]{\ifnum#1=58 %
\hatcurLBiiMxxxxxA
\else
\ifnum#1=59 %
\hatcurLBiiMxxxxxB
\else
\ifnum#1=60 %
\hatcurLBiiMxxxxxC
\else
\ifnum#1=61 %
\hatcurLBiiMxxxxxD
\else
\ifnum#1=62 %
\hatcurLBiiMxxxxxE
\else
\ifnum#1=63 %
\hatcurLBiiMxxxxxF
\else
\ifnum#1=64 %
\hatcurLBiiMxxxxxG
\else
??????\fi
\fi
\fi
\fi
\fi
\fi
\fi
}
\newcommand{\hatcurLBiir}[1]{\ifnum#1=58 %
\hatcurLBiirxxxxxA
\else
\ifnum#1=59 %
\hatcurLBiirxxxxxB
\else
\ifnum#1=60 %
\hatcurLBiirxxxxxC
\else
\ifnum#1=61 %
\hatcurLBiirxxxxxD
\else
\ifnum#1=62 %
\hatcurLBiirxxxxxE
\else
\ifnum#1=63 %
\hatcurLBiirxxxxxF
\else
\ifnum#1=64 %
\hatcurLBiirxxxxxG
\else
??????\fi
\fi
\fi
\fi
\fi
\fi
\fi
}
\newcommand{\hatcurLBiiR}[1]{\ifnum#1=58 %
\hatcurLBiiRxxxxxA
\else
\ifnum#1=59 %
\hatcurLBiiRxxxxxB
\else
\ifnum#1=60 %
\hatcurLBiiRxxxxxC
\else
\ifnum#1=61 %
\hatcurLBiiRxxxxxD
\else
\ifnum#1=62 %
\hatcurLBiiRxxxxxE
\else
\ifnum#1=63 %
\hatcurLBiiRxxxxxF
\else
\ifnum#1=64 %
\hatcurLBiiRxxxxxG
\else
??????\fi
\fi
\fi
\fi
\fi
\fi
\fi
}
\newcommand{\hatcurLBiiSfour}[1]{\ifnum#1=58 %
\hatcurLBiiSfourxxxxxA
\else
\ifnum#1=59 %
\hatcurLBiiSfourxxxxxB
\else
\ifnum#1=60 %
\hatcurLBiiSfourxxxxxC
\else
\ifnum#1=61 %
\hatcurLBiiSfourxxxxxD
\else
\ifnum#1=62 %
\hatcurLBiiSfourxxxxxE
\else
\ifnum#1=63 %
\hatcurLBiiSfourxxxxxF
\else
\ifnum#1=64 %
\hatcurLBiiSfourxxxxxG
\else
??????\fi
\fi
\fi
\fi
\fi
\fi
\fi
}
\newcommand{\hatcurLBiiSone}[1]{\ifnum#1=58 %
\hatcurLBiiSonexxxxxA
\else
\ifnum#1=59 %
\hatcurLBiiSonexxxxxB
\else
\ifnum#1=60 %
\hatcurLBiiSonexxxxxC
\else
\ifnum#1=61 %
\hatcurLBiiSonexxxxxD
\else
\ifnum#1=62 %
\hatcurLBiiSonexxxxxE
\else
\ifnum#1=63 %
\hatcurLBiiSonexxxxxF
\else
\ifnum#1=64 %
\hatcurLBiiSonexxxxxG
\else
??????\fi
\fi
\fi
\fi
\fi
\fi
\fi
}
\newcommand{\hatcurLBiiSthree}[1]{\ifnum#1=58 %
\hatcurLBiiSthreexxxxxA
\else
\ifnum#1=59 %
\hatcurLBiiSthreexxxxxB
\else
\ifnum#1=60 %
\hatcurLBiiSthreexxxxxC
\else
\ifnum#1=61 %
\hatcurLBiiSthreexxxxxD
\else
\ifnum#1=62 %
\hatcurLBiiSthreexxxxxE
\else
\ifnum#1=63 %
\hatcurLBiiSthreexxxxxF
\else
\ifnum#1=64 %
\hatcurLBiiSthreexxxxxG
\else
??????\fi
\fi
\fi
\fi
\fi
\fi
\fi
}
\newcommand{\hatcurLBiiStwo}[1]{\ifnum#1=58 %
\hatcurLBiiStwoxxxxxA
\else
\ifnum#1=59 %
\hatcurLBiiStwoxxxxxB
\else
\ifnum#1=60 %
\hatcurLBiiStwoxxxxxC
\else
\ifnum#1=61 %
\hatcurLBiiStwoxxxxxD
\else
\ifnum#1=62 %
\hatcurLBiiStwoxxxxxE
\else
\ifnum#1=63 %
\hatcurLBiiStwoxxxxxF
\else
\ifnum#1=64 %
\hatcurLBiiStwoxxxxxG
\else
??????\fi
\fi
\fi
\fi
\fi
\fi
\fi
}
\newcommand{\hatcurLBiiT}[1]{\ifnum#1=58 %
\hatcurLBiiTxxxxxA
\else
\ifnum#1=59 %
\hatcurLBiiTxxxxxB
\else
\ifnum#1=60 %
\hatcurLBiiTxxxxxC
\else
\ifnum#1=61 %
\hatcurLBiiTxxxxxD
\else
\ifnum#1=62 %
\hatcurLBiiTxxxxxE
\else
\ifnum#1=63 %
\hatcurLBiiTxxxxxF
\else
\ifnum#1=64 %
\hatcurLBiiTxxxxxG
\else
??????\fi
\fi
\fi
\fi
\fi
\fi
\fi
}
\newcommand{\hatcurLBiiu}[1]{\ifnum#1=58 %
\hatcurLBiiuxxxxxA
\else
\ifnum#1=59 %
\hatcurLBiiuxxxxxB
\else
\ifnum#1=60 %
\hatcurLBiiuxxxxxC
\else
\ifnum#1=61 %
\hatcurLBiiuxxxxxD
\else
\ifnum#1=62 %
\hatcurLBiiuxxxxxE
\else
\ifnum#1=63 %
\hatcurLBiiuxxxxxF
\else
\ifnum#1=64 %
\hatcurLBiiuxxxxxG
\else
??????\fi
\fi
\fi
\fi
\fi
\fi
\fi
}
\newcommand{\hatcurLBiiV}[1]{\ifnum#1=58 %
\hatcurLBiiVxxxxxA
\else
\ifnum#1=59 %
\hatcurLBiiVxxxxxB
\else
\ifnum#1=60 %
\hatcurLBiiVxxxxxC
\else
\ifnum#1=61 %
\hatcurLBiiVxxxxxD
\else
\ifnum#1=62 %
\hatcurLBiiVxxxxxE
\else
\ifnum#1=63 %
\hatcurLBiiVxxxxxF
\else
\ifnum#1=64 %
\hatcurLBiiVxxxxxG
\else
??????\fi
\fi
\fi
\fi
\fi
\fi
\fi
}
\newcommand{\hatcurLBiiz}[1]{\ifnum#1=58 %
\hatcurLBiizxxxxxA
\else
\ifnum#1=59 %
\hatcurLBiizxxxxxB
\else
\ifnum#1=60 %
\hatcurLBiizxxxxxC
\else
\ifnum#1=61 %
\hatcurLBiizxxxxxD
\else
\ifnum#1=62 %
\hatcurLBiizxxxxxE
\else
\ifnum#1=63 %
\hatcurLBiizxxxxxF
\else
\ifnum#1=64 %
\hatcurLBiizxxxxxG
\else
??????\fi
\fi
\fi
\fi
\fi
\fi
\fi
}
\newcommand{\hatcurLBiJ}[1]{\ifnum#1=58 %
\hatcurLBiJxxxxxA
\else
\ifnum#1=59 %
\hatcurLBiJxxxxxB
\else
\ifnum#1=60 %
\hatcurLBiJxxxxxC
\else
\ifnum#1=61 %
\hatcurLBiJxxxxxD
\else
\ifnum#1=62 %
\hatcurLBiJxxxxxE
\else
\ifnum#1=63 %
\hatcurLBiJxxxxxF
\else
\ifnum#1=64 %
\hatcurLBiJxxxxxG
\else
??????\fi
\fi
\fi
\fi
\fi
\fi
\fi
}
\newcommand{\hatcurLBiK}[1]{\ifnum#1=58 %
\hatcurLBiKxxxxxA
\else
\ifnum#1=59 %
\hatcurLBiKxxxxxB
\else
\ifnum#1=60 %
\hatcurLBiKxxxxxC
\else
\ifnum#1=61 %
\hatcurLBiKxxxxxD
\else
\ifnum#1=62 %
\hatcurLBiKxxxxxE
\else
\ifnum#1=63 %
\hatcurLBiKxxxxxF
\else
\ifnum#1=64 %
\hatcurLBiKxxxxxG
\else
??????\fi
\fi
\fi
\fi
\fi
\fi
\fi
}
\newcommand{\hatcurLBikep}[1]{\ifnum#1=58 %
\hatcurLBikepxxxxxA
\else
\ifnum#1=59 %
\hatcurLBikepxxxxxB
\else
\ifnum#1=60 %
\hatcurLBikepxxxxxC
\else
\ifnum#1=61 %
\hatcurLBikepxxxxxD
\else
\ifnum#1=62 %
\hatcurLBikepxxxxxE
\else
\ifnum#1=63 %
\hatcurLBikepxxxxxF
\else
\ifnum#1=64 %
\hatcurLBikepxxxxxG
\else
??????\fi
\fi
\fi
\fi
\fi
\fi
\fi
}
\newcommand{\hatcurLBiM}[1]{\ifnum#1=58 %
\hatcurLBiMxxxxxA
\else
\ifnum#1=59 %
\hatcurLBiMxxxxxB
\else
\ifnum#1=60 %
\hatcurLBiMxxxxxC
\else
\ifnum#1=61 %
\hatcurLBiMxxxxxD
\else
\ifnum#1=62 %
\hatcurLBiMxxxxxE
\else
\ifnum#1=63 %
\hatcurLBiMxxxxxF
\else
\ifnum#1=64 %
\hatcurLBiMxxxxxG
\else
??????\fi
\fi
\fi
\fi
\fi
\fi
\fi
}
\newcommand{\hatcurLBir}[1]{\ifnum#1=58 %
\hatcurLBirxxxxxA
\else
\ifnum#1=59 %
\hatcurLBirxxxxxB
\else
\ifnum#1=60 %
\hatcurLBirxxxxxC
\else
\ifnum#1=61 %
\hatcurLBirxxxxxD
\else
\ifnum#1=62 %
\hatcurLBirxxxxxE
\else
\ifnum#1=63 %
\hatcurLBirxxxxxF
\else
\ifnum#1=64 %
\hatcurLBirxxxxxG
\else
??????\fi
\fi
\fi
\fi
\fi
\fi
\fi
}
\newcommand{\hatcurLBiR}[1]{\ifnum#1=58 %
\hatcurLBiRxxxxxA
\else
\ifnum#1=59 %
\hatcurLBiRxxxxxB
\else
\ifnum#1=60 %
\hatcurLBiRxxxxxC
\else
\ifnum#1=61 %
\hatcurLBiRxxxxxD
\else
\ifnum#1=62 %
\hatcurLBiRxxxxxE
\else
\ifnum#1=63 %
\hatcurLBiRxxxxxF
\else
\ifnum#1=64 %
\hatcurLBiRxxxxxG
\else
??????\fi
\fi
\fi
\fi
\fi
\fi
\fi
}
\newcommand{\hatcurLBiSfour}[1]{\ifnum#1=58 %
\hatcurLBiSfourxxxxxA
\else
\ifnum#1=59 %
\hatcurLBiSfourxxxxxB
\else
\ifnum#1=60 %
\hatcurLBiSfourxxxxxC
\else
\ifnum#1=61 %
\hatcurLBiSfourxxxxxD
\else
\ifnum#1=62 %
\hatcurLBiSfourxxxxxE
\else
\ifnum#1=63 %
\hatcurLBiSfourxxxxxF
\else
\ifnum#1=64 %
\hatcurLBiSfourxxxxxG
\else
??????\fi
\fi
\fi
\fi
\fi
\fi
\fi
}
\newcommand{\hatcurLBiSone}[1]{\ifnum#1=58 %
\hatcurLBiSonexxxxxA
\else
\ifnum#1=59 %
\hatcurLBiSonexxxxxB
\else
\ifnum#1=60 %
\hatcurLBiSonexxxxxC
\else
\ifnum#1=61 %
\hatcurLBiSonexxxxxD
\else
\ifnum#1=62 %
\hatcurLBiSonexxxxxE
\else
\ifnum#1=63 %
\hatcurLBiSonexxxxxF
\else
\ifnum#1=64 %
\hatcurLBiSonexxxxxG
\else
??????\fi
\fi
\fi
\fi
\fi
\fi
\fi
}
\newcommand{\hatcurLBiSthree}[1]{\ifnum#1=58 %
\hatcurLBiSthreexxxxxA
\else
\ifnum#1=59 %
\hatcurLBiSthreexxxxxB
\else
\ifnum#1=60 %
\hatcurLBiSthreexxxxxC
\else
\ifnum#1=61 %
\hatcurLBiSthreexxxxxD
\else
\ifnum#1=62 %
\hatcurLBiSthreexxxxxE
\else
\ifnum#1=63 %
\hatcurLBiSthreexxxxxF
\else
\ifnum#1=64 %
\hatcurLBiSthreexxxxxG
\else
??????\fi
\fi
\fi
\fi
\fi
\fi
\fi
}
\newcommand{\hatcurLBiStwo}[1]{\ifnum#1=58 %
\hatcurLBiStwoxxxxxA
\else
\ifnum#1=59 %
\hatcurLBiStwoxxxxxB
\else
\ifnum#1=60 %
\hatcurLBiStwoxxxxxC
\else
\ifnum#1=61 %
\hatcurLBiStwoxxxxxD
\else
\ifnum#1=62 %
\hatcurLBiStwoxxxxxE
\else
\ifnum#1=63 %
\hatcurLBiStwoxxxxxF
\else
\ifnum#1=64 %
\hatcurLBiStwoxxxxxG
\else
??????\fi
\fi
\fi
\fi
\fi
\fi
\fi
}
\newcommand{\hatcurLBiT}[1]{\ifnum#1=58 %
\hatcurLBiTxxxxxA
\else
\ifnum#1=59 %
\hatcurLBiTxxxxxB
\else
\ifnum#1=60 %
\hatcurLBiTxxxxxC
\else
\ifnum#1=61 %
\hatcurLBiTxxxxxD
\else
\ifnum#1=62 %
\hatcurLBiTxxxxxE
\else
\ifnum#1=63 %
\hatcurLBiTxxxxxF
\else
\ifnum#1=64 %
\hatcurLBiTxxxxxG
\else
??????\fi
\fi
\fi
\fi
\fi
\fi
\fi
}
\newcommand{\hatcurLBiu}[1]{\ifnum#1=58 %
\hatcurLBiuxxxxxA
\else
\ifnum#1=59 %
\hatcurLBiuxxxxxB
\else
\ifnum#1=60 %
\hatcurLBiuxxxxxC
\else
\ifnum#1=61 %
\hatcurLBiuxxxxxD
\else
\ifnum#1=62 %
\hatcurLBiuxxxxxE
\else
\ifnum#1=63 %
\hatcurLBiuxxxxxF
\else
\ifnum#1=64 %
\hatcurLBiuxxxxxG
\else
??????\fi
\fi
\fi
\fi
\fi
\fi
\fi
}
\newcommand{\hatcurLBiV}[1]{\ifnum#1=58 %
\hatcurLBiVxxxxxA
\else
\ifnum#1=59 %
\hatcurLBiVxxxxxB
\else
\ifnum#1=60 %
\hatcurLBiVxxxxxC
\else
\ifnum#1=61 %
\hatcurLBiVxxxxxD
\else
\ifnum#1=62 %
\hatcurLBiVxxxxxE
\else
\ifnum#1=63 %
\hatcurLBiVxxxxxF
\else
\ifnum#1=64 %
\hatcurLBiVxxxxxG
\else
??????\fi
\fi
\fi
\fi
\fi
\fi
\fi
}
\newcommand{\hatcurLBiz}[1]{\ifnum#1=58 %
\hatcurLBizxxxxxA
\else
\ifnum#1=59 %
\hatcurLBizxxxxxB
\else
\ifnum#1=60 %
\hatcurLBizxxxxxC
\else
\ifnum#1=61 %
\hatcurLBizxxxxxD
\else
\ifnum#1=62 %
\hatcurLBizxxxxxE
\else
\ifnum#1=63 %
\hatcurLBizxxxxxF
\else
\ifnum#1=64 %
\hatcurLBizxxxxxG
\else
??????\fi
\fi
\fi
\fi
\fi
\fi
\fi
}
\newcommand{\hatcurLCbsq}[1]{\ifnum#1=58 %
\hatcurLCbsqxxxxxA
\else
\ifnum#1=59 %
\hatcurLCbsqxxxxxB
\else
\ifnum#1=60 %
\hatcurLCbsqxxxxxC
\else
\ifnum#1=61 %
\hatcurLCbsqxxxxxD
\else
\ifnum#1=62 %
\hatcurLCbsqxxxxxE
\else
\ifnum#1=63 %
\hatcurLCbsqxxxxxF
\else
\ifnum#1=64 %
\hatcurLCbsqxxxxxG
\else
??????\fi
\fi
\fi
\fi
\fi
\fi
\fi
}
\newcommand{\hatcurLCdip}[1]{\ifnum#1=58 %
\hatcurLCdipxxxxxA
\else
\ifnum#1=59 %
\hatcurLCdipxxxxxB
\else
\ifnum#1=60 %
\hatcurLCdipxxxxxC
\else
\ifnum#1=61 %
\hatcurLCdipxxxxxD
\else
\ifnum#1=62 %
\hatcurLCdipxxxxxE
\else
\ifnum#1=63 %
\hatcurLCdipxxxxxF
\else
\ifnum#1=64 %
\hatcurLCdipxxxxxG
\else
??????\fi
\fi
\fi
\fi
\fi
\fi
\fi
}
\newcommand{\hatcurLCdur}[1]{\ifnum#1=58 %
\hatcurLCdurxxxxxA
\else
\ifnum#1=59 %
\hatcurLCdurxxxxxB
\else
\ifnum#1=60 %
\hatcurLCdurxxxxxC
\else
\ifnum#1=61 %
\hatcurLCdurxxxxxD
\else
\ifnum#1=62 %
\hatcurLCdurxxxxxE
\else
\ifnum#1=63 %
\hatcurLCdurxxxxxF
\else
\ifnum#1=64 %
\hatcurLCdurxxxxxG
\else
??????\fi
\fi
\fi
\fi
\fi
\fi
\fi
}
\newcommand{\hatcurLCdurhr}[1]{\ifnum#1=58 %
\hatcurLCdurhrxxxxxA
\else
\ifnum#1=59 %
\hatcurLCdurhrxxxxxB
\else
\ifnum#1=60 %
\hatcurLCdurhrxxxxxC
\else
\ifnum#1=61 %
\hatcurLCdurhrxxxxxD
\else
\ifnum#1=62 %
\hatcurLCdurhrxxxxxE
\else
\ifnum#1=63 %
\hatcurLCdurhrxxxxxF
\else
\ifnum#1=64 %
\hatcurLCdurhrxxxxxG
\else
??????\fi
\fi
\fi
\fi
\fi
\fi
\fi
}
\newcommand{\hatcurLCdurhrshort}[1]{\ifnum#1=58 %
\hatcurLCdurhrshortxxxxxA
\else
\ifnum#1=59 %
\hatcurLCdurhrshortxxxxxB
\else
\ifnum#1=60 %
\hatcurLCdurhrshortxxxxxC
\else
\ifnum#1=61 %
\hatcurLCdurhrshortxxxxxD
\else
\ifnum#1=62 %
\hatcurLCdurhrshortxxxxxE
\else
\ifnum#1=63 %
\hatcurLCdurhrshortxxxxxF
\else
\ifnum#1=64 %
\hatcurLCdurhrshortxxxxxG
\else
??????\fi
\fi
\fi
\fi
\fi
\fi
\fi
}
\newcommand{\hatcurLCdurshort}[1]{\ifnum#1=58 %
\hatcurLCdurshortxxxxxA
\else
\ifnum#1=59 %
\hatcurLCdurshortxxxxxB
\else
\ifnum#1=60 %
\hatcurLCdurshortxxxxxC
\else
\ifnum#1=61 %
\hatcurLCdurshortxxxxxD
\else
\ifnum#1=62 %
\hatcurLCdurshortxxxxxE
\else
\ifnum#1=63 %
\hatcurLCdurshortxxxxxF
\else
\ifnum#1=64 %
\hatcurLCdurshortxxxxxG
\else
??????\fi
\fi
\fi
\fi
\fi
\fi
\fi
}
\newcommand{\hatcurLChatnetm}[1]{\ifnum#1=62 %
\hatcurLChatnetmxxxxxE
\else
??????\fi
}
\newcommand{\hatcurLChatnetmA}[1]{\ifnum#1=58 %
\hatcurLChatnetmAxxxxxA
\else
\ifnum#1=59 %
\hatcurLChatnetmAxxxxxB
\else
\ifnum#1=60 %
\hatcurLChatnetmAxxxxxC
\else
\ifnum#1=61 %
\hatcurLChatnetmAxxxxxD
\else
\ifnum#1=64 %
\hatcurLChatnetmAxxxxxG
\else
??????\fi
\fi
\fi
\fi
\fi
}
\newcommand{\hatcurLChatnetmB}[1]{\ifnum#1=58 %
\hatcurLChatnetmBxxxxxA
\else
\ifnum#1=59 %
\hatcurLChatnetmBxxxxxB
\else
\ifnum#1=60 %
\hatcurLChatnetmBxxxxxC
\else
\ifnum#1=61 %
\hatcurLChatnetmBxxxxxD
\else
\ifnum#1=64 %
\hatcurLChatnetmBxxxxxG
\else
??????\fi
\fi
\fi
\fi
\fi
}
\newcommand{\hatcurLChatnetmC}[1]{\ifnum#1=58 %
\hatcurLChatnetmCxxxxxA
\else
\ifnum#1=59 %
\hatcurLChatnetmCxxxxxB
\else
\ifnum#1=61 %
\hatcurLChatnetmCxxxxxD
\else
??????\fi
\fi
\fi
}
\newcommand{\hatcurLChatnetmD}[1]{\ifnum#1=59 %
\hatcurLChatnetmDxxxxxB
\else
??????\fi
}
\newcommand{\hatcurLChatnetmE}[1]{\ifnum#1=59 %
\hatcurLChatnetmExxxxxB
\else
??????\fi
}
\newcommand{\hatcurLChatnetmF}[1]{\ifnum#1=59 %
\hatcurLChatnetmFxxxxxB
\else
??????\fi
}
\newcommand{\hatcurLChatnetmG}[1]{\ifnum#1=59 %
\hatcurLChatnetmGxxxxxB
\else
??????\fi
}
\newcommand{\hatcurLChatnetmH}[1]{\ifnum#1=59 %
\hatcurLChatnetmHxxxxxB
\else
??????\fi
}
\newcommand{\hatcurLCiblend}[1]{\ifnum#1=62 %
\hatcurLCiblendxxxxxE
\else
??????\fi
}
\newcommand{\hatcurLCiblendA}[1]{\ifnum#1=58 %
\hatcurLCiblendAxxxxxA
\else
\ifnum#1=59 %
\hatcurLCiblendAxxxxxB
\else
\ifnum#1=60 %
\hatcurLCiblendAxxxxxC
\else
\ifnum#1=61 %
\hatcurLCiblendAxxxxxD
\else
\ifnum#1=64 %
\hatcurLCiblendAxxxxxG
\else
??????\fi
\fi
\fi
\fi
\fi
}
\newcommand{\hatcurLCiblendB}[1]{\ifnum#1=58 %
\hatcurLCiblendBxxxxxA
\else
\ifnum#1=59 %
\hatcurLCiblendBxxxxxB
\else
\ifnum#1=60 %
\hatcurLCiblendBxxxxxC
\else
\ifnum#1=61 %
\hatcurLCiblendBxxxxxD
\else
\ifnum#1=64 %
\hatcurLCiblendBxxxxxG
\else
??????\fi
\fi
\fi
\fi
\fi
}
\newcommand{\hatcurLCiblendC}[1]{\ifnum#1=58 %
\hatcurLCiblendCxxxxxA
\else
\ifnum#1=59 %
\hatcurLCiblendCxxxxxB
\else
\ifnum#1=61 %
\hatcurLCiblendCxxxxxD
\else
??????\fi
\fi
\fi
}
\newcommand{\hatcurLCiblendD}[1]{\ifnum#1=59 %
\hatcurLCiblendDxxxxxB
\else
??????\fi
}
\newcommand{\hatcurLCiblendE}[1]{\ifnum#1=59 %
\hatcurLCiblendExxxxxB
\else
??????\fi
}
\newcommand{\hatcurLCiblendF}[1]{\ifnum#1=59 %
\hatcurLCiblendFxxxxxB
\else
??????\fi
}
\newcommand{\hatcurLCiblendG}[1]{\ifnum#1=59 %
\hatcurLCiblendGxxxxxB
\else
??????\fi
}
\newcommand{\hatcurLCiblendH}[1]{\ifnum#1=59 %
\hatcurLCiblendHxxxxxB
\else
??????\fi
}
\newcommand{\hatcurLCimp}[1]{\ifnum#1=58 %
\hatcurLCimpxxxxxA
\else
\ifnum#1=59 %
\hatcurLCimpxxxxxB
\else
\ifnum#1=60 %
\hatcurLCimpxxxxxC
\else
\ifnum#1=61 %
\hatcurLCimpxxxxxD
\else
\ifnum#1=62 %
\hatcurLCimpxxxxxE
\else
\ifnum#1=63 %
\hatcurLCimpxxxxxF
\else
\ifnum#1=64 %
\hatcurLCimpxxxxxG
\else
??????\fi
\fi
\fi
\fi
\fi
\fi
\fi
}
\newcommand{\hatcurLCingdur}[1]{\ifnum#1=58 %
\hatcurLCingdurxxxxxA
\else
\ifnum#1=59 %
\hatcurLCingdurxxxxxB
\else
\ifnum#1=60 %
\hatcurLCingdurxxxxxC
\else
\ifnum#1=61 %
\hatcurLCingdurxxxxxD
\else
\ifnum#1=62 %
\hatcurLCingdurxxxxxE
\else
\ifnum#1=63 %
\hatcurLCingdurxxxxxF
\else
\ifnum#1=64 %
\hatcurLCingdurxxxxxG
\else
??????\fi
\fi
\fi
\fi
\fi
\fi
\fi
}
\newcommand{\hatcurLCP}[1]{\ifnum#1=58 %
\hatcurLCPxxxxxA
\else
\ifnum#1=59 %
\hatcurLCPxxxxxB
\else
\ifnum#1=60 %
\hatcurLCPxxxxxC
\else
\ifnum#1=61 %
\hatcurLCPxxxxxD
\else
\ifnum#1=62 %
\hatcurLCPxxxxxE
\else
\ifnum#1=63 %
\hatcurLCPxxxxxF
\else
\ifnum#1=64 %
\hatcurLCPxxxxxG
\else
??????\fi
\fi
\fi
\fi
\fi
\fi
\fi
}
\newcommand{\hatcurLCPprec}[1]{\ifnum#1=58 %
\hatcurLCPprecxxxxxA
\else
\ifnum#1=59 %
\hatcurLCPprecxxxxxB
\else
\ifnum#1=60 %
\hatcurLCPprecxxxxxC
\else
\ifnum#1=61 %
\hatcurLCPprecxxxxxD
\else
\ifnum#1=62 %
\hatcurLCPprecxxxxxE
\else
\ifnum#1=63 %
\hatcurLCPprecxxxxxF
\else
\ifnum#1=64 %
\hatcurLCPprecxxxxxG
\else
??????\fi
\fi
\fi
\fi
\fi
\fi
\fi
}
\newcommand{\hatcurLCPshort}[1]{\ifnum#1=58 %
\hatcurLCPshortxxxxxA
\else
\ifnum#1=59 %
\hatcurLCPshortxxxxxB
\else
\ifnum#1=60 %
\hatcurLCPshortxxxxxC
\else
\ifnum#1=61 %
\hatcurLCPshortxxxxxD
\else
\ifnum#1=62 %
\hatcurLCPshortxxxxxE
\else
\ifnum#1=63 %
\hatcurLCPshortxxxxxF
\else
\ifnum#1=64 %
\hatcurLCPshortxxxxxG
\else
??????\fi
\fi
\fi
\fi
\fi
\fi
\fi
}
\newcommand{\hatcurLCq}[1]{\ifnum#1=58 %
\hatcurLCqxxxxxA
\else
\ifnum#1=59 %
\hatcurLCqxxxxxB
\else
\ifnum#1=60 %
\hatcurLCqxxxxxC
\else
\ifnum#1=61 %
\hatcurLCqxxxxxD
\else
\ifnum#1=62 %
\hatcurLCqxxxxxE
\else
\ifnum#1=63 %
\hatcurLCqxxxxxF
\else
\ifnum#1=64 %
\hatcurLCqxxxxxG
\else
??????\fi
\fi
\fi
\fi
\fi
\fi
\fi
}
\newcommand{\hatcurLCqshort}[1]{\ifnum#1=58 %
\hatcurLCqshortxxxxxA
\else
\ifnum#1=59 %
\hatcurLCqshortxxxxxB
\else
\ifnum#1=60 %
\hatcurLCqshortxxxxxC
\else
\ifnum#1=61 %
\hatcurLCqshortxxxxxD
\else
\ifnum#1=62 %
\hatcurLCqshortxxxxxE
\else
\ifnum#1=63 %
\hatcurLCqshortxxxxxF
\else
\ifnum#1=64 %
\hatcurLCqshortxxxxxG
\else
??????\fi
\fi
\fi
\fi
\fi
\fi
\fi
}
\newcommand{\hatcurLCrho}[1]{\ifnum#1=58 %
\hatcurLCrhoxxxxxA
\else
\ifnum#1=59 %
\hatcurLCrhoxxxxxB
\else
\ifnum#1=60 %
\hatcurLCrhoxxxxxC
\else
\ifnum#1=61 %
\hatcurLCrhoxxxxxD
\else
\ifnum#1=62 %
\hatcurLCrhoxxxxxE
\else
\ifnum#1=63 %
\hatcurLCrhoxxxxxF
\else
\ifnum#1=64 %
\hatcurLCrhoxxxxxG
\else
??????\fi
\fi
\fi
\fi
\fi
\fi
\fi
}
\newcommand{\hatcurLCrprstar}[1]{\ifnum#1=58 %
\hatcurLCrprstarxxxxxA
\else
\ifnum#1=59 %
\hatcurLCrprstarxxxxxB
\else
\ifnum#1=60 %
\hatcurLCrprstarxxxxxC
\else
\ifnum#1=61 %
\hatcurLCrprstarxxxxxD
\else
\ifnum#1=62 %
\hatcurLCrprstarxxxxxE
\else
\ifnum#1=63 %
\hatcurLCrprstarxxxxxF
\else
\ifnum#1=64 %
\hatcurLCrprstarxxxxxG
\else
??????\fi
\fi
\fi
\fi
\fi
\fi
\fi
}
\newcommand{\hatcurLCT}[1]{\ifnum#1=58 %
\hatcurLCTxxxxxA
\else
\ifnum#1=59 %
\hatcurLCTxxxxxB
\else
\ifnum#1=60 %
\hatcurLCTxxxxxC
\else
\ifnum#1=61 %
\hatcurLCTxxxxxD
\else
\ifnum#1=62 %
\hatcurLCTxxxxxE
\else
\ifnum#1=63 %
\hatcurLCTxxxxxF
\else
\ifnum#1=64 %
\hatcurLCTxxxxxG
\else
??????\fi
\fi
\fi
\fi
\fi
\fi
\fi
}
\newcommand{\hatcurLCTA}[1]{\ifnum#1=58 %
\hatcurLCTAxxxxxA
\else
\ifnum#1=59 %
\hatcurLCTAxxxxxB
\else
\ifnum#1=60 %
\hatcurLCTAxxxxxC
\else
\ifnum#1=61 %
\hatcurLCTAxxxxxD
\else
\ifnum#1=62 %
\hatcurLCTAxxxxxE
\else
\ifnum#1=63 %
\hatcurLCTAxxxxxF
\else
\ifnum#1=64 %
\hatcurLCTAxxxxxG
\else
??????\fi
\fi
\fi
\fi
\fi
\fi
\fi
}
\newcommand{\hatcurLCTB}[1]{\ifnum#1=58 %
\hatcurLCTBxxxxxA
\else
\ifnum#1=59 %
\hatcurLCTBxxxxxB
\else
\ifnum#1=60 %
\hatcurLCTBxxxxxC
\else
\ifnum#1=61 %
\hatcurLCTBxxxxxD
\else
\ifnum#1=62 %
\hatcurLCTBxxxxxE
\else
\ifnum#1=63 %
\hatcurLCTBxxxxxF
\else
\ifnum#1=64 %
\hatcurLCTBxxxxxG
\else
??????\fi
\fi
\fi
\fi
\fi
\fi
\fi
}
\newcommand{\hatcurLCTshort}[1]{\ifnum#1=58 %
\hatcurLCTshortxxxxxA
\else
\ifnum#1=59 %
\hatcurLCTshortxxxxxB
\else
\ifnum#1=60 %
\hatcurLCTshortxxxxxC
\else
\ifnum#1=61 %
\hatcurLCTshortxxxxxD
\else
\ifnum#1=62 %
\hatcurLCTshortxxxxxE
\else
\ifnum#1=63 %
\hatcurLCTshortxxxxxF
\else
\ifnum#1=64 %
\hatcurLCTshortxxxxxG
\else
??????\fi
\fi
\fi
\fi
\fi
\fi
\fi
}
\newcommand{\hatcurLCzeta}[1]{\ifnum#1=58 %
\hatcurLCzetaxxxxxA
\else
\ifnum#1=59 %
\hatcurLCzetaxxxxxB
\else
\ifnum#1=60 %
\hatcurLCzetaxxxxxC
\else
\ifnum#1=61 %
\hatcurLCzetaxxxxxD
\else
\ifnum#1=62 %
\hatcurLCzetaxxxxxE
\else
\ifnum#1=63 %
\hatcurLCzetaxxxxxF
\else
\ifnum#1=64 %
\hatcurLCzetaxxxxxG
\else
??????\fi
\fi
\fi
\fi
\fi
\fi
\fi
}
\newcommand{\hatcurPPaequiv}[1]{\ifnum#1=58 %
\hatcurPPaequivxxxxxA
\else
\ifnum#1=59 %
\hatcurPPaequivxxxxxB
\else
\ifnum#1=60 %
\hatcurPPaequivxxxxxC
\else
\ifnum#1=61 %
\hatcurPPaequivxxxxxD
\else
\ifnum#1=62 %
\hatcurPPaequivxxxxxE
\else
\ifnum#1=63 %
\hatcurPPaequivxxxxxF
\else
\ifnum#1=64 %
\hatcurPPaequivxxxxxG
\else
??????\fi
\fi
\fi
\fi
\fi
\fi
\fi
}
\newcommand{\hatcurPPar}[1]{\ifnum#1=58 %
\hatcurPParxxxxxA
\else
\ifnum#1=59 %
\hatcurPParxxxxxB
\else
\ifnum#1=60 %
\hatcurPParxxxxxC
\else
\ifnum#1=61 %
\hatcurPParxxxxxD
\else
\ifnum#1=62 %
\hatcurPParxxxxxE
\else
\ifnum#1=63 %
\hatcurPParxxxxxF
\else
\ifnum#1=64 %
\hatcurPParxxxxxG
\else
??????\fi
\fi
\fi
\fi
\fi
\fi
\fi
}
\newcommand{\hatcurPParel}[1]{\ifnum#1=58 %
\hatcurPParelxxxxxA
\else
\ifnum#1=59 %
\hatcurPParelxxxxxB
\else
\ifnum#1=60 %
\hatcurPParelxxxxxC
\else
\ifnum#1=61 %
\hatcurPParelxxxxxD
\else
\ifnum#1=62 %
\hatcurPParelxxxxxE
\else
\ifnum#1=63 %
\hatcurPParelxxxxxF
\else
\ifnum#1=64 %
\hatcurPParelxxxxxG
\else
??????\fi
\fi
\fi
\fi
\fi
\fi
\fi
}
\newcommand{\hatcurPPfluxap}[1]{\ifnum#1=58 %
\hatcurPPfluxapxxxxxA
\else
\ifnum#1=59 %
\hatcurPPfluxapxxxxxB
\else
\ifnum#1=60 %
\hatcurPPfluxapxxxxxC
\else
\ifnum#1=61 %
\hatcurPPfluxapxxxxxD
\else
\ifnum#1=62 %
\hatcurPPfluxapxxxxxE
\else
\ifnum#1=63 %
\hatcurPPfluxapxxxxxF
\else
\ifnum#1=64 %
\hatcurPPfluxapxxxxxG
\else
??????\fi
\fi
\fi
\fi
\fi
\fi
\fi
}
\newcommand{\hatcurPPfluxapdim}[1]{\ifnum#1=58 %
\hatcurPPfluxapdimxxxxxA
\else
\ifnum#1=59 %
\hatcurPPfluxapdimxxxxxB
\else
\ifnum#1=60 %
\hatcurPPfluxapdimxxxxxC
\else
\ifnum#1=61 %
\hatcurPPfluxapdimxxxxxD
\else
\ifnum#1=62 %
\hatcurPPfluxapdimxxxxxE
\else
\ifnum#1=63 %
\hatcurPPfluxapdimxxxxxF
\else
\ifnum#1=64 %
\hatcurPPfluxapdimxxxxxG
\else
??????\fi
\fi
\fi
\fi
\fi
\fi
\fi
}
\newcommand{\hatcurPPfluxavg}[1]{\ifnum#1=58 %
\hatcurPPfluxavgxxxxxA
\else
\ifnum#1=59 %
\hatcurPPfluxavgxxxxxB
\else
\ifnum#1=60 %
\hatcurPPfluxavgxxxxxC
\else
\ifnum#1=61 %
\hatcurPPfluxavgxxxxxD
\else
\ifnum#1=62 %
\hatcurPPfluxavgxxxxxE
\else
\ifnum#1=63 %
\hatcurPPfluxavgxxxxxF
\else
\ifnum#1=64 %
\hatcurPPfluxavgxxxxxG
\else
??????\fi
\fi
\fi
\fi
\fi
\fi
\fi
}
\newcommand{\hatcurPPfluxavgdim}[1]{\ifnum#1=58 %
\hatcurPPfluxavgdimxxxxxA
\else
\ifnum#1=59 %
\hatcurPPfluxavgdimxxxxxB
\else
\ifnum#1=60 %
\hatcurPPfluxavgdimxxxxxC
\else
\ifnum#1=61 %
\hatcurPPfluxavgdimxxxxxD
\else
\ifnum#1=62 %
\hatcurPPfluxavgdimxxxxxE
\else
\ifnum#1=63 %
\hatcurPPfluxavgdimxxxxxF
\else
\ifnum#1=64 %
\hatcurPPfluxavgdimxxxxxG
\else
??????\fi
\fi
\fi
\fi
\fi
\fi
\fi
}
\newcommand{\hatcurPPfluxavglog}[1]{\ifnum#1=58 %
\hatcurPPfluxavglogxxxxxA
\else
\ifnum#1=59 %
\hatcurPPfluxavglogxxxxxB
\else
\ifnum#1=60 %
\hatcurPPfluxavglogxxxxxC
\else
\ifnum#1=61 %
\hatcurPPfluxavglogxxxxxD
\else
\ifnum#1=62 %
\hatcurPPfluxavglogxxxxxE
\else
\ifnum#1=63 %
\hatcurPPfluxavglogxxxxxF
\else
\ifnum#1=64 %
\hatcurPPfluxavglogxxxxxG
\else
??????\fi
\fi
\fi
\fi
\fi
\fi
\fi
}
\newcommand{\hatcurPPfluxperi}[1]{\ifnum#1=58 %
\hatcurPPfluxperixxxxxA
\else
\ifnum#1=59 %
\hatcurPPfluxperixxxxxB
\else
\ifnum#1=60 %
\hatcurPPfluxperixxxxxC
\else
\ifnum#1=61 %
\hatcurPPfluxperixxxxxD
\else
\ifnum#1=62 %
\hatcurPPfluxperixxxxxE
\else
\ifnum#1=63 %
\hatcurPPfluxperixxxxxF
\else
\ifnum#1=64 %
\hatcurPPfluxperixxxxxG
\else
??????\fi
\fi
\fi
\fi
\fi
\fi
\fi
}
\newcommand{\hatcurPPfluxperidim}[1]{\ifnum#1=58 %
\hatcurPPfluxperidimxxxxxA
\else
\ifnum#1=59 %
\hatcurPPfluxperidimxxxxxB
\else
\ifnum#1=60 %
\hatcurPPfluxperidimxxxxxC
\else
\ifnum#1=61 %
\hatcurPPfluxperidimxxxxxD
\else
\ifnum#1=62 %
\hatcurPPfluxperidimxxxxxE
\else
\ifnum#1=63 %
\hatcurPPfluxperidimxxxxxF
\else
\ifnum#1=64 %
\hatcurPPfluxperidimxxxxxG
\else
??????\fi
\fi
\fi
\fi
\fi
\fi
\fi
}
\newcommand{\hatcurPPg}[1]{\ifnum#1=58 %
\hatcurPPgxxxxxA
\else
\ifnum#1=59 %
\hatcurPPgxxxxxB
\else
\ifnum#1=60 %
\hatcurPPgxxxxxC
\else
\ifnum#1=61 %
\hatcurPPgxxxxxD
\else
\ifnum#1=62 %
\hatcurPPgxxxxxE
\else
\ifnum#1=63 %
\hatcurPPgxxxxxF
\else
\ifnum#1=64 %
\hatcurPPgxxxxxG
\else
??????\fi
\fi
\fi
\fi
\fi
\fi
\fi
}
\newcommand{\hatcurPPi}[1]{\ifnum#1=58 %
\hatcurPPixxxxxA
\else
\ifnum#1=59 %
\hatcurPPixxxxxB
\else
\ifnum#1=60 %
\hatcurPPixxxxxC
\else
\ifnum#1=61 %
\hatcurPPixxxxxD
\else
\ifnum#1=62 %
\hatcurPPixxxxxE
\else
\ifnum#1=63 %
\hatcurPPixxxxxF
\else
\ifnum#1=64 %
\hatcurPPixxxxxG
\else
??????\fi
\fi
\fi
\fi
\fi
\fi
\fi
}
\newcommand{\hatcurPPlogg}[1]{\ifnum#1=58 %
\hatcurPPloggxxxxxA
\else
\ifnum#1=59 %
\hatcurPPloggxxxxxB
\else
\ifnum#1=60 %
\hatcurPPloggxxxxxC
\else
\ifnum#1=61 %
\hatcurPPloggxxxxxD
\else
\ifnum#1=62 %
\hatcurPPloggxxxxxE
\else
\ifnum#1=63 %
\hatcurPPloggxxxxxF
\else
\ifnum#1=64 %
\hatcurPPloggxxxxxG
\else
??????\fi
\fi
\fi
\fi
\fi
\fi
\fi
}
\newcommand{\hatcurPPm}[1]{\ifnum#1=58 %
\hatcurPPmxxxxxA
\else
\ifnum#1=59 %
\hatcurPPmxxxxxB
\else
\ifnum#1=60 %
\hatcurPPmxxxxxC
\else
\ifnum#1=61 %
\hatcurPPmxxxxxD
\else
\ifnum#1=62 %
\hatcurPPmxxxxxE
\else
\ifnum#1=63 %
\hatcurPPmxxxxxF
\else
\ifnum#1=64 %
\hatcurPPmxxxxxG
\else
??????\fi
\fi
\fi
\fi
\fi
\fi
\fi
}
\newcommand{\hatcurPPme}[1]{\ifnum#1=58 %
\hatcurPPmexxxxxA
\else
\ifnum#1=59 %
\hatcurPPmexxxxxB
\else
\ifnum#1=60 %
\hatcurPPmexxxxxC
\else
\ifnum#1=61 %
\hatcurPPmexxxxxD
\else
\ifnum#1=62 %
\hatcurPPmexxxxxE
\else
\ifnum#1=63 %
\hatcurPPmexxxxxF
\else
\ifnum#1=64 %
\hatcurPPmexxxxxG
\else
??????\fi
\fi
\fi
\fi
\fi
\fi
\fi
}
\newcommand{\hatcurPPmelong}[1]{\ifnum#1=58 %
\hatcurPPmelongxxxxxA
\else
\ifnum#1=59 %
\hatcurPPmelongxxxxxB
\else
\ifnum#1=60 %
\hatcurPPmelongxxxxxC
\else
\ifnum#1=61 %
\hatcurPPmelongxxxxxD
\else
\ifnum#1=62 %
\hatcurPPmelongxxxxxE
\else
\ifnum#1=63 %
\hatcurPPmelongxxxxxF
\else
\ifnum#1=64 %
\hatcurPPmelongxxxxxG
\else
??????\fi
\fi
\fi
\fi
\fi
\fi
\fi
}
\newcommand{\hatcurPPmeshort}[1]{\ifnum#1=58 %
\hatcurPPmeshortxxxxxA
\else
\ifnum#1=59 %
\hatcurPPmeshortxxxxxB
\else
\ifnum#1=60 %
\hatcurPPmeshortxxxxxC
\else
\ifnum#1=61 %
\hatcurPPmeshortxxxxxD
\else
\ifnum#1=62 %
\hatcurPPmeshortxxxxxE
\else
\ifnum#1=63 %
\hatcurPPmeshortxxxxxF
\else
\ifnum#1=64 %
\hatcurPPmeshortxxxxxG
\else
??????\fi
\fi
\fi
\fi
\fi
\fi
\fi
}
\newcommand{\hatcurPPmlong}[1]{\ifnum#1=58 %
\hatcurPPmlongxxxxxA
\else
\ifnum#1=59 %
\hatcurPPmlongxxxxxB
\else
\ifnum#1=60 %
\hatcurPPmlongxxxxxC
\else
\ifnum#1=61 %
\hatcurPPmlongxxxxxD
\else
\ifnum#1=62 %
\hatcurPPmlongxxxxxE
\else
\ifnum#1=63 %
\hatcurPPmlongxxxxxF
\else
\ifnum#1=64 %
\hatcurPPmlongxxxxxG
\else
??????\fi
\fi
\fi
\fi
\fi
\fi
\fi
}
\newcommand{\hatcurPPmrcorr}[1]{\ifnum#1=58 %
\hatcurPPmrcorrxxxxxA
\else
\ifnum#1=59 %
\hatcurPPmrcorrxxxxxB
\else
\ifnum#1=60 %
\hatcurPPmrcorrxxxxxC
\else
\ifnum#1=61 %
\hatcurPPmrcorrxxxxxD
\else
\ifnum#1=62 %
\hatcurPPmrcorrxxxxxE
\else
\ifnum#1=63 %
\hatcurPPmrcorrxxxxxF
\else
\ifnum#1=64 %
\hatcurPPmrcorrxxxxxG
\else
??????\fi
\fi
\fi
\fi
\fi
\fi
\fi
}
\newcommand{\hatcurPPmshort}[1]{\ifnum#1=58 %
\hatcurPPmshortxxxxxA
\else
\ifnum#1=59 %
\hatcurPPmshortxxxxxB
\else
\ifnum#1=60 %
\hatcurPPmshortxxxxxC
\else
\ifnum#1=61 %
\hatcurPPmshortxxxxxD
\else
\ifnum#1=62 %
\hatcurPPmshortxxxxxE
\else
\ifnum#1=63 %
\hatcurPPmshortxxxxxF
\else
\ifnum#1=64 %
\hatcurPPmshortxxxxxG
\else
??????\fi
\fi
\fi
\fi
\fi
\fi
\fi
}
\newcommand{\hatcurPPperi}[1]{\ifnum#1=58 %
\hatcurPPperixxxxxA
\else
\ifnum#1=59 %
\hatcurPPperixxxxxB
\else
\ifnum#1=60 %
\hatcurPPperixxxxxC
\else
\ifnum#1=61 %
\hatcurPPperixxxxxD
\else
\ifnum#1=62 %
\hatcurPPperixxxxxE
\else
\ifnum#1=63 %
\hatcurPPperixxxxxF
\else
\ifnum#1=64 %
\hatcurPPperixxxxxG
\else
??????\fi
\fi
\fi
\fi
\fi
\fi
\fi
}
\newcommand{\hatcurPPphiconj}[1]{\ifnum#1=58 %
\hatcurPPphiconjxxxxxA
\else
\ifnum#1=59 %
\hatcurPPphiconjxxxxxB
\else
\ifnum#1=60 %
\hatcurPPphiconjxxxxxC
\else
\ifnum#1=61 %
\hatcurPPphiconjxxxxxD
\else
\ifnum#1=62 %
\hatcurPPphiconjxxxxxE
\else
\ifnum#1=63 %
\hatcurPPphiconjxxxxxF
\else
\ifnum#1=64 %
\hatcurPPphiconjxxxxxG
\else
??????\fi
\fi
\fi
\fi
\fi
\fi
\fi
}
\newcommand{\hatcurPPr}[1]{\ifnum#1=58 %
\hatcurPPrxxxxxA
\else
\ifnum#1=59 %
\hatcurPPrxxxxxB
\else
\ifnum#1=60 %
\hatcurPPrxxxxxC
\else
\ifnum#1=61 %
\hatcurPPrxxxxxD
\else
\ifnum#1=62 %
\hatcurPPrxxxxxE
\else
\ifnum#1=63 %
\hatcurPPrxxxxxF
\else
\ifnum#1=64 %
\hatcurPPrxxxxxG
\else
??????\fi
\fi
\fi
\fi
\fi
\fi
\fi
}
\newcommand{\hatcurPPre}[1]{\ifnum#1=58 %
\hatcurPPrexxxxxA
\else
\ifnum#1=59 %
\hatcurPPrexxxxxB
\else
\ifnum#1=60 %
\hatcurPPrexxxxxC
\else
\ifnum#1=61 %
\hatcurPPrexxxxxD
\else
\ifnum#1=62 %
\hatcurPPrexxxxxE
\else
\ifnum#1=63 %
\hatcurPPrexxxxxF
\else
\ifnum#1=64 %
\hatcurPPrexxxxxG
\else
??????\fi
\fi
\fi
\fi
\fi
\fi
\fi
}
\newcommand{\hatcurPPrelong}[1]{\ifnum#1=58 %
\hatcurPPrelongxxxxxA
\else
\ifnum#1=59 %
\hatcurPPrelongxxxxxB
\else
\ifnum#1=60 %
\hatcurPPrelongxxxxxC
\else
\ifnum#1=61 %
\hatcurPPrelongxxxxxD
\else
\ifnum#1=62 %
\hatcurPPrelongxxxxxE
\else
\ifnum#1=63 %
\hatcurPPrelongxxxxxF
\else
\ifnum#1=64 %
\hatcurPPrelongxxxxxG
\else
??????\fi
\fi
\fi
\fi
\fi
\fi
\fi
}
\newcommand{\hatcurPPreshort}[1]{\ifnum#1=58 %
\hatcurPPreshortxxxxxA
\else
\ifnum#1=59 %
\hatcurPPreshortxxxxxB
\else
\ifnum#1=60 %
\hatcurPPreshortxxxxxC
\else
\ifnum#1=61 %
\hatcurPPreshortxxxxxD
\else
\ifnum#1=62 %
\hatcurPPreshortxxxxxE
\else
\ifnum#1=63 %
\hatcurPPreshortxxxxxF
\else
\ifnum#1=64 %
\hatcurPPreshortxxxxxG
\else
??????\fi
\fi
\fi
\fi
\fi
\fi
\fi
}
\newcommand{\hatcurPPrho}[1]{\ifnum#1=58 %
\hatcurPPrhoxxxxxA
\else
\ifnum#1=59 %
\hatcurPPrhoxxxxxB
\else
\ifnum#1=60 %
\hatcurPPrhoxxxxxC
\else
\ifnum#1=61 %
\hatcurPPrhoxxxxxD
\else
\ifnum#1=62 %
\hatcurPPrhoxxxxxE
\else
\ifnum#1=63 %
\hatcurPPrhoxxxxxF
\else
\ifnum#1=64 %
\hatcurPPrhoxxxxxG
\else
??????\fi
\fi
\fi
\fi
\fi
\fi
\fi
}
\newcommand{\hatcurPPrlong}[1]{\ifnum#1=58 %
\hatcurPPrlongxxxxxA
\else
\ifnum#1=59 %
\hatcurPPrlongxxxxxB
\else
\ifnum#1=60 %
\hatcurPPrlongxxxxxC
\else
\ifnum#1=61 %
\hatcurPPrlongxxxxxD
\else
\ifnum#1=62 %
\hatcurPPrlongxxxxxE
\else
\ifnum#1=63 %
\hatcurPPrlongxxxxxF
\else
\ifnum#1=64 %
\hatcurPPrlongxxxxxG
\else
??????\fi
\fi
\fi
\fi
\fi
\fi
\fi
}
\newcommand{\hatcurPPrshort}[1]{\ifnum#1=58 %
\hatcurPPrshortxxxxxA
\else
\ifnum#1=59 %
\hatcurPPrshortxxxxxB
\else
\ifnum#1=60 %
\hatcurPPrshortxxxxxC
\else
\ifnum#1=61 %
\hatcurPPrshortxxxxxD
\else
\ifnum#1=62 %
\hatcurPPrshortxxxxxE
\else
\ifnum#1=63 %
\hatcurPPrshortxxxxxF
\else
\ifnum#1=64 %
\hatcurPPrshortxxxxxG
\else
??????\fi
\fi
\fi
\fi
\fi
\fi
\fi
}
\newcommand{\hatcurPPtcirc}[1]{\ifnum#1=58 %
\hatcurPPtcircxxxxxA
\else
\ifnum#1=59 %
\hatcurPPtcircxxxxxB
\else
\ifnum#1=60 %
\hatcurPPtcircxxxxxC
\else
\ifnum#1=61 %
\hatcurPPtcircxxxxxD
\else
\ifnum#1=62 %
\hatcurPPtcircxxxxxE
\else
\ifnum#1=63 %
\hatcurPPtcircxxxxxF
\else
\ifnum#1=64 %
\hatcurPPtcircxxxxxG
\else
??????\fi
\fi
\fi
\fi
\fi
\fi
\fi
}
\newcommand{\hatcurPPteff}[1]{\ifnum#1=58 %
\hatcurPPteffxxxxxA
\else
\ifnum#1=59 %
\hatcurPPteffxxxxxB
\else
\ifnum#1=60 %
\hatcurPPteffxxxxxC
\else
\ifnum#1=61 %
\hatcurPPteffxxxxxD
\else
\ifnum#1=62 %
\hatcurPPteffxxxxxE
\else
\ifnum#1=63 %
\hatcurPPteffxxxxxF
\else
\ifnum#1=64 %
\hatcurPPteffxxxxxG
\else
??????\fi
\fi
\fi
\fi
\fi
\fi
\fi
}
\newcommand{\hatcurPPtheta}[1]{\ifnum#1=58 %
\hatcurPPthetaxxxxxA
\else
\ifnum#1=59 %
\hatcurPPthetaxxxxxB
\else
\ifnum#1=60 %
\hatcurPPthetaxxxxxC
\else
\ifnum#1=61 %
\hatcurPPthetaxxxxxD
\else
\ifnum#1=62 %
\hatcurPPthetaxxxxxE
\else
\ifnum#1=63 %
\hatcurPPthetaxxxxxF
\else
\ifnum#1=64 %
\hatcurPPthetaxxxxxG
\else
??????\fi
\fi
\fi
\fi
\fi
\fi
\fi
}
\newcommand{\hatcurPPtinfall}[1]{\ifnum#1=58 %
\hatcurPPtinfallxxxxxA
\else
\ifnum#1=59 %
\hatcurPPtinfallxxxxxB
\else
\ifnum#1=60 %
\hatcurPPtinfallxxxxxC
\else
\ifnum#1=61 %
\hatcurPPtinfallxxxxxD
\else
\ifnum#1=62 %
\hatcurPPtinfallxxxxxE
\else
\ifnum#1=63 %
\hatcurPPtinfallxxxxxF
\else
\ifnum#1=64 %
\hatcurPPtinfallxxxxxG
\else
??????\fi
\fi
\fi
\fi
\fi
\fi
\fi
}
\newcommand{\hatcurRVeccen}[1]{\ifnum#1=58 %
\hatcurRVeccenxxxxxA
\else
\ifnum#1=59 %
\hatcurRVeccenxxxxxB
\else
\ifnum#1=60 %
\hatcurRVeccenxxxxxC
\else
\ifnum#1=61 %
\hatcurRVeccenxxxxxD
\else
\ifnum#1=62 %
\hatcurRVeccenxxxxxE
\else
\ifnum#1=63 %
\hatcurRVeccenxxxxxF
\else
\ifnum#1=64 %
\hatcurRVeccenxxxxxG
\else
??????\fi
\fi
\fi
\fi
\fi
\fi
\fi
}
\newcommand{\hatcurRVeccentwosiglim}[1]{\ifnum#1=58 %
\hatcurRVeccentwosiglimxxxxxA
\else
\ifnum#1=59 %
\hatcurRVeccentwosiglimxxxxxB
\else
\ifnum#1=60 %
\hatcurRVeccentwosiglimxxxxxC
\else
\ifnum#1=61 %
\hatcurRVeccentwosiglimxxxxxD
\else
\ifnum#1=62 %
\hatcurRVeccentwosiglimxxxxxE
\else
\ifnum#1=63 %
\hatcurRVeccentwosiglimxxxxxF
\else
\ifnum#1=64 %
\hatcurRVeccentwosiglimxxxxxG
\else
??????\fi
\fi
\fi
\fi
\fi
\fi
\fi
}
\newcommand{\hatcurRVfitrms}[1]{\ifnum#1=58 %
\hatcurRVfitrmsxxxxxA
\else
\ifnum#1=62 %
\hatcurRVfitrmsxxxxxE
\else
??????\fi
\fi
}
\newcommand{\hatcurRVfitrmsA}[1]{\ifnum#1=59 %
\hatcurRVfitrmsAxxxxxB
\else
\ifnum#1=60 %
\hatcurRVfitrmsAxxxxxC
\else
\ifnum#1=61 %
\hatcurRVfitrmsAxxxxxD
\else
\ifnum#1=63 %
\hatcurRVfitrmsAxxxxxF
\else
\ifnum#1=64 %
\hatcurRVfitrmsAxxxxxG
\else
??????\fi
\fi
\fi
\fi
\fi
}
\newcommand{\hatcurRVfitrmsB}[1]{\ifnum#1=59 %
\hatcurRVfitrmsBxxxxxB
\else
\ifnum#1=60 %
\hatcurRVfitrmsBxxxxxC
\else
\ifnum#1=61 %
\hatcurRVfitrmsBxxxxxD
\else
\ifnum#1=63 %
\hatcurRVfitrmsBxxxxxF
\else
\ifnum#1=64 %
\hatcurRVfitrmsBxxxxxG
\else
??????\fi
\fi
\fi
\fi
\fi
}
\newcommand{\hatcurRVfitrmsC}[1]{\ifnum#1=60 %
\hatcurRVfitrmsCxxxxxC
\else
\ifnum#1=61 %
\hatcurRVfitrmsCxxxxxD
\else
\ifnum#1=63 %
\hatcurRVfitrmsCxxxxxF
\else
??????\fi
\fi
\fi
}
\newcommand{\hatcurRVfitrmsD}[1]{\ifnum#1=60 %
\hatcurRVfitrmsDxxxxxC
\else
??????\fi
}
\newcommand{\hatcurRVfitrmsE}[1]{\ifnum#1=60 %
\hatcurRVfitrmsExxxxxC
\else
??????\fi
}
\newcommand{\hatcurRVgamma}[1]{\ifnum#1=58 %
\hatcurRVgammaxxxxxA
\else
\ifnum#1=62 %
\hatcurRVgammaxxxxxE
\else
??????\fi
\fi
}
\newcommand{\hatcurRVgammaA}[1]{\ifnum#1=59 %
\hatcurRVgammaAxxxxxB
\else
\ifnum#1=60 %
\hatcurRVgammaAxxxxxC
\else
\ifnum#1=61 %
\hatcurRVgammaAxxxxxD
\else
\ifnum#1=63 %
\hatcurRVgammaAxxxxxF
\else
\ifnum#1=64 %
\hatcurRVgammaAxxxxxG
\else
??????\fi
\fi
\fi
\fi
\fi
}
\newcommand{\hatcurRVgammaB}[1]{\ifnum#1=59 %
\hatcurRVgammaBxxxxxB
\else
\ifnum#1=60 %
\hatcurRVgammaBxxxxxC
\else
\ifnum#1=61 %
\hatcurRVgammaBxxxxxD
\else
\ifnum#1=63 %
\hatcurRVgammaBxxxxxF
\else
\ifnum#1=64 %
\hatcurRVgammaBxxxxxG
\else
??????\fi
\fi
\fi
\fi
\fi
}
\newcommand{\hatcurRVgammaC}[1]{\ifnum#1=60 %
\hatcurRVgammaCxxxxxC
\else
\ifnum#1=61 %
\hatcurRVgammaCxxxxxD
\else
\ifnum#1=63 %
\hatcurRVgammaCxxxxxF
\else
??????\fi
\fi
\fi
}
\newcommand{\hatcurRVgammaD}[1]{\ifnum#1=60 %
\hatcurRVgammaDxxxxxC
\else
??????\fi
}
\newcommand{\hatcurRVgammaE}[1]{\ifnum#1=60 %
\hatcurRVgammaExxxxxC
\else
??????\fi
}
\newcommand{\hatcurRVh}[1]{\ifnum#1=58 %
\hatcurRVhxxxxxA
\else
\ifnum#1=59 %
\hatcurRVhxxxxxB
\else
\ifnum#1=60 %
\hatcurRVhxxxxxC
\else
\ifnum#1=61 %
\hatcurRVhxxxxxD
\else
\ifnum#1=62 %
\hatcurRVhxxxxxE
\else
\ifnum#1=63 %
\hatcurRVhxxxxxF
\else
\ifnum#1=64 %
\hatcurRVhxxxxxG
\else
??????\fi
\fi
\fi
\fi
\fi
\fi
\fi
}
\newcommand{\hatcurRVjitter}[1]{\ifnum#1=58 %
\hatcurRVjitterxxxxxA
\else
\ifnum#1=62 %
\hatcurRVjitterxxxxxE
\else
??????\fi
\fi
}
\newcommand{\hatcurRVjitterA}[1]{\ifnum#1=59 %
\hatcurRVjitterAxxxxxB
\else
\ifnum#1=60 %
\hatcurRVjitterAxxxxxC
\else
\ifnum#1=61 %
\hatcurRVjitterAxxxxxD
\else
\ifnum#1=63 %
\hatcurRVjitterAxxxxxF
\else
\ifnum#1=64 %
\hatcurRVjitterAxxxxxG
\else
??????\fi
\fi
\fi
\fi
\fi
}
\newcommand{\hatcurRVjitterB}[1]{\ifnum#1=59 %
\hatcurRVjitterBxxxxxB
\else
\ifnum#1=60 %
\hatcurRVjitterBxxxxxC
\else
\ifnum#1=61 %
\hatcurRVjitterBxxxxxD
\else
\ifnum#1=63 %
\hatcurRVjitterBxxxxxF
\else
\ifnum#1=64 %
\hatcurRVjitterBxxxxxG
\else
??????\fi
\fi
\fi
\fi
\fi
}
\newcommand{\hatcurRVjitterC}[1]{\ifnum#1=60 %
\hatcurRVjitterCxxxxxC
\else
\ifnum#1=61 %
\hatcurRVjitterCxxxxxD
\else
\ifnum#1=63 %
\hatcurRVjitterCxxxxxF
\else
??????\fi
\fi
\fi
}
\newcommand{\hatcurRVjitterD}[1]{\ifnum#1=60 %
\hatcurRVjitterDxxxxxC
\else
??????\fi
}
\newcommand{\hatcurRVjitterE}[1]{\ifnum#1=60 %
\hatcurRVjitterExxxxxC
\else
??????\fi
}
\newcommand{\hatcurRVjittertwosiglim}[1]{\ifnum#1=58 %
\hatcurRVjittertwosiglimxxxxxA
\else
\ifnum#1=62 %
\hatcurRVjittertwosiglimxxxxxE
\else
??????\fi
\fi
}
\newcommand{\hatcurRVjittertwosiglimA}[1]{\ifnum#1=59 %
\hatcurRVjittertwosiglimAxxxxxB
\else
\ifnum#1=60 %
\hatcurRVjittertwosiglimAxxxxxC
\else
\ifnum#1=61 %
\hatcurRVjittertwosiglimAxxxxxD
\else
\ifnum#1=63 %
\hatcurRVjittertwosiglimAxxxxxF
\else
\ifnum#1=64 %
\hatcurRVjittertwosiglimAxxxxxG
\else
??????\fi
\fi
\fi
\fi
\fi
}
\newcommand{\hatcurRVjittertwosiglimB}[1]{\ifnum#1=59 %
\hatcurRVjittertwosiglimBxxxxxB
\else
\ifnum#1=60 %
\hatcurRVjittertwosiglimBxxxxxC
\else
\ifnum#1=61 %
\hatcurRVjittertwosiglimBxxxxxD
\else
\ifnum#1=63 %
\hatcurRVjittertwosiglimBxxxxxF
\else
\ifnum#1=64 %
\hatcurRVjittertwosiglimBxxxxxG
\else
??????\fi
\fi
\fi
\fi
\fi
}
\newcommand{\hatcurRVjittertwosiglimC}[1]{\ifnum#1=60 %
\hatcurRVjittertwosiglimCxxxxxC
\else
\ifnum#1=61 %
\hatcurRVjittertwosiglimCxxxxxD
\else
\ifnum#1=63 %
\hatcurRVjittertwosiglimCxxxxxF
\else
??????\fi
\fi
\fi
}
\newcommand{\hatcurRVjittertwosiglimD}[1]{\ifnum#1=60 %
\hatcurRVjittertwosiglimDxxxxxC
\else
??????\fi
}
\newcommand{\hatcurRVjittertwosiglimE}[1]{\ifnum#1=60 %
\hatcurRVjittertwosiglimExxxxxC
\else
??????\fi
}
\newcommand{\hatcurRVk}[1]{\ifnum#1=58 %
\hatcurRVkxxxxxA
\else
\ifnum#1=59 %
\hatcurRVkxxxxxB
\else
\ifnum#1=60 %
\hatcurRVkxxxxxC
\else
\ifnum#1=61 %
\hatcurRVkxxxxxD
\else
\ifnum#1=62 %
\hatcurRVkxxxxxE
\else
\ifnum#1=63 %
\hatcurRVkxxxxxF
\else
\ifnum#1=64 %
\hatcurRVkxxxxxG
\else
??????\fi
\fi
\fi
\fi
\fi
\fi
\fi
}
\newcommand{\hatcurRVK}[1]{\ifnum#1=58 %
\hatcurRVKxxxxxA
\else
\ifnum#1=59 %
\hatcurRVKxxxxxB
\else
\ifnum#1=60 %
\hatcurRVKxxxxxC
\else
\ifnum#1=61 %
\hatcurRVKxxxxxD
\else
\ifnum#1=62 %
\hatcurRVKxxxxxE
\else
\ifnum#1=63 %
\hatcurRVKxxxxxF
\else
\ifnum#1=64 %
\hatcurRVKxxxxxG
\else
??????\fi
\fi
\fi
\fi
\fi
\fi
\fi
}
\newcommand{\hatcurRVomega}[1]{\ifnum#1=58 %
\hatcurRVomegaxxxxxA
\else
\ifnum#1=59 %
\hatcurRVomegaxxxxxB
\else
\ifnum#1=60 %
\hatcurRVomegaxxxxxC
\else
\ifnum#1=61 %
\hatcurRVomegaxxxxxD
\else
\ifnum#1=62 %
\hatcurRVomegaxxxxxE
\else
\ifnum#1=63 %
\hatcurRVomegaxxxxxF
\else
\ifnum#1=64 %
\hatcurRVomegaxxxxxG
\else
??????\fi
\fi
\fi
\fi
\fi
\fi
\fi
}
\newcommand{\hatcurRVrh}[1]{\ifnum#1=58 %
\hatcurRVrhxxxxxA
\else
\ifnum#1=59 %
\hatcurRVrhxxxxxB
\else
\ifnum#1=60 %
\hatcurRVrhxxxxxC
\else
\ifnum#1=61 %
\hatcurRVrhxxxxxD
\else
\ifnum#1=62 %
\hatcurRVrhxxxxxE
\else
\ifnum#1=63 %
\hatcurRVrhxxxxxF
\else
\ifnum#1=64 %
\hatcurRVrhxxxxxG
\else
??????\fi
\fi
\fi
\fi
\fi
\fi
\fi
}
\newcommand{\hatcurRVrk}[1]{\ifnum#1=58 %
\hatcurRVrkxxxxxA
\else
\ifnum#1=59 %
\hatcurRVrkxxxxxB
\else
\ifnum#1=60 %
\hatcurRVrkxxxxxC
\else
\ifnum#1=61 %
\hatcurRVrkxxxxxD
\else
\ifnum#1=62 %
\hatcurRVrkxxxxxE
\else
\ifnum#1=63 %
\hatcurRVrkxxxxxF
\else
\ifnum#1=64 %
\hatcurRVrkxxxxxG
\else
??????\fi
\fi
\fi
\fi
\fi
\fi
\fi
}
\newcommand{\hatcurRVtrone}[1]{\ifnum#1=58 %
\hatcurRVtronexxxxxA
\else
\ifnum#1=59 %
\hatcurRVtronexxxxxB
\else
\ifnum#1=60 %
\hatcurRVtronexxxxxC
\else
\ifnum#1=61 %
\hatcurRVtronexxxxxD
\else
\ifnum#1=62 %
\hatcurRVtronexxxxxE
\else
\ifnum#1=63 %
\hatcurRVtronexxxxxF
\else
\ifnum#1=64 %
\hatcurRVtronexxxxxG
\else
??????\fi
\fi
\fi
\fi
\fi
\fi
\fi
}
\newcommand{\hatcurRVtrtwo}[1]{\ifnum#1=58 %
\hatcurRVtrtwoxxxxxA
\else
\ifnum#1=59 %
\hatcurRVtrtwoxxxxxB
\else
\ifnum#1=60 %
\hatcurRVtrtwoxxxxxC
\else
\ifnum#1=61 %
\hatcurRVtrtwoxxxxxD
\else
\ifnum#1=62 %
\hatcurRVtrtwoxxxxxE
\else
\ifnum#1=63 %
\hatcurRVtrtwoxxxxxF
\else
\ifnum#1=64 %
\hatcurRVtrtwoxxxxxG
\else
??????\fi
\fi
\fi
\fi
\fi
\fi
\fi
}
\newcommand{\hatcurSMEiilogg}[1]{\ifnum#1=58 %
\hatcurSMEiiloggxxxxxA
\else
\ifnum#1=59 %
\hatcurSMEiiloggxxxxxB
\else
\ifnum#1=60 %
\hatcurSMEiiloggxxxxxC
\else
\ifnum#1=61 %
\hatcurSMEiiloggxxxxxD
\else
\ifnum#1=62 %
\hatcurSMEiiloggxxxxxE
\else
??????\fi
\fi
\fi
\fi
\fi
}
\newcommand{\hatcurSMEiiteff}[1]{\ifnum#1=58 %
\hatcurSMEiiteffxxxxxA
\else
\ifnum#1=59 %
\hatcurSMEiiteffxxxxxB
\else
\ifnum#1=60 %
\hatcurSMEiiteffxxxxxC
\else
\ifnum#1=61 %
\hatcurSMEiiteffxxxxxD
\else
\ifnum#1=62 %
\hatcurSMEiiteffxxxxxE
\else
??????\fi
\fi
\fi
\fi
\fi
}
\newcommand{\hatcurSMEiivsin}[1]{\ifnum#1=58 %
\hatcurSMEiivsinxxxxxA
\else
\ifnum#1=59 %
\hatcurSMEiivsinxxxxxB
\else
\ifnum#1=60 %
\hatcurSMEiivsinxxxxxC
\else
\ifnum#1=61 %
\hatcurSMEiivsinxxxxxD
\else
\ifnum#1=62 %
\hatcurSMEiivsinxxxxxE
\else
??????\fi
\fi
\fi
\fi
\fi
}
\newcommand{\hatcurSMEiizfeh}[1]{\ifnum#1=58 %
\hatcurSMEiizfehxxxxxA
\else
\ifnum#1=59 %
\hatcurSMEiizfehxxxxxB
\else
\ifnum#1=60 %
\hatcurSMEiizfehxxxxxC
\else
\ifnum#1=61 %
\hatcurSMEiizfehxxxxxD
\else
\ifnum#1=62 %
\hatcurSMEiizfehxxxxxE
\else
??????\fi
\fi
\fi
\fi
\fi
}
\newcommand{\hatcurSMEiizfehshort}[1]{\ifnum#1=58 %
\hatcurSMEiizfehshortxxxxxA
\else
\ifnum#1=59 %
\hatcurSMEiizfehshortxxxxxB
\else
\ifnum#1=60 %
\hatcurSMEiizfehshortxxxxxC
\else
\ifnum#1=61 %
\hatcurSMEiizfehshortxxxxxD
\else
\ifnum#1=62 %
\hatcurSMEiizfehshortxxxxxE
\else
??????\fi
\fi
\fi
\fi
\fi
}
\newcommand{\hatcurSMEilogg}[1]{\ifnum#1=58 %
\hatcurSMEiloggxxxxxA
\else
\ifnum#1=59 %
\hatcurSMEiloggxxxxxB
\else
\ifnum#1=60 %
\hatcurSMEiloggxxxxxC
\else
\ifnum#1=61 %
\hatcurSMEiloggxxxxxD
\else
\ifnum#1=62 %
\hatcurSMEiloggxxxxxE
\else
\ifnum#1=63 %
\hatcurSMEiloggxxxxxF
\else
\ifnum#1=64 %
\hatcurSMEiloggxxxxxG
\else
??????\fi
\fi
\fi
\fi
\fi
\fi
\fi
}
\newcommand{\hatcurSMEiteff}[1]{\ifnum#1=58 %
\hatcurSMEiteffxxxxxA
\else
\ifnum#1=59 %
\hatcurSMEiteffxxxxxB
\else
\ifnum#1=60 %
\hatcurSMEiteffxxxxxC
\else
\ifnum#1=61 %
\hatcurSMEiteffxxxxxD
\else
\ifnum#1=62 %
\hatcurSMEiteffxxxxxE
\else
\ifnum#1=63 %
\hatcurSMEiteffxxxxxF
\else
\ifnum#1=64 %
\hatcurSMEiteffxxxxxG
\else
??????\fi
\fi
\fi
\fi
\fi
\fi
\fi
}
\newcommand{\hatcurSMEivmac}[1]{\ifnum#1=58 %
\hatcurSMEivmacxxxxxA
\else
\ifnum#1=59 %
\hatcurSMEivmacxxxxxB
\else
\ifnum#1=60 %
\hatcurSMEivmacxxxxxC
\else
\ifnum#1=61 %
\hatcurSMEivmacxxxxxD
\else
\ifnum#1=62 %
\hatcurSMEivmacxxxxxE
\else
\ifnum#1=63 %
\hatcurSMEivmacxxxxxF
\else
\ifnum#1=64 %
\hatcurSMEivmacxxxxxG
\else
??????\fi
\fi
\fi
\fi
\fi
\fi
\fi
}
\newcommand{\hatcurSMEivmic}[1]{\ifnum#1=58 %
\hatcurSMEivmicxxxxxA
\else
\ifnum#1=59 %
\hatcurSMEivmicxxxxxB
\else
\ifnum#1=60 %
\hatcurSMEivmicxxxxxC
\else
\ifnum#1=61 %
\hatcurSMEivmicxxxxxD
\else
\ifnum#1=62 %
\hatcurSMEivmicxxxxxE
\else
\ifnum#1=63 %
\hatcurSMEivmicxxxxxF
\else
\ifnum#1=64 %
\hatcurSMEivmicxxxxxG
\else
??????\fi
\fi
\fi
\fi
\fi
\fi
\fi
}
\newcommand{\hatcurSMEivsin}[1]{\ifnum#1=58 %
\hatcurSMEivsinxxxxxA
\else
\ifnum#1=59 %
\hatcurSMEivsinxxxxxB
\else
\ifnum#1=60 %
\hatcurSMEivsinxxxxxC
\else
\ifnum#1=61 %
\hatcurSMEivsinxxxxxD
\else
\ifnum#1=62 %
\hatcurSMEivsinxxxxxE
\else
\ifnum#1=63 %
\hatcurSMEivsinxxxxxF
\else
\ifnum#1=64 %
\hatcurSMEivsinxxxxxG
\else
??????\fi
\fi
\fi
\fi
\fi
\fi
\fi
}
\newcommand{\hatcurSMEizfeh}[1]{\ifnum#1=58 %
\hatcurSMEizfehxxxxxA
\else
\ifnum#1=59 %
\hatcurSMEizfehxxxxxB
\else
\ifnum#1=60 %
\hatcurSMEizfehxxxxxC
\else
\ifnum#1=61 %
\hatcurSMEizfehxxxxxD
\else
\ifnum#1=62 %
\hatcurSMEizfehxxxxxE
\else
\ifnum#1=63 %
\hatcurSMEizfehxxxxxF
\else
\ifnum#1=64 %
\hatcurSMEizfehxxxxxG
\else
??????\fi
\fi
\fi
\fi
\fi
\fi
\fi
}
\newcommand{\hatcurSMEizfehshort}[1]{\ifnum#1=58 %
\hatcurSMEizfehshortxxxxxA
\else
\ifnum#1=59 %
\hatcurSMEizfehshortxxxxxB
\else
\ifnum#1=60 %
\hatcurSMEizfehshortxxxxxC
\else
\ifnum#1=61 %
\hatcurSMEizfehshortxxxxxD
\else
\ifnum#1=62 %
\hatcurSMEizfehshortxxxxxE
\else
\ifnum#1=63 %
\hatcurSMEizfehshortxxxxxF
\else
\ifnum#1=64 %
\hatcurSMEizfehshortxxxxxG
\else
??????\fi
\fi
\fi
\fi
\fi
\fi
\fi
}
\newcommand{\hatcurXAv}[1]{\ifnum#1=58 %
\hatcurXAvxxxxxA
\else
\ifnum#1=59 %
\hatcurXAvxxxxxB
\else
\ifnum#1=60 %
\hatcurXAvxxxxxC
\else
\ifnum#1=61 %
\hatcurXAvxxxxxD
\else
\ifnum#1=62 %
\hatcurXAvxxxxxE
\else
\ifnum#1=63 %
\hatcurXAvxxxxxF
\else
\ifnum#1=64 %
\hatcurXAvxxxxxG
\else
??????\fi
\fi
\fi
\fi
\fi
\fi
\fi
}
\newcommand{\hatcurXdist}[1]{\ifnum#1=58 %
\hatcurXdistxxxxxA
\else
\ifnum#1=59 %
\hatcurXdistxxxxxB
\else
\ifnum#1=60 %
\hatcurXdistxxxxxC
\else
\ifnum#1=61 %
\hatcurXdistxxxxxD
\else
\ifnum#1=62 %
\hatcurXdistxxxxxE
\else
\ifnum#1=63 %
\hatcurXdistxxxxxF
\else
\ifnum#1=64 %
\hatcurXdistxxxxxG
\else
??????\fi
\fi
\fi
\fi
\fi
\fi
\fi
}
\newcommand{\hatcurXdistred}[1]{\ifnum#1=58 %
\hatcurXdistredxxxxxA
\else
\ifnum#1=59 %
\hatcurXdistredxxxxxB
\else
\ifnum#1=60 %
\hatcurXdistredxxxxxC
\else
\ifnum#1=61 %
\hatcurXdistredxxxxxD
\else
\ifnum#1=62 %
\hatcurXdistredxxxxxE
\else
\ifnum#1=63 %
\hatcurXdistredxxxxxF
\else
\ifnum#1=64 %
\hatcurXdistredxxxxxG
\else
??????\fi
\fi
\fi
\fi
\fi
\fi
\fi
}
\newcommand{\hatcurXEBV}[1]{\ifnum#1=58 %
\hatcurXEBVxxxxxA
\else
\ifnum#1=59 %
\hatcurXEBVxxxxxB
\else
\ifnum#1=60 %
\hatcurXEBVxxxxxC
\else
\ifnum#1=61 %
\hatcurXEBVxxxxxD
\else
\ifnum#1=62 %
\hatcurXEBVxxxxxE
\else
\ifnum#1=63 %
\hatcurXEBVxxxxxF
\else
\ifnum#1=64 %
\hatcurXEBVxxxxxG
\else
??????\fi
\fi
\fi
\fi
\fi
\fi
\fi
}
\newcommand{\hatcurXsecdur}[1]{\ifnum#1=58 %
\hatcurXsecdurxxxxxA
\else
\ifnum#1=59 %
\hatcurXsecdurxxxxxB
\else
\ifnum#1=60 %
\hatcurXsecdurxxxxxC
\else
\ifnum#1=61 %
\hatcurXsecdurxxxxxD
\else
\ifnum#1=62 %
\hatcurXsecdurxxxxxE
\else
\ifnum#1=63 %
\hatcurXsecdurxxxxxF
\else
\ifnum#1=64 %
\hatcurXsecdurxxxxxG
\else
??????\fi
\fi
\fi
\fi
\fi
\fi
\fi
}
\newcommand{\hatcurXsecingdur}[1]{\ifnum#1=58 %
\hatcurXsecingdurxxxxxA
\else
\ifnum#1=59 %
\hatcurXsecingdurxxxxxB
\else
\ifnum#1=60 %
\hatcurXsecingdurxxxxxC
\else
\ifnum#1=61 %
\hatcurXsecingdurxxxxxD
\else
\ifnum#1=62 %
\hatcurXsecingdurxxxxxE
\else
\ifnum#1=63 %
\hatcurXsecingdurxxxxxF
\else
\ifnum#1=64 %
\hatcurXsecingdurxxxxxG
\else
??????\fi
\fi
\fi
\fi
\fi
\fi
\fi
}
\newcommand{\hatcurXsecondary}[1]{\ifnum#1=58 %
\hatcurXsecondaryxxxxxA
\else
\ifnum#1=59 %
\hatcurXsecondaryxxxxxB
\else
\ifnum#1=60 %
\hatcurXsecondaryxxxxxC
\else
\ifnum#1=61 %
\hatcurXsecondaryxxxxxD
\else
\ifnum#1=62 %
\hatcurXsecondaryxxxxxE
\else
\ifnum#1=63 %
\hatcurXsecondaryxxxxxF
\else
\ifnum#1=64 %
\hatcurXsecondaryxxxxxG
\else
??????\fi
\fi
\fi
\fi
\fi
\fi
\fi
}
\newcommand{\hatcurXsecphase}[1]{\ifnum#1=58 %
\hatcurXsecphasexxxxxA
\else
\ifnum#1=59 %
\hatcurXsecphasexxxxxB
\else
\ifnum#1=60 %
\hatcurXsecphasexxxxxC
\else
\ifnum#1=61 %
\hatcurXsecphasexxxxxD
\else
\ifnum#1=62 %
\hatcurXsecphasexxxxxE
\else
\ifnum#1=63 %
\hatcurXsecphasexxxxxF
\else
\ifnum#1=64 %
\hatcurXsecphasexxxxxG
\else
??????\fi
\fi
\fi
\fi
\fi
\fi
\fi
}

%% file: HTR062-011.tex
\newcommand{\hatcurhtrxxxxxA}{HTR062-011}                       
\newcommand{\hatcurfieldxxxxxA}{\ensuremath{string}}            
\newcommand{\hatcurCCraxxxxxA}{\ensuremath{04^{\mathrm h}35^{\mathrm m}23.1828{\mathrm s}}}                   
\newcommand{\hatcurCCdecxxxxxA}{\ensuremath{+56{\arcdeg}52{\arcmin}05.5848{\arcsec}}}                 
\newcommand{\hatcurCCmagxxxxxA}{12.971}                         
\newcommand{\hatcurCCtwomassxxxxxA}{2MASS~04352318+5652055}     
\newcommand{\hatcurCCgscxxxxxA}{GSC~3740-01482}                 
\newcommand{\hatcurCCgaiaxxxxxA}{GAIA~277493610746748800}       
\newcommand{\hatcurCCgaiadrtwoxxxxxA}{277493615044741376} 
\newcommand{\hatcurCCtassmvxxxxxA}{\ensuremath{12.971\pm0.073}} 
\newcommand{\hatcurCCtassmvshortxxxxxA}{\ensuremath{13.0}}      
\newcommand{\hatcurCCtassmBxxxxxA}{\ensuremath{13.690\pm0.089}} 
\newcommand{\hatcurCCtassmBshortxxxxxA}{\ensuremath{13.7}}      
\newcommand{\hatcurCCtassmIxxxxxA}{\ensuremath{nff\pmnff}}      
\newcommand{\hatcurCCtassmIshortxxxxxA}{\ensuremath{0.0}}       
\newcommand{\hatcurCCtassmgxxxxxA}{\ensuremath{13.28\pm0.12}}   
\newcommand{\hatcurCCtassmgshortxxxxxA}{\ensuremath{13.3}}      
\newcommand{\hatcurCCtassmrxxxxxA}{\ensuremath{12.74\pm0.12}}   
\newcommand{\hatcurCCtassmrshortxxxxxA}{\ensuremath{12.7}}      
\newcommand{\hatcurCCtassmixxxxxA}{\ensuremath{12.50\pm0.12}}   
\newcommand{\hatcurCCtassmishortxxxxxA}{\ensuremath{12.5}}      
\newcommand{\hatcurCCparallaxxxxxxA}{\ensuremath{1.912\pm0.047}} 
\newcommand{\hatcurCCgaiamGxxxxxA}{\ensuremath{12.72020\pm0.00020}} 
\newcommand{\hatcurCCgaiamBPxxxxxA}{\ensuremath{13.1422\pm0.0016}} 
\newcommand{\hatcurCCgaiamRPxxxxxA}{\ensuremath{12.13470\pm0.00090}} 
\newcommand{\hatcurCCtwomassJmagxxxxxA}{\ensuremath{11.429\pm0.022}} 
\newcommand{\hatcurCCtwomassHmagxxxxxA}{\ensuremath{11.075\pm0.020}} 
\newcommand{\hatcurCCtwomassKmagxxxxxA}{\ensuremath{10.978\pm0.023}} 
\newcommand{\hatcurCCcitJmagxxxxxA}{\ensuremath{11.441\pm0.022}} 
\newcommand{\hatcurCCcitHmagxxxxxA}{\ensuremath{11.069\pm0.020}} 
\newcommand{\hatcurCCcitKmagxxxxxA}{\ensuremath{11.002\pm0.023}} 
\newcommand{\hatcurCCbbJmagxxxxxA}{\ensuremath{11.498\pm0.024}} 
\newcommand{\hatcurCCbbHmagxxxxxA}{\ensuremath{11.091\pm0.021}} 
\newcommand{\hatcurCCbbKmagxxxxxA}{\ensuremath{11.022\pm0.023}} 
\newcommand{\hatcurCCesoJmagxxxxxA}{\ensuremath{11.501\pm0.026}} 
\newcommand{\hatcurCCesoHmagxxxxxA}{\ensuremath{11.087\pm0.026}} 
\newcommand{\hatcurCCesoKmagxxxxxA}{\ensuremath{11.021\pm0.024}} 
\newcommand{\hatcurCCesoJHmagxxxxxA}{\ensuremath{0.413\pm0.034}} 
\newcommand{\hatcurCCesoJKmagxxxxxA}{\ensuremath{0.481\pm0.035}} 
\newcommand{\hatcurCCesoHKmagxxxxxA}{\ensuremath{0.067\pm0.035}} 
\newcommand{\hatcurCCWonemagxxxxxA}{\ensuremath{10.856\pm0.022}} 
\newcommand{\hatcurCCWtwomagxxxxxA}{\ensuremath{10.906\pm0.021}} 
\newcommand{\hatcurCCWthreemagxxxxxA}{\ensuremath{10.359\pm0.061}} 
\newcommand{\hatcurCCWfourmagxxxxxA}{\ensuremath{0\pm0}}        
\newcommand{\hatcurLCdipxxxxxA}{\ensuremath{7.4}}               
\newcommand{\hatcurLCrprstarxxxxxA}{\ensuremath{0.0895\pm0.0017}} 
\newcommand{\hatcurLCbsqxxxxxA}{\ensuremath{0.285_{-0.032}^{+0.033}}} 
\newcommand{\hatcurLCimpxxxxxA}{\ensuremath{0.534_{-0.031}^{+0.030}}} 
\newcommand{\hatcurLCzetaxxxxxA}{\ensuremath{13.00\pm0.10}}     
\newcommand{\hatcurLCdurxxxxxA}{\ensuremath{0.1729\pm0.0015}}   
\newcommand{\hatcurLCdurshortxxxxxA}{\ensuremath{0.1729}}       
\newcommand{\hatcurLCdurhrxxxxxA}{\ensuremath{4.150\pm0.035}}   
\newcommand{\hatcurLCdurhrshortxxxxxA}{\ensuremath{4.150}}      
\newcommand{\hatcurLCqxxxxxA}{\ensuremath{0.04310\pm0.00036}}   
\newcommand{\hatcurLCqshortxxxxxA}{\ensuremath{0.043}}          
\newcommand{\hatcurLCingdurxxxxxA}{\ensuremath{0.0193\pm0.0010}} 
\newcommand{\hatcurLCPxxxxxA}{\ensuremath{4.0138379\pm0.0000024}} 
\newcommand{\hatcurLCPprecxxxxxA}{\ensuremath{4.0138379}}       
\newcommand{\hatcurLCPshortxxxxxA}{\ensuremath{4.0138}}         
\newcommand{\hatcurLCTshortxxxxxA}{\ensuremath{7369.03094\pm0.00056}} 
\newcommand{\hatcurLCTxxxxxA}{\ensuremath{2457369.03094\pm0.00056}} 
\newcommand{\hatcurLCTAxxxxxA}{\ensuremath{2456132.76881\pm0.00081}} 
\newcommand{\hatcurLCTBxxxxxA}{\ensuremath{2458838.0956\pm0.0011}} 
\newcommand{\hatcurLChatnetmAxxxxxA}{\ensuremath{12.85295\pm0.00012}} 
\newcommand{\hatcurLCiblendAxxxxxA}{\ensuremath{1\pm0}}         
\newcommand{\hatcurLChatnetmBxxxxxA}{\ensuremath{0.000040\pm0.000036}} 
\newcommand{\hatcurLCiblendBxxxxxA}{\ensuremath{0.940\pm0.038}} 
\newcommand{\hatcurLChatnetmCxxxxxA}{\ensuremath{10.49539\pm0.00068}} 
\newcommand{\hatcurLCiblendCxxxxxA}{\ensuremath{-0.01137\pm0.00083}} 
\newcommand{\hatcurLCrhoxxxxxA}{\ensuremath{0.405\pm0.026}}     
\newcommand{\hatcurSMEiteffxxxxxA}{\ensuremath{5947\pm50}}      
\newcommand{\hatcurSMEizfehxxxxxA}{\ensuremath{0.032\pm0.080}}  
\newcommand{\hatcurSMEizfehshortxxxxxA}{\ensuremath{0.03}}      
\newcommand{\hatcurSMEiloggxxxxxA}{\ensuremath{4.24\pm0.10}}    
\newcommand{\hatcurSMEivsinxxxxxA}{\ensuremath{4.81\pm0.50}}    
\newcommand{\hatcurSMEivmacxxxxxA}{\ensuremath{nff\pmnff}}      
\newcommand{\hatcurSMEivmicxxxxxA}{\ensuremath{nff\pmnff}}      
\newcommand{\hatcurSMEiiteffxxxxxA}{\ensuremath{5931\pm50}}     
\newcommand{\hatcurSMEiizfehxxxxxA}{\ensuremath{0.012\pm0.080}} 
\newcommand{\hatcurSMEiizfehshortxxxxxA}{\ensuremath{0.012}}    
\newcommand{\hatcurSMEiiloggxxxxxA}{\ensuremath{4.183\pm0.036}} 
\newcommand{\hatcurSMEiivsinxxxxxA}{\ensuremath{4.91\pm0.50}}   
\newcommand{\hatcurLBiBxxxxxA}{\ensuremath{0.6762}}             
\newcommand{\hatcurLBiiBxxxxxA}{\ensuremath{0.1438}}            
\newcommand{\hatcurLBiVxxxxxA}{\ensuremath{0.5276}}             
\newcommand{\hatcurLBiiVxxxxxA}{\ensuremath{0.1912}}            
\newcommand{\hatcurLBiRxxxxxA}{\ensuremath{0.4418}}             
\newcommand{\hatcurLBiiRxxxxxA}{\ensuremath{0.1930}}            
\newcommand{\hatcurLBiIxxxxxA}{\ensuremath{0.3492}}             
\newcommand{\hatcurLBiiIxxxxxA}{\ensuremath{0.1958}}            
\newcommand{\hatcurLBiuxxxxxA}{\ensuremath{0.8162}}             
\newcommand{\hatcurLBiiuxxxxxA}{\ensuremath{0.0195}}            
\newcommand{\hatcurLBigxxxxxA}{\ensuremath{0.6254}}             
\newcommand{\hatcurLBiigxxxxxA}{\ensuremath{0.1537}}            
\newcommand{\hatcurLBirxxxxxA}{\ensuremath{0.23\pm0.14}}        
\newcommand{\hatcurLBiirxxxxxA}{\ensuremath{0.25\pm0.17}}       
\newcommand{\hatcurLBiixxxxxA}{\ensuremath{0.18\pm0.10}}        
\newcommand{\hatcurLBiiixxxxxA}{\ensuremath{0.14\pm0.15}}       
\newcommand{\hatcurLBizxxxxxA}{\ensuremath{0.3049}}             
\newcommand{\hatcurLBiizxxxxxA}{\ensuremath{0.1982}}            
\newcommand{\hatcurLBiJxxxxxA}{\ensuremath{0.2016}}             
\newcommand{\hatcurLBiiJxxxxxA}{\ensuremath{0.2236}}            
\newcommand{\hatcurLBiHxxxxxA}{\ensuremath{0.1199}}             
\newcommand{\hatcurLBiiHxxxxxA}{\ensuremath{0.2524}}            
\newcommand{\hatcurLBiKxxxxxA}{\ensuremath{0.1167}}             
\newcommand{\hatcurLBiiKxxxxxA}{\ensuremath{0.1968}}            
\newcommand{\hatcurLBiTxxxxxA}{\ensuremath{0.26\pm0.13}}        
\newcommand{\hatcurLBiiTxxxxxA}{\ensuremath{0.24\pm0.16}}       
\newcommand{\hatcurLBikepxxxxxA}{\ensuremath{0.4627}}           
\newcommand{\hatcurLBiikepxxxxxA}{\ensuremath{0.1998}}          
\newcommand{\hatcurLBiCxxxxxA}{\ensuremath{0.4546}}             
\newcommand{\hatcurLBiiCxxxxxA}{\ensuremath{0.1978}}            
\newcommand{\hatcurLBiMxxxxxA}{\ensuremath{0.5418}}             
\newcommand{\hatcurLBiiMxxxxxA}{\ensuremath{0.1801}}            
\newcommand{\hatcurLBiSonexxxxxA}{\ensuremath{0.0933}}            
\newcommand{\hatcurLBiiSonexxxxxA}{\ensuremath{0.1187}}           
\newcommand{\hatcurLBiStwoxxxxxA}{\ensuremath{0.0782}}            
\newcommand{\hatcurLBiiStwoxxxxxA}{\ensuremath{0.1011}}           
\newcommand{\hatcurLBiSthreexxxxxA}{\ensuremath{0.0662}}            
\newcommand{\hatcurLBiiSthreexxxxxA}{\ensuremath{0.0830}}           
\newcommand{\hatcurLBiSfourxxxxxA}{\ensuremath{0.0500}}            
\newcommand{\hatcurLBiiSfourxxxxxA}{\ensuremath{0.0658}}           
\newcommand{\hatcurISOmxxxxxA}{\ensuremath{1.031\pm0.028}}      
\newcommand{\hatcurISOmshortxxxxxA}{\ensuremath{1.03}}          
\newcommand{\hatcurISOmlongxxxxxA}{\ensuremath{1.031\pm0.028}}  
\newcommand{\hatcurISOrxxxxxA}{\ensuremath{1.530\pm0.034}}      
\newcommand{\hatcurISOrshortxxxxxA}{\ensuremath{1.53}}          
\newcommand{\hatcurISOrlongxxxxxA}{\ensuremath{1.530\pm0.034}}  
\newcommand{\hatcurISOrhoxxxxxA}{\ensuremath{0.405\pm0.026}}    
\newcommand{\hatcurISOrholongxxxxxA}{\ensuremath{0.405\pm0.026}} 
\newcommand{\hatcurISOloggxxxxxA}{\ensuremath{4.082\pm0.020}}   
\newcommand{\hatcurISOlumxxxxxA}{\ensuremath{2.86\pm0.15}}      
\newcommand{\hatcurISOlumshortxxxxxA}{\ensuremath{2.86}}        
\newcommand{\hatcurISOteffxxxxxA}{\ensuremath{6078\pm48}}       
\newcommand{\hatcurISOzfehxxxxxA}{\ensuremath{-0.224\pm0.057}}  
\newcommand{\hatcurISOagexxxxxA}{\ensuremath{7.11_{-0.72}^{+0.27}}} 
\newcommand{\hatcurISOspecxxxxxA}{G}                            
\newcommand{\hatcurRVKxxxxxA}{\ensuremath{46.4\pm3.6}}          
\newcommand{\hatcurRVrkxxxxxA}{\ensuremath{0\pm0}}              
\newcommand{\hatcurRVrhxxxxxA}{\ensuremath{0\pm0}}              
\newcommand{\hatcurRVkxxxxxA}{\ensuremath{0\pm0}}               
\newcommand{\hatcurRVhxxxxxA}{\ensuremath{0\pm0}}               
\newcommand{\hatcurRVtronexxxxxA}{\ensuremath{0\pm0}}           
\newcommand{\hatcurRVtrtwoxxxxxA}{\ensuremath{0\pm0}}           
\newcommand{\hatcurRVgammaxxxxxA}{\ensuremath{2.5\pm2.9}}       
\newcommand{\hatcurRVjitterxxxxxA}{\ensuremath{4.6\pm4.1}}      
\newcommand{\hatcurRVjittertwosiglimxxxxxA}{\ensuremath{<12.6}} 
\newcommand{\hatcurRVfitrmsxxxxxA}{\ensuremath{.1fym}}          %
\newcommand{\hatcurRVeccenxxxxxA}{\ensuremath{0\pm0}}           
\newcommand{\hatcurRVeccentwosiglimxxxxxA}{\ensuremath{<0.000}} 
\newcommand{\hatcurRVomegaxxxxxA}{\ensuremath{0\pm0}}           
\newcommand{\hatcurPPixxxxxA}{\ensuremath{85.64\pm0.34}}        
\newcommand{\hatcurPPgxxxxxA}{\ensuremath{5.18\pm0.53}}         
\newcommand{\hatcurPPloggxxxxxA}{\ensuremath{2.714\pm0.045}}    
\newcommand{\hatcurPParxxxxxA}{\ensuremath{7.02\pm0.15}}        
\newcommand{\hatcurPParelxxxxxA}{\ensuremath{0.04994\pm0.00044}} 
\newcommand{\hatcurPPrhoxxxxxA}{\ensuremath{0.194\pm0.024}}     
\newcommand{\hatcurPPmxxxxxA}{\ensuremath{0.372\pm0.030}}       
\newcommand{\hatcurPPmshortxxxxxA}{\ensuremath{0.37}}           
\newcommand{\hatcurPPmlongxxxxxA}{\ensuremath{0.372\pm0.030}}   
\newcommand{\hatcurPPmexxxxxA}{\ensuremath{118.2\pm9.5}}        
\newcommand{\hatcurPPmeshortxxxxxA}{\ensuremath{118.2}}         
\newcommand{\hatcurPPmelongxxxxxA}{\ensuremath{118.2\pm9.5}}    
\newcommand{\hatcurPPrxxxxxA}{\ensuremath{1.332\pm0.043}}       
\newcommand{\hatcurPPrshortxxxxxA}{\ensuremath{1.33}}           
\newcommand{\hatcurPPrlongxxxxxA}{\ensuremath{1.332\pm0.043}}   
\newcommand{\hatcurPPrexxxxxA}{\ensuremath{14.93\pm0.48}}       
\newcommand{\hatcurPPreshortxxxxxA}{\ensuremath{14.9}}          
\newcommand{\hatcurPPrelongxxxxxA}{\ensuremath{14.93\pm0.48}}   
\newcommand{\hatcurPPmrcorrxxxxxA}{\ensuremath{0.02}}           
\newcommand{\hatcurPPteffxxxxxA}{\ensuremath{1622\pm18}}        
\newcommand{\hatcurPPthetaxxxxxA}{\ensuremath{0.0269\pm0.0023}} 
\newcommand{\hatcurPPfluxperixxxxxA}{\ensuremath{1.561\pm0.068}} 
\newcommand{\hatcurPPfluxperidimxxxxxA}{\ensuremath{9}}         
\newcommand{\hatcurPPfluxapxxxxxA}{\ensuremath{1.561\pm0.068}}  
\newcommand{\hatcurPPfluxapdimxxxxxA}{\ensuremath{9}}           
\newcommand{\hatcurPPfluxavgxxxxxA}{\ensuremath{1.561\pm0.068}} 
\newcommand{\hatcurPPfluxavgdimxxxxxA}{\ensuremath{9}}          
\newcommand{\hatcurPPfluxavglogxxxxxA}{\ensuremath{9.193\pm0.019}} 
\newcommand{\hatcurXsecphasexxxxxA}{\ensuremath{0\pm0}}         
\newcommand{\hatcurXsecondaryxxxxxA}{\ensuremath{2457371.03786\pm0.00056}} 
\newcommand{\hatcurXsecdurxxxxxA}{\ensuremath{0.1729\pm0.0015}} 
\newcommand{\hatcurXsecingdurxxxxxA}{\ensuremath{0.0193\pm0.0010}} 
\newcommand{\hatcurPPphiconjxxxxxA}{\ensuremath{0\pm0}}         
\newcommand{\hatcurPPperixxxxxA}{\ensuremath{2457368.02748\pm0.00056}} 
\newcommand{\hatcurPPaequivxxxxxA}{\ensuremath{0.02950\pm0.00064}} 
\newcommand{\hatcurPPtcircxxxxxA}{\ensuremath{113\pm20}}        
\newcommand{\hatcurPPtinfallxxxxxA}{\ensuremath{1800\pm250}}    
\newcommand{\hatcurXdistxxxxxA}{\ensuremath{519\pm11}}          
\newcommand{\hatcurXAvxxxxxA}{\ensuremath{0.737\pm0.034}}       
\newcommand{\hatcurXdistredxxxxxA}{\ensuremath{519\pm11}}       
\newcommand{\hatcurXEBVxxxxxA}{\ensuremath{0.238\pm0.011}}      
\newcommand{\hatcurCCpmraxxxxxA}{\ensuremath{-10.883\pm0.072}}  
\newcommand{\hatcurCCpmdecxxxxxA}{\ensuremath{11.862\pm0.064}}  
\newcommand{\hatcurCCpmxxxxxA}{\ensuremath{16.098\pm0.096}}     

%% file: HTR080-003.tex
\newcommand{\hatcurhtrxxxxxB}{HTR080-003}                       
\newcommand{\hatcurfieldxxxxxB}{\ensuremath{string}}            
\newcommand{\hatcurCCraxxxxxB}{\ensuremath{19^{\mathrm h}29^{\mathrm m}50.0701{\mathrm s}}}                   
\newcommand{\hatcurCCdecxxxxxB}{\ensuremath{+62{\arcdeg}31{\arcmin}45.1751{\arcsec}}}                 
\newcommand{\hatcurCCmagxxxxxB}{11.883}                         
\newcommand{\hatcurCCtwomassxxxxxB}{2MASS~19295008+6231452}     
\newcommand{\hatcurCCgscxxxxxB}{GSC~4234-02195}                 
\newcommand{\hatcurCCgaiaxxxxxB}{GAIA~2241743199301759488}      
\newcommand{\hatcurCCgaiadrtwoxxxxxB}{2241743203599727744} 
\newcommand{\hatcurCCtassmvxxxxxB}{\ensuremath{11.883\pm0.065}} 
\newcommand{\hatcurCCtassmvshortxxxxxB}{\ensuremath{11.9}}      
\newcommand{\hatcurCCtassmBxxxxxB}{\ensuremath{12.581\pm0.094}} 
\newcommand{\hatcurCCtassmBshortxxxxxB}{\ensuremath{12.6}}      
\newcommand{\hatcurCCtassmIxxxxxB}{\ensuremath{11.073\pm0.078}} 
\newcommand{\hatcurCCtassmIshortxxxxxB}{\ensuremath{11.1}}      
\newcommand{\hatcurCCtassmgxxxxxB}{\ensuremath{12.16\pm0.10}}   
\newcommand{\hatcurCCtassmgshortxxxxxB}{\ensuremath{12.2}}      
\newcommand{\hatcurCCtassmrxxxxxB}{\ensuremath{11.650\pm0.050}} 
\newcommand{\hatcurCCtassmrshortxxxxxB}{\ensuremath{11.7}}      
\newcommand{\hatcurCCtassmixxxxxB}{\ensuremath{11.478\pm0.050}} 
\newcommand{\hatcurCCtassmishortxxxxxB}{\ensuremath{11.5}}      
\newcommand{\hatcurCCparallaxxxxxxB}{\ensuremath{3.738\pm0.019}} 
\newcommand{\hatcurCCgaiamGxxxxxB}{\ensuremath{11.67870\pm0.00030}} 
\newcommand{\hatcurCCgaiamBPxxxxxB}{\ensuremath{12.0587\pm0.0011}} 
\newcommand{\hatcurCCgaiamRPxxxxxB}{\ensuremath{11.15850\pm0.00050}} 
\newcommand{\hatcurCCtwomassJmagxxxxxB}{\ensuremath{10.581\pm0.020}} 
\newcommand{\hatcurCCtwomassHmagxxxxxB}{\ensuremath{10.268\pm0.018}} 
\newcommand{\hatcurCCtwomassKmagxxxxxB}{\ensuremath{10.208\pm0.022}} 
\newcommand{\hatcurCCcitJmagxxxxxB}{\ensuremath{10.598\pm0.020}} 
\newcommand{\hatcurCCcitHmagxxxxxB}{\ensuremath{10.263\pm0.019}} 
\newcommand{\hatcurCCcitKmagxxxxxB}{\ensuremath{10.232\pm0.022}} 
\newcommand{\hatcurCCbbJmagxxxxxB}{\ensuremath{10.647\pm0.022}} 
\newcommand{\hatcurCCbbHmagxxxxxB}{\ensuremath{10.284\pm0.019}} 
\newcommand{\hatcurCCbbKmagxxxxxB}{\ensuremath{10.252\pm0.022}} 
\newcommand{\hatcurCCesoJmagxxxxxB}{\ensuremath{10.649\pm0.024}} 
\newcommand{\hatcurCCesoHmagxxxxxB}{\ensuremath{10.278\pm0.022}} 
\newcommand{\hatcurCCesoKmagxxxxxB}{\ensuremath{10.251\pm0.023}} 
\newcommand{\hatcurCCesoJHmagxxxxxB}{\ensuremath{0.370\pm0.030}} 
\newcommand{\hatcurCCesoJKmagxxxxxB}{\ensuremath{0.399\pm0.032}} 
\newcommand{\hatcurCCesoHKmagxxxxxB}{\ensuremath{0.028\pm0.032}} 
\newcommand{\hatcurCCWonemagxxxxxB}{\ensuremath{10.144\pm0.023}} 
\newcommand{\hatcurCCWtwomagxxxxxB}{\ensuremath{10.211\pm0.020}} 
\newcommand{\hatcurCCWthreemagxxxxxB}{\ensuremath{10.191\pm0.037}} 
\newcommand{\hatcurCCWfourmagxxxxxB}{\ensuremath{0\pm0}}        
\newcommand{\hatcurLCdipxxxxxB}{\ensuremath{12.8}}              
\newcommand{\hatcurLCrprstarxxxxxB}{\ensuremath{0.10452\pm0.00096}} 
\newcommand{\hatcurLCbsqxxxxxB}{\ensuremath{0.689_{-0.015}^{+0.013}}} 
\newcommand{\hatcurLCimpxxxxxB}{\ensuremath{0.8299_{-0.0089}^{+0.0077}}} 
\newcommand{\hatcurLCzetaxxxxxB}{\ensuremath{26.81\pm0.41}}     
\newcommand{\hatcurLCdurxxxxxB}{\ensuremath{0.09747\pm0.00097}} 
\newcommand{\hatcurLCdurshortxxxxxB}{\ensuremath{0.0975}}       
\newcommand{\hatcurLCdurhrxxxxxB}{\ensuremath{2.339\pm0.023}}   
\newcommand{\hatcurLCdurhrshortxxxxxB}{\ensuremath{2.339}}      
\newcommand{\hatcurLCqxxxxxB}{\ensuremath{0.02350\pm0.00024}}   
\newcommand{\hatcurLCqshortxxxxxB}{\ensuremath{0.024}}          
\newcommand{\hatcurLCingdurxxxxxB}{\ensuremath{0.02624\pm0.00099}} 
\newcommand{\hatcurLCPxxxxxB}{\ensuremath{4.1419771\pm0.0000012}} 
\newcommand{\hatcurLCPprecxxxxxB}{\ensuremath{4.1419771}}       
\newcommand{\hatcurLCPshortxxxxxB}{\ensuremath{4.1420}}         
\newcommand{\hatcurLCTshortxxxxxB}{\ensuremath{8618.54088\pm0.00021}} 
\newcommand{\hatcurLCTxxxxxB}{\ensuremath{2458618.54088\pm0.00021}} 
\newcommand{\hatcurLCTAxxxxxB}{\ensuremath{2456129.21263\pm0.00068}} 
\newcommand{\hatcurLCTBxxxxxB}{\ensuremath{2458896.05334\pm0.00024}} 
\newcommand{\hatcurLChatnetmAxxxxxB}{\ensuremath{11.727050\pm0.000095}} 
\newcommand{\hatcurLCiblendAxxxxxB}{\ensuremath{1\pm0}}         
\newcommand{\hatcurLChatnetmBxxxxxB}{\ensuremath{0.000020\pm0.000024}} 
\newcommand{\hatcurLCiblendBxxxxxB}{\ensuremath{0.997\pm0.012}} 
\newcommand{\hatcurLChatnetmCxxxxxB}{\ensuremath{0.000030\pm0.000030}} 
\newcommand{\hatcurLCiblendCxxxxxB}{\ensuremath{0.9993\pm0.0018}} 
\newcommand{\hatcurLChatnetmDxxxxxB}{\ensuremath{0.000020\pm0.000023}} 
\newcommand{\hatcurLCiblendDxxxxxB}{\ensuremath{0.9982\pm0.0047}} 
\newcommand{\hatcurLChatnetmExxxxxB}{\ensuremath{0.000020\pm0.000024}} 
\newcommand{\hatcurLCiblendExxxxxB}{\ensuremath{0.9989\pm0.0038}} 
\newcommand{\hatcurLChatnetmFxxxxxB}{\ensuremath{-0.000010\pm0.000024}} 
\newcommand{\hatcurLCiblendFxxxxxB}{\ensuremath{0.996\pm0.023}} 
\newcommand{\hatcurLChatnetmGxxxxxB}{\ensuremath{-0.000020\pm0.000020}} 
\newcommand{\hatcurLCiblendGxxxxxB}{\ensuremath{0.996\pm0.012}} 
\newcommand{\hatcurLChatnetmHxxxxxB}{\ensuremath{-0.000010\pm0.000023}} 
\newcommand{\hatcurLCiblendHxxxxxB}{\ensuremath{0.9993\pm0.0018}} 
\newcommand{\hatcurLCrhoxxxxxB}{\ensuremath{1.059\pm0.038}}     
\newcommand{\hatcurSMEiteffxxxxxB}{\ensuremath{5678\pm50}}      
\newcommand{\hatcurSMEizfehxxxxxB}{\ensuremath{0.420\pm0.080}}  
\newcommand{\hatcurSMEizfehshortxxxxxB}{\ensuremath{0.42}}      
\newcommand{\hatcurSMEiloggxxxxxB}{\ensuremath{4.40\pm0.10}}    
\newcommand{\hatcurSMEivsinxxxxxB}{\ensuremath{3.05\pm0.50}}    
\newcommand{\hatcurSMEivmacxxxxxB}{\ensuremath{nff\pmnff}}      
\newcommand{\hatcurSMEivmicxxxxxB}{\ensuremath{nff\pmnff}}      
\newcommand{\hatcurSMEiiteffxxxxxB}{\ensuremath{5665\pm50}}     
\newcommand{\hatcurSMEiizfehxxxxxB}{\ensuremath{0.409\pm0.080}} 
\newcommand{\hatcurSMEiizfehshortxxxxxB}{\ensuremath{0.409}}    
\newcommand{\hatcurSMEiiloggxxxxxB}{\ensuremath{4.368\pm0.033}} 
\newcommand{\hatcurSMEiivsinxxxxxB}{\ensuremath{3.04\pm0.50}}   
\newcommand{\hatcurLBiBxxxxxB}{\ensuremath{0.7351}}             
\newcommand{\hatcurLBiiBxxxxxB}{\ensuremath{0.0925}}            
\newcommand{\hatcurLBiVxxxxxB}{\ensuremath{0.5691}}             
\newcommand{\hatcurLBiiVxxxxxB}{\ensuremath{0.1652}}            
\newcommand{\hatcurLBiRxxxxxB}{\ensuremath{0.4701}}             
\newcommand{\hatcurLBiiRxxxxxB}{\ensuremath{0.1808}}            
\newcommand{\hatcurLBiIxxxxxB}{\ensuremath{0.3726}}             
\newcommand{\hatcurLBiiIxxxxxB}{\ensuremath{0.1906}}            
\newcommand{\hatcurLBiuxxxxxB}{\ensuremath{0.9092}}             
\newcommand{\hatcurLBiiuxxxxxB}{\ensuremath{-0.0704}}           
\newcommand{\hatcurLBigxxxxxB}{\ensuremath{0.6783}}             
\newcommand{\hatcurLBiigxxxxxB}{\ensuremath{0.1108}}            
\newcommand{\hatcurLBirxxxxxB}{\ensuremath{0.43\pm0.16}}        
\newcommand{\hatcurLBiirxxxxxB}{\ensuremath{0.36\pm0.17}}       
\newcommand{\hatcurLBiixxxxxB}{\ensuremath{0.32\pm0.14}}        
\newcommand{\hatcurLBiiixxxxxB}{\ensuremath{0.21\pm0.15}}       
\newcommand{\hatcurLBizxxxxxB}{\ensuremath{0.3274}}             
\newcommand{\hatcurLBiizxxxxxB}{\ensuremath{0.1939}}            
\newcommand{\hatcurLBiJxxxxxB}{\ensuremath{0.2154}}             
\newcommand{\hatcurLBiiJxxxxxB}{\ensuremath{0.2260}}            
\newcommand{\hatcurLBiHxxxxxB}{\ensuremath{0.1252}}             
\newcommand{\hatcurLBiiHxxxxxB}{\ensuremath{0.2621}}            
\newcommand{\hatcurLBiKxxxxxB}{\ensuremath{0.1182}}             
\newcommand{\hatcurLBiiKxxxxxB}{\ensuremath{0.2081}}            
\newcommand{\hatcurLBiTxxxxxB}{\ensuremath{0.16\pm0.11}}        
\newcommand{\hatcurLBiiTxxxxxB}{\ensuremath{0.29\pm0.14}}       
\newcommand{\hatcurLBikepxxxxxB}{\ensuremath{0.4898}}           
\newcommand{\hatcurLBiikepxxxxxB}{\ensuremath{0.1829}}          
\newcommand{\hatcurLBiCxxxxxB}{\ensuremath{0.4801}}             
\newcommand{\hatcurLBiiCxxxxxB}{\ensuremath{0.1811}}            
\newcommand{\hatcurLBiMxxxxxB}{\ensuremath{0.5749}}             
\newcommand{\hatcurLBiiMxxxxxB}{\ensuremath{0.1555}}            
\newcommand{\hatcurLBiSonexxxxxB}{\ensuremath{0.0981}}            
\newcommand{\hatcurLBiiSonexxxxxB}{\ensuremath{0.1252}}           
\newcommand{\hatcurLBiStwoxxxxxB}{\ensuremath{0.0828}}            
\newcommand{\hatcurLBiiStwoxxxxxB}{\ensuremath{0.1060}}           
\newcommand{\hatcurLBiSthreexxxxxB}{\ensuremath{0.0697}}            
\newcommand{\hatcurLBiiSthreexxxxxB}{\ensuremath{0.0866}}           
\newcommand{\hatcurLBiSfourxxxxxB}{\ensuremath{0.0541}}            
\newcommand{\hatcurLBiiSfourxxxxxB}{\ensuremath{0.0685}}           
\newcommand{\hatcurISOmxxxxxB}{\ensuremath{1.008\pm0.022}}      
\newcommand{\hatcurISOmshortxxxxxB}{\ensuremath{1.01}}          
\newcommand{\hatcurISOmlongxxxxxB}{\ensuremath{1.008\pm0.022}}  
\newcommand{\hatcurISOrxxxxxB}{\ensuremath{1.1038\pm0.0073}}    
\newcommand{\hatcurISOrshortxxxxxB}{\ensuremath{1.10}}          
\newcommand{\hatcurISOrlongxxxxxB}{\ensuremath{1.1038\pm0.0073}} 
\newcommand{\hatcurISOrhoxxxxxB}{\ensuremath{1.059\pm0.038}}    
\newcommand{\hatcurISOrholongxxxxxB}{\ensuremath{1.059\pm0.038}} 
\newcommand{\hatcurISOloggxxxxxB}{\ensuremath{4.356\pm0.013}}   
\newcommand{\hatcurISOlumxxxxxB}{\ensuremath{1.132\pm0.015}}    
\newcommand{\hatcurISOlumshortxxxxxB}{\ensuremath{1.13}}        
\newcommand{\hatcurISOteffxxxxxB}{\ensuremath{5678\pm16}}       
\newcommand{\hatcurISOzfehxxxxxB}{\ensuremath{0.217\pm0.049}}   
\newcommand{\hatcurISOagexxxxxB}{\ensuremath{7.3\pm1.0}}        
\newcommand{\hatcurISOspecxxxxxB}{G}                            
\newcommand{\hatcurRVKxxxxxB}{\ensuremath{192.6\pm7.7}}         
\newcommand{\hatcurRVrkxxxxxB}{\ensuremath{0\pm0}}              
\newcommand{\hatcurRVrhxxxxxB}{\ensuremath{0\pm0}}              
\newcommand{\hatcurRVkxxxxxB}{\ensuremath{0\pm0}}               
\newcommand{\hatcurRVhxxxxxB}{\ensuremath{0\pm0}}               
\newcommand{\hatcurRVtronexxxxxB}{\ensuremath{0\pm0}}           
\newcommand{\hatcurRVtrtwoxxxxxB}{\ensuremath{0\pm0}}           
\newcommand{\hatcurRVgammaAxxxxxB}{\ensuremath{-152.3\pm8.3}}   
\newcommand{\hatcurRVjitterAxxxxxB}{\ensuremath{13\pm13}}       
\newcommand{\hatcurRVjittertwosiglimAxxxxxB}{\ensuremath{<38.4}} 
\newcommand{\hatcurRVfitrmsAxxxxxB}{\ensuremath{0.0}}           
\newcommand{\hatcurRVgammaBxxxxxB}{\ensuremath{-21157.5\pm6.0}} 
\newcommand{\hatcurRVjitterBxxxxxB}{\ensuremath{0.1\pm7.0}}     
\newcommand{\hatcurRVjittertwosiglimBxxxxxB}{\ensuremath{<17.6}} 
\newcommand{\hatcurRVfitrmsBxxxxxB}{\ensuremath{0.0}}           
\newcommand{\hatcurRVeccenxxxxxB}{\ensuremath{0\pm0}}           
\newcommand{\hatcurRVeccentwosiglimxxxxxB}{\ensuremath{<0.000}} 
\newcommand{\hatcurRVomegaxxxxxB}{\ensuremath{0\pm0}}           
\newcommand{\hatcurPPixxxxxB}{\ensuremath{85.180\pm0.100}}      
\newcommand{\hatcurPPgxxxxxB}{\ensuremath{30.2\pm1.6}}          
\newcommand{\hatcurPPloggxxxxxB}{\ensuremath{3.481\pm0.023}}    
\newcommand{\hatcurPParxxxxxB}{\ensuremath{9.87\pm0.12}}        
\newcommand{\hatcurPParelxxxxxB}{\ensuremath{0.05064\pm0.00037}} 
\newcommand{\hatcurPPrhoxxxxxB}{\ensuremath{1.347\pm0.081}}     
\newcommand{\hatcurPPmxxxxxB}{\ensuremath{1.540\pm0.067}}       
\newcommand{\hatcurPPmshortxxxxxB}{\ensuremath{1.54}}           
\newcommand{\hatcurPPmlongxxxxxB}{\ensuremath{1.540\pm0.067}}   
\newcommand{\hatcurPPmexxxxxB}{\ensuremath{489\pm21}}           
\newcommand{\hatcurPPmeshortxxxxxB}{\ensuremath{489.3}}         
\newcommand{\hatcurPPmelongxxxxxB}{\ensuremath{489\pm21}}       
\newcommand{\hatcurPPrxxxxxB}{\ensuremath{1.123\pm0.013}}       
\newcommand{\hatcurPPrshortxxxxxB}{\ensuremath{1.12}}           
\newcommand{\hatcurPPrlongxxxxxB}{\ensuremath{1.123\pm0.013}}   
\newcommand{\hatcurPPrexxxxxB}{\ensuremath{12.58\pm0.14}}       
\newcommand{\hatcurPPreshortxxxxxB}{\ensuremath{12.6}}          
\newcommand{\hatcurPPrelongxxxxxB}{\ensuremath{12.58\pm0.14}}   
\newcommand{\hatcurPPmrcorrxxxxxB}{\ensuremath{-0.16}}          
\newcommand{\hatcurPPteffxxxxxB}{\ensuremath{1277.8\pm6.5}}     
\newcommand{\hatcurPPthetaxxxxxB}{\ensuremath{0.1367\pm0.0058}} 
\newcommand{\hatcurPPfluxperixxxxxB}{\ensuremath{6.01\pm0.12}}  
\newcommand{\hatcurPPfluxperidimxxxxxB}{\ensuremath{8}}         
\newcommand{\hatcurPPfluxapxxxxxB}{\ensuremath{6.01\pm0.12}}    
\newcommand{\hatcurPPfluxapdimxxxxxB}{\ensuremath{8}}           
\newcommand{\hatcurPPfluxavgxxxxxB}{\ensuremath{6.01\pm0.12}}   
\newcommand{\hatcurPPfluxavgdimxxxxxB}{\ensuremath{8}}          
\newcommand{\hatcurPPfluxavglogxxxxxB}{\ensuremath{8.7787\pm0.0088}} 
\newcommand{\hatcurXsecphasexxxxxB}{\ensuremath{0\pm0}}         
\newcommand{\hatcurXsecondaryxxxxxB}{\ensuremath{2458620.61187\pm0.00021}} 
\newcommand{\hatcurXsecdurxxxxxB}{\ensuremath{0.09747\pm0.00097}} 
\newcommand{\hatcurXsecingdurxxxxxB}{\ensuremath{0.02624\pm0.00099}} 
\newcommand{\hatcurPPphiconjxxxxxB}{\ensuremath{0\pm0}}         
\newcommand{\hatcurPPperixxxxxB}{\ensuremath{2458617.50538\pm0.00021}} 
\newcommand{\hatcurPPaequivxxxxxB}{\ensuremath{0.04760\pm0.00048}} 
\newcommand{\hatcurPPtcircxxxxxB}{\ensuremath{1250\pm110}}      
\newcommand{\hatcurPPtinfallxxxxxB}{\ensuremath{2420\pm180}}    
\newcommand{\hatcurXdistxxxxxB}{\ensuremath{267.3\pm1.3}}       
\newcommand{\hatcurXAvxxxxxB}{\ensuremath{0.048\pm0.011}}       
\newcommand{\hatcurXdistredxxxxxB}{\ensuremath{267.3\pm1.3}}    
\newcommand{\hatcurXEBVxxxxxB}{\ensuremath{0.0160_{-0.0040}^{+0.0030}}} 
\newcommand{\hatcurCCpmraxxxxxB}{\ensuremath{20.957\pm0.046}}   
\newcommand{\hatcurCCpmdecxxxxxB}{\ensuremath{-6.056\pm0.043}}  
\newcommand{\hatcurCCpmxxxxxB}{\ensuremath{21.814\pm0.063}}     

%% file: HTR089-018.tex
\newcommand{\hatcurhtrxxxxxC}{HTR089-018}                       
\newcommand{\hatcurfieldxxxxxC}{\ensuremath{string}}            
\newcommand{\hatcurCCraxxxxxC}{\ensuremath{01^{\mathrm h}53^{\mathrm m}07.7727{\mathrm s}}}                   
\newcommand{\hatcurCCdecxxxxxC}{\ensuremath{+52{\arcdeg}03{\arcmin}14.01977{\arcsec}}}                
\newcommand{\hatcurCCmagxxxxxC}{9.710}                          
\newcommand{\hatcurCCtwomassxxxxxC}{2MASS~01530777+5203140}     
\newcommand{\hatcurCCgscxxxxxC}{GSC~3292-01330}                 
\newcommand{\hatcurCCgaiaxxxxxC}{GAIA~359678187913760384}       
\newcommand{\hatcurCCgaiadrtwoxxxxxC}{359678187913760384} 
\newcommand{\hatcurCCtassmvxxxxxC}{\ensuremath{9.710\pm0.050}}  
\newcommand{\hatcurCCtassmvshortxxxxxC}{\ensuremath{9.7}}       
\newcommand{\hatcurCCtassmBxxxxxC}{\ensuremath{10.230\pm0.040}} 
\newcommand{\hatcurCCtassmBshortxxxxxC}{\ensuremath{10.2}}      
\newcommand{\hatcurCCtassmIxxxxxC}{\ensuremath{9.077\pm0.042}}  
\newcommand{\hatcurCCtassmIshortxxxxxC}{\ensuremath{9.1}}       
\newcommand{\hatcurCCtassmgxxxxxC}{\ensuremath{nff\pmnff}}      
\newcommand{\hatcurCCtassmgshortxxxxxC}{\ensuremath{0.0}}       
\newcommand{\hatcurCCtassmrxxxxxC}{\ensuremath{nff\pmnff}}      
\newcommand{\hatcurCCtassmrshortxxxxxC}{\ensuremath{0.0}}       
\newcommand{\hatcurCCtassmixxxxxC}{\ensuremath{9.421\pm0.040}}  
\newcommand{\hatcurCCtassmishortxxxxxC}{\ensuremath{9.4}}       
\newcommand{\hatcurCCparallaxxxxxxC}{\ensuremath{4.260\pm0.049}} 
\newcommand{\hatcurCCgaiamGxxxxxC}{\ensuremath{9.56360\pm0.00030}} 
\newcommand{\hatcurCCgaiamBPxxxxxC}{\ensuremath{9.8631\pm0.0012}} 
\newcommand{\hatcurCCgaiamRPxxxxxC}{\ensuremath{9.1320\pm0.0013}} 
\newcommand{\hatcurCCtwomassJmagxxxxxC}{\ensuremath{8.677\pm0.052}} 
\newcommand{\hatcurCCtwomassHmagxxxxxC}{\ensuremath{8.396\pm0.029}} 
\newcommand{\hatcurCCtwomassKmagxxxxxC}{\ensuremath{8.368\pm0.031}} 
\newcommand{\hatcurCCcitJmagxxxxxC}{\ensuremath{8.697\pm0.050}} 
\newcommand{\hatcurCCcitHmagxxxxxC}{\ensuremath{8.392\pm0.029}} 
\newcommand{\hatcurCCcitKmagxxxxxC}{\ensuremath{8.392\pm0.031}} 
\newcommand{\hatcurCCbbJmagxxxxxC}{\ensuremath{8.742\pm0.055}}  
\newcommand{\hatcurCCbbHmagxxxxxC}{\ensuremath{8.412\pm0.030}}  
\newcommand{\hatcurCCbbKmagxxxxxC}{\ensuremath{8.412\pm0.031}}  
\newcommand{\hatcurCCesoJmagxxxxxC}{\ensuremath{8.742\pm0.055}} 
\newcommand{\hatcurCCesoHmagxxxxxC}{\ensuremath{8.406\pm0.033}} 
\newcommand{\hatcurCCesoKmagxxxxxC}{\ensuremath{8.411\pm0.032}} 
\newcommand{\hatcurCCesoJHmagxxxxxC}{\ensuremath{0.337\pm0.063}} 
\newcommand{\hatcurCCesoJKmagxxxxxC}{\ensuremath{0.332\pm0.064}} 
\newcommand{\hatcurCCesoHKmagxxxxxC}{\ensuremath{-0.005\pm0.047}} 
\newcommand{\hatcurCCWonemagxxxxxC}{\ensuremath{8.308\pm0.026}} 
\newcommand{\hatcurCCWtwomagxxxxxC}{\ensuremath{8.320\pm0.026}} 
\newcommand{\hatcurCCWthreemagxxxxxC}{\ensuremath{8.260\pm0.028}} 
\newcommand{\hatcurCCWfourmagxxxxxC}{\ensuremath{0\pm0}}        
\newcommand{\hatcurLCdipxxxxxC}{\ensuremath{5.8}}               
\newcommand{\hatcurLCrprstarxxxxxC}{\ensuremath{0.07622\pm0.00055}} 
\newcommand{\hatcurLCbsqxxxxxC}{\ensuremath{0.446_{-0.018}^{+0.014}}} 
\newcommand{\hatcurLCimpxxxxxC}{\ensuremath{0.668_{-0.014}^{+0.011}}} 
\newcommand{\hatcurLCzetaxxxxxC}{\ensuremath{10.81\pm0.10}}     
\newcommand{\hatcurLCdurxxxxxC}{\ensuremath{0.2098\pm0.0015}}   
\newcommand{\hatcurLCdurshortxxxxxC}{\ensuremath{0.2098}}       
\newcommand{\hatcurLCdurhrxxxxxC}{\ensuremath{5.036\pm0.037}}   
\newcommand{\hatcurLCdurhrshortxxxxxC}{\ensuremath{5.036}}      
\newcommand{\hatcurLCqxxxxxC}{\ensuremath{0.04380\pm0.00032}}   
\newcommand{\hatcurLCqshortxxxxxC}{\ensuremath{0.044}}          
\newcommand{\hatcurLCingdurxxxxxC}{\ensuremath{0.02557\pm0.00073}} 
\newcommand{\hatcurLCPxxxxxC}{\ensuremath{4.7947813\pm0.0000024}} 
\newcommand{\hatcurLCPprecxxxxxC}{\ensuremath{4.7947813}}       
\newcommand{\hatcurLCPshortxxxxxC}{\ensuremath{4.7948}}         
\newcommand{\hatcurLCTshortxxxxxC}{\ensuremath{8360.94029\pm0.00056}} 
\newcommand{\hatcurLCTxxxxxC}{\ensuremath{2458360.94029\pm0.00056}} 
\newcommand{\hatcurLCTAxxxxxC}{\ensuremath{2454764.8543\pm0.0017}} 
\newcommand{\hatcurLCTBxxxxxC}{\ensuremath{2458811.64971\pm0.00066}} 
\newcommand{\hatcurLChatnetmAxxxxxC}{\ensuremath{9.745750\pm0.000048}} 
\newcommand{\hatcurLCiblendAxxxxxC}{\ensuremath{0.872\pm0.015}} 
\newcommand{\hatcurLChatnetmBxxxxxC}{\ensuremath{0.000030\pm0.000013}} 
\newcommand{\hatcurLCiblendBxxxxxC}{\ensuremath{0.9957\pm0.0018}} 
\newcommand{\hatcurLCrhoxxxxxC}{\ensuremath{0.1909_{-0.0071}^{+0.0054}}} 
\newcommand{\hatcurSMEiteffxxxxxC}{\ensuremath{6223\pm50}}      
\newcommand{\hatcurSMEizfehxxxxxC}{\ensuremath{-0.404\pm0.080}} 
\newcommand{\hatcurSMEizfehshortxxxxxC}{\ensuremath{-0.40}}     
\newcommand{\hatcurSMEiloggxxxxxC}{\ensuremath{3.59\pm0.10}}    
\newcommand{\hatcurSMEivsinxxxxxC}{\ensuremath{10.71\pm0.50}}   
\newcommand{\hatcurSMEivmacxxxxxC}{\ensuremath{nff\pmnff}}      
\newcommand{\hatcurSMEivmicxxxxxC}{\ensuremath{nff\pmnff}}      
\newcommand{\hatcurSMEiiteffxxxxxC}{\ensuremath{6462\pm50}}     
\newcommand{\hatcurSMEiizfehxxxxxC}{\ensuremath{-0.237\pm0.080}} 
\newcommand{\hatcurSMEiizfehshortxxxxxC}{\ensuremath{-0.237}}   
\newcommand{\hatcurSMEiiloggxxxxxC}{\ensuremath{4.052\pm0.049}} 
\newcommand{\hatcurSMEiivsinxxxxxC}{\ensuremath{10.42\pm0.50}}  
\newcommand{\hatcurLBiBxxxxxC}{\ensuremath{0.5594}}             
\newcommand{\hatcurLBiiBxxxxxC}{\ensuremath{0.2417}}            
\newcommand{\hatcurLBiVxxxxxC}{\ensuremath{0.4555}}             
\newcommand{\hatcurLBiiVxxxxxC}{\ensuremath{0.2362}}            
\newcommand{\hatcurLBiRxxxxxC}{\ensuremath{0.3896}}             
\newcommand{\hatcurLBiiRxxxxxC}{\ensuremath{0.2179}}            
\newcommand{\hatcurLBiIxxxxxC}{\ensuremath{0.3076}}             
\newcommand{\hatcurLBiiIxxxxxC}{\ensuremath{0.2060}}            
\newcommand{\hatcurLBiuxxxxxC}{\ensuremath{0.6429}}             
\newcommand{\hatcurLBiiuxxxxxC}{\ensuremath{0.1541}}            
\newcommand{\hatcurLBigxxxxxC}{\ensuremath{0.5364}}             
\newcommand{\hatcurLBiigxxxxxC}{\ensuremath{0.2206}}            
\newcommand{\hatcurLBirxxxxxC}{\ensuremath{0.47\pm0.15}}        
\newcommand{\hatcurLBiirxxxxxC}{\ensuremath{0.28\pm0.15}}       
\newcommand{\hatcurLBiixxxxxC}{\ensuremath{0.18\pm0.12}}        
\newcommand{\hatcurLBiiixxxxxC}{\ensuremath{0.09\pm0.14}}       
\newcommand{\hatcurLBizxxxxxC}{\ensuremath{0.137_{-0.096}^{+0.129}}} 
\newcommand{\hatcurLBiizxxxxxC}{\ensuremath{0.12\pm0.15}}       
\newcommand{\hatcurLBiJxxxxxC}{\ensuremath{0.1789}}             
\newcommand{\hatcurLBiiJxxxxxC}{\ensuremath{0.2175}}            
\newcommand{\hatcurLBiHxxxxxC}{\ensuremath{0.1144}}             
\newcommand{\hatcurLBiiHxxxxxC}{\ensuremath{0.2251}}            
\newcommand{\hatcurLBiKxxxxxC}{\ensuremath{0.1120}}             
\newcommand{\hatcurLBiiKxxxxxC}{\ensuremath{0.1747}}            
\newcommand{\hatcurLBiTxxxxxC}{\ensuremath{0.19\pm0.11}}        
\newcommand{\hatcurLBiiTxxxxxC}{\ensuremath{0.23\pm0.15}}       
\newcommand{\hatcurLBikepxxxxxC}{\ensuremath{0.4383}}           
\newcommand{\hatcurLBiikepxxxxxC}{\ensuremath{0.2054}}          
\newcommand{\hatcurLBiCxxxxxC}{\ensuremath{0.4303}}             
\newcommand{\hatcurLBiiCxxxxxC}{\ensuremath{0.2057}}            
\newcommand{\hatcurLBiMxxxxxC}{\ensuremath{0.5086}}             
\newcommand{\hatcurLBiiMxxxxxC}{\ensuremath{0.1973}}            
\newcommand{\hatcurLBiSonexxxxxC}{\ensuremath{0.0824}}            
\newcommand{\hatcurLBiiSonexxxxxC}{\ensuremath{0.1090}}           
\newcommand{\hatcurLBiStwoxxxxxC}{\ensuremath{0.0699}}            
\newcommand{\hatcurLBiiStwoxxxxxC}{\ensuremath{0.0921}}           
\newcommand{\hatcurLBiSthreexxxxxC}{\ensuremath{0.0573}}            
\newcommand{\hatcurLBiiSthreexxxxxC}{\ensuremath{0.0757}}           
\newcommand{\hatcurLBiSfourxxxxxC}{\ensuremath{0.0435}}            
\newcommand{\hatcurLBiiSfourxxxxxC}{\ensuremath{0.0614}}           
\newcommand{\hatcurISOmxxxxxC}{\ensuremath{1.435\pm0.012}}      
\newcommand{\hatcurISOmshortxxxxxC}{\ensuremath{1.43}}          
\newcommand{\hatcurISOmlongxxxxxC}{\ensuremath{1.435\pm0.012}}  
\newcommand{\hatcurISOrxxxxxC}{\ensuremath{2.197_{-0.020}^{+0.027}}} 
\newcommand{\hatcurISOrshortxxxxxC}{\ensuremath{2.20}}          
\newcommand{\hatcurISOrlongxxxxxC}{\ensuremath{2.197_{-0.020}^{+0.027}}} 
\newcommand{\hatcurISOrhoxxxxxC}{\ensuremath{0.1909_{-0.0071}^{+0.0054}}} 
\newcommand{\hatcurISOrholongxxxxxC}{\ensuremath{0.1909_{-0.0071}^{+0.0054}}} 
\newcommand{\hatcurISOloggxxxxxC}{\ensuremath{3.9114\pm0.0097}} 
\newcommand{\hatcurISOlumxxxxxC}{\ensuremath{6.44\pm0.17}}      
\newcommand{\hatcurISOlumshortxxxxxC}{\ensuremath{6.44}}        
\newcommand{\hatcurISOteffxxxxxC}{\ensuremath{6212\pm26}}       
\newcommand{\hatcurISOzfehxxxxxC}{\ensuremath{0.037\pm0.037}}   
\newcommand{\hatcurISOagexxxxxC}{\ensuremath{2.765_{-0.056}^{+0.042}}} 
\newcommand{\hatcurISOspecxxxxxC}{F}                            
\newcommand{\hatcurRVKxxxxxC}{\ensuremath{54.1\pm3.5}}          
\newcommand{\hatcurRVrkxxxxxC}{\ensuremath{0\pm0}}              
\newcommand{\hatcurRVrhxxxxxC}{\ensuremath{0\pm0}}              
\newcommand{\hatcurRVkxxxxxC}{\ensuremath{0\pm0}}               
\newcommand{\hatcurRVhxxxxxC}{\ensuremath{0\pm0}}               
\newcommand{\hatcurRVtronexxxxxC}{\ensuremath{0\pm0}}           
\newcommand{\hatcurRVtrtwoxxxxxC}{\ensuremath{0\pm0}}           
\newcommand{\hatcurRVgammaAxxxxxC}{\ensuremath{10.0\pm3.8}}     
\newcommand{\hatcurRVjitterAxxxxxC}{\ensuremath{12.3\pm3.7}}    
\newcommand{\hatcurRVjittertwosiglimAxxxxxC}{\ensuremath{<20.1}} 
\newcommand{\hatcurRVfitrmsAxxxxxC}{\ensuremath{0.0}}           
\newcommand{\hatcurRVgammaBxxxxxC}{\ensuremath{56.0\pm6.3}}     
\newcommand{\hatcurRVjitterBxxxxxC}{\ensuremath{0.3\pm4.7}}     
\newcommand{\hatcurRVjittertwosiglimBxxxxxC}{\ensuremath{<12.4}} 
\newcommand{\hatcurRVfitrmsBxxxxxC}{\ensuremath{0.0}}           
\newcommand{\hatcurRVgammaCxxxxxC}{\ensuremath{6027.0\pm8.0}}   
\newcommand{\hatcurRVjitterCxxxxxC}{\ensuremath{6.8\pm5.4}}     
\newcommand{\hatcurRVjittertwosiglimCxxxxxC}{\ensuremath{<15.0}} 
\newcommand{\hatcurRVfitrmsCxxxxxC}{\ensuremath{0.0}}           
\newcommand{\hatcurRVgammaDxxxxxC}{\ensuremath{5923\pm15}}      
\newcommand{\hatcurRVjitterDxxxxxC}{\ensuremath{56\pm13}}       
\newcommand{\hatcurRVjittertwosiglimDxxxxxC}{\ensuremath{<79.4}} 
\newcommand{\hatcurRVfitrmsDxxxxxC}{\ensuremath{.1fym}}         %
\newcommand{\hatcurRVgammaExxxxxC}{\ensuremath{5842\pm13}}      
\newcommand{\hatcurRVjitterExxxxxC}{\ensuremath{22\pm19}}       
\newcommand{\hatcurRVjittertwosiglimExxxxxC}{\ensuremath{<54.9}} 
\newcommand{\hatcurRVfitrmsExxxxxC}{\ensuremath{.1fym}}         %
\newcommand{\hatcurRVeccenxxxxxC}{\ensuremath{0\pm0}}           
\newcommand{\hatcurRVeccentwosiglimxxxxxC}{\ensuremath{<0.000}} 
\newcommand{\hatcurRVomegaxxxxxC}{\ensuremath{0\pm0}}           
\newcommand{\hatcurPPixxxxxC}{\ensuremath{83.75\pm0.17}}        
\newcommand{\hatcurPPgxxxxxC}{\ensuremath{5.37\pm0.39}}         
\newcommand{\hatcurPPloggxxxxxC}{\ensuremath{2.730\pm0.032}}    
\newcommand{\hatcurPParxxxxxC}{\ensuremath{6.146_{-0.077}^{+0.057}}} 
\newcommand{\hatcurPParelxxxxxC}{\ensuremath{0.06277\pm0.00017}} 
\newcommand{\hatcurPPrhoxxxxxC}{\ensuremath{0.164\pm0.013}}     
\newcommand{\hatcurPPmxxxxxC}{\ensuremath{0.574\pm0.038}}       
\newcommand{\hatcurPPmshortxxxxxC}{\ensuremath{0.57}}           
\newcommand{\hatcurPPmlongxxxxxC}{\ensuremath{0.574\pm0.038}}   
\newcommand{\hatcurPPmexxxxxC}{\ensuremath{182\pm12}}           
\newcommand{\hatcurPPmeshortxxxxxC}{\ensuremath{182.5}}         
\newcommand{\hatcurPPmelongxxxxxC}{\ensuremath{182\pm12}}       
\newcommand{\hatcurPPrxxxxxC}{\ensuremath{1.631\pm0.024}}       
\newcommand{\hatcurPPrshortxxxxxC}{\ensuremath{1.63}}           
\newcommand{\hatcurPPrlongxxxxxC}{\ensuremath{1.631\pm0.024}}   
\newcommand{\hatcurPPrexxxxxC}{\ensuremath{18.28\pm0.26}}       
\newcommand{\hatcurPPreshortxxxxxC}{\ensuremath{18.3}}          
\newcommand{\hatcurPPrelongxxxxxC}{\ensuremath{18.28\pm0.26}}   
\newcommand{\hatcurPPmrcorrxxxxxC}{\ensuremath{-0.02}}          
\newcommand{\hatcurPPteffxxxxxC}{\ensuremath{1772\pm12}}        
\newcommand{\hatcurPPthetaxxxxxC}{\ensuremath{0.0306\pm0.0020}} 
\newcommand{\hatcurPPfluxperixxxxxC}{\ensuremath{2.225\pm0.062}} 
\newcommand{\hatcurPPfluxperidimxxxxxC}{\ensuremath{9}}         
\newcommand{\hatcurPPfluxapxxxxxC}{\ensuremath{2.225\pm0.062}}  
\newcommand{\hatcurPPfluxapdimxxxxxC}{\ensuremath{9}}           
\newcommand{\hatcurPPfluxavgxxxxxC}{\ensuremath{2.225\pm0.062}} 
\newcommand{\hatcurPPfluxavgdimxxxxxC}{\ensuremath{9}}          
\newcommand{\hatcurPPfluxavglogxxxxxC}{\ensuremath{9.347\pm0.012}} 
\newcommand{\hatcurXsecphasexxxxxC}{\ensuremath{0\pm0}}         
\newcommand{\hatcurXsecondaryxxxxxC}{\ensuremath{2458363.33768\pm0.00056}} 
\newcommand{\hatcurXsecdurxxxxxC}{\ensuremath{0.2098\pm0.0015}} 
\newcommand{\hatcurXsecingdurxxxxxC}{\ensuremath{0.02557\pm0.00073}} 
\newcommand{\hatcurPPphiconjxxxxxC}{\ensuremath{0\pm0}}         
\newcommand{\hatcurPPperixxxxxC}{\ensuremath{2458359.74160\pm0.00056}} 
\newcommand{\hatcurPPaequivxxxxxC}{\ensuremath{0.02470_{-0.00030}^{+0.00040}}} 
\newcommand{\hatcurPPtcircxxxxxC}{\ensuremath{172\pm17}}        
\newcommand{\hatcurPPtinfallxxxxxC}{\ensuremath{995\pm85}}      
\newcommand{\hatcurXdistxxxxxC}{\ensuremath{235.4\pm2.3}}       
\newcommand{\hatcurXAvxxxxxC}{\ensuremath{0.120\pm0.019}}       
\newcommand{\hatcurXdistredxxxxxC}{\ensuremath{235.4\pm2.3}}    
\newcommand{\hatcurXEBVxxxxxC}{\ensuremath{0.0390\pm0.0063}}    
\newcommand{\hatcurCCpmraxxxxxC}{\ensuremath{26.51\pm0.11}}     
\newcommand{\hatcurCCpmdecxxxxxC}{\ensuremath{6.165\pm0.075}}   
\newcommand{\hatcurCCpmxxxxxC}{\ensuremath{27.22\pm0.13}}       

%% file: HTR094-024.tex
\newcommand{\hatcurhtrxxxxxD}{HTR094-024}                       
\newcommand{\hatcurfieldxxxxxD}{\ensuremath{string}}            
\newcommand{\hatcurCCraxxxxxD}{\ensuremath{05^{\mathrm h}01^{\mathrm m}55.2577{\mathrm s}}}                   
\newcommand{\hatcurCCdecxxxxxD}{\ensuremath{+50{\arcdeg}07{\arcmin}52.5746{\arcsec}}}                 
\newcommand{\hatcurCCmagxxxxxD}{13.188}                         
\newcommand{\hatcurCCtwomassxxxxxD}{2MASS~05015525+5007526}     
\newcommand{\hatcurCCgscxxxxxD}{GSC~3352-00595}                 
\newcommand{\hatcurCCgaiaxxxxxD}{GAIA~256580178035425152}       
\newcommand{\hatcurCCgaiadrtwoxxxxxD}{256580182331399296} 
\newcommand{\hatcurCCtassmvxxxxxD}{\ensuremath{13.188\pm0.029}} 
\newcommand{\hatcurCCtassmvshortxxxxxD}{\ensuremath{13.2}}      
\newcommand{\hatcurCCtassmBxxxxxD}{\ensuremath{14.040\pm0.056}} 
\newcommand{\hatcurCCtassmBshortxxxxxD}{\ensuremath{14.0}}      
\newcommand{\hatcurCCtassmIxxxxxD}{\ensuremath{12.05\pm0.12}}   
\newcommand{\hatcurCCtassmIshortxxxxxD}{\ensuremath{12.1}}      
\newcommand{\hatcurCCtassmgxxxxxD}{\ensuremath{13.550\pm0.045}} 
\newcommand{\hatcurCCtassmgshortxxxxxD}{\ensuremath{13.6}}      
\newcommand{\hatcurCCtassmrxxxxxD}{\ensuremath{12.915\pm0.033}} 
\newcommand{\hatcurCCtassmrshortxxxxxD}{\ensuremath{12.9}}      
\newcommand{\hatcurCCtassmixxxxxD}{\ensuremath{12.675\pm0.051}} 
\newcommand{\hatcurCCtassmishortxxxxxD}{\ensuremath{12.7}}      
\newcommand{\hatcurCCparallaxxxxxxD}{\ensuremath{2.923\pm0.035}} 
\newcommand{\hatcurCCgaiamGxxxxxD}{\ensuremath{12.93860\pm0.00040}} 
\newcommand{\hatcurCCgaiamBPxxxxxD}{\ensuremath{13.4067\pm0.0018}} 
\newcommand{\hatcurCCgaiamRPxxxxxD}{\ensuremath{12.33560\pm0.00080}} 
\newcommand{\hatcurCCtwomassJmagxxxxxD}{\ensuremath{11.598\pm0.021}} 
\newcommand{\hatcurCCtwomassHmagxxxxxD}{\ensuremath{11.263\pm0.029}} 
\newcommand{\hatcurCCtwomassKmagxxxxxD}{\ensuremath{11.141\pm0.020}} 
\newcommand{\hatcurCCcitJmagxxxxxD}{\ensuremath{11.610\pm0.021}} 
\newcommand{\hatcurCCcitHmagxxxxxD}{\ensuremath{11.257\pm0.029}} 
\newcommand{\hatcurCCcitKmagxxxxxD}{\ensuremath{11.165\pm0.020}} 
\newcommand{\hatcurCCbbJmagxxxxxD}{\ensuremath{11.667\pm0.023}} 
\newcommand{\hatcurCCbbHmagxxxxxD}{\ensuremath{11.279\pm0.030}} 
\newcommand{\hatcurCCbbKmagxxxxxD}{\ensuremath{11.185\pm0.020}} 
\newcommand{\hatcurCCesoJmagxxxxxD}{\ensuremath{11.670\pm0.025}} 
\newcommand{\hatcurCCesoHmagxxxxxD}{\ensuremath{11.276\pm0.035}} 
\newcommand{\hatcurCCesoKmagxxxxxD}{\ensuremath{11.184\pm0.021}} 
\newcommand{\hatcurCCesoJHmagxxxxxD}{\ensuremath{0.393\pm0.041}} 
\newcommand{\hatcurCCesoJKmagxxxxxD}{\ensuremath{0.487\pm0.032}} 
\newcommand{\hatcurCCesoHKmagxxxxxD}{\ensuremath{0.093\pm0.041}} 
\newcommand{\hatcurCCWonemagxxxxxD}{\ensuremath{11.089\pm0.022}} 
\newcommand{\hatcurCCWtwomagxxxxxD}{\ensuremath{11.176\pm0.022}} 
\newcommand{\hatcurCCWthreemagxxxxxD}{\ensuremath{0\pm0}}       
\newcommand{\hatcurCCWfourmagxxxxxD}{\ensuremath{0\pm0}}        
\newcommand{\hatcurLCdipxxxxxD}{\ensuremath{12.7}}              
\newcommand{\hatcurLCrprstarxxxxxD}{\ensuremath{0.0984\pm0.0025}} 
\newcommand{\hatcurLCbsqxxxxxD}{\ensuremath{0.589_{-0.026}^{+0.024}}} 
\newcommand{\hatcurLCimpxxxxxD}{\ensuremath{0.767_{-0.017}^{+0.015}}} 
\newcommand{\hatcurLCzetaxxxxxD}{\ensuremath{35.46\pm0.88}}     
\newcommand{\hatcurLCdurxxxxxD}{\ensuremath{0.0691\pm0.0012}}   
\newcommand{\hatcurLCdurshortxxxxxD}{\ensuremath{0.0691}}       
\newcommand{\hatcurLCdurhrxxxxxD}{\ensuremath{1.658\pm0.030}}   
\newcommand{\hatcurLCdurhrshortxxxxxD}{\ensuremath{1.658}}      
\newcommand{\hatcurLCqxxxxxD}{\ensuremath{0.03630\pm0.00065}}   
\newcommand{\hatcurLCqshortxxxxxD}{\ensuremath{0.036}}          
\newcommand{\hatcurLCingdurxxxxxD}{\ensuremath{0.01372\pm0.00085}} 
\newcommand{\hatcurLCPxxxxxD}{\ensuremath{1.90231289\pm0.00000077}} 
\newcommand{\hatcurLCPprecxxxxxD}{\ensuremath{1.9023129}}       
\newcommand{\hatcurLCPshortxxxxxD}{\ensuremath{1.9023}}         
\newcommand{\hatcurLCTshortxxxxxD}{\ensuremath{7851.21119\pm0.00047}} 
\newcommand{\hatcurLCTxxxxxD}{\ensuremath{2457851.21119\pm0.00047}} 
\newcommand{\hatcurLCTAxxxxxD}{\ensuremath{2454392.8064\pm0.0013}} 
\newcommand{\hatcurLCTBxxxxxD}{\ensuremath{2458840.41390\pm0.00071}} 
\newcommand{\hatcurLChatnetmAxxxxxD}{\ensuremath{12.61785\pm0.00019}} 
\newcommand{\hatcurLCiblendAxxxxxD}{\ensuremath{0.87\pm0.10}}   
\newcommand{\hatcurLChatnetmBxxxxxD}{\ensuremath{12.80532\pm0.00011}} 
\newcommand{\hatcurLCiblendBxxxxxD}{\ensuremath{0.915\pm0.065}} 
\newcommand{\hatcurLChatnetmCxxxxxD}{\ensuremath{0.000010\pm0.000032}} 
\newcommand{\hatcurLCiblendCxxxxxD}{\ensuremath{0.694\pm0.040}} 
\newcommand{\hatcurLCrhoxxxxxD}{\ensuremath{1.715\pm0.094}}     
\newcommand{\hatcurSMEiteffxxxxxD}{\ensuremath{5546\pm50}}      
\newcommand{\hatcurSMEizfehxxxxxD}{\ensuremath{0.399\pm0.080}}  
\newcommand{\hatcurSMEizfehshortxxxxxD}{\ensuremath{0.40}}      
\newcommand{\hatcurSMEiloggxxxxxD}{\ensuremath{4.43\pm0.10}}    
\newcommand{\hatcurSMEivsinxxxxxD}{\ensuremath{3.72\pm0.50}}    
\newcommand{\hatcurSMEivmacxxxxxD}{\ensuremath{nff\pmnff}}      
\newcommand{\hatcurSMEivmicxxxxxD}{\ensuremath{nff\pmnff}}      
\newcommand{\hatcurSMEiiteffxxxxxD}{\ensuremath{5551\pm50}}     
\newcommand{\hatcurSMEiizfehxxxxxD}{\ensuremath{0.396\pm0.080}} 
\newcommand{\hatcurSMEiizfehshortxxxxxD}{\ensuremath{0.396}}    
\newcommand{\hatcurSMEiiloggxxxxxD}{\ensuremath{4.468\pm0.035}} 
\newcommand{\hatcurSMEiivsinxxxxxD}{\ensuremath{3.69\pm0.50}}   
\newcommand{\hatcurLBiBxxxxxD}{\ensuremath{0.7596}}             
\newcommand{\hatcurLBiiBxxxxxD}{\ensuremath{0.0707}}            
\newcommand{\hatcurLBiVxxxxxD}{\ensuremath{0.5876}}             
\newcommand{\hatcurLBiiVxxxxxD}{\ensuremath{0.1536}}            
\newcommand{\hatcurLBiRxxxxxD}{\ensuremath{0.38\pm0.16}}        
\newcommand{\hatcurLBiiRxxxxxD}{\ensuremath{0.36\pm0.17}}       
\newcommand{\hatcurLBiIxxxxxD}{\ensuremath{0.3836}}             
\newcommand{\hatcurLBiiIxxxxxD}{\ensuremath{0.1879}}            
\newcommand{\hatcurLBiuxxxxxD}{\ensuremath{0.9484}}             
\newcommand{\hatcurLBiiuxxxxxD}{\ensuremath{-0.1150}}           
\newcommand{\hatcurLBigxxxxxD}{\ensuremath{0.7007}}             
\newcommand{\hatcurLBiigxxxxxD}{\ensuremath{0.0918}}            
\newcommand{\hatcurLBirxxxxxD}{\ensuremath{0.40\pm0.16}}        
\newcommand{\hatcurLBiirxxxxxD}{\ensuremath{0.39\pm0.16}}       
\newcommand{\hatcurLBiixxxxxD}{\ensuremath{0.32\pm0.16}}        
\newcommand{\hatcurLBiiixxxxxD}{\ensuremath{0.30\pm0.16}}       
\newcommand{\hatcurLBizxxxxxD}{\ensuremath{0.3380}}             
\newcommand{\hatcurLBiizxxxxxD}{\ensuremath{0.1913}}            
\newcommand{\hatcurLBiJxxxxxD}{\ensuremath{0.2216}}             
\newcommand{\hatcurLBiiJxxxxxD}{\ensuremath{0.2275}}            
\newcommand{\hatcurLBiHxxxxxD}{\ensuremath{0.1268}}             
\newcommand{\hatcurLBiiHxxxxxD}{\ensuremath{0.2668}}            
\newcommand{\hatcurLBiKxxxxxD}{\ensuremath{0.1187}}             
\newcommand{\hatcurLBiiKxxxxxD}{\ensuremath{0.2135}}            
\newcommand{\hatcurLBiTxxxxxD}{\ensuremath{0.31\pm0.14}}        
\newcommand{\hatcurLBiiTxxxxxD}{\ensuremath{0.29\pm0.16}}       
\newcommand{\hatcurLBikepxxxxxD}{\ensuremath{0.4997}}           
\newcommand{\hatcurLBiikepxxxxxD}{\ensuremath{0.1776}}          
\newcommand{\hatcurLBiCxxxxxD}{\ensuremath{0.4891}}             
\newcommand{\hatcurLBiiCxxxxxD}{\ensuremath{0.1761}}            
\newcommand{\hatcurLBiMxxxxxD}{\ensuremath{0.5864}}             
\newcommand{\hatcurLBiiMxxxxxD}{\ensuremath{0.1476}}            
\newcommand{\hatcurLBiSonexxxxxD}{\ensuremath{0.1004}}            
\newcommand{\hatcurLBiiSonexxxxxD}{\ensuremath{0.1280}}           
\newcommand{\hatcurLBiStwoxxxxxD}{\ensuremath{0.0854}}            
\newcommand{\hatcurLBiiStwoxxxxxD}{\ensuremath{0.1068}}           
\newcommand{\hatcurLBiSthreexxxxxD}{\ensuremath{0.0715}}            
\newcommand{\hatcurLBiiSthreexxxxxD}{\ensuremath{0.0878}}           
\newcommand{\hatcurLBiSfourxxxxxD}{\ensuremath{0.0563}}            
\newcommand{\hatcurLBiiSfourxxxxxD}{\ensuremath{0.0694}}           
\newcommand{\hatcurISOmxxxxxD}{\ensuremath{1.004\pm0.033}}      
\newcommand{\hatcurISOmshortxxxxxD}{\ensuremath{1.00}}          
\newcommand{\hatcurISOmlongxxxxxD}{\ensuremath{1.004\pm0.033}}  
\newcommand{\hatcurISOrxxxxxD}{\ensuremath{0.938\pm0.011}}      
\newcommand{\hatcurISOrshortxxxxxD}{\ensuremath{0.94}}          
\newcommand{\hatcurISOrlongxxxxxD}{\ensuremath{0.938\pm0.011}}  
\newcommand{\hatcurISOrhoxxxxxD}{\ensuremath{1.715\pm0.094}}    
\newcommand{\hatcurISOrholongxxxxxD}{\ensuremath{1.715\pm0.094}} 
\newcommand{\hatcurISOloggxxxxxD}{\ensuremath{4.496\pm0.021}}   
\newcommand{\hatcurISOlumxxxxxD}{\ensuremath{0.767\pm0.031}}    
\newcommand{\hatcurISOlumshortxxxxxD}{\ensuremath{0.77}}        
\newcommand{\hatcurISOteffxxxxxD}{\ensuremath{5587\pm45}}       
\newcommand{\hatcurISOzfehxxxxxD}{\ensuremath{0.194\pm0.060}}   
\newcommand{\hatcurISOagexxxxxD}{\ensuremath{2.6\pm2.0}}        
\newcommand{\hatcurISOspecxxxxxD}{G}                            
\newcommand{\hatcurRVKxxxxxD}{\ensuremath{173\pm11}}            
\newcommand{\hatcurRVrkxxxxxD}{\ensuremath{0\pm0}}              
\newcommand{\hatcurRVrhxxxxxD}{\ensuremath{0\pm0}}              
\newcommand{\hatcurRVkxxxxxD}{\ensuremath{0\pm0}}               
\newcommand{\hatcurRVhxxxxxD}{\ensuremath{0\pm0}}               
\newcommand{\hatcurRVtronexxxxxD}{\ensuremath{0\pm0}}           
\newcommand{\hatcurRVtrtwoxxxxxD}{\ensuremath{0\pm0}}           
\newcommand{\hatcurRVgammaAxxxxxD}{\ensuremath{-57\pm15}}       
\newcommand{\hatcurRVjitterAxxxxxD}{\ensuremath{17\pm21}}       
\newcommand{\hatcurRVjittertwosiglimAxxxxxD}{\ensuremath{<61.3}} 
\newcommand{\hatcurRVfitrmsAxxxxxD}{\ensuremath{0.0}}           
\newcommand{\hatcurRVgammaBxxxxxD}{\ensuremath{211\pm12}}       
\newcommand{\hatcurRVjitterBxxxxxD}{\ensuremath{39\pm11}}       
\newcommand{\hatcurRVjittertwosiglimBxxxxxD}{\ensuremath{<58.4}} 
\newcommand{\hatcurRVfitrmsBxxxxxD}{\ensuremath{0.0}}           
\newcommand{\hatcurRVgammaCxxxxxD}{\ensuremath{0\pm230000000000}} 
\newcommand{\hatcurRVjitterCxxxxxD}{\ensuremath{0\pm57}}        
\newcommand{\hatcurRVjittertwosiglimCxxxxxD}{\ensuremath{<81.3}} 
\newcommand{\hatcurRVfitrmsCxxxxxD}{\ensuremath{0.0}}           
\newcommand{\hatcurRVeccenxxxxxD}{\ensuremath{0\pm0}}           
\newcommand{\hatcurRVeccentwosiglimxxxxxD}{\ensuremath{<0.000}} 
\newcommand{\hatcurRVomegaxxxxxD}{\ensuremath{0\pm0}}           
\newcommand{\hatcurPPixxxxxD}{\ensuremath{83.62\pm0.24}}        
\newcommand{\hatcurPPgxxxxxD}{\ensuremath{32.4\pm2.9}}          
\newcommand{\hatcurPPloggxxxxxD}{\ensuremath{3.510\pm0.040}}    
\newcommand{\hatcurPParxxxxxD}{\ensuremath{6.90\pm0.13}}        
\newcommand{\hatcurPParelxxxxxD}{\ensuremath{0.03010\pm0.00034}} 
\newcommand{\hatcurPPrhoxxxxxD}{\ensuremath{1.80\pm0.20}}       
\newcommand{\hatcurPPmxxxxxD}{\ensuremath{1.057\pm0.070}}       
\newcommand{\hatcurPPmshortxxxxxD}{\ensuremath{1.06}}           
\newcommand{\hatcurPPmlongxxxxxD}{\ensuremath{1.057\pm0.070}}   
\newcommand{\hatcurPPmexxxxxD}{\ensuremath{336\pm22}}           
\newcommand{\hatcurPPmeshortxxxxxD}{\ensuremath{335.8}}         
\newcommand{\hatcurPPmelongxxxxxD}{\ensuremath{336\pm22}}       
\newcommand{\hatcurPPrxxxxxD}{\ensuremath{0.899\pm0.027}}       
\newcommand{\hatcurPPrshortxxxxxD}{\ensuremath{0.90}}           
\newcommand{\hatcurPPrlongxxxxxD}{\ensuremath{0.899\pm0.027}}   
\newcommand{\hatcurPPrexxxxxD}{\ensuremath{10.08\pm0.30}}       
\newcommand{\hatcurPPreshortxxxxxD}{\ensuremath{10.1}}          
\newcommand{\hatcurPPrelongxxxxxD}{\ensuremath{10.08\pm0.30}}   
\newcommand{\hatcurPPmrcorrxxxxxD}{\ensuremath{-0.04}}          
\newcommand{\hatcurPPteffxxxxxD}{\ensuremath{1505\pm16}}        
\newcommand{\hatcurPPthetaxxxxxD}{\ensuremath{0.0702\pm0.0047}} 
\newcommand{\hatcurPPfluxperixxxxxD}{\ensuremath{1.156\pm0.049}} 
\newcommand{\hatcurPPfluxperidimxxxxxD}{\ensuremath{9}}         
\newcommand{\hatcurPPfluxapxxxxxD}{\ensuremath{1.156\pm0.049}}  
\newcommand{\hatcurPPfluxapdimxxxxxD}{\ensuremath{9}}           
\newcommand{\hatcurPPfluxavgxxxxxD}{\ensuremath{1.156\pm0.049}} 
\newcommand{\hatcurPPfluxavgdimxxxxxD}{\ensuremath{9}}          
\newcommand{\hatcurPPfluxavglogxxxxxD}{\ensuremath{9.063\pm0.018}} 
\newcommand{\hatcurXsecphasexxxxxD}{\ensuremath{0\pm0}}         
\newcommand{\hatcurXsecondaryxxxxxD}{\ensuremath{2457852.16234\pm0.00047}} 
\newcommand{\hatcurXsecdurxxxxxD}{\ensuremath{0.0691\pm0.0012}} 
\newcommand{\hatcurXsecingdurxxxxxD}{\ensuremath{0.01372\pm0.00085}} 
\newcommand{\hatcurPPphiconjxxxxxD}{\ensuremath{0\pm0}}         
\newcommand{\hatcurPPperixxxxxD}{\ensuremath{2457850.73561\pm0.00047}} 
\newcommand{\hatcurPPaequivxxxxxD}{\ensuremath{0.03430\pm0.00072}} 
\newcommand{\hatcurPPtcircxxxxxD}{\ensuremath{90\pm16}}         
\newcommand{\hatcurPPtinfallxxxxxD}{\ensuremath{270\pm31}}      
\newcommand{\hatcurXdistxxxxxD}{\ensuremath{343.2\pm3.9}}       
\newcommand{\hatcurXAvxxxxxD}{\ensuremath{0.389\pm0.043}}       
\newcommand{\hatcurXdistredxxxxxD}{\ensuremath{343.2\pm3.9}}    
\newcommand{\hatcurXEBVxxxxxD}{\ensuremath{0.126\pm0.014}}      
\newcommand{\hatcurCCpmraxxxxxD}{\ensuremath{-11.021\pm0.067}}  
\newcommand{\hatcurCCpmdecxxxxxD}{\ensuremath{-21.440\pm0.063}} 
\newcommand{\hatcurCCpmxxxxxD}{\ensuremath{24.107\pm0.092}}     

%% file: HTR130-001.tex
\newcommand{\hatcurhtrxxxxxE}{HTR130-001}                       
\newcommand{\hatcurfieldxxxxxE}{\ensuremath{string}}            
\newcommand{\hatcurCCraxxxxxE}{\ensuremath{04^{\mathrm h}58^{\mathrm m}01.0287{\mathrm s}}}                   
\newcommand{\hatcurCCdecxxxxxE}{\ensuremath{+48{\arcdeg}18{\arcmin}03.7570{\arcsec}}}                 
\newcommand{\hatcurCCmagxxxxxE}{\ensuremath{string}}            
\newcommand{\hatcurCCtwomassxxxxxE}{2MASS~04580102+4818038}     
\newcommand{\hatcurCCgscxxxxxE}{GSC~3348-01101}                 
\newcommand{\hatcurCCgaiaxxxxxE}{GAIA~255397137880340480}       
\newcommand{\hatcurCCgaiadrtwoxxxxxE}{255397142179844224} 
\newcommand{\hatcurCCtassmvxxxxxE}{\ensuremath{0\pm0}}          
\newcommand{\hatcurCCtassmvshortxxxxxE}{\ensuremath{.1fym}}     
\newcommand{\hatcurCCtassmBxxxxxE}{\ensuremath{0\pm0}}          
\newcommand{\hatcurCCtassmBshortxxxxxE}{\ensuremath{.1fym}}     
\newcommand{\hatcurCCtassmIxxxxxE}{\ensuremath{0\pm0}}          
\newcommand{\hatcurCCtassmIshortxxxxxE}{\ensuremath{.1fym}}     
\newcommand{\hatcurCCtassmgxxxxxE}{\ensuremath{0\pm0}}          
\newcommand{\hatcurCCtassmgshortxxxxxE}{\ensuremath{.1fym}}     
\newcommand{\hatcurCCtassmrxxxxxE}{\ensuremath{0\pm0}}          
\newcommand{\hatcurCCtassmrshortxxxxxE}{\ensuremath{.1fym}}     
\newcommand{\hatcurCCtassmixxxxxE}{\ensuremath{0\pm0}}          
\newcommand{\hatcurCCtassmishortxxxxxE}{\ensuremath{.1fym}}     
\newcommand{\hatcurCCparallaxxxxxxE}{\ensuremath{2.839\pm0.040}} 
\newcommand{\hatcurCCgaiamGxxxxxE}{\ensuremath{12.43620\pm0.00030}} 
\newcommand{\hatcurCCgaiamBPxxxxxE}{\ensuremath{12.8932\pm0.0013}} 
\newcommand{\hatcurCCgaiamRPxxxxxE}{\ensuremath{11.84000\pm0.00090}} 
\newcommand{\hatcurCCtwomassJmagxxxxxE}{\ensuremath{11.144\pm0.029}} 
\newcommand{\hatcurCCtwomassHmagxxxxxE}{\ensuremath{10.803\pm0.041}} 
\newcommand{\hatcurCCtwomassKmagxxxxxE}{\ensuremath{10.701\pm0.026}} 
\newcommand{\hatcurCCcitJmagxxxxxE}{\ensuremath{11.157\pm0.029}} 
\newcommand{\hatcurCCcitHmagxxxxxE}{\ensuremath{10.797\pm0.040}} 
\newcommand{\hatcurCCcitKmagxxxxxE}{\ensuremath{10.725\pm0.026}} 
\newcommand{\hatcurCCbbJmagxxxxxE}{\ensuremath{11.212\pm0.031}} 
\newcommand{\hatcurCCbbHmagxxxxxE}{\ensuremath{10.819\pm0.042}} 
\newcommand{\hatcurCCbbKmagxxxxxE}{\ensuremath{10.745\pm0.026}} 
\newcommand{\hatcurCCesoJmagxxxxxE}{\ensuremath{11.215\pm0.033}} 
\newcommand{\hatcurCCesoHmagxxxxxE}{\ensuremath{10.815\pm0.047}} 
\newcommand{\hatcurCCesoKmagxxxxxE}{\ensuremath{10.744\pm0.027}} 
\newcommand{\hatcurCCesoJHmagxxxxxE}{\ensuremath{0.399\pm0.055}} 
\newcommand{\hatcurCCesoJKmagxxxxxE}{\ensuremath{0.472\pm0.042}} 
\newcommand{\hatcurCCesoHKmagxxxxxE}{\ensuremath{0.073\pm0.054}} 
\newcommand{\hatcurCCWonemagxxxxxE}{\ensuremath{0\pm0}}         
\newcommand{\hatcurCCWtwomagxxxxxE}{\ensuremath{0\pm0}}         
\newcommand{\hatcurCCWthreemagxxxxxE}{\ensuremath{0\pm0}}       
\newcommand{\hatcurCCWfourmagxxxxxE}{\ensuremath{0\pm0}}        
\newcommand{\hatcurLCdipxxxxxE}{\ensuremath{9.3}}               
\newcommand{\hatcurLCrprstarxxxxxE}{\ensuremath{0.0942\pm0.0019}} 
\newcommand{\hatcurLCbsqxxxxxE}{\ensuremath{0.063_{-0.035}^{+0.036}}} 
\newcommand{\hatcurLCimpxxxxxE}{\ensuremath{0.250_{-0.084}^{+0.064}}} 
\newcommand{\hatcurLCzetaxxxxxE}{\ensuremath{17.02\pm0.17}}     
\newcommand{\hatcurLCdurxxxxxE}{\ensuremath{0.1293\pm0.0012}}   
\newcommand{\hatcurLCdurshortxxxxxE}{\ensuremath{0.1293}}       
\newcommand{\hatcurLCdurhrxxxxxE}{\ensuremath{3.104\pm0.028}}   
\newcommand{\hatcurLCdurhrshortxxxxxE}{\ensuremath{3.104}}      
\newcommand{\hatcurLCqxxxxxE}{\ensuremath{0.04890\pm0.00045}}   
\newcommand{\hatcurLCqshortxxxxxE}{\ensuremath{0.049}}          
\newcommand{\hatcurLCingdurxxxxxE}{\ensuremath{0.01183\pm0.00050}} 
\newcommand{\hatcurLCPxxxxxE}{\ensuremath{2.6453235\pm0.0000039}} 
\newcommand{\hatcurLCPprecxxxxxE}{\ensuremath{2.6453235}}       
\newcommand{\hatcurLCPshortxxxxxE}{\ensuremath{2.6453}}         
\newcommand{\hatcurLCTshortxxxxxE}{\ensuremath{7118.38979\pm0.00044}} 
\newcommand{\hatcurLCTxxxxxE}{\ensuremath{2457118.38979\pm0.00044}} 
\newcommand{\hatcurLCTAxxxxxE}{\ensuremath{2456131.6841\pm0.0014}} 
\newcommand{\hatcurLCTBxxxxxE}{\ensuremath{2457292.98114\pm0.00058}} 
\newcommand{\hatcurLChatnetmxxxxxE}{\ensuremath{12.379220\pm0.000097}} 
\newcommand{\hatcurLCiblendxxxxxE}{\ensuremath{0.839\pm0.055}}  
\newcommand{\hatcurLCrhoxxxxxE}{\ensuremath{0.901\pm0.042}}     
\newcommand{\hatcurSMEiteffxxxxxE}{\ensuremath{5617\pm50}}      
\newcommand{\hatcurSMEizfehxxxxxE}{\ensuremath{0.461\pm0.080}}  
\newcommand{\hatcurSMEizfehshortxxxxxE}{\ensuremath{0.46}}      
\newcommand{\hatcurSMEiloggxxxxxE}{\ensuremath{4.31\pm0.10}}    
\newcommand{\hatcurSMEivsinxxxxxE}{\ensuremath{3.53\pm0.50}}    
\newcommand{\hatcurSMEivmacxxxxxE}{\ensuremath{nff\pmnff}}      
\newcommand{\hatcurSMEivmicxxxxxE}{\ensuremath{nff\pmnff}}      
\newcommand{\hatcurSMEiiteffxxxxxE}{\ensuremath{5601\pm50}}     
\newcommand{\hatcurSMEiizfehxxxxxE}{\ensuremath{0.449\pm0.080}} 
\newcommand{\hatcurSMEiizfehshortxxxxxE}{\ensuremath{0.449}}    
\newcommand{\hatcurSMEiiloggxxxxxE}{\ensuremath{4.305\pm0.030}} 
\newcommand{\hatcurSMEiivsinxxxxxE}{\ensuremath{3.55\pm0.50}}   
\newcommand{\hatcurLBiBxxxxxE}{\ensuremath{0.7609}}             
\newcommand{\hatcurLBiiBxxxxxE}{\ensuremath{0.0700}}            
\newcommand{\hatcurLBiVxxxxxE}{\ensuremath{0.5874}}             
\newcommand{\hatcurLBiiVxxxxxE}{\ensuremath{0.1516}}            
\newcommand{\hatcurLBiRxxxxxE}{\ensuremath{0.4820}}             
\newcommand{\hatcurLBiiRxxxxxE}{\ensuremath{0.1735}}            
\newcommand{\hatcurLBiIxxxxxE}{\ensuremath{0.3806}}             
\newcommand{\hatcurLBiiIxxxxxE}{\ensuremath{0.1873}}            
\newcommand{\hatcurLBiuxxxxxE}{\ensuremath{0.9339}}             
\newcommand{\hatcurLBiiuxxxxxE}{\ensuremath{-0.0938}}           
\newcommand{\hatcurLBigxxxxxE}{\ensuremath{0.7007}}             
\newcommand{\hatcurLBiigxxxxxE}{\ensuremath{0.0918}}            
\newcommand{\hatcurLBirxxxxxE}{\ensuremath{0.36\pm0.15}}        
\newcommand{\hatcurLBiirxxxxxE}{\ensuremath{0.31\pm0.18}}       
\newcommand{\hatcurLBiixxxxxE}{\ensuremath{0.33\pm0.10}}        
\newcommand{\hatcurLBiiixxxxxE}{\ensuremath{0.22\pm0.16}}       
\newcommand{\hatcurLBizxxxxxE}{\ensuremath{0.35\pm0.16}}        
\newcommand{\hatcurLBiizxxxxxE}{\ensuremath{0.35\pm0.17}}       
\newcommand{\hatcurLBiJxxxxxE}{\ensuremath{0.2219}}             
\newcommand{\hatcurLBiiJxxxxxE}{\ensuremath{0.2241}}            
\newcommand{\hatcurLBiHxxxxxE}{\ensuremath{0.1309}}             
\newcommand{\hatcurLBiiHxxxxxE}{\ensuremath{0.2607}}            
\newcommand{\hatcurLBiKxxxxxE}{\ensuremath{0.1205}}             
\newcommand{\hatcurLBiiKxxxxxE}{\ensuremath{0.2088}}            
\newcommand{\hatcurLBiTxxxxxE}{\ensuremath{0.3799}}             
\newcommand{\hatcurLBiiTxxxxxE}{\ensuremath{0.2072}}            
\newcommand{\hatcurLBikepxxxxxE}{\ensuremath{0.4960}}           
\newcommand{\hatcurLBiikepxxxxxE}{\ensuremath{0.1811}}          
\newcommand{\hatcurLBiCxxxxxE}{\ensuremath{0.4858}}             
\newcommand{\hatcurLBiiCxxxxxE}{\ensuremath{0.1794}}            
\newcommand{\hatcurLBiMxxxxxE}{\ensuremath{0.5828}}             
\newcommand{\hatcurLBiiMxxxxxE}{\ensuremath{0.1517}}            
\newcommand{\hatcurLBiSonexxxxxE}{\ensuremath{0.0977}}            
\newcommand{\hatcurLBiiSonexxxxxE}{\ensuremath{0.1285}}           
\newcommand{\hatcurLBiStwoxxxxxE}{\ensuremath{0.0835}}            
\newcommand{\hatcurLBiiStwoxxxxxE}{\ensuremath{0.1074}}           
\newcommand{\hatcurLBiSthreexxxxxE}{\ensuremath{0.0701}}            
\newcommand{\hatcurLBiiSthreexxxxxE}{\ensuremath{0.0881}}           
\newcommand{\hatcurLBiSfourxxxxxE}{\ensuremath{0.0548}}            
\newcommand{\hatcurLBiiSfourxxxxxE}{\ensuremath{0.0697}}           
\newcommand{\hatcurISOmxxxxxE}{\ensuremath{1.023\pm0.020}}      
\newcommand{\hatcurISOmshortxxxxxE}{\ensuremath{1.02}}          
\newcommand{\hatcurISOmlongxxxxxE}{\ensuremath{1.023\pm0.020}}  
\newcommand{\hatcurISOrxxxxxE}{\ensuremath{1.170\pm0.016}}      
\newcommand{\hatcurISOrshortxxxxxE}{\ensuremath{1.17}}          
\newcommand{\hatcurISOrlongxxxxxE}{\ensuremath{1.170\pm0.016}}  
\newcommand{\hatcurISOrhoxxxxxE}{\ensuremath{0.901\pm0.042}}    
\newcommand{\hatcurISOrholongxxxxxE}{\ensuremath{0.901\pm0.042}} 
\newcommand{\hatcurISOloggxxxxxE}{\ensuremath{4.312\pm0.015}}   
\newcommand{\hatcurISOlumxxxxxE}{\ensuremath{1.232\pm0.053}}    
\newcommand{\hatcurISOlumshortxxxxxE}{\ensuremath{1.23}}        
\newcommand{\hatcurISOteffxxxxxE}{\ensuremath{5629\pm48}}       
\newcommand{\hatcurISOzfehxxxxxE}{\ensuremath{0.414\pm0.090}}   
\newcommand{\hatcurISOagexxxxxE}{\ensuremath{8.1\pm1.1}}        
\newcommand{\hatcurISOspecxxxxxE}{G}                            
\newcommand{\hatcurRVKxxxxxE}{\ensuremath{110\pm13}}            
\newcommand{\hatcurRVrkxxxxxE}{\ensuremath{0\pm0}}              
\newcommand{\hatcurRVrhxxxxxE}{\ensuremath{0\pm0}}              
\newcommand{\hatcurRVkxxxxxE}{\ensuremath{0\pm0}}               
\newcommand{\hatcurRVhxxxxxE}{\ensuremath{0\pm0}}               
\newcommand{\hatcurRVtronexxxxxE}{\ensuremath{0\pm0}}           
\newcommand{\hatcurRVtrtwoxxxxxE}{\ensuremath{0\pm0}}           
\newcommand{\hatcurRVgammaxxxxxE}{\ensuremath{127\pm11}}        
\newcommand{\hatcurRVjitterxxxxxE}{\ensuremath{33\pm11}}        
\newcommand{\hatcurRVjittertwosiglimxxxxxE}{\ensuremath{<53.6}} 
\newcommand{\hatcurRVfitrmsxxxxxE}{\ensuremath{.1fym}}          %
\newcommand{\hatcurRVeccenxxxxxE}{\ensuremath{0\pm0}}           
\newcommand{\hatcurRVeccentwosiglimxxxxxE}{\ensuremath{<0.000}} 
\newcommand{\hatcurRVomegaxxxxxE}{\ensuremath{0\pm0}}           
\newcommand{\hatcurPPixxxxxE}{\ensuremath{87.93\pm0.64}}        
\newcommand{\hatcurPPgxxxxxE}{\ensuremath{16.4\pm2.1}}          
\newcommand{\hatcurPPloggxxxxxE}{\ensuremath{3.214\pm0.056}}    
\newcommand{\hatcurPParxxxxxE}{\ensuremath{6.93\pm0.11}}        
\newcommand{\hatcurPParelxxxxxE}{\ensuremath{0.03772\pm0.00024}} 
\newcommand{\hatcurPPrhoxxxxxE}{\ensuremath{0.77\pm0.11}}       
\newcommand{\hatcurPPmxxxxxE}{\ensuremath{0.761\pm0.088}}       
\newcommand{\hatcurPPmshortxxxxxE}{\ensuremath{0.76}}           
\newcommand{\hatcurPPmlongxxxxxE}{\ensuremath{0.761\pm0.088}}   
\newcommand{\hatcurPPmexxxxxE}{\ensuremath{242\pm28}}           
\newcommand{\hatcurPPmeshortxxxxxE}{\ensuremath{241.9}}         
\newcommand{\hatcurPPmelongxxxxxE}{\ensuremath{242\pm28}}       
\newcommand{\hatcurPPrxxxxxE}{\ensuremath{1.073\pm0.029}}       
\newcommand{\hatcurPPrshortxxxxxE}{\ensuremath{1.07}}           
\newcommand{\hatcurPPrlongxxxxxE}{\ensuremath{1.073\pm0.029}}   
\newcommand{\hatcurPPrexxxxxE}{\ensuremath{12.03\pm0.32}}       
\newcommand{\hatcurPPreshortxxxxxE}{\ensuremath{12.0}}          
\newcommand{\hatcurPPrelongxxxxxE}{\ensuremath{12.03\pm0.32}}   
\newcommand{\hatcurPPmrcorrxxxxxE}{\ensuremath{0.02}}           
\newcommand{\hatcurPPteffxxxxxE}{\ensuremath{1512\pm13}}        
\newcommand{\hatcurPPthetaxxxxxE}{\ensuremath{0.0522\pm0.0061}} 
\newcommand{\hatcurPPfluxperixxxxxE}{\ensuremath{1.179\pm0.040}} 
\newcommand{\hatcurPPfluxperidimxxxxxE}{\ensuremath{9}}         
\newcommand{\hatcurPPfluxapxxxxxE}{\ensuremath{1.179\pm0.040}}  
\newcommand{\hatcurPPfluxapdimxxxxxE}{\ensuremath{9}}           
\newcommand{\hatcurPPfluxavgxxxxxE}{\ensuremath{1.179\pm0.040}} 
\newcommand{\hatcurPPfluxavgdimxxxxxE}{\ensuremath{9}}          
\newcommand{\hatcurPPfluxavglogxxxxxE}{\ensuremath{9.072\pm0.015}} 
\newcommand{\hatcurXsecphasexxxxxE}{\ensuremath{0\pm0}}         
\newcommand{\hatcurXsecondaryxxxxxE}{\ensuremath{2457119.71245\pm0.00044}} 
\newcommand{\hatcurXsecdurxxxxxE}{\ensuremath{0.1293\pm0.0012}} 
\newcommand{\hatcurXsecingdurxxxxxE}{\ensuremath{0.01183\pm0.00050}} 
\newcommand{\hatcurPPphiconjxxxxxE}{\ensuremath{0\pm0}}         
\newcommand{\hatcurPPperixxxxxE}{\ensuremath{2457117.72846\pm0.00044}} 
\newcommand{\hatcurPPaequivxxxxxE}{\ensuremath{0.03400\pm0.00058}} 
\newcommand{\hatcurPPtcircxxxxxE}{\ensuremath{112\pm20}}        
\newcommand{\hatcurPPtinfallxxxxxE}{\ensuremath{540\pm79}}      
\newcommand{\hatcurXdistxxxxxE}{\ensuremath{353.1\pm4.4}}       
\newcommand{\hatcurXAvxxxxxE}{\ensuremath{0.339\pm0.046}}       
\newcommand{\hatcurXdistredxxxxxE}{\ensuremath{353.1\pm4.4}}    
\newcommand{\hatcurXEBVxxxxxE}{\ensuremath{0.109\pm0.015}}      
\newcommand{\hatcurCCpmraxxxxxE}{\ensuremath{14.732\pm0.080}}   
\newcommand{\hatcurCCpmdecxxxxxE}{\ensuremath{-43.776\pm0.061}} 
\newcommand{\hatcurCCpmxxxxxE}{\ensuremath{46.19\pm0.10}}       

%% file: HTR384-014.tex
\newcommand{\hatcurhtrxxxxxF}{HTR384-014}                       
\newcommand{\hatcurfieldxxxxxF}{\ensuremath{string}}            
\newcommand{\hatcurCCraxxxxxF}{\ensuremath{17^{\mathrm h}58^{\mathrm m}17.3121{\mathrm s}}}                   
\newcommand{\hatcurCCdecxxxxxF}{\ensuremath{+05{\arcdeg}45{\arcmin}40.9400{\arcsec}}}                 
\newcommand{\hatcurCCmagxxxxxF}{13.753}                         
\newcommand{\hatcurCCtwomassxxxxxF}{2MASS~17581730+0545409}     
\newcommand{\hatcurCCgscxxxxxF}{GSC~0429-01697}                 
\newcommand{\hatcurCCgaiaxxxxxF}{GAIA~4474644332250439552}      
\newcommand{\hatcurCCgaiadrtwoxxxxxF}{4474644332250439552} 
\newcommand{\hatcurCCtassmvxxxxxF}{\ensuremath{13.753\pm0.065}} 
\newcommand{\hatcurCCtassmvshortxxxxxF}{\ensuremath{13.8}}      
\newcommand{\hatcurCCtassmBxxxxxF}{\ensuremath{14.729\pm0.066}} 
\newcommand{\hatcurCCtassmBshortxxxxxF}{\ensuremath{14.7}}      
\newcommand{\hatcurCCtassmIxxxxxF}{\ensuremath{nff\pmnff}}      
\newcommand{\hatcurCCtassmIshortxxxxxF}{\ensuremath{0.0}}       
\newcommand{\hatcurCCtassmgxxxxxF}{\ensuremath{14.258\pm0.026}} 
\newcommand{\hatcurCCtassmgshortxxxxxF}{\ensuremath{14.3}}      
\newcommand{\hatcurCCtassmrxxxxxF}{\ensuremath{13.418\pm0.093}} 
\newcommand{\hatcurCCtassmrshortxxxxxF}{\ensuremath{13.4}}      
\newcommand{\hatcurCCtassmixxxxxF}{\ensuremath{13.136\pm0.086}} 
\newcommand{\hatcurCCtassmishortxxxxxF}{\ensuremath{13.1}}      
\newcommand{\hatcurCCparallaxxxxxxF}{\ensuremath{2.450\pm0.024}} 
\newcommand{\hatcurCCgaiamGxxxxxF}{\ensuremath{13.51060\pm0.00030}} 
\newcommand{\hatcurCCgaiamBPxxxxxF}{\ensuremath{14.0381\pm0.0012}} 
\newcommand{\hatcurCCgaiamRPxxxxxF}{\ensuremath{12.84760\pm0.00080}} 
\newcommand{\hatcurCCtwomassJmagxxxxxF}{\ensuremath{12.021\pm0.021}} 
\newcommand{\hatcurCCtwomassHmagxxxxxF}{\ensuremath{11.630\pm0.028}} 
\newcommand{\hatcurCCtwomassKmagxxxxxF}{\ensuremath{11.512\pm0.023}} 
\newcommand{\hatcurCCcitJmagxxxxxF}{\ensuremath{12.030\pm0.022}} 
\newcommand{\hatcurCCcitHmagxxxxxF}{\ensuremath{11.623\pm0.028}} 
\newcommand{\hatcurCCcitKmagxxxxxF}{\ensuremath{11.536\pm0.023}} 
\newcommand{\hatcurCCbbJmagxxxxxF}{\ensuremath{12.091\pm0.024}} 
\newcommand{\hatcurCCbbHmagxxxxxF}{\ensuremath{11.646\pm0.029}} 
\newcommand{\hatcurCCbbKmagxxxxxF}{\ensuremath{11.556\pm0.023}} 
\newcommand{\hatcurCCesoJmagxxxxxF}{\ensuremath{12.095\pm0.026}} 
\newcommand{\hatcurCCesoHmagxxxxxF}{\ensuremath{11.643\pm0.034}} 
\newcommand{\hatcurCCesoKmagxxxxxF}{\ensuremath{11.554\pm0.024}} 
\newcommand{\hatcurCCesoJHmagxxxxxF}{\ensuremath{0.452\pm0.040}} 
\newcommand{\hatcurCCesoJKmagxxxxxF}{\ensuremath{0.541\pm0.034}} 
\newcommand{\hatcurCCesoHKmagxxxxxF}{\ensuremath{0.089\pm0.042}} 
\newcommand{\hatcurCCWonemagxxxxxF}{\ensuremath{11.453\pm0.022}} 
\newcommand{\hatcurCCWtwomagxxxxxF}{\ensuremath{11.514\pm0.022}} 
\newcommand{\hatcurCCWthreemagxxxxxF}{\ensuremath{0\pm0}}       
\newcommand{\hatcurCCWfourmagxxxxxF}{\ensuremath{0\pm0}}        
\newcommand{\hatcurLCdipxxxxxF}{\ensuremath{17.6}}              
\newcommand{\hatcurLCrprstarxxxxxF}{\ensuremath{0.1191\pm0.0032}} 
\newcommand{\hatcurLCbsqxxxxxF}{\ensuremath{0.069_{-0.030}^{+0.040}}} 
\newcommand{\hatcurLCimpxxxxxF}{\ensuremath{0.262_{-0.066}^{+0.068}}} 
\newcommand{\hatcurLCzetaxxxxxF}{\ensuremath{18.45\pm0.27}}     
\newcommand{\hatcurLCdurxxxxxF}{\ensuremath{0.1222\pm0.0016}}   
\newcommand{\hatcurLCdurshortxxxxxF}{\ensuremath{0.1222}}       
\newcommand{\hatcurLCdurhrxxxxxF}{\ensuremath{2.933\pm0.037}}   
\newcommand{\hatcurLCdurhrshortxxxxxF}{\ensuremath{2.933}}      
\newcommand{\hatcurLCqxxxxxF}{\ensuremath{0.03620\pm0.00046}}   
\newcommand{\hatcurLCqshortxxxxxF}{\ensuremath{0.036}}          
\newcommand{\hatcurLCingdurxxxxxF}{\ensuremath{0.01392\pm0.00058}} 
\newcommand{\hatcurLCPxxxxxF}{\ensuremath{3.377728\pm0.000013}} 
\newcommand{\hatcurLCPprecxxxxxF}{\ensuremath{3.3777279}}       
\newcommand{\hatcurLCPshortxxxxxF}{\ensuremath{3.3777}}         
\newcommand{\hatcurLCTshortxxxxxF}{\ensuremath{6382.94256\pm0.00053}} 
\newcommand{\hatcurLCTxxxxxF}{\ensuremath{2456382.94256\pm0.00053}} 
\newcommand{\hatcurLCTAxxxxxF}{\ensuremath{2454964.2969\pm0.0055}} 
\newcommand{\hatcurLCTBxxxxxF}{\ensuremath{2456399.83120\pm0.00054}} 
\newcommand{\hatcurLCrhoxxxxxF}{\ensuremath{1.448\pm0.061}}     
\newcommand{\hatcurSMEiteffxxxxxF}{\ensuremath{5365\pm50}}      
\newcommand{\hatcurSMEizfehxxxxxF}{\ensuremath{0.428\pm0.080}}  
\newcommand{\hatcurSMEizfehshortxxxxxF}{\ensuremath{0.43}}      
\newcommand{\hatcurSMEiloggxxxxxF}{\ensuremath{4.40\pm0.10}}    
\newcommand{\hatcurSMEivsinxxxxxF}{\ensuremath{3.22\pm0.50}}    
\newcommand{\hatcurSMEivmacxxxxxF}{\ensuremath{nff\pmnff}}      
\newcommand{\hatcurSMEivmicxxxxxF}{\ensuremath{nff\pmnff}}      
\newcommand{\hatcurLBiBxxxxxF}{\ensuremath{0.8183}}             
\newcommand{\hatcurLBiiBxxxxxF}{\ensuremath{0.0186}}            
\newcommand{\hatcurLBiVxxxxxF}{\ensuremath{0.6325}}             
\newcommand{\hatcurLBiiVxxxxxF}{\ensuremath{0.1208}}            
\newcommand{\hatcurLBiRxxxxxF}{\ensuremath{0.5132}}             
\newcommand{\hatcurLBiiRxxxxxF}{\ensuremath{0.1589}}            
\newcommand{\hatcurLBiIxxxxxF}{\ensuremath{0.4073}}             
\newcommand{\hatcurLBiiIxxxxxF}{\ensuremath{0.1780}}            
\newcommand{\hatcurLBiuxxxxxF}{\ensuremath{1.0107}}             
\newcommand{\hatcurLBiiuxxxxxF}{\ensuremath{-0.1834}}           
\newcommand{\hatcurLBigxxxxxF}{\ensuremath{0.7517}}             
\newcommand{\hatcurLBiigxxxxxF}{\ensuremath{0.0481}}            
\newcommand{\hatcurLBirxxxxxF}{\ensuremath{0.44\pm0.15}}        
\newcommand{\hatcurLBiirxxxxxF}{\ensuremath{0.38\pm0.16}}       
\newcommand{\hatcurLBiixxxxxF}{\ensuremath{0.47\pm0.13}}        
\newcommand{\hatcurLBiiixxxxxF}{\ensuremath{0.41\pm0.14}}       
\newcommand{\hatcurLBizxxxxxF}{\ensuremath{0.3593}}             
\newcommand{\hatcurLBiizxxxxxF}{\ensuremath{0.1834}}            
\newcommand{\hatcurLBiJxxxxxF}{\ensuremath{0.2380}}             
\newcommand{\hatcurLBiiJxxxxxF}{\ensuremath{0.2252}}            
\newcommand{\hatcurLBiHxxxxxF}{\ensuremath{0.1368}}             
\newcommand{\hatcurLBiiHxxxxxF}{\ensuremath{0.2704}}            
\newcommand{\hatcurLBiKxxxxxF}{\ensuremath{0.1239}}             
\newcommand{\hatcurLBiiKxxxxxF}{\ensuremath{0.2193}}            
\newcommand{\hatcurLBiTxxxxxF}{\ensuremath{0.3989}}             
\newcommand{\hatcurLBiiTxxxxxF}{\ensuremath{0.2029}}            
\newcommand{\hatcurLBikepxxxxxF}{\ensuremath{0.5196}}           
\newcommand{\hatcurLBiikepxxxxxF}{\ensuremath{0.1699}}          
\newcommand{\hatcurLBiCxxxxxF}{\ensuremath{0.5069}}             
\newcommand{\hatcurLBiiCxxxxxF}{\ensuremath{0.1693}}            
\newcommand{\hatcurLBiMxxxxxF}{\ensuremath{0.6102}}             
\newcommand{\hatcurLBiiMxxxxxF}{\ensuremath{0.1350}}            
\newcommand{\hatcurLBiSonexxxxxF}{\ensuremath{0.1019}}            
\newcommand{\hatcurLBiiSonexxxxxF}{\ensuremath{0.1351}}           
\newcommand{\hatcurLBiStwoxxxxxF}{\ensuremath{0.0886}}            
\newcommand{\hatcurLBiiStwoxxxxxF}{\ensuremath{0.1097}}           
\newcommand{\hatcurLBiSthreexxxxxF}{\ensuremath{0.0735}}            
\newcommand{\hatcurLBiiSthreexxxxxF}{\ensuremath{0.0906}}           
\newcommand{\hatcurLBiSfourxxxxxF}{\ensuremath{0.0584}}            
\newcommand{\hatcurLBiiSfourxxxxxF}{\ensuremath{0.0727}}           
\newcommand{\hatcurISOmxxxxxF}{\ensuremath{0.925\pm0.023}}      
\newcommand{\hatcurISOmshortxxxxxF}{\ensuremath{0.92}}          
\newcommand{\hatcurISOmlongxxxxxF}{\ensuremath{0.925\pm0.023}}  
\newcommand{\hatcurISOrxxxxxF}{\ensuremath{0.9661_{-0.0082}^{+0.0110}}} 
\newcommand{\hatcurISOrshortxxxxxF}{\ensuremath{0.97}}          
\newcommand{\hatcurISOrlongxxxxxF}{\ensuremath{0.9661_{-0.0082}^{+0.0110}}} 
\newcommand{\hatcurISOrhoxxxxxF}{\ensuremath{1.448\pm0.061}}    
\newcommand{\hatcurISOrholongxxxxxF}{\ensuremath{1.448\pm0.061}} 
\newcommand{\hatcurISOloggxxxxxF}{\ensuremath{4.435\pm0.015}}   
\newcommand{\hatcurISOlumxxxxxF}{\ensuremath{0.714\pm0.028}}    
\newcommand{\hatcurISOlumshortxxxxxF}{\ensuremath{0.71}}        
\newcommand{\hatcurISOteffxxxxxF}{\ensuremath{5400_{-39}^{+55}}} 
\newcommand{\hatcurISOzfehxxxxxF}{\ensuremath{0.251\pm0.061}}   
\newcommand{\hatcurISOagexxxxxF}{\ensuremath{9.0\pm1.7}}        
\newcommand{\hatcurISOspecxxxxxF}{G}                            
\newcommand{\hatcurRVKxxxxxF}{\ensuremath{87.3\pm3.2}}          
\newcommand{\hatcurRVrkxxxxxF}{\ensuremath{0\pm0}}              
\newcommand{\hatcurRVrhxxxxxF}{\ensuremath{0\pm0}}              
\newcommand{\hatcurRVkxxxxxF}{\ensuremath{0\pm0}}               
\newcommand{\hatcurRVhxxxxxF}{\ensuremath{0\pm0}}               
\newcommand{\hatcurRVtronexxxxxF}{\ensuremath{0\pm0}}           
\newcommand{\hatcurRVtrtwoxxxxxF}{\ensuremath{0\pm0}}           
\newcommand{\hatcurRVgammaAxxxxxF}{\ensuremath{6.7\pm2.4}}      
\newcommand{\hatcurRVjitterAxxxxxF}{\ensuremath{0.01\pm0.91}}   
\newcommand{\hatcurRVjittertwosiglimAxxxxxF}{\ensuremath{<2.4}} 
\newcommand{\hatcurRVfitrmsAxxxxxF}{\ensuremath{0.0}}           
\newcommand{\hatcurRVgammaBxxxxxF}{\ensuremath{-105\pm27}}      
\newcommand{\hatcurRVjitterBxxxxxF}{\ensuremath{0.03\pm0.74}}   
\newcommand{\hatcurRVjittertwosiglimBxxxxxF}{\ensuremath{<1.9}} 
\newcommand{\hatcurRVfitrmsBxxxxxF}{\ensuremath{0.0}}           
\newcommand{\hatcurRVgammaCxxxxxF}{\ensuremath{-69596.7\pm7.9}} 
\newcommand{\hatcurRVjitterCxxxxxF}{\ensuremath{16\pm10}}       
\newcommand{\hatcurRVjittertwosiglimCxxxxxF}{\ensuremath{<35.7}} 
\newcommand{\hatcurRVfitrmsCxxxxxF}{\ensuremath{0.0}}           
\newcommand{\hatcurRVeccenxxxxxF}{\ensuremath{0\pm0}}           
\newcommand{\hatcurRVeccentwosiglimxxxxxF}{\ensuremath{<0.000}} 
\newcommand{\hatcurRVomegaxxxxxF}{\ensuremath{0\pm0}}           
\newcommand{\hatcurPPixxxxxF}{\ensuremath{88.45\pm0.44}}        
\newcommand{\hatcurPPgxxxxxF}{\ensuremath{12.08\pm0.93}}        
\newcommand{\hatcurPPloggxxxxxF}{\ensuremath{3.082\pm0.034}}    
\newcommand{\hatcurPParxxxxxF}{\ensuremath{9.56\pm0.14}}        
\newcommand{\hatcurPParelxxxxxF}{\ensuremath{0.04294\pm0.00035}} 
\newcommand{\hatcurPPrhoxxxxxF}{\ensuremath{0.540\pm0.055}}     
\newcommand{\hatcurPPmxxxxxF}{\ensuremath{0.614\pm0.024}}       
\newcommand{\hatcurPPmshortxxxxxF}{\ensuremath{0.61}}           
\newcommand{\hatcurPPmlongxxxxxF}{\ensuremath{0.614\pm0.024}}   
\newcommand{\hatcurPPmexxxxxF}{\ensuremath{195.3\pm7.6}}        
\newcommand{\hatcurPPmeshortxxxxxF}{\ensuremath{195.3}}         
\newcommand{\hatcurPPmelongxxxxxF}{\ensuremath{195.3\pm7.6}}    
\newcommand{\hatcurPPrxxxxxF}{\ensuremath{1.119\pm0.033}}       
\newcommand{\hatcurPPrshortxxxxxF}{\ensuremath{1.12}}           
\newcommand{\hatcurPPrlongxxxxxF}{\ensuremath{1.119\pm0.033}}   
\newcommand{\hatcurPPrexxxxxF}{\ensuremath{12.55\pm0.37}}       
\newcommand{\hatcurPPreshortxxxxxF}{\ensuremath{12.5}}          
\newcommand{\hatcurPPrelongxxxxxF}{\ensuremath{12.55\pm0.37}}   
\newcommand{\hatcurPPmrcorrxxxxxF}{\ensuremath{-0.25}}          
\newcommand{\hatcurPPteffxxxxxF}{\ensuremath{1237\pm11}}        
\newcommand{\hatcurPPthetaxxxxxF}{\ensuremath{0.0506\pm0.0026}} 
\newcommand{\hatcurPPfluxperixxxxxF}{\ensuremath{5.27\pm0.18}}  
\newcommand{\hatcurPPfluxperidimxxxxxF}{\ensuremath{8}}         
\newcommand{\hatcurPPfluxapxxxxxF}{\ensuremath{5.27\pm0.18}}    
\newcommand{\hatcurPPfluxapdimxxxxxF}{\ensuremath{8}}           
\newcommand{\hatcurPPfluxavgxxxxxF}{\ensuremath{5.27\pm0.18}}   
\newcommand{\hatcurPPfluxavgdimxxxxxF}{\ensuremath{8}}          
\newcommand{\hatcurPPfluxavglogxxxxxF}{\ensuremath{8.722\pm0.015}} 
\newcommand{\hatcurXsecphasexxxxxF}{\ensuremath{0\pm0}}         
\newcommand{\hatcurXsecondaryxxxxxF}{\ensuremath{2456384.63142\pm0.00053}} 
\newcommand{\hatcurXsecdurxxxxxF}{\ensuremath{0.1222\pm0.0016}} 
\newcommand{\hatcurXsecingdurxxxxxF}{\ensuremath{0.01392\pm0.00058}} 
\newcommand{\hatcurPPphiconjxxxxxF}{\ensuremath{0\pm0}}         
\newcommand{\hatcurPPperixxxxxF}{\ensuremath{2456382.09813\pm0.00053}} 
\newcommand{\hatcurPPaequivxxxxxF}{\ensuremath{0.05080\pm0.00088}} 
\newcommand{\hatcurPPtcircxxxxxF}{\ensuremath{197\pm31}}        
\newcommand{\hatcurPPtinfallxxxxxF}{\ensuremath{3910\pm330}}    
\newcommand{\hatcurXdistxxxxxF}{\ensuremath{408.0\pm4.1}}       
\newcommand{\hatcurXAvxxxxxF}{\ensuremath{0.506\pm0.046}}       
\newcommand{\hatcurXdistredxxxxxF}{\ensuremath{408.0\pm4.0}}    
\newcommand{\hatcurXEBVxxxxxF}{\ensuremath{0.163\pm0.015}}      
\newcommand{\hatcurCCpmraxxxxxF}{\ensuremath{-14.871\pm0.036}}  
\newcommand{\hatcurCCpmdecxxxxxF}{\ensuremath{-0.301\pm0.039}}  
\newcommand{\hatcurCCpmxxxxxF}{\ensuremath{14.874\pm0.053}}     

%% file: HTR405-001.tex
\newcommand{\hatcurhtrxxxxxG}{HTR405-001}                       
\newcommand{\hatcurfieldxxxxxG}{\ensuremath{string}}            
\newcommand{\hatcurCCraxxxxxG}{\ensuremath{04^{\mathrm h}35^{\mathrm m}53.8469{\mathrm s}}}                   
\newcommand{\hatcurCCdecxxxxxG}{\ensuremath{+02{\arcdeg}25{\arcmin}52.6434{\arcsec}}}                 
\newcommand{\hatcurCCmagxxxxxG}{12.771}                         
\newcommand{\hatcurCCtwomassxxxxxG}{2MASS~04355384+0225526}     
\newcommand{\hatcurCCgscxxxxxG}{GSC~0086-00341}                 
\newcommand{\hatcurCCgaiaxxxxxG}{GAIA~3279418602369232000}      
\newcommand{\hatcurCCgaiadrtwoxxxxxG}{3279418602369232000} 
\newcommand{\hatcurCCtassmvxxxxxG}{\ensuremath{12.771\pm0.010}} 
\newcommand{\hatcurCCtassmvshortxxxxxG}{\ensuremath{12.8}}      
\newcommand{\hatcurCCtassmBxxxxxG}{\ensuremath{13.446\pm0.011}} 
\newcommand{\hatcurCCtassmBshortxxxxxG}{\ensuremath{13.4}}      
\newcommand{\hatcurCCtassmIxxxxxG}{\ensuremath{12.105\pm0.070}} 
\newcommand{\hatcurCCtassmIshortxxxxxG}{\ensuremath{12.1}}      
\newcommand{\hatcurCCtassmgxxxxxG}{\ensuremath{13.062\pm0.013}} 
\newcommand{\hatcurCCtassmgshortxxxxxG}{\ensuremath{13.1}}      
\newcommand{\hatcurCCtassmrxxxxxG}{\ensuremath{12.553\pm0.075}} 
\newcommand{\hatcurCCtassmrshortxxxxxG}{\ensuremath{12.6}}      
\newcommand{\hatcurCCtassmixxxxxG}{\ensuremath{12.440\pm0.037}} 
\newcommand{\hatcurCCtassmishortxxxxxG}{\ensuremath{12.4}}      
\newcommand{\hatcurCCparallaxxxxxxG}{\ensuremath{1.505\pm0.035}} 
\newcommand{\hatcurCCgaiamGxxxxxG}{\ensuremath{12.62210\pm0.00010}} 
\newcommand{\hatcurCCgaiamBPxxxxxG}{\ensuremath{12.98580\pm0.00080}} 
\newcommand{\hatcurCCgaiamRPxxxxxG}{\ensuremath{12.09530\pm0.00070}} 
\newcommand{\hatcurCCtwomassJmagxxxxxG}{\ensuremath{11.485\pm0.026}} 
\newcommand{\hatcurCCtwomassHmagxxxxxG}{\ensuremath{11.225\pm0.025}} 
\newcommand{\hatcurCCtwomassKmagxxxxxG}{\ensuremath{11.123\pm0.021}} 
\newcommand{\hatcurCCcitJmagxxxxxG}{\ensuremath{11.502\pm0.026}} 
\newcommand{\hatcurCCcitHmagxxxxxG}{\ensuremath{11.219\pm0.025}} 
\newcommand{\hatcurCCcitKmagxxxxxG}{\ensuremath{11.147\pm0.021}} 
\newcommand{\hatcurCCbbJmagxxxxxG}{\ensuremath{11.551\pm0.028}} 
\newcommand{\hatcurCCbbHmagxxxxxG}{\ensuremath{11.241\pm0.026}} 
\newcommand{\hatcurCCbbKmagxxxxxG}{\ensuremath{11.167\pm0.021}} 
\newcommand{\hatcurCCesoJmagxxxxxG}{\ensuremath{11.553\pm0.029}} 
\newcommand{\hatcurCCesoHmagxxxxxG}{\ensuremath{11.238\pm0.030}} 
\newcommand{\hatcurCCesoKmagxxxxxG}{\ensuremath{11.166\pm0.022}} 
\newcommand{\hatcurCCesoJHmagxxxxxG}{\ensuremath{0.315\pm0.040}} 
\newcommand{\hatcurCCesoJKmagxxxxxG}{\ensuremath{0.387\pm0.036}} 
\newcommand{\hatcurCCesoHKmagxxxxxG}{\ensuremath{0.072\pm0.038}} 
\newcommand{\hatcurCCWonemagxxxxxG}{\ensuremath{11.058\pm0.021}} 
\newcommand{\hatcurCCWtwomagxxxxxG}{\ensuremath{11.055\pm0.021}} 
\newcommand{\hatcurCCWthreemagxxxxxG}{\ensuremath{0\pm0}}       
\newcommand{\hatcurCCWfourmagxxxxxG}{\ensuremath{0\pm0}}        
\newcommand{\hatcurLCdipxxxxxG}{\ensuremath{11.2}}              
\newcommand{\hatcurLCrprstarxxxxxG}{\ensuremath{0.1007\pm0.0034}} 
\newcommand{\hatcurLCbsqxxxxxG}{\ensuremath{0.054_{-0.030}^{+0.046}}} 
\newcommand{\hatcurLCimpxxxxxG}{\ensuremath{0.232_{-0.079}^{+0.085}}} 
\newcommand{\hatcurLCzetaxxxxxG}{\ensuremath{10.79\pm0.11}}     
\newcommand{\hatcurLCdurxxxxxG}{\ensuremath{0.2052\pm0.0020}}   
\newcommand{\hatcurLCdurshortxxxxxG}{\ensuremath{0.2052}}       
\newcommand{\hatcurLCdurhrxxxxxG}{\ensuremath{4.926\pm0.048}}   
\newcommand{\hatcurLCdurhrshortxxxxxG}{\ensuremath{4.926}}      
\newcommand{\hatcurLCqxxxxxG}{\ensuremath{0.05120\pm0.00050}}   
\newcommand{\hatcurLCqshortxxxxxG}{\ensuremath{0.051}}          
\newcommand{\hatcurLCingdurxxxxxG}{\ensuremath{0.0199\pm0.0010}} 
\newcommand{\hatcurLCPxxxxxG}{\ensuremath{4.0072320\pm0.0000017}} 
\newcommand{\hatcurLCPprecxxxxxG}{\ensuremath{4.0072320}}       
\newcommand{\hatcurLCPshortxxxxxG}{\ensuremath{4.0072}}         
\newcommand{\hatcurLCTshortxxxxxG}{\ensuremath{7751.46354\pm0.00063}} 
\newcommand{\hatcurLCTxxxxxG}{\ensuremath{2457751.46354\pm0.00063}} 
\newcommand{\hatcurLCTAxxxxxG}{\ensuremath{2455078.6398\pm0.0011}} 
\newcommand{\hatcurLCTBxxxxxG}{\ensuremath{2458460.74358\pm0.00077}} 
\newcommand{\hatcurLChatnetmAxxxxxG}{\ensuremath{12.52213\pm0.00014}} 
\newcommand{\hatcurLCiblendAxxxxxG}{\ensuremath{0.748\pm0.072}} 
\newcommand{\hatcurLChatnetmBxxxxxG}{\ensuremath{0.000040\pm0.000035}} 
\newcommand{\hatcurLCiblendBxxxxxG}{\ensuremath{0.871\pm0.062}} 
\newcommand{\hatcurLCrhoxxxxxG}{\ensuremath{0.350\pm0.018}}     
\newcommand{\hatcurSMEiteffxxxxxG}{\ensuremath{6302\pm50}}      
\newcommand{\hatcurSMEizfehxxxxxG}{\ensuremath{-0.010\pm0.080}} 
\newcommand{\hatcurSMEizfehshortxxxxxG}{\ensuremath{-0.01}}     
\newcommand{\hatcurSMEiloggxxxxxG}{\ensuremath{3.99\pm0.10}}    
\newcommand{\hatcurSMEivsinxxxxxG}{\ensuremath{12.70\pm0.50}}   
\newcommand{\hatcurSMEivmacxxxxxG}{\ensuremath{nff\pmnff}}      
\newcommand{\hatcurSMEivmicxxxxxG}{\ensuremath{nff\pmnff}}      
\newcommand{\hatcurLBiBxxxxxG}{\ensuremath{0.5991}}             
\newcommand{\hatcurLBiiBxxxxxG}{\ensuremath{0.2117}}            
\newcommand{\hatcurLBiVxxxxxG}{\ensuremath{0.4809}}             
\newcommand{\hatcurLBiiVxxxxxG}{\ensuremath{0.2207}}            
\newcommand{\hatcurLBiRxxxxxG}{\ensuremath{0.4086}}             
\newcommand{\hatcurLBiiRxxxxxG}{\ensuremath{0.2085}}            
\newcommand{\hatcurLBiIxxxxxG}{\ensuremath{0.3221}}             
\newcommand{\hatcurLBiiIxxxxxG}{\ensuremath{0.2018}}            
\newcommand{\hatcurLBiuxxxxxG}{\ensuremath{0.6931}}             
\newcommand{\hatcurLBiiuxxxxxG}{\ensuremath{0.1202}}            
\newcommand{\hatcurLBigxxxxxG}{\ensuremath{0.5665}}             
\newcommand{\hatcurLBiigxxxxxG}{\ensuremath{0.1994}}            
\newcommand{\hatcurLBirxxxxxG}{\ensuremath{0.26\pm0.14}}        
\newcommand{\hatcurLBiirxxxxxG}{\ensuremath{0.26\pm0.17}}       
\newcommand{\hatcurLBiixxxxxG}{\ensuremath{0.26\pm0.14}}        
\newcommand{\hatcurLBiiixxxxxG}{\ensuremath{0.27\pm0.17}}       
\newcommand{\hatcurLBizxxxxxG}{\ensuremath{0.2782}}             
\newcommand{\hatcurLBiizxxxxxG}{\ensuremath{0.2037}}            
\newcommand{\hatcurLBiJxxxxxG}{\ensuremath{0.1869}}             
\newcommand{\hatcurLBiiJxxxxxG}{\ensuremath{0.2193}}            
\newcommand{\hatcurLBiHxxxxxG}{\ensuremath{0.1166}}             
\newcommand{\hatcurLBiiHxxxxxG}{\ensuremath{0.2345}}            
\newcommand{\hatcurLBiKxxxxxG}{\ensuremath{0.1148}}             
\newcommand{\hatcurLBiiKxxxxxG}{\ensuremath{0.1811}}            
\newcommand{\hatcurLBiTxxxxxG}{\ensuremath{0.29\pm0.11}}        
\newcommand{\hatcurLBiiTxxxxxG}{\ensuremath{0.37\pm0.16}}       
\newcommand{\hatcurLBikepxxxxxG}{\ensuremath{0.4509}}           
\newcommand{\hatcurLBiikepxxxxxG}{\ensuremath{0.1998}}          
\newcommand{\hatcurLBiCxxxxxG}{\ensuremath{0.4428}}             
\newcommand{\hatcurLBiiCxxxxxG}{\ensuremath{0.1995}}            
\newcommand{\hatcurLBiMxxxxxG}{\ensuremath{0.5248}}             
\newcommand{\hatcurLBiiMxxxxxG}{\ensuremath{0.1878}}            
\newcommand{\hatcurLBiSonexxxxxG}{\ensuremath{0.0852}}            
\newcommand{\hatcurLBiiSonexxxxxG}{\ensuremath{0.1124}}           
\newcommand{\hatcurLBiStwoxxxxxG}{\ensuremath{0.0722}}            
\newcommand{\hatcurLBiiStwoxxxxxG}{\ensuremath{0.0951}}           
\newcommand{\hatcurLBiSthreexxxxxG}{\ensuremath{0.0596}}            
\newcommand{\hatcurLBiiSthreexxxxxG}{\ensuremath{0.0788}}           
\newcommand{\hatcurLBiSfourxxxxxG}{\ensuremath{0.0452}}            
\newcommand{\hatcurLBiiSfourxxxxxG}{\ensuremath{0.0629}}           
\newcommand{\hatcurISOmxxxxxG}{\ensuremath{1.298\pm0.021}}      
\newcommand{\hatcurISOmshortxxxxxG}{\ensuremath{1.30}}          
\newcommand{\hatcurISOmlongxxxxxG}{\ensuremath{1.298\pm0.021}}  
\newcommand{\hatcurISOrxxxxxG}{\ensuremath{1.735_{-0.028}^{+0.041}}} 
\newcommand{\hatcurISOrshortxxxxxG}{\ensuremath{1.73}}          
\newcommand{\hatcurISOrlongxxxxxG}{\ensuremath{1.735_{-0.028}^{+0.041}}} 
\newcommand{\hatcurISOrhoxxxxxG}{\ensuremath{0.350\pm0.018}}    
\newcommand{\hatcurISOrholongxxxxxG}{\ensuremath{0.350\pm0.018}} 
\newcommand{\hatcurISOloggxxxxxG}{\ensuremath{4.072\pm0.015}}   
\newcommand{\hatcurISOlumxxxxxG}{\ensuremath{4.66_{-0.17}^{+0.29}}} 
\newcommand{\hatcurISOlumshortxxxxxG}{\ensuremath{4.66}}        
\newcommand{\hatcurISOteffxxxxxG}{\ensuremath{6457\pm29}}       
\newcommand{\hatcurISOzfehxxxxxG}{\ensuremath{-0.113_{-0.056}^{+0.027}}} 
\newcommand{\hatcurISOagexxxxxG}{\ensuremath{2.88\pm0.13}}      
\newcommand{\hatcurISOspecxxxxxG}{F}                            
\newcommand{\hatcurRVKxxxxxG}{\ensuremath{62\pm18}}             
\newcommand{\hatcurRVrkxxxxxG}{\ensuremath{0\pm0}}              
\newcommand{\hatcurRVrhxxxxxG}{\ensuremath{0\pm0}}              
\newcommand{\hatcurRVkxxxxxG}{\ensuremath{0\pm0}}               
\newcommand{\hatcurRVhxxxxxG}{\ensuremath{0\pm0}}               
\newcommand{\hatcurRVtronexxxxxG}{\ensuremath{0\pm0}}           
\newcommand{\hatcurRVtrtwoxxxxxG}{\ensuremath{0\pm0}}           
\newcommand{\hatcurRVgammaAxxxxxG}{\ensuremath{12\pm11}}        
\newcommand{\hatcurRVjitterAxxxxxG}{\ensuremath{21\pm10}}       
\newcommand{\hatcurRVjittertwosiglimAxxxxxG}{\ensuremath{<42.9}} 
\newcommand{\hatcurRVfitrmsAxxxxxG}{\ensuremath{0.0}}           
\newcommand{\hatcurRVgammaBxxxxxG}{\ensuremath{24486\pm20}}     
\newcommand{\hatcurRVjitterBxxxxxG}{\ensuremath{3\pm29}}        
\newcommand{\hatcurRVjittertwosiglimBxxxxxG}{\ensuremath{<69.2}} 
\newcommand{\hatcurRVfitrmsBxxxxxG}{\ensuremath{0.0}}           
\newcommand{\hatcurRVeccenxxxxxG}{\ensuremath{0\pm0}}           
\newcommand{\hatcurRVeccentwosiglimxxxxxG}{\ensuremath{<0.000}} 
\newcommand{\hatcurRVomegaxxxxxG}{\ensuremath{0\pm0}}           
\newcommand{\hatcurPPixxxxxG}{\ensuremath{88.01\pm0.70}}        
\newcommand{\hatcurPPgxxxxxG}{\ensuremath{4.9\pm1.5}}           
\newcommand{\hatcurPPloggxxxxxG}{\ensuremath{2.69\pm0.12}}      
\newcommand{\hatcurPParxxxxxG}{\ensuremath{6.67\pm0.12}}        
\newcommand{\hatcurPParelxxxxxG}{\ensuremath{0.05387\pm0.00030}} 
\newcommand{\hatcurPPrhoxxxxxG}{\ensuremath{0.144_{-0.035}^{+0.046}}} 
\newcommand{\hatcurPPmxxxxxG}{\ensuremath{0.58_{-0.13}^{+0.18}}} 
\newcommand{\hatcurPPmshortxxxxxG}{\ensuremath{0.58}}           
\newcommand{\hatcurPPmlongxxxxxG}{\ensuremath{0.58_{-0.13}^{+0.18}}} 
\newcommand{\hatcurPPmexxxxxG}{\ensuremath{183_{-41}^{+57}}}    
\newcommand{\hatcurPPmeshortxxxxxG}{\ensuremath{183.4}}         
\newcommand{\hatcurPPmelongxxxxxG}{\ensuremath{183_{-41}^{+57}}} 
\newcommand{\hatcurPPrxxxxxG}{\ensuremath{1.703\pm0.070}}       
\newcommand{\hatcurPPrshortxxxxxG}{\ensuremath{1.70}}           
\newcommand{\hatcurPPrlongxxxxxG}{\ensuremath{1.703\pm0.070}}   
\newcommand{\hatcurPPrexxxxxG}{\ensuremath{19.09\pm0.79}}       
\newcommand{\hatcurPPreshortxxxxxG}{\ensuremath{19.1}}          
\newcommand{\hatcurPPrelongxxxxxG}{\ensuremath{19.09\pm0.79}}   
\newcommand{\hatcurPPmrcorrxxxxxG}{\ensuremath{0.06}}           
\newcommand{\hatcurPPteffxxxxxG}{\ensuremath{1766_{-16}^{+22}}} 
\newcommand{\hatcurPPthetaxxxxxG}{\ensuremath{0.0281_{-0.0064}^{+0.0084}}} 
\newcommand{\hatcurPPfluxperixxxxxG}{\ensuremath{2.193_{-0.079}^{+0.108}}} 
\newcommand{\hatcurPPfluxperidimxxxxxG}{\ensuremath{9}}         
\newcommand{\hatcurPPfluxapxxxxxG}{\ensuremath{2.193_{-0.079}^{+0.108}}} 
\newcommand{\hatcurPPfluxapdimxxxxxG}{\ensuremath{9}}           
\newcommand{\hatcurPPfluxavgxxxxxG}{\ensuremath{2.193_{-0.079}^{+0.108}}} 
\newcommand{\hatcurPPfluxavgdimxxxxxG}{\ensuremath{9}}          
\newcommand{\hatcurPPfluxavglogxxxxxG}{\ensuremath{9.341_{-0.016}^{+0.021}}} 
\newcommand{\hatcurXsecphasexxxxxG}{\ensuremath{0\pm0}}         
\newcommand{\hatcurXsecondaryxxxxxG}{\ensuremath{2457753.46715\pm0.00063}} 
\newcommand{\hatcurXsecdurxxxxxG}{\ensuremath{0.2052\pm0.0020}} 
\newcommand{\hatcurXsecingdurxxxxxG}{\ensuremath{0.0199\pm0.0010}} 
\newcommand{\hatcurPPphiconjxxxxxG}{\ensuremath{0\pm0}}         
\newcommand{\hatcurPPperixxxxxG}{\ensuremath{2457750.46173\pm0.00063}} 
\newcommand{\hatcurPPaequivxxxxxG}{\ensuremath{0.02490\pm0.00050}} 
\newcommand{\hatcurPPtcircxxxxxG}{\ensuremath{59_{-16}^{+23}}}  
\newcommand{\hatcurPPtinfallxxxxxG}{\ensuremath{1130\pm360}}    
\newcommand{\hatcurXdistxxxxxG}{\ensuremath{655_{-11}^{+17}}}   
\newcommand{\hatcurXAvxxxxxG}{\ensuremath{0.650_{-0.021}^{+0.014}}} 
\newcommand{\hatcurXdistredxxxxxG}{\ensuremath{655_{-11}^{+17}}} 
\newcommand{\hatcurXEBVxxxxxG}{\ensuremath{0.2100_{-0.0070}^{+0.0040}}} 
\newcommand{\hatcurCCpmraxxxxxG}{\ensuremath{7.784\pm0.079}}    
\newcommand{\hatcurCCpmdecxxxxxG}{\ensuremath{-3.863\pm0.044}}  
\newcommand{\hatcurCCpmxxxxxG}{\ensuremath{8.690\pm0.090}}      

%% file: planetinfo_multi_eccen.tex
\input{HTR062-011.eccen.tex}
\input{HTR080-003.eccen.tex}
\input{HTR089-018.eccen.tex}
\input{HTR094-024.eccen.tex}
\input{HTR130-001.eccen.tex}
\input{HTR384-014.eccen.tex}
\input{HTR405-001.eccen.tex}
\newcommand{\hatcurCCbbHmageccen}[1]{\ifnum#1=58 %
\hatcurCCbbHmageccenxxxxxA
\else
\ifnum#1=59 %
\hatcurCCbbHmageccenxxxxxB
\else
\ifnum#1=60 %
\hatcurCCbbHmageccenxxxxxC
\else
\ifnum#1=61 %
\hatcurCCbbHmageccenxxxxxD
\else
\ifnum#1=62 %
\hatcurCCbbHmageccenxxxxxE
\else
\ifnum#1=63 %
\hatcurCCbbHmageccenxxxxxF
\else
\ifnum#1=64 %
\hatcurCCbbHmageccenxxxxxG
\else
??????\fi
\fi
\fi
\fi
\fi
\fi
\fi
}
\newcommand{\hatcurCCbbJmageccen}[1]{\ifnum#1=58 %
\hatcurCCbbJmageccenxxxxxA
\else
\ifnum#1=59 %
\hatcurCCbbJmageccenxxxxxB
\else
\ifnum#1=60 %
\hatcurCCbbJmageccenxxxxxC
\else
\ifnum#1=61 %
\hatcurCCbbJmageccenxxxxxD
\else
\ifnum#1=62 %
\hatcurCCbbJmageccenxxxxxE
\else
\ifnum#1=63 %
\hatcurCCbbJmageccenxxxxxF
\else
\ifnum#1=64 %
\hatcurCCbbJmageccenxxxxxG
\else
??????\fi
\fi
\fi
\fi
\fi
\fi
\fi
}
\newcommand{\hatcurCCbbKmageccen}[1]{\ifnum#1=58 %
\hatcurCCbbKmageccenxxxxxA
\else
\ifnum#1=59 %
\hatcurCCbbKmageccenxxxxxB
\else
\ifnum#1=60 %
\hatcurCCbbKmageccenxxxxxC
\else
\ifnum#1=61 %
\hatcurCCbbKmageccenxxxxxD
\else
\ifnum#1=62 %
\hatcurCCbbKmageccenxxxxxE
\else
\ifnum#1=63 %
\hatcurCCbbKmageccenxxxxxF
\else
\ifnum#1=64 %
\hatcurCCbbKmageccenxxxxxG
\else
??????\fi
\fi
\fi
\fi
\fi
\fi
\fi
}
\newcommand{\hatcurCCcitHmageccen}[1]{\ifnum#1=58 %
\hatcurCCcitHmageccenxxxxxA
\else
\ifnum#1=59 %
\hatcurCCcitHmageccenxxxxxB
\else
\ifnum#1=60 %
\hatcurCCcitHmageccenxxxxxC
\else
\ifnum#1=61 %
\hatcurCCcitHmageccenxxxxxD
\else
\ifnum#1=62 %
\hatcurCCcitHmageccenxxxxxE
\else
\ifnum#1=63 %
\hatcurCCcitHmageccenxxxxxF
\else
\ifnum#1=64 %
\hatcurCCcitHmageccenxxxxxG
\else
??????\fi
\fi
\fi
\fi
\fi
\fi
\fi
}
\newcommand{\hatcurCCcitJmageccen}[1]{\ifnum#1=58 %
\hatcurCCcitJmageccenxxxxxA
\else
\ifnum#1=59 %
\hatcurCCcitJmageccenxxxxxB
\else
\ifnum#1=60 %
\hatcurCCcitJmageccenxxxxxC
\else
\ifnum#1=61 %
\hatcurCCcitJmageccenxxxxxD
\else
\ifnum#1=62 %
\hatcurCCcitJmageccenxxxxxE
\else
\ifnum#1=63 %
\hatcurCCcitJmageccenxxxxxF
\else
\ifnum#1=64 %
\hatcurCCcitJmageccenxxxxxG
\else
??????\fi
\fi
\fi
\fi
\fi
\fi
\fi
}
\newcommand{\hatcurCCcitKmageccen}[1]{\ifnum#1=58 %
\hatcurCCcitKmageccenxxxxxA
\else
\ifnum#1=59 %
\hatcurCCcitKmageccenxxxxxB
\else
\ifnum#1=60 %
\hatcurCCcitKmageccenxxxxxC
\else
\ifnum#1=61 %
\hatcurCCcitKmageccenxxxxxD
\else
\ifnum#1=62 %
\hatcurCCcitKmageccenxxxxxE
\else
\ifnum#1=63 %
\hatcurCCcitKmageccenxxxxxF
\else
\ifnum#1=64 %
\hatcurCCcitKmageccenxxxxxG
\else
??????\fi
\fi
\fi
\fi
\fi
\fi
\fi
}
\newcommand{\hatcurCCdececcen}[1]{\ifnum#1=58 %
\hatcurCCdececcenxxxxxA
\else
\ifnum#1=59 %
\hatcurCCdececcenxxxxxB
\else
\ifnum#1=60 %
\hatcurCCdececcenxxxxxC
\else
\ifnum#1=61 %
\hatcurCCdececcenxxxxxD
\else
\ifnum#1=62 %
\hatcurCCdececcenxxxxxE
\else
\ifnum#1=63 %
\hatcurCCdececcenxxxxxF
\else
\ifnum#1=64 %
\hatcurCCdececcenxxxxxG
\else
??????\fi
\fi
\fi
\fi
\fi
\fi
\fi
}
\newcommand{\hatcurCCesoHKmageccen}[1]{\ifnum#1=58 %
\hatcurCCesoHKmageccenxxxxxA
\else
\ifnum#1=59 %
\hatcurCCesoHKmageccenxxxxxB
\else
\ifnum#1=60 %
\hatcurCCesoHKmageccenxxxxxC
\else
\ifnum#1=61 %
\hatcurCCesoHKmageccenxxxxxD
\else
\ifnum#1=62 %
\hatcurCCesoHKmageccenxxxxxE
\else
\ifnum#1=63 %
\hatcurCCesoHKmageccenxxxxxF
\else
\ifnum#1=64 %
\hatcurCCesoHKmageccenxxxxxG
\else
??????\fi
\fi
\fi
\fi
\fi
\fi
\fi
}
\newcommand{\hatcurCCesoHmageccen}[1]{\ifnum#1=58 %
\hatcurCCesoHmageccenxxxxxA
\else
\ifnum#1=59 %
\hatcurCCesoHmageccenxxxxxB
\else
\ifnum#1=60 %
\hatcurCCesoHmageccenxxxxxC
\else
\ifnum#1=61 %
\hatcurCCesoHmageccenxxxxxD
\else
\ifnum#1=62 %
\hatcurCCesoHmageccenxxxxxE
\else
\ifnum#1=63 %
\hatcurCCesoHmageccenxxxxxF
\else
\ifnum#1=64 %
\hatcurCCesoHmageccenxxxxxG
\else
??????\fi
\fi
\fi
\fi
\fi
\fi
\fi
}
\newcommand{\hatcurCCesoJHmageccen}[1]{\ifnum#1=58 %
\hatcurCCesoJHmageccenxxxxxA
\else
\ifnum#1=59 %
\hatcurCCesoJHmageccenxxxxxB
\else
\ifnum#1=60 %
\hatcurCCesoJHmageccenxxxxxC
\else
\ifnum#1=61 %
\hatcurCCesoJHmageccenxxxxxD
\else
\ifnum#1=62 %
\hatcurCCesoJHmageccenxxxxxE
\else
\ifnum#1=63 %
\hatcurCCesoJHmageccenxxxxxF
\else
\ifnum#1=64 %
\hatcurCCesoJHmageccenxxxxxG
\else
??????\fi
\fi
\fi
\fi
\fi
\fi
\fi
}
\newcommand{\hatcurCCesoJKmageccen}[1]{\ifnum#1=58 %
\hatcurCCesoJKmageccenxxxxxA
\else
\ifnum#1=59 %
\hatcurCCesoJKmageccenxxxxxB
\else
\ifnum#1=60 %
\hatcurCCesoJKmageccenxxxxxC
\else
\ifnum#1=61 %
\hatcurCCesoJKmageccenxxxxxD
\else
\ifnum#1=62 %
\hatcurCCesoJKmageccenxxxxxE
\else
\ifnum#1=63 %
\hatcurCCesoJKmageccenxxxxxF
\else
\ifnum#1=64 %
\hatcurCCesoJKmageccenxxxxxG
\else
??????\fi
\fi
\fi
\fi
\fi
\fi
\fi
}
\newcommand{\hatcurCCesoJmageccen}[1]{\ifnum#1=58 %
\hatcurCCesoJmageccenxxxxxA
\else
\ifnum#1=59 %
\hatcurCCesoJmageccenxxxxxB
\else
\ifnum#1=60 %
\hatcurCCesoJmageccenxxxxxC
\else
\ifnum#1=61 %
\hatcurCCesoJmageccenxxxxxD
\else
\ifnum#1=62 %
\hatcurCCesoJmageccenxxxxxE
\else
\ifnum#1=63 %
\hatcurCCesoJmageccenxxxxxF
\else
\ifnum#1=64 %
\hatcurCCesoJmageccenxxxxxG
\else
??????\fi
\fi
\fi
\fi
\fi
\fi
\fi
}
\newcommand{\hatcurCCesoKmageccen}[1]{\ifnum#1=58 %
\hatcurCCesoKmageccenxxxxxA
\else
\ifnum#1=59 %
\hatcurCCesoKmageccenxxxxxB
\else
\ifnum#1=60 %
\hatcurCCesoKmageccenxxxxxC
\else
\ifnum#1=61 %
\hatcurCCesoKmageccenxxxxxD
\else
\ifnum#1=62 %
\hatcurCCesoKmageccenxxxxxE
\else
\ifnum#1=63 %
\hatcurCCesoKmageccenxxxxxF
\else
\ifnum#1=64 %
\hatcurCCesoKmageccenxxxxxG
\else
??????\fi
\fi
\fi
\fi
\fi
\fi
\fi
}
\newcommand{\hatcurCCgaiadrtwoeccen}[1]{\ifnum#1=58 %
\hatcurCCgaiadrtwoeccenxxxxxA
\else
\ifnum#1=59 %
\hatcurCCgaiadrtwoeccenxxxxxB
\else
\ifnum#1=60 %
\hatcurCCgaiadrtwoeccenxxxxxC
\else
\ifnum#1=61 %
\hatcurCCgaiadrtwoeccenxxxxxD
\else
\ifnum#1=62 %
\hatcurCCgaiadrtwoeccenxxxxxE
\else
\ifnum#1=63 %
\hatcurCCgaiadrtwoeccenxxxxxF
\else
\ifnum#1=64 %
\hatcurCCgaiadrtwoeccenxxxxxG
\else
??????\fi
\fi
\fi
\fi
\fi
\fi
\fi
}
\newcommand{\hatcurCCgaiaeccen}[1]{\ifnum#1=58 %
\hatcurCCgaiaeccenxxxxxA
\else
\ifnum#1=59 %
\hatcurCCgaiaeccenxxxxxB
\else
\ifnum#1=60 %
\hatcurCCgaiaeccenxxxxxC
\else
\ifnum#1=61 %
\hatcurCCgaiaeccenxxxxxD
\else
\ifnum#1=62 %
\hatcurCCgaiaeccenxxxxxE
\else
\ifnum#1=63 %
\hatcurCCgaiaeccenxxxxxF
\else
\ifnum#1=64 %
\hatcurCCgaiaeccenxxxxxG
\else
??????\fi
\fi
\fi
\fi
\fi
\fi
\fi
}
\newcommand{\hatcurCCgaiamBPeccen}[1]{\ifnum#1=58 %
\hatcurCCgaiamBPeccenxxxxxA
\else
\ifnum#1=59 %
\hatcurCCgaiamBPeccenxxxxxB
\else
\ifnum#1=60 %
\hatcurCCgaiamBPeccenxxxxxC
\else
\ifnum#1=61 %
\hatcurCCgaiamBPeccenxxxxxD
\else
\ifnum#1=62 %
\hatcurCCgaiamBPeccenxxxxxE
\else
\ifnum#1=63 %
\hatcurCCgaiamBPeccenxxxxxF
\else
\ifnum#1=64 %
\hatcurCCgaiamBPeccenxxxxxG
\else
??????\fi
\fi
\fi
\fi
\fi
\fi
\fi
}
\newcommand{\hatcurCCgaiamGeccen}[1]{\ifnum#1=58 %
\hatcurCCgaiamGeccenxxxxxA
\else
\ifnum#1=59 %
\hatcurCCgaiamGeccenxxxxxB
\else
\ifnum#1=60 %
\hatcurCCgaiamGeccenxxxxxC
\else
\ifnum#1=61 %
\hatcurCCgaiamGeccenxxxxxD
\else
\ifnum#1=62 %
\hatcurCCgaiamGeccenxxxxxE
\else
\ifnum#1=63 %
\hatcurCCgaiamGeccenxxxxxF
\else
\ifnum#1=64 %
\hatcurCCgaiamGeccenxxxxxG
\else
??????\fi
\fi
\fi
\fi
\fi
\fi
\fi
}
\newcommand{\hatcurCCgaiamRPeccen}[1]{\ifnum#1=58 %
\hatcurCCgaiamRPeccenxxxxxA
\else
\ifnum#1=59 %
\hatcurCCgaiamRPeccenxxxxxB
\else
\ifnum#1=60 %
\hatcurCCgaiamRPeccenxxxxxC
\else
\ifnum#1=61 %
\hatcurCCgaiamRPeccenxxxxxD
\else
\ifnum#1=62 %
\hatcurCCgaiamRPeccenxxxxxE
\else
\ifnum#1=63 %
\hatcurCCgaiamRPeccenxxxxxF
\else
\ifnum#1=64 %
\hatcurCCgaiamRPeccenxxxxxG
\else
??????\fi
\fi
\fi
\fi
\fi
\fi
\fi
}
\newcommand{\hatcurCCgsceccen}[1]{\ifnum#1=58 %
\hatcurCCgsceccenxxxxxA
\else
\ifnum#1=59 %
\hatcurCCgsceccenxxxxxB
\else
\ifnum#1=60 %
\hatcurCCgsceccenxxxxxC
\else
\ifnum#1=61 %
\hatcurCCgsceccenxxxxxD
\else
\ifnum#1=62 %
\hatcurCCgsceccenxxxxxE
\else
\ifnum#1=63 %
\hatcurCCgsceccenxxxxxF
\else
\ifnum#1=64 %
\hatcurCCgsceccenxxxxxG
\else
??????\fi
\fi
\fi
\fi
\fi
\fi
\fi
}
\newcommand{\hatcurCCmageccen}[1]{\ifnum#1=58 %
\hatcurCCmageccenxxxxxA
\else
\ifnum#1=59 %
\hatcurCCmageccenxxxxxB
\else
\ifnum#1=60 %
\hatcurCCmageccenxxxxxC
\else
\ifnum#1=61 %
\hatcurCCmageccenxxxxxD
\else
\ifnum#1=62 %
\hatcurCCmageccenxxxxxE
\else
\ifnum#1=63 %
\hatcurCCmageccenxxxxxF
\else
\ifnum#1=64 %
\hatcurCCmageccenxxxxxG
\else
??????\fi
\fi
\fi
\fi
\fi
\fi
\fi
}
\newcommand{\hatcurCCparallaxeccen}[1]{\ifnum#1=58 %
\hatcurCCparallaxeccenxxxxxA
\else
\ifnum#1=59 %
\hatcurCCparallaxeccenxxxxxB
\else
\ifnum#1=60 %
\hatcurCCparallaxeccenxxxxxC
\else
\ifnum#1=61 %
\hatcurCCparallaxeccenxxxxxD
\else
\ifnum#1=62 %
\hatcurCCparallaxeccenxxxxxE
\else
\ifnum#1=63 %
\hatcurCCparallaxeccenxxxxxF
\else
\ifnum#1=64 %
\hatcurCCparallaxeccenxxxxxG
\else
??????\fi
\fi
\fi
\fi
\fi
\fi
\fi
}
\newcommand{\hatcurCCpmdececcen}[1]{\ifnum#1=58 %
\hatcurCCpmdececcenxxxxxA
\else
\ifnum#1=59 %
\hatcurCCpmdececcenxxxxxB
\else
\ifnum#1=60 %
\hatcurCCpmdececcenxxxxxC
\else
\ifnum#1=61 %
\hatcurCCpmdececcenxxxxxD
\else
\ifnum#1=62 %
\hatcurCCpmdececcenxxxxxE
\else
\ifnum#1=63 %
\hatcurCCpmdececcenxxxxxF
\else
\ifnum#1=64 %
\hatcurCCpmdececcenxxxxxG
\else
??????\fi
\fi
\fi
\fi
\fi
\fi
\fi
}
\newcommand{\hatcurCCpmeccen}[1]{\ifnum#1=58 %
\hatcurCCpmeccenxxxxxA
\else
\ifnum#1=59 %
\hatcurCCpmeccenxxxxxB
\else
\ifnum#1=60 %
\hatcurCCpmeccenxxxxxC
\else
\ifnum#1=61 %
\hatcurCCpmeccenxxxxxD
\else
\ifnum#1=62 %
\hatcurCCpmeccenxxxxxE
\else
\ifnum#1=63 %
\hatcurCCpmeccenxxxxxF
\else
\ifnum#1=64 %
\hatcurCCpmeccenxxxxxG
\else
??????\fi
\fi
\fi
\fi
\fi
\fi
\fi
}
\newcommand{\hatcurCCpmraeccen}[1]{\ifnum#1=58 %
\hatcurCCpmraeccenxxxxxA
\else
\ifnum#1=59 %
\hatcurCCpmraeccenxxxxxB
\else
\ifnum#1=60 %
\hatcurCCpmraeccenxxxxxC
\else
\ifnum#1=61 %
\hatcurCCpmraeccenxxxxxD
\else
\ifnum#1=62 %
\hatcurCCpmraeccenxxxxxE
\else
\ifnum#1=63 %
\hatcurCCpmraeccenxxxxxF
\else
\ifnum#1=64 %
\hatcurCCpmraeccenxxxxxG
\else
??????\fi
\fi
\fi
\fi
\fi
\fi
\fi
}
\newcommand{\hatcurCCraeccen}[1]{\ifnum#1=58 %
\hatcurCCraeccenxxxxxA
\else
\ifnum#1=59 %
\hatcurCCraeccenxxxxxB
\else
\ifnum#1=60 %
\hatcurCCraeccenxxxxxC
\else
\ifnum#1=61 %
\hatcurCCraeccenxxxxxD
\else
\ifnum#1=62 %
\hatcurCCraeccenxxxxxE
\else
\ifnum#1=63 %
\hatcurCCraeccenxxxxxF
\else
\ifnum#1=64 %
\hatcurCCraeccenxxxxxG
\else
??????\fi
\fi
\fi
\fi
\fi
\fi
\fi
}
\newcommand{\hatcurCCtassmBeccen}[1]{\ifnum#1=58 %
\hatcurCCtassmBeccenxxxxxA
\else
\ifnum#1=59 %
\hatcurCCtassmBeccenxxxxxB
\else
\ifnum#1=60 %
\hatcurCCtassmBeccenxxxxxC
\else
\ifnum#1=61 %
\hatcurCCtassmBeccenxxxxxD
\else
\ifnum#1=62 %
\hatcurCCtassmBeccenxxxxxE
\else
\ifnum#1=63 %
\hatcurCCtassmBeccenxxxxxF
\else
\ifnum#1=64 %
\hatcurCCtassmBeccenxxxxxG
\else
??????\fi
\fi
\fi
\fi
\fi
\fi
\fi
}
\newcommand{\hatcurCCtassmBshorteccen}[1]{\ifnum#1=58 %
\hatcurCCtassmBshorteccenxxxxxA
\else
\ifnum#1=59 %
\hatcurCCtassmBshorteccenxxxxxB
\else
\ifnum#1=60 %
\hatcurCCtassmBshorteccenxxxxxC
\else
\ifnum#1=61 %
\hatcurCCtassmBshorteccenxxxxxD
\else
\ifnum#1=62 %
\hatcurCCtassmBshorteccenxxxxxE
\else
\ifnum#1=63 %
\hatcurCCtassmBshorteccenxxxxxF
\else
\ifnum#1=64 %
\hatcurCCtassmBshorteccenxxxxxG
\else
??????\fi
\fi
\fi
\fi
\fi
\fi
\fi
}
\newcommand{\hatcurCCtassmgeccen}[1]{\ifnum#1=58 %
\hatcurCCtassmgeccenxxxxxA
\else
\ifnum#1=59 %
\hatcurCCtassmgeccenxxxxxB
\else
\ifnum#1=60 %
\hatcurCCtassmgeccenxxxxxC
\else
\ifnum#1=61 %
\hatcurCCtassmgeccenxxxxxD
\else
\ifnum#1=62 %
\hatcurCCtassmgeccenxxxxxE
\else
\ifnum#1=63 %
\hatcurCCtassmgeccenxxxxxF
\else
\ifnum#1=64 %
\hatcurCCtassmgeccenxxxxxG
\else
??????\fi
\fi
\fi
\fi
\fi
\fi
\fi
}
\newcommand{\hatcurCCtassmgshorteccen}[1]{\ifnum#1=58 %
\hatcurCCtassmgshorteccenxxxxxA
\else
\ifnum#1=59 %
\hatcurCCtassmgshorteccenxxxxxB
\else
\ifnum#1=60 %
\hatcurCCtassmgshorteccenxxxxxC
\else
\ifnum#1=61 %
\hatcurCCtassmgshorteccenxxxxxD
\else
\ifnum#1=62 %
\hatcurCCtassmgshorteccenxxxxxE
\else
\ifnum#1=63 %
\hatcurCCtassmgshorteccenxxxxxF
\else
\ifnum#1=64 %
\hatcurCCtassmgshorteccenxxxxxG
\else
??????\fi
\fi
\fi
\fi
\fi
\fi
\fi
}
\newcommand{\hatcurCCtassmieccen}[1]{\ifnum#1=58 %
\hatcurCCtassmieccenxxxxxA
\else
\ifnum#1=59 %
\hatcurCCtassmieccenxxxxxB
\else
\ifnum#1=60 %
\hatcurCCtassmieccenxxxxxC
\else
\ifnum#1=61 %
\hatcurCCtassmieccenxxxxxD
\else
\ifnum#1=62 %
\hatcurCCtassmieccenxxxxxE
\else
\ifnum#1=63 %
\hatcurCCtassmieccenxxxxxF
\else
\ifnum#1=64 %
\hatcurCCtassmieccenxxxxxG
\else
??????\fi
\fi
\fi
\fi
\fi
\fi
\fi
}
\newcommand{\hatcurCCtassmIeccen}[1]{\ifnum#1=58 %
\hatcurCCtassmIeccenxxxxxA
\else
\ifnum#1=59 %
\hatcurCCtassmIeccenxxxxxB
\else
\ifnum#1=60 %
\hatcurCCtassmIeccenxxxxxC
\else
\ifnum#1=61 %
\hatcurCCtassmIeccenxxxxxD
\else
\ifnum#1=62 %
\hatcurCCtassmIeccenxxxxxE
\else
\ifnum#1=63 %
\hatcurCCtassmIeccenxxxxxF
\else
\ifnum#1=64 %
\hatcurCCtassmIeccenxxxxxG
\else
??????\fi
\fi
\fi
\fi
\fi
\fi
\fi
}
\newcommand{\hatcurCCtassmishorteccen}[1]{\ifnum#1=58 %
\hatcurCCtassmishorteccenxxxxxA
\else
\ifnum#1=59 %
\hatcurCCtassmishorteccenxxxxxB
\else
\ifnum#1=60 %
\hatcurCCtassmishorteccenxxxxxC
\else
\ifnum#1=61 %
\hatcurCCtassmishorteccenxxxxxD
\else
\ifnum#1=62 %
\hatcurCCtassmishorteccenxxxxxE
\else
\ifnum#1=63 %
\hatcurCCtassmishorteccenxxxxxF
\else
\ifnum#1=64 %
\hatcurCCtassmishorteccenxxxxxG
\else
??????\fi
\fi
\fi
\fi
\fi
\fi
\fi
}
\newcommand{\hatcurCCtassmIshorteccen}[1]{\ifnum#1=58 %
\hatcurCCtassmIshorteccenxxxxxA
\else
\ifnum#1=59 %
\hatcurCCtassmIshorteccenxxxxxB
\else
\ifnum#1=60 %
\hatcurCCtassmIshorteccenxxxxxC
\else
\ifnum#1=61 %
\hatcurCCtassmIshorteccenxxxxxD
\else
\ifnum#1=62 %
\hatcurCCtassmIshorteccenxxxxxE
\else
\ifnum#1=63 %
\hatcurCCtassmIshorteccenxxxxxF
\else
\ifnum#1=64 %
\hatcurCCtassmIshorteccenxxxxxG
\else
??????\fi
\fi
\fi
\fi
\fi
\fi
\fi
}
\newcommand{\hatcurCCtassmreccen}[1]{\ifnum#1=58 %
\hatcurCCtassmreccenxxxxxA
\else
\ifnum#1=59 %
\hatcurCCtassmreccenxxxxxB
\else
\ifnum#1=60 %
\hatcurCCtassmreccenxxxxxC
\else
\ifnum#1=61 %
\hatcurCCtassmreccenxxxxxD
\else
\ifnum#1=62 %
\hatcurCCtassmreccenxxxxxE
\else
\ifnum#1=63 %
\hatcurCCtassmreccenxxxxxF
\else
\ifnum#1=64 %
\hatcurCCtassmreccenxxxxxG
\else
??????\fi
\fi
\fi
\fi
\fi
\fi
\fi
}
\newcommand{\hatcurCCtassmrshorteccen}[1]{\ifnum#1=58 %
\hatcurCCtassmrshorteccenxxxxxA
\else
\ifnum#1=59 %
\hatcurCCtassmrshorteccenxxxxxB
\else
\ifnum#1=60 %
\hatcurCCtassmrshorteccenxxxxxC
\else
\ifnum#1=61 %
\hatcurCCtassmrshorteccenxxxxxD
\else
\ifnum#1=62 %
\hatcurCCtassmrshorteccenxxxxxE
\else
\ifnum#1=63 %
\hatcurCCtassmrshorteccenxxxxxF
\else
\ifnum#1=64 %
\hatcurCCtassmrshorteccenxxxxxG
\else
??????\fi
\fi
\fi
\fi
\fi
\fi
\fi
}
\newcommand{\hatcurCCtassmveccen}[1]{\ifnum#1=58 %
\hatcurCCtassmveccenxxxxxA
\else
\ifnum#1=59 %
\hatcurCCtassmveccenxxxxxB
\else
\ifnum#1=60 %
\hatcurCCtassmveccenxxxxxC
\else
\ifnum#1=61 %
\hatcurCCtassmveccenxxxxxD
\else
\ifnum#1=62 %
\hatcurCCtassmveccenxxxxxE
\else
\ifnum#1=63 %
\hatcurCCtassmveccenxxxxxF
\else
\ifnum#1=64 %
\hatcurCCtassmveccenxxxxxG
\else
??????\fi
\fi
\fi
\fi
\fi
\fi
\fi
}
\newcommand{\hatcurCCtassmvshorteccen}[1]{\ifnum#1=58 %
\hatcurCCtassmvshorteccenxxxxxA
\else
\ifnum#1=59 %
\hatcurCCtassmvshorteccenxxxxxB
\else
\ifnum#1=60 %
\hatcurCCtassmvshorteccenxxxxxC
\else
\ifnum#1=61 %
\hatcurCCtassmvshorteccenxxxxxD
\else
\ifnum#1=62 %
\hatcurCCtassmvshorteccenxxxxxE
\else
\ifnum#1=63 %
\hatcurCCtassmvshorteccenxxxxxF
\else
\ifnum#1=64 %
\hatcurCCtassmvshorteccenxxxxxG
\else
??????\fi
\fi
\fi
\fi
\fi
\fi
\fi
}
\newcommand{\hatcurCCtwomasseccen}[1]{\ifnum#1=58 %
\hatcurCCtwomasseccenxxxxxA
\else
\ifnum#1=59 %
\hatcurCCtwomasseccenxxxxxB
\else
\ifnum#1=60 %
\hatcurCCtwomasseccenxxxxxC
\else
\ifnum#1=61 %
\hatcurCCtwomasseccenxxxxxD
\else
\ifnum#1=62 %
\hatcurCCtwomasseccenxxxxxE
\else
\ifnum#1=63 %
\hatcurCCtwomasseccenxxxxxF
\else
\ifnum#1=64 %
\hatcurCCtwomasseccenxxxxxG
\else
??????\fi
\fi
\fi
\fi
\fi
\fi
\fi
}
\newcommand{\hatcurCCtwomassHmageccen}[1]{\ifnum#1=58 %
\hatcurCCtwomassHmageccenxxxxxA
\else
\ifnum#1=59 %
\hatcurCCtwomassHmageccenxxxxxB
\else
\ifnum#1=60 %
\hatcurCCtwomassHmageccenxxxxxC
\else
\ifnum#1=61 %
\hatcurCCtwomassHmageccenxxxxxD
\else
\ifnum#1=62 %
\hatcurCCtwomassHmageccenxxxxxE
\else
\ifnum#1=63 %
\hatcurCCtwomassHmageccenxxxxxF
\else
\ifnum#1=64 %
\hatcurCCtwomassHmageccenxxxxxG
\else
??????\fi
\fi
\fi
\fi
\fi
\fi
\fi
}
\newcommand{\hatcurCCtwomassJmageccen}[1]{\ifnum#1=58 %
\hatcurCCtwomassJmageccenxxxxxA
\else
\ifnum#1=59 %
\hatcurCCtwomassJmageccenxxxxxB
\else
\ifnum#1=60 %
\hatcurCCtwomassJmageccenxxxxxC
\else
\ifnum#1=61 %
\hatcurCCtwomassJmageccenxxxxxD
\else
\ifnum#1=62 %
\hatcurCCtwomassJmageccenxxxxxE
\else
\ifnum#1=63 %
\hatcurCCtwomassJmageccenxxxxxF
\else
\ifnum#1=64 %
\hatcurCCtwomassJmageccenxxxxxG
\else
??????\fi
\fi
\fi
\fi
\fi
\fi
\fi
}
\newcommand{\hatcurCCtwomassKmageccen}[1]{\ifnum#1=58 %
\hatcurCCtwomassKmageccenxxxxxA
\else
\ifnum#1=59 %
\hatcurCCtwomassKmageccenxxxxxB
\else
\ifnum#1=60 %
\hatcurCCtwomassKmageccenxxxxxC
\else
\ifnum#1=61 %
\hatcurCCtwomassKmageccenxxxxxD
\else
\ifnum#1=62 %
\hatcurCCtwomassKmageccenxxxxxE
\else
\ifnum#1=63 %
\hatcurCCtwomassKmageccenxxxxxF
\else
\ifnum#1=64 %
\hatcurCCtwomassKmageccenxxxxxG
\else
??????\fi
\fi
\fi
\fi
\fi
\fi
\fi
}
\newcommand{\hatcurCCWfourmageccen}[1]{\ifnum#1=58 %
\hatcurCCWfourmageccenxxxxxA
\else
\ifnum#1=59 %
\hatcurCCWfourmageccenxxxxxB
\else
\ifnum#1=60 %
\hatcurCCWfourmageccenxxxxxC
\else
\ifnum#1=61 %
\hatcurCCWfourmageccenxxxxxD
\else
\ifnum#1=62 %
\hatcurCCWfourmageccenxxxxxE
\else
\ifnum#1=63 %
\hatcurCCWfourmageccenxxxxxF
\else
\ifnum#1=64 %
\hatcurCCWfourmageccenxxxxxG
\else
??????\fi
\fi
\fi
\fi
\fi
\fi
\fi
}
\newcommand{\hatcurCCWonemageccen}[1]{\ifnum#1=58 %
\hatcurCCWonemageccenxxxxxA
\else
\ifnum#1=59 %
\hatcurCCWonemageccenxxxxxB
\else
\ifnum#1=60 %
\hatcurCCWonemageccenxxxxxC
\else
\ifnum#1=61 %
\hatcurCCWonemageccenxxxxxD
\else
\ifnum#1=62 %
\hatcurCCWonemageccenxxxxxE
\else
\ifnum#1=63 %
\hatcurCCWonemageccenxxxxxF
\else
\ifnum#1=64 %
\hatcurCCWonemageccenxxxxxG
\else
??????\fi
\fi
\fi
\fi
\fi
\fi
\fi
}
\newcommand{\hatcurCCWthreemageccen}[1]{\ifnum#1=58 %
\hatcurCCWthreemageccenxxxxxA
\else
\ifnum#1=59 %
\hatcurCCWthreemageccenxxxxxB
\else
\ifnum#1=60 %
\hatcurCCWthreemageccenxxxxxC
\else
\ifnum#1=61 %
\hatcurCCWthreemageccenxxxxxD
\else
\ifnum#1=62 %
\hatcurCCWthreemageccenxxxxxE
\else
\ifnum#1=63 %
\hatcurCCWthreemageccenxxxxxF
\else
\ifnum#1=64 %
\hatcurCCWthreemageccenxxxxxG
\else
??????\fi
\fi
\fi
\fi
\fi
\fi
\fi
}
\newcommand{\hatcurCCWtwomageccen}[1]{\ifnum#1=58 %
\hatcurCCWtwomageccenxxxxxA
\else
\ifnum#1=59 %
\hatcurCCWtwomageccenxxxxxB
\else
\ifnum#1=60 %
\hatcurCCWtwomageccenxxxxxC
\else
\ifnum#1=61 %
\hatcurCCWtwomageccenxxxxxD
\else
\ifnum#1=62 %
\hatcurCCWtwomageccenxxxxxE
\else
\ifnum#1=63 %
\hatcurCCWtwomageccenxxxxxF
\else
\ifnum#1=64 %
\hatcurCCWtwomageccenxxxxxG
\else
??????\fi
\fi
\fi
\fi
\fi
\fi
\fi
}
\newcommand{\hatcurfieldeccen}[1]{\ifnum#1=58 %
\hatcurfieldeccenxxxxxA
\else
\ifnum#1=59 %
\hatcurfieldeccenxxxxxB
\else
\ifnum#1=60 %
\hatcurfieldeccenxxxxxC
\else
\ifnum#1=61 %
\hatcurfieldeccenxxxxxD
\else
\ifnum#1=62 %
\hatcurfieldeccenxxxxxE
\else
\ifnum#1=63 %
\hatcurfieldeccenxxxxxF
\else
\ifnum#1=64 %
\hatcurfieldeccenxxxxxG
\else
??????\fi
\fi
\fi
\fi
\fi
\fi
\fi
}
\newcommand{\hatcurhtreccen}[1]{\ifnum#1=58 %
\hatcurhtreccenxxxxxA
\else
\ifnum#1=59 %
\hatcurhtreccenxxxxxB
\else
\ifnum#1=60 %
\hatcurhtreccenxxxxxC
\else
\ifnum#1=61 %
\hatcurhtreccenxxxxxD
\else
\ifnum#1=62 %
\hatcurhtreccenxxxxxE
\else
\ifnum#1=63 %
\hatcurhtreccenxxxxxF
\else
\ifnum#1=64 %
\hatcurhtreccenxxxxxG
\else
??????\fi
\fi
\fi
\fi
\fi
\fi
\fi
}
\newcommand{\hatcurISOageeccen}[1]{\ifnum#1=58 %
\hatcurISOageeccenxxxxxA
\else
\ifnum#1=59 %
\hatcurISOageeccenxxxxxB
\else
\ifnum#1=60 %
\hatcurISOageeccenxxxxxC
\else
\ifnum#1=61 %
\hatcurISOageeccenxxxxxD
\else
\ifnum#1=62 %
\hatcurISOageeccenxxxxxE
\else
\ifnum#1=63 %
\hatcurISOageeccenxxxxxF
\else
\ifnum#1=64 %
\hatcurISOageeccenxxxxxG
\else
??????\fi
\fi
\fi
\fi
\fi
\fi
\fi
}
\newcommand{\hatcurISOloggeccen}[1]{\ifnum#1=58 %
\hatcurISOloggeccenxxxxxA
\else
\ifnum#1=59 %
\hatcurISOloggeccenxxxxxB
\else
\ifnum#1=60 %
\hatcurISOloggeccenxxxxxC
\else
\ifnum#1=61 %
\hatcurISOloggeccenxxxxxD
\else
\ifnum#1=62 %
\hatcurISOloggeccenxxxxxE
\else
\ifnum#1=63 %
\hatcurISOloggeccenxxxxxF
\else
\ifnum#1=64 %
\hatcurISOloggeccenxxxxxG
\else
??????\fi
\fi
\fi
\fi
\fi
\fi
\fi
}
\newcommand{\hatcurISOlumeccen}[1]{\ifnum#1=58 %
\hatcurISOlumeccenxxxxxA
\else
\ifnum#1=59 %
\hatcurISOlumeccenxxxxxB
\else
\ifnum#1=60 %
\hatcurISOlumeccenxxxxxC
\else
\ifnum#1=61 %
\hatcurISOlumeccenxxxxxD
\else
\ifnum#1=62 %
\hatcurISOlumeccenxxxxxE
\else
\ifnum#1=63 %
\hatcurISOlumeccenxxxxxF
\else
\ifnum#1=64 %
\hatcurISOlumeccenxxxxxG
\else
??????\fi
\fi
\fi
\fi
\fi
\fi
\fi
}
\newcommand{\hatcurISOlumshorteccen}[1]{\ifnum#1=58 %
\hatcurISOlumshorteccenxxxxxA
\else
\ifnum#1=59 %
\hatcurISOlumshorteccenxxxxxB
\else
\ifnum#1=60 %
\hatcurISOlumshorteccenxxxxxC
\else
\ifnum#1=61 %
\hatcurISOlumshorteccenxxxxxD
\else
\ifnum#1=62 %
\hatcurISOlumshorteccenxxxxxE
\else
\ifnum#1=63 %
\hatcurISOlumshorteccenxxxxxF
\else
\ifnum#1=64 %
\hatcurISOlumshorteccenxxxxxG
\else
??????\fi
\fi
\fi
\fi
\fi
\fi
\fi
}
\newcommand{\hatcurISOmeccen}[1]{\ifnum#1=58 %
\hatcurISOmeccenxxxxxA
\else
\ifnum#1=59 %
\hatcurISOmeccenxxxxxB
\else
\ifnum#1=60 %
\hatcurISOmeccenxxxxxC
\else
\ifnum#1=61 %
\hatcurISOmeccenxxxxxD
\else
\ifnum#1=62 %
\hatcurISOmeccenxxxxxE
\else
\ifnum#1=63 %
\hatcurISOmeccenxxxxxF
\else
\ifnum#1=64 %
\hatcurISOmeccenxxxxxG
\else
??????\fi
\fi
\fi
\fi
\fi
\fi
\fi
}
\newcommand{\hatcurISOmlongeccen}[1]{\ifnum#1=58 %
\hatcurISOmlongeccenxxxxxA
\else
\ifnum#1=59 %
\hatcurISOmlongeccenxxxxxB
\else
\ifnum#1=60 %
\hatcurISOmlongeccenxxxxxC
\else
\ifnum#1=61 %
\hatcurISOmlongeccenxxxxxD
\else
\ifnum#1=62 %
\hatcurISOmlongeccenxxxxxE
\else
\ifnum#1=63 %
\hatcurISOmlongeccenxxxxxF
\else
\ifnum#1=64 %
\hatcurISOmlongeccenxxxxxG
\else
??????\fi
\fi
\fi
\fi
\fi
\fi
\fi
}
\newcommand{\hatcurISOmshorteccen}[1]{\ifnum#1=58 %
\hatcurISOmshorteccenxxxxxA
\else
\ifnum#1=59 %
\hatcurISOmshorteccenxxxxxB
\else
\ifnum#1=60 %
\hatcurISOmshorteccenxxxxxC
\else
\ifnum#1=61 %
\hatcurISOmshorteccenxxxxxD
\else
\ifnum#1=62 %
\hatcurISOmshorteccenxxxxxE
\else
\ifnum#1=63 %
\hatcurISOmshorteccenxxxxxF
\else
\ifnum#1=64 %
\hatcurISOmshorteccenxxxxxG
\else
??????\fi
\fi
\fi
\fi
\fi
\fi
\fi
}
\newcommand{\hatcurISOreccen}[1]{\ifnum#1=58 %
\hatcurISOreccenxxxxxA
\else
\ifnum#1=59 %
\hatcurISOreccenxxxxxB
\else
\ifnum#1=60 %
\hatcurISOreccenxxxxxC
\else
\ifnum#1=61 %
\hatcurISOreccenxxxxxD
\else
\ifnum#1=62 %
\hatcurISOreccenxxxxxE
\else
\ifnum#1=63 %
\hatcurISOreccenxxxxxF
\else
\ifnum#1=64 %
\hatcurISOreccenxxxxxG
\else
??????\fi
\fi
\fi
\fi
\fi
\fi
\fi
}
\newcommand{\hatcurISOrhoeccen}[1]{\ifnum#1=58 %
\hatcurISOrhoeccenxxxxxA
\else
\ifnum#1=59 %
\hatcurISOrhoeccenxxxxxB
\else
\ifnum#1=60 %
\hatcurISOrhoeccenxxxxxC
\else
\ifnum#1=61 %
\hatcurISOrhoeccenxxxxxD
\else
\ifnum#1=62 %
\hatcurISOrhoeccenxxxxxE
\else
\ifnum#1=63 %
\hatcurISOrhoeccenxxxxxF
\else
\ifnum#1=64 %
\hatcurISOrhoeccenxxxxxG
\else
??????\fi
\fi
\fi
\fi
\fi
\fi
\fi
}
\newcommand{\hatcurISOrholongeccen}[1]{\ifnum#1=58 %
\hatcurISOrholongeccenxxxxxA
\else
\ifnum#1=59 %
\hatcurISOrholongeccenxxxxxB
\else
\ifnum#1=60 %
\hatcurISOrholongeccenxxxxxC
\else
\ifnum#1=61 %
\hatcurISOrholongeccenxxxxxD
\else
\ifnum#1=62 %
\hatcurISOrholongeccenxxxxxE
\else
\ifnum#1=63 %
\hatcurISOrholongeccenxxxxxF
\else
\ifnum#1=64 %
\hatcurISOrholongeccenxxxxxG
\else
??????\fi
\fi
\fi
\fi
\fi
\fi
\fi
}
\newcommand{\hatcurISOrlongeccen}[1]{\ifnum#1=58 %
\hatcurISOrlongeccenxxxxxA
\else
\ifnum#1=59 %
\hatcurISOrlongeccenxxxxxB
\else
\ifnum#1=60 %
\hatcurISOrlongeccenxxxxxC
\else
\ifnum#1=61 %
\hatcurISOrlongeccenxxxxxD
\else
\ifnum#1=62 %
\hatcurISOrlongeccenxxxxxE
\else
\ifnum#1=63 %
\hatcurISOrlongeccenxxxxxF
\else
\ifnum#1=64 %
\hatcurISOrlongeccenxxxxxG
\else
??????\fi
\fi
\fi
\fi
\fi
\fi
\fi
}
\newcommand{\hatcurISOrshorteccen}[1]{\ifnum#1=58 %
\hatcurISOrshorteccenxxxxxA
\else
\ifnum#1=59 %
\hatcurISOrshorteccenxxxxxB
\else
\ifnum#1=60 %
\hatcurISOrshorteccenxxxxxC
\else
\ifnum#1=61 %
\hatcurISOrshorteccenxxxxxD
\else
\ifnum#1=62 %
\hatcurISOrshorteccenxxxxxE
\else
\ifnum#1=63 %
\hatcurISOrshorteccenxxxxxF
\else
\ifnum#1=64 %
\hatcurISOrshorteccenxxxxxG
\else
??????\fi
\fi
\fi
\fi
\fi
\fi
\fi
}
\newcommand{\hatcurISOspececcen}[1]{\ifnum#1=58 %
\hatcurISOspececcenxxxxxA
\else
\ifnum#1=59 %
\hatcurISOspececcenxxxxxB
\else
\ifnum#1=60 %
\hatcurISOspececcenxxxxxC
\else
\ifnum#1=61 %
\hatcurISOspececcenxxxxxD
\else
\ifnum#1=62 %
\hatcurISOspececcenxxxxxE
\else
\ifnum#1=63 %
\hatcurISOspececcenxxxxxF
\else
\ifnum#1=64 %
\hatcurISOspececcenxxxxxG
\else
??????\fi
\fi
\fi
\fi
\fi
\fi
\fi
}
\newcommand{\hatcurISOteffeccen}[1]{\ifnum#1=58 %
\hatcurISOteffeccenxxxxxA
\else
\ifnum#1=59 %
\hatcurISOteffeccenxxxxxB
\else
\ifnum#1=60 %
\hatcurISOteffeccenxxxxxC
\else
\ifnum#1=61 %
\hatcurISOteffeccenxxxxxD
\else
\ifnum#1=62 %
\hatcurISOteffeccenxxxxxE
\else
\ifnum#1=63 %
\hatcurISOteffeccenxxxxxF
\else
\ifnum#1=64 %
\hatcurISOteffeccenxxxxxG
\else
??????\fi
\fi
\fi
\fi
\fi
\fi
\fi
}
\newcommand{\hatcurISOzfeheccen}[1]{\ifnum#1=58 %
\hatcurISOzfeheccenxxxxxA
\else
\ifnum#1=59 %
\hatcurISOzfeheccenxxxxxB
\else
\ifnum#1=60 %
\hatcurISOzfeheccenxxxxxC
\else
\ifnum#1=61 %
\hatcurISOzfeheccenxxxxxD
\else
\ifnum#1=62 %
\hatcurISOzfeheccenxxxxxE
\else
\ifnum#1=63 %
\hatcurISOzfeheccenxxxxxF
\else
\ifnum#1=64 %
\hatcurISOzfeheccenxxxxxG
\else
??????\fi
\fi
\fi
\fi
\fi
\fi
\fi
}
\newcommand{\hatcurLBiBeccen}[1]{\ifnum#1=58 %
\hatcurLBiBeccenxxxxxA
\else
\ifnum#1=59 %
\hatcurLBiBeccenxxxxxB
\else
\ifnum#1=60 %
\hatcurLBiBeccenxxxxxC
\else
\ifnum#1=61 %
\hatcurLBiBeccenxxxxxD
\else
\ifnum#1=62 %
\hatcurLBiBeccenxxxxxE
\else
\ifnum#1=63 %
\hatcurLBiBeccenxxxxxF
\else
\ifnum#1=64 %
\hatcurLBiBeccenxxxxxG
\else
??????\fi
\fi
\fi
\fi
\fi
\fi
\fi
}
\newcommand{\hatcurLBiCeccen}[1]{\ifnum#1=58 %
\hatcurLBiCeccenxxxxxA
\else
\ifnum#1=59 %
\hatcurLBiCeccenxxxxxB
\else
\ifnum#1=60 %
\hatcurLBiCeccenxxxxxC
\else
\ifnum#1=61 %
\hatcurLBiCeccenxxxxxD
\else
\ifnum#1=62 %
\hatcurLBiCeccenxxxxxE
\else
\ifnum#1=63 %
\hatcurLBiCeccenxxxxxF
\else
\ifnum#1=64 %
\hatcurLBiCeccenxxxxxG
\else
??????\fi
\fi
\fi
\fi
\fi
\fi
\fi
}
\newcommand{\hatcurLBigeccen}[1]{\ifnum#1=58 %
\hatcurLBigeccenxxxxxA
\else
\ifnum#1=59 %
\hatcurLBigeccenxxxxxB
\else
\ifnum#1=60 %
\hatcurLBigeccenxxxxxC
\else
\ifnum#1=61 %
\hatcurLBigeccenxxxxxD
\else
\ifnum#1=62 %
\hatcurLBigeccenxxxxxE
\else
\ifnum#1=63 %
\hatcurLBigeccenxxxxxF
\else
\ifnum#1=64 %
\hatcurLBigeccenxxxxxG
\else
??????\fi
\fi
\fi
\fi
\fi
\fi
\fi
}
\newcommand{\hatcurLBiHeccen}[1]{\ifnum#1=58 %
\hatcurLBiHeccenxxxxxA
\else
\ifnum#1=59 %
\hatcurLBiHeccenxxxxxB
\else
\ifnum#1=60 %
\hatcurLBiHeccenxxxxxC
\else
\ifnum#1=61 %
\hatcurLBiHeccenxxxxxD
\else
\ifnum#1=62 %
\hatcurLBiHeccenxxxxxE
\else
\ifnum#1=63 %
\hatcurLBiHeccenxxxxxF
\else
\ifnum#1=64 %
\hatcurLBiHeccenxxxxxG
\else
??????\fi
\fi
\fi
\fi
\fi
\fi
\fi
}
\newcommand{\hatcurLBiiBeccen}[1]{\ifnum#1=58 %
\hatcurLBiiBeccenxxxxxA
\else
\ifnum#1=59 %
\hatcurLBiiBeccenxxxxxB
\else
\ifnum#1=60 %
\hatcurLBiiBeccenxxxxxC
\else
\ifnum#1=61 %
\hatcurLBiiBeccenxxxxxD
\else
\ifnum#1=62 %
\hatcurLBiiBeccenxxxxxE
\else
\ifnum#1=63 %
\hatcurLBiiBeccenxxxxxF
\else
\ifnum#1=64 %
\hatcurLBiiBeccenxxxxxG
\else
??????\fi
\fi
\fi
\fi
\fi
\fi
\fi
}
\newcommand{\hatcurLBiiCeccen}[1]{\ifnum#1=58 %
\hatcurLBiiCeccenxxxxxA
\else
\ifnum#1=59 %
\hatcurLBiiCeccenxxxxxB
\else
\ifnum#1=60 %
\hatcurLBiiCeccenxxxxxC
\else
\ifnum#1=61 %
\hatcurLBiiCeccenxxxxxD
\else
\ifnum#1=62 %
\hatcurLBiiCeccenxxxxxE
\else
\ifnum#1=63 %
\hatcurLBiiCeccenxxxxxF
\else
\ifnum#1=64 %
\hatcurLBiiCeccenxxxxxG
\else
??????\fi
\fi
\fi
\fi
\fi
\fi
\fi
}
\newcommand{\hatcurLBiieccen}[1]{\ifnum#1=58 %
\hatcurLBiieccenxxxxxA
\else
\ifnum#1=59 %
\hatcurLBiieccenxxxxxB
\else
\ifnum#1=60 %
\hatcurLBiieccenxxxxxC
\else
\ifnum#1=61 %
\hatcurLBiieccenxxxxxD
\else
\ifnum#1=62 %
\hatcurLBiieccenxxxxxE
\else
\ifnum#1=63 %
\hatcurLBiieccenxxxxxF
\else
\ifnum#1=64 %
\hatcurLBiieccenxxxxxG
\else
??????\fi
\fi
\fi
\fi
\fi
\fi
\fi
}
\newcommand{\hatcurLBiIeccen}[1]{\ifnum#1=58 %
\hatcurLBiIeccenxxxxxA
\else
\ifnum#1=59 %
\hatcurLBiIeccenxxxxxB
\else
\ifnum#1=60 %
\hatcurLBiIeccenxxxxxC
\else
\ifnum#1=61 %
\hatcurLBiIeccenxxxxxD
\else
\ifnum#1=62 %
\hatcurLBiIeccenxxxxxE
\else
\ifnum#1=63 %
\hatcurLBiIeccenxxxxxF
\else
\ifnum#1=64 %
\hatcurLBiIeccenxxxxxG
\else
??????\fi
\fi
\fi
\fi
\fi
\fi
\fi
}
\newcommand{\hatcurLBiigeccen}[1]{\ifnum#1=58 %
\hatcurLBiigeccenxxxxxA
\else
\ifnum#1=59 %
\hatcurLBiigeccenxxxxxB
\else
\ifnum#1=60 %
\hatcurLBiigeccenxxxxxC
\else
\ifnum#1=61 %
\hatcurLBiigeccenxxxxxD
\else
\ifnum#1=62 %
\hatcurLBiigeccenxxxxxE
\else
\ifnum#1=63 %
\hatcurLBiigeccenxxxxxF
\else
\ifnum#1=64 %
\hatcurLBiigeccenxxxxxG
\else
??????\fi
\fi
\fi
\fi
\fi
\fi
\fi
}
\newcommand{\hatcurLBiiHeccen}[1]{\ifnum#1=58 %
\hatcurLBiiHeccenxxxxxA
\else
\ifnum#1=59 %
\hatcurLBiiHeccenxxxxxB
\else
\ifnum#1=60 %
\hatcurLBiiHeccenxxxxxC
\else
\ifnum#1=61 %
\hatcurLBiiHeccenxxxxxD
\else
\ifnum#1=62 %
\hatcurLBiiHeccenxxxxxE
\else
\ifnum#1=63 %
\hatcurLBiiHeccenxxxxxF
\else
\ifnum#1=64 %
\hatcurLBiiHeccenxxxxxG
\else
??????\fi
\fi
\fi
\fi
\fi
\fi
\fi
}
\newcommand{\hatcurLBiiieccen}[1]{\ifnum#1=58 %
\hatcurLBiiieccenxxxxxA
\else
\ifnum#1=59 %
\hatcurLBiiieccenxxxxxB
\else
\ifnum#1=60 %
\hatcurLBiiieccenxxxxxC
\else
\ifnum#1=61 %
\hatcurLBiiieccenxxxxxD
\else
\ifnum#1=62 %
\hatcurLBiiieccenxxxxxE
\else
\ifnum#1=63 %
\hatcurLBiiieccenxxxxxF
\else
\ifnum#1=64 %
\hatcurLBiiieccenxxxxxG
\else
??????\fi
\fi
\fi
\fi
\fi
\fi
\fi
}
\newcommand{\hatcurLBiiIeccen}[1]{\ifnum#1=58 %
\hatcurLBiiIeccenxxxxxA
\else
\ifnum#1=59 %
\hatcurLBiiIeccenxxxxxB
\else
\ifnum#1=60 %
\hatcurLBiiIeccenxxxxxC
\else
\ifnum#1=61 %
\hatcurLBiiIeccenxxxxxD
\else
\ifnum#1=62 %
\hatcurLBiiIeccenxxxxxE
\else
\ifnum#1=63 %
\hatcurLBiiIeccenxxxxxF
\else
\ifnum#1=64 %
\hatcurLBiiIeccenxxxxxG
\else
??????\fi
\fi
\fi
\fi
\fi
\fi
\fi
}
\newcommand{\hatcurLBiiJeccen}[1]{\ifnum#1=58 %
\hatcurLBiiJeccenxxxxxA
\else
\ifnum#1=59 %
\hatcurLBiiJeccenxxxxxB
\else
\ifnum#1=60 %
\hatcurLBiiJeccenxxxxxC
\else
\ifnum#1=61 %
\hatcurLBiiJeccenxxxxxD
\else
\ifnum#1=62 %
\hatcurLBiiJeccenxxxxxE
\else
\ifnum#1=63 %
\hatcurLBiiJeccenxxxxxF
\else
\ifnum#1=64 %
\hatcurLBiiJeccenxxxxxG
\else
??????\fi
\fi
\fi
\fi
\fi
\fi
\fi
}
\newcommand{\hatcurLBiiKeccen}[1]{\ifnum#1=58 %
\hatcurLBiiKeccenxxxxxA
\else
\ifnum#1=59 %
\hatcurLBiiKeccenxxxxxB
\else
\ifnum#1=60 %
\hatcurLBiiKeccenxxxxxC
\else
\ifnum#1=61 %
\hatcurLBiiKeccenxxxxxD
\else
\ifnum#1=62 %
\hatcurLBiiKeccenxxxxxE
\else
\ifnum#1=63 %
\hatcurLBiiKeccenxxxxxF
\else
\ifnum#1=64 %
\hatcurLBiiKeccenxxxxxG
\else
??????\fi
\fi
\fi
\fi
\fi
\fi
\fi
}
\newcommand{\hatcurLBiikepeccen}[1]{\ifnum#1=58 %
\hatcurLBiikepeccenxxxxxA
\else
\ifnum#1=59 %
\hatcurLBiikepeccenxxxxxB
\else
\ifnum#1=60 %
\hatcurLBiikepeccenxxxxxC
\else
\ifnum#1=61 %
\hatcurLBiikepeccenxxxxxD
\else
\ifnum#1=62 %
\hatcurLBiikepeccenxxxxxE
\else
\ifnum#1=63 %
\hatcurLBiikepeccenxxxxxF
\else
\ifnum#1=64 %
\hatcurLBiikepeccenxxxxxG
\else
??????\fi
\fi
\fi
\fi
\fi
\fi
\fi
}
\newcommand{\hatcurLBiiMeccen}[1]{\ifnum#1=58 %
\hatcurLBiiMeccenxxxxxA
\else
\ifnum#1=59 %
\hatcurLBiiMeccenxxxxxB
\else
\ifnum#1=60 %
\hatcurLBiiMeccenxxxxxC
\else
\ifnum#1=61 %
\hatcurLBiiMeccenxxxxxD
\else
\ifnum#1=62 %
\hatcurLBiiMeccenxxxxxE
\else
\ifnum#1=63 %
\hatcurLBiiMeccenxxxxxF
\else
\ifnum#1=64 %
\hatcurLBiiMeccenxxxxxG
\else
??????\fi
\fi
\fi
\fi
\fi
\fi
\fi
}
\newcommand{\hatcurLBiireccen}[1]{\ifnum#1=58 %
\hatcurLBiireccenxxxxxA
\else
\ifnum#1=59 %
\hatcurLBiireccenxxxxxB
\else
\ifnum#1=60 %
\hatcurLBiireccenxxxxxC
\else
\ifnum#1=61 %
\hatcurLBiireccenxxxxxD
\else
\ifnum#1=62 %
\hatcurLBiireccenxxxxxE
\else
\ifnum#1=63 %
\hatcurLBiireccenxxxxxF
\else
\ifnum#1=64 %
\hatcurLBiireccenxxxxxG
\else
??????\fi
\fi
\fi
\fi
\fi
\fi
\fi
}
\newcommand{\hatcurLBiiReccen}[1]{\ifnum#1=58 %
\hatcurLBiiReccenxxxxxA
\else
\ifnum#1=59 %
\hatcurLBiiReccenxxxxxB
\else
\ifnum#1=60 %
\hatcurLBiiReccenxxxxxC
\else
\ifnum#1=61 %
\hatcurLBiiReccenxxxxxD
\else
\ifnum#1=62 %
\hatcurLBiiReccenxxxxxE
\else
\ifnum#1=63 %
\hatcurLBiiReccenxxxxxF
\else
\ifnum#1=64 %
\hatcurLBiiReccenxxxxxG
\else
??????\fi
\fi
\fi
\fi
\fi
\fi
\fi
}
\newcommand{\hatcurLBiiSfoureccen}[1]{\ifnum#1=58 %
\hatcurLBiiSfoureccenxxxxxA
\else
\ifnum#1=59 %
\hatcurLBiiSfoureccenxxxxxB
\else
\ifnum#1=60 %
\hatcurLBiiSfoureccenxxxxxC
\else
\ifnum#1=61 %
\hatcurLBiiSfoureccenxxxxxD
\else
\ifnum#1=62 %
\hatcurLBiiSfoureccenxxxxxE
\else
\ifnum#1=63 %
\hatcurLBiiSfoureccenxxxxxF
\else
\ifnum#1=64 %
\hatcurLBiiSfoureccenxxxxxG
\else
??????\fi
\fi
\fi
\fi
\fi
\fi
\fi
}
\newcommand{\hatcurLBiiSoneeccen}[1]{\ifnum#1=58 %
\hatcurLBiiSoneeccenxxxxxA
\else
\ifnum#1=59 %
\hatcurLBiiSoneeccenxxxxxB
\else
\ifnum#1=60 %
\hatcurLBiiSoneeccenxxxxxC
\else
\ifnum#1=61 %
\hatcurLBiiSoneeccenxxxxxD
\else
\ifnum#1=62 %
\hatcurLBiiSoneeccenxxxxxE
\else
\ifnum#1=63 %
\hatcurLBiiSoneeccenxxxxxF
\else
\ifnum#1=64 %
\hatcurLBiiSoneeccenxxxxxG
\else
??????\fi
\fi
\fi
\fi
\fi
\fi
\fi
}
\newcommand{\hatcurLBiiSthreeeccen}[1]{\ifnum#1=58 %
\hatcurLBiiSthreeeccenxxxxxA
\else
\ifnum#1=59 %
\hatcurLBiiSthreeeccenxxxxxB
\else
\ifnum#1=60 %
\hatcurLBiiSthreeeccenxxxxxC
\else
\ifnum#1=61 %
\hatcurLBiiSthreeeccenxxxxxD
\else
\ifnum#1=62 %
\hatcurLBiiSthreeeccenxxxxxE
\else
\ifnum#1=63 %
\hatcurLBiiSthreeeccenxxxxxF
\else
\ifnum#1=64 %
\hatcurLBiiSthreeeccenxxxxxG
\else
??????\fi
\fi
\fi
\fi
\fi
\fi
\fi
}
\newcommand{\hatcurLBiiStwoeccen}[1]{\ifnum#1=58 %
\hatcurLBiiStwoeccenxxxxxA
\else
\ifnum#1=59 %
\hatcurLBiiStwoeccenxxxxxB
\else
\ifnum#1=60 %
\hatcurLBiiStwoeccenxxxxxC
\else
\ifnum#1=61 %
\hatcurLBiiStwoeccenxxxxxD
\else
\ifnum#1=62 %
\hatcurLBiiStwoeccenxxxxxE
\else
\ifnum#1=63 %
\hatcurLBiiStwoeccenxxxxxF
\else
\ifnum#1=64 %
\hatcurLBiiStwoeccenxxxxxG
\else
??????\fi
\fi
\fi
\fi
\fi
\fi
\fi
}
\newcommand{\hatcurLBiiTeccen}[1]{\ifnum#1=58 %
\hatcurLBiiTeccenxxxxxA
\else
\ifnum#1=59 %
\hatcurLBiiTeccenxxxxxB
\else
\ifnum#1=60 %
\hatcurLBiiTeccenxxxxxC
\else
\ifnum#1=61 %
\hatcurLBiiTeccenxxxxxD
\else
\ifnum#1=62 %
\hatcurLBiiTeccenxxxxxE
\else
\ifnum#1=63 %
\hatcurLBiiTeccenxxxxxF
\else
\ifnum#1=64 %
\hatcurLBiiTeccenxxxxxG
\else
??????\fi
\fi
\fi
\fi
\fi
\fi
\fi
}
\newcommand{\hatcurLBiiueccen}[1]{\ifnum#1=58 %
\hatcurLBiiueccenxxxxxA
\else
\ifnum#1=59 %
\hatcurLBiiueccenxxxxxB
\else
\ifnum#1=60 %
\hatcurLBiiueccenxxxxxC
\else
\ifnum#1=61 %
\hatcurLBiiueccenxxxxxD
\else
\ifnum#1=62 %
\hatcurLBiiueccenxxxxxE
\else
\ifnum#1=63 %
\hatcurLBiiueccenxxxxxF
\else
\ifnum#1=64 %
\hatcurLBiiueccenxxxxxG
\else
??????\fi
\fi
\fi
\fi
\fi
\fi
\fi
}
\newcommand{\hatcurLBiiVeccen}[1]{\ifnum#1=58 %
\hatcurLBiiVeccenxxxxxA
\else
\ifnum#1=59 %
\hatcurLBiiVeccenxxxxxB
\else
\ifnum#1=60 %
\hatcurLBiiVeccenxxxxxC
\else
\ifnum#1=61 %
\hatcurLBiiVeccenxxxxxD
\else
\ifnum#1=62 %
\hatcurLBiiVeccenxxxxxE
\else
\ifnum#1=63 %
\hatcurLBiiVeccenxxxxxF
\else
\ifnum#1=64 %
\hatcurLBiiVeccenxxxxxG
\else
??????\fi
\fi
\fi
\fi
\fi
\fi
\fi
}
\newcommand{\hatcurLBiizeccen}[1]{\ifnum#1=58 %
\hatcurLBiizeccenxxxxxA
\else
\ifnum#1=59 %
\hatcurLBiizeccenxxxxxB
\else
\ifnum#1=60 %
\hatcurLBiizeccenxxxxxC
\else
\ifnum#1=61 %
\hatcurLBiizeccenxxxxxD
\else
\ifnum#1=62 %
\hatcurLBiizeccenxxxxxE
\else
\ifnum#1=63 %
\hatcurLBiizeccenxxxxxF
\else
\ifnum#1=64 %
\hatcurLBiizeccenxxxxxG
\else
??????\fi
\fi
\fi
\fi
\fi
\fi
\fi
}
\newcommand{\hatcurLBiJeccen}[1]{\ifnum#1=58 %
\hatcurLBiJeccenxxxxxA
\else
\ifnum#1=59 %
\hatcurLBiJeccenxxxxxB
\else
\ifnum#1=60 %
\hatcurLBiJeccenxxxxxC
\else
\ifnum#1=61 %
\hatcurLBiJeccenxxxxxD
\else
\ifnum#1=62 %
\hatcurLBiJeccenxxxxxE
\else
\ifnum#1=63 %
\hatcurLBiJeccenxxxxxF
\else
\ifnum#1=64 %
\hatcurLBiJeccenxxxxxG
\else
??????\fi
\fi
\fi
\fi
\fi
\fi
\fi
}
\newcommand{\hatcurLBiKeccen}[1]{\ifnum#1=58 %
\hatcurLBiKeccenxxxxxA
\else
\ifnum#1=59 %
\hatcurLBiKeccenxxxxxB
\else
\ifnum#1=60 %
\hatcurLBiKeccenxxxxxC
\else
\ifnum#1=61 %
\hatcurLBiKeccenxxxxxD
\else
\ifnum#1=62 %
\hatcurLBiKeccenxxxxxE
\else
\ifnum#1=63 %
\hatcurLBiKeccenxxxxxF
\else
\ifnum#1=64 %
\hatcurLBiKeccenxxxxxG
\else
??????\fi
\fi
\fi
\fi
\fi
\fi
\fi
}
\newcommand{\hatcurLBikepeccen}[1]{\ifnum#1=58 %
\hatcurLBikepeccenxxxxxA
\else
\ifnum#1=59 %
\hatcurLBikepeccenxxxxxB
\else
\ifnum#1=60 %
\hatcurLBikepeccenxxxxxC
\else
\ifnum#1=61 %
\hatcurLBikepeccenxxxxxD
\else
\ifnum#1=62 %
\hatcurLBikepeccenxxxxxE
\else
\ifnum#1=63 %
\hatcurLBikepeccenxxxxxF
\else
\ifnum#1=64 %
\hatcurLBikepeccenxxxxxG
\else
??????\fi
\fi
\fi
\fi
\fi
\fi
\fi
}
\newcommand{\hatcurLBiMeccen}[1]{\ifnum#1=58 %
\hatcurLBiMeccenxxxxxA
\else
\ifnum#1=59 %
\hatcurLBiMeccenxxxxxB
\else
\ifnum#1=60 %
\hatcurLBiMeccenxxxxxC
\else
\ifnum#1=61 %
\hatcurLBiMeccenxxxxxD
\else
\ifnum#1=62 %
\hatcurLBiMeccenxxxxxE
\else
\ifnum#1=63 %
\hatcurLBiMeccenxxxxxF
\else
\ifnum#1=64 %
\hatcurLBiMeccenxxxxxG
\else
??????\fi
\fi
\fi
\fi
\fi
\fi
\fi
}
\newcommand{\hatcurLBireccen}[1]{\ifnum#1=58 %
\hatcurLBireccenxxxxxA
\else
\ifnum#1=59 %
\hatcurLBireccenxxxxxB
\else
\ifnum#1=60 %
\hatcurLBireccenxxxxxC
\else
\ifnum#1=61 %
\hatcurLBireccenxxxxxD
\else
\ifnum#1=62 %
\hatcurLBireccenxxxxxE
\else
\ifnum#1=63 %
\hatcurLBireccenxxxxxF
\else
\ifnum#1=64 %
\hatcurLBireccenxxxxxG
\else
??????\fi
\fi
\fi
\fi
\fi
\fi
\fi
}
\newcommand{\hatcurLBiReccen}[1]{\ifnum#1=58 %
\hatcurLBiReccenxxxxxA
\else
\ifnum#1=59 %
\hatcurLBiReccenxxxxxB
\else
\ifnum#1=60 %
\hatcurLBiReccenxxxxxC
\else
\ifnum#1=61 %
\hatcurLBiReccenxxxxxD
\else
\ifnum#1=62 %
\hatcurLBiReccenxxxxxE
\else
\ifnum#1=63 %
\hatcurLBiReccenxxxxxF
\else
\ifnum#1=64 %
\hatcurLBiReccenxxxxxG
\else
??????\fi
\fi
\fi
\fi
\fi
\fi
\fi
}
\newcommand{\hatcurLBiSfoureccen}[1]{\ifnum#1=58 %
\hatcurLBiSfoureccenxxxxxA
\else
\ifnum#1=59 %
\hatcurLBiSfoureccenxxxxxB
\else
\ifnum#1=60 %
\hatcurLBiSfoureccenxxxxxC
\else
\ifnum#1=61 %
\hatcurLBiSfoureccenxxxxxD
\else
\ifnum#1=62 %
\hatcurLBiSfoureccenxxxxxE
\else
\ifnum#1=63 %
\hatcurLBiSfoureccenxxxxxF
\else
\ifnum#1=64 %
\hatcurLBiSfoureccenxxxxxG
\else
??????\fi
\fi
\fi
\fi
\fi
\fi
\fi
}
\newcommand{\hatcurLBiSoneeccen}[1]{\ifnum#1=58 %
\hatcurLBiSoneeccenxxxxxA
\else
\ifnum#1=59 %
\hatcurLBiSoneeccenxxxxxB
\else
\ifnum#1=60 %
\hatcurLBiSoneeccenxxxxxC
\else
\ifnum#1=61 %
\hatcurLBiSoneeccenxxxxxD
\else
\ifnum#1=62 %
\hatcurLBiSoneeccenxxxxxE
\else
\ifnum#1=63 %
\hatcurLBiSoneeccenxxxxxF
\else
\ifnum#1=64 %
\hatcurLBiSoneeccenxxxxxG
\else
??????\fi
\fi
\fi
\fi
\fi
\fi
\fi
}
\newcommand{\hatcurLBiSthreeeccen}[1]{\ifnum#1=58 %
\hatcurLBiSthreeeccenxxxxxA
\else
\ifnum#1=59 %
\hatcurLBiSthreeeccenxxxxxB
\else
\ifnum#1=60 %
\hatcurLBiSthreeeccenxxxxxC
\else
\ifnum#1=61 %
\hatcurLBiSthreeeccenxxxxxD
\else
\ifnum#1=62 %
\hatcurLBiSthreeeccenxxxxxE
\else
\ifnum#1=63 %
\hatcurLBiSthreeeccenxxxxxF
\else
\ifnum#1=64 %
\hatcurLBiSthreeeccenxxxxxG
\else
??????\fi
\fi
\fi
\fi
\fi
\fi
\fi
}
\newcommand{\hatcurLBiStwoeccen}[1]{\ifnum#1=58 %
\hatcurLBiStwoeccenxxxxxA
\else
\ifnum#1=59 %
\hatcurLBiStwoeccenxxxxxB
\else
\ifnum#1=60 %
\hatcurLBiStwoeccenxxxxxC
\else
\ifnum#1=61 %
\hatcurLBiStwoeccenxxxxxD
\else
\ifnum#1=62 %
\hatcurLBiStwoeccenxxxxxE
\else
\ifnum#1=63 %
\hatcurLBiStwoeccenxxxxxF
\else
\ifnum#1=64 %
\hatcurLBiStwoeccenxxxxxG
\else
??????\fi
\fi
\fi
\fi
\fi
\fi
\fi
}
\newcommand{\hatcurLBiTeccen}[1]{\ifnum#1=58 %
\hatcurLBiTeccenxxxxxA
\else
\ifnum#1=59 %
\hatcurLBiTeccenxxxxxB
\else
\ifnum#1=60 %
\hatcurLBiTeccenxxxxxC
\else
\ifnum#1=61 %
\hatcurLBiTeccenxxxxxD
\else
\ifnum#1=62 %
\hatcurLBiTeccenxxxxxE
\else
\ifnum#1=63 %
\hatcurLBiTeccenxxxxxF
\else
\ifnum#1=64 %
\hatcurLBiTeccenxxxxxG
\else
??????\fi
\fi
\fi
\fi
\fi
\fi
\fi
}
\newcommand{\hatcurLBiueccen}[1]{\ifnum#1=58 %
\hatcurLBiueccenxxxxxA
\else
\ifnum#1=59 %
\hatcurLBiueccenxxxxxB
\else
\ifnum#1=60 %
\hatcurLBiueccenxxxxxC
\else
\ifnum#1=61 %
\hatcurLBiueccenxxxxxD
\else
\ifnum#1=62 %
\hatcurLBiueccenxxxxxE
\else
\ifnum#1=63 %
\hatcurLBiueccenxxxxxF
\else
\ifnum#1=64 %
\hatcurLBiueccenxxxxxG
\else
??????\fi
\fi
\fi
\fi
\fi
\fi
\fi
}
\newcommand{\hatcurLBiVeccen}[1]{\ifnum#1=58 %
\hatcurLBiVeccenxxxxxA
\else
\ifnum#1=59 %
\hatcurLBiVeccenxxxxxB
\else
\ifnum#1=60 %
\hatcurLBiVeccenxxxxxC
\else
\ifnum#1=61 %
\hatcurLBiVeccenxxxxxD
\else
\ifnum#1=62 %
\hatcurLBiVeccenxxxxxE
\else
\ifnum#1=63 %
\hatcurLBiVeccenxxxxxF
\else
\ifnum#1=64 %
\hatcurLBiVeccenxxxxxG
\else
??????\fi
\fi
\fi
\fi
\fi
\fi
\fi
}
\newcommand{\hatcurLBizeccen}[1]{\ifnum#1=58 %
\hatcurLBizeccenxxxxxA
\else
\ifnum#1=59 %
\hatcurLBizeccenxxxxxB
\else
\ifnum#1=60 %
\hatcurLBizeccenxxxxxC
\else
\ifnum#1=61 %
\hatcurLBizeccenxxxxxD
\else
\ifnum#1=62 %
\hatcurLBizeccenxxxxxE
\else
\ifnum#1=63 %
\hatcurLBizeccenxxxxxF
\else
\ifnum#1=64 %
\hatcurLBizeccenxxxxxG
\else
??????\fi
\fi
\fi
\fi
\fi
\fi
\fi
}
\newcommand{\hatcurLCbsqeccen}[1]{\ifnum#1=58 %
\hatcurLCbsqeccenxxxxxA
\else
\ifnum#1=59 %
\hatcurLCbsqeccenxxxxxB
\else
\ifnum#1=60 %
\hatcurLCbsqeccenxxxxxC
\else
\ifnum#1=61 %
\hatcurLCbsqeccenxxxxxD
\else
\ifnum#1=62 %
\hatcurLCbsqeccenxxxxxE
\else
\ifnum#1=63 %
\hatcurLCbsqeccenxxxxxF
\else
\ifnum#1=64 %
\hatcurLCbsqeccenxxxxxG
\else
??????\fi
\fi
\fi
\fi
\fi
\fi
\fi
}
\newcommand{\hatcurLCdipeccen}[1]{\ifnum#1=58 %
\hatcurLCdipeccenxxxxxA
\else
\ifnum#1=59 %
\hatcurLCdipeccenxxxxxB
\else
\ifnum#1=60 %
\hatcurLCdipeccenxxxxxC
\else
\ifnum#1=61 %
\hatcurLCdipeccenxxxxxD
\else
\ifnum#1=62 %
\hatcurLCdipeccenxxxxxE
\else
\ifnum#1=63 %
\hatcurLCdipeccenxxxxxF
\else
\ifnum#1=64 %
\hatcurLCdipeccenxxxxxG
\else
??????\fi
\fi
\fi
\fi
\fi
\fi
\fi
}
\newcommand{\hatcurLCdureccen}[1]{\ifnum#1=58 %
\hatcurLCdureccenxxxxxA
\else
\ifnum#1=59 %
\hatcurLCdureccenxxxxxB
\else
\ifnum#1=60 %
\hatcurLCdureccenxxxxxC
\else
\ifnum#1=61 %
\hatcurLCdureccenxxxxxD
\else
\ifnum#1=62 %
\hatcurLCdureccenxxxxxE
\else
\ifnum#1=63 %
\hatcurLCdureccenxxxxxF
\else
\ifnum#1=64 %
\hatcurLCdureccenxxxxxG
\else
??????\fi
\fi
\fi
\fi
\fi
\fi
\fi
}
\newcommand{\hatcurLCdurhreccen}[1]{\ifnum#1=58 %
\hatcurLCdurhreccenxxxxxA
\else
\ifnum#1=59 %
\hatcurLCdurhreccenxxxxxB
\else
\ifnum#1=60 %
\hatcurLCdurhreccenxxxxxC
\else
\ifnum#1=61 %
\hatcurLCdurhreccenxxxxxD
\else
\ifnum#1=62 %
\hatcurLCdurhreccenxxxxxE
\else
\ifnum#1=63 %
\hatcurLCdurhreccenxxxxxF
\else
\ifnum#1=64 %
\hatcurLCdurhreccenxxxxxG
\else
??????\fi
\fi
\fi
\fi
\fi
\fi
\fi
}
\newcommand{\hatcurLCdurhrshorteccen}[1]{\ifnum#1=58 %
\hatcurLCdurhrshorteccenxxxxxA
\else
\ifnum#1=59 %
\hatcurLCdurhrshorteccenxxxxxB
\else
\ifnum#1=60 %
\hatcurLCdurhrshorteccenxxxxxC
\else
\ifnum#1=61 %
\hatcurLCdurhrshorteccenxxxxxD
\else
\ifnum#1=62 %
\hatcurLCdurhrshorteccenxxxxxE
\else
\ifnum#1=63 %
\hatcurLCdurhrshorteccenxxxxxF
\else
\ifnum#1=64 %
\hatcurLCdurhrshorteccenxxxxxG
\else
??????\fi
\fi
\fi
\fi
\fi
\fi
\fi
}
\newcommand{\hatcurLCdurshorteccen}[1]{\ifnum#1=58 %
\hatcurLCdurshorteccenxxxxxA
\else
\ifnum#1=59 %
\hatcurLCdurshorteccenxxxxxB
\else
\ifnum#1=60 %
\hatcurLCdurshorteccenxxxxxC
\else
\ifnum#1=61 %
\hatcurLCdurshorteccenxxxxxD
\else
\ifnum#1=62 %
\hatcurLCdurshorteccenxxxxxE
\else
\ifnum#1=63 %
\hatcurLCdurshorteccenxxxxxF
\else
\ifnum#1=64 %
\hatcurLCdurshorteccenxxxxxG
\else
??????\fi
\fi
\fi
\fi
\fi
\fi
\fi
}
\newcommand{\hatcurLChatnetmAeccen}[1]{\ifnum#1=58 %
\hatcurLChatnetmAeccenxxxxxA
\else
\ifnum#1=59 %
\hatcurLChatnetmAeccenxxxxxB
\else
\ifnum#1=60 %
\hatcurLChatnetmAeccenxxxxxC
\else
\ifnum#1=61 %
\hatcurLChatnetmAeccenxxxxxD
\else
\ifnum#1=64 %
\hatcurLChatnetmAeccenxxxxxG
\else
??????\fi
\fi
\fi
\fi
\fi
}
\newcommand{\hatcurLChatnetmBeccen}[1]{\ifnum#1=58 %
\hatcurLChatnetmBeccenxxxxxA
\else
\ifnum#1=59 %
\hatcurLChatnetmBeccenxxxxxB
\else
\ifnum#1=60 %
\hatcurLChatnetmBeccenxxxxxC
\else
\ifnum#1=61 %
\hatcurLChatnetmBeccenxxxxxD
\else
\ifnum#1=64 %
\hatcurLChatnetmBeccenxxxxxG
\else
??????\fi
\fi
\fi
\fi
\fi
}
\newcommand{\hatcurLChatnetmCeccen}[1]{\ifnum#1=59 %
\hatcurLChatnetmCeccenxxxxxB
\else
\ifnum#1=61 %
\hatcurLChatnetmCeccenxxxxxD
\else
??????\fi
\fi
}
\newcommand{\hatcurLChatnetmDeccen}[1]{\ifnum#1=59 %
\hatcurLChatnetmDeccenxxxxxB
\else
??????\fi
}
\newcommand{\hatcurLChatnetmeccen}[1]{\ifnum#1=62 %
\hatcurLChatnetmeccenxxxxxE
\else
??????\fi
}
\newcommand{\hatcurLChatnetmEeccen}[1]{\ifnum#1=59 %
\hatcurLChatnetmEeccenxxxxxB
\else
??????\fi
}
\newcommand{\hatcurLChatnetmFeccen}[1]{\ifnum#1=59 %
\hatcurLChatnetmFeccenxxxxxB
\else
??????\fi
}
\newcommand{\hatcurLChatnetmGeccen}[1]{\ifnum#1=59 %
\hatcurLChatnetmGeccenxxxxxB
\else
??????\fi
}
\newcommand{\hatcurLChatnetmHeccen}[1]{\ifnum#1=59 %
\hatcurLChatnetmHeccenxxxxxB
\else
??????\fi
}
\newcommand{\hatcurLCiblendAeccen}[1]{\ifnum#1=58 %
\hatcurLCiblendAeccenxxxxxA
\else
\ifnum#1=59 %
\hatcurLCiblendAeccenxxxxxB
\else
\ifnum#1=60 %
\hatcurLCiblendAeccenxxxxxC
\else
\ifnum#1=61 %
\hatcurLCiblendAeccenxxxxxD
\else
\ifnum#1=64 %
\hatcurLCiblendAeccenxxxxxG
\else
??????\fi
\fi
\fi
\fi
\fi
}
\newcommand{\hatcurLCiblendBeccen}[1]{\ifnum#1=58 %
\hatcurLCiblendBeccenxxxxxA
\else
\ifnum#1=59 %
\hatcurLCiblendBeccenxxxxxB
\else
\ifnum#1=60 %
\hatcurLCiblendBeccenxxxxxC
\else
\ifnum#1=61 %
\hatcurLCiblendBeccenxxxxxD
\else
\ifnum#1=64 %
\hatcurLCiblendBeccenxxxxxG
\else
??????\fi
\fi
\fi
\fi
\fi
}
\newcommand{\hatcurLCiblendCeccen}[1]{\ifnum#1=59 %
\hatcurLCiblendCeccenxxxxxB
\else
\ifnum#1=61 %
\hatcurLCiblendCeccenxxxxxD
\else
??????\fi
\fi
}
\newcommand{\hatcurLCiblendDeccen}[1]{\ifnum#1=59 %
\hatcurLCiblendDeccenxxxxxB
\else
??????\fi
}
\newcommand{\hatcurLCiblendeccen}[1]{\ifnum#1=62 %
\hatcurLCiblendeccenxxxxxE
\else
??????\fi
}
\newcommand{\hatcurLCiblendEeccen}[1]{\ifnum#1=59 %
\hatcurLCiblendEeccenxxxxxB
\else
??????\fi
}
\newcommand{\hatcurLCiblendFeccen}[1]{\ifnum#1=59 %
\hatcurLCiblendFeccenxxxxxB
\else
??????\fi
}
\newcommand{\hatcurLCiblendGeccen}[1]{\ifnum#1=59 %
\hatcurLCiblendGeccenxxxxxB
\else
??????\fi
}
\newcommand{\hatcurLCiblendHeccen}[1]{\ifnum#1=59 %
\hatcurLCiblendHeccenxxxxxB
\else
??????\fi
}
\newcommand{\hatcurLCimpeccen}[1]{\ifnum#1=58 %
\hatcurLCimpeccenxxxxxA
\else
\ifnum#1=59 %
\hatcurLCimpeccenxxxxxB
\else
\ifnum#1=60 %
\hatcurLCimpeccenxxxxxC
\else
\ifnum#1=61 %
\hatcurLCimpeccenxxxxxD
\else
\ifnum#1=62 %
\hatcurLCimpeccenxxxxxE
\else
\ifnum#1=63 %
\hatcurLCimpeccenxxxxxF
\else
\ifnum#1=64 %
\hatcurLCimpeccenxxxxxG
\else
??????\fi
\fi
\fi
\fi
\fi
\fi
\fi
}
\newcommand{\hatcurLCingdureccen}[1]{\ifnum#1=58 %
\hatcurLCingdureccenxxxxxA
\else
\ifnum#1=59 %
\hatcurLCingdureccenxxxxxB
\else
\ifnum#1=60 %
\hatcurLCingdureccenxxxxxC
\else
\ifnum#1=61 %
\hatcurLCingdureccenxxxxxD
\else
\ifnum#1=62 %
\hatcurLCingdureccenxxxxxE
\else
\ifnum#1=63 %
\hatcurLCingdureccenxxxxxF
\else
\ifnum#1=64 %
\hatcurLCingdureccenxxxxxG
\else
??????\fi
\fi
\fi
\fi
\fi
\fi
\fi
}
\newcommand{\hatcurLCPeccen}[1]{\ifnum#1=58 %
\hatcurLCPeccenxxxxxA
\else
\ifnum#1=59 %
\hatcurLCPeccenxxxxxB
\else
\ifnum#1=60 %
\hatcurLCPeccenxxxxxC
\else
\ifnum#1=61 %
\hatcurLCPeccenxxxxxD
\else
\ifnum#1=62 %
\hatcurLCPeccenxxxxxE
\else
\ifnum#1=63 %
\hatcurLCPeccenxxxxxF
\else
\ifnum#1=64 %
\hatcurLCPeccenxxxxxG
\else
??????\fi
\fi
\fi
\fi
\fi
\fi
\fi
}
\newcommand{\hatcurLCPprececcen}[1]{\ifnum#1=58 %
\hatcurLCPprececcenxxxxxA
\else
\ifnum#1=59 %
\hatcurLCPprececcenxxxxxB
\else
\ifnum#1=60 %
\hatcurLCPprececcenxxxxxC
\else
\ifnum#1=61 %
\hatcurLCPprececcenxxxxxD
\else
\ifnum#1=62 %
\hatcurLCPprececcenxxxxxE
\else
\ifnum#1=63 %
\hatcurLCPprececcenxxxxxF
\else
\ifnum#1=64 %
\hatcurLCPprececcenxxxxxG
\else
??????\fi
\fi
\fi
\fi
\fi
\fi
\fi
}
\newcommand{\hatcurLCPshorteccen}[1]{\ifnum#1=58 %
\hatcurLCPshorteccenxxxxxA
\else
\ifnum#1=59 %
\hatcurLCPshorteccenxxxxxB
\else
\ifnum#1=60 %
\hatcurLCPshorteccenxxxxxC
\else
\ifnum#1=61 %
\hatcurLCPshorteccenxxxxxD
\else
\ifnum#1=62 %
\hatcurLCPshorteccenxxxxxE
\else
\ifnum#1=63 %
\hatcurLCPshorteccenxxxxxF
\else
\ifnum#1=64 %
\hatcurLCPshorteccenxxxxxG
\else
??????\fi
\fi
\fi
\fi
\fi
\fi
\fi
}
\newcommand{\hatcurLCqeccen}[1]{\ifnum#1=58 %
\hatcurLCqeccenxxxxxA
\else
\ifnum#1=59 %
\hatcurLCqeccenxxxxxB
\else
\ifnum#1=60 %
\hatcurLCqeccenxxxxxC
\else
\ifnum#1=61 %
\hatcurLCqeccenxxxxxD
\else
\ifnum#1=62 %
\hatcurLCqeccenxxxxxE
\else
\ifnum#1=63 %
\hatcurLCqeccenxxxxxF
\else
\ifnum#1=64 %
\hatcurLCqeccenxxxxxG
\else
??????\fi
\fi
\fi
\fi
\fi
\fi
\fi
}
\newcommand{\hatcurLCqshorteccen}[1]{\ifnum#1=58 %
\hatcurLCqshorteccenxxxxxA
\else
\ifnum#1=59 %
\hatcurLCqshorteccenxxxxxB
\else
\ifnum#1=60 %
\hatcurLCqshorteccenxxxxxC
\else
\ifnum#1=61 %
\hatcurLCqshorteccenxxxxxD
\else
\ifnum#1=62 %
\hatcurLCqshorteccenxxxxxE
\else
\ifnum#1=63 %
\hatcurLCqshorteccenxxxxxF
\else
\ifnum#1=64 %
\hatcurLCqshorteccenxxxxxG
\else
??????\fi
\fi
\fi
\fi
\fi
\fi
\fi
}
\newcommand{\hatcurLCrhoeccen}[1]{\ifnum#1=58 %
\hatcurLCrhoeccenxxxxxA
\else
\ifnum#1=59 %
\hatcurLCrhoeccenxxxxxB
\else
\ifnum#1=60 %
\hatcurLCrhoeccenxxxxxC
\else
\ifnum#1=61 %
\hatcurLCrhoeccenxxxxxD
\else
\ifnum#1=62 %
\hatcurLCrhoeccenxxxxxE
\else
\ifnum#1=63 %
\hatcurLCrhoeccenxxxxxF
\else
\ifnum#1=64 %
\hatcurLCrhoeccenxxxxxG
\else
??????\fi
\fi
\fi
\fi
\fi
\fi
\fi
}
\newcommand{\hatcurLCrprstareccen}[1]{\ifnum#1=58 %
\hatcurLCrprstareccenxxxxxA
\else
\ifnum#1=59 %
\hatcurLCrprstareccenxxxxxB
\else
\ifnum#1=60 %
\hatcurLCrprstareccenxxxxxC
\else
\ifnum#1=61 %
\hatcurLCrprstareccenxxxxxD
\else
\ifnum#1=62 %
\hatcurLCrprstareccenxxxxxE
\else
\ifnum#1=63 %
\hatcurLCrprstareccenxxxxxF
\else
\ifnum#1=64 %
\hatcurLCrprstareccenxxxxxG
\else
??????\fi
\fi
\fi
\fi
\fi
\fi
\fi
}
\newcommand{\hatcurLCTAeccen}[1]{\ifnum#1=58 %
\hatcurLCTAeccenxxxxxA
\else
\ifnum#1=59 %
\hatcurLCTAeccenxxxxxB
\else
\ifnum#1=60 %
\hatcurLCTAeccenxxxxxC
\else
\ifnum#1=61 %
\hatcurLCTAeccenxxxxxD
\else
\ifnum#1=62 %
\hatcurLCTAeccenxxxxxE
\else
\ifnum#1=63 %
\hatcurLCTAeccenxxxxxF
\else
\ifnum#1=64 %
\hatcurLCTAeccenxxxxxG
\else
??????\fi
\fi
\fi
\fi
\fi
\fi
\fi
}
\newcommand{\hatcurLCTBeccen}[1]{\ifnum#1=58 %
\hatcurLCTBeccenxxxxxA
\else
\ifnum#1=59 %
\hatcurLCTBeccenxxxxxB
\else
\ifnum#1=60 %
\hatcurLCTBeccenxxxxxC
\else
\ifnum#1=61 %
\hatcurLCTBeccenxxxxxD
\else
\ifnum#1=62 %
\hatcurLCTBeccenxxxxxE
\else
\ifnum#1=63 %
\hatcurLCTBeccenxxxxxF
\else
\ifnum#1=64 %
\hatcurLCTBeccenxxxxxG
\else
??????\fi
\fi
\fi
\fi
\fi
\fi
\fi
}
\newcommand{\hatcurLCTeccen}[1]{\ifnum#1=58 %
\hatcurLCTeccenxxxxxA
\else
\ifnum#1=59 %
\hatcurLCTeccenxxxxxB
\else
\ifnum#1=60 %
\hatcurLCTeccenxxxxxC
\else
\ifnum#1=61 %
\hatcurLCTeccenxxxxxD
\else
\ifnum#1=62 %
\hatcurLCTeccenxxxxxE
\else
\ifnum#1=63 %
\hatcurLCTeccenxxxxxF
\else
\ifnum#1=64 %
\hatcurLCTeccenxxxxxG
\else
??????\fi
\fi
\fi
\fi
\fi
\fi
\fi
}
\newcommand{\hatcurLCzetaeccen}[1]{\ifnum#1=58 %
\hatcurLCzetaeccenxxxxxA
\else
\ifnum#1=59 %
\hatcurLCzetaeccenxxxxxB
\else
\ifnum#1=60 %
\hatcurLCzetaeccenxxxxxC
\else
\ifnum#1=61 %
\hatcurLCzetaeccenxxxxxD
\else
\ifnum#1=62 %
\hatcurLCzetaeccenxxxxxE
\else
\ifnum#1=63 %
\hatcurLCzetaeccenxxxxxF
\else
\ifnum#1=64 %
\hatcurLCzetaeccenxxxxxG
\else
??????\fi
\fi
\fi
\fi
\fi
\fi
\fi
}
\newcommand{\hatcurPPaequiveccen}[1]{\ifnum#1=58 %
\hatcurPPaequiveccenxxxxxA
\else
\ifnum#1=59 %
\hatcurPPaequiveccenxxxxxB
\else
\ifnum#1=60 %
\hatcurPPaequiveccenxxxxxC
\else
\ifnum#1=61 %
\hatcurPPaequiveccenxxxxxD
\else
\ifnum#1=62 %
\hatcurPPaequiveccenxxxxxE
\else
\ifnum#1=63 %
\hatcurPPaequiveccenxxxxxF
\else
\ifnum#1=64 %
\hatcurPPaequiveccenxxxxxG
\else
??????\fi
\fi
\fi
\fi
\fi
\fi
\fi
}
\newcommand{\hatcurPPareccen}[1]{\ifnum#1=58 %
\hatcurPPareccenxxxxxA
\else
\ifnum#1=59 %
\hatcurPPareccenxxxxxB
\else
\ifnum#1=60 %
\hatcurPPareccenxxxxxC
\else
\ifnum#1=61 %
\hatcurPPareccenxxxxxD
\else
\ifnum#1=62 %
\hatcurPPareccenxxxxxE
\else
\ifnum#1=63 %
\hatcurPPareccenxxxxxF
\else
\ifnum#1=64 %
\hatcurPPareccenxxxxxG
\else
??????\fi
\fi
\fi
\fi
\fi
\fi
\fi
}
\newcommand{\hatcurPPareleccen}[1]{\ifnum#1=58 %
\hatcurPPareleccenxxxxxA
\else
\ifnum#1=59 %
\hatcurPPareleccenxxxxxB
\else
\ifnum#1=60 %
\hatcurPPareleccenxxxxxC
\else
\ifnum#1=61 %
\hatcurPPareleccenxxxxxD
\else
\ifnum#1=62 %
\hatcurPPareleccenxxxxxE
\else
\ifnum#1=63 %
\hatcurPPareleccenxxxxxF
\else
\ifnum#1=64 %
\hatcurPPareleccenxxxxxG
\else
??????\fi
\fi
\fi
\fi
\fi
\fi
\fi
}
\newcommand{\hatcurPPfluxapdimeccen}[1]{\ifnum#1=58 %
\hatcurPPfluxapdimeccenxxxxxA
\else
\ifnum#1=59 %
\hatcurPPfluxapdimeccenxxxxxB
\else
\ifnum#1=60 %
\hatcurPPfluxapdimeccenxxxxxC
\else
\ifnum#1=61 %
\hatcurPPfluxapdimeccenxxxxxD
\else
\ifnum#1=62 %
\hatcurPPfluxapdimeccenxxxxxE
\else
\ifnum#1=63 %
\hatcurPPfluxapdimeccenxxxxxF
\else
\ifnum#1=64 %
\hatcurPPfluxapdimeccenxxxxxG
\else
??????\fi
\fi
\fi
\fi
\fi
\fi
\fi
}
\newcommand{\hatcurPPfluxapeccen}[1]{\ifnum#1=58 %
\hatcurPPfluxapeccenxxxxxA
\else
\ifnum#1=59 %
\hatcurPPfluxapeccenxxxxxB
\else
\ifnum#1=60 %
\hatcurPPfluxapeccenxxxxxC
\else
\ifnum#1=61 %
\hatcurPPfluxapeccenxxxxxD
\else
\ifnum#1=62 %
\hatcurPPfluxapeccenxxxxxE
\else
\ifnum#1=63 %
\hatcurPPfluxapeccenxxxxxF
\else
\ifnum#1=64 %
\hatcurPPfluxapeccenxxxxxG
\else
??????\fi
\fi
\fi
\fi
\fi
\fi
\fi
}
\newcommand{\hatcurPPfluxavgdimeccen}[1]{\ifnum#1=58 %
\hatcurPPfluxavgdimeccenxxxxxA
\else
\ifnum#1=59 %
\hatcurPPfluxavgdimeccenxxxxxB
\else
\ifnum#1=60 %
\hatcurPPfluxavgdimeccenxxxxxC
\else
\ifnum#1=61 %
\hatcurPPfluxavgdimeccenxxxxxD
\else
\ifnum#1=62 %
\hatcurPPfluxavgdimeccenxxxxxE
\else
\ifnum#1=63 %
\hatcurPPfluxavgdimeccenxxxxxF
\else
\ifnum#1=64 %
\hatcurPPfluxavgdimeccenxxxxxG
\else
??????\fi
\fi
\fi
\fi
\fi
\fi
\fi
}
\newcommand{\hatcurPPfluxavgeccen}[1]{\ifnum#1=58 %
\hatcurPPfluxavgeccenxxxxxA
\else
\ifnum#1=59 %
\hatcurPPfluxavgeccenxxxxxB
\else
\ifnum#1=60 %
\hatcurPPfluxavgeccenxxxxxC
\else
\ifnum#1=61 %
\hatcurPPfluxavgeccenxxxxxD
\else
\ifnum#1=62 %
\hatcurPPfluxavgeccenxxxxxE
\else
\ifnum#1=63 %
\hatcurPPfluxavgeccenxxxxxF
\else
\ifnum#1=64 %
\hatcurPPfluxavgeccenxxxxxG
\else
??????\fi
\fi
\fi
\fi
\fi
\fi
\fi
}
\newcommand{\hatcurPPfluxavglogeccen}[1]{\ifnum#1=58 %
\hatcurPPfluxavglogeccenxxxxxA
\else
\ifnum#1=59 %
\hatcurPPfluxavglogeccenxxxxxB
\else
\ifnum#1=60 %
\hatcurPPfluxavglogeccenxxxxxC
\else
\ifnum#1=61 %
\hatcurPPfluxavglogeccenxxxxxD
\else
\ifnum#1=62 %
\hatcurPPfluxavglogeccenxxxxxE
\else
\ifnum#1=63 %
\hatcurPPfluxavglogeccenxxxxxF
\else
\ifnum#1=64 %
\hatcurPPfluxavglogeccenxxxxxG
\else
??????\fi
\fi
\fi
\fi
\fi
\fi
\fi
}
\newcommand{\hatcurPPfluxperidimeccen}[1]{\ifnum#1=58 %
\hatcurPPfluxperidimeccenxxxxxA
\else
\ifnum#1=59 %
\hatcurPPfluxperidimeccenxxxxxB
\else
\ifnum#1=60 %
\hatcurPPfluxperidimeccenxxxxxC
\else
\ifnum#1=61 %
\hatcurPPfluxperidimeccenxxxxxD
\else
\ifnum#1=62 %
\hatcurPPfluxperidimeccenxxxxxE
\else
\ifnum#1=63 %
\hatcurPPfluxperidimeccenxxxxxF
\else
\ifnum#1=64 %
\hatcurPPfluxperidimeccenxxxxxG
\else
??????\fi
\fi
\fi
\fi
\fi
\fi
\fi
}
\newcommand{\hatcurPPfluxperieccen}[1]{\ifnum#1=58 %
\hatcurPPfluxperieccenxxxxxA
\else
\ifnum#1=59 %
\hatcurPPfluxperieccenxxxxxB
\else
\ifnum#1=60 %
\hatcurPPfluxperieccenxxxxxC
\else
\ifnum#1=61 %
\hatcurPPfluxperieccenxxxxxD
\else
\ifnum#1=62 %
\hatcurPPfluxperieccenxxxxxE
\else
\ifnum#1=63 %
\hatcurPPfluxperieccenxxxxxF
\else
\ifnum#1=64 %
\hatcurPPfluxperieccenxxxxxG
\else
??????\fi
\fi
\fi
\fi
\fi
\fi
\fi
}
\newcommand{\hatcurPPgeccen}[1]{\ifnum#1=58 %
\hatcurPPgeccenxxxxxA
\else
\ifnum#1=59 %
\hatcurPPgeccenxxxxxB
\else
\ifnum#1=60 %
\hatcurPPgeccenxxxxxC
\else
\ifnum#1=61 %
\hatcurPPgeccenxxxxxD
\else
\ifnum#1=62 %
\hatcurPPgeccenxxxxxE
\else
\ifnum#1=63 %
\hatcurPPgeccenxxxxxF
\else
\ifnum#1=64 %
\hatcurPPgeccenxxxxxG
\else
??????\fi
\fi
\fi
\fi
\fi
\fi
\fi
}
\newcommand{\hatcurPPieccen}[1]{\ifnum#1=58 %
\hatcurPPieccenxxxxxA
\else
\ifnum#1=59 %
\hatcurPPieccenxxxxxB
\else
\ifnum#1=60 %
\hatcurPPieccenxxxxxC
\else
\ifnum#1=61 %
\hatcurPPieccenxxxxxD
\else
\ifnum#1=62 %
\hatcurPPieccenxxxxxE
\else
\ifnum#1=63 %
\hatcurPPieccenxxxxxF
\else
\ifnum#1=64 %
\hatcurPPieccenxxxxxG
\else
??????\fi
\fi
\fi
\fi
\fi
\fi
\fi
}
\newcommand{\hatcurPPloggeccen}[1]{\ifnum#1=58 %
\hatcurPPloggeccenxxxxxA
\else
\ifnum#1=59 %
\hatcurPPloggeccenxxxxxB
\else
\ifnum#1=60 %
\hatcurPPloggeccenxxxxxC
\else
\ifnum#1=61 %
\hatcurPPloggeccenxxxxxD
\else
\ifnum#1=62 %
\hatcurPPloggeccenxxxxxE
\else
\ifnum#1=63 %
\hatcurPPloggeccenxxxxxF
\else
\ifnum#1=64 %
\hatcurPPloggeccenxxxxxG
\else
??????\fi
\fi
\fi
\fi
\fi
\fi
\fi
}
\newcommand{\hatcurPPmeccen}[1]{\ifnum#1=58 %
\hatcurPPmeccenxxxxxA
\else
\ifnum#1=59 %
\hatcurPPmeccenxxxxxB
\else
\ifnum#1=60 %
\hatcurPPmeccenxxxxxC
\else
\ifnum#1=61 %
\hatcurPPmeccenxxxxxD
\else
\ifnum#1=62 %
\hatcurPPmeccenxxxxxE
\else
\ifnum#1=63 %
\hatcurPPmeccenxxxxxF
\else
\ifnum#1=64 %
\hatcurPPmeccenxxxxxG
\else
??????\fi
\fi
\fi
\fi
\fi
\fi
\fi
}
\newcommand{\hatcurPPmeeccen}[1]{\ifnum#1=58 %
\hatcurPPmeeccenxxxxxA
\else
\ifnum#1=59 %
\hatcurPPmeeccenxxxxxB
\else
\ifnum#1=60 %
\hatcurPPmeeccenxxxxxC
\else
\ifnum#1=61 %
\hatcurPPmeeccenxxxxxD
\else
\ifnum#1=62 %
\hatcurPPmeeccenxxxxxE
\else
\ifnum#1=63 %
\hatcurPPmeeccenxxxxxF
\else
\ifnum#1=64 %
\hatcurPPmeeccenxxxxxG
\else
??????\fi
\fi
\fi
\fi
\fi
\fi
\fi
}
\newcommand{\hatcurPPmelongeccen}[1]{\ifnum#1=58 %
\hatcurPPmelongeccenxxxxxA
\else
\ifnum#1=59 %
\hatcurPPmelongeccenxxxxxB
\else
\ifnum#1=60 %
\hatcurPPmelongeccenxxxxxC
\else
\ifnum#1=61 %
\hatcurPPmelongeccenxxxxxD
\else
\ifnum#1=62 %
\hatcurPPmelongeccenxxxxxE
\else
\ifnum#1=63 %
\hatcurPPmelongeccenxxxxxF
\else
\ifnum#1=64 %
\hatcurPPmelongeccenxxxxxG
\else
??????\fi
\fi
\fi
\fi
\fi
\fi
\fi
}
\newcommand{\hatcurPPmeshorteccen}[1]{\ifnum#1=58 %
\hatcurPPmeshorteccenxxxxxA
\else
\ifnum#1=59 %
\hatcurPPmeshorteccenxxxxxB
\else
\ifnum#1=60 %
\hatcurPPmeshorteccenxxxxxC
\else
\ifnum#1=61 %
\hatcurPPmeshorteccenxxxxxD
\else
\ifnum#1=62 %
\hatcurPPmeshorteccenxxxxxE
\else
\ifnum#1=63 %
\hatcurPPmeshorteccenxxxxxF
\else
\ifnum#1=64 %
\hatcurPPmeshorteccenxxxxxG
\else
??????\fi
\fi
\fi
\fi
\fi
\fi
\fi
}
\newcommand{\hatcurPPmlongeccen}[1]{\ifnum#1=58 %
\hatcurPPmlongeccenxxxxxA
\else
\ifnum#1=59 %
\hatcurPPmlongeccenxxxxxB
\else
\ifnum#1=60 %
\hatcurPPmlongeccenxxxxxC
\else
\ifnum#1=61 %
\hatcurPPmlongeccenxxxxxD
\else
\ifnum#1=62 %
\hatcurPPmlongeccenxxxxxE
\else
\ifnum#1=63 %
\hatcurPPmlongeccenxxxxxF
\else
\ifnum#1=64 %
\hatcurPPmlongeccenxxxxxG
\else
??????\fi
\fi
\fi
\fi
\fi
\fi
\fi
}
\newcommand{\hatcurPPmrcorreccen}[1]{\ifnum#1=58 %
\hatcurPPmrcorreccenxxxxxA
\else
\ifnum#1=59 %
\hatcurPPmrcorreccenxxxxxB
\else
\ifnum#1=60 %
\hatcurPPmrcorreccenxxxxxC
\else
\ifnum#1=61 %
\hatcurPPmrcorreccenxxxxxD
\else
\ifnum#1=62 %
\hatcurPPmrcorreccenxxxxxE
\else
\ifnum#1=63 %
\hatcurPPmrcorreccenxxxxxF
\else
\ifnum#1=64 %
\hatcurPPmrcorreccenxxxxxG
\else
??????\fi
\fi
\fi
\fi
\fi
\fi
\fi
}
\newcommand{\hatcurPPmshorteccen}[1]{\ifnum#1=58 %
\hatcurPPmshorteccenxxxxxA
\else
\ifnum#1=59 %
\hatcurPPmshorteccenxxxxxB
\else
\ifnum#1=60 %
\hatcurPPmshorteccenxxxxxC
\else
\ifnum#1=61 %
\hatcurPPmshorteccenxxxxxD
\else
\ifnum#1=62 %
\hatcurPPmshorteccenxxxxxE
\else
\ifnum#1=63 %
\hatcurPPmshorteccenxxxxxF
\else
\ifnum#1=64 %
\hatcurPPmshorteccenxxxxxG
\else
??????\fi
\fi
\fi
\fi
\fi
\fi
\fi
}
\newcommand{\hatcurPPperieccen}[1]{\ifnum#1=58 %
\hatcurPPperieccenxxxxxA
\else
\ifnum#1=59 %
\hatcurPPperieccenxxxxxB
\else
\ifnum#1=60 %
\hatcurPPperieccenxxxxxC
\else
\ifnum#1=61 %
\hatcurPPperieccenxxxxxD
\else
\ifnum#1=62 %
\hatcurPPperieccenxxxxxE
\else
\ifnum#1=63 %
\hatcurPPperieccenxxxxxF
\else
\ifnum#1=64 %
\hatcurPPperieccenxxxxxG
\else
??????\fi
\fi
\fi
\fi
\fi
\fi
\fi
}
\newcommand{\hatcurPPphiconjeccen}[1]{\ifnum#1=58 %
\hatcurPPphiconjeccenxxxxxA
\else
\ifnum#1=59 %
\hatcurPPphiconjeccenxxxxxB
\else
\ifnum#1=60 %
\hatcurPPphiconjeccenxxxxxC
\else
\ifnum#1=61 %
\hatcurPPphiconjeccenxxxxxD
\else
\ifnum#1=62 %
\hatcurPPphiconjeccenxxxxxE
\else
\ifnum#1=63 %
\hatcurPPphiconjeccenxxxxxF
\else
\ifnum#1=64 %
\hatcurPPphiconjeccenxxxxxG
\else
??????\fi
\fi
\fi
\fi
\fi
\fi
\fi
}
\newcommand{\hatcurPPreccen}[1]{\ifnum#1=58 %
\hatcurPPreccenxxxxxA
\else
\ifnum#1=59 %
\hatcurPPreccenxxxxxB
\else
\ifnum#1=60 %
\hatcurPPreccenxxxxxC
\else
\ifnum#1=61 %
\hatcurPPreccenxxxxxD
\else
\ifnum#1=62 %
\hatcurPPreccenxxxxxE
\else
\ifnum#1=63 %
\hatcurPPreccenxxxxxF
\else
\ifnum#1=64 %
\hatcurPPreccenxxxxxG
\else
??????\fi
\fi
\fi
\fi
\fi
\fi
\fi
}
\newcommand{\hatcurPPreeccen}[1]{\ifnum#1=58 %
\hatcurPPreeccenxxxxxA
\else
\ifnum#1=59 %
\hatcurPPreeccenxxxxxB
\else
\ifnum#1=60 %
\hatcurPPreeccenxxxxxC
\else
\ifnum#1=61 %
\hatcurPPreeccenxxxxxD
\else
\ifnum#1=62 %
\hatcurPPreeccenxxxxxE
\else
\ifnum#1=63 %
\hatcurPPreeccenxxxxxF
\else
\ifnum#1=64 %
\hatcurPPreeccenxxxxxG
\else
??????\fi
\fi
\fi
\fi
\fi
\fi
\fi
}
\newcommand{\hatcurPPrelongeccen}[1]{\ifnum#1=58 %
\hatcurPPrelongeccenxxxxxA
\else
\ifnum#1=59 %
\hatcurPPrelongeccenxxxxxB
\else
\ifnum#1=60 %
\hatcurPPrelongeccenxxxxxC
\else
\ifnum#1=61 %
\hatcurPPrelongeccenxxxxxD
\else
\ifnum#1=62 %
\hatcurPPrelongeccenxxxxxE
\else
\ifnum#1=63 %
\hatcurPPrelongeccenxxxxxF
\else
\ifnum#1=64 %
\hatcurPPrelongeccenxxxxxG
\else
??????\fi
\fi
\fi
\fi
\fi
\fi
\fi
}
\newcommand{\hatcurPPreshorteccen}[1]{\ifnum#1=58 %
\hatcurPPreshorteccenxxxxxA
\else
\ifnum#1=59 %
\hatcurPPreshorteccenxxxxxB
\else
\ifnum#1=60 %
\hatcurPPreshorteccenxxxxxC
\else
\ifnum#1=61 %
\hatcurPPreshorteccenxxxxxD
\else
\ifnum#1=62 %
\hatcurPPreshorteccenxxxxxE
\else
\ifnum#1=63 %
\hatcurPPreshorteccenxxxxxF
\else
\ifnum#1=64 %
\hatcurPPreshorteccenxxxxxG
\else
??????\fi
\fi
\fi
\fi
\fi
\fi
\fi
}
\newcommand{\hatcurPPrhoeccen}[1]{\ifnum#1=58 %
\hatcurPPrhoeccenxxxxxA
\else
\ifnum#1=59 %
\hatcurPPrhoeccenxxxxxB
\else
\ifnum#1=60 %
\hatcurPPrhoeccenxxxxxC
\else
\ifnum#1=61 %
\hatcurPPrhoeccenxxxxxD
\else
\ifnum#1=62 %
\hatcurPPrhoeccenxxxxxE
\else
\ifnum#1=63 %
\hatcurPPrhoeccenxxxxxF
\else
\ifnum#1=64 %
\hatcurPPrhoeccenxxxxxG
\else
??????\fi
\fi
\fi
\fi
\fi
\fi
\fi
}
\newcommand{\hatcurPPrlongeccen}[1]{\ifnum#1=58 %
\hatcurPPrlongeccenxxxxxA
\else
\ifnum#1=59 %
\hatcurPPrlongeccenxxxxxB
\else
\ifnum#1=60 %
\hatcurPPrlongeccenxxxxxC
\else
\ifnum#1=61 %
\hatcurPPrlongeccenxxxxxD
\else
\ifnum#1=62 %
\hatcurPPrlongeccenxxxxxE
\else
\ifnum#1=63 %
\hatcurPPrlongeccenxxxxxF
\else
\ifnum#1=64 %
\hatcurPPrlongeccenxxxxxG
\else
??????\fi
\fi
\fi
\fi
\fi
\fi
\fi
}
\newcommand{\hatcurPPrshorteccen}[1]{\ifnum#1=58 %
\hatcurPPrshorteccenxxxxxA
\else
\ifnum#1=59 %
\hatcurPPrshorteccenxxxxxB
\else
\ifnum#1=60 %
\hatcurPPrshorteccenxxxxxC
\else
\ifnum#1=61 %
\hatcurPPrshorteccenxxxxxD
\else
\ifnum#1=62 %
\hatcurPPrshorteccenxxxxxE
\else
\ifnum#1=63 %
\hatcurPPrshorteccenxxxxxF
\else
\ifnum#1=64 %
\hatcurPPrshorteccenxxxxxG
\else
??????\fi
\fi
\fi
\fi
\fi
\fi
\fi
}
\newcommand{\hatcurPPtcirceccen}[1]{\ifnum#1=58 %
\hatcurPPtcirceccenxxxxxA
\else
\ifnum#1=59 %
\hatcurPPtcirceccenxxxxxB
\else
\ifnum#1=60 %
\hatcurPPtcirceccenxxxxxC
\else
\ifnum#1=61 %
\hatcurPPtcirceccenxxxxxD
\else
\ifnum#1=62 %
\hatcurPPtcirceccenxxxxxE
\else
\ifnum#1=63 %
\hatcurPPtcirceccenxxxxxF
\else
\ifnum#1=64 %
\hatcurPPtcirceccenxxxxxG
\else
??????\fi
\fi
\fi
\fi
\fi
\fi
\fi
}
\newcommand{\hatcurPPteffeccen}[1]{\ifnum#1=58 %
\hatcurPPteffeccenxxxxxA
\else
\ifnum#1=59 %
\hatcurPPteffeccenxxxxxB
\else
\ifnum#1=60 %
\hatcurPPteffeccenxxxxxC
\else
\ifnum#1=61 %
\hatcurPPteffeccenxxxxxD
\else
\ifnum#1=62 %
\hatcurPPteffeccenxxxxxE
\else
\ifnum#1=63 %
\hatcurPPteffeccenxxxxxF
\else
\ifnum#1=64 %
\hatcurPPteffeccenxxxxxG
\else
??????\fi
\fi
\fi
\fi
\fi
\fi
\fi
}
\newcommand{\hatcurPPthetaeccen}[1]{\ifnum#1=58 %
\hatcurPPthetaeccenxxxxxA
\else
\ifnum#1=59 %
\hatcurPPthetaeccenxxxxxB
\else
\ifnum#1=60 %
\hatcurPPthetaeccenxxxxxC
\else
\ifnum#1=61 %
\hatcurPPthetaeccenxxxxxD
\else
\ifnum#1=62 %
\hatcurPPthetaeccenxxxxxE
\else
\ifnum#1=63 %
\hatcurPPthetaeccenxxxxxF
\else
\ifnum#1=64 %
\hatcurPPthetaeccenxxxxxG
\else
??????\fi
\fi
\fi
\fi
\fi
\fi
\fi
}
\newcommand{\hatcurPPtinfalleccen}[1]{\ifnum#1=58 %
\hatcurPPtinfalleccenxxxxxA
\else
\ifnum#1=59 %
\hatcurPPtinfalleccenxxxxxB
\else
\ifnum#1=60 %
\hatcurPPtinfalleccenxxxxxC
\else
\ifnum#1=61 %
\hatcurPPtinfalleccenxxxxxD
\else
\ifnum#1=62 %
\hatcurPPtinfalleccenxxxxxE
\else
\ifnum#1=63 %
\hatcurPPtinfalleccenxxxxxF
\else
\ifnum#1=64 %
\hatcurPPtinfalleccenxxxxxG
\else
??????\fi
\fi
\fi
\fi
\fi
\fi
\fi
}
\newcommand{\hatcurRVecceneccen}[1]{\ifnum#1=58 %
\hatcurRVecceneccenxxxxxA
\else
\ifnum#1=59 %
\hatcurRVecceneccenxxxxxB
\else
\ifnum#1=60 %
\hatcurRVecceneccenxxxxxC
\else
\ifnum#1=61 %
\hatcurRVecceneccenxxxxxD
\else
\ifnum#1=62 %
\hatcurRVecceneccenxxxxxE
\else
\ifnum#1=63 %
\hatcurRVecceneccenxxxxxF
\else
\ifnum#1=64 %
\hatcurRVecceneccenxxxxxG
\else
??????\fi
\fi
\fi
\fi
\fi
\fi
\fi
}
\newcommand{\hatcurRVeccentwosiglimeccen}[1]{\ifnum#1=58 %
\hatcurRVeccentwosiglimeccenxxxxxA
\else
\ifnum#1=59 %
\hatcurRVeccentwosiglimeccenxxxxxB
\else
\ifnum#1=60 %
\hatcurRVeccentwosiglimeccenxxxxxC
\else
\ifnum#1=61 %
\hatcurRVeccentwosiglimeccenxxxxxD
\else
\ifnum#1=62 %
\hatcurRVeccentwosiglimeccenxxxxxE
\else
\ifnum#1=63 %
\hatcurRVeccentwosiglimeccenxxxxxF
\else
\ifnum#1=64 %
\hatcurRVeccentwosiglimeccenxxxxxG
\else
??????\fi
\fi
\fi
\fi
\fi
\fi
\fi
}
\newcommand{\hatcurRVfitrmsAeccen}[1]{\ifnum#1=59 %
\hatcurRVfitrmsAeccenxxxxxB
\else
\ifnum#1=60 %
\hatcurRVfitrmsAeccenxxxxxC
\else
\ifnum#1=61 %
\hatcurRVfitrmsAeccenxxxxxD
\else
\ifnum#1=63 %
\hatcurRVfitrmsAeccenxxxxxF
\else
\ifnum#1=64 %
\hatcurRVfitrmsAeccenxxxxxG
\else
??????\fi
\fi
\fi
\fi
\fi
}
\newcommand{\hatcurRVfitrmsBeccen}[1]{\ifnum#1=59 %
\hatcurRVfitrmsBeccenxxxxxB
\else
\ifnum#1=60 %
\hatcurRVfitrmsBeccenxxxxxC
\else
\ifnum#1=61 %
\hatcurRVfitrmsBeccenxxxxxD
\else
\ifnum#1=63 %
\hatcurRVfitrmsBeccenxxxxxF
\else
\ifnum#1=64 %
\hatcurRVfitrmsBeccenxxxxxG
\else
??????\fi
\fi
\fi
\fi
\fi
}
\newcommand{\hatcurRVfitrmsCeccen}[1]{\ifnum#1=60 %
\hatcurRVfitrmsCeccenxxxxxC
\else
\ifnum#1=61 %
\hatcurRVfitrmsCeccenxxxxxD
\else
\ifnum#1=63 %
\hatcurRVfitrmsCeccenxxxxxF
\else
??????\fi
\fi
\fi
}
\newcommand{\hatcurRVfitrmsDeccen}[1]{\ifnum#1=60 %
\hatcurRVfitrmsDeccenxxxxxC
\else
??????\fi
}
\newcommand{\hatcurRVfitrmseccen}[1]{\ifnum#1=58 %
\hatcurRVfitrmseccenxxxxxA
\else
\ifnum#1=62 %
\hatcurRVfitrmseccenxxxxxE
\else
??????\fi
\fi
}
\newcommand{\hatcurRVfitrmsEeccen}[1]{\ifnum#1=60 %
\hatcurRVfitrmsEeccenxxxxxC
\else
??????\fi
}
\newcommand{\hatcurRVgammaAeccen}[1]{\ifnum#1=59 %
\hatcurRVgammaAeccenxxxxxB
\else
\ifnum#1=60 %
\hatcurRVgammaAeccenxxxxxC
\else
\ifnum#1=61 %
\hatcurRVgammaAeccenxxxxxD
\else
\ifnum#1=63 %
\hatcurRVgammaAeccenxxxxxF
\else
\ifnum#1=64 %
\hatcurRVgammaAeccenxxxxxG
\else
??????\fi
\fi
\fi
\fi
\fi
}
\newcommand{\hatcurRVgammaBeccen}[1]{\ifnum#1=59 %
\hatcurRVgammaBeccenxxxxxB
\else
\ifnum#1=60 %
\hatcurRVgammaBeccenxxxxxC
\else
\ifnum#1=61 %
\hatcurRVgammaBeccenxxxxxD
\else
\ifnum#1=63 %
\hatcurRVgammaBeccenxxxxxF
\else
\ifnum#1=64 %
\hatcurRVgammaBeccenxxxxxG
\else
??????\fi
\fi
\fi
\fi
\fi
}
\newcommand{\hatcurRVgammaCeccen}[1]{\ifnum#1=60 %
\hatcurRVgammaCeccenxxxxxC
\else
\ifnum#1=61 %
\hatcurRVgammaCeccenxxxxxD
\else
\ifnum#1=63 %
\hatcurRVgammaCeccenxxxxxF
\else
??????\fi
\fi
\fi
}
\newcommand{\hatcurRVgammaDeccen}[1]{\ifnum#1=60 %
\hatcurRVgammaDeccenxxxxxC
\else
??????\fi
}
\newcommand{\hatcurRVgammaeccen}[1]{\ifnum#1=58 %
\hatcurRVgammaeccenxxxxxA
\else
\ifnum#1=62 %
\hatcurRVgammaeccenxxxxxE
\else
??????\fi
\fi
}
\newcommand{\hatcurRVgammaEeccen}[1]{\ifnum#1=60 %
\hatcurRVgammaEeccenxxxxxC
\else
??????\fi
}
\newcommand{\hatcurRVheccen}[1]{\ifnum#1=58 %
\hatcurRVheccenxxxxxA
\else
\ifnum#1=59 %
\hatcurRVheccenxxxxxB
\else
\ifnum#1=60 %
\hatcurRVheccenxxxxxC
\else
\ifnum#1=61 %
\hatcurRVheccenxxxxxD
\else
\ifnum#1=62 %
\hatcurRVheccenxxxxxE
\else
\ifnum#1=63 %
\hatcurRVheccenxxxxxF
\else
\ifnum#1=64 %
\hatcurRVheccenxxxxxG
\else
??????\fi
\fi
\fi
\fi
\fi
\fi
\fi
}
\newcommand{\hatcurRVjitterAeccen}[1]{\ifnum#1=59 %
\hatcurRVjitterAeccenxxxxxB
\else
\ifnum#1=60 %
\hatcurRVjitterAeccenxxxxxC
\else
\ifnum#1=61 %
\hatcurRVjitterAeccenxxxxxD
\else
\ifnum#1=63 %
\hatcurRVjitterAeccenxxxxxF
\else
\ifnum#1=64 %
\hatcurRVjitterAeccenxxxxxG
\else
??????\fi
\fi
\fi
\fi
\fi
}
\newcommand{\hatcurRVjitterBeccen}[1]{\ifnum#1=59 %
\hatcurRVjitterBeccenxxxxxB
\else
\ifnum#1=60 %
\hatcurRVjitterBeccenxxxxxC
\else
\ifnum#1=61 %
\hatcurRVjitterBeccenxxxxxD
\else
\ifnum#1=63 %
\hatcurRVjitterBeccenxxxxxF
\else
\ifnum#1=64 %
\hatcurRVjitterBeccenxxxxxG
\else
??????\fi
\fi
\fi
\fi
\fi
}
\newcommand{\hatcurRVjitterCeccen}[1]{\ifnum#1=60 %
\hatcurRVjitterCeccenxxxxxC
\else
\ifnum#1=61 %
\hatcurRVjitterCeccenxxxxxD
\else
\ifnum#1=63 %
\hatcurRVjitterCeccenxxxxxF
\else
??????\fi
\fi
\fi
}
\newcommand{\hatcurRVjitterDeccen}[1]{\ifnum#1=60 %
\hatcurRVjitterDeccenxxxxxC
\else
??????\fi
}
\newcommand{\hatcurRVjittereccen}[1]{\ifnum#1=58 %
\hatcurRVjittereccenxxxxxA
\else
\ifnum#1=62 %
\hatcurRVjittereccenxxxxxE
\else
??????\fi
\fi
}
\newcommand{\hatcurRVjitterEeccen}[1]{\ifnum#1=60 %
\hatcurRVjitterEeccenxxxxxC
\else
??????\fi
}
\newcommand{\hatcurRVjittertwosiglimAeccen}[1]{\ifnum#1=59 %
\hatcurRVjittertwosiglimAeccenxxxxxB
\else
\ifnum#1=60 %
\hatcurRVjittertwosiglimAeccenxxxxxC
\else
\ifnum#1=61 %
\hatcurRVjittertwosiglimAeccenxxxxxD
\else
\ifnum#1=63 %
\hatcurRVjittertwosiglimAeccenxxxxxF
\else
\ifnum#1=64 %
\hatcurRVjittertwosiglimAeccenxxxxxG
\else
??????\fi
\fi
\fi
\fi
\fi
}
\newcommand{\hatcurRVjittertwosiglimBeccen}[1]{\ifnum#1=59 %
\hatcurRVjittertwosiglimBeccenxxxxxB
\else
\ifnum#1=60 %
\hatcurRVjittertwosiglimBeccenxxxxxC
\else
\ifnum#1=61 %
\hatcurRVjittertwosiglimBeccenxxxxxD
\else
\ifnum#1=63 %
\hatcurRVjittertwosiglimBeccenxxxxxF
\else
\ifnum#1=64 %
\hatcurRVjittertwosiglimBeccenxxxxxG
\else
??????\fi
\fi
\fi
\fi
\fi
}
\newcommand{\hatcurRVjittertwosiglimCeccen}[1]{\ifnum#1=60 %
\hatcurRVjittertwosiglimCeccenxxxxxC
\else
\ifnum#1=61 %
\hatcurRVjittertwosiglimCeccenxxxxxD
\else
\ifnum#1=63 %
\hatcurRVjittertwosiglimCeccenxxxxxF
\else
??????\fi
\fi
\fi
}
\newcommand{\hatcurRVjittertwosiglimDeccen}[1]{\ifnum#1=60 %
\hatcurRVjittertwosiglimDeccenxxxxxC
\else
??????\fi
}
\newcommand{\hatcurRVjittertwosiglimeccen}[1]{\ifnum#1=58 %
\hatcurRVjittertwosiglimeccenxxxxxA
\else
\ifnum#1=62 %
\hatcurRVjittertwosiglimeccenxxxxxE
\else
??????\fi
\fi
}
\newcommand{\hatcurRVjittertwosiglimEeccen}[1]{\ifnum#1=60 %
\hatcurRVjittertwosiglimEeccenxxxxxC
\else
??????\fi
}
\newcommand{\hatcurRVkeccen}[1]{\ifnum#1=58 %
\hatcurRVkeccenxxxxxA
\else
\ifnum#1=59 %
\hatcurRVkeccenxxxxxB
\else
\ifnum#1=60 %
\hatcurRVkeccenxxxxxC
\else
\ifnum#1=61 %
\hatcurRVkeccenxxxxxD
\else
\ifnum#1=62 %
\hatcurRVkeccenxxxxxE
\else
\ifnum#1=63 %
\hatcurRVkeccenxxxxxF
\else
\ifnum#1=64 %
\hatcurRVkeccenxxxxxG
\else
??????\fi
\fi
\fi
\fi
\fi
\fi
\fi
}
\newcommand{\hatcurRVKeccen}[1]{\ifnum#1=58 %
\hatcurRVKeccenxxxxxA
\else
\ifnum#1=59 %
\hatcurRVKeccenxxxxxB
\else
\ifnum#1=60 %
\hatcurRVKeccenxxxxxC
\else
\ifnum#1=61 %
\hatcurRVKeccenxxxxxD
\else
\ifnum#1=62 %
\hatcurRVKeccenxxxxxE
\else
\ifnum#1=63 %
\hatcurRVKeccenxxxxxF
\else
\ifnum#1=64 %
\hatcurRVKeccenxxxxxG
\else
??????\fi
\fi
\fi
\fi
\fi
\fi
\fi
}
\newcommand{\hatcurRVomegaeccen}[1]{\ifnum#1=58 %
\hatcurRVomegaeccenxxxxxA
\else
\ifnum#1=59 %
\hatcurRVomegaeccenxxxxxB
\else
\ifnum#1=60 %
\hatcurRVomegaeccenxxxxxC
\else
\ifnum#1=61 %
\hatcurRVomegaeccenxxxxxD
\else
\ifnum#1=62 %
\hatcurRVomegaeccenxxxxxE
\else
\ifnum#1=63 %
\hatcurRVomegaeccenxxxxxF
\else
\ifnum#1=64 %
\hatcurRVomegaeccenxxxxxG
\else
??????\fi
\fi
\fi
\fi
\fi
\fi
\fi
}
\newcommand{\hatcurRVrheccen}[1]{\ifnum#1=58 %
\hatcurRVrheccenxxxxxA
\else
\ifnum#1=59 %
\hatcurRVrheccenxxxxxB
\else
\ifnum#1=60 %
\hatcurRVrheccenxxxxxC
\else
\ifnum#1=61 %
\hatcurRVrheccenxxxxxD
\else
\ifnum#1=62 %
\hatcurRVrheccenxxxxxE
\else
\ifnum#1=63 %
\hatcurRVrheccenxxxxxF
\else
\ifnum#1=64 %
\hatcurRVrheccenxxxxxG
\else
??????\fi
\fi
\fi
\fi
\fi
\fi
\fi
}
\newcommand{\hatcurRVrkeccen}[1]{\ifnum#1=58 %
\hatcurRVrkeccenxxxxxA
\else
\ifnum#1=59 %
\hatcurRVrkeccenxxxxxB
\else
\ifnum#1=60 %
\hatcurRVrkeccenxxxxxC
\else
\ifnum#1=61 %
\hatcurRVrkeccenxxxxxD
\else
\ifnum#1=62 %
\hatcurRVrkeccenxxxxxE
\else
\ifnum#1=63 %
\hatcurRVrkeccenxxxxxF
\else
\ifnum#1=64 %
\hatcurRVrkeccenxxxxxG
\else
??????\fi
\fi
\fi
\fi
\fi
\fi
\fi
}
\newcommand{\hatcurRVtroneeccen}[1]{\ifnum#1=58 %
\hatcurRVtroneeccenxxxxxA
\else
\ifnum#1=59 %
\hatcurRVtroneeccenxxxxxB
\else
\ifnum#1=60 %
\hatcurRVtroneeccenxxxxxC
\else
\ifnum#1=61 %
\hatcurRVtroneeccenxxxxxD
\else
\ifnum#1=62 %
\hatcurRVtroneeccenxxxxxE
\else
\ifnum#1=63 %
\hatcurRVtroneeccenxxxxxF
\else
\ifnum#1=64 %
\hatcurRVtroneeccenxxxxxG
\else
??????\fi
\fi
\fi
\fi
\fi
\fi
\fi
}
\newcommand{\hatcurRVtrtwoeccen}[1]{\ifnum#1=58 %
\hatcurRVtrtwoeccenxxxxxA
\else
\ifnum#1=59 %
\hatcurRVtrtwoeccenxxxxxB
\else
\ifnum#1=60 %
\hatcurRVtrtwoeccenxxxxxC
\else
\ifnum#1=61 %
\hatcurRVtrtwoeccenxxxxxD
\else
\ifnum#1=62 %
\hatcurRVtrtwoeccenxxxxxE
\else
\ifnum#1=63 %
\hatcurRVtrtwoeccenxxxxxF
\else
\ifnum#1=64 %
\hatcurRVtrtwoeccenxxxxxG
\else
??????\fi
\fi
\fi
\fi
\fi
\fi
\fi
}
\newcommand{\hatcurSMEiiloggeccen}[1]{\ifnum#1=58 %
\hatcurSMEiiloggeccenxxxxxA
\else
\ifnum#1=59 %
\hatcurSMEiiloggeccenxxxxxB
\else
\ifnum#1=60 %
\hatcurSMEiiloggeccenxxxxxC
\else
\ifnum#1=61 %
\hatcurSMEiiloggeccenxxxxxD
\else
\ifnum#1=62 %
\hatcurSMEiiloggeccenxxxxxE
\else
??????\fi
\fi
\fi
\fi
\fi
}
\newcommand{\hatcurSMEiiteffeccen}[1]{\ifnum#1=58 %
\hatcurSMEiiteffeccenxxxxxA
\else
\ifnum#1=59 %
\hatcurSMEiiteffeccenxxxxxB
\else
\ifnum#1=60 %
\hatcurSMEiiteffeccenxxxxxC
\else
\ifnum#1=61 %
\hatcurSMEiiteffeccenxxxxxD
\else
\ifnum#1=62 %
\hatcurSMEiiteffeccenxxxxxE
\else
??????\fi
\fi
\fi
\fi
\fi
}
\newcommand{\hatcurSMEiivsineccen}[1]{\ifnum#1=58 %
\hatcurSMEiivsineccenxxxxxA
\else
\ifnum#1=59 %
\hatcurSMEiivsineccenxxxxxB
\else
\ifnum#1=60 %
\hatcurSMEiivsineccenxxxxxC
\else
\ifnum#1=61 %
\hatcurSMEiivsineccenxxxxxD
\else
\ifnum#1=62 %
\hatcurSMEiivsineccenxxxxxE
\else
??????\fi
\fi
\fi
\fi
\fi
}
\newcommand{\hatcurSMEiizfeheccen}[1]{\ifnum#1=58 %
\hatcurSMEiizfeheccenxxxxxA
\else
\ifnum#1=59 %
\hatcurSMEiizfeheccenxxxxxB
\else
\ifnum#1=60 %
\hatcurSMEiizfeheccenxxxxxC
\else
\ifnum#1=61 %
\hatcurSMEiizfeheccenxxxxxD
\else
\ifnum#1=62 %
\hatcurSMEiizfeheccenxxxxxE
\else
??????\fi
\fi
\fi
\fi
\fi
}
\newcommand{\hatcurSMEiizfehshorteccen}[1]{\ifnum#1=58 %
\hatcurSMEiizfehshorteccenxxxxxA
\else
\ifnum#1=59 %
\hatcurSMEiizfehshorteccenxxxxxB
\else
\ifnum#1=60 %
\hatcurSMEiizfehshorteccenxxxxxC
\else
\ifnum#1=61 %
\hatcurSMEiizfehshorteccenxxxxxD
\else
\ifnum#1=62 %
\hatcurSMEiizfehshorteccenxxxxxE
\else
??????\fi
\fi
\fi
\fi
\fi
}
\newcommand{\hatcurSMEiloggeccen}[1]{\ifnum#1=58 %
\hatcurSMEiloggeccenxxxxxA
\else
\ifnum#1=59 %
\hatcurSMEiloggeccenxxxxxB
\else
\ifnum#1=60 %
\hatcurSMEiloggeccenxxxxxC
\else
\ifnum#1=61 %
\hatcurSMEiloggeccenxxxxxD
\else
\ifnum#1=62 %
\hatcurSMEiloggeccenxxxxxE
\else
\ifnum#1=63 %
\hatcurSMEiloggeccenxxxxxF
\else
\ifnum#1=64 %
\hatcurSMEiloggeccenxxxxxG
\else
??????\fi
\fi
\fi
\fi
\fi
\fi
\fi
}
\newcommand{\hatcurSMEiteffeccen}[1]{\ifnum#1=58 %
\hatcurSMEiteffeccenxxxxxA
\else
\ifnum#1=59 %
\hatcurSMEiteffeccenxxxxxB
\else
\ifnum#1=60 %
\hatcurSMEiteffeccenxxxxxC
\else
\ifnum#1=61 %
\hatcurSMEiteffeccenxxxxxD
\else
\ifnum#1=62 %
\hatcurSMEiteffeccenxxxxxE
\else
\ifnum#1=63 %
\hatcurSMEiteffeccenxxxxxF
\else
\ifnum#1=64 %
\hatcurSMEiteffeccenxxxxxG
\else
??????\fi
\fi
\fi
\fi
\fi
\fi
\fi
}
\newcommand{\hatcurSMEivmaceccen}[1]{\ifnum#1=58 %
\hatcurSMEivmaceccenxxxxxA
\else
\ifnum#1=59 %
\hatcurSMEivmaceccenxxxxxB
\else
\ifnum#1=60 %
\hatcurSMEivmaceccenxxxxxC
\else
\ifnum#1=61 %
\hatcurSMEivmaceccenxxxxxD
\else
\ifnum#1=62 %
\hatcurSMEivmaceccenxxxxxE
\else
\ifnum#1=63 %
\hatcurSMEivmaceccenxxxxxF
\else
\ifnum#1=64 %
\hatcurSMEivmaceccenxxxxxG
\else
??????\fi
\fi
\fi
\fi
\fi
\fi
\fi
}
\newcommand{\hatcurSMEivmiceccen}[1]{\ifnum#1=58 %
\hatcurSMEivmiceccenxxxxxA
\else
\ifnum#1=59 %
\hatcurSMEivmiceccenxxxxxB
\else
\ifnum#1=60 %
\hatcurSMEivmiceccenxxxxxC
\else
\ifnum#1=61 %
\hatcurSMEivmiceccenxxxxxD
\else
\ifnum#1=62 %
\hatcurSMEivmiceccenxxxxxE
\else
\ifnum#1=63 %
\hatcurSMEivmiceccenxxxxxF
\else
\ifnum#1=64 %
\hatcurSMEivmiceccenxxxxxG
\else
??????\fi
\fi
\fi
\fi
\fi
\fi
\fi
}
\newcommand{\hatcurSMEivsineccen}[1]{\ifnum#1=58 %
\hatcurSMEivsineccenxxxxxA
\else
\ifnum#1=59 %
\hatcurSMEivsineccenxxxxxB
\else
\ifnum#1=60 %
\hatcurSMEivsineccenxxxxxC
\else
\ifnum#1=61 %
\hatcurSMEivsineccenxxxxxD
\else
\ifnum#1=62 %
\hatcurSMEivsineccenxxxxxE
\else
\ifnum#1=63 %
\hatcurSMEivsineccenxxxxxF
\else
\ifnum#1=64 %
\hatcurSMEivsineccenxxxxxG
\else
??????\fi
\fi
\fi
\fi
\fi
\fi
\fi
}
\newcommand{\hatcurSMEizfeheccen}[1]{\ifnum#1=58 %
\hatcurSMEizfeheccenxxxxxA
\else
\ifnum#1=59 %
\hatcurSMEizfeheccenxxxxxB
\else
\ifnum#1=60 %
\hatcurSMEizfeheccenxxxxxC
\else
\ifnum#1=61 %
\hatcurSMEizfeheccenxxxxxD
\else
\ifnum#1=62 %
\hatcurSMEizfeheccenxxxxxE
\else
\ifnum#1=63 %
\hatcurSMEizfeheccenxxxxxF
\else
\ifnum#1=64 %
\hatcurSMEizfeheccenxxxxxG
\else
??????\fi
\fi
\fi
\fi
\fi
\fi
\fi
}
\newcommand{\hatcurSMEizfehshorteccen}[1]{\ifnum#1=58 %
\hatcurSMEizfehshorteccenxxxxxA
\else
\ifnum#1=59 %
\hatcurSMEizfehshorteccenxxxxxB
\else
\ifnum#1=60 %
\hatcurSMEizfehshorteccenxxxxxC
\else
\ifnum#1=61 %
\hatcurSMEizfehshorteccenxxxxxD
\else
\ifnum#1=62 %
\hatcurSMEizfehshorteccenxxxxxE
\else
\ifnum#1=63 %
\hatcurSMEizfehshorteccenxxxxxF
\else
\ifnum#1=64 %
\hatcurSMEizfehshorteccenxxxxxG
\else
??????\fi
\fi
\fi
\fi
\fi
\fi
\fi
}
\newcommand{\hatcurXAveccen}[1]{\ifnum#1=58 %
\hatcurXAveccenxxxxxA
\else
\ifnum#1=59 %
\hatcurXAveccenxxxxxB
\else
\ifnum#1=60 %
\hatcurXAveccenxxxxxC
\else
\ifnum#1=61 %
\hatcurXAveccenxxxxxD
\else
\ifnum#1=62 %
\hatcurXAveccenxxxxxE
\else
\ifnum#1=63 %
\hatcurXAveccenxxxxxF
\else
\ifnum#1=64 %
\hatcurXAveccenxxxxxG
\else
??????\fi
\fi
\fi
\fi
\fi
\fi
\fi
}
\newcommand{\hatcurXdisteccen}[1]{\ifnum#1=58 %
\hatcurXdisteccenxxxxxA
\else
\ifnum#1=59 %
\hatcurXdisteccenxxxxxB
\else
\ifnum#1=60 %
\hatcurXdisteccenxxxxxC
\else
\ifnum#1=61 %
\hatcurXdisteccenxxxxxD
\else
\ifnum#1=62 %
\hatcurXdisteccenxxxxxE
\else
\ifnum#1=63 %
\hatcurXdisteccenxxxxxF
\else
\ifnum#1=64 %
\hatcurXdisteccenxxxxxG
\else
??????\fi
\fi
\fi
\fi
\fi
\fi
\fi
}
\newcommand{\hatcurXdistredeccen}[1]{\ifnum#1=58 %
\hatcurXdistredeccenxxxxxA
\else
\ifnum#1=59 %
\hatcurXdistredeccenxxxxxB
\else
\ifnum#1=60 %
\hatcurXdistredeccenxxxxxC
\else
\ifnum#1=61 %
\hatcurXdistredeccenxxxxxD
\else
\ifnum#1=62 %
\hatcurXdistredeccenxxxxxE
\else
\ifnum#1=63 %
\hatcurXdistredeccenxxxxxF
\else
\ifnum#1=64 %
\hatcurXdistredeccenxxxxxG
\else
??????\fi
\fi
\fi
\fi
\fi
\fi
\fi
}
\newcommand{\hatcurXEBVeccen}[1]{\ifnum#1=58 %
\hatcurXEBVeccenxxxxxA
\else
\ifnum#1=59 %
\hatcurXEBVeccenxxxxxB
\else
\ifnum#1=60 %
\hatcurXEBVeccenxxxxxC
\else
\ifnum#1=61 %
\hatcurXEBVeccenxxxxxD
\else
\ifnum#1=62 %
\hatcurXEBVeccenxxxxxE
\else
\ifnum#1=63 %
\hatcurXEBVeccenxxxxxF
\else
\ifnum#1=64 %
\hatcurXEBVeccenxxxxxG
\else
??????\fi
\fi
\fi
\fi
\fi
\fi
\fi
}
\newcommand{\hatcurXsecdureccen}[1]{\ifnum#1=58 %
\hatcurXsecdureccenxxxxxA
\else
\ifnum#1=59 %
\hatcurXsecdureccenxxxxxB
\else
\ifnum#1=60 %
\hatcurXsecdureccenxxxxxC
\else
\ifnum#1=61 %
\hatcurXsecdureccenxxxxxD
\else
\ifnum#1=62 %
\hatcurXsecdureccenxxxxxE
\else
\ifnum#1=63 %
\hatcurXsecdureccenxxxxxF
\else
\ifnum#1=64 %
\hatcurXsecdureccenxxxxxG
\else
??????\fi
\fi
\fi
\fi
\fi
\fi
\fi
}
\newcommand{\hatcurXsecingdureccen}[1]{\ifnum#1=58 %
\hatcurXsecingdureccenxxxxxA
\else
\ifnum#1=59 %
\hatcurXsecingdureccenxxxxxB
\else
\ifnum#1=60 %
\hatcurXsecingdureccenxxxxxC
\else
\ifnum#1=61 %
\hatcurXsecingdureccenxxxxxD
\else
\ifnum#1=62 %
\hatcurXsecingdureccenxxxxxE
\else
\ifnum#1=63 %
\hatcurXsecingdureccenxxxxxF
\else
\ifnum#1=64 %
\hatcurXsecingdureccenxxxxxG
\else
??????\fi
\fi
\fi
\fi
\fi
\fi
\fi
}
\newcommand{\hatcurXsecondaryeccen}[1]{\ifnum#1=58 %
\hatcurXsecondaryeccenxxxxxA
\else
\ifnum#1=59 %
\hatcurXsecondaryeccenxxxxxB
\else
\ifnum#1=60 %
\hatcurXsecondaryeccenxxxxxC
\else
\ifnum#1=61 %
\hatcurXsecondaryeccenxxxxxD
\else
\ifnum#1=62 %
\hatcurXsecondaryeccenxxxxxE
\else
\ifnum#1=63 %
\hatcurXsecondaryeccenxxxxxF
\else
\ifnum#1=64 %
\hatcurXsecondaryeccenxxxxxG
\else
??????\fi
\fi
\fi
\fi
\fi
\fi
\fi
}
\newcommand{\hatcurXsecphaseeccen}[1]{\ifnum#1=58 %
\hatcurXsecphaseeccenxxxxxA
\else
\ifnum#1=59 %
\hatcurXsecphaseeccenxxxxxB
\else
\ifnum#1=60 %
\hatcurXsecphaseeccenxxxxxC
\else
\ifnum#1=61 %
\hatcurXsecphaseeccenxxxxxD
\else
\ifnum#1=62 %
\hatcurXsecphaseeccenxxxxxE
\else
\ifnum#1=63 %
\hatcurXsecphaseeccenxxxxxF
\else
\ifnum#1=64 %
\hatcurXsecphaseeccenxxxxxG
\else
??????\fi
\fi
\fi
\fi
\fi
\fi
\fi
}

%% file: HTR062-011.eccen.tex
\newcommand{\hatcurhtreccenxxxxxA}{HTR062-011}                       
\newcommand{\hatcurfieldeccenxxxxxA}{\ensuremath{string}}            
\newcommand{\hatcurCCraeccenxxxxxA}{\ensuremath{04^{\mathrm h}35^{\mathrm m}23.1828{\mathrm s}}}                   
\newcommand{\hatcurCCdececcenxxxxxA}{\ensuremath{+56{\arcdeg}52{\arcmin}05.5848{\arcsec}}}                 
\newcommand{\hatcurCCmageccenxxxxxA}{12.971}                         
\newcommand{\hatcurCCtwomasseccenxxxxxA}{2MASS~04352318+5652055}     
\newcommand{\hatcurCCgsceccenxxxxxA}{GSC~3740-01482}                 
\newcommand{\hatcurCCgaiaeccenxxxxxA}{GAIA~277493610746748800}       
\newcommand{\hatcurCCgaiadrtwoeccenxxxxxA}{GAIA~DR2~277493615044741376} 
\newcommand{\hatcurCCtassmveccenxxxxxA}{\ensuremath{12.971\pm0.073}} 
\newcommand{\hatcurCCtassmvshorteccenxxxxxA}{\ensuremath{13.0}}      
\newcommand{\hatcurCCtassmBeccenxxxxxA}{\ensuremath{13.690\pm0.089}} 
\newcommand{\hatcurCCtassmBshorteccenxxxxxA}{\ensuremath{13.7}}      
\newcommand{\hatcurCCtassmIeccenxxxxxA}{\ensuremath{nff\pmnff}}      
\newcommand{\hatcurCCtassmIshorteccenxxxxxA}{\ensuremath{0.0}}       
\newcommand{\hatcurCCtassmgeccenxxxxxA}{\ensuremath{13.28\pm0.12}}   
\newcommand{\hatcurCCtassmgshorteccenxxxxxA}{\ensuremath{13.3}}      
\newcommand{\hatcurCCtassmreccenxxxxxA}{\ensuremath{12.74\pm0.12}}   
\newcommand{\hatcurCCtassmrshorteccenxxxxxA}{\ensuremath{12.7}}      
\newcommand{\hatcurCCtassmieccenxxxxxA}{\ensuremath{12.50\pm0.12}}   
\newcommand{\hatcurCCtassmishorteccenxxxxxA}{\ensuremath{12.5}}      
\newcommand{\hatcurCCparallaxeccenxxxxxA}{\ensuremath{1.912\pm0.047}} 
\newcommand{\hatcurCCgaiamGeccenxxxxxA}{\ensuremath{12.72020\pm0.00020}} 
\newcommand{\hatcurCCgaiamBPeccenxxxxxA}{\ensuremath{13.1422\pm0.0016}} 
\newcommand{\hatcurCCgaiamRPeccenxxxxxA}{\ensuremath{12.13470\pm0.00090}} 
\newcommand{\hatcurCCtwomassJmageccenxxxxxA}{\ensuremath{11.429\pm0.022}} 
\newcommand{\hatcurCCtwomassHmageccenxxxxxA}{\ensuremath{11.075\pm0.020}} 
\newcommand{\hatcurCCtwomassKmageccenxxxxxA}{\ensuremath{10.978\pm0.023}} 
\newcommand{\hatcurCCcitJmageccenxxxxxA}{\ensuremath{11.441\pm0.022}} 
\newcommand{\hatcurCCcitHmageccenxxxxxA}{\ensuremath{11.069\pm0.020}} 
\newcommand{\hatcurCCcitKmageccenxxxxxA}{\ensuremath{11.002\pm0.023}} 
\newcommand{\hatcurCCbbJmageccenxxxxxA}{\ensuremath{11.498\pm0.024}} 
\newcommand{\hatcurCCbbHmageccenxxxxxA}{\ensuremath{11.091\pm0.021}} 
\newcommand{\hatcurCCbbKmageccenxxxxxA}{\ensuremath{11.022\pm0.023}} 
\newcommand{\hatcurCCesoJmageccenxxxxxA}{\ensuremath{11.501\pm0.026}} 
\newcommand{\hatcurCCesoHmageccenxxxxxA}{\ensuremath{11.087\pm0.026}} 
\newcommand{\hatcurCCesoKmageccenxxxxxA}{\ensuremath{11.021\pm0.024}} 
\newcommand{\hatcurCCesoJHmageccenxxxxxA}{\ensuremath{0.413\pm0.034}} 
\newcommand{\hatcurCCesoJKmageccenxxxxxA}{\ensuremath{0.481\pm0.035}} 
\newcommand{\hatcurCCesoHKmageccenxxxxxA}{\ensuremath{0.067\pm0.035}} 
\newcommand{\hatcurCCWonemageccenxxxxxA}{\ensuremath{10.856\pm0.022}} 
\newcommand{\hatcurCCWtwomageccenxxxxxA}{\ensuremath{10.906\pm0.021}} 
\newcommand{\hatcurCCWthreemageccenxxxxxA}{\ensuremath{10.359\pm0.061}} 
\newcommand{\hatcurCCWfourmageccenxxxxxA}{\ensuremath{0\pm0}}        
\newcommand{\hatcurLCdipeccenxxxxxA}{\ensuremath{7.4}}               
\newcommand{\hatcurLCrprstareccenxxxxxA}{\ensuremath{0.0891\pm0.0018}} 
\newcommand{\hatcurLCbsqeccenxxxxxA}{\ensuremath{0.273_{-0.040}^{+0.041}}} 
\newcommand{\hatcurLCimpeccenxxxxxA}{\ensuremath{0.523_{-0.040}^{+0.037}}} 
\newcommand{\hatcurLCzetaeccenxxxxxA}{\ensuremath{13.02\pm0.10}}     
\newcommand{\hatcurLCdureccenxxxxxA}{\ensuremath{0.1723\pm0.0016}}   
\newcommand{\hatcurLCdurshorteccenxxxxxA}{\ensuremath{0.1723}}       
\newcommand{\hatcurLCdurhreccenxxxxxA}{\ensuremath{4.136\pm0.039}}   
\newcommand{\hatcurLCdurhrshorteccenxxxxxA}{\ensuremath{4.136}}      
\newcommand{\hatcurLCqeccenxxxxxA}{\ensuremath{0.04290\pm0.00041}}   
\newcommand{\hatcurLCqshorteccenxxxxxA}{\ensuremath{0.043}}          
\newcommand{\hatcurLCingdureccenxxxxxA}{\ensuremath{0.0190\pm0.0012}} 
\newcommand{\hatcurLCPeccenxxxxxA}{\ensuremath{4.0138372\pm0.0000021}} 
\newcommand{\hatcurLCPprececcenxxxxxA}{\ensuremath{4.0138372}}       
\newcommand{\hatcurLCPshorteccenxxxxxA}{\ensuremath{4.0138}}         
\newcommand{\hatcurLCTeccenxxxxxA}{\ensuremath{2457533.59819\pm0.00056}} 
\newcommand{\hatcurLCTAeccenxxxxxA}{\ensuremath{2456132.76900\pm0.00077}} 
\newcommand{\hatcurLCTBeccenxxxxxA}{\ensuremath{2458838.0952\pm0.0010}} 
\newcommand{\hatcurLChatnetmAeccenxxxxxA}{\ensuremath{12.85298\pm0.00011}} 
\newcommand{\hatcurLCiblendAeccenxxxxxA}{\ensuremath{1\pm0}}         
\newcommand{\hatcurLChatnetmBeccenxxxxxA}{\ensuremath{0.000050\pm0.000041}} 
\newcommand{\hatcurLCiblendBeccenxxxxxA}{\ensuremath{0.944\pm0.037}} 
\newcommand{\hatcurLCrhoeccenxxxxxA}{\ensuremath{0.410\pm0.031}}     
\newcommand{\hatcurSMEiteffeccenxxxxxA}{\ensuremath{5947\pm50}}      
\newcommand{\hatcurSMEizfeheccenxxxxxA}{\ensuremath{0.032\pm0.080}}  
\newcommand{\hatcurSMEizfehshorteccenxxxxxA}{\ensuremath{0.03}}      
\newcommand{\hatcurSMEiloggeccenxxxxxA}{\ensuremath{4.24\pm0.10}}    
\newcommand{\hatcurSMEivsineccenxxxxxA}{\ensuremath{4.81\pm0.50}}    
\newcommand{\hatcurSMEivmaceccenxxxxxA}{\ensuremath{nff\pmnff}}      
\newcommand{\hatcurSMEivmiceccenxxxxxA}{\ensuremath{nff\pmnff}}      
\newcommand{\hatcurSMEiiteffeccenxxxxxA}{\ensuremath{5931\pm50}}     
\newcommand{\hatcurSMEiizfeheccenxxxxxA}{\ensuremath{0.012\pm0.080}} 
\newcommand{\hatcurSMEiizfehshorteccenxxxxxA}{\ensuremath{0.012}}    
\newcommand{\hatcurSMEiiloggeccenxxxxxA}{\ensuremath{4.183\pm0.036}} 
\newcommand{\hatcurSMEiivsineccenxxxxxA}{\ensuremath{4.91\pm0.50}}   
\newcommand{\hatcurLBiBeccenxxxxxA}{\ensuremath{0.6762}}             
\newcommand{\hatcurLBiiBeccenxxxxxA}{\ensuremath{0.1438}}            
\newcommand{\hatcurLBiVeccenxxxxxA}{\ensuremath{0.5276}}             
\newcommand{\hatcurLBiiVeccenxxxxxA}{\ensuremath{0.1912}}            
\newcommand{\hatcurLBiReccenxxxxxA}{\ensuremath{0.4418}}             
\newcommand{\hatcurLBiiReccenxxxxxA}{\ensuremath{0.1930}}            
\newcommand{\hatcurLBiIeccenxxxxxA}{\ensuremath{0.3492}}             
\newcommand{\hatcurLBiiIeccenxxxxxA}{\ensuremath{0.1958}}            
\newcommand{\hatcurLBiueccenxxxxxA}{\ensuremath{0.8162}}             
\newcommand{\hatcurLBiiueccenxxxxxA}{\ensuremath{0.0195}}            
\newcommand{\hatcurLBigeccenxxxxxA}{\ensuremath{0.6254}}             
\newcommand{\hatcurLBiigeccenxxxxxA}{\ensuremath{0.1537}}            
\newcommand{\hatcurLBireccenxxxxxA}{\ensuremath{0.24\pm0.14}}        
\newcommand{\hatcurLBiireccenxxxxxA}{\ensuremath{0.27\pm0.18}}       
\newcommand{\hatcurLBiieccenxxxxxA}{\ensuremath{0.18\pm0.11}}        
\newcommand{\hatcurLBiiieccenxxxxxA}{\ensuremath{0.10\pm0.15}}       
\newcommand{\hatcurLBizeccenxxxxxA}{\ensuremath{0.3049}}             
\newcommand{\hatcurLBiizeccenxxxxxA}{\ensuremath{0.1982}}            
\newcommand{\hatcurLBiJeccenxxxxxA}{\ensuremath{0.2016}}             
\newcommand{\hatcurLBiiJeccenxxxxxA}{\ensuremath{0.2236}}            
\newcommand{\hatcurLBiHeccenxxxxxA}{\ensuremath{0.1199}}             
\newcommand{\hatcurLBiiHeccenxxxxxA}{\ensuremath{0.2524}}            
\newcommand{\hatcurLBiKeccenxxxxxA}{\ensuremath{0.1167}}             
\newcommand{\hatcurLBiiKeccenxxxxxA}{\ensuremath{0.1968}}            
\newcommand{\hatcurLBiTeccenxxxxxA}{\ensuremath{0.23\pm0.11}}        
\newcommand{\hatcurLBiiTeccenxxxxxA}{\ensuremath{0.26\pm0.15}}       
\newcommand{\hatcurLBikepeccenxxxxxA}{\ensuremath{0.4627}}           
\newcommand{\hatcurLBiikepeccenxxxxxA}{\ensuremath{0.1998}}          
\newcommand{\hatcurLBiCeccenxxxxxA}{\ensuremath{0.4546}}             
\newcommand{\hatcurLBiiCeccenxxxxxA}{\ensuremath{0.1978}}            
\newcommand{\hatcurLBiMeccenxxxxxA}{\ensuremath{0.5418}}             
\newcommand{\hatcurLBiiMeccenxxxxxA}{\ensuremath{0.1801}}            
\newcommand{\hatcurLBiSoneeccenxxxxxA}{\ensuremath{0.0933}}            
\newcommand{\hatcurLBiiSoneeccenxxxxxA}{\ensuremath{0.1187}}           
\newcommand{\hatcurLBiStwoeccenxxxxxA}{\ensuremath{0.0782}}            
\newcommand{\hatcurLBiiStwoeccenxxxxxA}{\ensuremath{0.1011}}           
\newcommand{\hatcurLBiSthreeeccenxxxxxA}{\ensuremath{0.0662}}            
\newcommand{\hatcurLBiiSthreeeccenxxxxxA}{\ensuremath{0.0830}}           
\newcommand{\hatcurLBiSfoureccenxxxxxA}{\ensuremath{0.0500}}            
\newcommand{\hatcurLBiiSfoureccenxxxxxA}{\ensuremath{0.0658}}           
\newcommand{\hatcurISOmeccenxxxxxA}{\ensuremath{1.035_{-0.018}^{+0.028}}} 
\newcommand{\hatcurISOmshorteccenxxxxxA}{\ensuremath{1.03}}          
\newcommand{\hatcurISOmlongeccenxxxxxA}{\ensuremath{1.035_{-0.018}^{+0.028}}} 
\newcommand{\hatcurISOreccenxxxxxA}{\ensuremath{1.526\pm0.038}}      
\newcommand{\hatcurISOrshorteccenxxxxxA}{\ensuremath{1.53}}          
\newcommand{\hatcurISOrlongeccenxxxxxA}{\ensuremath{1.526\pm0.038}}  
\newcommand{\hatcurISOrhoeccenxxxxxA}{\ensuremath{0.410\pm0.031}}    
\newcommand{\hatcurISOrholongeccenxxxxxA}{\ensuremath{0.410\pm0.031}} 
\newcommand{\hatcurISOloggeccenxxxxxA}{\ensuremath{4.086\pm0.023}}   
\newcommand{\hatcurISOlumeccenxxxxxA}{\ensuremath{2.83\pm0.17}}      
\newcommand{\hatcurISOlumshorteccenxxxxxA}{\ensuremath{2.83}}        
\newcommand{\hatcurISOteffeccenxxxxxA}{\ensuremath{6074\pm47}}       
\newcommand{\hatcurISOzfeheccenxxxxxA}{\ensuremath{-0.203\pm0.062}}  
\newcommand{\hatcurISOageeccenxxxxxA}{\ensuremath{7.10_{-0.72}^{+0.17}}} 
\newcommand{\hatcurISOspececcenxxxxxA}{G}                            
\newcommand{\hatcurRVKeccenxxxxxA}{\ensuremath{46.7\pm3.0}}          
\newcommand{\hatcurRVrkeccenxxxxxA}{\ensuremath{-0.07\pm0.12}}       
\newcommand{\hatcurRVrheccenxxxxxA}{\ensuremath{0.028\pm0.100}}      
\newcommand{\hatcurRVkeccenxxxxxA}{\ensuremath{-0.008_{-0.029}^{+0.016}}} 
\newcommand{\hatcurRVheccenxxxxxA}{\ensuremath{0.003_{-0.014}^{+0.020}}} 
\newcommand{\hatcurRVtroneeccenxxxxxA}{\ensuremath{0\pm0}}           
\newcommand{\hatcurRVtrtwoeccenxxxxxA}{\ensuremath{0\pm0}}           
\newcommand{\hatcurRVgammaeccenxxxxxA}{\ensuremath{2.5\pm2.8}}       
\newcommand{\hatcurRVjittereccenxxxxxA}{\ensuremath{4.9\pm3.6}}      
\newcommand{\hatcurRVjittertwosiglimeccenxxxxxA}{\ensuremath{<11.8}} 
\newcommand{\hatcurRVfitrmseccenxxxxxA}{\ensuremath{.1fym}}          %
\newcommand{\hatcurRVecceneccenxxxxxA}{\ensuremath{0.024\pm0.023}}   
\newcommand{\hatcurRVeccentwosiglimeccenxxxxxA}{\ensuremath{<0.073}} 
\newcommand{\hatcurRVomegaeccenxxxxxA}{\ensuremath{162\pm84}}        
\newcommand{\hatcurPPieccenxxxxxA}{\ensuremath{85.72\pm0.38}}        
\newcommand{\hatcurPPgeccenxxxxxA}{\ensuremath{5.29\pm0.50}}         
\newcommand{\hatcurPPloggeccenxxxxxA}{\ensuremath{2.724\pm0.041}}    
\newcommand{\hatcurPPareccenxxxxxA}{\ensuremath{7.04\pm0.18}}        
\newcommand{\hatcurPPareleccenxxxxxA}{\ensuremath{0.05000_{-0.00030}^{+0.00045}}} 
\newcommand{\hatcurPPrhoeccenxxxxxA}{\ensuremath{0.199\pm0.024}}     
\newcommand{\hatcurPPmeccenxxxxxA}{\ensuremath{0.374\pm0.025}}       
\newcommand{\hatcurPPmshorteccenxxxxxA}{\ensuremath{0.37}}           
\newcommand{\hatcurPPmlongeccenxxxxxA}{\ensuremath{0.374\pm0.025}}   
\newcommand{\hatcurPPmeeccenxxxxxA}{\ensuremath{119.0\pm7.9}}        
\newcommand{\hatcurPPmeshorteccenxxxxxA}{\ensuremath{119.0}}         
\newcommand{\hatcurPPmelongeccenxxxxxA}{\ensuremath{119.0\pm7.9}}    
\newcommand{\hatcurPPreccenxxxxxA}{\ensuremath{1.328\pm0.047}}       
\newcommand{\hatcurPPrshorteccenxxxxxA}{\ensuremath{1.33}}           
\newcommand{\hatcurPPrlongeccenxxxxxA}{\ensuremath{1.328\pm0.047}}   
\newcommand{\hatcurPPreeccenxxxxxA}{\ensuremath{14.88\pm0.52}}       
\newcommand{\hatcurPPreshorteccenxxxxxA}{\ensuremath{14.9}}          
\newcommand{\hatcurPPrelongeccenxxxxxA}{\ensuremath{14.88\pm0.52}}   
\newcommand{\hatcurPPmrcorreccenxxxxxA}{\ensuremath{0.04}}           
\newcommand{\hatcurPPteffeccenxxxxxA}{\ensuremath{1618\pm22}}        
\newcommand{\hatcurPPthetaeccenxxxxxA}{\ensuremath{0.0273\pm0.0020}} 
\newcommand{\hatcurPPfluxperieccenxxxxxA}{\ensuremath{1.63\pm0.13}}  
\newcommand{\hatcurPPfluxperidimeccenxxxxxA}{\ensuremath{9}}         
\newcommand{\hatcurPPfluxapeccenxxxxxA}{\ensuremath{1.463\pm0.094}}  
\newcommand{\hatcurPPfluxapdimeccenxxxxxA}{\ensuremath{9}}           
\newcommand{\hatcurPPfluxavgeccenxxxxxA}{\ensuremath{1.545\pm0.082}} 
\newcommand{\hatcurPPfluxavgdimeccenxxxxxA}{\ensuremath{9}}          
\newcommand{\hatcurPPfluxavglogeccenxxxxxA}{\ensuremath{9.189\pm0.023}} 
\newcommand{\hatcurXsecphaseeccenxxxxxA}{\ensuremath{0.495\pm0.018}} 
\newcommand{\hatcurXsecondaryeccenxxxxxA}{\ensuremath{2457535.585\pm0.071}} 
\newcommand{\hatcurXsecdureccenxxxxxA}{\ensuremath{0.1733\pm0.0045}} 
\newcommand{\hatcurXsecingdureccenxxxxxA}{\ensuremath{0.0192\pm0.0012}} 
\newcommand{\hatcurPPphiconjeccenxxxxxA}{\ensuremath{-0.12_{-0.18}^{+0.30}}} 
\newcommand{\hatcurPPperieccenxxxxxA}{\ensuremath{2457534.09\pm0.94}} 
\newcommand{\hatcurPPaequiveccenxxxxxA}{\ensuremath{0.02970\pm0.00079}} 
\newcommand{\hatcurPPtcirceccenxxxxxA}{\ensuremath{116_{-20}^{+26}}} 
\newcommand{\hatcurPPtinfalleccenxxxxxA}{\ensuremath{1830\pm270}}    
\newcommand{\hatcurXdisteccenxxxxxA}{\ensuremath{518\pm13}}          
\newcommand{\hatcurXAveccenxxxxxA}{\ensuremath{0.734\pm0.033}}       
\newcommand{\hatcurXdistredeccenxxxxxA}{\ensuremath{518\pm13}}       
\newcommand{\hatcurXEBVeccenxxxxxA}{\ensuremath{0.237\pm0.011}}      
\newcommand{\hatcurCCpmraeccenxxxxxA}{\ensuremath{-10.883\pm0.072}}  
\newcommand{\hatcurCCpmdececcenxxxxxA}{\ensuremath{11.862\pm0.064}}  
\newcommand{\hatcurCCpmeccenxxxxxA}{\ensuremath{16.098\pm0.096}}     

%% file: HTR080-003.eccen.tex
\newcommand{\hatcurhtreccenxxxxxB}{HTR080-003}                       
\newcommand{\hatcurfieldeccenxxxxxB}{\ensuremath{string}}            
\newcommand{\hatcurCCraeccenxxxxxB}{\ensuremath{19^{\mathrm h}29^{\mathrm m}50.0701{\mathrm s}}}                   
\newcommand{\hatcurCCdececcenxxxxxB}{\ensuremath{+62{\arcdeg}31{\arcmin}45.1751{\arcsec}}}                 
\newcommand{\hatcurCCmageccenxxxxxB}{11.883}                         
\newcommand{\hatcurCCtwomasseccenxxxxxB}{2MASS~19295008+6231452}     
\newcommand{\hatcurCCgsceccenxxxxxB}{GSC~4234-02195}                 
\newcommand{\hatcurCCgaiaeccenxxxxxB}{GAIA~2241743199301759488}      
\newcommand{\hatcurCCgaiadrtwoeccenxxxxxB}{GAIA~DR2~2241743203599727744} 
\newcommand{\hatcurCCtassmveccenxxxxxB}{\ensuremath{11.883\pm0.065}} 
\newcommand{\hatcurCCtassmvshorteccenxxxxxB}{\ensuremath{11.9}}      
\newcommand{\hatcurCCtassmBeccenxxxxxB}{\ensuremath{12.581\pm0.094}} 
\newcommand{\hatcurCCtassmBshorteccenxxxxxB}{\ensuremath{12.6}}      
\newcommand{\hatcurCCtassmIeccenxxxxxB}{\ensuremath{11.073\pm0.078}} 
\newcommand{\hatcurCCtassmIshorteccenxxxxxB}{\ensuremath{11.1}}      
\newcommand{\hatcurCCtassmgeccenxxxxxB}{\ensuremath{12.16\pm0.10}}   
\newcommand{\hatcurCCtassmgshorteccenxxxxxB}{\ensuremath{12.2}}      
\newcommand{\hatcurCCtassmreccenxxxxxB}{\ensuremath{11.650\pm0.050}} 
\newcommand{\hatcurCCtassmrshorteccenxxxxxB}{\ensuremath{11.7}}      
\newcommand{\hatcurCCtassmieccenxxxxxB}{\ensuremath{11.478\pm0.050}} 
\newcommand{\hatcurCCtassmishorteccenxxxxxB}{\ensuremath{11.5}}      
\newcommand{\hatcurCCparallaxeccenxxxxxB}{\ensuremath{3.738\pm0.019}} 
\newcommand{\hatcurCCgaiamGeccenxxxxxB}{\ensuremath{11.67870\pm0.00030}} 
\newcommand{\hatcurCCgaiamBPeccenxxxxxB}{\ensuremath{12.0587\pm0.0011}} 
\newcommand{\hatcurCCgaiamRPeccenxxxxxB}{\ensuremath{11.15850\pm0.00050}} 
\newcommand{\hatcurCCtwomassJmageccenxxxxxB}{\ensuremath{10.581\pm0.020}} 
\newcommand{\hatcurCCtwomassHmageccenxxxxxB}{\ensuremath{10.268\pm0.018}} 
\newcommand{\hatcurCCtwomassKmageccenxxxxxB}{\ensuremath{10.208\pm0.022}} 
\newcommand{\hatcurCCcitJmageccenxxxxxB}{\ensuremath{10.598\pm0.020}} 
\newcommand{\hatcurCCcitHmageccenxxxxxB}{\ensuremath{10.263\pm0.019}} 
\newcommand{\hatcurCCcitKmageccenxxxxxB}{\ensuremath{10.232\pm0.022}} 
\newcommand{\hatcurCCbbJmageccenxxxxxB}{\ensuremath{10.647\pm0.022}} 
\newcommand{\hatcurCCbbHmageccenxxxxxB}{\ensuremath{10.284\pm0.019}} 
\newcommand{\hatcurCCbbKmageccenxxxxxB}{\ensuremath{10.252\pm0.022}} 
\newcommand{\hatcurCCesoJmageccenxxxxxB}{\ensuremath{10.649\pm0.024}} 
\newcommand{\hatcurCCesoHmageccenxxxxxB}{\ensuremath{10.278\pm0.022}} 
\newcommand{\hatcurCCesoKmageccenxxxxxB}{\ensuremath{10.251\pm0.023}} 
\newcommand{\hatcurCCesoJHmageccenxxxxxB}{\ensuremath{0.370\pm0.030}} 
\newcommand{\hatcurCCesoJKmageccenxxxxxB}{\ensuremath{0.399\pm0.032}} 
\newcommand{\hatcurCCesoHKmageccenxxxxxB}{\ensuremath{0.028\pm0.032}} 
\newcommand{\hatcurCCWonemageccenxxxxxB}{\ensuremath{10.144\pm0.023}} 
\newcommand{\hatcurCCWtwomageccenxxxxxB}{\ensuremath{10.211\pm0.020}} 
\newcommand{\hatcurCCWthreemageccenxxxxxB}{\ensuremath{10.191\pm0.037}} 
\newcommand{\hatcurCCWfourmageccenxxxxxB}{\ensuremath{0\pm0}}        
\newcommand{\hatcurLCdipeccenxxxxxB}{\ensuremath{12.8}}              
\newcommand{\hatcurLCrprstareccenxxxxxB}{\ensuremath{0.1045\pm0.0011}} 
\newcommand{\hatcurLCbsqeccenxxxxxB}{\ensuremath{0.687_{-0.017}^{+0.016}}} 
\newcommand{\hatcurLCimpeccenxxxxxB}{\ensuremath{0.8291_{-0.0101}^{+0.0096}}} 
\newcommand{\hatcurLCzetaeccenxxxxxB}{\ensuremath{26.78\pm0.44}}     
\newcommand{\hatcurLCdureccenxxxxxB}{\ensuremath{0.0975\pm0.0010}}   
\newcommand{\hatcurLCdurshorteccenxxxxxB}{\ensuremath{0.0975}}       
\newcommand{\hatcurLCdurhreccenxxxxxB}{\ensuremath{2.340\pm0.025}}   
\newcommand{\hatcurLCdurhrshorteccenxxxxxB}{\ensuremath{2.340}}      
\newcommand{\hatcurLCqeccenxxxxxB}{\ensuremath{0.02350\pm0.00025}}   
\newcommand{\hatcurLCqshorteccenxxxxxB}{\ensuremath{0.024}}          
\newcommand{\hatcurLCingdureccenxxxxxB}{\ensuremath{0.0261\pm0.0011}} 
\newcommand{\hatcurLCPeccenxxxxxB}{\ensuremath{4.1419769\pm0.0000010}} 
\newcommand{\hatcurLCPprececcenxxxxxB}{\ensuremath{4.1419769}}       
\newcommand{\hatcurLCPshorteccenxxxxxB}{\ensuremath{4.1420}}         
\newcommand{\hatcurLCTeccenxxxxxB}{\ensuremath{2458618.54086\pm0.00021}} 
\newcommand{\hatcurLCTAeccenxxxxxB}{\ensuremath{2456129.21270\pm0.00061}} 
\newcommand{\hatcurLCTBeccenxxxxxB}{\ensuremath{2458896.05332\pm0.00023}} 
\newcommand{\hatcurLChatnetmAeccenxxxxxB}{\ensuremath{11.727050\pm0.000095}} 
\newcommand{\hatcurLCiblendAeccenxxxxxB}{\ensuremath{1\pm0}}         
\newcommand{\hatcurLChatnetmBeccenxxxxxB}{\ensuremath{0.000020\pm0.000023}} 
\newcommand{\hatcurLCiblendBeccenxxxxxB}{\ensuremath{0.9978\pm0.0031}} 
\newcommand{\hatcurLChatnetmCeccenxxxxxB}{\ensuremath{0.000030\pm0.000032}} 
\newcommand{\hatcurLCiblendCeccenxxxxxB}{\ensuremath{0.9988\pm0.0018}} 
\newcommand{\hatcurLChatnetmDeccenxxxxxB}{\ensuremath{0.000020\pm0.000024}} 
\newcommand{\hatcurLCiblendDeccenxxxxxB}{\ensuremath{0.9955\pm0.0046}} 
\newcommand{\hatcurLChatnetmEeccenxxxxxB}{\ensuremath{0.000030\pm0.000027}} 
\newcommand{\hatcurLCiblendEeccenxxxxxB}{\ensuremath{0.9961\pm0.0043}} 
\newcommand{\hatcurLChatnetmFeccenxxxxxB}{\ensuremath{-0.000010\pm0.000024}} 
\newcommand{\hatcurLCiblendFeccenxxxxxB}{\ensuremath{0.9949\pm0.0086}} 
\newcommand{\hatcurLChatnetmGeccenxxxxxB}{\ensuremath{-0.000020\pm0.000022}} 
\newcommand{\hatcurLCiblendGeccenxxxxxB}{\ensuremath{0.99946\pm0.00092}} 
\newcommand{\hatcurLChatnetmHeccenxxxxxB}{\ensuremath{-0.000010\pm0.000023}} 
\newcommand{\hatcurLCiblendHeccenxxxxxB}{\ensuremath{0.9963\pm0.0052}} 
\newcommand{\hatcurLCrhoeccenxxxxxB}{\ensuremath{1.054_{-0.032}^{+0.047}}} 
\newcommand{\hatcurSMEiteffeccenxxxxxB}{\ensuremath{5678\pm50}}      
\newcommand{\hatcurSMEizfeheccenxxxxxB}{\ensuremath{0.420\pm0.080}}  
\newcommand{\hatcurSMEizfehshorteccenxxxxxB}{\ensuremath{0.42}}      
\newcommand{\hatcurSMEiloggeccenxxxxxB}{\ensuremath{4.40\pm0.10}}    
\newcommand{\hatcurSMEivsineccenxxxxxB}{\ensuremath{3.05\pm0.50}}    
\newcommand{\hatcurSMEivmaceccenxxxxxB}{\ensuremath{nff\pmnff}}      
\newcommand{\hatcurSMEivmiceccenxxxxxB}{\ensuremath{nff\pmnff}}      
\newcommand{\hatcurSMEiiteffeccenxxxxxB}{\ensuremath{5665\pm50}}     
\newcommand{\hatcurSMEiizfeheccenxxxxxB}{\ensuremath{0.409\pm0.080}} 
\newcommand{\hatcurSMEiizfehshorteccenxxxxxB}{\ensuremath{0.409}}    
\newcommand{\hatcurSMEiiloggeccenxxxxxB}{\ensuremath{4.368\pm0.033}} 
\newcommand{\hatcurSMEiivsineccenxxxxxB}{\ensuremath{3.04\pm0.50}}   
\newcommand{\hatcurLBiBeccenxxxxxB}{\ensuremath{0.7351}}             
\newcommand{\hatcurLBiiBeccenxxxxxB}{\ensuremath{0.0925}}            
\newcommand{\hatcurLBiVeccenxxxxxB}{\ensuremath{0.5691}}             
\newcommand{\hatcurLBiiVeccenxxxxxB}{\ensuremath{0.1652}}            
\newcommand{\hatcurLBiReccenxxxxxB}{\ensuremath{0.4701}}             
\newcommand{\hatcurLBiiReccenxxxxxB}{\ensuremath{0.1808}}            
\newcommand{\hatcurLBiIeccenxxxxxB}{\ensuremath{0.3726}}             
\newcommand{\hatcurLBiiIeccenxxxxxB}{\ensuremath{0.1906}}            
\newcommand{\hatcurLBiueccenxxxxxB}{\ensuremath{0.9092}}             
\newcommand{\hatcurLBiiueccenxxxxxB}{\ensuremath{-0.0704}}           
\newcommand{\hatcurLBigeccenxxxxxB}{\ensuremath{0.6783}}             
\newcommand{\hatcurLBiigeccenxxxxxB}{\ensuremath{0.1108}}            
\newcommand{\hatcurLBireccenxxxxxB}{\ensuremath{0.44\pm0.16}}        
\newcommand{\hatcurLBiireccenxxxxxB}{\ensuremath{0.34\pm0.15}}       
\newcommand{\hatcurLBiieccenxxxxxB}{\ensuremath{0.33\pm0.14}}        
\newcommand{\hatcurLBiiieccenxxxxxB}{\ensuremath{0.21\pm0.17}}       
\newcommand{\hatcurLBizeccenxxxxxB}{\ensuremath{0.3274}}             
\newcommand{\hatcurLBiizeccenxxxxxB}{\ensuremath{0.1939}}            
\newcommand{\hatcurLBiJeccenxxxxxB}{\ensuremath{0.2154}}             
\newcommand{\hatcurLBiiJeccenxxxxxB}{\ensuremath{0.2260}}            
\newcommand{\hatcurLBiHeccenxxxxxB}{\ensuremath{0.1252}}             
\newcommand{\hatcurLBiiHeccenxxxxxB}{\ensuremath{0.2621}}            
\newcommand{\hatcurLBiKeccenxxxxxB}{\ensuremath{0.1182}}             
\newcommand{\hatcurLBiiKeccenxxxxxB}{\ensuremath{0.2081}}            
\newcommand{\hatcurLBiTeccenxxxxxB}{\ensuremath{0.17\pm0.11}}        
\newcommand{\hatcurLBiiTeccenxxxxxB}{\ensuremath{0.27\pm0.15}}       
\newcommand{\hatcurLBikepeccenxxxxxB}{\ensuremath{0.4898}}           
\newcommand{\hatcurLBiikepeccenxxxxxB}{\ensuremath{0.1829}}          
\newcommand{\hatcurLBiCeccenxxxxxB}{\ensuremath{0.4801}}             
\newcommand{\hatcurLBiiCeccenxxxxxB}{\ensuremath{0.1811}}            
\newcommand{\hatcurLBiMeccenxxxxxB}{\ensuremath{0.5749}}             
\newcommand{\hatcurLBiiMeccenxxxxxB}{\ensuremath{0.1555}}            
\newcommand{\hatcurLBiSoneeccenxxxxxB}{\ensuremath{0.0981}}            
\newcommand{\hatcurLBiiSoneeccenxxxxxB}{\ensuremath{0.1252}}           
\newcommand{\hatcurLBiStwoeccenxxxxxB}{\ensuremath{0.0828}}            
\newcommand{\hatcurLBiiStwoeccenxxxxxB}{\ensuremath{0.1060}}           
\newcommand{\hatcurLBiSthreeeccenxxxxxB}{\ensuremath{0.0697}}            
\newcommand{\hatcurLBiiSthreeeccenxxxxxB}{\ensuremath{0.0866}}           
\newcommand{\hatcurLBiSfoureccenxxxxxB}{\ensuremath{0.0541}}            
\newcommand{\hatcurLBiiSfoureccenxxxxxB}{\ensuremath{0.0685}}           
\newcommand{\hatcurISOmeccenxxxxxB}{\ensuremath{1.008\pm0.022}}      
\newcommand{\hatcurISOmshorteccenxxxxxB}{\ensuremath{1.01}}          
\newcommand{\hatcurISOmlongeccenxxxxxB}{\ensuremath{1.008\pm0.022}}  
\newcommand{\hatcurISOreccenxxxxxB}{\ensuremath{1.1034\pm0.0077}}    
\newcommand{\hatcurISOrshorteccenxxxxxB}{\ensuremath{1.10}}          
\newcommand{\hatcurISOrlongeccenxxxxxB}{\ensuremath{1.1034\pm0.0077}} 
\newcommand{\hatcurISOrhoeccenxxxxxB}{\ensuremath{1.054_{-0.032}^{+0.047}}} 
\newcommand{\hatcurISOrholongeccenxxxxxB}{\ensuremath{1.054_{-0.032}^{+0.047}}} 
\newcommand{\hatcurISOloggeccenxxxxxB}{\ensuremath{4.355\pm0.014}}   
\newcommand{\hatcurISOlumeccenxxxxxB}{\ensuremath{1.130\pm0.016}}    
\newcommand{\hatcurISOlumshorteccenxxxxxB}{\ensuremath{1.13}}        
\newcommand{\hatcurISOteffeccenxxxxxB}{\ensuremath{5676\pm17}}       
\newcommand{\hatcurISOzfeheccenxxxxxB}{\ensuremath{0.221\pm0.051}}   
\newcommand{\hatcurISOageeccenxxxxxB}{\ensuremath{7.36_{-1.25}^{+0.92}}} 
\newcommand{\hatcurISOspececcenxxxxxB}{G}                            
\newcommand{\hatcurRVKeccenxxxxxB}{\ensuremath{192.0\pm8.6}}         
\newcommand{\hatcurRVrkeccenxxxxxB}{\ensuremath{0.042\pm0.070}}      
\newcommand{\hatcurRVrheccenxxxxxB}{\ensuremath{0.008\pm0.081}}      
\newcommand{\hatcurRVkeccenxxxxxB}{\ensuremath{0.0037_{-0.0069}^{+0.0102}}} 
\newcommand{\hatcurRVheccenxxxxxB}{\ensuremath{0.000\pm0.011}}       
\newcommand{\hatcurRVtroneeccenxxxxxB}{\ensuremath{0\pm0}}           
\newcommand{\hatcurRVtrtwoeccenxxxxxB}{\ensuremath{0\pm0}}           
\newcommand{\hatcurRVgammaAeccenxxxxxB}{\ensuremath{-151.1\pm9.0}}   
\newcommand{\hatcurRVjitterAeccenxxxxxB}{\ensuremath{9\pm14}}        
\newcommand{\hatcurRVjittertwosiglimAeccenxxxxxB}{\ensuremath{<36.7}} 
\newcommand{\hatcurRVfitrmsAeccenxxxxxB}{\ensuremath{0.0}}           
\newcommand{\hatcurRVgammaBeccenxxxxxB}{\ensuremath{-21157.3\pm4.7}} 
\newcommand{\hatcurRVjitterBeccenxxxxxB}{\ensuremath{0.1\pm7.0}}     
\newcommand{\hatcurRVjittertwosiglimBeccenxxxxxB}{\ensuremath{<12.7}} 
\newcommand{\hatcurRVfitrmsBeccenxxxxxB}{\ensuremath{0.0}}           
\newcommand{\hatcurRVecceneccenxxxxxB}{\ensuremath{0.0100\pm0.0094}} 
\newcommand{\hatcurRVeccentwosiglimeccenxxxxxB}{\ensuremath{<0.030}} 
\newcommand{\hatcurRVomegaeccenxxxxxB}{\ensuremath{160\pm120}}       
\newcommand{\hatcurPPieccenxxxxxB}{\ensuremath{85.17\pm0.11}}        
\newcommand{\hatcurPPgeccenxxxxxB}{\ensuremath{30.3\pm1.7}}          
\newcommand{\hatcurPPloggeccenxxxxxB}{\ensuremath{3.481\pm0.025}}    
\newcommand{\hatcurPPareccenxxxxxB}{\ensuremath{9.855_{-0.099}^{+0.146}}} 
\newcommand{\hatcurPPareleccenxxxxxB}{\ensuremath{0.05064\pm0.00037}} 
\newcommand{\hatcurPPrhoeccenxxxxxB}{\ensuremath{1.351\pm0.087}}     
\newcommand{\hatcurPPmeccenxxxxxB}{\ensuremath{1.530\pm0.072}}       
\newcommand{\hatcurPPmshorteccenxxxxxB}{\ensuremath{1.53}}           
\newcommand{\hatcurPPmlongeccenxxxxxB}{\ensuremath{1.530\pm0.072}}   
\newcommand{\hatcurPPmeeccenxxxxxB}{\ensuremath{486\pm23}}           
\newcommand{\hatcurPPmeshorteccenxxxxxB}{\ensuremath{486.4}}         
\newcommand{\hatcurPPmelongeccenxxxxxB}{\ensuremath{486\pm23}}       
\newcommand{\hatcurPPreccenxxxxxB}{\ensuremath{1.121\pm0.012}}       
\newcommand{\hatcurPPrshorteccenxxxxxB}{\ensuremath{1.12}}           
\newcommand{\hatcurPPrlongeccenxxxxxB}{\ensuremath{1.121\pm0.012}}   
\newcommand{\hatcurPPreeccenxxxxxB}{\ensuremath{12.56\pm0.14}}       
\newcommand{\hatcurPPreshorteccenxxxxxB}{\ensuremath{12.6}}          
\newcommand{\hatcurPPrelongeccenxxxxxB}{\ensuremath{12.56\pm0.14}}   
\newcommand{\hatcurPPmrcorreccenxxxxxB}{\ensuremath{-0.25}}          
\newcommand{\hatcurPPteffeccenxxxxxB}{\ensuremath{1277.9\pm6.6}}     
\newcommand{\hatcurPPthetaeccenxxxxxB}{\ensuremath{0.1363\pm0.0065}} 
\newcommand{\hatcurPPfluxperieccenxxxxxB}{\ensuremath{6.15\pm0.17}}  
\newcommand{\hatcurPPfluxperidimeccenxxxxxB}{\ensuremath{8}}         
\newcommand{\hatcurPPfluxapeccenxxxxxB}{\ensuremath{5.86\pm0.16}}    
\newcommand{\hatcurPPfluxapdimeccenxxxxxB}{\ensuremath{8}}           
\newcommand{\hatcurPPfluxavgeccenxxxxxB}{\ensuremath{6.01\pm0.12}}   
\newcommand{\hatcurPPfluxavgdimeccenxxxxxB}{\ensuremath{8}}          
\newcommand{\hatcurPPfluxavglogeccenxxxxxB}{\ensuremath{8.7789\pm0.0090}} 
\newcommand{\hatcurXsecphaseeccenxxxxxB}{\ensuremath{0.5023\pm0.0059}} 
\newcommand{\hatcurXsecondaryeccenxxxxxB}{\ensuremath{2458620.621\pm0.025}} 
\newcommand{\hatcurXsecdureccenxxxxxB}{\ensuremath{0.0974\pm0.0013}} 
\newcommand{\hatcurXsecingdureccenxxxxxB}{\ensuremath{0.0261\pm0.0022}} 
\newcommand{\hatcurPPphiconjeccenxxxxxB}{\ensuremath{0.12_{-0.33}^{+0.23}}} 
\newcommand{\hatcurPPperieccenxxxxxB}{\ensuremath{2458618.1\pm1.1}}  
\newcommand{\hatcurPPaequiveccenxxxxxB}{\ensuremath{0.04760\pm0.00049}} 
\newcommand{\hatcurPPtcirceccenxxxxxB}{\ensuremath{1250\pm110}}      
\newcommand{\hatcurPPtinfalleccenxxxxxB}{\ensuremath{2420\pm200}}    
\newcommand{\hatcurXdisteccenxxxxxB}{\ensuremath{267.0\pm1.4}}       
\newcommand{\hatcurXAveccenxxxxxB}{\ensuremath{0.050\pm0.012}}       
\newcommand{\hatcurXdistredeccenxxxxxB}{\ensuremath{267.0\pm1.4}}    
\newcommand{\hatcurXEBVeccenxxxxxB}{\ensuremath{0.0160\pm0.0039}}    
\newcommand{\hatcurCCpmraeccenxxxxxB}{\ensuremath{20.957\pm0.046}}   
\newcommand{\hatcurCCpmdececcenxxxxxB}{\ensuremath{-6.056\pm0.043}}  
\newcommand{\hatcurCCpmeccenxxxxxB}{\ensuremath{21.814\pm0.063}}     

%% file: HTR089-018.eccen.tex
\newcommand{\hatcurhtreccenxxxxxC}{HTR089-018}                       
\newcommand{\hatcurfieldeccenxxxxxC}{\ensuremath{string}}            
\newcommand{\hatcurCCraeccenxxxxxC}{\ensuremath{01^{\mathrm h}53^{\mathrm m}07.7727{\mathrm s}}}                   
\newcommand{\hatcurCCdececcenxxxxxC}{\ensuremath{+52{\arcdeg}03{\arcmin}14.01977{\arcsec}}}                
\newcommand{\hatcurCCmageccenxxxxxC}{9.710}                          
\newcommand{\hatcurCCtwomasseccenxxxxxC}{2MASS~01530777+5203140}     
\newcommand{\hatcurCCgsceccenxxxxxC}{GSC~3292-01330}                 
\newcommand{\hatcurCCgaiaeccenxxxxxC}{GAIA~359678187913760384}       
\newcommand{\hatcurCCgaiadrtwoeccenxxxxxC}{GAIA~DR2~359678187913760384} 
\newcommand{\hatcurCCtassmveccenxxxxxC}{\ensuremath{9.710\pm0.050}}  
\newcommand{\hatcurCCtassmvshorteccenxxxxxC}{\ensuremath{9.7}}       
\newcommand{\hatcurCCtassmBeccenxxxxxC}{\ensuremath{10.230\pm0.040}} 
\newcommand{\hatcurCCtassmBshorteccenxxxxxC}{\ensuremath{10.2}}      
\newcommand{\hatcurCCtassmIeccenxxxxxC}{\ensuremath{9.077\pm0.042}}  
\newcommand{\hatcurCCtassmIshorteccenxxxxxC}{\ensuremath{9.1}}       
\newcommand{\hatcurCCtassmgeccenxxxxxC}{\ensuremath{nff\pmnff}}      
\newcommand{\hatcurCCtassmgshorteccenxxxxxC}{\ensuremath{0.0}}       
\newcommand{\hatcurCCtassmreccenxxxxxC}{\ensuremath{nff\pmnff}}      
\newcommand{\hatcurCCtassmrshorteccenxxxxxC}{\ensuremath{0.0}}       
\newcommand{\hatcurCCtassmieccenxxxxxC}{\ensuremath{9.421\pm0.040}}  
\newcommand{\hatcurCCtassmishorteccenxxxxxC}{\ensuremath{9.4}}       
\newcommand{\hatcurCCparallaxeccenxxxxxC}{\ensuremath{4.260\pm0.049}} 
\newcommand{\hatcurCCgaiamGeccenxxxxxC}{\ensuremath{9.56360\pm0.00030}} 
\newcommand{\hatcurCCgaiamBPeccenxxxxxC}{\ensuremath{9.8631\pm0.0012}} 
\newcommand{\hatcurCCgaiamRPeccenxxxxxC}{\ensuremath{9.1320\pm0.0013}} 
\newcommand{\hatcurCCtwomassJmageccenxxxxxC}{\ensuremath{8.677\pm0.052}} 
\newcommand{\hatcurCCtwomassHmageccenxxxxxC}{\ensuremath{8.396\pm0.029}} 
\newcommand{\hatcurCCtwomassKmageccenxxxxxC}{\ensuremath{8.368\pm0.031}} 
\newcommand{\hatcurCCcitJmageccenxxxxxC}{\ensuremath{8.697\pm0.050}} 
\newcommand{\hatcurCCcitHmageccenxxxxxC}{\ensuremath{8.392\pm0.029}} 
\newcommand{\hatcurCCcitKmageccenxxxxxC}{\ensuremath{8.392\pm0.031}} 
\newcommand{\hatcurCCbbJmageccenxxxxxC}{\ensuremath{8.742\pm0.055}}  
\newcommand{\hatcurCCbbHmageccenxxxxxC}{\ensuremath{8.412\pm0.030}}  
\newcommand{\hatcurCCbbKmageccenxxxxxC}{\ensuremath{8.412\pm0.031}}  
\newcommand{\hatcurCCesoJmageccenxxxxxC}{\ensuremath{8.742\pm0.055}} 
\newcommand{\hatcurCCesoHmageccenxxxxxC}{\ensuremath{8.406\pm0.033}} 
\newcommand{\hatcurCCesoKmageccenxxxxxC}{\ensuremath{8.411\pm0.032}} 
\newcommand{\hatcurCCesoJHmageccenxxxxxC}{\ensuremath{0.337\pm0.063}} 
\newcommand{\hatcurCCesoJKmageccenxxxxxC}{\ensuremath{0.332\pm0.064}} 
\newcommand{\hatcurCCesoHKmageccenxxxxxC}{\ensuremath{-0.005\pm0.047}} 
\newcommand{\hatcurCCWonemageccenxxxxxC}{\ensuremath{8.308\pm0.026}} 
\newcommand{\hatcurCCWtwomageccenxxxxxC}{\ensuremath{8.320\pm0.026}} 
\newcommand{\hatcurCCWthreemageccenxxxxxC}{\ensuremath{8.260\pm0.028}} 
\newcommand{\hatcurCCWfourmageccenxxxxxC}{\ensuremath{0\pm0}}        
\newcommand{\hatcurLCdipeccenxxxxxC}{\ensuremath{5.8}}               
\newcommand{\hatcurLCrprstareccenxxxxxC}{\ensuremath{0.0752\pm0.0014}} 
\newcommand{\hatcurLCbsqeccenxxxxxC}{\ensuremath{0.357_{-0.133}^{+0.068}}} 
\newcommand{\hatcurLCimpeccenxxxxxC}{\ensuremath{0.597_{-0.124}^{+0.054}}} 
\newcommand{\hatcurLCzetaeccenxxxxxC}{\ensuremath{10.76\pm0.11}}     
\newcommand{\hatcurLCdureccenxxxxxC}{\ensuremath{0.2071\pm0.0028}}   
\newcommand{\hatcurLCdurshorteccenxxxxxC}{\ensuremath{0.2071}}       
\newcommand{\hatcurLCdurhreccenxxxxxC}{\ensuremath{4.970\pm0.067}}   
\newcommand{\hatcurLCdurhrshorteccenxxxxxC}{\ensuremath{4.970}}      
\newcommand{\hatcurLCqeccenxxxxxC}{\ensuremath{0.04320\pm0.00059}}   
\newcommand{\hatcurLCqshorteccenxxxxxC}{\ensuremath{0.043}}          
\newcommand{\hatcurLCingdureccenxxxxxC}{\ensuremath{0.0218\pm0.0033}} 
\newcommand{\hatcurLCPeccenxxxxxC}{\ensuremath{4.7947817\pm0.0000024}} 
\newcommand{\hatcurLCPprececcenxxxxxC}{\ensuremath{4.7947817}}       
\newcommand{\hatcurLCPshorteccenxxxxxC}{\ensuremath{4.7948}}         
\newcommand{\hatcurLCTeccenxxxxxC}{\ensuremath{2458365.73565\pm0.00049}} 
\newcommand{\hatcurLCTAeccenxxxxxC}{\ensuremath{2454764.8546\pm0.0017}} 
\newcommand{\hatcurLCTBeccenxxxxxC}{\ensuremath{2458811.65038\pm0.00059}} 
\newcommand{\hatcurLChatnetmAeccenxxxxxC}{\ensuremath{9.745730\pm0.000052}} 
\newcommand{\hatcurLCiblendAeccenxxxxxC}{\ensuremath{1\pm0}}         
\newcommand{\hatcurLChatnetmBeccenxxxxxC}{\ensuremath{0.000040\pm0.000012}} 
\newcommand{\hatcurLCiblendBeccenxxxxxC}{\ensuremath{0.9947\pm0.0047}} 
\newcommand{\hatcurLCrhoeccenxxxxxC}{\ensuremath{0.1854\pm0.0093}}   
\newcommand{\hatcurSMEiteffeccenxxxxxC}{\ensuremath{6223\pm50}}      
\newcommand{\hatcurSMEizfeheccenxxxxxC}{\ensuremath{-0.404\pm0.080}} 
\newcommand{\hatcurSMEizfehshorteccenxxxxxC}{\ensuremath{-0.40}}     
\newcommand{\hatcurSMEiloggeccenxxxxxC}{\ensuremath{3.59\pm0.10}}    
\newcommand{\hatcurSMEivsineccenxxxxxC}{\ensuremath{10.71\pm0.50}}   
\newcommand{\hatcurSMEivmaceccenxxxxxC}{\ensuremath{nff\pmnff}}      
\newcommand{\hatcurSMEivmiceccenxxxxxC}{\ensuremath{nff\pmnff}}      
\newcommand{\hatcurSMEiiteffeccenxxxxxC}{\ensuremath{6462\pm50}}     
\newcommand{\hatcurSMEiizfeheccenxxxxxC}{\ensuremath{-0.237\pm0.080}} 
\newcommand{\hatcurSMEiizfehshorteccenxxxxxC}{\ensuremath{-0.237}}   
\newcommand{\hatcurSMEiiloggeccenxxxxxC}{\ensuremath{4.052\pm0.049}} 
\newcommand{\hatcurSMEiivsineccenxxxxxC}{\ensuremath{10.42\pm0.50}}  
\newcommand{\hatcurLBiBeccenxxxxxC}{\ensuremath{0.5594}}             
\newcommand{\hatcurLBiiBeccenxxxxxC}{\ensuremath{0.2417}}            
\newcommand{\hatcurLBiVeccenxxxxxC}{\ensuremath{0.4555}}             
\newcommand{\hatcurLBiiVeccenxxxxxC}{\ensuremath{0.2362}}            
\newcommand{\hatcurLBiReccenxxxxxC}{\ensuremath{0.3896}}             
\newcommand{\hatcurLBiiReccenxxxxxC}{\ensuremath{0.2179}}            
\newcommand{\hatcurLBiIeccenxxxxxC}{\ensuremath{0.3076}}             
\newcommand{\hatcurLBiiIeccenxxxxxC}{\ensuremath{0.2060}}            
\newcommand{\hatcurLBiueccenxxxxxC}{\ensuremath{0.6429}}             
\newcommand{\hatcurLBiiueccenxxxxxC}{\ensuremath{0.1541}}            
\newcommand{\hatcurLBigeccenxxxxxC}{\ensuremath{0.5364}}             
\newcommand{\hatcurLBiigeccenxxxxxC}{\ensuremath{0.2206}}            
\newcommand{\hatcurLBireccenxxxxxC}{\ensuremath{0.46\pm0.16}}        
\newcommand{\hatcurLBiireccenxxxxxC}{\ensuremath{0.22\pm0.16}}       
\newcommand{\hatcurLBiieccenxxxxxC}{\ensuremath{0.18\pm0.11}}        
\newcommand{\hatcurLBiiieccenxxxxxC}{\ensuremath{0.17\pm0.15}}       
\newcommand{\hatcurLBizeccenxxxxxC}{\ensuremath{0.103_{-0.074}^{+0.134}}} 
\newcommand{\hatcurLBiizeccenxxxxxC}{\ensuremath{0.20\pm0.14}}       
\newcommand{\hatcurLBiJeccenxxxxxC}{\ensuremath{0.1789}}             
\newcommand{\hatcurLBiiJeccenxxxxxC}{\ensuremath{0.2175}}            
\newcommand{\hatcurLBiHeccenxxxxxC}{\ensuremath{0.1144}}             
\newcommand{\hatcurLBiiHeccenxxxxxC}{\ensuremath{0.2251}}            
\newcommand{\hatcurLBiKeccenxxxxxC}{\ensuremath{0.1120}}             
\newcommand{\hatcurLBiiKeccenxxxxxC}{\ensuremath{0.1747}}            
\newcommand{\hatcurLBiTeccenxxxxxC}{\ensuremath{0.25\pm0.12}}        
\newcommand{\hatcurLBiiTeccenxxxxxC}{\ensuremath{0.21\pm0.16}}       
\newcommand{\hatcurLBikepeccenxxxxxC}{\ensuremath{0.4383}}           
\newcommand{\hatcurLBiikepeccenxxxxxC}{\ensuremath{0.2054}}          
\newcommand{\hatcurLBiCeccenxxxxxC}{\ensuremath{0.4303}}             
\newcommand{\hatcurLBiiCeccenxxxxxC}{\ensuremath{0.2057}}            
\newcommand{\hatcurLBiMeccenxxxxxC}{\ensuremath{0.5086}}             
\newcommand{\hatcurLBiiMeccenxxxxxC}{\ensuremath{0.1973}}            
\newcommand{\hatcurLBiSoneeccenxxxxxC}{\ensuremath{0.0824}}            
\newcommand{\hatcurLBiiSoneeccenxxxxxC}{\ensuremath{0.1090}}           
\newcommand{\hatcurLBiStwoeccenxxxxxC}{\ensuremath{0.0699}}            
\newcommand{\hatcurLBiiStwoeccenxxxxxC}{\ensuremath{0.0921}}           
\newcommand{\hatcurLBiSthreeeccenxxxxxC}{\ensuremath{0.0573}}            
\newcommand{\hatcurLBiiSthreeeccenxxxxxC}{\ensuremath{0.0757}}           
\newcommand{\hatcurLBiSfoureccenxxxxxC}{\ensuremath{0.0435}}            
\newcommand{\hatcurLBiiSfoureccenxxxxxC}{\ensuremath{0.0614}}           
\newcommand{\hatcurISOmeccenxxxxxC}{\ensuremath{1.388_{-0.023}^{+0.039}}} 
\newcommand{\hatcurISOmshorteccenxxxxxC}{\ensuremath{1.39}}          
\newcommand{\hatcurISOmlongeccenxxxxxC}{\ensuremath{1.388_{-0.023}^{+0.039}}} 
\newcommand{\hatcurISOreccenxxxxxC}{\ensuremath{2.193_{-0.025}^{+0.033}}} 
\newcommand{\hatcurISOrshorteccenxxxxxC}{\ensuremath{2.19}}          
\newcommand{\hatcurISOrlongeccenxxxxxC}{\ensuremath{2.193_{-0.025}^{+0.033}}} 
\newcommand{\hatcurISOrhoeccenxxxxxC}{\ensuremath{0.1854\pm0.0093}}  
\newcommand{\hatcurISOrholongeccenxxxxxC}{\ensuremath{0.1854\pm0.0093}} 
\newcommand{\hatcurISOloggeccenxxxxxC}{\ensuremath{3.898\pm0.017}}   
\newcommand{\hatcurISOlumeccenxxxxxC}{\ensuremath{6.46\pm0.26}}      
\newcommand{\hatcurISOlumshorteccenxxxxxC}{\ensuremath{6.46}}        
\newcommand{\hatcurISOteffeccenxxxxxC}{\ensuremath{6220\pm44}}       
\newcommand{\hatcurISOzfeheccenxxxxxC}{\ensuremath{-0.035\pm0.062}}  
\newcommand{\hatcurISOageeccenxxxxxC}{\ensuremath{3.00_{-0.25}^{+0.12}}} 
\newcommand{\hatcurISOspececcenxxxxxC}{F}                            
\newcommand{\hatcurRVKeccenxxxxxC}{\ensuremath{37.9\pm6.0}}          
\newcommand{\hatcurRVrkeccenxxxxxC}{\ensuremath{0.14\pm0.13}}        
\newcommand{\hatcurRVrheccenxxxxxC}{\ensuremath{0.26\pm0.11}}        
\newcommand{\hatcurRVkeccenxxxxxC}{\ensuremath{0.039\pm0.053}}       
\newcommand{\hatcurRVheccenxxxxxC}{\ensuremath{0.080_{-0.048}^{+0.082}}} 
\newcommand{\hatcurRVtroneeccenxxxxxC}{\ensuremath{0\pm0}}           
\newcommand{\hatcurRVtrtwoeccenxxxxxC}{\ensuremath{0\pm0}}           
\newcommand{\hatcurRVgammaAeccenxxxxxC}{\ensuremath{-6.9\pm6.1}}     
\newcommand{\hatcurRVjitterAeccenxxxxxC}{\ensuremath{17.8\pm6.2}}    
\newcommand{\hatcurRVjittertwosiglimAeccenxxxxxC}{\ensuremath{<29.9}} 
\newcommand{\hatcurRVfitrmsAeccenxxxxxC}{\ensuremath{0.0}}           
\newcommand{\hatcurRVgammaBeccenxxxxxC}{\ensuremath{73.2\pm7.7}}     
\newcommand{\hatcurRVjitterBeccenxxxxxC}{\ensuremath{52\pm21}}       
\newcommand{\hatcurRVjittertwosiglimBeccenxxxxxC}{\ensuremath{<93.5}} 
\newcommand{\hatcurRVfitrmsBeccenxxxxxC}{\ensuremath{0.0}}           
\newcommand{\hatcurRVgammaCeccenxxxxxC}{\ensuremath{6019.8\pm8.4}}   
\newcommand{\hatcurRVjitterCeccenxxxxxC}{\ensuremath{11.7\pm4.5}}    
\newcommand{\hatcurRVjittertwosiglimCeccenxxxxxC}{\ensuremath{<18.8}} 
\newcommand{\hatcurRVfitrmsCeccenxxxxxC}{\ensuremath{0.0}}           
\newcommand{\hatcurRVgammaDeccenxxxxxC}{\ensuremath{5951\pm21}}      
\newcommand{\hatcurRVjitterDeccenxxxxxC}{\ensuremath{81\pm16}}       
\newcommand{\hatcurRVjittertwosiglimDeccenxxxxxC}{\ensuremath{<107.5}} 
\newcommand{\hatcurRVfitrmsDeccenxxxxxC}{\ensuremath{.1fym}}         %
\newcommand{\hatcurRVgammaEeccenxxxxxC}{\ensuremath{5784\pm20}}      
\newcommand{\hatcurRVjitterEeccenxxxxxC}{\ensuremath{106\pm21}}      
\newcommand{\hatcurRVjittertwosiglimEeccenxxxxxC}{\ensuremath{<144.0}} 
\newcommand{\hatcurRVfitrmsEeccenxxxxxC}{\ensuremath{.1fym}}         %
\newcommand{\hatcurRVecceneccenxxxxxC}{\ensuremath{0.104\pm0.066}}   
\newcommand{\hatcurRVeccentwosiglimeccenxxxxxC}{\ensuremath{<0.250}} 
\newcommand{\hatcurRVomegaeccenxxxxxC}{\ensuremath{64\pm41}}         
\newcommand{\hatcurPPieccenxxxxxC}{\ensuremath{83.89_{-0.39}^{+0.87}}} 
\newcommand{\hatcurPPgeccenxxxxxC}{\ensuremath{3.74\pm0.66}}         
\newcommand{\hatcurPPloggeccenxxxxxC}{\ensuremath{2.572\pm0.079}}    
\newcommand{\hatcurPPareccenxxxxxC}{\ensuremath{6.08\pm0.10}}        
\newcommand{\hatcurPPareleccenxxxxxC}{\ensuremath{0.06208_{-0.00034}^{+0.00058}}} 
\newcommand{\hatcurPPrhoeccenxxxxxC}{\ensuremath{0.116\pm0.023}}     
\newcommand{\hatcurPPmeccenxxxxxC}{\ensuremath{0.388\pm0.058}}       
\newcommand{\hatcurPPmshorteccenxxxxxC}{\ensuremath{0.39}}           
\newcommand{\hatcurPPmlongeccenxxxxxC}{\ensuremath{0.388\pm0.058}}   
\newcommand{\hatcurPPmeeccenxxxxxC}{\ensuremath{123\pm19}}           
\newcommand{\hatcurPPmeshorteccenxxxxxC}{\ensuremath{123.2}}         
\newcommand{\hatcurPPmelongeccenxxxxxC}{\ensuremath{123\pm19}}       
\newcommand{\hatcurPPreccenxxxxxC}{\ensuremath{1.607\pm0.043}}       
\newcommand{\hatcurPPrshorteccenxxxxxC}{\ensuremath{1.61}}           
\newcommand{\hatcurPPrlongeccenxxxxxC}{\ensuremath{1.607\pm0.043}}   
\newcommand{\hatcurPPreeccenxxxxxC}{\ensuremath{18.01\pm0.48}}       
\newcommand{\hatcurPPreshorteccenxxxxxC}{\ensuremath{18.0}}          
\newcommand{\hatcurPPrelongeccenxxxxxC}{\ensuremath{18.01\pm0.48}}   
\newcommand{\hatcurPPmrcorreccenxxxxxC}{\ensuremath{-0.34}}          
\newcommand{\hatcurPPteffeccenxxxxxC}{\ensuremath{1786\pm19}}        
\newcommand{\hatcurPPthetaeccenxxxxxC}{\ensuremath{0.0217\pm0.0036}} 
\newcommand{\hatcurPPfluxperieccenxxxxxC}{\ensuremath{2.85_{-0.31}^{+0.55}}} 
\newcommand{\hatcurPPfluxperidimeccenxxxxxC}{\ensuremath{9}}         
\newcommand{\hatcurPPfluxapeccenxxxxxC}{\ensuremath{1.87\pm0.22}}    
\newcommand{\hatcurPPfluxapdimeccenxxxxxC}{\ensuremath{9}}           
\newcommand{\hatcurPPfluxavgeccenxxxxxC}{\ensuremath{2.294\pm0.099}} 
\newcommand{\hatcurPPfluxavgdimeccenxxxxxC}{\ensuremath{9}}          
\newcommand{\hatcurPPfluxavglogeccenxxxxxC}{\ensuremath{9.361\pm0.018}} 
\newcommand{\hatcurXsecphaseeccenxxxxxC}{\ensuremath{0.525\pm0.034}} 
\newcommand{\hatcurXsecondaryeccenxxxxxC}{\ensuremath{2458368.25\pm0.16}} 
\newcommand{\hatcurXsecdureccenxxxxxC}{\ensuremath{0.222\pm0.029}}   
\newcommand{\hatcurXsecingdureccenxxxxxC}{\ensuremath{0.0281\pm0.0022}} 
\newcommand{\hatcurPPphiconjeccenxxxxxC}{\ensuremath{0.058\pm0.063}} 
\newcommand{\hatcurPPperieccenxxxxxC}{\ensuremath{2458365.46\pm0.30}} 
\newcommand{\hatcurPPaequiveccenxxxxxC}{\ensuremath{0.02450\pm0.00050}} 
\newcommand{\hatcurPPtcirceccenxxxxxC}{\ensuremath{109\pm21}}        
\newcommand{\hatcurPPtinfalleccenxxxxxC}{\ensuremath{1370_{-180}^{+280}}} 
\newcommand{\hatcurXdisteccenxxxxxC}{\ensuremath{233.6_{-2.2}^{+3.1}}} 
\newcommand{\hatcurXAveccenxxxxxC}{\ensuremath{0.136\pm0.030}}       
\newcommand{\hatcurXdistredeccenxxxxxC}{\ensuremath{233.6_{-2.2}^{+3.1}}} 
\newcommand{\hatcurXEBVeccenxxxxxC}{\ensuremath{0.0440\pm0.0097}}    
\newcommand{\hatcurCCpmraeccenxxxxxC}{\ensuremath{26.51\pm0.11}}     
\newcommand{\hatcurCCpmdececcenxxxxxC}{\ensuremath{6.165\pm0.075}}   
\newcommand{\hatcurCCpmeccenxxxxxC}{\ensuremath{27.22\pm0.13}}       

%% file: HTR094-024.eccen.tex
\newcommand{\hatcurhtreccenxxxxxD}{HTR094-024}                       
\newcommand{\hatcurfieldeccenxxxxxD}{\ensuremath{string}}            
\newcommand{\hatcurCCraeccenxxxxxD}{\ensuremath{05^{\mathrm h}01^{\mathrm m}55.2577{\mathrm s}}}                   
\newcommand{\hatcurCCdececcenxxxxxD}{\ensuremath{+50{\arcdeg}07{\arcmin}52.5746{\arcsec}}}                 
\newcommand{\hatcurCCmageccenxxxxxD}{13.188}                         
\newcommand{\hatcurCCtwomasseccenxxxxxD}{2MASS~05015525+5007526}     
\newcommand{\hatcurCCgsceccenxxxxxD}{GSC~3352-00595}                 
\newcommand{\hatcurCCgaiaeccenxxxxxD}{GAIA~256580178035425152}       
\newcommand{\hatcurCCgaiadrtwoeccenxxxxxD}{GAIA~DR2~256580182331399296} 
\newcommand{\hatcurCCtassmveccenxxxxxD}{\ensuremath{13.188\pm0.029}} 
\newcommand{\hatcurCCtassmvshorteccenxxxxxD}{\ensuremath{13.2}}      
\newcommand{\hatcurCCtassmBeccenxxxxxD}{\ensuremath{14.040\pm0.056}} 
\newcommand{\hatcurCCtassmBshorteccenxxxxxD}{\ensuremath{14.0}}      
\newcommand{\hatcurCCtassmIeccenxxxxxD}{\ensuremath{12.05\pm0.12}}   
\newcommand{\hatcurCCtassmIshorteccenxxxxxD}{\ensuremath{12.1}}      
\newcommand{\hatcurCCtassmgeccenxxxxxD}{\ensuremath{13.550\pm0.045}} 
\newcommand{\hatcurCCtassmgshorteccenxxxxxD}{\ensuremath{13.6}}      
\newcommand{\hatcurCCtassmreccenxxxxxD}{\ensuremath{12.915\pm0.033}} 
\newcommand{\hatcurCCtassmrshorteccenxxxxxD}{\ensuremath{12.9}}      
\newcommand{\hatcurCCtassmieccenxxxxxD}{\ensuremath{12.675\pm0.051}} 
\newcommand{\hatcurCCtassmishorteccenxxxxxD}{\ensuremath{12.7}}      
\newcommand{\hatcurCCparallaxeccenxxxxxD}{\ensuremath{2.923\pm0.035}} 
\newcommand{\hatcurCCgaiamGeccenxxxxxD}{\ensuremath{12.93860\pm0.00040}} 
\newcommand{\hatcurCCgaiamBPeccenxxxxxD}{\ensuremath{13.4067\pm0.0018}} 
\newcommand{\hatcurCCgaiamRPeccenxxxxxD}{\ensuremath{12.33560\pm0.00080}} 
\newcommand{\hatcurCCtwomassJmageccenxxxxxD}{\ensuremath{11.598\pm0.021}} 
\newcommand{\hatcurCCtwomassHmageccenxxxxxD}{\ensuremath{11.263\pm0.029}} 
\newcommand{\hatcurCCtwomassKmageccenxxxxxD}{\ensuremath{11.141\pm0.020}} 
\newcommand{\hatcurCCcitJmageccenxxxxxD}{\ensuremath{11.610\pm0.021}} 
\newcommand{\hatcurCCcitHmageccenxxxxxD}{\ensuremath{11.257\pm0.029}} 
\newcommand{\hatcurCCcitKmageccenxxxxxD}{\ensuremath{11.165\pm0.020}} 
\newcommand{\hatcurCCbbJmageccenxxxxxD}{\ensuremath{11.667\pm0.023}} 
\newcommand{\hatcurCCbbHmageccenxxxxxD}{\ensuremath{11.279\pm0.030}} 
\newcommand{\hatcurCCbbKmageccenxxxxxD}{\ensuremath{11.185\pm0.020}} 
\newcommand{\hatcurCCesoJmageccenxxxxxD}{\ensuremath{11.670\pm0.025}} 
\newcommand{\hatcurCCesoHmageccenxxxxxD}{\ensuremath{11.276\pm0.035}} 
\newcommand{\hatcurCCesoKmageccenxxxxxD}{\ensuremath{11.184\pm0.021}} 
\newcommand{\hatcurCCesoJHmageccenxxxxxD}{\ensuremath{0.393\pm0.041}} 
\newcommand{\hatcurCCesoJKmageccenxxxxxD}{\ensuremath{0.487\pm0.032}} 
\newcommand{\hatcurCCesoHKmageccenxxxxxD}{\ensuremath{0.093\pm0.041}} 
\newcommand{\hatcurCCWonemageccenxxxxxD}{\ensuremath{11.089\pm0.022}} 
\newcommand{\hatcurCCWtwomageccenxxxxxD}{\ensuremath{11.176\pm0.022}} 
\newcommand{\hatcurCCWthreemageccenxxxxxD}{\ensuremath{0\pm0}}       
\newcommand{\hatcurCCWfourmageccenxxxxxD}{\ensuremath{0\pm0}}        
\newcommand{\hatcurLCdipeccenxxxxxD}{\ensuremath{12.7}}              
\newcommand{\hatcurLCrprstareccenxxxxxD}{\ensuremath{0.0999\pm0.0026}} 
\newcommand{\hatcurLCbsqeccenxxxxxD}{\ensuremath{0.637_{-0.037}^{+0.033}}} 
\newcommand{\hatcurLCimpeccenxxxxxD}{\ensuremath{0.798_{-0.023}^{+0.020}}} 
\newcommand{\hatcurLCzetaeccenxxxxxD}{\ensuremath{35.84\pm0.88}}     
\newcommand{\hatcurLCdureccenxxxxxD}{\ensuremath{0.0702\pm0.0015}}   
\newcommand{\hatcurLCdurshorteccenxxxxxD}{\ensuremath{0.0702}}       
\newcommand{\hatcurLCdurhreccenxxxxxD}{\ensuremath{1.684\pm0.036}}   
\newcommand{\hatcurLCdurhrshorteccenxxxxxD}{\ensuremath{1.684}}      
\newcommand{\hatcurLCqeccenxxxxxD}{\ensuremath{0.03690\pm0.00078}}   
\newcommand{\hatcurLCqshorteccenxxxxxD}{\ensuremath{0.037}}          
\newcommand{\hatcurLCingdureccenxxxxxD}{\ensuremath{0.0159\pm0.0016}} 
\newcommand{\hatcurLCPeccenxxxxxD}{\ensuremath{1.90231300\pm0.00000074}} 
\newcommand{\hatcurLCPprececcenxxxxxD}{\ensuremath{1.9023130}}       
\newcommand{\hatcurLCPshorteccenxxxxxD}{\ensuremath{1.9023}}         
\newcommand{\hatcurLCTeccenxxxxxD}{\ensuremath{2457912.08504\pm0.00045}} 
\newcommand{\hatcurLCTAeccenxxxxxD}{\ensuremath{2454392.8059\pm0.0013}} 
\newcommand{\hatcurLCTBeccenxxxxxD}{\ensuremath{2458840.41377\pm0.00067}} 
\newcommand{\hatcurLChatnetmAeccenxxxxxD}{\ensuremath{12.61782\pm0.00020}} 
\newcommand{\hatcurLCiblendAeccenxxxxxD}{\ensuremath{0.886\pm0.098}} 
\newcommand{\hatcurLChatnetmBeccenxxxxxD}{\ensuremath{12.80531\pm0.00011}} 
\newcommand{\hatcurLCiblendBeccenxxxxxD}{\ensuremath{0.918\pm0.064}} 
\newcommand{\hatcurLChatnetmCeccenxxxxxD}{\ensuremath{0.000010\pm0.000033}} 
\newcommand{\hatcurLCiblendCeccenxxxxxD}{\ensuremath{0.691\pm0.045}} 
\newcommand{\hatcurLCrhoeccenxxxxxD}{\ensuremath{1.717\pm0.094}}     
\newcommand{\hatcurSMEiteffeccenxxxxxD}{\ensuremath{5546\pm50}}      
\newcommand{\hatcurSMEizfeheccenxxxxxD}{\ensuremath{0.399\pm0.080}}  
\newcommand{\hatcurSMEizfehshorteccenxxxxxD}{\ensuremath{0.40}}      
\newcommand{\hatcurSMEiloggeccenxxxxxD}{\ensuremath{4.43\pm0.10}}    
\newcommand{\hatcurSMEivsineccenxxxxxD}{\ensuremath{3.72\pm0.50}}    
\newcommand{\hatcurSMEivmaceccenxxxxxD}{\ensuremath{nff\pmnff}}      
\newcommand{\hatcurSMEivmiceccenxxxxxD}{\ensuremath{nff\pmnff}}      
\newcommand{\hatcurSMEiiteffeccenxxxxxD}{\ensuremath{5551\pm50}}     
\newcommand{\hatcurSMEiizfeheccenxxxxxD}{\ensuremath{0.396\pm0.080}} 
\newcommand{\hatcurSMEiizfehshorteccenxxxxxD}{\ensuremath{0.396}}    
\newcommand{\hatcurSMEiiloggeccenxxxxxD}{\ensuremath{4.468\pm0.035}} 
\newcommand{\hatcurSMEiivsineccenxxxxxD}{\ensuremath{3.69\pm0.50}}   
\newcommand{\hatcurLBiBeccenxxxxxD}{\ensuremath{0.7596}}             
\newcommand{\hatcurLBiiBeccenxxxxxD}{\ensuremath{0.0707}}            
\newcommand{\hatcurLBiVeccenxxxxxD}{\ensuremath{0.5876}}             
\newcommand{\hatcurLBiiVeccenxxxxxD}{\ensuremath{0.1536}}            
\newcommand{\hatcurLBiReccenxxxxxD}{\ensuremath{0.36\pm0.17}}        
\newcommand{\hatcurLBiiReccenxxxxxD}{\ensuremath{0.36\pm0.17}}       
\newcommand{\hatcurLBiIeccenxxxxxD}{\ensuremath{0.3836}}             
\newcommand{\hatcurLBiiIeccenxxxxxD}{\ensuremath{0.1879}}            
\newcommand{\hatcurLBiueccenxxxxxD}{\ensuremath{0.9484}}             
\newcommand{\hatcurLBiiueccenxxxxxD}{\ensuremath{-0.1150}}           
\newcommand{\hatcurLBigeccenxxxxxD}{\ensuremath{0.7007}}             
\newcommand{\hatcurLBiigeccenxxxxxD}{\ensuremath{0.0918}}            
\newcommand{\hatcurLBireccenxxxxxD}{\ensuremath{0.38\pm0.16}}        
\newcommand{\hatcurLBiireccenxxxxxD}{\ensuremath{0.38\pm0.16}}       
\newcommand{\hatcurLBiieccenxxxxxD}{\ensuremath{0.30\pm0.15}}        
\newcommand{\hatcurLBiiieccenxxxxxD}{\ensuremath{0.27\pm0.16}}       
\newcommand{\hatcurLBizeccenxxxxxD}{\ensuremath{0.3380}}             
\newcommand{\hatcurLBiizeccenxxxxxD}{\ensuremath{0.1913}}            
\newcommand{\hatcurLBiJeccenxxxxxD}{\ensuremath{0.2216}}             
\newcommand{\hatcurLBiiJeccenxxxxxD}{\ensuremath{0.2275}}            
\newcommand{\hatcurLBiHeccenxxxxxD}{\ensuremath{0.1268}}             
\newcommand{\hatcurLBiiHeccenxxxxxD}{\ensuremath{0.2668}}            
\newcommand{\hatcurLBiKeccenxxxxxD}{\ensuremath{0.1187}}             
\newcommand{\hatcurLBiiKeccenxxxxxD}{\ensuremath{0.2135}}            
\newcommand{\hatcurLBiTeccenxxxxxD}{\ensuremath{0.30\pm0.15}}        
\newcommand{\hatcurLBiiTeccenxxxxxD}{\ensuremath{0.30\pm0.18}}       
\newcommand{\hatcurLBikepeccenxxxxxD}{\ensuremath{0.4997}}           
\newcommand{\hatcurLBiikepeccenxxxxxD}{\ensuremath{0.1776}}          
\newcommand{\hatcurLBiCeccenxxxxxD}{\ensuremath{0.4891}}             
\newcommand{\hatcurLBiiCeccenxxxxxD}{\ensuremath{0.1761}}            
\newcommand{\hatcurLBiMeccenxxxxxD}{\ensuremath{0.5864}}             
\newcommand{\hatcurLBiiMeccenxxxxxD}{\ensuremath{0.1476}}            
\newcommand{\hatcurLBiSoneeccenxxxxxD}{\ensuremath{0.1004}}            
\newcommand{\hatcurLBiiSoneeccenxxxxxD}{\ensuremath{0.1280}}           
\newcommand{\hatcurLBiStwoeccenxxxxxD}{\ensuremath{0.0854}}            
\newcommand{\hatcurLBiiStwoeccenxxxxxD}{\ensuremath{0.1068}}           
\newcommand{\hatcurLBiSthreeeccenxxxxxD}{\ensuremath{0.0715}}            
\newcommand{\hatcurLBiiSthreeeccenxxxxxD}{\ensuremath{0.0878}}           
\newcommand{\hatcurLBiSfoureccenxxxxxD}{\ensuremath{0.0563}}            
\newcommand{\hatcurLBiiSfoureccenxxxxxD}{\ensuremath{0.0694}}           
\newcommand{\hatcurISOmeccenxxxxxD}{\ensuremath{1.000\pm0.034}}      
\newcommand{\hatcurISOmshorteccenxxxxxD}{\ensuremath{1.00}}          
\newcommand{\hatcurISOmlongeccenxxxxxD}{\ensuremath{1.000\pm0.034}}  
\newcommand{\hatcurISOreccenxxxxxD}{\ensuremath{0.937\pm0.011}}      
\newcommand{\hatcurISOrshorteccenxxxxxD}{\ensuremath{0.94}}          
\newcommand{\hatcurISOrlongeccenxxxxxD}{\ensuremath{0.937\pm0.011}}  
\newcommand{\hatcurISOrhoeccenxxxxxD}{\ensuremath{1.717\pm0.094}}    
\newcommand{\hatcurISOrholongeccenxxxxxD}{\ensuremath{1.717\pm0.094}} 
\newcommand{\hatcurISOloggeccenxxxxxD}{\ensuremath{4.495\pm0.020}}   
\newcommand{\hatcurISOlumeccenxxxxxD}{\ensuremath{0.765\pm0.031}}    
\newcommand{\hatcurISOlumshorteccenxxxxxD}{\ensuremath{0.76}}        
\newcommand{\hatcurISOteffeccenxxxxxD}{\ensuremath{5584\pm45}}       
\newcommand{\hatcurISOzfeheccenxxxxxD}{\ensuremath{0.196\pm0.055}}   
\newcommand{\hatcurISOageeccenxxxxxD}{\ensuremath{2.7_{-1.8}^{+2.5}}} 
\newcommand{\hatcurISOspececcenxxxxxD}{G}                            
\newcommand{\hatcurRVKeccenxxxxxD}{\ensuremath{178.3\pm6.1}}         
\newcommand{\hatcurRVrkeccenxxxxxD}{\ensuremath{-0.122_{-0.084}^{+0.123}}} 
\newcommand{\hatcurRVrheccenxxxxxD}{\ensuremath{-0.214_{-0.078}^{+0.173}}} 
\newcommand{\hatcurRVkeccenxxxxxD}{\ensuremath{-0.033_{-0.025}^{+0.033}}} 
\newcommand{\hatcurRVheccenxxxxxD}{\ensuremath{-0.059_{-0.032}^{+0.054}}} 
\newcommand{\hatcurRVtroneeccenxxxxxD}{\ensuremath{0\pm0}}           
\newcommand{\hatcurRVtrtwoeccenxxxxxD}{\ensuremath{0\pm0}}           
\newcommand{\hatcurRVgammaAeccenxxxxxD}{\ensuremath{-54.3\pm7.5}}    
\newcommand{\hatcurRVjitterAeccenxxxxxD}{\ensuremath{1.7\pm8.8}}     
\newcommand{\hatcurRVjittertwosiglimAeccenxxxxxD}{\ensuremath{<21.5}} 
\newcommand{\hatcurRVfitrmsAeccenxxxxxD}{\ensuremath{0.0}}           
\newcommand{\hatcurRVgammaBeccenxxxxxD}{\ensuremath{209\pm12}}       
\newcommand{\hatcurRVjitterBeccenxxxxxD}{\ensuremath{41\pm14}}       
\newcommand{\hatcurRVjittertwosiglimBeccenxxxxxD}{\ensuremath{<65.3}} 
\newcommand{\hatcurRVfitrmsBeccenxxxxxD}{\ensuremath{0.0}}           
\newcommand{\hatcurRVgammaCeccenxxxxxD}{\ensuremath{6000\pm4500}}    
\newcommand{\hatcurRVjitterCeccenxxxxxD}{\ensuremath{0\pm27}}        
\newcommand{\hatcurRVjittertwosiglimCeccenxxxxxD}{\ensuremath{<58.0}} 
\newcommand{\hatcurRVfitrmsCeccenxxxxxD}{\ensuremath{0.0}}           
\newcommand{\hatcurRVecceneccenxxxxxD}{\ensuremath{0.081\pm0.034}}   
\newcommand{\hatcurRVeccentwosiglimeccenxxxxxD}{\ensuremath{<0.113}} 
\newcommand{\hatcurRVomegaeccenxxxxxD}{\ensuremath{239\pm59}}        
\newcommand{\hatcurPPieccenxxxxxD}{\ensuremath{83.70\pm0.25}}        
\newcommand{\hatcurPPgeccenxxxxxD}{\ensuremath{32.3\pm2.4}}          
\newcommand{\hatcurPPloggeccenxxxxxD}{\ensuremath{3.510\pm0.032}}    
\newcommand{\hatcurPPareccenxxxxxD}{\ensuremath{6.90_{-0.15}^{+0.11}}} 
\newcommand{\hatcurPPareleccenxxxxxD}{\ensuremath{0.03006_{-0.00041}^{+0.00031}}} 
\newcommand{\hatcurPPrhoeccenxxxxxD}{\ensuremath{1.77\pm0.18}}       
\newcommand{\hatcurPPmeccenxxxxxD}{\ensuremath{1.086\pm0.046}}       
\newcommand{\hatcurPPmshorteccenxxxxxD}{\ensuremath{1.09}}           
\newcommand{\hatcurPPmlongeccenxxxxxD}{\ensuremath{1.086\pm0.046}}   
\newcommand{\hatcurPPmeeccenxxxxxD}{\ensuremath{345\pm14}}           
\newcommand{\hatcurPPmeshorteccenxxxxxD}{\ensuremath{345.2}}         
\newcommand{\hatcurPPmelongeccenxxxxxD}{\ensuremath{345\pm14}}       
\newcommand{\hatcurPPreccenxxxxxD}{\ensuremath{0.910\pm0.027}}       
\newcommand{\hatcurPPrshorteccenxxxxxD}{\ensuremath{0.91}}           
\newcommand{\hatcurPPrlongeccenxxxxxD}{\ensuremath{0.910\pm0.027}}   
\newcommand{\hatcurPPreeccenxxxxxD}{\ensuremath{10.21\pm0.30}}       
\newcommand{\hatcurPPreshorteccenxxxxxD}{\ensuremath{10.2}}          
\newcommand{\hatcurPPrelongeccenxxxxxD}{\ensuremath{10.21\pm0.30}}   
\newcommand{\hatcurPPmrcorreccenxxxxxD}{\ensuremath{-0.05}}          
\newcommand{\hatcurPPteffeccenxxxxxD}{\ensuremath{1505\pm15}}        
\newcommand{\hatcurPPthetaeccenxxxxxD}{\ensuremath{0.0713\pm0.0031}} 
\newcommand{\hatcurPPfluxperieccenxxxxxD}{\ensuremath{1.347\pm0.100}} 
\newcommand{\hatcurPPfluxperidimeccenxxxxxD}{\ensuremath{9}}         
\newcommand{\hatcurPPfluxapeccenxxxxxD}{\ensuremath{9.95_{-0.64}^{+1.02}}} 
\newcommand{\hatcurPPfluxapdimeccenxxxxxD}{\ensuremath{8}}           
\newcommand{\hatcurPPfluxavgeccenxxxxxD}{\ensuremath{1.157\pm0.045}} 
\newcommand{\hatcurPPfluxavgdimeccenxxxxxD}{\ensuremath{9}}          
\newcommand{\hatcurPPfluxavglogeccenxxxxxD}{\ensuremath{9.063\pm0.017}} 
\newcommand{\hatcurXsecphaseeccenxxxxxD}{\ensuremath{0.479\pm0.018}} 
\newcommand{\hatcurXsecondaryeccenxxxxxD}{\ensuremath{2457912.996\pm0.035}} 
\newcommand{\hatcurXsecdureccenxxxxxD}{\ensuremath{0.0692\pm0.0012}} 
\newcommand{\hatcurXsecingdureccenxxxxxD}{\ensuremath{0.0119\pm0.0016}} 
\newcommand{\hatcurPPphiconjeccenxxxxxD}{\ensuremath{-0.375_{-0.066}^{+0.379}}} 
\newcommand{\hatcurPPperieccenxxxxxD}{\ensuremath{2457912.80\pm0.56}} 
\newcommand{\hatcurPPaequiveccenxxxxxD}{\ensuremath{0.03430_{-0.00060}^{+0.00080}}} 
\newcommand{\hatcurPPtcirceccenxxxxxD}{\ensuremath{82\pm14}}         
\newcommand{\hatcurPPtinfalleccenxxxxxD}{\ensuremath{262_{-30}^{+23}}} 
\newcommand{\hatcurXdisteccenxxxxxD}{\ensuremath{342.7\pm3.9}}       
\newcommand{\hatcurXAveccenxxxxxD}{\ensuremath{0.385\pm0.042}}       
\newcommand{\hatcurXdistredeccenxxxxxD}{\ensuremath{342.7\pm3.9}}    
\newcommand{\hatcurXEBVeccenxxxxxD}{\ensuremath{0.124\pm0.014}}      
\newcommand{\hatcurCCpmraeccenxxxxxD}{\ensuremath{-11.021\pm0.067}}  
\newcommand{\hatcurCCpmdececcenxxxxxD}{\ensuremath{-21.440\pm0.063}} 
\newcommand{\hatcurCCpmeccenxxxxxD}{\ensuremath{24.107\pm0.092}}     

%% file: HTR130-001.eccen.tex
\newcommand{\hatcurhtreccenxxxxxE}{HTR130-001}                       
\newcommand{\hatcurfieldeccenxxxxxE}{\ensuremath{string}}            
\newcommand{\hatcurCCraeccenxxxxxE}{\ensuremath{04^{\mathrm h}58^{\mathrm m}01.0287{\mathrm s}}}                   
\newcommand{\hatcurCCdececcenxxxxxE}{\ensuremath{+48{\arcdeg}18{\arcmin}03.7570{\arcsec}}}                 
\newcommand{\hatcurCCmageccenxxxxxE}{\ensuremath{string}}            
\newcommand{\hatcurCCtwomasseccenxxxxxE}{2MASS~04580102+4818038}     
\newcommand{\hatcurCCgsceccenxxxxxE}{GSC~3348-01101}                 
\newcommand{\hatcurCCgaiaeccenxxxxxE}{GAIA~255397137880340480}       
\newcommand{\hatcurCCgaiadrtwoeccenxxxxxE}{GAIA~DR2~255397142179844224} 
\newcommand{\hatcurCCtassmveccenxxxxxE}{\ensuremath{0\pm0}}          
\newcommand{\hatcurCCtassmvshorteccenxxxxxE}{\ensuremath{.1fym}}     
\newcommand{\hatcurCCtassmBeccenxxxxxE}{\ensuremath{0\pm0}}          
\newcommand{\hatcurCCtassmBshorteccenxxxxxE}{\ensuremath{.1fym}}     
\newcommand{\hatcurCCtassmIeccenxxxxxE}{\ensuremath{0\pm0}}          
\newcommand{\hatcurCCtassmIshorteccenxxxxxE}{\ensuremath{.1fym}}     
\newcommand{\hatcurCCtassmgeccenxxxxxE}{\ensuremath{0\pm0}}          
\newcommand{\hatcurCCtassmgshorteccenxxxxxE}{\ensuremath{.1fym}}     
\newcommand{\hatcurCCtassmreccenxxxxxE}{\ensuremath{0\pm0}}          
\newcommand{\hatcurCCtassmrshorteccenxxxxxE}{\ensuremath{.1fym}}     
\newcommand{\hatcurCCtassmieccenxxxxxE}{\ensuremath{0\pm0}}          
\newcommand{\hatcurCCtassmishorteccenxxxxxE}{\ensuremath{.1fym}}     
\newcommand{\hatcurCCparallaxeccenxxxxxE}{\ensuremath{2.839\pm0.040}} 
\newcommand{\hatcurCCgaiamGeccenxxxxxE}{\ensuremath{12.43620\pm0.00030}} 
\newcommand{\hatcurCCgaiamBPeccenxxxxxE}{\ensuremath{12.8932\pm0.0013}} 
\newcommand{\hatcurCCgaiamRPeccenxxxxxE}{\ensuremath{11.84000\pm0.00090}} 
\newcommand{\hatcurCCtwomassJmageccenxxxxxE}{\ensuremath{11.144\pm0.029}} 
\newcommand{\hatcurCCtwomassHmageccenxxxxxE}{\ensuremath{10.803\pm0.041}} 
\newcommand{\hatcurCCtwomassKmageccenxxxxxE}{\ensuremath{10.701\pm0.026}} 
\newcommand{\hatcurCCcitJmageccenxxxxxE}{\ensuremath{11.157\pm0.029}} 
\newcommand{\hatcurCCcitHmageccenxxxxxE}{\ensuremath{10.797\pm0.040}} 
\newcommand{\hatcurCCcitKmageccenxxxxxE}{\ensuremath{10.725\pm0.026}} 
\newcommand{\hatcurCCbbJmageccenxxxxxE}{\ensuremath{11.212\pm0.031}} 
\newcommand{\hatcurCCbbHmageccenxxxxxE}{\ensuremath{10.819\pm0.042}} 
\newcommand{\hatcurCCbbKmageccenxxxxxE}{\ensuremath{10.745\pm0.026}} 
\newcommand{\hatcurCCesoJmageccenxxxxxE}{\ensuremath{11.215\pm0.033}} 
\newcommand{\hatcurCCesoHmageccenxxxxxE}{\ensuremath{10.815\pm0.047}} 
\newcommand{\hatcurCCesoKmageccenxxxxxE}{\ensuremath{10.744\pm0.027}} 
\newcommand{\hatcurCCesoJHmageccenxxxxxE}{\ensuremath{0.399\pm0.055}} 
\newcommand{\hatcurCCesoJKmageccenxxxxxE}{\ensuremath{0.472\pm0.042}} 
\newcommand{\hatcurCCesoHKmageccenxxxxxE}{\ensuremath{0.073\pm0.054}} 
\newcommand{\hatcurCCWonemageccenxxxxxE}{\ensuremath{0\pm0}}         
\newcommand{\hatcurCCWtwomageccenxxxxxE}{\ensuremath{0\pm0}}         
\newcommand{\hatcurCCWthreemageccenxxxxxE}{\ensuremath{0\pm0}}       
\newcommand{\hatcurCCWfourmageccenxxxxxE}{\ensuremath{0\pm0}}        
\newcommand{\hatcurLCdipeccenxxxxxE}{\ensuremath{9.3}}               
\newcommand{\hatcurLCrprstareccenxxxxxE}{\ensuremath{0.0936\pm0.0019}} 
\newcommand{\hatcurLCbsqeccenxxxxxE}{\ensuremath{0.067_{-0.041}^{+0.051}}} 
\newcommand{\hatcurLCimpeccenxxxxxE}{\ensuremath{0.259_{-0.099}^{+0.085}}} 
\newcommand{\hatcurLCzetaeccenxxxxxE}{\ensuremath{17.04\pm0.17}}     
\newcommand{\hatcurLCdureccenxxxxxE}{\ensuremath{0.1292\pm0.0013}}   
\newcommand{\hatcurLCdurshorteccenxxxxxE}{\ensuremath{0.1292}}       
\newcommand{\hatcurLCdurhreccenxxxxxE}{\ensuremath{3.102\pm0.032}}   
\newcommand{\hatcurLCdurhrshorteccenxxxxxE}{\ensuremath{3.102}}      
\newcommand{\hatcurLCqeccenxxxxxE}{\ensuremath{0.04890\pm0.00051}}   
\newcommand{\hatcurLCqshorteccenxxxxxE}{\ensuremath{0.049}}          
\newcommand{\hatcurLCingdureccenxxxxxE}{\ensuremath{0.01177\pm0.00075}} 
\newcommand{\hatcurLCPeccenxxxxxE}{\ensuremath{2.6453232\pm0.0000039}} 
\newcommand{\hatcurLCPprececcenxxxxxE}{\ensuremath{2.6453232}}       
\newcommand{\hatcurLCPshorteccenxxxxxE}{\ensuremath{2.6453}}         
\newcommand{\hatcurLCTeccenxxxxxE}{\ensuremath{2457084.00047\pm0.00043}} 
\newcommand{\hatcurLCTAeccenxxxxxE}{\ensuremath{2456131.6841\pm0.0013}} 
\newcommand{\hatcurLCTBeccenxxxxxE}{\ensuremath{2457292.98098\pm0.00059}} 
\newcommand{\hatcurLChatnetmeccenxxxxxE}{\ensuremath{12.379230\pm0.000097}} 
\newcommand{\hatcurLCiblendeccenxxxxxE}{\ensuremath{0.858\pm0.055}}  
\newcommand{\hatcurLCrhoeccenxxxxxE}{\ensuremath{0.912\pm0.047}}     
\newcommand{\hatcurSMEiteffeccenxxxxxE}{\ensuremath{5617\pm50}}      
\newcommand{\hatcurSMEizfeheccenxxxxxE}{\ensuremath{0.461\pm0.080}}  
\newcommand{\hatcurSMEizfehshorteccenxxxxxE}{\ensuremath{0.46}}      
\newcommand{\hatcurSMEiloggeccenxxxxxE}{\ensuremath{4.31\pm0.10}}    
\newcommand{\hatcurSMEivsineccenxxxxxE}{\ensuremath{3.53\pm0.50}}    
\newcommand{\hatcurSMEivmaceccenxxxxxE}{\ensuremath{nff\pmnff}}      
\newcommand{\hatcurSMEivmiceccenxxxxxE}{\ensuremath{nff\pmnff}}      
\newcommand{\hatcurSMEiiteffeccenxxxxxE}{\ensuremath{5601\pm50}}     
\newcommand{\hatcurSMEiizfeheccenxxxxxE}{\ensuremath{0.449\pm0.080}} 
\newcommand{\hatcurSMEiizfehshorteccenxxxxxE}{\ensuremath{0.449}}    
\newcommand{\hatcurSMEiiloggeccenxxxxxE}{\ensuremath{4.305\pm0.030}} 
\newcommand{\hatcurSMEiivsineccenxxxxxE}{\ensuremath{3.55\pm0.50}}   
\newcommand{\hatcurLBiBeccenxxxxxE}{\ensuremath{0.7609}}             
\newcommand{\hatcurLBiiBeccenxxxxxE}{\ensuremath{0.0700}}            
\newcommand{\hatcurLBiVeccenxxxxxE}{\ensuremath{0.5874}}             
\newcommand{\hatcurLBiiVeccenxxxxxE}{\ensuremath{0.1516}}            
\newcommand{\hatcurLBiReccenxxxxxE}{\ensuremath{0.4820}}             
\newcommand{\hatcurLBiiReccenxxxxxE}{\ensuremath{0.1735}}            
\newcommand{\hatcurLBiIeccenxxxxxE}{\ensuremath{0.3806}}             
\newcommand{\hatcurLBiiIeccenxxxxxE}{\ensuremath{0.1873}}            
\newcommand{\hatcurLBiueccenxxxxxE}{\ensuremath{0.9339}}             
\newcommand{\hatcurLBiiueccenxxxxxE}{\ensuremath{-0.0938}}           
\newcommand{\hatcurLBigeccenxxxxxE}{\ensuremath{0.7007}}             
\newcommand{\hatcurLBiigeccenxxxxxE}{\ensuremath{0.0918}}            
\newcommand{\hatcurLBireccenxxxxxE}{\ensuremath{0.34\pm0.15}}        
\newcommand{\hatcurLBiireccenxxxxxE}{\ensuremath{0.29\pm0.17}}       
\newcommand{\hatcurLBiieccenxxxxxE}{\ensuremath{0.33\pm0.10}}        
\newcommand{\hatcurLBiiieccenxxxxxE}{\ensuremath{0.21\pm0.15}}       
\newcommand{\hatcurLBizeccenxxxxxE}{\ensuremath{0.36\pm0.16}}        
\newcommand{\hatcurLBiizeccenxxxxxE}{\ensuremath{0.34\pm0.17}}       
\newcommand{\hatcurLBiJeccenxxxxxE}{\ensuremath{0.2219}}             
\newcommand{\hatcurLBiiJeccenxxxxxE}{\ensuremath{0.2241}}            
\newcommand{\hatcurLBiHeccenxxxxxE}{\ensuremath{0.1309}}             
\newcommand{\hatcurLBiiHeccenxxxxxE}{\ensuremath{0.2607}}            
\newcommand{\hatcurLBiKeccenxxxxxE}{\ensuremath{0.1205}}             
\newcommand{\hatcurLBiiKeccenxxxxxE}{\ensuremath{0.2088}}            
\newcommand{\hatcurLBiTeccenxxxxxE}{\ensuremath{0.3799}}             
\newcommand{\hatcurLBiiTeccenxxxxxE}{\ensuremath{0.2072}}            
\newcommand{\hatcurLBikepeccenxxxxxE}{\ensuremath{0.4960}}           
\newcommand{\hatcurLBiikepeccenxxxxxE}{\ensuremath{0.1811}}          
\newcommand{\hatcurLBiCeccenxxxxxE}{\ensuremath{0.4858}}             
\newcommand{\hatcurLBiiCeccenxxxxxE}{\ensuremath{0.1794}}            
\newcommand{\hatcurLBiMeccenxxxxxE}{\ensuremath{0.5828}}             
\newcommand{\hatcurLBiiMeccenxxxxxE}{\ensuremath{0.1517}}            
\newcommand{\hatcurLBiSoneeccenxxxxxE}{\ensuremath{0.0977}}            
\newcommand{\hatcurLBiiSoneeccenxxxxxE}{\ensuremath{0.1285}}           
\newcommand{\hatcurLBiStwoeccenxxxxxE}{\ensuremath{0.0835}}            
\newcommand{\hatcurLBiiStwoeccenxxxxxE}{\ensuremath{0.1074}}           
\newcommand{\hatcurLBiSthreeeccenxxxxxE}{\ensuremath{0.0701}}            
\newcommand{\hatcurLBiiSthreeeccenxxxxxE}{\ensuremath{0.0881}}           
\newcommand{\hatcurLBiSfoureccenxxxxxE}{\ensuremath{0.0548}}            
\newcommand{\hatcurLBiiSfoureccenxxxxxE}{\ensuremath{0.0697}}           
\newcommand{\hatcurISOmeccenxxxxxE}{\ensuremath{1.023\pm0.020}}      
\newcommand{\hatcurISOmshorteccenxxxxxE}{\ensuremath{1.02}}          
\newcommand{\hatcurISOmlongeccenxxxxxE}{\ensuremath{1.023\pm0.020}}  
\newcommand{\hatcurISOreccenxxxxxE}{\ensuremath{1.165\pm0.018}}      
\newcommand{\hatcurISOrshorteccenxxxxxE}{\ensuremath{1.17}}          
\newcommand{\hatcurISOrlongeccenxxxxxE}{\ensuremath{1.165\pm0.018}}  
\newcommand{\hatcurISOrhoeccenxxxxxE}{\ensuremath{0.912\pm0.047}}    
\newcommand{\hatcurISOrholongeccenxxxxxE}{\ensuremath{0.912\pm0.047}} 
\newcommand{\hatcurISOloggeccenxxxxxE}{\ensuremath{4.315\pm0.017}}   
\newcommand{\hatcurISOlumeccenxxxxxE}{\ensuremath{1.227\pm0.054}}    
\newcommand{\hatcurISOlumshorteccenxxxxxE}{\ensuremath{1.23}}        
\newcommand{\hatcurISOteffeccenxxxxxE}{\ensuremath{5634\pm48}}       
\newcommand{\hatcurISOzfeheccenxxxxxE}{\ensuremath{0.413\pm0.092}}   
\newcommand{\hatcurISOageeccenxxxxxE}{\ensuremath{8.0\pm1.1}}        
\newcommand{\hatcurISOspececcenxxxxxE}{G}                            
\newcommand{\hatcurRVKeccenxxxxxE}{\ensuremath{113\pm11}}            
\newcommand{\hatcurRVrkeccenxxxxxE}{\ensuremath{-0.00\pm0.16}}       
\newcommand{\hatcurRVrheccenxxxxxE}{\ensuremath{-0.03\pm0.11}}       
\newcommand{\hatcurRVkeccenxxxxxE}{\ensuremath{-0.000\pm0.043}}      
\newcommand{\hatcurRVheccenxxxxxE}{\ensuremath{-0.003_{-0.029}^{+0.015}}} 
\newcommand{\hatcurRVtroneeccenxxxxxE}{\ensuremath{0\pm0}}           
\newcommand{\hatcurRVtrtwoeccenxxxxxE}{\ensuremath{0\pm0}}           
\newcommand{\hatcurRVgammaeccenxxxxxE}{\ensuremath{127.7\pm8.8}}     
\newcommand{\hatcurRVjittereccenxxxxxE}{\ensuremath{31.9\pm8.3}}     
\newcommand{\hatcurRVjittertwosiglimeccenxxxxxE}{\ensuremath{<47.7}} 
\newcommand{\hatcurRVfitrmseccenxxxxxE}{\ensuremath{.1fym}}          %
\newcommand{\hatcurRVecceneccenxxxxxE}{\ensuremath{0.030\pm0.034}}   
\newcommand{\hatcurRVeccentwosiglimeccenxxxxxE}{\ensuremath{<0.101}} 
\newcommand{\hatcurRVomegaeccenxxxxxE}{\ensuremath{200\pm110}}       
\newcommand{\hatcurPPieccenxxxxxE}{\ensuremath{87.87\pm0.72}}        
\newcommand{\hatcurPPgeccenxxxxxE}{\ensuremath{17.1\pm2.0}}          
\newcommand{\hatcurPPloggeccenxxxxxE}{\ensuremath{3.234_{-0.059}^{+0.042}}} 
\newcommand{\hatcurPPareccenxxxxxE}{\ensuremath{6.96\pm0.12}}        
\newcommand{\hatcurPPareleccenxxxxxE}{\ensuremath{0.03773\pm0.00024}} 
\newcommand{\hatcurPPrhoeccenxxxxxE}{\ensuremath{0.81\pm0.11}}       
\newcommand{\hatcurPPmeccenxxxxxE}{\ensuremath{0.780_{-0.088}^{+0.068}}} 
\newcommand{\hatcurPPmshorteccenxxxxxE}{\ensuremath{0.78}}           
\newcommand{\hatcurPPmlongeccenxxxxxE}{\ensuremath{0.780_{-0.088}^{+0.068}}} 
\newcommand{\hatcurPPmeeccenxxxxxE}{\ensuremath{248_{-28}^{+22}}}    
\newcommand{\hatcurPPmeshorteccenxxxxxE}{\ensuremath{247.9}}         
\newcommand{\hatcurPPmelongeccenxxxxxE}{\ensuremath{248_{-28}^{+22}}} 
\newcommand{\hatcurPPreccenxxxxxE}{\ensuremath{1.062\pm0.027}}       
\newcommand{\hatcurPPrshorteccenxxxxxE}{\ensuremath{1.06}}           
\newcommand{\hatcurPPrlongeccenxxxxxE}{\ensuremath{1.062\pm0.027}}   
\newcommand{\hatcurPPreeccenxxxxxE}{\ensuremath{11.90\pm0.30}}       
\newcommand{\hatcurPPreshorteccenxxxxxE}{\ensuremath{11.9}}          
\newcommand{\hatcurPPrelongeccenxxxxxE}{\ensuremath{11.90\pm0.30}}   
\newcommand{\hatcurPPmrcorreccenxxxxxE}{\ensuremath{-0.09}}          
\newcommand{\hatcurPPteffeccenxxxxxE}{\ensuremath{1510\pm14}}        
\newcommand{\hatcurPPthetaeccenxxxxxE}{\ensuremath{0.0541_{-0.0062}^{+0.0047}}} 
\newcommand{\hatcurPPfluxperieccenxxxxxE}{\ensuremath{1.257_{-0.074}^{+0.102}}} 
\newcommand{\hatcurPPfluxperidimeccenxxxxxE}{\ensuremath{9}}         
\newcommand{\hatcurPPfluxapeccenxxxxxE}{\ensuremath{1.099\pm0.079}}  
\newcommand{\hatcurPPfluxapdimeccenxxxxxE}{\ensuremath{9}}           
\newcommand{\hatcurPPfluxavgeccenxxxxxE}{\ensuremath{1.173\pm0.044}} 
\newcommand{\hatcurPPfluxavgdimeccenxxxxxE}{\ensuremath{9}}          
\newcommand{\hatcurPPfluxavglogeccenxxxxxE}{\ensuremath{9.069\pm0.016}} 
\newcommand{\hatcurXsecphaseeccenxxxxxE}{\ensuremath{0.500\pm0.027}} 
\newcommand{\hatcurXsecondaryeccenxxxxxE}{\ensuremath{2457085.323\pm0.072}} 
\newcommand{\hatcurXsecdureccenxxxxxE}{\ensuremath{0.1280\pm0.0054}} 
\newcommand{\hatcurXsecingdureccenxxxxxE}{\ensuremath{0.01161\pm0.00048}} 
\newcommand{\hatcurPPphiconjeccenxxxxxE}{\ensuremath{-0.01\pm0.30}}  
\newcommand{\hatcurPPperieccenxxxxxE}{\ensuremath{2457084.02\pm0.78}} 
\newcommand{\hatcurPPaequiveccenxxxxxE}{\ensuremath{0.03410\pm0.00063}} 
\newcommand{\hatcurPPtcirceccenxxxxxE}{\ensuremath{118\pm21}}        
\newcommand{\hatcurPPtinfalleccenxxxxxE}{\ensuremath{543_{-58}^{+75}}} 
\newcommand{\hatcurXdisteccenxxxxxE}{\ensuremath{351.7\pm4.5}}       
\newcommand{\hatcurXAveccenxxxxxE}{\ensuremath{0.343\pm0.047}}       
\newcommand{\hatcurXdistredeccenxxxxxE}{\ensuremath{351.7\pm4.5}}    
\newcommand{\hatcurXEBVeccenxxxxxE}{\ensuremath{0.111\pm0.015}}      
\newcommand{\hatcurCCpmraeccenxxxxxE}{\ensuremath{14.732\pm0.080}}   
\newcommand{\hatcurCCpmdececcenxxxxxE}{\ensuremath{-43.776\pm0.061}} 
\newcommand{\hatcurCCpmeccenxxxxxE}{\ensuremath{46.19\pm0.10}}       

%% file: HTR384-014.eccen.tex
\newcommand{\hatcurhtreccenxxxxxF}{HTR384-014}                       
\newcommand{\hatcurfieldeccenxxxxxF}{\ensuremath{string}}            
\newcommand{\hatcurCCraeccenxxxxxF}{\ensuremath{17^{\mathrm h}58^{\mathrm m}17.3121{\mathrm s}}}                   
\newcommand{\hatcurCCdececcenxxxxxF}{\ensuremath{+05{\arcdeg}45{\arcmin}40.9400{\arcsec}}}                 
\newcommand{\hatcurCCmageccenxxxxxF}{13.753}                         
\newcommand{\hatcurCCtwomasseccenxxxxxF}{2MASS~17581730+0545409}     
\newcommand{\hatcurCCgsceccenxxxxxF}{GSC~0429-01697}                 
\newcommand{\hatcurCCgaiaeccenxxxxxF}{GAIA~4474644332250439552}      
\newcommand{\hatcurCCgaiadrtwoeccenxxxxxF}{GAIA~DR2~4474644332250439552} 
\newcommand{\hatcurCCtassmveccenxxxxxF}{\ensuremath{13.753\pm0.065}} 
\newcommand{\hatcurCCtassmvshorteccenxxxxxF}{\ensuremath{13.8}}      
\newcommand{\hatcurCCtassmBeccenxxxxxF}{\ensuremath{14.729\pm0.066}} 
\newcommand{\hatcurCCtassmBshorteccenxxxxxF}{\ensuremath{14.7}}      
\newcommand{\hatcurCCtassmIeccenxxxxxF}{\ensuremath{nff\pmnff}}      
\newcommand{\hatcurCCtassmIshorteccenxxxxxF}{\ensuremath{0.0}}       
\newcommand{\hatcurCCtassmgeccenxxxxxF}{\ensuremath{14.258\pm0.026}} 
\newcommand{\hatcurCCtassmgshorteccenxxxxxF}{\ensuremath{14.3}}      
\newcommand{\hatcurCCtassmreccenxxxxxF}{\ensuremath{13.418\pm0.093}} 
\newcommand{\hatcurCCtassmrshorteccenxxxxxF}{\ensuremath{13.4}}      
\newcommand{\hatcurCCtassmieccenxxxxxF}{\ensuremath{13.136\pm0.086}} 
\newcommand{\hatcurCCtassmishorteccenxxxxxF}{\ensuremath{13.1}}      
\newcommand{\hatcurCCparallaxeccenxxxxxF}{\ensuremath{2.450\pm0.024}} 
\newcommand{\hatcurCCgaiamGeccenxxxxxF}{\ensuremath{13.51060\pm0.00030}} 
\newcommand{\hatcurCCgaiamBPeccenxxxxxF}{\ensuremath{14.0381\pm0.0012}} 
\newcommand{\hatcurCCgaiamRPeccenxxxxxF}{\ensuremath{12.84760\pm0.00080}} 
\newcommand{\hatcurCCtwomassJmageccenxxxxxF}{\ensuremath{12.021\pm0.021}} 
\newcommand{\hatcurCCtwomassHmageccenxxxxxF}{\ensuremath{11.630\pm0.028}} 
\newcommand{\hatcurCCtwomassKmageccenxxxxxF}{\ensuremath{11.512\pm0.023}} 
\newcommand{\hatcurCCcitJmageccenxxxxxF}{\ensuremath{12.030\pm0.022}} 
\newcommand{\hatcurCCcitHmageccenxxxxxF}{\ensuremath{11.623\pm0.028}} 
\newcommand{\hatcurCCcitKmageccenxxxxxF}{\ensuremath{11.536\pm0.023}} 
\newcommand{\hatcurCCbbJmageccenxxxxxF}{\ensuremath{12.091\pm0.024}} 
\newcommand{\hatcurCCbbHmageccenxxxxxF}{\ensuremath{11.646\pm0.029}} 
\newcommand{\hatcurCCbbKmageccenxxxxxF}{\ensuremath{11.556\pm0.023}} 
\newcommand{\hatcurCCesoJmageccenxxxxxF}{\ensuremath{12.095\pm0.026}} 
\newcommand{\hatcurCCesoHmageccenxxxxxF}{\ensuremath{11.643\pm0.034}} 
\newcommand{\hatcurCCesoKmageccenxxxxxF}{\ensuremath{11.554\pm0.024}} 
\newcommand{\hatcurCCesoJHmageccenxxxxxF}{\ensuremath{0.452\pm0.040}} 
\newcommand{\hatcurCCesoJKmageccenxxxxxF}{\ensuremath{0.541\pm0.034}} 
\newcommand{\hatcurCCesoHKmageccenxxxxxF}{\ensuremath{0.089\pm0.042}} 
\newcommand{\hatcurCCWonemageccenxxxxxF}{\ensuremath{11.453\pm0.022}} 
\newcommand{\hatcurCCWtwomageccenxxxxxF}{\ensuremath{11.514\pm0.022}} 
\newcommand{\hatcurCCWthreemageccenxxxxxF}{\ensuremath{0\pm0}}       
\newcommand{\hatcurCCWfourmageccenxxxxxF}{\ensuremath{0\pm0}}        
\newcommand{\hatcurLCdipeccenxxxxxF}{\ensuremath{17.6}}              
\newcommand{\hatcurLCrprstareccenxxxxxF}{\ensuremath{0.1225\pm0.0039}} 
\newcommand{\hatcurLCbsqeccenxxxxxF}{\ensuremath{0.130_{-0.058}^{+0.054}}} 
\newcommand{\hatcurLCimpeccenxxxxxF}{\ensuremath{0.360_{-0.091}^{+0.069}}} 
\newcommand{\hatcurLCzetaeccenxxxxxF}{\ensuremath{18.77_{-0.35}^{+0.26}}} 
\newcommand{\hatcurLCdureccenxxxxxF}{\ensuremath{0.1215\pm0.0017}}   
\newcommand{\hatcurLCdurshorteccenxxxxxF}{\ensuremath{0.1215}}       
\newcommand{\hatcurLCdurhreccenxxxxxF}{\ensuremath{2.917\pm0.041}}   
\newcommand{\hatcurLCdurhrshorteccenxxxxxF}{\ensuremath{2.917}}      
\newcommand{\hatcurLCqeccenxxxxxF}{\ensuremath{0.03600\pm0.00050}}   
\newcommand{\hatcurLCqshorteccenxxxxxF}{\ensuremath{0.036}}          
\newcommand{\hatcurLCingdureccenxxxxxF}{\ensuremath{0.0151\pm0.0011}} 
\newcommand{\hatcurLCPeccenxxxxxF}{\ensuremath{3.377729\pm0.000011}} 
\newcommand{\hatcurLCPprececcenxxxxxF}{\ensuremath{3.3777293}}       
\newcommand{\hatcurLCPshorteccenxxxxxF}{\ensuremath{3.3777}}         
\newcommand{\hatcurLCTeccenxxxxxF}{\ensuremath{2456376.18692\pm0.00061}} 
\newcommand{\hatcurLCTAeccenxxxxxF}{\ensuremath{2454964.2962\pm0.0048}} 
\newcommand{\hatcurLCTBeccenxxxxxF}{\ensuremath{2456399.83104\pm0.00061}} 
\newcommand{\hatcurLCrhoeccenxxxxxF}{\ensuremath{1.447_{-0.065}^{+0.087}}} 
\newcommand{\hatcurSMEiteffeccenxxxxxF}{\ensuremath{5365\pm50}}      
\newcommand{\hatcurSMEizfeheccenxxxxxF}{\ensuremath{0.428\pm0.080}}  
\newcommand{\hatcurSMEizfehshorteccenxxxxxF}{\ensuremath{0.43}}      
\newcommand{\hatcurSMEiloggeccenxxxxxF}{\ensuremath{4.40\pm0.10}}    
\newcommand{\hatcurSMEivsineccenxxxxxF}{\ensuremath{3.22\pm0.50}}    
\newcommand{\hatcurSMEivmaceccenxxxxxF}{\ensuremath{nff\pmnff}}      
\newcommand{\hatcurSMEivmiceccenxxxxxF}{\ensuremath{nff\pmnff}}      
\newcommand{\hatcurLBiBeccenxxxxxF}{\ensuremath{0.8183}}             
\newcommand{\hatcurLBiiBeccenxxxxxF}{\ensuremath{0.0186}}            
\newcommand{\hatcurLBiVeccenxxxxxF}{\ensuremath{0.6325}}             
\newcommand{\hatcurLBiiVeccenxxxxxF}{\ensuremath{0.1208}}            
\newcommand{\hatcurLBiReccenxxxxxF}{\ensuremath{0.5132}}             
\newcommand{\hatcurLBiiReccenxxxxxF}{\ensuremath{0.1589}}            
\newcommand{\hatcurLBiIeccenxxxxxF}{\ensuremath{0.4073}}             
\newcommand{\hatcurLBiiIeccenxxxxxF}{\ensuremath{0.1780}}            
\newcommand{\hatcurLBiueccenxxxxxF}{\ensuremath{1.0107}}             
\newcommand{\hatcurLBiiueccenxxxxxF}{\ensuremath{-0.1834}}           
\newcommand{\hatcurLBigeccenxxxxxF}{\ensuremath{0.7517}}             
\newcommand{\hatcurLBiigeccenxxxxxF}{\ensuremath{0.0481}}            
\newcommand{\hatcurLBireccenxxxxxF}{\ensuremath{0.52\pm0.16}}        
\newcommand{\hatcurLBiireccenxxxxxF}{\ensuremath{0.30\pm0.16}}       
\newcommand{\hatcurLBiieccenxxxxxF}{\ensuremath{0.42\pm0.14}}        
\newcommand{\hatcurLBiiieccenxxxxxF}{\ensuremath{0.37\pm0.17}}       
\newcommand{\hatcurLBizeccenxxxxxF}{\ensuremath{0.3593}}             
\newcommand{\hatcurLBiizeccenxxxxxF}{\ensuremath{0.1834}}            
\newcommand{\hatcurLBiJeccenxxxxxF}{\ensuremath{0.2380}}             
\newcommand{\hatcurLBiiJeccenxxxxxF}{\ensuremath{0.2252}}            
\newcommand{\hatcurLBiHeccenxxxxxF}{\ensuremath{0.1368}}             
\newcommand{\hatcurLBiiHeccenxxxxxF}{\ensuremath{0.2704}}            
\newcommand{\hatcurLBiKeccenxxxxxF}{\ensuremath{0.1239}}             
\newcommand{\hatcurLBiiKeccenxxxxxF}{\ensuremath{0.2193}}            
\newcommand{\hatcurLBiTeccenxxxxxF}{\ensuremath{0.3989}}             
\newcommand{\hatcurLBiiTeccenxxxxxF}{\ensuremath{0.2029}}            
\newcommand{\hatcurLBikepeccenxxxxxF}{\ensuremath{0.5196}}           
\newcommand{\hatcurLBiikepeccenxxxxxF}{\ensuremath{0.1699}}          
\newcommand{\hatcurLBiCeccenxxxxxF}{\ensuremath{0.5069}}             
\newcommand{\hatcurLBiiCeccenxxxxxF}{\ensuremath{0.1693}}            
\newcommand{\hatcurLBiMeccenxxxxxF}{\ensuremath{0.6102}}             
\newcommand{\hatcurLBiiMeccenxxxxxF}{\ensuremath{0.1350}}            
\newcommand{\hatcurLBiSoneeccenxxxxxF}{\ensuremath{0.1019}}            
\newcommand{\hatcurLBiiSoneeccenxxxxxF}{\ensuremath{0.1351}}           
\newcommand{\hatcurLBiStwoeccenxxxxxF}{\ensuremath{0.0886}}            
\newcommand{\hatcurLBiiStwoeccenxxxxxF}{\ensuremath{0.1097}}           
\newcommand{\hatcurLBiSthreeeccenxxxxxF}{\ensuremath{0.0735}}            
\newcommand{\hatcurLBiiSthreeeccenxxxxxF}{\ensuremath{0.0906}}           
\newcommand{\hatcurLBiSfoureccenxxxxxF}{\ensuremath{0.0584}}            
\newcommand{\hatcurLBiiSfoureccenxxxxxF}{\ensuremath{0.0727}}           
\newcommand{\hatcurISOmeccenxxxxxF}{\ensuremath{0.927\pm0.023}}      
\newcommand{\hatcurISOmshorteccenxxxxxF}{\ensuremath{0.93}}          
\newcommand{\hatcurISOmlongeccenxxxxxF}{\ensuremath{0.927\pm0.023}}  
\newcommand{\hatcurISOreccenxxxxxF}{\ensuremath{0.966\pm0.010}}      
\newcommand{\hatcurISOrshorteccenxxxxxF}{\ensuremath{0.97}}          
\newcommand{\hatcurISOrlongeccenxxxxxF}{\ensuremath{0.966\pm0.010}}  
\newcommand{\hatcurISOrhoeccenxxxxxF}{\ensuremath{1.447_{-0.065}^{+0.087}}} 
\newcommand{\hatcurISOrholongeccenxxxxxF}{\ensuremath{1.447_{-0.065}^{+0.087}}} 
\newcommand{\hatcurISOloggeccenxxxxxF}{\ensuremath{4.435\pm0.018}}   
\newcommand{\hatcurISOlumeccenxxxxxF}{\ensuremath{0.709\pm0.027}}    
\newcommand{\hatcurISOlumshorteccenxxxxxF}{\ensuremath{0.71}}        
\newcommand{\hatcurISOteffeccenxxxxxF}{\ensuremath{5395\pm46}}       
\newcommand{\hatcurISOzfeheccenxxxxxF}{\ensuremath{0.270\pm0.065}}   
\newcommand{\hatcurISOageeccenxxxxxF}{\ensuremath{9.0\pm1.9}}        
\newcommand{\hatcurISOspececcenxxxxxF}{G}                            
\newcommand{\hatcurRVKeccenxxxxxF}{\ensuremath{87.9\pm3.1}}          
\newcommand{\hatcurRVrkeccenxxxxxF}{\ensuremath{-0.087\pm0.087}}     
\newcommand{\hatcurRVrheccenxxxxxF}{\ensuremath{-0.105\pm0.091}}     
\newcommand{\hatcurRVkeccenxxxxxF}{\ensuremath{-0.013_{-0.024}^{+0.015}}} 
\newcommand{\hatcurRVheccenxxxxxF}{\ensuremath{-0.016_{-0.023}^{+0.015}}} 
\newcommand{\hatcurRVtroneeccenxxxxxF}{\ensuremath{0\pm0}}           
\newcommand{\hatcurRVtrtwoeccenxxxxxF}{\ensuremath{0\pm0}}           
\newcommand{\hatcurRVgammaAeccenxxxxxF}{\ensuremath{7.7\pm2.8}}      
\newcommand{\hatcurRVjitterAeccenxxxxxF}{\ensuremath{0.1\pm1.5}}     
\newcommand{\hatcurRVjittertwosiglimAeccenxxxxxF}{\ensuremath{<3.6}} 
\newcommand{\hatcurRVfitrmsAeccenxxxxxF}{\ensuremath{0.0}}           
\newcommand{\hatcurRVgammaBeccenxxxxxF}{\ensuremath{-94\pm29}}       
\newcommand{\hatcurRVjitterBeccenxxxxxF}{\ensuremath{0.02\pm0.75}}   
\newcommand{\hatcurRVjittertwosiglimBeccenxxxxxF}{\ensuremath{<1.5}} 
\newcommand{\hatcurRVfitrmsBeccenxxxxxF}{\ensuremath{0.0}}           
\newcommand{\hatcurRVgammaCeccenxxxxxF}{\ensuremath{-69595.7\pm7.5}} 
\newcommand{\hatcurRVjitterCeccenxxxxxF}{\ensuremath{16\pm10}}       
\newcommand{\hatcurRVjittertwosiglimCeccenxxxxxF}{\ensuremath{<34.5}} 
\newcommand{\hatcurRVfitrmsCeccenxxxxxF}{\ensuremath{0.0}}           
\newcommand{\hatcurRVecceneccenxxxxxF}{\ensuremath{0.028\pm0.021}}   
\newcommand{\hatcurRVeccentwosiglimeccenxxxxxF}{\ensuremath{<0.069}} 
\newcommand{\hatcurRVomegaeccenxxxxxF}{\ensuremath{229\pm59}}        
\newcommand{\hatcurPPieccenxxxxxF}{\ensuremath{87.87_{-0.38}^{+0.53}}} 
\newcommand{\hatcurPPgeccenxxxxxF}{\ensuremath{11.46_{-0.75}^{+1.25}}} 
\newcommand{\hatcurPPloggeccenxxxxxF}{\ensuremath{3.059_{-0.029}^{+0.045}}} 
\newcommand{\hatcurPPareccenxxxxxF}{\ensuremath{9.56\pm0.16}}        
\newcommand{\hatcurPPareleccenxxxxxF}{\ensuremath{0.04297\pm0.00036}} 
\newcommand{\hatcurPPrhoeccenxxxxxF}{\ensuremath{0.500_{-0.046}^{+0.073}}} 
\newcommand{\hatcurPPmeccenxxxxxF}{\ensuremath{0.617\pm0.024}}       
\newcommand{\hatcurPPmshorteccenxxxxxF}{\ensuremath{0.62}}           
\newcommand{\hatcurPPmlongeccenxxxxxF}{\ensuremath{0.617\pm0.024}}   
\newcommand{\hatcurPPmeeccenxxxxxF}{\ensuremath{196.0\pm7.7}}        
\newcommand{\hatcurPPmeshorteccenxxxxxF}{\ensuremath{196.0}}         
\newcommand{\hatcurPPmelongeccenxxxxxF}{\ensuremath{196.0\pm7.7}}    
\newcommand{\hatcurPPreccenxxxxxF}{\ensuremath{1.151\pm0.041}}       
\newcommand{\hatcurPPrshorteccenxxxxxF}{\ensuremath{1.15}}           
\newcommand{\hatcurPPrlongeccenxxxxxF}{\ensuremath{1.151\pm0.041}}   
\newcommand{\hatcurPPreeccenxxxxxF}{\ensuremath{12.90\pm0.46}}       
\newcommand{\hatcurPPreshorteccenxxxxxF}{\ensuremath{12.9}}          
\newcommand{\hatcurPPrelongeccenxxxxxF}{\ensuremath{12.90\pm0.46}}   
\newcommand{\hatcurPPmrcorreccenxxxxxF}{\ensuremath{-0.17}}          
\newcommand{\hatcurPPteffeccenxxxxxF}{\ensuremath{1234\pm11}}        
\newcommand{\hatcurPPthetaeccenxxxxxF}{\ensuremath{0.0494_{-0.0021}^{+0.0029}}} 
\newcommand{\hatcurPPfluxperieccenxxxxxF}{\ensuremath{5.53\pm0.31}}  
\newcommand{\hatcurPPfluxperidimeccenxxxxxF}{\ensuremath{8}}         
\newcommand{\hatcurPPfluxapeccenxxxxxF}{\ensuremath{4.93\pm0.28}}    
\newcommand{\hatcurPPfluxapdimeccenxxxxxF}{\ensuremath{8}}           
\newcommand{\hatcurPPfluxavgeccenxxxxxF}{\ensuremath{5.22\pm0.19}}   
\newcommand{\hatcurPPfluxavgdimeccenxxxxxF}{\ensuremath{8}}          
\newcommand{\hatcurPPfluxavglogeccenxxxxxF}{\ensuremath{8.718\pm0.016}} 
\newcommand{\hatcurXsecphaseeccenxxxxxF}{\ensuremath{0.492\pm0.012}} 
\newcommand{\hatcurXsecondaryeccenxxxxxF}{\ensuremath{2456377.848\pm0.041}} 
\newcommand{\hatcurXsecdureccenxxxxxF}{\ensuremath{0.1180\pm0.0042}} 
\newcommand{\hatcurXsecingdureccenxxxxxF}{\ensuremath{0.01439\pm0.00074}} 
\newcommand{\hatcurPPphiconjeccenxxxxxF}{\ensuremath{-0.337_{-0.091}^{+0.551}}} 
\newcommand{\hatcurPPperieccenxxxxxF}{\ensuremath{2456377.3\pm1.1}}  
\newcommand{\hatcurPPaequiveccenxxxxxF}{\ensuremath{0.05110\pm0.00094}} 
\newcommand{\hatcurPPtcirceccenxxxxxF}{\ensuremath{171_{-26}^{+42}}} 
\newcommand{\hatcurPPtinfalleccenxxxxxF}{\ensuremath{3880\pm400}}    
\newcommand{\hatcurXdisteccenxxxxxF}{\ensuremath{408.7\pm3.5}}       
\newcommand{\hatcurXAveccenxxxxxF}{\ensuremath{0.490_{-0.042}^{+0.056}}} 
\newcommand{\hatcurXdistredeccenxxxxxF}{\ensuremath{408.7\pm3.5}}    
\newcommand{\hatcurXEBVeccenxxxxxF}{\ensuremath{0.158\pm0.016}}      
\newcommand{\hatcurCCpmraeccenxxxxxF}{\ensuremath{-14.871\pm0.036}}  
\newcommand{\hatcurCCpmdececcenxxxxxF}{\ensuremath{-0.301\pm0.039}}  
\newcommand{\hatcurCCpmeccenxxxxxF}{\ensuremath{14.874\pm0.053}}     

%% file: HTR405-001.eccen.tex
\newcommand{\hatcurhtreccenxxxxxG}{HTR405-001}                       
\newcommand{\hatcurfieldeccenxxxxxG}{\ensuremath{string}}            
\newcommand{\hatcurCCraeccenxxxxxG}{\ensuremath{04^{\mathrm h}35^{\mathrm m}53.8469{\mathrm s}}}                   
\newcommand{\hatcurCCdececcenxxxxxG}{\ensuremath{+02{\arcdeg}25{\arcmin}52.6434{\arcsec}}}                 
\newcommand{\hatcurCCmageccenxxxxxG}{12.771}                         
\newcommand{\hatcurCCtwomasseccenxxxxxG}{2MASS~04355384+0225526}     
\newcommand{\hatcurCCgsceccenxxxxxG}{GSC~0086-00341}                 
\newcommand{\hatcurCCgaiaeccenxxxxxG}{GAIA~3279418602369232000}      
\newcommand{\hatcurCCgaiadrtwoeccenxxxxxG}{GAIA~DR2~3279418602369232000} 
\newcommand{\hatcurCCtassmveccenxxxxxG}{\ensuremath{12.771\pm0.010}} 
\newcommand{\hatcurCCtassmvshorteccenxxxxxG}{\ensuremath{12.8}}      
\newcommand{\hatcurCCtassmBeccenxxxxxG}{\ensuremath{13.446\pm0.011}} 
\newcommand{\hatcurCCtassmBshorteccenxxxxxG}{\ensuremath{13.4}}      
\newcommand{\hatcurCCtassmIeccenxxxxxG}{\ensuremath{12.105\pm0.070}} 
\newcommand{\hatcurCCtassmIshorteccenxxxxxG}{\ensuremath{12.1}}      
\newcommand{\hatcurCCtassmgeccenxxxxxG}{\ensuremath{13.062\pm0.013}} 
\newcommand{\hatcurCCtassmgshorteccenxxxxxG}{\ensuremath{13.1}}      
\newcommand{\hatcurCCtassmreccenxxxxxG}{\ensuremath{12.553\pm0.075}} 
\newcommand{\hatcurCCtassmrshorteccenxxxxxG}{\ensuremath{12.6}}      
\newcommand{\hatcurCCtassmieccenxxxxxG}{\ensuremath{12.440\pm0.037}} 
\newcommand{\hatcurCCtassmishorteccenxxxxxG}{\ensuremath{12.4}}      
\newcommand{\hatcurCCparallaxeccenxxxxxG}{\ensuremath{1.505\pm0.035}} 
\newcommand{\hatcurCCgaiamGeccenxxxxxG}{\ensuremath{12.62210\pm0.00010}} 
\newcommand{\hatcurCCgaiamBPeccenxxxxxG}{\ensuremath{12.98580\pm0.00080}} 
\newcommand{\hatcurCCgaiamRPeccenxxxxxG}{\ensuremath{12.09530\pm0.00070}} 
\newcommand{\hatcurCCtwomassJmageccenxxxxxG}{\ensuremath{11.485\pm0.026}} 
\newcommand{\hatcurCCtwomassHmageccenxxxxxG}{\ensuremath{11.225\pm0.025}} 
\newcommand{\hatcurCCtwomassKmageccenxxxxxG}{\ensuremath{11.123\pm0.021}} 
\newcommand{\hatcurCCcitJmageccenxxxxxG}{\ensuremath{11.502\pm0.026}} 
\newcommand{\hatcurCCcitHmageccenxxxxxG}{\ensuremath{11.219\pm0.025}} 
\newcommand{\hatcurCCcitKmageccenxxxxxG}{\ensuremath{11.147\pm0.021}} 
\newcommand{\hatcurCCbbJmageccenxxxxxG}{\ensuremath{11.551\pm0.028}} 
\newcommand{\hatcurCCbbHmageccenxxxxxG}{\ensuremath{11.241\pm0.026}} 
\newcommand{\hatcurCCbbKmageccenxxxxxG}{\ensuremath{11.167\pm0.021}} 
\newcommand{\hatcurCCesoJmageccenxxxxxG}{\ensuremath{11.553\pm0.029}} 
\newcommand{\hatcurCCesoHmageccenxxxxxG}{\ensuremath{11.238\pm0.030}} 
\newcommand{\hatcurCCesoKmageccenxxxxxG}{\ensuremath{11.166\pm0.022}} 
\newcommand{\hatcurCCesoJHmageccenxxxxxG}{\ensuremath{0.315\pm0.040}} 
\newcommand{\hatcurCCesoJKmageccenxxxxxG}{\ensuremath{0.387\pm0.036}} 
\newcommand{\hatcurCCesoHKmageccenxxxxxG}{\ensuremath{0.072\pm0.038}} 
\newcommand{\hatcurCCWonemageccenxxxxxG}{\ensuremath{11.058\pm0.021}} 
\newcommand{\hatcurCCWtwomageccenxxxxxG}{\ensuremath{11.055\pm0.021}} 
\newcommand{\hatcurCCWthreemageccenxxxxxG}{\ensuremath{0\pm0}}       
\newcommand{\hatcurCCWfourmageccenxxxxxG}{\ensuremath{0\pm0}}        
\newcommand{\hatcurLCdipeccenxxxxxG}{\ensuremath{11.2}}              
\newcommand{\hatcurLCrprstareccenxxxxxG}{\ensuremath{0.1031\pm0.0038}} 
\newcommand{\hatcurLCbsqeccenxxxxxG}{\ensuremath{0.025_{-0.018}^{+0.030}}} 
\newcommand{\hatcurLCimpeccenxxxxxG}{\ensuremath{0.157_{-0.073}^{+0.077}}} 
\newcommand{\hatcurLCzetaeccenxxxxxG}{\ensuremath{10.83\pm0.11}}     
\newcommand{\hatcurLCdureccenxxxxxG}{\ensuremath{0.2043\pm0.0023}}   
\newcommand{\hatcurLCdurshorteccenxxxxxG}{\ensuremath{0.2043}}       
\newcommand{\hatcurLCdurhreccenxxxxxG}{\ensuremath{4.902\pm0.056}}   
\newcommand{\hatcurLCdurhrshorteccenxxxxxG}{\ensuremath{4.902}}      
\newcommand{\hatcurLCqeccenxxxxxG}{\ensuremath{0.05100\pm0.00058}}   
\newcommand{\hatcurLCqshorteccenxxxxxG}{\ensuremath{0.051}}          
\newcommand{\hatcurLCingdureccenxxxxxG}{\ensuremath{0.01961\pm0.00098}} 
\newcommand{\hatcurLCPeccenxxxxxG}{\ensuremath{4.0072318\pm0.0000016}} 
\newcommand{\hatcurLCPprececcenxxxxxG}{\ensuremath{4.0072318}}       
\newcommand{\hatcurLCPshorteccenxxxxxG}{\ensuremath{4.0072}}         
\newcommand{\hatcurLCTeccenxxxxxG}{\ensuremath{2457430.88497\pm0.00062}} 
\newcommand{\hatcurLCTAeccenxxxxxG}{\ensuremath{2455078.6400\pm0.0010}} 
\newcommand{\hatcurLCTBeccenxxxxxG}{\ensuremath{2458460.74353\pm0.00079}} 
\newcommand{\hatcurLChatnetmAeccenxxxxxG}{\ensuremath{12.52213\pm0.00012}} 
\newcommand{\hatcurLCiblendAeccenxxxxxG}{\ensuremath{0.720\pm0.066}} 
\newcommand{\hatcurLChatnetmBeccenxxxxxG}{\ensuremath{0.000060\pm0.000021}} 
\newcommand{\hatcurLCiblendBeccenxxxxxG}{\ensuremath{0.828\pm0.066}} 
\newcommand{\hatcurLCrhoeccenxxxxxG}{\ensuremath{0.328\pm0.020}}     
\newcommand{\hatcurSMEiteffeccenxxxxxG}{\ensuremath{6302\pm50}}      
\newcommand{\hatcurSMEizfeheccenxxxxxG}{\ensuremath{-0.010\pm0.080}} 
\newcommand{\hatcurSMEizfehshorteccenxxxxxG}{\ensuremath{-0.01}}     
\newcommand{\hatcurSMEiloggeccenxxxxxG}{\ensuremath{3.99\pm0.10}}    
\newcommand{\hatcurSMEivsineccenxxxxxG}{\ensuremath{12.70\pm0.50}}   
\newcommand{\hatcurSMEivmaceccenxxxxxG}{\ensuremath{nff\pmnff}}      
\newcommand{\hatcurSMEivmiceccenxxxxxG}{\ensuremath{nff\pmnff}}      
\newcommand{\hatcurLBiBeccenxxxxxG}{\ensuremath{0.5991}}             
\newcommand{\hatcurLBiiBeccenxxxxxG}{\ensuremath{0.2117}}            
\newcommand{\hatcurLBiVeccenxxxxxG}{\ensuremath{0.4809}}             
\newcommand{\hatcurLBiiVeccenxxxxxG}{\ensuremath{0.2207}}            
\newcommand{\hatcurLBiReccenxxxxxG}{\ensuremath{0.4086}}             
\newcommand{\hatcurLBiiReccenxxxxxG}{\ensuremath{0.2085}}            
\newcommand{\hatcurLBiIeccenxxxxxG}{\ensuremath{0.3221}}             
\newcommand{\hatcurLBiiIeccenxxxxxG}{\ensuremath{0.2018}}            
\newcommand{\hatcurLBiueccenxxxxxG}{\ensuremath{0.6931}}             
\newcommand{\hatcurLBiiueccenxxxxxG}{\ensuremath{0.1202}}            
\newcommand{\hatcurLBigeccenxxxxxG}{\ensuremath{0.5665}}             
\newcommand{\hatcurLBiigeccenxxxxxG}{\ensuremath{0.1994}}            
\newcommand{\hatcurLBireccenxxxxxG}{\ensuremath{0.21\pm0.13}}        
\newcommand{\hatcurLBiireccenxxxxxG}{\ensuremath{0.30\pm0.19}}       
\newcommand{\hatcurLBiieccenxxxxxG}{\ensuremath{0.30\pm0.13}}        
\newcommand{\hatcurLBiiieccenxxxxxG}{\ensuremath{0.26\pm0.15}}       
\newcommand{\hatcurLBizeccenxxxxxG}{\ensuremath{0.2782}}             
\newcommand{\hatcurLBiizeccenxxxxxG}{\ensuremath{0.2037}}            
\newcommand{\hatcurLBiJeccenxxxxxG}{\ensuremath{0.1869}}             
\newcommand{\hatcurLBiiJeccenxxxxxG}{\ensuremath{0.2193}}            
\newcommand{\hatcurLBiHeccenxxxxxG}{\ensuremath{0.1166}}             
\newcommand{\hatcurLBiiHeccenxxxxxG}{\ensuremath{0.2345}}            
\newcommand{\hatcurLBiKeccenxxxxxG}{\ensuremath{0.1148}}             
\newcommand{\hatcurLBiiKeccenxxxxxG}{\ensuremath{0.1811}}            
\newcommand{\hatcurLBiTeccenxxxxxG}{\ensuremath{0.27\pm0.10}}        
\newcommand{\hatcurLBiiTeccenxxxxxG}{\ensuremath{0.37\pm0.17}}       
\newcommand{\hatcurLBikepeccenxxxxxG}{\ensuremath{0.4509}}           
\newcommand{\hatcurLBiikepeccenxxxxxG}{\ensuremath{0.1998}}          
\newcommand{\hatcurLBiCeccenxxxxxG}{\ensuremath{0.4428}}             
\newcommand{\hatcurLBiiCeccenxxxxxG}{\ensuremath{0.1995}}            
\newcommand{\hatcurLBiMeccenxxxxxG}{\ensuremath{0.5248}}             
\newcommand{\hatcurLBiiMeccenxxxxxG}{\ensuremath{0.1878}}            
\newcommand{\hatcurLBiSoneeccenxxxxxG}{\ensuremath{0.0852}}            
\newcommand{\hatcurLBiiSoneeccenxxxxxG}{\ensuremath{0.1124}}           
\newcommand{\hatcurLBiStwoeccenxxxxxG}{\ensuremath{0.0722}}            
\newcommand{\hatcurLBiiStwoeccenxxxxxG}{\ensuremath{0.0951}}           
\newcommand{\hatcurLBiSthreeeccenxxxxxG}{\ensuremath{0.0596}}            
\newcommand{\hatcurLBiiSthreeeccenxxxxxG}{\ensuremath{0.0788}}           
\newcommand{\hatcurLBiSfoureccenxxxxxG}{\ensuremath{0.0452}}            
\newcommand{\hatcurLBiiSfoureccenxxxxxG}{\ensuremath{0.0629}}           
\newcommand{\hatcurISOmeccenxxxxxG}{\ensuremath{1.299\pm0.029}}      
\newcommand{\hatcurISOmshorteccenxxxxxG}{\ensuremath{1.30}}          
\newcommand{\hatcurISOmlongeccenxxxxxG}{\ensuremath{1.299\pm0.029}}  
\newcommand{\hatcurISOreccenxxxxxG}{\ensuremath{1.777\pm0.038}}      
\newcommand{\hatcurISOrshorteccenxxxxxG}{\ensuremath{1.78}}          
\newcommand{\hatcurISOrlongeccenxxxxxG}{\ensuremath{1.777\pm0.038}}  
\newcommand{\hatcurISOrhoeccenxxxxxG}{\ensuremath{0.328\pm0.020}}    
\newcommand{\hatcurISOrholongeccenxxxxxG}{\ensuremath{0.328\pm0.020}} 
\newcommand{\hatcurISOloggeccenxxxxxG}{\ensuremath{4.054\pm0.018}}   
\newcommand{\hatcurISOlumeccenxxxxxG}{\ensuremath{4.79\pm0.24}}      
\newcommand{\hatcurISOlumshorteccenxxxxxG}{\ensuremath{4.79}}        
\newcommand{\hatcurISOteffeccenxxxxxG}{\ensuremath{6416\pm42}}       
\newcommand{\hatcurISOzfeheccenxxxxxG}{\ensuremath{-0.116_{-0.055}^{+0.035}}} 
\newcommand{\hatcurISOageeccenxxxxxG}{\ensuremath{2.97_{-0.22}^{+0.14}}} 
\newcommand{\hatcurISOspececcenxxxxxG}{F}                            
\newcommand{\hatcurRVKeccenxxxxxG}{\ensuremath{54.1\pm5.8}}          
\newcommand{\hatcurRVrkeccenxxxxxG}{\ensuremath{-0.00\pm0.14}}       
\newcommand{\hatcurRVrheccenxxxxxG}{\ensuremath{0.181_{-0.092}^{+0.059}}} 
\newcommand{\hatcurRVkeccenxxxxxG}{\ensuremath{-0.000_{-0.026}^{+0.036}}} 
\newcommand{\hatcurRVheccenxxxxxG}{\ensuremath{0.040\pm0.023}}       
\newcommand{\hatcurRVtroneeccenxxxxxG}{\ensuremath{0\pm0}}           
\newcommand{\hatcurRVtrtwoeccenxxxxxG}{\ensuremath{0\pm0}}           
\newcommand{\hatcurRVgammaAeccenxxxxxG}{\ensuremath{2.6\pm6.1}}      
\newcommand{\hatcurRVjitterAeccenxxxxxG}{\ensuremath{18.0\pm8.4}}    
\newcommand{\hatcurRVjittertwosiglimAeccenxxxxxG}{\ensuremath{<35.2}} 
\newcommand{\hatcurRVfitrmsAeccenxxxxxG}{\ensuremath{0.0}}           
\newcommand{\hatcurRVgammaBeccenxxxxxG}{\ensuremath{24488\pm15}}     
\newcommand{\hatcurRVjitterBeccenxxxxxG}{\ensuremath{2\pm19}}        
\newcommand{\hatcurRVjittertwosiglimBeccenxxxxxG}{\ensuremath{<44.9}} 
\newcommand{\hatcurRVfitrmsBeccenxxxxxG}{\ensuremath{0.0}}           
\newcommand{\hatcurRVecceneccenxxxxxG}{\ensuremath{0.050\pm0.031}}   
\newcommand{\hatcurRVeccentwosiglimeccenxxxxxG}{\ensuremath{<0.101}} 
\newcommand{\hatcurRVomegaeccenxxxxxG}{\ensuremath{92\pm46}}         
\newcommand{\hatcurPPieccenxxxxxG}{\ensuremath{88.58\pm0.68}}        
\newcommand{\hatcurPPgeccenxxxxxG}{\ensuremath{3.90_{-0.39}^{+0.56}}} 
\newcommand{\hatcurPPloggeccenxxxxxG}{\ensuremath{2.591\pm0.057}}    
\newcommand{\hatcurPPareccenxxxxxG}{\ensuremath{6.53\pm0.13}}        
\newcommand{\hatcurPPareleccenxxxxxG}{\ensuremath{0.05388\pm0.00041}} 
\newcommand{\hatcurPPrhoeccenxxxxxG}{\ensuremath{0.110_{-0.014}^{+0.020}}} 
\newcommand{\hatcurPPmeccenxxxxxG}{\ensuremath{0.503\pm0.052}}       
\newcommand{\hatcurPPmshorteccenxxxxxG}{\ensuremath{0.50}}           
\newcommand{\hatcurPPmlongeccenxxxxxG}{\ensuremath{0.503\pm0.052}}   
\newcommand{\hatcurPPmeeccenxxxxxG}{\ensuremath{160\pm16}}           
\newcommand{\hatcurPPmeshorteccenxxxxxG}{\ensuremath{160.0}}         
\newcommand{\hatcurPPmelongeccenxxxxxG}{\ensuremath{160\pm16}}       
\newcommand{\hatcurPPreccenxxxxxG}{\ensuremath{1.780\pm0.077}}       
\newcommand{\hatcurPPrshorteccenxxxxxG}{\ensuremath{1.78}}           
\newcommand{\hatcurPPrlongeccenxxxxxG}{\ensuremath{1.780\pm0.077}}   
\newcommand{\hatcurPPreeccenxxxxxG}{\ensuremath{19.95\pm0.86}}       
\newcommand{\hatcurPPreshorteccenxxxxxG}{\ensuremath{19.9}}          
\newcommand{\hatcurPPrelongeccenxxxxxG}{\ensuremath{19.95\pm0.86}}   
\newcommand{\hatcurPPmrcorreccenxxxxxG}{\ensuremath{0.04}}           
\newcommand{\hatcurPPteffeccenxxxxxG}{\ensuremath{1778\pm20}}        
\newcommand{\hatcurPPthetaeccenxxxxxG}{\ensuremath{0.0233_{-0.0021}^{+0.0028}}} 
\newcommand{\hatcurPPfluxperieccenxxxxxG}{\ensuremath{2.50\pm0.24}}  
\newcommand{\hatcurPPfluxperidimeccenxxxxxG}{\ensuremath{9}}         
\newcommand{\hatcurPPfluxapeccenxxxxxG}{\ensuremath{2.04\pm0.11}}    
\newcommand{\hatcurPPfluxapdimeccenxxxxxG}{\ensuremath{9}}           
\newcommand{\hatcurPPfluxavgeccenxxxxxG}{\ensuremath{2.25\pm0.10}}   
\newcommand{\hatcurPPfluxavgdimeccenxxxxxG}{\ensuremath{9}}          
\newcommand{\hatcurPPfluxavglogeccenxxxxxG}{\ensuremath{9.353\pm0.020}} 
\newcommand{\hatcurXsecphaseeccenxxxxxG}{\ensuremath{0.500\pm0.025}} 
\newcommand{\hatcurXsecondaryeccenxxxxxG}{\ensuremath{2457432.89\pm0.10}} 
\newcommand{\hatcurXsecdureccenxxxxxG}{\ensuremath{0.2208\pm0.0095}} 
\newcommand{\hatcurXsecingdureccenxxxxxG}{\ensuremath{0.0213\pm0.0012}} 
\newcommand{\hatcurPPphiconjeccenxxxxxG}{\ensuremath{-0.00\pm0.10}}  
\newcommand{\hatcurPPperieccenxxxxxG}{\ensuremath{2457430.89\pm0.42}} 
\newcommand{\hatcurPPaequiveccenxxxxxG}{\ensuremath{0.02460\pm0.00055}} 
\newcommand{\hatcurPPtcirceccenxxxxxG}{\ensuremath{40.6_{-7.9}^{+11.7}}} 
\newcommand{\hatcurPPtinfalleccenxxxxxG}{\ensuremath{1170\pm170}}    
\newcommand{\hatcurXdisteccenxxxxxG}{\ensuremath{671\pm14}}          
\newcommand{\hatcurXAveccenxxxxxG}{\ensuremath{0.623\pm0.028}}       
\newcommand{\hatcurXdistredeccenxxxxxG}{\ensuremath{671\pm14}}       
\newcommand{\hatcurXEBVeccenxxxxxG}{\ensuremath{0.2010\pm0.0090}}    
\newcommand{\hatcurCCpmraeccenxxxxxG}{\ensuremath{7.784\pm0.079}}    
\newcommand{\hatcurCCpmdececcenxxxxxG}{\ensuremath{-3.863\pm0.044}}  
\newcommand{\hatcurCCpmeccenxxxxxG}{\ensuremath{8.690\pm0.090}}      

%% file: planetinfo_multi_orig.tex
\input{HTR062-011.orig.tex}
\input{HTR080-003.orig.tex}
\input{HTR089-018.orig.tex}
\input{HTR094-024.orig.tex}
\input{HTR130-001.orig.tex}
\input{HTR384-014.orig.tex}
\input{HTR405-001.orig.tex}
\newcommand{\hatcurCCbbHmagorig}[1]{\ifnum#1=58 %
\hatcurCCbbHmagorigxxxxxA
\else
\ifnum#1=59 %
\hatcurCCbbHmagorigxxxxxB
\else
\ifnum#1=60 %
\hatcurCCbbHmagorigxxxxxC
\else
\ifnum#1=61 %
\hatcurCCbbHmagorigxxxxxD
\else
\ifnum#1=62 %
\hatcurCCbbHmagorigxxxxxE
\else
\ifnum#1=63 %
\hatcurCCbbHmagorigxxxxxF
\else
\ifnum#1=64 %
\hatcurCCbbHmagorigxxxxxG
\else
??????\fi
\fi
\fi
\fi
\fi
\fi
\fi
}
\newcommand{\hatcurCCbbJmagorig}[1]{\ifnum#1=58 %
\hatcurCCbbJmagorigxxxxxA
\else
\ifnum#1=59 %
\hatcurCCbbJmagorigxxxxxB
\else
\ifnum#1=60 %
\hatcurCCbbJmagorigxxxxxC
\else
\ifnum#1=61 %
\hatcurCCbbJmagorigxxxxxD
\else
\ifnum#1=62 %
\hatcurCCbbJmagorigxxxxxE
\else
\ifnum#1=63 %
\hatcurCCbbJmagorigxxxxxF
\else
\ifnum#1=64 %
\hatcurCCbbJmagorigxxxxxG
\else
??????\fi
\fi
\fi
\fi
\fi
\fi
\fi
}
\newcommand{\hatcurCCbbKmagorig}[1]{\ifnum#1=58 %
\hatcurCCbbKmagorigxxxxxA
\else
\ifnum#1=59 %
\hatcurCCbbKmagorigxxxxxB
\else
\ifnum#1=60 %
\hatcurCCbbKmagorigxxxxxC
\else
\ifnum#1=61 %
\hatcurCCbbKmagorigxxxxxD
\else
\ifnum#1=62 %
\hatcurCCbbKmagorigxxxxxE
\else
\ifnum#1=63 %
\hatcurCCbbKmagorigxxxxxF
\else
\ifnum#1=64 %
\hatcurCCbbKmagorigxxxxxG
\else
??????\fi
\fi
\fi
\fi
\fi
\fi
\fi
}
\newcommand{\hatcurCCcitHmagorig}[1]{\ifnum#1=58 %
\hatcurCCcitHmagorigxxxxxA
\else
\ifnum#1=59 %
\hatcurCCcitHmagorigxxxxxB
\else
\ifnum#1=60 %
\hatcurCCcitHmagorigxxxxxC
\else
\ifnum#1=61 %
\hatcurCCcitHmagorigxxxxxD
\else
\ifnum#1=62 %
\hatcurCCcitHmagorigxxxxxE
\else
\ifnum#1=63 %
\hatcurCCcitHmagorigxxxxxF
\else
\ifnum#1=64 %
\hatcurCCcitHmagorigxxxxxG
\else
??????\fi
\fi
\fi
\fi
\fi
\fi
\fi
}
\newcommand{\hatcurCCcitJmagorig}[1]{\ifnum#1=58 %
\hatcurCCcitJmagorigxxxxxA
\else
\ifnum#1=59 %
\hatcurCCcitJmagorigxxxxxB
\else
\ifnum#1=60 %
\hatcurCCcitJmagorigxxxxxC
\else
\ifnum#1=61 %
\hatcurCCcitJmagorigxxxxxD
\else
\ifnum#1=62 %
\hatcurCCcitJmagorigxxxxxE
\else
\ifnum#1=63 %
\hatcurCCcitJmagorigxxxxxF
\else
\ifnum#1=64 %
\hatcurCCcitJmagorigxxxxxG
\else
??????\fi
\fi
\fi
\fi
\fi
\fi
\fi
}
\newcommand{\hatcurCCcitKmagorig}[1]{\ifnum#1=58 %
\hatcurCCcitKmagorigxxxxxA
\else
\ifnum#1=59 %
\hatcurCCcitKmagorigxxxxxB
\else
\ifnum#1=60 %
\hatcurCCcitKmagorigxxxxxC
\else
\ifnum#1=61 %
\hatcurCCcitKmagorigxxxxxD
\else
\ifnum#1=62 %
\hatcurCCcitKmagorigxxxxxE
\else
\ifnum#1=63 %
\hatcurCCcitKmagorigxxxxxF
\else
\ifnum#1=64 %
\hatcurCCcitKmagorigxxxxxG
\else
??????\fi
\fi
\fi
\fi
\fi
\fi
\fi
}
\newcommand{\hatcurCCdecorig}[1]{\ifnum#1=58 %
\hatcurCCdecorigxxxxxA
\else
\ifnum#1=59 %
\hatcurCCdecorigxxxxxB
\else
\ifnum#1=60 %
\hatcurCCdecorigxxxxxC
\else
\ifnum#1=61 %
\hatcurCCdecorigxxxxxD
\else
\ifnum#1=62 %
\hatcurCCdecorigxxxxxE
\else
\ifnum#1=63 %
\hatcurCCdecorigxxxxxF
\else
\ifnum#1=64 %
\hatcurCCdecorigxxxxxG
\else
??????\fi
\fi
\fi
\fi
\fi
\fi
\fi
}
\newcommand{\hatcurCCesoHKmagorig}[1]{\ifnum#1=58 %
\hatcurCCesoHKmagorigxxxxxA
\else
\ifnum#1=59 %
\hatcurCCesoHKmagorigxxxxxB
\else
\ifnum#1=60 %
\hatcurCCesoHKmagorigxxxxxC
\else
\ifnum#1=61 %
\hatcurCCesoHKmagorigxxxxxD
\else
\ifnum#1=62 %
\hatcurCCesoHKmagorigxxxxxE
\else
\ifnum#1=63 %
\hatcurCCesoHKmagorigxxxxxF
\else
\ifnum#1=64 %
\hatcurCCesoHKmagorigxxxxxG
\else
??????\fi
\fi
\fi
\fi
\fi
\fi
\fi
}
\newcommand{\hatcurCCesoHmagorig}[1]{\ifnum#1=58 %
\hatcurCCesoHmagorigxxxxxA
\else
\ifnum#1=59 %
\hatcurCCesoHmagorigxxxxxB
\else
\ifnum#1=60 %
\hatcurCCesoHmagorigxxxxxC
\else
\ifnum#1=61 %
\hatcurCCesoHmagorigxxxxxD
\else
\ifnum#1=62 %
\hatcurCCesoHmagorigxxxxxE
\else
\ifnum#1=63 %
\hatcurCCesoHmagorigxxxxxF
\else
\ifnum#1=64 %
\hatcurCCesoHmagorigxxxxxG
\else
??????\fi
\fi
\fi
\fi
\fi
\fi
\fi
}
\newcommand{\hatcurCCesoJHmagorig}[1]{\ifnum#1=58 %
\hatcurCCesoJHmagorigxxxxxA
\else
\ifnum#1=59 %
\hatcurCCesoJHmagorigxxxxxB
\else
\ifnum#1=60 %
\hatcurCCesoJHmagorigxxxxxC
\else
\ifnum#1=61 %
\hatcurCCesoJHmagorigxxxxxD
\else
\ifnum#1=62 %
\hatcurCCesoJHmagorigxxxxxE
\else
\ifnum#1=63 %
\hatcurCCesoJHmagorigxxxxxF
\else
\ifnum#1=64 %
\hatcurCCesoJHmagorigxxxxxG
\else
??????\fi
\fi
\fi
\fi
\fi
\fi
\fi
}
\newcommand{\hatcurCCesoJKmagorig}[1]{\ifnum#1=58 %
\hatcurCCesoJKmagorigxxxxxA
\else
\ifnum#1=59 %
\hatcurCCesoJKmagorigxxxxxB
\else
\ifnum#1=60 %
\hatcurCCesoJKmagorigxxxxxC
\else
\ifnum#1=61 %
\hatcurCCesoJKmagorigxxxxxD
\else
\ifnum#1=62 %
\hatcurCCesoJKmagorigxxxxxE
\else
\ifnum#1=63 %
\hatcurCCesoJKmagorigxxxxxF
\else
\ifnum#1=64 %
\hatcurCCesoJKmagorigxxxxxG
\else
??????\fi
\fi
\fi
\fi
\fi
\fi
\fi
}
\newcommand{\hatcurCCesoJmagorig}[1]{\ifnum#1=58 %
\hatcurCCesoJmagorigxxxxxA
\else
\ifnum#1=59 %
\hatcurCCesoJmagorigxxxxxB
\else
\ifnum#1=60 %
\hatcurCCesoJmagorigxxxxxC
\else
\ifnum#1=61 %
\hatcurCCesoJmagorigxxxxxD
\else
\ifnum#1=62 %
\hatcurCCesoJmagorigxxxxxE
\else
\ifnum#1=63 %
\hatcurCCesoJmagorigxxxxxF
\else
\ifnum#1=64 %
\hatcurCCesoJmagorigxxxxxG
\else
??????\fi
\fi
\fi
\fi
\fi
\fi
\fi
}
\newcommand{\hatcurCCesoKmagorig}[1]{\ifnum#1=58 %
\hatcurCCesoKmagorigxxxxxA
\else
\ifnum#1=59 %
\hatcurCCesoKmagorigxxxxxB
\else
\ifnum#1=60 %
\hatcurCCesoKmagorigxxxxxC
\else
\ifnum#1=61 %
\hatcurCCesoKmagorigxxxxxD
\else
\ifnum#1=62 %
\hatcurCCesoKmagorigxxxxxE
\else
\ifnum#1=63 %
\hatcurCCesoKmagorigxxxxxF
\else
\ifnum#1=64 %
\hatcurCCesoKmagorigxxxxxG
\else
??????\fi
\fi
\fi
\fi
\fi
\fi
\fi
}
\newcommand{\hatcurCCgscorig}[1]{\ifnum#1=58 %
\hatcurCCgscorigxxxxxA
\else
\ifnum#1=59 %
\hatcurCCgscorigxxxxxB
\else
\ifnum#1=60 %
\hatcurCCgscorigxxxxxC
\else
\ifnum#1=61 %
\hatcurCCgscorigxxxxxD
\else
\ifnum#1=62 %
\hatcurCCgscorigxxxxxE
\else
\ifnum#1=63 %
\hatcurCCgscorigxxxxxF
\else
\ifnum#1=64 %
\hatcurCCgscorigxxxxxG
\else
??????\fi
\fi
\fi
\fi
\fi
\fi
\fi
}
\newcommand{\hatcurCCmagorig}[1]{\ifnum#1=58 %
\hatcurCCmagorigxxxxxA
\else
\ifnum#1=59 %
\hatcurCCmagorigxxxxxB
\else
\ifnum#1=60 %
\hatcurCCmagorigxxxxxC
\else
\ifnum#1=61 %
\hatcurCCmagorigxxxxxD
\else
\ifnum#1=62 %
\hatcurCCmagorigxxxxxE
\else
\ifnum#1=63 %
\hatcurCCmagorigxxxxxF
\else
\ifnum#1=64 %
\hatcurCCmagorigxxxxxG
\else
??????\fi
\fi
\fi
\fi
\fi
\fi
\fi
}
\newcommand{\hatcurCCpmdecorig}[1]{\ifnum#1=58 %
\hatcurCCpmdecorigxxxxxA
\else
\ifnum#1=59 %
\hatcurCCpmdecorigxxxxxB
\else
\ifnum#1=60 %
\hatcurCCpmdecorigxxxxxC
\else
\ifnum#1=61 %
\hatcurCCpmdecorigxxxxxD
\else
\ifnum#1=62 %
\hatcurCCpmdecorigxxxxxE
\else
\ifnum#1=63 %
\hatcurCCpmdecorigxxxxxF
\else
\ifnum#1=64 %
\hatcurCCpmdecorigxxxxxG
\else
??????\fi
\fi
\fi
\fi
\fi
\fi
\fi
}
\newcommand{\hatcurCCpmorig}[1]{\ifnum#1=58 %
\hatcurCCpmorigxxxxxA
\else
\ifnum#1=59 %
\hatcurCCpmorigxxxxxB
\else
\ifnum#1=60 %
\hatcurCCpmorigxxxxxC
\else
\ifnum#1=61 %
\hatcurCCpmorigxxxxxD
\else
\ifnum#1=62 %
\hatcurCCpmorigxxxxxE
\else
\ifnum#1=63 %
\hatcurCCpmorigxxxxxF
\else
\ifnum#1=64 %
\hatcurCCpmorigxxxxxG
\else
??????\fi
\fi
\fi
\fi
\fi
\fi
\fi
}
\newcommand{\hatcurCCpmraorig}[1]{\ifnum#1=58 %
\hatcurCCpmraorigxxxxxA
\else
\ifnum#1=59 %
\hatcurCCpmraorigxxxxxB
\else
\ifnum#1=60 %
\hatcurCCpmraorigxxxxxC
\else
\ifnum#1=61 %
\hatcurCCpmraorigxxxxxD
\else
\ifnum#1=62 %
\hatcurCCpmraorigxxxxxE
\else
\ifnum#1=63 %
\hatcurCCpmraorigxxxxxF
\else
\ifnum#1=64 %
\hatcurCCpmraorigxxxxxG
\else
??????\fi
\fi
\fi
\fi
\fi
\fi
\fi
}
\newcommand{\hatcurCCraorig}[1]{\ifnum#1=58 %
\hatcurCCraorigxxxxxA
\else
\ifnum#1=59 %
\hatcurCCraorigxxxxxB
\else
\ifnum#1=60 %
\hatcurCCraorigxxxxxC
\else
\ifnum#1=61 %
\hatcurCCraorigxxxxxD
\else
\ifnum#1=62 %
\hatcurCCraorigxxxxxE
\else
\ifnum#1=63 %
\hatcurCCraorigxxxxxF
\else
\ifnum#1=64 %
\hatcurCCraorigxxxxxG
\else
??????\fi
\fi
\fi
\fi
\fi
\fi
\fi
}
\newcommand{\hatcurCCtassmBorig}[1]{\ifnum#1=58 %
\hatcurCCtassmBorigxxxxxA
\else
\ifnum#1=59 %
\hatcurCCtassmBorigxxxxxB
\else
\ifnum#1=60 %
\hatcurCCtassmBorigxxxxxC
\else
\ifnum#1=61 %
\hatcurCCtassmBorigxxxxxD
\else
\ifnum#1=62 %
\hatcurCCtassmBorigxxxxxE
\else
\ifnum#1=63 %
\hatcurCCtassmBorigxxxxxF
\else
\ifnum#1=64 %
\hatcurCCtassmBorigxxxxxG
\else
??????\fi
\fi
\fi
\fi
\fi
\fi
\fi
}
\newcommand{\hatcurCCtassmBshortorig}[1]{\ifnum#1=58 %
\hatcurCCtassmBshortorigxxxxxA
\else
\ifnum#1=59 %
\hatcurCCtassmBshortorigxxxxxB
\else
\ifnum#1=60 %
\hatcurCCtassmBshortorigxxxxxC
\else
\ifnum#1=61 %
\hatcurCCtassmBshortorigxxxxxD
\else
\ifnum#1=62 %
\hatcurCCtassmBshortorigxxxxxE
\else
\ifnum#1=63 %
\hatcurCCtassmBshortorigxxxxxF
\else
\ifnum#1=64 %
\hatcurCCtassmBshortorigxxxxxG
\else
??????\fi
\fi
\fi
\fi
\fi
\fi
\fi
}
\newcommand{\hatcurCCtassmgorig}[1]{\ifnum#1=58 %
\hatcurCCtassmgorigxxxxxA
\else
\ifnum#1=59 %
\hatcurCCtassmgorigxxxxxB
\else
\ifnum#1=60 %
\hatcurCCtassmgorigxxxxxC
\else
\ifnum#1=61 %
\hatcurCCtassmgorigxxxxxD
\else
\ifnum#1=62 %
\hatcurCCtassmgorigxxxxxE
\else
\ifnum#1=63 %
\hatcurCCtassmgorigxxxxxF
\else
\ifnum#1=64 %
\hatcurCCtassmgorigxxxxxG
\else
??????\fi
\fi
\fi
\fi
\fi
\fi
\fi
}
\newcommand{\hatcurCCtassmgshortorig}[1]{\ifnum#1=58 %
\hatcurCCtassmgshortorigxxxxxA
\else
\ifnum#1=59 %
\hatcurCCtassmgshortorigxxxxxB
\else
\ifnum#1=60 %
\hatcurCCtassmgshortorigxxxxxC
\else
\ifnum#1=61 %
\hatcurCCtassmgshortorigxxxxxD
\else
\ifnum#1=62 %
\hatcurCCtassmgshortorigxxxxxE
\else
\ifnum#1=63 %
\hatcurCCtassmgshortorigxxxxxF
\else
\ifnum#1=64 %
\hatcurCCtassmgshortorigxxxxxG
\else
??????\fi
\fi
\fi
\fi
\fi
\fi
\fi
}
\newcommand{\hatcurCCtassmiorig}[1]{\ifnum#1=58 %
\hatcurCCtassmiorigxxxxxA
\else
\ifnum#1=59 %
\hatcurCCtassmiorigxxxxxB
\else
\ifnum#1=60 %
\hatcurCCtassmiorigxxxxxC
\else
\ifnum#1=61 %
\hatcurCCtassmiorigxxxxxD
\else
\ifnum#1=62 %
\hatcurCCtassmiorigxxxxxE
\else
\ifnum#1=63 %
\hatcurCCtassmiorigxxxxxF
\else
\ifnum#1=64 %
\hatcurCCtassmiorigxxxxxG
\else
??????\fi
\fi
\fi
\fi
\fi
\fi
\fi
}
\newcommand{\hatcurCCtassmIorig}[1]{\ifnum#1=58 %
\hatcurCCtassmIorigxxxxxA
\else
\ifnum#1=59 %
\hatcurCCtassmIorigxxxxxB
\else
\ifnum#1=60 %
\hatcurCCtassmIorigxxxxxC
\else
\ifnum#1=61 %
\hatcurCCtassmIorigxxxxxD
\else
\ifnum#1=62 %
\hatcurCCtassmIorigxxxxxE
\else
\ifnum#1=63 %
\hatcurCCtassmIorigxxxxxF
\else
\ifnum#1=64 %
\hatcurCCtassmIorigxxxxxG
\else
??????\fi
\fi
\fi
\fi
\fi
\fi
\fi
}
\newcommand{\hatcurCCtassmishortorig}[1]{\ifnum#1=58 %
\hatcurCCtassmishortorigxxxxxA
\else
\ifnum#1=59 %
\hatcurCCtassmishortorigxxxxxB
\else
\ifnum#1=60 %
\hatcurCCtassmishortorigxxxxxC
\else
\ifnum#1=61 %
\hatcurCCtassmishortorigxxxxxD
\else
\ifnum#1=62 %
\hatcurCCtassmishortorigxxxxxE
\else
\ifnum#1=63 %
\hatcurCCtassmishortorigxxxxxF
\else
\ifnum#1=64 %
\hatcurCCtassmishortorigxxxxxG
\else
??????\fi
\fi
\fi
\fi
\fi
\fi
\fi
}
\newcommand{\hatcurCCtassmIshortorig}[1]{\ifnum#1=58 %
\hatcurCCtassmIshortorigxxxxxA
\else
\ifnum#1=59 %
\hatcurCCtassmIshortorigxxxxxB
\else
\ifnum#1=60 %
\hatcurCCtassmIshortorigxxxxxC
\else
\ifnum#1=61 %
\hatcurCCtassmIshortorigxxxxxD
\else
\ifnum#1=62 %
\hatcurCCtassmIshortorigxxxxxE
\else
\ifnum#1=63 %
\hatcurCCtassmIshortorigxxxxxF
\else
\ifnum#1=64 %
\hatcurCCtassmIshortorigxxxxxG
\else
??????\fi
\fi
\fi
\fi
\fi
\fi
\fi
}
\newcommand{\hatcurCCtassmrorig}[1]{\ifnum#1=58 %
\hatcurCCtassmrorigxxxxxA
\else
\ifnum#1=59 %
\hatcurCCtassmrorigxxxxxB
\else
\ifnum#1=60 %
\hatcurCCtassmrorigxxxxxC
\else
\ifnum#1=61 %
\hatcurCCtassmrorigxxxxxD
\else
\ifnum#1=62 %
\hatcurCCtassmrorigxxxxxE
\else
\ifnum#1=63 %
\hatcurCCtassmrorigxxxxxF
\else
\ifnum#1=64 %
\hatcurCCtassmrorigxxxxxG
\else
??????\fi
\fi
\fi
\fi
\fi
\fi
\fi
}
\newcommand{\hatcurCCtassmrshortorig}[1]{\ifnum#1=58 %
\hatcurCCtassmrshortorigxxxxxA
\else
\ifnum#1=59 %
\hatcurCCtassmrshortorigxxxxxB
\else
\ifnum#1=60 %
\hatcurCCtassmrshortorigxxxxxC
\else
\ifnum#1=61 %
\hatcurCCtassmrshortorigxxxxxD
\else
\ifnum#1=62 %
\hatcurCCtassmrshortorigxxxxxE
\else
\ifnum#1=63 %
\hatcurCCtassmrshortorigxxxxxF
\else
\ifnum#1=64 %
\hatcurCCtassmrshortorigxxxxxG
\else
??????\fi
\fi
\fi
\fi
\fi
\fi
\fi
}
\newcommand{\hatcurCCtassmvorig}[1]{\ifnum#1=58 %
\hatcurCCtassmvorigxxxxxA
\else
\ifnum#1=59 %
\hatcurCCtassmvorigxxxxxB
\else
\ifnum#1=60 %
\hatcurCCtassmvorigxxxxxC
\else
\ifnum#1=61 %
\hatcurCCtassmvorigxxxxxD
\else
\ifnum#1=62 %
\hatcurCCtassmvorigxxxxxE
\else
\ifnum#1=63 %
\hatcurCCtassmvorigxxxxxF
\else
\ifnum#1=64 %
\hatcurCCtassmvorigxxxxxG
\else
??????\fi
\fi
\fi
\fi
\fi
\fi
\fi
}
\newcommand{\hatcurCCtassmvshortorig}[1]{\ifnum#1=58 %
\hatcurCCtassmvshortorigxxxxxA
\else
\ifnum#1=59 %
\hatcurCCtassmvshortorigxxxxxB
\else
\ifnum#1=60 %
\hatcurCCtassmvshortorigxxxxxC
\else
\ifnum#1=61 %
\hatcurCCtassmvshortorigxxxxxD
\else
\ifnum#1=62 %
\hatcurCCtassmvshortorigxxxxxE
\else
\ifnum#1=63 %
\hatcurCCtassmvshortorigxxxxxF
\else
\ifnum#1=64 %
\hatcurCCtassmvshortorigxxxxxG
\else
??????\fi
\fi
\fi
\fi
\fi
\fi
\fi
}
\newcommand{\hatcurCCtwomassHmagorig}[1]{\ifnum#1=58 %
\hatcurCCtwomassHmagorigxxxxxA
\else
\ifnum#1=59 %
\hatcurCCtwomassHmagorigxxxxxB
\else
\ifnum#1=60 %
\hatcurCCtwomassHmagorigxxxxxC
\else
\ifnum#1=61 %
\hatcurCCtwomassHmagorigxxxxxD
\else
\ifnum#1=62 %
\hatcurCCtwomassHmagorigxxxxxE
\else
\ifnum#1=63 %
\hatcurCCtwomassHmagorigxxxxxF
\else
\ifnum#1=64 %
\hatcurCCtwomassHmagorigxxxxxG
\else
??????\fi
\fi
\fi
\fi
\fi
\fi
\fi
}
\newcommand{\hatcurCCtwomassJmagorig}[1]{\ifnum#1=58 %
\hatcurCCtwomassJmagorigxxxxxA
\else
\ifnum#1=59 %
\hatcurCCtwomassJmagorigxxxxxB
\else
\ifnum#1=60 %
\hatcurCCtwomassJmagorigxxxxxC
\else
\ifnum#1=61 %
\hatcurCCtwomassJmagorigxxxxxD
\else
\ifnum#1=62 %
\hatcurCCtwomassJmagorigxxxxxE
\else
\ifnum#1=63 %
\hatcurCCtwomassJmagorigxxxxxF
\else
\ifnum#1=64 %
\hatcurCCtwomassJmagorigxxxxxG
\else
??????\fi
\fi
\fi
\fi
\fi
\fi
\fi
}
\newcommand{\hatcurCCtwomassKmagorig}[1]{\ifnum#1=58 %
\hatcurCCtwomassKmagorigxxxxxA
\else
\ifnum#1=59 %
\hatcurCCtwomassKmagorigxxxxxB
\else
\ifnum#1=60 %
\hatcurCCtwomassKmagorigxxxxxC
\else
\ifnum#1=61 %
\hatcurCCtwomassKmagorigxxxxxD
\else
\ifnum#1=62 %
\hatcurCCtwomassKmagorigxxxxxE
\else
\ifnum#1=63 %
\hatcurCCtwomassKmagorigxxxxxF
\else
\ifnum#1=64 %
\hatcurCCtwomassKmagorigxxxxxG
\else
??????\fi
\fi
\fi
\fi
\fi
\fi
\fi
}
\newcommand{\hatcurCCtwomassorig}[1]{\ifnum#1=58 %
\hatcurCCtwomassorigxxxxxA
\else
\ifnum#1=59 %
\hatcurCCtwomassorigxxxxxB
\else
\ifnum#1=60 %
\hatcurCCtwomassorigxxxxxC
\else
\ifnum#1=61 %
\hatcurCCtwomassorigxxxxxD
\else
\ifnum#1=62 %
\hatcurCCtwomassorigxxxxxE
\else
\ifnum#1=63 %
\hatcurCCtwomassorigxxxxxF
\else
\ifnum#1=64 %
\hatcurCCtwomassorigxxxxxG
\else
??????\fi
\fi
\fi
\fi
\fi
\fi
\fi
}
\newcommand{\hatcurfieldorig}[1]{\ifnum#1=58 %
\hatcurfieldorigxxxxxA
\else
\ifnum#1=59 %
\hatcurfieldorigxxxxxB
\else
\ifnum#1=60 %
\hatcurfieldorigxxxxxC
\else
\ifnum#1=61 %
\hatcurfieldorigxxxxxD
\else
\ifnum#1=62 %
\hatcurfieldorigxxxxxE
\else
\ifnum#1=63 %
\hatcurfieldorigxxxxxF
\else
\ifnum#1=64 %
\hatcurfieldorigxxxxxG
\else
??????\fi
\fi
\fi
\fi
\fi
\fi
\fi
}
\newcommand{\hatcurhtrorig}[1]{\ifnum#1=58 %
\hatcurhtrorigxxxxxA
\else
\ifnum#1=59 %
\hatcurhtrorigxxxxxB
\else
\ifnum#1=60 %
\hatcurhtrorigxxxxxC
\else
\ifnum#1=61 %
\hatcurhtrorigxxxxxD
\else
\ifnum#1=62 %
\hatcurhtrorigxxxxxE
\else
\ifnum#1=63 %
\hatcurhtrorigxxxxxF
\else
\ifnum#1=64 %
\hatcurhtrorigxxxxxG
\else
??????\fi
\fi
\fi
\fi
\fi
\fi
\fi
}
\newcommand{\hatcurISOageorig}[1]{\ifnum#1=58 %
\hatcurISOageorigxxxxxA
\else
\ifnum#1=59 %
\hatcurISOageorigxxxxxB
\else
\ifnum#1=60 %
\hatcurISOageorigxxxxxC
\else
\ifnum#1=61 %
\hatcurISOageorigxxxxxD
\else
\ifnum#1=62 %
\hatcurISOageorigxxxxxE
\else
\ifnum#1=63 %
\hatcurISOageorigxxxxxF
\else
\ifnum#1=64 %
\hatcurISOageorigxxxxxG
\else
??????\fi
\fi
\fi
\fi
\fi
\fi
\fi
}
\newcommand{\hatcurISOJKorig}[1]{\ifnum#1=58 %
\hatcurISOJKorigxxxxxA
\else
\ifnum#1=59 %
\hatcurISOJKorigxxxxxB
\else
\ifnum#1=60 %
\hatcurISOJKorigxxxxxC
\else
\ifnum#1=61 %
\hatcurISOJKorigxxxxxD
\else
\ifnum#1=62 %
\hatcurISOJKorigxxxxxE
\else
\ifnum#1=63 %
\hatcurISOJKorigxxxxxF
\else
\ifnum#1=64 %
\hatcurISOJKorigxxxxxG
\else
??????\fi
\fi
\fi
\fi
\fi
\fi
\fi
}
\newcommand{\hatcurISOloggorig}[1]{\ifnum#1=58 %
\hatcurISOloggorigxxxxxA
\else
\ifnum#1=59 %
\hatcurISOloggorigxxxxxB
\else
\ifnum#1=60 %
\hatcurISOloggorigxxxxxC
\else
\ifnum#1=61 %
\hatcurISOloggorigxxxxxD
\else
\ifnum#1=62 %
\hatcurISOloggorigxxxxxE
\else
\ifnum#1=63 %
\hatcurISOloggorigxxxxxF
\else
\ifnum#1=64 %
\hatcurISOloggorigxxxxxG
\else
??????\fi
\fi
\fi
\fi
\fi
\fi
\fi
}
\newcommand{\hatcurISOlumorig}[1]{\ifnum#1=58 %
\hatcurISOlumorigxxxxxA
\else
\ifnum#1=59 %
\hatcurISOlumorigxxxxxB
\else
\ifnum#1=60 %
\hatcurISOlumorigxxxxxC
\else
\ifnum#1=61 %
\hatcurISOlumorigxxxxxD
\else
\ifnum#1=62 %
\hatcurISOlumorigxxxxxE
\else
\ifnum#1=63 %
\hatcurISOlumorigxxxxxF
\else
\ifnum#1=64 %
\hatcurISOlumorigxxxxxG
\else
??????\fi
\fi
\fi
\fi
\fi
\fi
\fi
}
\newcommand{\hatcurISOlumshortorig}[1]{\ifnum#1=58 %
\hatcurISOlumshortorigxxxxxA
\else
\ifnum#1=59 %
\hatcurISOlumshortorigxxxxxB
\else
\ifnum#1=60 %
\hatcurISOlumshortorigxxxxxC
\else
\ifnum#1=61 %
\hatcurISOlumshortorigxxxxxD
\else
\ifnum#1=62 %
\hatcurISOlumshortorigxxxxxE
\else
\ifnum#1=63 %
\hatcurISOlumshortorigxxxxxF
\else
\ifnum#1=64 %
\hatcurISOlumshortorigxxxxxG
\else
??????\fi
\fi
\fi
\fi
\fi
\fi
\fi
}
\newcommand{\hatcurISOMHorig}[1]{\ifnum#1=58 %
\hatcurISOMHorigxxxxxA
\else
\ifnum#1=59 %
\hatcurISOMHorigxxxxxB
\else
\ifnum#1=60 %
\hatcurISOMHorigxxxxxC
\else
\ifnum#1=61 %
\hatcurISOMHorigxxxxxD
\else
\ifnum#1=62 %
\hatcurISOMHorigxxxxxE
\else
\ifnum#1=63 %
\hatcurISOMHorigxxxxxF
\else
\ifnum#1=64 %
\hatcurISOMHorigxxxxxG
\else
??????\fi
\fi
\fi
\fi
\fi
\fi
\fi
}
\newcommand{\hatcurISOMJorig}[1]{\ifnum#1=58 %
\hatcurISOMJorigxxxxxA
\else
\ifnum#1=59 %
\hatcurISOMJorigxxxxxB
\else
\ifnum#1=60 %
\hatcurISOMJorigxxxxxC
\else
\ifnum#1=61 %
\hatcurISOMJorigxxxxxD
\else
\ifnum#1=62 %
\hatcurISOMJorigxxxxxE
\else
\ifnum#1=63 %
\hatcurISOMJorigxxxxxF
\else
\ifnum#1=64 %
\hatcurISOMJorigxxxxxG
\else
??????\fi
\fi
\fi
\fi
\fi
\fi
\fi
}
\newcommand{\hatcurISOMKorig}[1]{\ifnum#1=58 %
\hatcurISOMKorigxxxxxA
\else
\ifnum#1=59 %
\hatcurISOMKorigxxxxxB
\else
\ifnum#1=60 %
\hatcurISOMKorigxxxxxC
\else
\ifnum#1=61 %
\hatcurISOMKorigxxxxxD
\else
\ifnum#1=62 %
\hatcurISOMKorigxxxxxE
\else
\ifnum#1=63 %
\hatcurISOMKorigxxxxxF
\else
\ifnum#1=64 %
\hatcurISOMKorigxxxxxG
\else
??????\fi
\fi
\fi
\fi
\fi
\fi
\fi
}
\newcommand{\hatcurISOmlongorig}[1]{\ifnum#1=58 %
\hatcurISOmlongorigxxxxxA
\else
\ifnum#1=59 %
\hatcurISOmlongorigxxxxxB
\else
\ifnum#1=60 %
\hatcurISOmlongorigxxxxxC
\else
\ifnum#1=61 %
\hatcurISOmlongorigxxxxxD
\else
\ifnum#1=62 %
\hatcurISOmlongorigxxxxxE
\else
\ifnum#1=63 %
\hatcurISOmlongorigxxxxxF
\else
\ifnum#1=64 %
\hatcurISOmlongorigxxxxxG
\else
??????\fi
\fi
\fi
\fi
\fi
\fi
\fi
}
\newcommand{\hatcurISOmorig}[1]{\ifnum#1=58 %
\hatcurISOmorigxxxxxA
\else
\ifnum#1=59 %
\hatcurISOmorigxxxxxB
\else
\ifnum#1=60 %
\hatcurISOmorigxxxxxC
\else
\ifnum#1=61 %
\hatcurISOmorigxxxxxD
\else
\ifnum#1=62 %
\hatcurISOmorigxxxxxE
\else
\ifnum#1=63 %
\hatcurISOmorigxxxxxF
\else
\ifnum#1=64 %
\hatcurISOmorigxxxxxG
\else
??????\fi
\fi
\fi
\fi
\fi
\fi
\fi
}
\newcommand{\hatcurISOmshortorig}[1]{\ifnum#1=58 %
\hatcurISOmshortorigxxxxxA
\else
\ifnum#1=59 %
\hatcurISOmshortorigxxxxxB
\else
\ifnum#1=60 %
\hatcurISOmshortorigxxxxxC
\else
\ifnum#1=61 %
\hatcurISOmshortorigxxxxxD
\else
\ifnum#1=62 %
\hatcurISOmshortorigxxxxxE
\else
\ifnum#1=63 %
\hatcurISOmshortorigxxxxxF
\else
\ifnum#1=64 %
\hatcurISOmshortorigxxxxxG
\else
??????\fi
\fi
\fi
\fi
\fi
\fi
\fi
}
\newcommand{\hatcurISOmvorig}[1]{\ifnum#1=58 %
\hatcurISOmvorigxxxxxA
\else
\ifnum#1=59 %
\hatcurISOmvorigxxxxxB
\else
\ifnum#1=60 %
\hatcurISOmvorigxxxxxC
\else
\ifnum#1=61 %
\hatcurISOmvorigxxxxxD
\else
\ifnum#1=62 %
\hatcurISOmvorigxxxxxE
\else
\ifnum#1=63 %
\hatcurISOmvorigxxxxxF
\else
\ifnum#1=64 %
\hatcurISOmvorigxxxxxG
\else
??????\fi
\fi
\fi
\fi
\fi
\fi
\fi
}
\newcommand{\hatcurISOrholongorig}[1]{\ifnum#1=58 %
\hatcurISOrholongorigxxxxxA
\else
\ifnum#1=59 %
\hatcurISOrholongorigxxxxxB
\else
\ifnum#1=60 %
\hatcurISOrholongorigxxxxxC
\else
\ifnum#1=61 %
\hatcurISOrholongorigxxxxxD
\else
\ifnum#1=62 %
\hatcurISOrholongorigxxxxxE
\else
\ifnum#1=63 %
\hatcurISOrholongorigxxxxxF
\else
\ifnum#1=64 %
\hatcurISOrholongorigxxxxxG
\else
??????\fi
\fi
\fi
\fi
\fi
\fi
\fi
}
\newcommand{\hatcurISOrhoorig}[1]{\ifnum#1=58 %
\hatcurISOrhoorigxxxxxA
\else
\ifnum#1=59 %
\hatcurISOrhoorigxxxxxB
\else
\ifnum#1=60 %
\hatcurISOrhoorigxxxxxC
\else
\ifnum#1=61 %
\hatcurISOrhoorigxxxxxD
\else
\ifnum#1=62 %
\hatcurISOrhoorigxxxxxE
\else
\ifnum#1=63 %
\hatcurISOrhoorigxxxxxF
\else
\ifnum#1=64 %
\hatcurISOrhoorigxxxxxG
\else
??????\fi
\fi
\fi
\fi
\fi
\fi
\fi
}
\newcommand{\hatcurISOrlongorig}[1]{\ifnum#1=58 %
\hatcurISOrlongorigxxxxxA
\else
\ifnum#1=59 %
\hatcurISOrlongorigxxxxxB
\else
\ifnum#1=60 %
\hatcurISOrlongorigxxxxxC
\else
\ifnum#1=61 %
\hatcurISOrlongorigxxxxxD
\else
\ifnum#1=62 %
\hatcurISOrlongorigxxxxxE
\else
\ifnum#1=63 %
\hatcurISOrlongorigxxxxxF
\else
\ifnum#1=64 %
\hatcurISOrlongorigxxxxxG
\else
??????\fi
\fi
\fi
\fi
\fi
\fi
\fi
}
\newcommand{\hatcurISOrorig}[1]{\ifnum#1=58 %
\hatcurISOrorigxxxxxA
\else
\ifnum#1=59 %
\hatcurISOrorigxxxxxB
\else
\ifnum#1=60 %
\hatcurISOrorigxxxxxC
\else
\ifnum#1=61 %
\hatcurISOrorigxxxxxD
\else
\ifnum#1=62 %
\hatcurISOrorigxxxxxE
\else
\ifnum#1=63 %
\hatcurISOrorigxxxxxF
\else
\ifnum#1=64 %
\hatcurISOrorigxxxxxG
\else
??????\fi
\fi
\fi
\fi
\fi
\fi
\fi
}
\newcommand{\hatcurISOrshortorig}[1]{\ifnum#1=58 %
\hatcurISOrshortorigxxxxxA
\else
\ifnum#1=59 %
\hatcurISOrshortorigxxxxxB
\else
\ifnum#1=60 %
\hatcurISOrshortorigxxxxxC
\else
\ifnum#1=61 %
\hatcurISOrshortorigxxxxxD
\else
\ifnum#1=62 %
\hatcurISOrshortorigxxxxxE
\else
\ifnum#1=63 %
\hatcurISOrshortorigxxxxxF
\else
\ifnum#1=64 %
\hatcurISOrshortorigxxxxxG
\else
??????\fi
\fi
\fi
\fi
\fi
\fi
\fi
}
\newcommand{\hatcurISOsigmaorig}[1]{\ifnum#1=58 %
\hatcurISOsigmaorigxxxxxA
\else
\ifnum#1=59 %
\hatcurISOsigmaorigxxxxxB
\else
\ifnum#1=60 %
\hatcurISOsigmaorigxxxxxC
\else
\ifnum#1=61 %
\hatcurISOsigmaorigxxxxxD
\else
\ifnum#1=62 %
\hatcurISOsigmaorigxxxxxE
\else
\ifnum#1=63 %
\hatcurISOsigmaorigxxxxxF
\else
\ifnum#1=64 %
\hatcurISOsigmaorigxxxxxG
\else
??????\fi
\fi
\fi
\fi
\fi
\fi
\fi
}
\newcommand{\hatcurISOspecorig}[1]{\ifnum#1=58 %
\hatcurISOspecorigxxxxxA
\else
\ifnum#1=59 %
\hatcurISOspecorigxxxxxB
\else
\ifnum#1=60 %
\hatcurISOspecorigxxxxxC
\else
\ifnum#1=61 %
\hatcurISOspecorigxxxxxD
\else
\ifnum#1=62 %
\hatcurISOspecorigxxxxxE
\else
\ifnum#1=63 %
\hatcurISOspecorigxxxxxF
\else
\ifnum#1=64 %
\hatcurISOspecorigxxxxxG
\else
??????\fi
\fi
\fi
\fi
\fi
\fi
\fi
}
\newcommand{\hatcurISOviorig}[1]{\ifnum#1=58 %
\hatcurISOviorigxxxxxA
\else
\ifnum#1=59 %
\hatcurISOviorigxxxxxB
\else
\ifnum#1=60 %
\hatcurISOviorigxxxxxC
\else
\ifnum#1=61 %
\hatcurISOviorigxxxxxD
\else
\ifnum#1=62 %
\hatcurISOviorigxxxxxE
\else
\ifnum#1=63 %
\hatcurISOviorigxxxxxF
\else
\ifnum#1=64 %
\hatcurISOviorigxxxxxG
\else
??????\fi
\fi
\fi
\fi
\fi
\fi
\fi
}
\newcommand{\hatcurLBigorig}[1]{\ifnum#1=58 %
\hatcurLBigorigxxxxxA
\else
\ifnum#1=59 %
\hatcurLBigorigxxxxxB
\else
\ifnum#1=60 %
\hatcurLBigorigxxxxxC
\else
\ifnum#1=61 %
\hatcurLBigorigxxxxxD
\else
\ifnum#1=62 %
\hatcurLBigorigxxxxxE
\else
\ifnum#1=63 %
\hatcurLBigorigxxxxxF
\else
\ifnum#1=64 %
\hatcurLBigorigxxxxxG
\else
??????\fi
\fi
\fi
\fi
\fi
\fi
\fi
}
\newcommand{\hatcurLBiigorig}[1]{\ifnum#1=58 %
\hatcurLBiigorigxxxxxA
\else
\ifnum#1=59 %
\hatcurLBiigorigxxxxxB
\else
\ifnum#1=60 %
\hatcurLBiigorigxxxxxC
\else
\ifnum#1=61 %
\hatcurLBiigorigxxxxxD
\else
\ifnum#1=62 %
\hatcurLBiigorigxxxxxE
\else
\ifnum#1=63 %
\hatcurLBiigorigxxxxxF
\else
\ifnum#1=64 %
\hatcurLBiigorigxxxxxG
\else
??????\fi
\fi
\fi
\fi
\fi
\fi
\fi
}
\newcommand{\hatcurLBiiiorig}[1]{\ifnum#1=58 %
\hatcurLBiiiorigxxxxxA
\else
\ifnum#1=59 %
\hatcurLBiiiorigxxxxxB
\else
\ifnum#1=60 %
\hatcurLBiiiorigxxxxxC
\else
\ifnum#1=61 %
\hatcurLBiiiorigxxxxxD
\else
\ifnum#1=62 %
\hatcurLBiiiorigxxxxxE
\else
\ifnum#1=63 %
\hatcurLBiiiorigxxxxxF
\else
\ifnum#1=64 %
\hatcurLBiiiorigxxxxxG
\else
??????\fi
\fi
\fi
\fi
\fi
\fi
\fi
}
\newcommand{\hatcurLBiiIorig}[1]{\ifnum#1=58 %
\hatcurLBiiIorigxxxxxA
\else
\ifnum#1=59 %
\hatcurLBiiIorigxxxxxB
\else
\ifnum#1=60 %
\hatcurLBiiIorigxxxxxC
\else
\ifnum#1=61 %
\hatcurLBiiIorigxxxxxD
\else
\ifnum#1=62 %
\hatcurLBiiIorigxxxxxE
\else
\ifnum#1=63 %
\hatcurLBiiIorigxxxxxF
\else
\ifnum#1=64 %
\hatcurLBiiIorigxxxxxG
\else
??????\fi
\fi
\fi
\fi
\fi
\fi
\fi
}
\newcommand{\hatcurLBiikeporig}[1]{\ifnum#1=58 %
\hatcurLBiikeporigxxxxxA
\else
\ifnum#1=59 %
\hatcurLBiikeporigxxxxxB
\else
\ifnum#1=60 %
\hatcurLBiikeporigxxxxxC
\else
\ifnum#1=61 %
\hatcurLBiikeporigxxxxxD
\else
\ifnum#1=62 %
\hatcurLBiikeporigxxxxxE
\else
\ifnum#1=63 %
\hatcurLBiikeporigxxxxxF
\else
\ifnum#1=64 %
\hatcurLBiikeporigxxxxxG
\else
??????\fi
\fi
\fi
\fi
\fi
\fi
\fi
}
\newcommand{\hatcurLBiiorig}[1]{\ifnum#1=58 %
\hatcurLBiiorigxxxxxA
\else
\ifnum#1=59 %
\hatcurLBiiorigxxxxxB
\else
\ifnum#1=60 %
\hatcurLBiiorigxxxxxC
\else
\ifnum#1=61 %
\hatcurLBiiorigxxxxxD
\else
\ifnum#1=62 %
\hatcurLBiiorigxxxxxE
\else
\ifnum#1=63 %
\hatcurLBiiorigxxxxxF
\else
\ifnum#1=64 %
\hatcurLBiiorigxxxxxG
\else
??????\fi
\fi
\fi
\fi
\fi
\fi
\fi
}
\newcommand{\hatcurLBiIorig}[1]{\ifnum#1=58 %
\hatcurLBiIorigxxxxxA
\else
\ifnum#1=59 %
\hatcurLBiIorigxxxxxB
\else
\ifnum#1=60 %
\hatcurLBiIorigxxxxxC
\else
\ifnum#1=61 %
\hatcurLBiIorigxxxxxD
\else
\ifnum#1=62 %
\hatcurLBiIorigxxxxxE
\else
\ifnum#1=63 %
\hatcurLBiIorigxxxxxF
\else
\ifnum#1=64 %
\hatcurLBiIorigxxxxxG
\else
??????\fi
\fi
\fi
\fi
\fi
\fi
\fi
}
\newcommand{\hatcurLBiirorig}[1]{\ifnum#1=58 %
\hatcurLBiirorigxxxxxA
\else
\ifnum#1=59 %
\hatcurLBiirorigxxxxxB
\else
\ifnum#1=60 %
\hatcurLBiirorigxxxxxC
\else
\ifnum#1=61 %
\hatcurLBiirorigxxxxxD
\else
\ifnum#1=62 %
\hatcurLBiirorigxxxxxE
\else
\ifnum#1=63 %
\hatcurLBiirorigxxxxxF
\else
\ifnum#1=64 %
\hatcurLBiirorigxxxxxG
\else
??????\fi
\fi
\fi
\fi
\fi
\fi
\fi
}
\newcommand{\hatcurLBiiRorig}[1]{\ifnum#1=58 %
\hatcurLBiiRorigxxxxxA
\else
\ifnum#1=59 %
\hatcurLBiiRorigxxxxxB
\else
\ifnum#1=60 %
\hatcurLBiiRorigxxxxxC
\else
\ifnum#1=61 %
\hatcurLBiiRorigxxxxxD
\else
\ifnum#1=62 %
\hatcurLBiiRorigxxxxxE
\else
\ifnum#1=63 %
\hatcurLBiiRorigxxxxxF
\else
\ifnum#1=64 %
\hatcurLBiiRorigxxxxxG
\else
??????\fi
\fi
\fi
\fi
\fi
\fi
\fi
}
\newcommand{\hatcurLBiizorig}[1]{\ifnum#1=58 %
\hatcurLBiizorigxxxxxA
\else
\ifnum#1=59 %
\hatcurLBiizorigxxxxxB
\else
\ifnum#1=60 %
\hatcurLBiizorigxxxxxC
\else
\ifnum#1=61 %
\hatcurLBiizorigxxxxxD
\else
\ifnum#1=62 %
\hatcurLBiizorigxxxxxE
\else
\ifnum#1=63 %
\hatcurLBiizorigxxxxxF
\else
\ifnum#1=64 %
\hatcurLBiizorigxxxxxG
\else
??????\fi
\fi
\fi
\fi
\fi
\fi
\fi
}
\newcommand{\hatcurLBikeporig}[1]{\ifnum#1=58 %
\hatcurLBikeporigxxxxxA
\else
\ifnum#1=59 %
\hatcurLBikeporigxxxxxB
\else
\ifnum#1=60 %
\hatcurLBikeporigxxxxxC
\else
\ifnum#1=61 %
\hatcurLBikeporigxxxxxD
\else
\ifnum#1=62 %
\hatcurLBikeporigxxxxxE
\else
\ifnum#1=63 %
\hatcurLBikeporigxxxxxF
\else
\ifnum#1=64 %
\hatcurLBikeporigxxxxxG
\else
??????\fi
\fi
\fi
\fi
\fi
\fi
\fi
}
\newcommand{\hatcurLBirorig}[1]{\ifnum#1=58 %
\hatcurLBirorigxxxxxA
\else
\ifnum#1=59 %
\hatcurLBirorigxxxxxB
\else
\ifnum#1=60 %
\hatcurLBirorigxxxxxC
\else
\ifnum#1=61 %
\hatcurLBirorigxxxxxD
\else
\ifnum#1=62 %
\hatcurLBirorigxxxxxE
\else
\ifnum#1=63 %
\hatcurLBirorigxxxxxF
\else
\ifnum#1=64 %
\hatcurLBirorigxxxxxG
\else
??????\fi
\fi
\fi
\fi
\fi
\fi
\fi
}
\newcommand{\hatcurLBiRorig}[1]{\ifnum#1=58 %
\hatcurLBiRorigxxxxxA
\else
\ifnum#1=59 %
\hatcurLBiRorigxxxxxB
\else
\ifnum#1=60 %
\hatcurLBiRorigxxxxxC
\else
\ifnum#1=61 %
\hatcurLBiRorigxxxxxD
\else
\ifnum#1=62 %
\hatcurLBiRorigxxxxxE
\else
\ifnum#1=63 %
\hatcurLBiRorigxxxxxF
\else
\ifnum#1=64 %
\hatcurLBiRorigxxxxxG
\else
??????\fi
\fi
\fi
\fi
\fi
\fi
\fi
}
\newcommand{\hatcurLBizorig}[1]{\ifnum#1=58 %
\hatcurLBizorigxxxxxA
\else
\ifnum#1=59 %
\hatcurLBizorigxxxxxB
\else
\ifnum#1=60 %
\hatcurLBizorigxxxxxC
\else
\ifnum#1=61 %
\hatcurLBizorigxxxxxD
\else
\ifnum#1=62 %
\hatcurLBizorigxxxxxE
\else
\ifnum#1=63 %
\hatcurLBizorigxxxxxF
\else
\ifnum#1=64 %
\hatcurLBizorigxxxxxG
\else
??????\fi
\fi
\fi
\fi
\fi
\fi
\fi
}
\newcommand{\hatcurLCbsqorig}[1]{\ifnum#1=58 %
\hatcurLCbsqorigxxxxxA
\else
\ifnum#1=59 %
\hatcurLCbsqorigxxxxxB
\else
\ifnum#1=60 %
\hatcurLCbsqorigxxxxxC
\else
\ifnum#1=61 %
\hatcurLCbsqorigxxxxxD
\else
\ifnum#1=62 %
\hatcurLCbsqorigxxxxxE
\else
\ifnum#1=63 %
\hatcurLCbsqorigxxxxxF
\else
\ifnum#1=64 %
\hatcurLCbsqorigxxxxxG
\else
??????\fi
\fi
\fi
\fi
\fi
\fi
\fi
}
\newcommand{\hatcurLCdiporig}[1]{\ifnum#1=58 %
\hatcurLCdiporigxxxxxA
\else
\ifnum#1=59 %
\hatcurLCdiporigxxxxxB
\else
\ifnum#1=60 %
\hatcurLCdiporigxxxxxC
\else
\ifnum#1=61 %
\hatcurLCdiporigxxxxxD
\else
\ifnum#1=62 %
\hatcurLCdiporigxxxxxE
\else
\ifnum#1=63 %
\hatcurLCdiporigxxxxxF
\else
\ifnum#1=64 %
\hatcurLCdiporigxxxxxG
\else
??????\fi
\fi
\fi
\fi
\fi
\fi
\fi
}
\newcommand{\hatcurLCdurhrorig}[1]{\ifnum#1=58 %
\hatcurLCdurhrorigxxxxxA
\else
\ifnum#1=59 %
\hatcurLCdurhrorigxxxxxB
\else
\ifnum#1=60 %
\hatcurLCdurhrorigxxxxxC
\else
\ifnum#1=61 %
\hatcurLCdurhrorigxxxxxD
\else
\ifnum#1=62 %
\hatcurLCdurhrorigxxxxxE
\else
\ifnum#1=63 %
\hatcurLCdurhrorigxxxxxF
\else
\ifnum#1=64 %
\hatcurLCdurhrorigxxxxxG
\else
??????\fi
\fi
\fi
\fi
\fi
\fi
\fi
}
\newcommand{\hatcurLCdurhrshortorig}[1]{\ifnum#1=58 %
\hatcurLCdurhrshortorigxxxxxA
\else
\ifnum#1=59 %
\hatcurLCdurhrshortorigxxxxxB
\else
\ifnum#1=60 %
\hatcurLCdurhrshortorigxxxxxC
\else
\ifnum#1=61 %
\hatcurLCdurhrshortorigxxxxxD
\else
\ifnum#1=62 %
\hatcurLCdurhrshortorigxxxxxE
\else
\ifnum#1=63 %
\hatcurLCdurhrshortorigxxxxxF
\else
\ifnum#1=64 %
\hatcurLCdurhrshortorigxxxxxG
\else
??????\fi
\fi
\fi
\fi
\fi
\fi
\fi
}
\newcommand{\hatcurLCdurorig}[1]{\ifnum#1=58 %
\hatcurLCdurorigxxxxxA
\else
\ifnum#1=59 %
\hatcurLCdurorigxxxxxB
\else
\ifnum#1=60 %
\hatcurLCdurorigxxxxxC
\else
\ifnum#1=61 %
\hatcurLCdurorigxxxxxD
\else
\ifnum#1=62 %
\hatcurLCdurorigxxxxxE
\else
\ifnum#1=63 %
\hatcurLCdurorigxxxxxF
\else
\ifnum#1=64 %
\hatcurLCdurorigxxxxxG
\else
??????\fi
\fi
\fi
\fi
\fi
\fi
\fi
}
\newcommand{\hatcurLCdurshortorig}[1]{\ifnum#1=58 %
\hatcurLCdurshortorigxxxxxA
\else
\ifnum#1=59 %
\hatcurLCdurshortorigxxxxxB
\else
\ifnum#1=60 %
\hatcurLCdurshortorigxxxxxC
\else
\ifnum#1=61 %
\hatcurLCdurshortorigxxxxxD
\else
\ifnum#1=62 %
\hatcurLCdurshortorigxxxxxE
\else
\ifnum#1=63 %
\hatcurLCdurshortorigxxxxxF
\else
\ifnum#1=64 %
\hatcurLCdurshortorigxxxxxG
\else
??????\fi
\fi
\fi
\fi
\fi
\fi
\fi
}
\newcommand{\hatcurLChatnetmAorig}[1]{\ifnum#1=61 %
\hatcurLChatnetmAorigxxxxxD
\else
??????\fi
}
\newcommand{\hatcurLChatnetmBorig}[1]{\ifnum#1=61 %
\hatcurLChatnetmBorigxxxxxD
\else
??????\fi
}
\newcommand{\hatcurLChatnetmorig}[1]{\ifnum#1=58 %
\hatcurLChatnetmorigxxxxxA
\else
\ifnum#1=59 %
\hatcurLChatnetmorigxxxxxB
\else
\ifnum#1=62 %
\hatcurLChatnetmorigxxxxxE
\else
\ifnum#1=63 %
\hatcurLChatnetmorigxxxxxF
\else
\ifnum#1=64 %
\hatcurLChatnetmorigxxxxxG
\else
??????\fi
\fi
\fi
\fi
\fi
}
\newcommand{\hatcurLCiblendAorig}[1]{\ifnum#1=61 %
\hatcurLCiblendAorigxxxxxD
\else
??????\fi
}
\newcommand{\hatcurLCiblendBorig}[1]{\ifnum#1=61 %
\hatcurLCiblendBorigxxxxxD
\else
??????\fi
}
\newcommand{\hatcurLCiblendorig}[1]{\ifnum#1=58 %
\hatcurLCiblendorigxxxxxA
\else
\ifnum#1=59 %
\hatcurLCiblendorigxxxxxB
\else
\ifnum#1=62 %
\hatcurLCiblendorigxxxxxE
\else
\ifnum#1=63 %
\hatcurLCiblendorigxxxxxF
\else
\ifnum#1=64 %
\hatcurLCiblendorigxxxxxG
\else
??????\fi
\fi
\fi
\fi
\fi
}
\newcommand{\hatcurLCimporig}[1]{\ifnum#1=58 %
\hatcurLCimporigxxxxxA
\else
\ifnum#1=59 %
\hatcurLCimporigxxxxxB
\else
\ifnum#1=60 %
\hatcurLCimporigxxxxxC
\else
\ifnum#1=61 %
\hatcurLCimporigxxxxxD
\else
\ifnum#1=62 %
\hatcurLCimporigxxxxxE
\else
\ifnum#1=63 %
\hatcurLCimporigxxxxxF
\else
\ifnum#1=64 %
\hatcurLCimporigxxxxxG
\else
??????\fi
\fi
\fi
\fi
\fi
\fi
\fi
}
\newcommand{\hatcurLCingdurorig}[1]{\ifnum#1=58 %
\hatcurLCingdurorigxxxxxA
\else
\ifnum#1=59 %
\hatcurLCingdurorigxxxxxB
\else
\ifnum#1=60 %
\hatcurLCingdurorigxxxxxC
\else
\ifnum#1=61 %
\hatcurLCingdurorigxxxxxD
\else
\ifnum#1=62 %
\hatcurLCingdurorigxxxxxE
\else
\ifnum#1=63 %
\hatcurLCingdurorigxxxxxF
\else
\ifnum#1=64 %
\hatcurLCingdurorigxxxxxG
\else
??????\fi
\fi
\fi
\fi
\fi
\fi
\fi
}
\newcommand{\hatcurLCPorig}[1]{\ifnum#1=58 %
\hatcurLCPorigxxxxxA
\else
\ifnum#1=59 %
\hatcurLCPorigxxxxxB
\else
\ifnum#1=60 %
\hatcurLCPorigxxxxxC
\else
\ifnum#1=61 %
\hatcurLCPorigxxxxxD
\else
\ifnum#1=62 %
\hatcurLCPorigxxxxxE
\else
\ifnum#1=63 %
\hatcurLCPorigxxxxxF
\else
\ifnum#1=64 %
\hatcurLCPorigxxxxxG
\else
??????\fi
\fi
\fi
\fi
\fi
\fi
\fi
}
\newcommand{\hatcurLCPprecorig}[1]{\ifnum#1=58 %
\hatcurLCPprecorigxxxxxA
\else
\ifnum#1=59 %
\hatcurLCPprecorigxxxxxB
\else
\ifnum#1=60 %
\hatcurLCPprecorigxxxxxC
\else
\ifnum#1=61 %
\hatcurLCPprecorigxxxxxD
\else
\ifnum#1=62 %
\hatcurLCPprecorigxxxxxE
\else
\ifnum#1=63 %
\hatcurLCPprecorigxxxxxF
\else
\ifnum#1=64 %
\hatcurLCPprecorigxxxxxG
\else
??????\fi
\fi
\fi
\fi
\fi
\fi
\fi
}
\newcommand{\hatcurLCPshortorig}[1]{\ifnum#1=58 %
\hatcurLCPshortorigxxxxxA
\else
\ifnum#1=59 %
\hatcurLCPshortorigxxxxxB
\else
\ifnum#1=60 %
\hatcurLCPshortorigxxxxxC
\else
\ifnum#1=61 %
\hatcurLCPshortorigxxxxxD
\else
\ifnum#1=62 %
\hatcurLCPshortorigxxxxxE
\else
\ifnum#1=63 %
\hatcurLCPshortorigxxxxxF
\else
\ifnum#1=64 %
\hatcurLCPshortorigxxxxxG
\else
??????\fi
\fi
\fi
\fi
\fi
\fi
\fi
}
\newcommand{\hatcurLCqorig}[1]{\ifnum#1=58 %
\hatcurLCqorigxxxxxA
\else
\ifnum#1=59 %
\hatcurLCqorigxxxxxB
\else
\ifnum#1=60 %
\hatcurLCqorigxxxxxC
\else
\ifnum#1=61 %
\hatcurLCqorigxxxxxD
\else
\ifnum#1=62 %
\hatcurLCqorigxxxxxE
\else
\ifnum#1=63 %
\hatcurLCqorigxxxxxF
\else
\ifnum#1=64 %
\hatcurLCqorigxxxxxG
\else
??????\fi
\fi
\fi
\fi
\fi
\fi
\fi
}
\newcommand{\hatcurLCqshortorig}[1]{\ifnum#1=58 %
\hatcurLCqshortorigxxxxxA
\else
\ifnum#1=59 %
\hatcurLCqshortorigxxxxxB
\else
\ifnum#1=60 %
\hatcurLCqshortorigxxxxxC
\else
\ifnum#1=61 %
\hatcurLCqshortorigxxxxxD
\else
\ifnum#1=62 %
\hatcurLCqshortorigxxxxxE
\else
\ifnum#1=63 %
\hatcurLCqshortorigxxxxxF
\else
\ifnum#1=64 %
\hatcurLCqshortorigxxxxxG
\else
??????\fi
\fi
\fi
\fi
\fi
\fi
\fi
}
\newcommand{\hatcurLCrhoorig}[1]{\ifnum#1=58 %
\hatcurLCrhoorigxxxxxA
\else
\ifnum#1=59 %
\hatcurLCrhoorigxxxxxB
\else
\ifnum#1=60 %
\hatcurLCrhoorigxxxxxC
\else
\ifnum#1=61 %
\hatcurLCrhoorigxxxxxD
\else
\ifnum#1=62 %
\hatcurLCrhoorigxxxxxE
\else
\ifnum#1=63 %
\hatcurLCrhoorigxxxxxF
\else
\ifnum#1=64 %
\hatcurLCrhoorigxxxxxG
\else
??????\fi
\fi
\fi
\fi
\fi
\fi
\fi
}
\newcommand{\hatcurLCrprstarorig}[1]{\ifnum#1=58 %
\hatcurLCrprstarorigxxxxxA
\else
\ifnum#1=59 %
\hatcurLCrprstarorigxxxxxB
\else
\ifnum#1=60 %
\hatcurLCrprstarorigxxxxxC
\else
\ifnum#1=61 %
\hatcurLCrprstarorigxxxxxD
\else
\ifnum#1=62 %
\hatcurLCrprstarorigxxxxxE
\else
\ifnum#1=63 %
\hatcurLCrprstarorigxxxxxF
\else
\ifnum#1=64 %
\hatcurLCrprstarorigxxxxxG
\else
??????\fi
\fi
\fi
\fi
\fi
\fi
\fi
}
\newcommand{\hatcurLCTAorig}[1]{\ifnum#1=58 %
\hatcurLCTAorigxxxxxA
\else
\ifnum#1=59 %
\hatcurLCTAorigxxxxxB
\else
\ifnum#1=60 %
\hatcurLCTAorigxxxxxC
\else
\ifnum#1=61 %
\hatcurLCTAorigxxxxxD
\else
\ifnum#1=62 %
\hatcurLCTAorigxxxxxE
\else
\ifnum#1=63 %
\hatcurLCTAorigxxxxxF
\else
\ifnum#1=64 %
\hatcurLCTAorigxxxxxG
\else
??????\fi
\fi
\fi
\fi
\fi
\fi
\fi
}
\newcommand{\hatcurLCTBorig}[1]{\ifnum#1=58 %
\hatcurLCTBorigxxxxxA
\else
\ifnum#1=59 %
\hatcurLCTBorigxxxxxB
\else
\ifnum#1=60 %
\hatcurLCTBorigxxxxxC
\else
\ifnum#1=61 %
\hatcurLCTBorigxxxxxD
\else
\ifnum#1=62 %
\hatcurLCTBorigxxxxxE
\else
\ifnum#1=63 %
\hatcurLCTBorigxxxxxF
\else
\ifnum#1=64 %
\hatcurLCTBorigxxxxxG
\else
??????\fi
\fi
\fi
\fi
\fi
\fi
\fi
}
\newcommand{\hatcurLCTorig}[1]{\ifnum#1=58 %
\hatcurLCTorigxxxxxA
\else
\ifnum#1=59 %
\hatcurLCTorigxxxxxB
\else
\ifnum#1=60 %
\hatcurLCTorigxxxxxC
\else
\ifnum#1=61 %
\hatcurLCTorigxxxxxD
\else
\ifnum#1=62 %
\hatcurLCTorigxxxxxE
\else
\ifnum#1=63 %
\hatcurLCTorigxxxxxF
\else
\ifnum#1=64 %
\hatcurLCTorigxxxxxG
\else
??????\fi
\fi
\fi
\fi
\fi
\fi
\fi
}
\newcommand{\hatcurLCTshortorig}[1]{\ifnum#1=58 %
\hatcurLCTshortorigxxxxxA
\else
\ifnum#1=59 %
\hatcurLCTshortorigxxxxxB
\else
\ifnum#1=60 %
\hatcurLCTshortorigxxxxxC
\else
\ifnum#1=61 %
\hatcurLCTshortorigxxxxxD
\else
\ifnum#1=62 %
\hatcurLCTshortorigxxxxxE
\else
\ifnum#1=63 %
\hatcurLCTshortorigxxxxxF
\else
\ifnum#1=64 %
\hatcurLCTshortorigxxxxxG
\else
??????\fi
\fi
\fi
\fi
\fi
\fi
\fi
}
\newcommand{\hatcurLCzetaorig}[1]{\ifnum#1=58 %
\hatcurLCzetaorigxxxxxA
\else
\ifnum#1=59 %
\hatcurLCzetaorigxxxxxB
\else
\ifnum#1=60 %
\hatcurLCzetaorigxxxxxC
\else
\ifnum#1=61 %
\hatcurLCzetaorigxxxxxD
\else
\ifnum#1=62 %
\hatcurLCzetaorigxxxxxE
\else
\ifnum#1=63 %
\hatcurLCzetaorigxxxxxF
\else
\ifnum#1=64 %
\hatcurLCzetaorigxxxxxG
\else
??????\fi
\fi
\fi
\fi
\fi
\fi
\fi
}
\newcommand{\hatcurPPaequivorig}[1]{\ifnum#1=58 %
\hatcurPPaequivorigxxxxxA
\else
\ifnum#1=59 %
\hatcurPPaequivorigxxxxxB
\else
\ifnum#1=60 %
\hatcurPPaequivorigxxxxxC
\else
\ifnum#1=61 %
\hatcurPPaequivorigxxxxxD
\else
\ifnum#1=62 %
\hatcurPPaequivorigxxxxxE
\else
\ifnum#1=63 %
\hatcurPPaequivorigxxxxxF
\else
\ifnum#1=64 %
\hatcurPPaequivorigxxxxxG
\else
??????\fi
\fi
\fi
\fi
\fi
\fi
\fi
}
\newcommand{\hatcurPParelorig}[1]{\ifnum#1=58 %
\hatcurPParelorigxxxxxA
\else
\ifnum#1=59 %
\hatcurPParelorigxxxxxB
\else
\ifnum#1=60 %
\hatcurPParelorigxxxxxC
\else
\ifnum#1=61 %
\hatcurPParelorigxxxxxD
\else
\ifnum#1=62 %
\hatcurPParelorigxxxxxE
\else
\ifnum#1=63 %
\hatcurPParelorigxxxxxF
\else
\ifnum#1=64 %
\hatcurPParelorigxxxxxG
\else
??????\fi
\fi
\fi
\fi
\fi
\fi
\fi
}
\newcommand{\hatcurPParorig}[1]{\ifnum#1=58 %
\hatcurPParorigxxxxxA
\else
\ifnum#1=59 %
\hatcurPParorigxxxxxB
\else
\ifnum#1=60 %
\hatcurPParorigxxxxxC
\else
\ifnum#1=61 %
\hatcurPParorigxxxxxD
\else
\ifnum#1=62 %
\hatcurPParorigxxxxxE
\else
\ifnum#1=63 %
\hatcurPParorigxxxxxF
\else
\ifnum#1=64 %
\hatcurPParorigxxxxxG
\else
??????\fi
\fi
\fi
\fi
\fi
\fi
\fi
}
\newcommand{\hatcurPPfluxapdimorig}[1]{\ifnum#1=58 %
\hatcurPPfluxapdimorigxxxxxA
\else
\ifnum#1=59 %
\hatcurPPfluxapdimorigxxxxxB
\else
\ifnum#1=60 %
\hatcurPPfluxapdimorigxxxxxC
\else
\ifnum#1=61 %
\hatcurPPfluxapdimorigxxxxxD
\else
\ifnum#1=62 %
\hatcurPPfluxapdimorigxxxxxE
\else
\ifnum#1=63 %
\hatcurPPfluxapdimorigxxxxxF
\else
\ifnum#1=64 %
\hatcurPPfluxapdimorigxxxxxG
\else
??????\fi
\fi
\fi
\fi
\fi
\fi
\fi
}
\newcommand{\hatcurPPfluxaporig}[1]{\ifnum#1=58 %
\hatcurPPfluxaporigxxxxxA
\else
\ifnum#1=59 %
\hatcurPPfluxaporigxxxxxB
\else
\ifnum#1=60 %
\hatcurPPfluxaporigxxxxxC
\else
\ifnum#1=61 %
\hatcurPPfluxaporigxxxxxD
\else
\ifnum#1=62 %
\hatcurPPfluxaporigxxxxxE
\else
\ifnum#1=63 %
\hatcurPPfluxaporigxxxxxF
\else
\ifnum#1=64 %
\hatcurPPfluxaporigxxxxxG
\else
??????\fi
\fi
\fi
\fi
\fi
\fi
\fi
}
\newcommand{\hatcurPPfluxavgdimorig}[1]{\ifnum#1=58 %
\hatcurPPfluxavgdimorigxxxxxA
\else
\ifnum#1=59 %
\hatcurPPfluxavgdimorigxxxxxB
\else
\ifnum#1=60 %
\hatcurPPfluxavgdimorigxxxxxC
\else
\ifnum#1=61 %
\hatcurPPfluxavgdimorigxxxxxD
\else
\ifnum#1=62 %
\hatcurPPfluxavgdimorigxxxxxE
\else
\ifnum#1=63 %
\hatcurPPfluxavgdimorigxxxxxF
\else
\ifnum#1=64 %
\hatcurPPfluxavgdimorigxxxxxG
\else
??????\fi
\fi
\fi
\fi
\fi
\fi
\fi
}
\newcommand{\hatcurPPfluxavglogorig}[1]{\ifnum#1=58 %
\hatcurPPfluxavglogorigxxxxxA
\else
\ifnum#1=59 %
\hatcurPPfluxavglogorigxxxxxB
\else
\ifnum#1=60 %
\hatcurPPfluxavglogorigxxxxxC
\else
\ifnum#1=61 %
\hatcurPPfluxavglogorigxxxxxD
\else
\ifnum#1=62 %
\hatcurPPfluxavglogorigxxxxxE
\else
\ifnum#1=63 %
\hatcurPPfluxavglogorigxxxxxF
\else
\ifnum#1=64 %
\hatcurPPfluxavglogorigxxxxxG
\else
??????\fi
\fi
\fi
\fi
\fi
\fi
\fi
}
\newcommand{\hatcurPPfluxavgorig}[1]{\ifnum#1=58 %
\hatcurPPfluxavgorigxxxxxA
\else
\ifnum#1=59 %
\hatcurPPfluxavgorigxxxxxB
\else
\ifnum#1=60 %
\hatcurPPfluxavgorigxxxxxC
\else
\ifnum#1=61 %
\hatcurPPfluxavgorigxxxxxD
\else
\ifnum#1=62 %
\hatcurPPfluxavgorigxxxxxE
\else
\ifnum#1=63 %
\hatcurPPfluxavgorigxxxxxF
\else
\ifnum#1=64 %
\hatcurPPfluxavgorigxxxxxG
\else
??????\fi
\fi
\fi
\fi
\fi
\fi
\fi
}
\newcommand{\hatcurPPfluxperidimorig}[1]{\ifnum#1=58 %
\hatcurPPfluxperidimorigxxxxxA
\else
\ifnum#1=59 %
\hatcurPPfluxperidimorigxxxxxB
\else
\ifnum#1=60 %
\hatcurPPfluxperidimorigxxxxxC
\else
\ifnum#1=61 %
\hatcurPPfluxperidimorigxxxxxD
\else
\ifnum#1=62 %
\hatcurPPfluxperidimorigxxxxxE
\else
\ifnum#1=63 %
\hatcurPPfluxperidimorigxxxxxF
\else
\ifnum#1=64 %
\hatcurPPfluxperidimorigxxxxxG
\else
??????\fi
\fi
\fi
\fi
\fi
\fi
\fi
}
\newcommand{\hatcurPPfluxperiorig}[1]{\ifnum#1=58 %
\hatcurPPfluxperiorigxxxxxA
\else
\ifnum#1=59 %
\hatcurPPfluxperiorigxxxxxB
\else
\ifnum#1=60 %
\hatcurPPfluxperiorigxxxxxC
\else
\ifnum#1=61 %
\hatcurPPfluxperiorigxxxxxD
\else
\ifnum#1=62 %
\hatcurPPfluxperiorigxxxxxE
\else
\ifnum#1=63 %
\hatcurPPfluxperiorigxxxxxF
\else
\ifnum#1=64 %
\hatcurPPfluxperiorigxxxxxG
\else
??????\fi
\fi
\fi
\fi
\fi
\fi
\fi
}
\newcommand{\hatcurPPgorig}[1]{\ifnum#1=58 %
\hatcurPPgorigxxxxxA
\else
\ifnum#1=59 %
\hatcurPPgorigxxxxxB
\else
\ifnum#1=60 %
\hatcurPPgorigxxxxxC
\else
\ifnum#1=61 %
\hatcurPPgorigxxxxxD
\else
\ifnum#1=62 %
\hatcurPPgorigxxxxxE
\else
\ifnum#1=63 %
\hatcurPPgorigxxxxxF
\else
\ifnum#1=64 %
\hatcurPPgorigxxxxxG
\else
??????\fi
\fi
\fi
\fi
\fi
\fi
\fi
}
\newcommand{\hatcurPPiorig}[1]{\ifnum#1=58 %
\hatcurPPiorigxxxxxA
\else
\ifnum#1=59 %
\hatcurPPiorigxxxxxB
\else
\ifnum#1=60 %
\hatcurPPiorigxxxxxC
\else
\ifnum#1=61 %
\hatcurPPiorigxxxxxD
\else
\ifnum#1=62 %
\hatcurPPiorigxxxxxE
\else
\ifnum#1=63 %
\hatcurPPiorigxxxxxF
\else
\ifnum#1=64 %
\hatcurPPiorigxxxxxG
\else
??????\fi
\fi
\fi
\fi
\fi
\fi
\fi
}
\newcommand{\hatcurPPloggorig}[1]{\ifnum#1=58 %
\hatcurPPloggorigxxxxxA
\else
\ifnum#1=59 %
\hatcurPPloggorigxxxxxB
\else
\ifnum#1=60 %
\hatcurPPloggorigxxxxxC
\else
\ifnum#1=61 %
\hatcurPPloggorigxxxxxD
\else
\ifnum#1=62 %
\hatcurPPloggorigxxxxxE
\else
\ifnum#1=63 %
\hatcurPPloggorigxxxxxF
\else
\ifnum#1=64 %
\hatcurPPloggorigxxxxxG
\else
??????\fi
\fi
\fi
\fi
\fi
\fi
\fi
}
\newcommand{\hatcurPPmelongorig}[1]{\ifnum#1=58 %
\hatcurPPmelongorigxxxxxA
\else
\ifnum#1=59 %
\hatcurPPmelongorigxxxxxB
\else
\ifnum#1=60 %
\hatcurPPmelongorigxxxxxC
\else
\ifnum#1=61 %
\hatcurPPmelongorigxxxxxD
\else
\ifnum#1=62 %
\hatcurPPmelongorigxxxxxE
\else
\ifnum#1=63 %
\hatcurPPmelongorigxxxxxF
\else
\ifnum#1=64 %
\hatcurPPmelongorigxxxxxG
\else
??????\fi
\fi
\fi
\fi
\fi
\fi
\fi
}
\newcommand{\hatcurPPmeorig}[1]{\ifnum#1=58 %
\hatcurPPmeorigxxxxxA
\else
\ifnum#1=59 %
\hatcurPPmeorigxxxxxB
\else
\ifnum#1=60 %
\hatcurPPmeorigxxxxxC
\else
\ifnum#1=61 %
\hatcurPPmeorigxxxxxD
\else
\ifnum#1=62 %
\hatcurPPmeorigxxxxxE
\else
\ifnum#1=63 %
\hatcurPPmeorigxxxxxF
\else
\ifnum#1=64 %
\hatcurPPmeorigxxxxxG
\else
??????\fi
\fi
\fi
\fi
\fi
\fi
\fi
}
\newcommand{\hatcurPPmeshortorig}[1]{\ifnum#1=58 %
\hatcurPPmeshortorigxxxxxA
\else
\ifnum#1=59 %
\hatcurPPmeshortorigxxxxxB
\else
\ifnum#1=60 %
\hatcurPPmeshortorigxxxxxC
\else
\ifnum#1=61 %
\hatcurPPmeshortorigxxxxxD
\else
\ifnum#1=62 %
\hatcurPPmeshortorigxxxxxE
\else
\ifnum#1=63 %
\hatcurPPmeshortorigxxxxxF
\else
\ifnum#1=64 %
\hatcurPPmeshortorigxxxxxG
\else
??????\fi
\fi
\fi
\fi
\fi
\fi
\fi
}
\newcommand{\hatcurPPmlongorig}[1]{\ifnum#1=58 %
\hatcurPPmlongorigxxxxxA
\else
\ifnum#1=59 %
\hatcurPPmlongorigxxxxxB
\else
\ifnum#1=60 %
\hatcurPPmlongorigxxxxxC
\else
\ifnum#1=61 %
\hatcurPPmlongorigxxxxxD
\else
\ifnum#1=62 %
\hatcurPPmlongorigxxxxxE
\else
\ifnum#1=63 %
\hatcurPPmlongorigxxxxxF
\else
\ifnum#1=64 %
\hatcurPPmlongorigxxxxxG
\else
??????\fi
\fi
\fi
\fi
\fi
\fi
\fi
}
\newcommand{\hatcurPPmorig}[1]{\ifnum#1=58 %
\hatcurPPmorigxxxxxA
\else
\ifnum#1=59 %
\hatcurPPmorigxxxxxB
\else
\ifnum#1=60 %
\hatcurPPmorigxxxxxC
\else
\ifnum#1=61 %
\hatcurPPmorigxxxxxD
\else
\ifnum#1=62 %
\hatcurPPmorigxxxxxE
\else
\ifnum#1=63 %
\hatcurPPmorigxxxxxF
\else
\ifnum#1=64 %
\hatcurPPmorigxxxxxG
\else
??????\fi
\fi
\fi
\fi
\fi
\fi
\fi
}
\newcommand{\hatcurPPmrcorrorig}[1]{\ifnum#1=58 %
\hatcurPPmrcorrorigxxxxxA
\else
\ifnum#1=59 %
\hatcurPPmrcorrorigxxxxxB
\else
\ifnum#1=60 %
\hatcurPPmrcorrorigxxxxxC
\else
\ifnum#1=61 %
\hatcurPPmrcorrorigxxxxxD
\else
\ifnum#1=62 %
\hatcurPPmrcorrorigxxxxxE
\else
\ifnum#1=63 %
\hatcurPPmrcorrorigxxxxxF
\else
\ifnum#1=64 %
\hatcurPPmrcorrorigxxxxxG
\else
??????\fi
\fi
\fi
\fi
\fi
\fi
\fi
}
\newcommand{\hatcurPPmshortorig}[1]{\ifnum#1=58 %
\hatcurPPmshortorigxxxxxA
\else
\ifnum#1=59 %
\hatcurPPmshortorigxxxxxB
\else
\ifnum#1=60 %
\hatcurPPmshortorigxxxxxC
\else
\ifnum#1=61 %
\hatcurPPmshortorigxxxxxD
\else
\ifnum#1=62 %
\hatcurPPmshortorigxxxxxE
\else
\ifnum#1=63 %
\hatcurPPmshortorigxxxxxF
\else
\ifnum#1=64 %
\hatcurPPmshortorigxxxxxG
\else
??????\fi
\fi
\fi
\fi
\fi
\fi
\fi
}
\newcommand{\hatcurPPperiorig}[1]{\ifnum#1=58 %
\hatcurPPperiorigxxxxxA
\else
\ifnum#1=59 %
\hatcurPPperiorigxxxxxB
\else
\ifnum#1=60 %
\hatcurPPperiorigxxxxxC
\else
\ifnum#1=61 %
\hatcurPPperiorigxxxxxD
\else
\ifnum#1=62 %
\hatcurPPperiorigxxxxxE
\else
\ifnum#1=63 %
\hatcurPPperiorigxxxxxF
\else
\ifnum#1=64 %
\hatcurPPperiorigxxxxxG
\else
??????\fi
\fi
\fi
\fi
\fi
\fi
\fi
}
\newcommand{\hatcurPPphiconjorig}[1]{\ifnum#1=58 %
\hatcurPPphiconjorigxxxxxA
\else
\ifnum#1=59 %
\hatcurPPphiconjorigxxxxxB
\else
\ifnum#1=60 %
\hatcurPPphiconjorigxxxxxC
\else
\ifnum#1=61 %
\hatcurPPphiconjorigxxxxxD
\else
\ifnum#1=62 %
\hatcurPPphiconjorigxxxxxE
\else
\ifnum#1=63 %
\hatcurPPphiconjorigxxxxxF
\else
\ifnum#1=64 %
\hatcurPPphiconjorigxxxxxG
\else
??????\fi
\fi
\fi
\fi
\fi
\fi
\fi
}
\newcommand{\hatcurPPrelongorig}[1]{\ifnum#1=58 %
\hatcurPPrelongorigxxxxxA
\else
\ifnum#1=59 %
\hatcurPPrelongorigxxxxxB
\else
\ifnum#1=60 %
\hatcurPPrelongorigxxxxxC
\else
\ifnum#1=61 %
\hatcurPPrelongorigxxxxxD
\else
\ifnum#1=62 %
\hatcurPPrelongorigxxxxxE
\else
\ifnum#1=63 %
\hatcurPPrelongorigxxxxxF
\else
\ifnum#1=64 %
\hatcurPPrelongorigxxxxxG
\else
??????\fi
\fi
\fi
\fi
\fi
\fi
\fi
}
\newcommand{\hatcurPPreorig}[1]{\ifnum#1=58 %
\hatcurPPreorigxxxxxA
\else
\ifnum#1=59 %
\hatcurPPreorigxxxxxB
\else
\ifnum#1=60 %
\hatcurPPreorigxxxxxC
\else
\ifnum#1=61 %
\hatcurPPreorigxxxxxD
\else
\ifnum#1=62 %
\hatcurPPreorigxxxxxE
\else
\ifnum#1=63 %
\hatcurPPreorigxxxxxF
\else
\ifnum#1=64 %
\hatcurPPreorigxxxxxG
\else
??????\fi
\fi
\fi
\fi
\fi
\fi
\fi
}
\newcommand{\hatcurPPreshortorig}[1]{\ifnum#1=58 %
\hatcurPPreshortorigxxxxxA
\else
\ifnum#1=59 %
\hatcurPPreshortorigxxxxxB
\else
\ifnum#1=60 %
\hatcurPPreshortorigxxxxxC
\else
\ifnum#1=61 %
\hatcurPPreshortorigxxxxxD
\else
\ifnum#1=62 %
\hatcurPPreshortorigxxxxxE
\else
\ifnum#1=63 %
\hatcurPPreshortorigxxxxxF
\else
\ifnum#1=64 %
\hatcurPPreshortorigxxxxxG
\else
??????\fi
\fi
\fi
\fi
\fi
\fi
\fi
}
\newcommand{\hatcurPPrhoorig}[1]{\ifnum#1=58 %
\hatcurPPrhoorigxxxxxA
\else
\ifnum#1=59 %
\hatcurPPrhoorigxxxxxB
\else
\ifnum#1=60 %
\hatcurPPrhoorigxxxxxC
\else
\ifnum#1=61 %
\hatcurPPrhoorigxxxxxD
\else
\ifnum#1=62 %
\hatcurPPrhoorigxxxxxE
\else
\ifnum#1=63 %
\hatcurPPrhoorigxxxxxF
\else
\ifnum#1=64 %
\hatcurPPrhoorigxxxxxG
\else
??????\fi
\fi
\fi
\fi
\fi
\fi
\fi
}
\newcommand{\hatcurPPrlongorig}[1]{\ifnum#1=58 %
\hatcurPPrlongorigxxxxxA
\else
\ifnum#1=59 %
\hatcurPPrlongorigxxxxxB
\else
\ifnum#1=60 %
\hatcurPPrlongorigxxxxxC
\else
\ifnum#1=61 %
\hatcurPPrlongorigxxxxxD
\else
\ifnum#1=62 %
\hatcurPPrlongorigxxxxxE
\else
\ifnum#1=63 %
\hatcurPPrlongorigxxxxxF
\else
\ifnum#1=64 %
\hatcurPPrlongorigxxxxxG
\else
??????\fi
\fi
\fi
\fi
\fi
\fi
\fi
}
\newcommand{\hatcurPProrig}[1]{\ifnum#1=58 %
\hatcurPProrigxxxxxA
\else
\ifnum#1=59 %
\hatcurPProrigxxxxxB
\else
\ifnum#1=60 %
\hatcurPProrigxxxxxC
\else
\ifnum#1=61 %
\hatcurPProrigxxxxxD
\else
\ifnum#1=62 %
\hatcurPProrigxxxxxE
\else
\ifnum#1=63 %
\hatcurPProrigxxxxxF
\else
\ifnum#1=64 %
\hatcurPProrigxxxxxG
\else
??????\fi
\fi
\fi
\fi
\fi
\fi
\fi
}
\newcommand{\hatcurPPrshortorig}[1]{\ifnum#1=58 %
\hatcurPPrshortorigxxxxxA
\else
\ifnum#1=59 %
\hatcurPPrshortorigxxxxxB
\else
\ifnum#1=60 %
\hatcurPPrshortorigxxxxxC
\else
\ifnum#1=61 %
\hatcurPPrshortorigxxxxxD
\else
\ifnum#1=62 %
\hatcurPPrshortorigxxxxxE
\else
\ifnum#1=63 %
\hatcurPPrshortorigxxxxxF
\else
\ifnum#1=64 %
\hatcurPPrshortorigxxxxxG
\else
??????\fi
\fi
\fi
\fi
\fi
\fi
\fi
}
\newcommand{\hatcurPPtcircorig}[1]{\ifnum#1=58 %
\hatcurPPtcircorigxxxxxA
\else
\ifnum#1=59 %
\hatcurPPtcircorigxxxxxB
\else
\ifnum#1=60 %
\hatcurPPtcircorigxxxxxC
\else
\ifnum#1=61 %
\hatcurPPtcircorigxxxxxD
\else
\ifnum#1=62 %
\hatcurPPtcircorigxxxxxE
\else
\ifnum#1=63 %
\hatcurPPtcircorigxxxxxF
\else
\ifnum#1=64 %
\hatcurPPtcircorigxxxxxG
\else
??????\fi
\fi
\fi
\fi
\fi
\fi
\fi
}
\newcommand{\hatcurPPtefforig}[1]{\ifnum#1=58 %
\hatcurPPtefforigxxxxxA
\else
\ifnum#1=59 %
\hatcurPPtefforigxxxxxB
\else
\ifnum#1=60 %
\hatcurPPtefforigxxxxxC
\else
\ifnum#1=61 %
\hatcurPPtefforigxxxxxD
\else
\ifnum#1=62 %
\hatcurPPtefforigxxxxxE
\else
\ifnum#1=63 %
\hatcurPPtefforigxxxxxF
\else
\ifnum#1=64 %
\hatcurPPtefforigxxxxxG
\else
??????\fi
\fi
\fi
\fi
\fi
\fi
\fi
}
\newcommand{\hatcurPPthetaorig}[1]{\ifnum#1=58 %
\hatcurPPthetaorigxxxxxA
\else
\ifnum#1=59 %
\hatcurPPthetaorigxxxxxB
\else
\ifnum#1=60 %
\hatcurPPthetaorigxxxxxC
\else
\ifnum#1=61 %
\hatcurPPthetaorigxxxxxD
\else
\ifnum#1=62 %
\hatcurPPthetaorigxxxxxE
\else
\ifnum#1=63 %
\hatcurPPthetaorigxxxxxF
\else
\ifnum#1=64 %
\hatcurPPthetaorigxxxxxG
\else
??????\fi
\fi
\fi
\fi
\fi
\fi
\fi
}
\newcommand{\hatcurPPtinfallorig}[1]{\ifnum#1=58 %
\hatcurPPtinfallorigxxxxxA
\else
\ifnum#1=59 %
\hatcurPPtinfallorigxxxxxB
\else
\ifnum#1=60 %
\hatcurPPtinfallorigxxxxxC
\else
\ifnum#1=61 %
\hatcurPPtinfallorigxxxxxD
\else
\ifnum#1=62 %
\hatcurPPtinfallorigxxxxxE
\else
\ifnum#1=63 %
\hatcurPPtinfallorigxxxxxF
\else
\ifnum#1=64 %
\hatcurPPtinfallorigxxxxxG
\else
??????\fi
\fi
\fi
\fi
\fi
\fi
\fi
}
\newcommand{\hatcurRVeccenorig}[1]{\ifnum#1=58 %
\hatcurRVeccenorigxxxxxA
\else
\ifnum#1=59 %
\hatcurRVeccenorigxxxxxB
\else
\ifnum#1=60 %
\hatcurRVeccenorigxxxxxC
\else
\ifnum#1=61 %
\hatcurRVeccenorigxxxxxD
\else
\ifnum#1=62 %
\hatcurRVeccenorigxxxxxE
\else
\ifnum#1=63 %
\hatcurRVeccenorigxxxxxF
\else
\ifnum#1=64 %
\hatcurRVeccenorigxxxxxG
\else
??????\fi
\fi
\fi
\fi
\fi
\fi
\fi
}
\newcommand{\hatcurRVeccentwosiglimorig}[1]{\ifnum#1=58 %
\hatcurRVeccentwosiglimorigxxxxxA
\else
\ifnum#1=59 %
\hatcurRVeccentwosiglimorigxxxxxB
\else
\ifnum#1=60 %
\hatcurRVeccentwosiglimorigxxxxxC
\else
\ifnum#1=61 %
\hatcurRVeccentwosiglimorigxxxxxD
\else
\ifnum#1=62 %
\hatcurRVeccentwosiglimorigxxxxxE
\else
\ifnum#1=63 %
\hatcurRVeccentwosiglimorigxxxxxF
\else
\ifnum#1=64 %
\hatcurRVeccentwosiglimorigxxxxxG
\else
??????\fi
\fi
\fi
\fi
\fi
\fi
\fi
}
\newcommand{\hatcurRVfitrmsAorig}[1]{\ifnum#1=59 %
\hatcurRVfitrmsAorigxxxxxB
\else
\ifnum#1=60 %
\hatcurRVfitrmsAorigxxxxxC
\else
\ifnum#1=61 %
\hatcurRVfitrmsAorigxxxxxD
\else
\ifnum#1=63 %
\hatcurRVfitrmsAorigxxxxxF
\else
\ifnum#1=64 %
\hatcurRVfitrmsAorigxxxxxG
\else
??????\fi
\fi
\fi
\fi
\fi
}
\newcommand{\hatcurRVfitrmsBorig}[1]{\ifnum#1=59 %
\hatcurRVfitrmsBorigxxxxxB
\else
\ifnum#1=60 %
\hatcurRVfitrmsBorigxxxxxC
\else
\ifnum#1=61 %
\hatcurRVfitrmsBorigxxxxxD
\else
\ifnum#1=63 %
\hatcurRVfitrmsBorigxxxxxF
\else
\ifnum#1=64 %
\hatcurRVfitrmsBorigxxxxxG
\else
??????\fi
\fi
\fi
\fi
\fi
}
\newcommand{\hatcurRVfitrmsCorig}[1]{\ifnum#1=60 %
\hatcurRVfitrmsCorigxxxxxC
\else
\ifnum#1=63 %
\hatcurRVfitrmsCorigxxxxxF
\else
??????\fi
\fi
}
\newcommand{\hatcurRVfitrmsorig}[1]{\ifnum#1=58 %
\hatcurRVfitrmsorigxxxxxA
\else
\ifnum#1=62 %
\hatcurRVfitrmsorigxxxxxE
\else
??????\fi
\fi
}
\newcommand{\hatcurRVgammaAorig}[1]{\ifnum#1=59 %
\hatcurRVgammaAorigxxxxxB
\else
\ifnum#1=60 %
\hatcurRVgammaAorigxxxxxC
\else
\ifnum#1=61 %
\hatcurRVgammaAorigxxxxxD
\else
\ifnum#1=63 %
\hatcurRVgammaAorigxxxxxF
\else
\ifnum#1=64 %
\hatcurRVgammaAorigxxxxxG
\else
??????\fi
\fi
\fi
\fi
\fi
}
\newcommand{\hatcurRVgammaBorig}[1]{\ifnum#1=59 %
\hatcurRVgammaBorigxxxxxB
\else
\ifnum#1=60 %
\hatcurRVgammaBorigxxxxxC
\else
\ifnum#1=61 %
\hatcurRVgammaBorigxxxxxD
\else
\ifnum#1=63 %
\hatcurRVgammaBorigxxxxxF
\else
\ifnum#1=64 %
\hatcurRVgammaBorigxxxxxG
\else
??????\fi
\fi
\fi
\fi
\fi
}
\newcommand{\hatcurRVgammaCorig}[1]{\ifnum#1=60 %
\hatcurRVgammaCorigxxxxxC
\else
\ifnum#1=63 %
\hatcurRVgammaCorigxxxxxF
\else
??????\fi
\fi
}
\newcommand{\hatcurRVgammaorig}[1]{\ifnum#1=58 %
\hatcurRVgammaorigxxxxxA
\else
\ifnum#1=62 %
\hatcurRVgammaorigxxxxxE
\else
??????\fi
\fi
}
\newcommand{\hatcurRVhorig}[1]{\ifnum#1=58 %
\hatcurRVhorigxxxxxA
\else
\ifnum#1=59 %
\hatcurRVhorigxxxxxB
\else
\ifnum#1=60 %
\hatcurRVhorigxxxxxC
\else
\ifnum#1=61 %
\hatcurRVhorigxxxxxD
\else
\ifnum#1=62 %
\hatcurRVhorigxxxxxE
\else
\ifnum#1=63 %
\hatcurRVhorigxxxxxF
\else
\ifnum#1=64 %
\hatcurRVhorigxxxxxG
\else
??????\fi
\fi
\fi
\fi
\fi
\fi
\fi
}
\newcommand{\hatcurRVjitterAorig}[1]{\ifnum#1=59 %
\hatcurRVjitterAorigxxxxxB
\else
\ifnum#1=60 %
\hatcurRVjitterAorigxxxxxC
\else
\ifnum#1=61 %
\hatcurRVjitterAorigxxxxxD
\else
\ifnum#1=63 %
\hatcurRVjitterAorigxxxxxF
\else
\ifnum#1=64 %
\hatcurRVjitterAorigxxxxxG
\else
??????\fi
\fi
\fi
\fi
\fi
}
\newcommand{\hatcurRVjitterBorig}[1]{\ifnum#1=59 %
\hatcurRVjitterBorigxxxxxB
\else
\ifnum#1=60 %
\hatcurRVjitterBorigxxxxxC
\else
\ifnum#1=61 %
\hatcurRVjitterBorigxxxxxD
\else
\ifnum#1=63 %
\hatcurRVjitterBorigxxxxxF
\else
\ifnum#1=64 %
\hatcurRVjitterBorigxxxxxG
\else
??????\fi
\fi
\fi
\fi
\fi
}
\newcommand{\hatcurRVjitterCorig}[1]{\ifnum#1=60 %
\hatcurRVjitterCorigxxxxxC
\else
\ifnum#1=63 %
\hatcurRVjitterCorigxxxxxF
\else
??????\fi
\fi
}
\newcommand{\hatcurRVjitterorig}[1]{\ifnum#1=58 %
\hatcurRVjitterorigxxxxxA
\else
\ifnum#1=62 %
\hatcurRVjitterorigxxxxxE
\else
??????\fi
\fi
}
\newcommand{\hatcurRVjittertwosiglimAorig}[1]{\ifnum#1=59 %
\hatcurRVjittertwosiglimAorigxxxxxB
\else
\ifnum#1=60 %
\hatcurRVjittertwosiglimAorigxxxxxC
\else
\ifnum#1=61 %
\hatcurRVjittertwosiglimAorigxxxxxD
\else
\ifnum#1=63 %
\hatcurRVjittertwosiglimAorigxxxxxF
\else
\ifnum#1=64 %
\hatcurRVjittertwosiglimAorigxxxxxG
\else
??????\fi
\fi
\fi
\fi
\fi
}
\newcommand{\hatcurRVjittertwosiglimBorig}[1]{\ifnum#1=59 %
\hatcurRVjittertwosiglimBorigxxxxxB
\else
\ifnum#1=60 %
\hatcurRVjittertwosiglimBorigxxxxxC
\else
\ifnum#1=61 %
\hatcurRVjittertwosiglimBorigxxxxxD
\else
\ifnum#1=63 %
\hatcurRVjittertwosiglimBorigxxxxxF
\else
\ifnum#1=64 %
\hatcurRVjittertwosiglimBorigxxxxxG
\else
??????\fi
\fi
\fi
\fi
\fi
}
\newcommand{\hatcurRVjittertwosiglimCorig}[1]{\ifnum#1=60 %
\hatcurRVjittertwosiglimCorigxxxxxC
\else
\ifnum#1=63 %
\hatcurRVjittertwosiglimCorigxxxxxF
\else
??????\fi
\fi
}
\newcommand{\hatcurRVjittertwosiglimorig}[1]{\ifnum#1=58 %
\hatcurRVjittertwosiglimorigxxxxxA
\else
\ifnum#1=62 %
\hatcurRVjittertwosiglimorigxxxxxE
\else
??????\fi
\fi
}
\newcommand{\hatcurRVkorig}[1]{\ifnum#1=58 %
\hatcurRVkorigxxxxxA
\else
\ifnum#1=59 %
\hatcurRVkorigxxxxxB
\else
\ifnum#1=60 %
\hatcurRVkorigxxxxxC
\else
\ifnum#1=61 %
\hatcurRVkorigxxxxxD
\else
\ifnum#1=62 %
\hatcurRVkorigxxxxxE
\else
\ifnum#1=63 %
\hatcurRVkorigxxxxxF
\else
\ifnum#1=64 %
\hatcurRVkorigxxxxxG
\else
??????\fi
\fi
\fi
\fi
\fi
\fi
\fi
}
\newcommand{\hatcurRVKorig}[1]{\ifnum#1=58 %
\hatcurRVKorigxxxxxA
\else
\ifnum#1=59 %
\hatcurRVKorigxxxxxB
\else
\ifnum#1=60 %
\hatcurRVKorigxxxxxC
\else
\ifnum#1=61 %
\hatcurRVKorigxxxxxD
\else
\ifnum#1=62 %
\hatcurRVKorigxxxxxE
\else
\ifnum#1=63 %
\hatcurRVKorigxxxxxF
\else
\ifnum#1=64 %
\hatcurRVKorigxxxxxG
\else
??????\fi
\fi
\fi
\fi
\fi
\fi
\fi
}
\newcommand{\hatcurRVomegaorig}[1]{\ifnum#1=58 %
\hatcurRVomegaorigxxxxxA
\else
\ifnum#1=59 %
\hatcurRVomegaorigxxxxxB
\else
\ifnum#1=60 %
\hatcurRVomegaorigxxxxxC
\else
\ifnum#1=61 %
\hatcurRVomegaorigxxxxxD
\else
\ifnum#1=62 %
\hatcurRVomegaorigxxxxxE
\else
\ifnum#1=63 %
\hatcurRVomegaorigxxxxxF
\else
\ifnum#1=64 %
\hatcurRVomegaorigxxxxxG
\else
??????\fi
\fi
\fi
\fi
\fi
\fi
\fi
}
\newcommand{\hatcurRVrhorig}[1]{\ifnum#1=58 %
\hatcurRVrhorigxxxxxA
\else
\ifnum#1=59 %
\hatcurRVrhorigxxxxxB
\else
\ifnum#1=60 %
\hatcurRVrhorigxxxxxC
\else
\ifnum#1=61 %
\hatcurRVrhorigxxxxxD
\else
\ifnum#1=62 %
\hatcurRVrhorigxxxxxE
\else
\ifnum#1=63 %
\hatcurRVrhorigxxxxxF
\else
\ifnum#1=64 %
\hatcurRVrhorigxxxxxG
\else
??????\fi
\fi
\fi
\fi
\fi
\fi
\fi
}
\newcommand{\hatcurRVrkorig}[1]{\ifnum#1=58 %
\hatcurRVrkorigxxxxxA
\else
\ifnum#1=59 %
\hatcurRVrkorigxxxxxB
\else
\ifnum#1=60 %
\hatcurRVrkorigxxxxxC
\else
\ifnum#1=61 %
\hatcurRVrkorigxxxxxD
\else
\ifnum#1=62 %
\hatcurRVrkorigxxxxxE
\else
\ifnum#1=63 %
\hatcurRVrkorigxxxxxF
\else
\ifnum#1=64 %
\hatcurRVrkorigxxxxxG
\else
??????\fi
\fi
\fi
\fi
\fi
\fi
\fi
}
\newcommand{\hatcurRVtroneorig}[1]{\ifnum#1=58 %
\hatcurRVtroneorigxxxxxA
\else
\ifnum#1=59 %
\hatcurRVtroneorigxxxxxB
\else
\ifnum#1=60 %
\hatcurRVtroneorigxxxxxC
\else
\ifnum#1=61 %
\hatcurRVtroneorigxxxxxD
\else
\ifnum#1=62 %
\hatcurRVtroneorigxxxxxE
\else
\ifnum#1=63 %
\hatcurRVtroneorigxxxxxF
\else
\ifnum#1=64 %
\hatcurRVtroneorigxxxxxG
\else
??????\fi
\fi
\fi
\fi
\fi
\fi
\fi
}
\newcommand{\hatcurRVtrtwoorig}[1]{\ifnum#1=58 %
\hatcurRVtrtwoorigxxxxxA
\else
\ifnum#1=59 %
\hatcurRVtrtwoorigxxxxxB
\else
\ifnum#1=60 %
\hatcurRVtrtwoorigxxxxxC
\else
\ifnum#1=61 %
\hatcurRVtrtwoorigxxxxxD
\else
\ifnum#1=62 %
\hatcurRVtrtwoorigxxxxxE
\else
\ifnum#1=63 %
\hatcurRVtrtwoorigxxxxxF
\else
\ifnum#1=64 %
\hatcurRVtrtwoorigxxxxxG
\else
??????\fi
\fi
\fi
\fi
\fi
\fi
\fi
}
\newcommand{\hatcurSMEiiloggorig}[1]{\ifnum#1=58 %
\hatcurSMEiiloggorigxxxxxA
\else
\ifnum#1=59 %
\hatcurSMEiiloggorigxxxxxB
\else
\ifnum#1=60 %
\hatcurSMEiiloggorigxxxxxC
\else
\ifnum#1=61 %
\hatcurSMEiiloggorigxxxxxD
\else
\ifnum#1=62 %
\hatcurSMEiiloggorigxxxxxE
\else
??????\fi
\fi
\fi
\fi
\fi
}
\newcommand{\hatcurSMEiitefforig}[1]{\ifnum#1=58 %
\hatcurSMEiitefforigxxxxxA
\else
\ifnum#1=59 %
\hatcurSMEiitefforigxxxxxB
\else
\ifnum#1=60 %
\hatcurSMEiitefforigxxxxxC
\else
\ifnum#1=61 %
\hatcurSMEiitefforigxxxxxD
\else
\ifnum#1=62 %
\hatcurSMEiitefforigxxxxxE
\else
??????\fi
\fi
\fi
\fi
\fi
}
\newcommand{\hatcurSMEiivsinorig}[1]{\ifnum#1=58 %
\hatcurSMEiivsinorigxxxxxA
\else
\ifnum#1=59 %
\hatcurSMEiivsinorigxxxxxB
\else
\ifnum#1=60 %
\hatcurSMEiivsinorigxxxxxC
\else
\ifnum#1=61 %
\hatcurSMEiivsinorigxxxxxD
\else
\ifnum#1=62 %
\hatcurSMEiivsinorigxxxxxE
\else
??????\fi
\fi
\fi
\fi
\fi
}
\newcommand{\hatcurSMEiizfehorig}[1]{\ifnum#1=58 %
\hatcurSMEiizfehorigxxxxxA
\else
\ifnum#1=59 %
\hatcurSMEiizfehorigxxxxxB
\else
\ifnum#1=60 %
\hatcurSMEiizfehorigxxxxxC
\else
\ifnum#1=61 %
\hatcurSMEiizfehorigxxxxxD
\else
\ifnum#1=62 %
\hatcurSMEiizfehorigxxxxxE
\else
??????\fi
\fi
\fi
\fi
\fi
}
\newcommand{\hatcurSMEiizfehshortorig}[1]{\ifnum#1=58 %
\hatcurSMEiizfehshortorigxxxxxA
\else
\ifnum#1=59 %
\hatcurSMEiizfehshortorigxxxxxB
\else
\ifnum#1=60 %
\hatcurSMEiizfehshortorigxxxxxC
\else
\ifnum#1=61 %
\hatcurSMEiizfehshortorigxxxxxD
\else
\ifnum#1=62 %
\hatcurSMEiizfehshortorigxxxxxE
\else
??????\fi
\fi
\fi
\fi
\fi
}
\newcommand{\hatcurSMEiloggorig}[1]{\ifnum#1=58 %
\hatcurSMEiloggorigxxxxxA
\else
\ifnum#1=59 %
\hatcurSMEiloggorigxxxxxB
\else
\ifnum#1=60 %
\hatcurSMEiloggorigxxxxxC
\else
\ifnum#1=61 %
\hatcurSMEiloggorigxxxxxD
\else
\ifnum#1=62 %
\hatcurSMEiloggorigxxxxxE
\else
\ifnum#1=63 %
\hatcurSMEiloggorigxxxxxF
\else
\ifnum#1=64 %
\hatcurSMEiloggorigxxxxxG
\else
??????\fi
\fi
\fi
\fi
\fi
\fi
\fi
}
\newcommand{\hatcurSMEitefforig}[1]{\ifnum#1=58 %
\hatcurSMEitefforigxxxxxA
\else
\ifnum#1=59 %
\hatcurSMEitefforigxxxxxB
\else
\ifnum#1=60 %
\hatcurSMEitefforigxxxxxC
\else
\ifnum#1=61 %
\hatcurSMEitefforigxxxxxD
\else
\ifnum#1=62 %
\hatcurSMEitefforigxxxxxE
\else
\ifnum#1=63 %
\hatcurSMEitefforigxxxxxF
\else
\ifnum#1=64 %
\hatcurSMEitefforigxxxxxG
\else
??????\fi
\fi
\fi
\fi
\fi
\fi
\fi
}
\newcommand{\hatcurSMEivmacorig}[1]{\ifnum#1=58 %
\hatcurSMEivmacorigxxxxxA
\else
\ifnum#1=59 %
\hatcurSMEivmacorigxxxxxB
\else
\ifnum#1=60 %
\hatcurSMEivmacorigxxxxxC
\else
\ifnum#1=61 %
\hatcurSMEivmacorigxxxxxD
\else
\ifnum#1=62 %
\hatcurSMEivmacorigxxxxxE
\else
\ifnum#1=63 %
\hatcurSMEivmacorigxxxxxF
\else
\ifnum#1=64 %
\hatcurSMEivmacorigxxxxxG
\else
??????\fi
\fi
\fi
\fi
\fi
\fi
\fi
}
\newcommand{\hatcurSMEivmicorig}[1]{\ifnum#1=58 %
\hatcurSMEivmicorigxxxxxA
\else
\ifnum#1=59 %
\hatcurSMEivmicorigxxxxxB
\else
\ifnum#1=60 %
\hatcurSMEivmicorigxxxxxC
\else
\ifnum#1=61 %
\hatcurSMEivmicorigxxxxxD
\else
\ifnum#1=62 %
\hatcurSMEivmicorigxxxxxE
\else
\ifnum#1=63 %
\hatcurSMEivmicorigxxxxxF
\else
\ifnum#1=64 %
\hatcurSMEivmicorigxxxxxG
\else
??????\fi
\fi
\fi
\fi
\fi
\fi
\fi
}
\newcommand{\hatcurSMEivsinorig}[1]{\ifnum#1=58 %
\hatcurSMEivsinorigxxxxxA
\else
\ifnum#1=59 %
\hatcurSMEivsinorigxxxxxB
\else
\ifnum#1=60 %
\hatcurSMEivsinorigxxxxxC
\else
\ifnum#1=61 %
\hatcurSMEivsinorigxxxxxD
\else
\ifnum#1=62 %
\hatcurSMEivsinorigxxxxxE
\else
\ifnum#1=63 %
\hatcurSMEivsinorigxxxxxF
\else
\ifnum#1=64 %
\hatcurSMEivsinorigxxxxxG
\else
??????\fi
\fi
\fi
\fi
\fi
\fi
\fi
}
\newcommand{\hatcurSMEizfehorig}[1]{\ifnum#1=58 %
\hatcurSMEizfehorigxxxxxA
\else
\ifnum#1=59 %
\hatcurSMEizfehorigxxxxxB
\else
\ifnum#1=60 %
\hatcurSMEizfehorigxxxxxC
\else
\ifnum#1=61 %
\hatcurSMEizfehorigxxxxxD
\else
\ifnum#1=62 %
\hatcurSMEizfehorigxxxxxE
\else
\ifnum#1=63 %
\hatcurSMEizfehorigxxxxxF
\else
\ifnum#1=64 %
\hatcurSMEizfehorigxxxxxG
\else
??????\fi
\fi
\fi
\fi
\fi
\fi
\fi
}
\newcommand{\hatcurSMEizfehshortorig}[1]{\ifnum#1=58 %
\hatcurSMEizfehshortorigxxxxxA
\else
\ifnum#1=59 %
\hatcurSMEizfehshortorigxxxxxB
\else
\ifnum#1=60 %
\hatcurSMEizfehshortorigxxxxxC
\else
\ifnum#1=61 %
\hatcurSMEizfehshortorigxxxxxD
\else
\ifnum#1=62 %
\hatcurSMEizfehshortorigxxxxxE
\else
\ifnum#1=63 %
\hatcurSMEizfehshortorigxxxxxF
\else
\ifnum#1=64 %
\hatcurSMEizfehshortorigxxxxxG
\else
??????\fi
\fi
\fi
\fi
\fi
\fi
\fi
}
\newcommand{\hatcurXAvorig}[1]{\ifnum#1=58 %
\hatcurXAvorigxxxxxA
\else
\ifnum#1=59 %
\hatcurXAvorigxxxxxB
\else
\ifnum#1=60 %
\hatcurXAvorigxxxxxC
\else
\ifnum#1=61 %
\hatcurXAvorigxxxxxD
\else
\ifnum#1=62 %
\hatcurXAvorigxxxxxE
\else
\ifnum#1=63 %
\hatcurXAvorigxxxxxF
\else
\ifnum#1=64 %
\hatcurXAvorigxxxxxG
\else
??????\fi
\fi
\fi
\fi
\fi
\fi
\fi
}
\newcommand{\hatcurXdistorig}[1]{\ifnum#1=58 %
\hatcurXdistorigxxxxxA
\else
\ifnum#1=59 %
\hatcurXdistorigxxxxxB
\else
\ifnum#1=60 %
\hatcurXdistorigxxxxxC
\else
\ifnum#1=61 %
\hatcurXdistorigxxxxxD
\else
\ifnum#1=62 %
\hatcurXdistorigxxxxxE
\else
\ifnum#1=63 %
\hatcurXdistorigxxxxxF
\else
\ifnum#1=64 %
\hatcurXdistorigxxxxxG
\else
??????\fi
\fi
\fi
\fi
\fi
\fi
\fi
}
\newcommand{\hatcurXdistredorig}[1]{\ifnum#1=58 %
\hatcurXdistredorigxxxxxA
\else
\ifnum#1=59 %
\hatcurXdistredorigxxxxxB
\else
\ifnum#1=60 %
\hatcurXdistredorigxxxxxC
\else
\ifnum#1=61 %
\hatcurXdistredorigxxxxxD
\else
\ifnum#1=62 %
\hatcurXdistredorigxxxxxE
\else
\ifnum#1=63 %
\hatcurXdistredorigxxxxxF
\else
\ifnum#1=64 %
\hatcurXdistredorigxxxxxG
\else
??????\fi
\fi
\fi
\fi
\fi
\fi
\fi
}
\newcommand{\hatcurXEBVorig}[1]{\ifnum#1=58 %
\hatcurXEBVorigxxxxxA
\else
\ifnum#1=59 %
\hatcurXEBVorigxxxxxB
\else
\ifnum#1=60 %
\hatcurXEBVorigxxxxxC
\else
\ifnum#1=61 %
\hatcurXEBVorigxxxxxD
\else
\ifnum#1=62 %
\hatcurXEBVorigxxxxxE
\else
\ifnum#1=63 %
\hatcurXEBVorigxxxxxF
\else
\ifnum#1=64 %
\hatcurXEBVorigxxxxxG
\else
??????\fi
\fi
\fi
\fi
\fi
\fi
\fi
}
\newcommand{\hatcurXjhisoredorig}[1]{\ifnum#1=58 %
\hatcurXjhisoredorigxxxxxA
\else
\ifnum#1=59 %
\hatcurXjhisoredorigxxxxxB
\else
\ifnum#1=60 %
\hatcurXjhisoredorigxxxxxC
\else
\ifnum#1=61 %
\hatcurXjhisoredorigxxxxxD
\else
\ifnum#1=62 %
\hatcurXjhisoredorigxxxxxE
\else
\ifnum#1=63 %
\hatcurXjhisoredorigxxxxxF
\else
\ifnum#1=64 %
\hatcurXjhisoredorigxxxxxG
\else
??????\fi
\fi
\fi
\fi
\fi
\fi
\fi
}
\newcommand{\hatcurXjkisoredorig}[1]{\ifnum#1=58 %
\hatcurXjkisoredorigxxxxxA
\else
\ifnum#1=59 %
\hatcurXjkisoredorigxxxxxB
\else
\ifnum#1=60 %
\hatcurXjkisoredorigxxxxxC
\else
\ifnum#1=61 %
\hatcurXjkisoredorigxxxxxD
\else
\ifnum#1=62 %
\hatcurXjkisoredorigxxxxxE
\else
\ifnum#1=63 %
\hatcurXjkisoredorigxxxxxF
\else
\ifnum#1=64 %
\hatcurXjkisoredorigxxxxxG
\else
??????\fi
\fi
\fi
\fi
\fi
\fi
\fi
}
\newcommand{\hatcurXmhisoredorig}[1]{\ifnum#1=58 %
\hatcurXmhisoredorigxxxxxA
\else
\ifnum#1=59 %
\hatcurXmhisoredorigxxxxxB
\else
\ifnum#1=60 %
\hatcurXmhisoredorigxxxxxC
\else
\ifnum#1=61 %
\hatcurXmhisoredorigxxxxxD
\else
\ifnum#1=62 %
\hatcurXmhisoredorigxxxxxE
\else
\ifnum#1=63 %
\hatcurXmhisoredorigxxxxxF
\else
\ifnum#1=64 %
\hatcurXmhisoredorigxxxxxG
\else
??????\fi
\fi
\fi
\fi
\fi
\fi
\fi
}
\newcommand{\hatcurXmiisoredorig}[1]{\ifnum#1=58 %
\hatcurXmiisoredorigxxxxxA
\else
\ifnum#1=59 %
\hatcurXmiisoredorigxxxxxB
\else
\ifnum#1=60 %
\hatcurXmiisoredorigxxxxxC
\else
\ifnum#1=61 %
\hatcurXmiisoredorigxxxxxD
\else
\ifnum#1=62 %
\hatcurXmiisoredorigxxxxxE
\else
\ifnum#1=63 %
\hatcurXmiisoredorigxxxxxF
\else
\ifnum#1=64 %
\hatcurXmiisoredorigxxxxxG
\else
??????\fi
\fi
\fi
\fi
\fi
\fi
\fi
}
\newcommand{\hatcurXmjisoredorig}[1]{\ifnum#1=58 %
\hatcurXmjisoredorigxxxxxA
\else
\ifnum#1=59 %
\hatcurXmjisoredorigxxxxxB
\else
\ifnum#1=60 %
\hatcurXmjisoredorigxxxxxC
\else
\ifnum#1=61 %
\hatcurXmjisoredorigxxxxxD
\else
\ifnum#1=62 %
\hatcurXmjisoredorigxxxxxE
\else
\ifnum#1=63 %
\hatcurXmjisoredorigxxxxxF
\else
\ifnum#1=64 %
\hatcurXmjisoredorigxxxxxG
\else
??????\fi
\fi
\fi
\fi
\fi
\fi
\fi
}
\newcommand{\hatcurXmkisoredorig}[1]{\ifnum#1=58 %
\hatcurXmkisoredorigxxxxxA
\else
\ifnum#1=59 %
\hatcurXmkisoredorigxxxxxB
\else
\ifnum#1=60 %
\hatcurXmkisoredorigxxxxxC
\else
\ifnum#1=61 %
\hatcurXmkisoredorigxxxxxD
\else
\ifnum#1=62 %
\hatcurXmkisoredorigxxxxxE
\else
\ifnum#1=63 %
\hatcurXmkisoredorigxxxxxF
\else
\ifnum#1=64 %
\hatcurXmkisoredorigxxxxxG
\else
??????\fi
\fi
\fi
\fi
\fi
\fi
\fi
}
\newcommand{\hatcurXmvisoredorig}[1]{\ifnum#1=58 %
\hatcurXmvisoredorigxxxxxA
\else
\ifnum#1=59 %
\hatcurXmvisoredorigxxxxxB
\else
\ifnum#1=60 %
\hatcurXmvisoredorigxxxxxC
\else
\ifnum#1=61 %
\hatcurXmvisoredorigxxxxxD
\else
\ifnum#1=62 %
\hatcurXmvisoredorigxxxxxE
\else
\ifnum#1=63 %
\hatcurXmvisoredorigxxxxxF
\else
\ifnum#1=64 %
\hatcurXmvisoredorigxxxxxG
\else
??????\fi
\fi
\fi
\fi
\fi
\fi
\fi
}
\newcommand{\hatcurXsecdurorig}[1]{\ifnum#1=58 %
\hatcurXsecdurorigxxxxxA
\else
\ifnum#1=59 %
\hatcurXsecdurorigxxxxxB
\else
\ifnum#1=60 %
\hatcurXsecdurorigxxxxxC
\else
\ifnum#1=61 %
\hatcurXsecdurorigxxxxxD
\else
\ifnum#1=62 %
\hatcurXsecdurorigxxxxxE
\else
\ifnum#1=63 %
\hatcurXsecdurorigxxxxxF
\else
\ifnum#1=64 %
\hatcurXsecdurorigxxxxxG
\else
??????\fi
\fi
\fi
\fi
\fi
\fi
\fi
}
\newcommand{\hatcurXsecingdurorig}[1]{\ifnum#1=58 %
\hatcurXsecingdurorigxxxxxA
\else
\ifnum#1=59 %
\hatcurXsecingdurorigxxxxxB
\else
\ifnum#1=60 %
\hatcurXsecingdurorigxxxxxC
\else
\ifnum#1=61 %
\hatcurXsecingdurorigxxxxxD
\else
\ifnum#1=62 %
\hatcurXsecingdurorigxxxxxE
\else
\ifnum#1=63 %
\hatcurXsecingdurorigxxxxxF
\else
\ifnum#1=64 %
\hatcurXsecingdurorigxxxxxG
\else
??????\fi
\fi
\fi
\fi
\fi
\fi
\fi
}
\newcommand{\hatcurXsecondaryorig}[1]{\ifnum#1=58 %
\hatcurXsecondaryorigxxxxxA
\else
\ifnum#1=59 %
\hatcurXsecondaryorigxxxxxB
\else
\ifnum#1=60 %
\hatcurXsecondaryorigxxxxxC
\else
\ifnum#1=61 %
\hatcurXsecondaryorigxxxxxD
\else
\ifnum#1=62 %
\hatcurXsecondaryorigxxxxxE
\else
\ifnum#1=63 %
\hatcurXsecondaryorigxxxxxF
\else
\ifnum#1=64 %
\hatcurXsecondaryorigxxxxxG
\else
??????\fi
\fi
\fi
\fi
\fi
\fi
\fi
}
\newcommand{\hatcurXsecphaseorig}[1]{\ifnum#1=58 %
\hatcurXsecphaseorigxxxxxA
\else
\ifnum#1=59 %
\hatcurXsecphaseorigxxxxxB
\else
\ifnum#1=60 %
\hatcurXsecphaseorigxxxxxC
\else
\ifnum#1=61 %
\hatcurXsecphaseorigxxxxxD
\else
\ifnum#1=62 %
\hatcurXsecphaseorigxxxxxE
\else
\ifnum#1=63 %
\hatcurXsecphaseorigxxxxxF
\else
\ifnum#1=64 %
\hatcurXsecphaseorigxxxxxG
\else
??????\fi
\fi
\fi
\fi
\fi
\fi
\fi
}
\newcommand{\hatcurXviisoredorig}[1]{\ifnum#1=58 %
\hatcurXviisoredorigxxxxxA
\else
\ifnum#1=59 %
\hatcurXviisoredorigxxxxxB
\else
\ifnum#1=60 %
\hatcurXviisoredorigxxxxxC
\else
\ifnum#1=61 %
\hatcurXviisoredorigxxxxxD
\else
\ifnum#1=62 %
\hatcurXviisoredorigxxxxxE
\else
\ifnum#1=63 %
\hatcurXviisoredorigxxxxxF
\else
\ifnum#1=64 %
\hatcurXviisoredorigxxxxxG
\else
??????\fi
\fi
\fi
\fi
\fi
\fi
\fi
}
\newcommand{\hatcurXvkisoredorig}[1]{\ifnum#1=58 %
\hatcurXvkisoredorigxxxxxA
\else
\ifnum#1=59 %
\hatcurXvkisoredorigxxxxxB
\else
\ifnum#1=60 %
\hatcurXvkisoredorigxxxxxC
\else
\ifnum#1=61 %
\hatcurXvkisoredorigxxxxxD
\else
\ifnum#1=62 %
\hatcurXvkisoredorigxxxxxE
\else
\ifnum#1=63 %
\hatcurXvkisoredorigxxxxxF
\else
\ifnum#1=64 %
\hatcurXvkisoredorigxxxxxG
\else
??????\fi
\fi
\fi
\fi
\fi
\fi
\fi
}

%% file: HTR062-011.orig.tex
\newcommand{\hatcurhtrorigxxxxxA}{HTR062-011}                              
\newcommand{\hatcurfieldorigxxxxxA}{\ensuremath{string}}                   
\newcommand{\hatcurCCraorigxxxxxA}{\ensuremath{04^{\mathrm h}35^{\mathrm m}23.17{\mathrm s}}}                            
\newcommand{\hatcurCCdecorigxxxxxA}{\ensuremath{+56{\arcdeg}52{\arcmin}05.5{\arcsec}}}                           
\newcommand{\hatcurCCmagorigxxxxxA}{12.971}                                
\newcommand{\hatcurCCtwomassorigxxxxxA}{2MASS~04352318+5652055}            
\newcommand{\hatcurCCgscorigxxxxxA}{GSC~3740-01482}                        
\newcommand{\hatcurCCtassmvorigxxxxxA}{\ensuremath{12.971\pm0.073}}        
\newcommand{\hatcurCCtassmvshortorigxxxxxA}{\ensuremath{13.0}}             
\newcommand{\hatcurCCtassmBorigxxxxxA}{\ensuremath{13.690\pm0.089}}        
\newcommand{\hatcurCCtassmBshortorigxxxxxA}{\ensuremath{13.7}}             
\newcommand{\hatcurCCtassmIorigxxxxxA}{\ensuremath{nff\pmnff}}             
\newcommand{\hatcurCCtassmIshortorigxxxxxA}{\ensuremath{0.0}}              
\newcommand{\hatcurCCtassmgorigxxxxxA}{\ensuremath{13.28\pm0.12}}          
\newcommand{\hatcurCCtassmgshortorigxxxxxA}{\ensuremath{13.3}}             
\newcommand{\hatcurCCtassmrorigxxxxxA}{\ensuremath{12.74\pm0.12}}          
\newcommand{\hatcurCCtassmrshortorigxxxxxA}{\ensuremath{12.7}}             
\newcommand{\hatcurCCtassmiorigxxxxxA}{\ensuremath{12.50\pm0.12}}          
\newcommand{\hatcurCCtassmishortorigxxxxxA}{\ensuremath{12.5}}             
\newcommand{\hatcurCCtwomassJmagorigxxxxxA}{\ensuremath{11.429\pm0.022}}   
\newcommand{\hatcurCCtwomassHmagorigxxxxxA}{\ensuremath{11.075\pm0.020}}   
\newcommand{\hatcurCCtwomassKmagorigxxxxxA}{\ensuremath{10.978\pm0.023}}   
\newcommand{\hatcurCCcitJmagorigxxxxxA}{\ensuremath{11.441\pm0.022}}       
\newcommand{\hatcurCCcitHmagorigxxxxxA}{\ensuremath{11.069\pm0.020}}       
\newcommand{\hatcurCCcitKmagorigxxxxxA}{\ensuremath{11.002\pm0.023}}       
\newcommand{\hatcurCCbbJmagorigxxxxxA}{\ensuremath{11.498\pm0.024}}        
\newcommand{\hatcurCCbbHmagorigxxxxxA}{\ensuremath{11.091\pm0.021}}        
\newcommand{\hatcurCCbbKmagorigxxxxxA}{\ensuremath{11.022\pm0.023}}        
\newcommand{\hatcurCCesoJmagorigxxxxxA}{\ensuremath{11.501\pm0.026}}       
\newcommand{\hatcurCCesoHmagorigxxxxxA}{\ensuremath{11.087\pm0.026}}       
\newcommand{\hatcurCCesoKmagorigxxxxxA}{\ensuremath{11.021\pm0.024}}       
\newcommand{\hatcurCCesoJHmagorigxxxxxA}{\ensuremath{0.413\pm0.034}}       
\newcommand{\hatcurCCesoJKmagorigxxxxxA}{\ensuremath{0.481\pm0.035}}       
\newcommand{\hatcurCCesoHKmagorigxxxxxA}{\ensuremath{0.067\pm0.035}}       
\newcommand{\hatcurLCdiporigxxxxxA}{\ensuremath{7.4}}                      
\newcommand{\hatcurLCrprstarorigxxxxxA}{\ensuremath{0.0877\pm0.0021}}      
\newcommand{\hatcurLCbsqorigxxxxxA}{\ensuremath{0.092_{-0.062}^{+0.086}}}  
\newcommand{\hatcurLCimporigxxxxxA}{\ensuremath{0.30_{-0.13}^{+0.12}}}     
\newcommand{\hatcurLCzetaorigxxxxxA}{\ensuremath{12.880\pm0.094}}          
\newcommand{\hatcurLCdurorigxxxxxA}{\ensuremath{0.1704\pm0.0020}}          
\newcommand{\hatcurLCdurshortorigxxxxxA}{\ensuremath{0.1704}}              
\newcommand{\hatcurLCdurhrorigxxxxxA}{\ensuremath{4.090\pm0.049}}          
\newcommand{\hatcurLCdurhrshortorigxxxxxA}{\ensuremath{4.090}}             
\newcommand{\hatcurLCqorigxxxxxA}{\ensuremath{0.04250\pm0.00051}}          
\newcommand{\hatcurLCqshortorigxxxxxA}{\ensuremath{0.043}}                 
\newcommand{\hatcurLCingdurorigxxxxxA}{\ensuremath{0.0150\pm0.0015}}       
\newcommand{\hatcurLCPorigxxxxxA}{\ensuremath{4.013848\pm0.000013}}        
\newcommand{\hatcurLCPprecorigxxxxxA}{\ensuremath{4.0138476}}              
\newcommand{\hatcurLCPshortorigxxxxxA}{\ensuremath{4.0138}}                
\newcommand{\hatcurLCTorigxxxxxA}{\ensuremath{2456787.02468\pm0.00066}}    
\newcommand{\hatcurLCTshortorigxxxxxA}{\ensuremath{6787.02468\pm0.00066}}    
\newcommand{\hatcurLCTAorigxxxxxA}{\ensuremath{2456132.7675\pm0.0020}}     
\newcommand{\hatcurLCTBorigxxxxxA}{\ensuremath{2456999.7586\pm0.0011}}     
\newcommand{\hatcurLChatnetmorigxxxxxA}{\ensuremath{12.85294\pm0.00012}}   
\newcommand{\hatcurLCiblendorigxxxxxA}{\ensuremath{1\pm0}}                 
\newcommand{\hatcurLCrhoorigxxxxxA}{\ensuremath{0.564\pm0.066}}            
\newcommand{\hatcurSMEitefforigxxxxxA}{\ensuremath{5947\pm50}}             
\newcommand{\hatcurSMEizfehorigxxxxxA}{\ensuremath{0.032\pm0.080}}         
\newcommand{\hatcurSMEizfehshortorigxxxxxA}{\ensuremath{0.03}}             
\newcommand{\hatcurSMEiloggorigxxxxxA}{\ensuremath{4.24\pm0.10}}           
\newcommand{\hatcurSMEivsinorigxxxxxA}{\ensuremath{4.81\pm0.50}}           
\newcommand{\hatcurSMEivmacorigxxxxxA}{\ensuremath{0.0}}                   
\newcommand{\hatcurSMEivmicorigxxxxxA}{\ensuremath{0.0}}                   
\newcommand{\hatcurSMEiitefforigxxxxxA}{\ensuremath{5931\pm50}}            
\newcommand{\hatcurSMEiizfehorigxxxxxA}{\ensuremath{0.012\pm0.080}}        
\newcommand{\hatcurSMEiizfehshortorigxxxxxA}{\ensuremath{0.012}}           
\newcommand{\hatcurSMEiiloggorigxxxxxA}{\ensuremath{4.183\pm0.036}}        
\newcommand{\hatcurSMEiivsinorigxxxxxA}{\ensuremath{4.91\pm0.50}}          
\newcommand{\hatcurLBizorigxxxxxA}{\ensuremath{0.1839}}                    
\newcommand{\hatcurLBiizorigxxxxxA}{\ensuremath{0.3408}}                   
\newcommand{\hatcurLBiiorigxxxxxA}{\ensuremath{0.2401}}                    
\newcommand{\hatcurLBiiiorigxxxxxA}{\ensuremath{0.3451}}                   
\newcommand{\hatcurLBiIorigxxxxxA}{\ensuremath{0.2206}}                    
\newcommand{\hatcurLBiiIorigxxxxxA}{\ensuremath{0.3449}}                   
\newcommand{\hatcurLBigorigxxxxxA}{\ensuremath{0.5087}}                    
\newcommand{\hatcurLBiigorigxxxxxA}{\ensuremath{0.2683}}                   
\newcommand{\hatcurLBirorigxxxxxA}{\ensuremath{0.3231}}                    
\newcommand{\hatcurLBiirorigxxxxxA}{\ensuremath{0.3457}}                   
\newcommand{\hatcurLBiRorigxxxxxA}{\ensuremath{0.2999}}                    
\newcommand{\hatcurLBiiRorigxxxxxA}{\ensuremath{0.3467}}                   
\newcommand{\hatcurLBikeporigxxxxxA}{\ensuremath{0.1000}}                  
\newcommand{\hatcurLBiikeporigxxxxxA}{\ensuremath{0.1000}}                 
\newcommand{\hatcurISOmorigxxxxxA}{\ensuremath{1.110_{-0.042}^{+0.080}}}   
\newcommand{\hatcurISOmshortorigxxxxxA}{\ensuremath{1.11}}                 
\newcommand{\hatcurISOmlongorigxxxxxA}{\ensuremath{1.110_{-0.042}^{+0.080}}} 
\newcommand{\hatcurISOrorigxxxxxA}{\ensuremath{1.410_{-0.057}^{+0.081}}}   
\newcommand{\hatcurISOrshortorigxxxxxA}{\ensuremath{1.41}}                 
\newcommand{\hatcurISOrlongorigxxxxxA}{\ensuremath{1.410_{-0.057}^{+0.081}}} 
\newcommand{\hatcurISOrhoorigxxxxxA}{\ensuremath{0.564\pm0.066}}           
\newcommand{\hatcurISOrholongorigxxxxxA}{\ensuremath{0.564\pm0.066}}       
\newcommand{\hatcurISOloggorigxxxxxA}{\ensuremath{4.189\pm0.035}}          
\newcommand{\hatcurISOlumorigxxxxxA}{\ensuremath{2.21_{-0.21}^{+0.28}}}    
\newcommand{\hatcurISOlumshortorigxxxxxA}{\ensuremath{2.21}}               
\newcommand{\hatcurISOmvorigxxxxxA}{\ensuremath{3.95\pm0.12}}              
\newcommand{\hatcurISOviorigxxxxxA}{\ensuremath{0.650\pm0.015}}            
\newcommand{\hatcurISOageorigxxxxxA}{\ensuremath{6.10_{-1.66}^{+0.82}}}    
\newcommand{\hatcurISOsigmaorigxxxxxA}{\ensuremath{0.000300\pm0.000055}}   
\newcommand{\hatcurISOMJorigxxxxxA}{\ensuremath{2.88\pm0.11}}              
\newcommand{\hatcurISOMHorigxxxxxA}{\ensuremath{2.56\pm0.11}}              
\newcommand{\hatcurISOMKorigxxxxxA}{\ensuremath{2.51\pm0.11}}              
\newcommand{\hatcurISOJKorigxxxxxA}{\ensuremath{0.370\pm0.010}}            
\newcommand{\hatcurISOspecorigxxxxxA}{G}                                   
\newcommand{\hatcurRVKorigxxxxxA}{\ensuremath{46.6\pm3.7}}                 
\newcommand{\hatcurRVrkorigxxxxxA}{\ensuremath{0\pm0}}                     
\newcommand{\hatcurRVrhorigxxxxxA}{\ensuremath{0\pm0}}                     
\newcommand{\hatcurRVkorigxxxxxA}{\ensuremath{0\pm0}}                      
\newcommand{\hatcurRVhorigxxxxxA}{\ensuremath{0\pm0}}                      
\newcommand{\hatcurRVtroneorigxxxxxA}{\ensuremath{0\pm0}}                  
\newcommand{\hatcurRVtrtwoorigxxxxxA}{\ensuremath{0\pm0}}                  
\newcommand{\hatcurRVgammaorigxxxxxA}{\ensuremath{2.4\pm3.0}}              
\newcommand{\hatcurRVjitterorigxxxxxA}{\ensuremath{4.4\pm4.3}}             
\newcommand{\hatcurRVjittertwosiglimorigxxxxxA}{\ensuremath{<12.3}}        
\newcommand{\hatcurRVfitrmsorigxxxxxA}{\ensuremath{.1fym}}                 %
\newcommand{\hatcurRVeccenorigxxxxxA}{\ensuremath{0\pm0}}                  
\newcommand{\hatcurRVeccentwosiglimorigxxxxxA}{\ensuremath{<0.000}}        
\newcommand{\hatcurRVomegaorigxxxxxA}{\ensuremath{0\pm0}}                  
\newcommand{\hatcurPPiorigxxxxxA}{\ensuremath{87.78\pm0.94}}               
\newcommand{\hatcurPPgorigxxxxxA}{\ensuremath{6.74\pm0.89}}                
\newcommand{\hatcurPPloggorigxxxxxA}{\ensuremath{2.829\pm0.059}}           
\newcommand{\hatcurPParorigxxxxxA}{\ensuremath{7.83_{-0.39}^{+0.28}}}      
\newcommand{\hatcurPParelorigxxxxxA}{\ensuremath{0.05117_{-0.00065}^{+0.00119}}} 
\newcommand{\hatcurPPrhoorigxxxxxA}{\ensuremath{0.281\pm0.050}}            
\newcommand{\hatcurPPmorigxxxxxA}{\ensuremath{0.394\pm0.034}}              
\newcommand{\hatcurPPmshortorigxxxxxA}{\ensuremath{0.39}}                  
\newcommand{\hatcurPPmlongorigxxxxxA}{\ensuremath{0.394\pm0.034}}          
\newcommand{\hatcurPPmeorigxxxxxA}{\ensuremath{125\pm11}}                  
\newcommand{\hatcurPPmeshortorigxxxxxA}{\ensuremath{125.1}}                
\newcommand{\hatcurPPmelongorigxxxxxA}{\ensuremath{125\pm11}}              
\newcommand{\hatcurPProrigxxxxxA}{\ensuremath{1.203_{-0.063}^{+0.090}}}    
\newcommand{\hatcurPPrshortorigxxxxxA}{\ensuremath{1.20}}                  
\newcommand{\hatcurPPrlongorigxxxxxA}{\ensuremath{1.203_{-0.063}^{+0.090}}} 
\newcommand{\hatcurPPreorigxxxxxA}{\ensuremath{13.48_{-0.71}^{+1.01}}}     
\newcommand{\hatcurPPreshortorigxxxxxA}{\ensuremath{13.5}}                 
\newcommand{\hatcurPPrelongorigxxxxxA}{\ensuremath{13.48_{-0.71}^{+1.01}}} 
\newcommand{\hatcurPPmrcorrorigxxxxxA}{\ensuremath{0.19}}                  
\newcommand{\hatcurPPtefforigxxxxxA}{\ensuremath{1500_{-29}^{+40}}}        
\newcommand{\hatcurPPthetaorigxxxxxA}{\ensuremath{0.0299\pm0.0029}}        
\newcommand{\hatcurPPfluxperiorigxxxxxA}{\ensuremath{1.141_{-0.086}^{+0.126}}} 
\newcommand{\hatcurPPfluxperidimorigxxxxxA}{\ensuremath{9}}                
\newcommand{\hatcurPPfluxaporigxxxxxA}{\ensuremath{1.141_{-0.086}^{+0.126}}} 
\newcommand{\hatcurPPfluxapdimorigxxxxxA}{\ensuremath{9}}                  
\newcommand{\hatcurPPfluxavgorigxxxxxA}{\ensuremath{1.141_{-0.086}^{+0.126}}} 
\newcommand{\hatcurPPfluxavgdimorigxxxxxA}{\ensuremath{9}}                 
\newcommand{\hatcurPPfluxavglogorigxxxxxA}{\ensuremath{9.057_{-0.034}^{+0.045}}} 
\newcommand{\hatcurXsecphaseorigxxxxxA}{\ensuremath{0\pm0}}                
\newcommand{\hatcurXsecondaryorigxxxxxA}{\ensuremath{2456789.03160\pm0.00066}} 
\newcommand{\hatcurXsecdurorigxxxxxA}{\ensuremath{0.1704\pm0.0020}}        
\newcommand{\hatcurXsecingdurorigxxxxxA}{\ensuremath{0.0150\pm0.0015}}     
\newcommand{\hatcurPPphiconjorigxxxxxA}{\ensuremath{0\pm0}}                
\newcommand{\hatcurPPperiorigxxxxxA}{\ensuremath{2456786.02122\pm0.00065}} 
\newcommand{\hatcurPPaequivorigxxxxxA}{\ensuremath{0.0346\pm0.0015}}       
\newcommand{\hatcurPPtcircorigxxxxxA}{\ensuremath{212\pm58}}               
\newcommand{\hatcurPPtinfallorigxxxxxA}{\ensuremath{3180\pm650}}           
\newcommand{\hatcurXdistorigxxxxxA}{\ensuremath{505_{-22}^{+30}}}          
\newcommand{\hatcurXAvorigxxxxxA}{\ensuremath{0.587\pm0.091}}              
\newcommand{\hatcurXdistredorigxxxxxA}{\ensuremath{488_{-21}^{+29}}}       
\newcommand{\hatcurXEBVorigxxxxxA}{\ensuremath{0.189\pm0.029}}             
\newcommand{\hatcurXmvisoredorigxxxxxA}{\ensuremath{12.981\pm0.067}}       
\newcommand{\hatcurXmiisoredorigxxxxxA}{\ensuremath{12.026\pm0.026}}       
\newcommand{\hatcurXmjisoredorigxxxxxA}{\ensuremath{11.483\pm0.015}}       
\newcommand{\hatcurXmhisoredorigxxxxxA}{\ensuremath{11.112\pm0.015}}       
\newcommand{\hatcurXmkisoredorigxxxxxA}{\ensuremath{11.014\pm0.017}}       
\newcommand{\hatcurXviisoredorigxxxxxA}{\ensuremath{0.956\pm0.044}}        
\newcommand{\hatcurXvkisoredorigxxxxxA}{\ensuremath{1.967\pm0.075}}        
\newcommand{\hatcurXjhisoredorigxxxxxA}{\ensuremath{0.3710\pm0.0090}}      
\newcommand{\hatcurXjkisoredorigxxxxxA}{\ensuremath{0.469\pm0.015}}        
\newcommand{\hatcurCCpmraorigxxxxxA}{\ensuremath{-10.3\pm1.3}}             
\newcommand{\hatcurCCpmdecorigxxxxxA}{\ensuremath{9.1\pm1.4}}              
\newcommand{\hatcurCCpmorigxxxxxA}{\ensuremath{13.7\pm1.9}}                

%% file: HTR080-003.orig.tex
\newcommand{\hatcurhtrorigxxxxxB}{HTR080-003}                              
\newcommand{\hatcurfieldorigxxxxxB}{\ensuremath{string}}                   
\newcommand{\hatcurCCraorigxxxxxB}{\ensuremath{19^{\mathrm h}29^{\mathrm m}49.92{\mathrm s}}}                            
\newcommand{\hatcurCCdecorigxxxxxB}{\ensuremath{+62{\arcdeg}31{\arcmin}45.2{\arcsec}}}                           
\newcommand{\hatcurCCmagorigxxxxxB}{11.883}                                
\newcommand{\hatcurCCtwomassorigxxxxxB}{2MASS~19295008+6231452}            
\newcommand{\hatcurCCgscorigxxxxxB}{GSC~4234-02195}                        
\newcommand{\hatcurCCtassmvorigxxxxxB}{\ensuremath{11.883\pm0.065}}        
\newcommand{\hatcurCCtassmvshortorigxxxxxB}{\ensuremath{11.9}}             
\newcommand{\hatcurCCtassmBorigxxxxxB}{\ensuremath{12.581\pm0.094}}        
\newcommand{\hatcurCCtassmBshortorigxxxxxB}{\ensuremath{12.6}}             
\newcommand{\hatcurCCtassmIorigxxxxxB}{\ensuremath{11.073\pm0.078}}        
\newcommand{\hatcurCCtassmIshortorigxxxxxB}{\ensuremath{11.1}}             
\newcommand{\hatcurCCtassmgorigxxxxxB}{\ensuremath{12.16\pm0.10}}          
\newcommand{\hatcurCCtassmgshortorigxxxxxB}{\ensuremath{12.2}}             
\newcommand{\hatcurCCtassmrorigxxxxxB}{\ensuremath{11.650\pm0.050}}        
\newcommand{\hatcurCCtassmrshortorigxxxxxB}{\ensuremath{11.7}}             
\newcommand{\hatcurCCtassmiorigxxxxxB}{\ensuremath{11.478\pm0.050}}        
\newcommand{\hatcurCCtassmishortorigxxxxxB}{\ensuremath{11.5}}             
\newcommand{\hatcurCCtwomassJmagorigxxxxxB}{\ensuremath{10.581\pm0.020}}   
\newcommand{\hatcurCCtwomassHmagorigxxxxxB}{\ensuremath{10.268\pm0.018}}   
\newcommand{\hatcurCCtwomassKmagorigxxxxxB}{\ensuremath{10.208\pm0.022}}   
\newcommand{\hatcurCCcitJmagorigxxxxxB}{\ensuremath{10.598\pm0.020}}       
\newcommand{\hatcurCCcitHmagorigxxxxxB}{\ensuremath{10.263\pm0.019}}       
\newcommand{\hatcurCCcitKmagorigxxxxxB}{\ensuremath{10.232\pm0.022}}       
\newcommand{\hatcurCCbbJmagorigxxxxxB}{\ensuremath{10.647\pm0.022}}        
\newcommand{\hatcurCCbbHmagorigxxxxxB}{\ensuremath{10.284\pm0.019}}        
\newcommand{\hatcurCCbbKmagorigxxxxxB}{\ensuremath{10.252\pm0.022}}        
\newcommand{\hatcurCCesoJmagorigxxxxxB}{\ensuremath{10.649\pm0.024}}       
\newcommand{\hatcurCCesoHmagorigxxxxxB}{\ensuremath{10.278\pm0.022}}       
\newcommand{\hatcurCCesoKmagorigxxxxxB}{\ensuremath{10.251\pm0.023}}       
\newcommand{\hatcurCCesoJHmagorigxxxxxB}{\ensuremath{0.370\pm0.030}}       
\newcommand{\hatcurCCesoJKmagorigxxxxxB}{\ensuremath{0.399\pm0.032}}       
\newcommand{\hatcurCCesoHKmagorigxxxxxB}{\ensuremath{0.028\pm0.032}}       
\newcommand{\hatcurLCdiporigxxxxxB}{\ensuremath{12.8}}                     
\newcommand{\hatcurLCrprstarorigxxxxxB}{\ensuremath{0.1023\pm0.0024}}      
\newcommand{\hatcurLCbsqorigxxxxxB}{\ensuremath{0.673_{-0.024}^{+0.026}}}  
\newcommand{\hatcurLCimporigxxxxxB}{\ensuremath{0.820_{-0.015}^{+0.016}}}  
\newcommand{\hatcurLCzetaorigxxxxxB}{\ensuremath{26.30_{-0.44}^{+0.29}}}   
\newcommand{\hatcurLCdurorigxxxxxB}{\ensuremath{0.0982\pm0.0020}}          
\newcommand{\hatcurLCdurshortorigxxxxxB}{\ensuremath{0.0982}}              
\newcommand{\hatcurLCdurhrorigxxxxxB}{\ensuremath{2.357\pm0.049}}          
\newcommand{\hatcurLCdurhrshortorigxxxxxB}{\ensuremath{2.357}}             
\newcommand{\hatcurLCqorigxxxxxB}{\ensuremath{0.02370\pm0.00049}}          
\newcommand{\hatcurLCqshortorigxxxxxB}{\ensuremath{0.024}}                 
\newcommand{\hatcurLCingdurorigxxxxxB}{\ensuremath{0.0248\pm0.0026}}       
\newcommand{\hatcurLCPorigxxxxxB}{\ensuremath{4.141950\pm0.000017}}        
\newcommand{\hatcurLCPprecorigxxxxxB}{\ensuremath{4.1419496}}              
\newcommand{\hatcurLCPshortorigxxxxxB}{\ensuremath{4.1419}}                
\newcommand{\hatcurLCTorigxxxxxB}{\ensuremath{2456754.65000\pm0.00053}}    
\newcommand{\hatcurLCTshortorigxxxxxB}{\ensuremath{6754.65000\pm0.00053}}    
\newcommand{\hatcurLCTAorigxxxxxB}{\ensuremath{2456129.2159\pm0.0024}}     
\newcommand{\hatcurLCTBorigxxxxxB}{\ensuremath{2456791.92753\pm0.00058}}   
\newcommand{\hatcurLChatnetmorigxxxxxB}{\ensuremath{11.727050\pm0.000087}} 
\newcommand{\hatcurLCiblendorigxxxxxB}{\ensuremath{1\pm0}}                 
\newcommand{\hatcurLCrhoorigxxxxxB}{\ensuremath{1.07\pm0.12}}              
\newcommand{\hatcurSMEitefforigxxxxxB}{\ensuremath{5678\pm50}}             
\newcommand{\hatcurSMEizfehorigxxxxxB}{\ensuremath{0.420\pm0.080}}         
\newcommand{\hatcurSMEizfehshortorigxxxxxB}{\ensuremath{0.42}}             
\newcommand{\hatcurSMEiloggorigxxxxxB}{\ensuremath{4.40\pm0.10}}           
\newcommand{\hatcurSMEivsinorigxxxxxB}{\ensuremath{3.05\pm0.50}}           
\newcommand{\hatcurSMEivmacorigxxxxxB}{\ensuremath{0.0}}                   
\newcommand{\hatcurSMEivmicorigxxxxxB}{\ensuremath{0.0}}                   
\newcommand{\hatcurSMEiitefforigxxxxxB}{\ensuremath{5665\pm50}}            
\newcommand{\hatcurSMEiizfehorigxxxxxB}{\ensuremath{0.409\pm0.080}}        
\newcommand{\hatcurSMEiizfehshortorigxxxxxB}{\ensuremath{0.409}}           
\newcommand{\hatcurSMEiiloggorigxxxxxB}{\ensuremath{4.368\pm0.033}}        
\newcommand{\hatcurSMEiivsinorigxxxxxB}{\ensuremath{3.04\pm0.50}}          
\newcommand{\hatcurLBizorigxxxxxB}{\ensuremath{0.2318}}                    
\newcommand{\hatcurLBiizorigxxxxxB}{\ensuremath{0.3284}}                   
\newcommand{\hatcurLBiiorigxxxxxB}{\ensuremath{0.3052}}                    
\newcommand{\hatcurLBiiiorigxxxxxB}{\ensuremath{0.3214}}                   
\newcommand{\hatcurLBiIorigxxxxxB}{\ensuremath{0.2807}}                    
\newcommand{\hatcurLBiiIorigxxxxxB}{\ensuremath{0.3245}}                   
\newcommand{\hatcurLBigorigxxxxxB}{\ensuremath{0.6266}}                    
\newcommand{\hatcurLBiigorigxxxxxB}{\ensuremath{0.1833}}                   
\newcommand{\hatcurLBirorigxxxxxB}{\ensuremath{0.4088}}                    
\newcommand{\hatcurLBiirorigxxxxxB}{\ensuremath{0.3012}}                   
\newcommand{\hatcurLBiRorigxxxxxB}{\ensuremath{0.3800}}                    
\newcommand{\hatcurLBiiRorigxxxxxB}{\ensuremath{0.3079}}                   
\newcommand{\hatcurLBikeporigxxxxxB}{\ensuremath{0.1000}}                  
\newcommand{\hatcurLBiikeporigxxxxxB}{\ensuremath{0.1000}}                 
\newcommand{\hatcurISOmorigxxxxxB}{\ensuremath{1.089\pm0.022}}             
\newcommand{\hatcurISOmshortorigxxxxxB}{\ensuremath{1.09}}                 
\newcommand{\hatcurISOmlongorigxxxxxB}{\ensuremath{1.089\pm0.022}}         
\newcommand{\hatcurISOrorigxxxxxB}{\ensuremath{1.127\pm0.046}}             
\newcommand{\hatcurISOrshortorigxxxxxB}{\ensuremath{1.13}}                 
\newcommand{\hatcurISOrlongorigxxxxxB}{\ensuremath{1.127\pm0.046}}         
\newcommand{\hatcurISOrhoorigxxxxxB}{\ensuremath{1.07\pm0.13}}             
\newcommand{\hatcurISOrholongorigxxxxxB}{\ensuremath{1.07\pm0.13}}         
\newcommand{\hatcurISOloggorigxxxxxB}{\ensuremath{4.369\pm0.032}}          
\newcommand{\hatcurISOlumorigxxxxxB}{\ensuremath{1.17\pm0.11}}             
\newcommand{\hatcurISOlumshortorigxxxxxB}{\ensuremath{1.17}}               
\newcommand{\hatcurISOmvorigxxxxxB}{\ensuremath{4.67\pm0.11}}              
\newcommand{\hatcurISOviorigxxxxxB}{\ensuremath{0.744\pm0.017}}            
\newcommand{\hatcurISOageorigxxxxxB}{\ensuremath{4.3\pm1.0}}               
\newcommand{\hatcurISOsigmaorigxxxxxB}{\ensuremath{0.00170\pm0.00015}}     
\newcommand{\hatcurISOMJorigxxxxxB}{\ensuremath{3.460\pm0.094}}            
\newcommand{\hatcurISOMHorigxxxxxB}{\ensuremath{3.113\pm0.090}}            
\newcommand{\hatcurISOMKorigxxxxxB}{\ensuremath{3.056\pm0.090}}            
\newcommand{\hatcurISOJKorigxxxxxB}{\ensuremath{0.400\pm0.010}}            
\newcommand{\hatcurISOspecorigxxxxxB}{G}                                   
\newcommand{\hatcurRVKorigxxxxxB}{\ensuremath{193.6\pm6.7}}                
\newcommand{\hatcurRVrkorigxxxxxB}{\ensuremath{0\pm0}}                     
\newcommand{\hatcurRVrhorigxxxxxB}{\ensuremath{0\pm0}}                     
\newcommand{\hatcurRVkorigxxxxxB}{\ensuremath{0\pm0}}                      
\newcommand{\hatcurRVhorigxxxxxB}{\ensuremath{0\pm0}}                      
\newcommand{\hatcurRVtroneorigxxxxxB}{\ensuremath{0\pm0}}                  
\newcommand{\hatcurRVtrtwoorigxxxxxB}{\ensuremath{0\pm0}}                  
\newcommand{\hatcurRVgammaAorigxxxxxB}{\ensuremath{-153.5\pm7.2}}          
\newcommand{\hatcurRVjitterAorigxxxxxB}{\ensuremath{2\pm13}}               
\newcommand{\hatcurRVjittertwosiglimAorigxxxxxB}{\ensuremath{<30.2}}       
\newcommand{\hatcurRVfitrmsAorigxxxxxB}{\ensuremath{0.0}}                  
\newcommand{\hatcurRVgammaBorigxxxxxB}{\ensuremath{-21157.4\pm4.3}}        
\newcommand{\hatcurRVjitterBorigxxxxxB}{\ensuremath{0.0\pm2.2}}            
\newcommand{\hatcurRVjittertwosiglimBorigxxxxxB}{\ensuremath{<4.0}}        
\newcommand{\hatcurRVfitrmsBorigxxxxxB}{\ensuremath{0.0}}                  
\newcommand{\hatcurRVeccenorigxxxxxB}{\ensuremath{0\pm0}}                  
\newcommand{\hatcurRVeccentwosiglimorigxxxxxB}{\ensuremath{<0.000}}        
\newcommand{\hatcurRVomegaorigxxxxxB}{\ensuremath{0\pm0}}                  
\newcommand{\hatcurPPiorigxxxxxB}{\ensuremath{85.24\pm0.27}}               
\newcommand{\hatcurPPgorigxxxxxB}{\ensuremath{31.9\pm4.2}}                 
\newcommand{\hatcurPPloggorigxxxxxB}{\ensuremath{3.503\pm0.054}}           
\newcommand{\hatcurPParorigxxxxxB}{\ensuremath{9.89\pm0.38}}               
\newcommand{\hatcurPParelorigxxxxxB}{\ensuremath{0.05193\pm0.00036}}       
\newcommand{\hatcurPPrhoorigxxxxxB}{\ensuremath{1.43\pm0.29}}              
\newcommand{\hatcurPPmorigxxxxxB}{\ensuremath{1.624\pm0.061}}              
\newcommand{\hatcurPPmshortorigxxxxxB}{\ensuremath{1.62}}                  
\newcommand{\hatcurPPmlongorigxxxxxB}{\ensuremath{1.624\pm0.061}}          
\newcommand{\hatcurPPmeorigxxxxxB}{\ensuremath{516\pm19}}                  
\newcommand{\hatcurPPmeshortorigxxxxxB}{\ensuremath{516.1}}                
\newcommand{\hatcurPPmelongorigxxxxxB}{\ensuremath{516\pm19}}              
\newcommand{\hatcurPProrigxxxxxB}{\ensuremath{1.121\pm0.066}}              
\newcommand{\hatcurPPrshortorigxxxxxB}{\ensuremath{1.12}}                  
\newcommand{\hatcurPPrlongorigxxxxxB}{\ensuremath{1.121\pm0.066}}          
\newcommand{\hatcurPPreorigxxxxxB}{\ensuremath{12.57\pm0.74}}              
\newcommand{\hatcurPPreshortorigxxxxxB}{\ensuremath{12.6}}                 
\newcommand{\hatcurPPrelongorigxxxxxB}{\ensuremath{12.57\pm0.74}}          
\newcommand{\hatcurPPmrcorrorigxxxxxB}{\ensuremath{0.04}}                  
\newcommand{\hatcurPPtefforigxxxxxB}{\ensuremath{1273\pm27}}               
\newcommand{\hatcurPPthetaorigxxxxxB}{\ensuremath{0.137\pm0.010}}          
\newcommand{\hatcurPPfluxperiorigxxxxxB}{\ensuremath{5.94\pm0.49}}         
\newcommand{\hatcurPPfluxperidimorigxxxxxB}{\ensuremath{8}}                
\newcommand{\hatcurPPfluxaporigxxxxxB}{\ensuremath{5.94\pm0.49}}           
\newcommand{\hatcurPPfluxapdimorigxxxxxB}{\ensuremath{8}}                  
\newcommand{\hatcurPPfluxavgorigxxxxxB}{\ensuremath{5.94\pm0.49}}          
\newcommand{\hatcurPPfluxavgdimorigxxxxxB}{\ensuremath{8}}                 
\newcommand{\hatcurPPfluxavglogorigxxxxxB}{\ensuremath{8.773\pm0.036}}     
\newcommand{\hatcurXsecphaseorigxxxxxB}{\ensuremath{0\pm0}}                
\newcommand{\hatcurXsecondaryorigxxxxxB}{\ensuremath{2456756.72097\pm0.00053}} 
\newcommand{\hatcurXsecdurorigxxxxxB}{\ensuremath{0.0982\pm0.0020}}        
\newcommand{\hatcurXsecingdurorigxxxxxB}{\ensuremath{0.0248\pm0.0026}}     
\newcommand{\hatcurPPphiconjorigxxxxxB}{\ensuremath{0\pm0}}                
\newcommand{\hatcurPPperiorigxxxxxB}{\ensuremath{2456753.61451\pm0.00052}} 
\newcommand{\hatcurPPaequivorigxxxxxB}{\ensuremath{0.0480\pm0.0020}}       
\newcommand{\hatcurPPtcircorigxxxxxB}{\ensuremath{1400\pm500}}             
\newcommand{\hatcurPPtinfallorigxxxxxB}{\ensuremath{2490\pm500}}           
\newcommand{\hatcurXdistorigxxxxxB}{\ensuremath{275\pm12}}                 
\newcommand{\hatcurXAvorigxxxxxB}{\ensuremath{0.024_{-0.024}^{+0.087}}}    
\newcommand{\hatcurXdistredorigxxxxxB}{\ensuremath{272\pm11}}              
\newcommand{\hatcurXEBVorigxxxxxB}{\ensuremath{0.0080_{-0.0080}^{+0.0280}}} 
\newcommand{\hatcurXmvisoredorigxxxxxB}{\ensuremath{11.882\pm0.043}}       
\newcommand{\hatcurXmiisoredorigxxxxxB}{\ensuremath{11.116\pm0.020}}       
\newcommand{\hatcurXmjisoredorigxxxxxB}{\ensuremath{10.645\pm0.014}}       
\newcommand{\hatcurXmhisoredorigxxxxxB}{\ensuremath{10.294\pm0.013}}       
\newcommand{\hatcurXmkisoredorigxxxxxB}{\ensuremath{10.233\pm0.015}}       
\newcommand{\hatcurXviisoredorigxxxxxB}{\ensuremath{0.764_{-0.020}^{+0.032}}} 
\newcommand{\hatcurXvkisoredorigxxxxxB}{\ensuremath{1.648\pm0.048}}        
\newcommand{\hatcurXjhisoredorigxxxxxB}{\ensuremath{0.3510\pm0.0091}}      
\newcommand{\hatcurXjkisoredorigxxxxxB}{\ensuremath{0.411\pm0.012}}        
\newcommand{\hatcurCCpmraorigxxxxxB}{\ensuremath{20.7\pm1.1}}              
\newcommand{\hatcurCCpmdecorigxxxxxB}{\ensuremath{-5.00\pm0.90}}           
\newcommand{\hatcurCCpmorigxxxxxB}{\ensuremath{21.3\pm1.4}}                

%% file: HTR089-018.orig.tex
\newcommand{\hatcurhtrorigxxxxxC}{HTR089-018}                       
\newcommand{\hatcurfieldorigxxxxxC}{\ensuremath{string}}            
\newcommand{\hatcurCCraorigxxxxxC}{\ensuremath{01^{\mathrm h}53^{\mathrm m}07.76{\mathrm s}}}                     
\newcommand{\hatcurCCdecorigxxxxxC}{\ensuremath{+52{\arcdeg}03{\arcmin}14.1{\arcsec}}}                    
\newcommand{\hatcurCCmagorigxxxxxC}{9.523}                          
\newcommand{\hatcurCCtwomassorigxxxxxC}{2MASS~01530777+5203140}     
\newcommand{\hatcurCCgscorigxxxxxC}{GSC~3292-01330}                 
\newcommand{\hatcurCCtassmvorigxxxxxC}{\ensuremath{9.5230\pm0.0010}} 
\newcommand{\hatcurCCtassmvshortorigxxxxxC}{\ensuremath{9.5}}       
\newcommand{\hatcurCCtassmBorigxxxxxC}{\ensuremath{10.18\pm0.11}}   
\newcommand{\hatcurCCtassmBshortorigxxxxxC}{\ensuremath{10.2}}      
\newcommand{\hatcurCCtassmIorigxxxxxC}{\ensuremath{9.077\pm0.042}}  
\newcommand{\hatcurCCtassmIshortorigxxxxxC}{\ensuremath{9.1}}       
\newcommand{\hatcurCCtassmgorigxxxxxC}{\ensuremath{9.7580\pm0.0010}} 
\newcommand{\hatcurCCtassmgshortorigxxxxxC}{\ensuremath{9.8}}       
\newcommand{\hatcurCCtassmrorigxxxxxC}{\ensuremath{9.5000\pm0.0010}} 
\newcommand{\hatcurCCtassmrshortorigxxxxxC}{\ensuremath{9.5}}       
\newcommand{\hatcurCCtassmiorigxxxxxC}{\ensuremath{9.3160\pm0.0010}} 
\newcommand{\hatcurCCtassmishortorigxxxxxC}{\ensuremath{9.3}}       
\newcommand{\hatcurCCtwomassJmagorigxxxxxC}{\ensuremath{8.677\pm0.052}} 
\newcommand{\hatcurCCtwomassHmagorigxxxxxC}{\ensuremath{8.396\pm0.029}} 
\newcommand{\hatcurCCtwomassKmagorigxxxxxC}{\ensuremath{8.368\pm0.031}} 
\newcommand{\hatcurCCcitJmagorigxxxxxC}{\ensuremath{8.697\pm0.050}} 
\newcommand{\hatcurCCcitHmagorigxxxxxC}{\ensuremath{8.392\pm0.029}} 
\newcommand{\hatcurCCcitKmagorigxxxxxC}{\ensuremath{8.392\pm0.031}} 
\newcommand{\hatcurCCbbJmagorigxxxxxC}{\ensuremath{8.742\pm0.055}}  
\newcommand{\hatcurCCbbHmagorigxxxxxC}{\ensuremath{8.412\pm0.030}}  
\newcommand{\hatcurCCbbKmagorigxxxxxC}{\ensuremath{8.412\pm0.031}}  
\newcommand{\hatcurCCesoJmagorigxxxxxC}{\ensuremath{8.742\pm0.055}} 
\newcommand{\hatcurCCesoHmagorigxxxxxC}{\ensuremath{8.406\pm0.033}} 
\newcommand{\hatcurCCesoKmagorigxxxxxC}{\ensuremath{8.411\pm0.032}} 
\newcommand{\hatcurCCesoJHmagorigxxxxxC}{\ensuremath{0.337\pm0.063}} 
\newcommand{\hatcurCCesoJKmagorigxxxxxC}{\ensuremath{0.332\pm0.064}} 
\newcommand{\hatcurCCesoHKmagorigxxxxxC}{\ensuremath{-0.005\pm0.047}} 
\newcommand{\hatcurLCdiporigxxxxxC}{\ensuremath{5.8}}               
\newcommand{\hatcurLCrprstarorigxxxxxC}{\ensuremath{0.0671\pm0.0021}} 
\newcommand{\hatcurLCbsqorigxxxxxC}{\ensuremath{0.119_{-0.090}^{+0.146}}} 
\newcommand{\hatcurLCimporigxxxxxC}{\ensuremath{0.35_{-0.17}^{+0.17}}} 
\newcommand{\hatcurLCzetaorigxxxxxC}{\ensuremath{10.593\pm0.100}}   
\newcommand{\hatcurLCdurorigxxxxxC}{\ensuremath{0.2035\pm0.0027}}   
\newcommand{\hatcurLCdurshortorigxxxxxC}{\ensuremath{0.2035}}       
\newcommand{\hatcurLCdurhrorigxxxxxC}{\ensuremath{4.885\pm0.064}}   
\newcommand{\hatcurLCdurhrshortorigxxxxxC}{\ensuremath{4.885}}      
\newcommand{\hatcurLCqorigxxxxxC}{\ensuremath{0.04240\pm0.00056}}   
\newcommand{\hatcurLCqshortorigxxxxxC}{\ensuremath{0.042}}          
\newcommand{\hatcurLCingdurorigxxxxxC}{\ensuremath{0.0143\pm0.0026}} 
\newcommand{\hatcurLCPorigxxxxxC}{\ensuremath{4.7947693\pm0.0000077}} 
\newcommand{\hatcurLCPprecorigxxxxxC}{\ensuremath{4.7947693}}       
\newcommand{\hatcurLCPshortorigxxxxxC}{\ensuremath{4.7948}}         
\newcommand{\hatcurLCTorigxxxxxC}{\ensuremath{2456721.12317\pm0.00095}} 
\newcommand{\hatcurLCTshortorigxxxxxC}{\ensuremath{6721.12317\pm0.00095}} 
\newcommand{\hatcurLCTAorigxxxxxC}{\ensuremath{2454764.8575\pm0.0032}} 
\newcommand{\hatcurLCTBorigxxxxxC}{\ensuremath{2456941.6825\pm0.0010}} 
\newcommand{\hatcurLCrhoorigxxxxxC}{\ensuremath{0.355_{-0.081}^{+0.055}}} 
\newcommand{\hatcurSMEitefforigxxxxxC}{\ensuremath{6223\pm50}}      
\newcommand{\hatcurSMEizfehorigxxxxxC}{\ensuremath{-0.404\pm0.080}} 
\newcommand{\hatcurSMEizfehshortorigxxxxxC}{\ensuremath{-0.40}}     
\newcommand{\hatcurSMEiloggorigxxxxxC}{\ensuremath{3.59\pm0.10}}    
\newcommand{\hatcurSMEivsinorigxxxxxC}{\ensuremath{10.71\pm0.50}}   
\newcommand{\hatcurSMEivmacorigxxxxxC}{\ensuremath{0.0}}            
\newcommand{\hatcurSMEivmicorigxxxxxC}{\ensuremath{0.0}}            
\newcommand{\hatcurSMEiitefforigxxxxxC}{\ensuremath{6462\pm50}}     
\newcommand{\hatcurSMEiizfehorigxxxxxC}{\ensuremath{-0.237\pm0.080}} 
\newcommand{\hatcurSMEiizfehshortorigxxxxxC}{\ensuremath{-0.237}}   
\newcommand{\hatcurSMEiiloggorigxxxxxC}{\ensuremath{4.052\pm0.049}} 
\newcommand{\hatcurSMEiivsinorigxxxxxC}{\ensuremath{10.42\pm0.50}}  
\newcommand{\hatcurLBizorigxxxxxC}{\ensuremath{0.1281}}             
\newcommand{\hatcurLBiizorigxxxxxC}{\ensuremath{0.3535}}            
\newcommand{\hatcurLBiiorigxxxxxC}{\ensuremath{0.1761}}             
\newcommand{\hatcurLBiiiorigxxxxxC}{\ensuremath{0.3644}}            
\newcommand{\hatcurLBiIorigxxxxxC}{\ensuremath{0.1586}}             
\newcommand{\hatcurLBiiIorigxxxxxC}{\ensuremath{0.3618}}            
\newcommand{\hatcurLBigorigxxxxxC}{\ensuremath{0.4024}}             
\newcommand{\hatcurLBiigorigxxxxxC}{\ensuremath{0.3343}}            
\newcommand{\hatcurLBirorigxxxxxC}{\ensuremath{0.2448}}             
\newcommand{\hatcurLBiirorigxxxxxC}{\ensuremath{0.3779}}            
\newcommand{\hatcurLBiRorigxxxxxC}{\ensuremath{0.2249}}             
\newcommand{\hatcurLBiiRorigxxxxxC}{\ensuremath{0.3760}}            
\newcommand{\hatcurLBikeporigxxxxxC}{\ensuremath{0.1000}}           
\newcommand{\hatcurLBiikeporigxxxxxC}{\ensuremath{0.1000}}          
\newcommand{\hatcurISOmorigxxxxxC}{\ensuremath{1.310\pm0.070}}      
\newcommand{\hatcurISOmshortorigxxxxxC}{\ensuremath{1.31}}          
\newcommand{\hatcurISOmlongorigxxxxxC}{\ensuremath{1.310\pm0.070}}  
\newcommand{\hatcurISOrorigxxxxxC}{\ensuremath{1.726_{-0.092}^{+0.178}}} 
\newcommand{\hatcurISOrshortorigxxxxxC}{\ensuremath{1.73}}          
\newcommand{\hatcurISOrlongorigxxxxxC}{\ensuremath{1.726_{-0.092}^{+0.178}}} 
\newcommand{\hatcurISOrhoorigxxxxxC}{\ensuremath{0.357_{-0.082}^{+0.055}}} 
\newcommand{\hatcurISOrholongorigxxxxxC}{\ensuremath{0.357_{-0.082}^{+0.055}}} 
\newcommand{\hatcurISOloggorigxxxxxC}{\ensuremath{4.078\pm0.056}}   
\newcommand{\hatcurISOlumorigxxxxxC}{\ensuremath{4.66_{-0.52}^{+1.03}}} 
\newcommand{\hatcurISOlumshortorigxxxxxC}{\ensuremath{4.66}}        
\newcommand{\hatcurISOmvorigxxxxxC}{\ensuremath{3.09\pm0.18}}       
\newcommand{\hatcurISOviorigxxxxxC}{\ensuremath{0.506\pm0.012}}     
\newcommand{\hatcurISOageorigxxxxxC}{\ensuremath{2.88\pm0.56}}      
\newcommand{\hatcurISOsigmaorigxxxxxC}{\ensuremath{0.000200\pm0.000046}} 
\newcommand{\hatcurISOMJorigxxxxxC}{\ensuremath{2.27\pm0.18}}       
\newcommand{\hatcurISOMHorigxxxxxC}{\ensuremath{2.04\pm0.17}}       
\newcommand{\hatcurISOMKorigxxxxxC}{\ensuremath{2.00\pm0.17}}       
\newcommand{\hatcurISOJKorigxxxxxC}{\ensuremath{0.270\pm0.010}}     
\newcommand{\hatcurISOspecorigxxxxxC}{F}                            
\newcommand{\hatcurRVKorigxxxxxC}{\ensuremath{49.7\pm4.7}}          
\newcommand{\hatcurRVrkorigxxxxxC}{\ensuremath{0\pm0}}              
\newcommand{\hatcurRVrhorigxxxxxC}{\ensuremath{0\pm0}}              
\newcommand{\hatcurRVkorigxxxxxC}{\ensuremath{0\pm0}}               
\newcommand{\hatcurRVhorigxxxxxC}{\ensuremath{0\pm0}}               
\newcommand{\hatcurRVtroneorigxxxxxC}{\ensuremath{0\pm0}}           
\newcommand{\hatcurRVtrtwoorigxxxxxC}{\ensuremath{0\pm0}}           
\newcommand{\hatcurRVgammaAorigxxxxxC}{\ensuremath{9.2\pm5.8}}      
\newcommand{\hatcurRVjitterAorigxxxxxC}{\ensuremath{12.2\pm5.1}}    
\newcommand{\hatcurRVjittertwosiglimAorigxxxxxC}{\ensuremath{<21.0}} 
\newcommand{\hatcurRVfitrmsAorigxxxxxC}{\ensuremath{0.0}}           
\newcommand{\hatcurRVgammaBorigxxxxxC}{\ensuremath{54.4\pm7.0}}     
\newcommand{\hatcurRVjitterBorigxxxxxC}{\ensuremath{0.1\pm3.3}}     
\newcommand{\hatcurRVjittertwosiglimBorigxxxxxC}{\ensuremath{<7.6}} 
\newcommand{\hatcurRVfitrmsBorigxxxxxC}{\ensuremath{0.0}}           
\newcommand{\hatcurRVgammaCorigxxxxxC}{\ensuremath{6027.2\pm7.2}}   
\newcommand{\hatcurRVjitterCorigxxxxxC}{\ensuremath{0.1\pm3.1}}     
\newcommand{\hatcurRVjittertwosiglimCorigxxxxxC}{\ensuremath{<5.8}} 
\newcommand{\hatcurRVfitrmsCorigxxxxxC}{\ensuremath{0.0}}           
\newcommand{\hatcurRVeccenorigxxxxxC}{\ensuremath{0\pm0}}           
\newcommand{\hatcurRVeccentwosiglimorigxxxxxC}{\ensuremath{<0.000}} 
\newcommand{\hatcurRVomegaorigxxxxxC}{\ensuremath{0\pm0}}           
\newcommand{\hatcurPPiorigxxxxxC}{\ensuremath{87.4\pm1.4}}          
\newcommand{\hatcurPPgorigxxxxxC}{\ensuremath{9.6_{-2.0}^{+1.5}}}   
\newcommand{\hatcurPPloggorigxxxxxC}{\ensuremath{2.984_{-0.100}^{+0.062}}} 
\newcommand{\hatcurPParorigxxxxxC}{\ensuremath{7.57_{-0.63}^{+0.37}}} 
\newcommand{\hatcurPParelorigxxxxxC}{\ensuremath{0.0609\pm0.0011}}  
\newcommand{\hatcurPPrhoorigxxxxxC}{\ensuremath{0.43\pm0.11}}       
\newcommand{\hatcurPPmorigxxxxxC}{\ensuremath{0.492\pm0.049}}       
\newcommand{\hatcurPPmshortorigxxxxxC}{\ensuremath{0.49}}           
\newcommand{\hatcurPPmlongorigxxxxxC}{\ensuremath{0.492\pm0.049}}   
\newcommand{\hatcurPPmeorigxxxxxC}{\ensuremath{156\pm16}}           
\newcommand{\hatcurPPmeshortorigxxxxxC}{\ensuremath{156.3}}         
\newcommand{\hatcurPPmelongorigxxxxxC}{\ensuremath{156\pm16}}       
\newcommand{\hatcurPProrigxxxxxC}{\ensuremath{1.116_{-0.064}^{+0.150}}} 
\newcommand{\hatcurPPrshortorigxxxxxC}{\ensuremath{1.12}}           
\newcommand{\hatcurPPrlongorigxxxxxC}{\ensuremath{1.116_{-0.064}^{+0.150}}} 
\newcommand{\hatcurPPreorigxxxxxC}{\ensuremath{12.51_{-0.72}^{+1.69}}} 
\newcommand{\hatcurPPreshortorigxxxxxC}{\ensuremath{12.5}}          
\newcommand{\hatcurPPrelongorigxxxxxC}{\ensuremath{12.51_{-0.72}^{+1.69}}} 
\newcommand{\hatcurPPmrcorrorigxxxxxC}{\ensuremath{0.16}}           
\newcommand{\hatcurPPtefforigxxxxxC}{\ensuremath{1662_{-42}^{+73}}} 
\newcommand{\hatcurPPthetaorigxxxxxC}{\ensuremath{0.0403\pm0.0054}} 
\newcommand{\hatcurPPfluxperiorigxxxxxC}{\ensuremath{1.72_{-0.17}^{+0.32}}} 
\newcommand{\hatcurPPfluxperidimorigxxxxxC}{\ensuremath{9}}         
\newcommand{\hatcurPPfluxaporigxxxxxC}{\ensuremath{1.72_{-0.17}^{+0.32}}} 
\newcommand{\hatcurPPfluxapdimorigxxxxxC}{\ensuremath{9}}           
\newcommand{\hatcurPPfluxavgorigxxxxxC}{\ensuremath{1.72_{-0.17}^{+0.32}}} 
\newcommand{\hatcurPPfluxavgdimorigxxxxxC}{\ensuremath{9}}          
\newcommand{\hatcurPPfluxavglogorigxxxxxC}{\ensuremath{9.237_{-0.044}^{+0.075}}} 
\newcommand{\hatcurXsecphaseorigxxxxxC}{\ensuremath{0\pm0}}         
\newcommand{\hatcurXsecondaryorigxxxxxC}{\ensuremath{2456723.52055\pm0.00095}} 
\newcommand{\hatcurXsecdurorigxxxxxC}{\ensuremath{0.2035\pm0.0027}} 
\newcommand{\hatcurXsecingdurorigxxxxxC}{\ensuremath{0.0143\pm0.0026}} 
\newcommand{\hatcurPPphiconjorigxxxxxC}{\ensuremath{0\pm0}}         
\newcommand{\hatcurPPperiorigxxxxxC}{\ensuremath{2456719.92447\pm0.00095}} 
\newcommand{\hatcurPPaequivorigxxxxxC}{\ensuremath{0.0281_{-0.0023}^{+0.0015}}} 
\newcommand{\hatcurPPtcircorigxxxxxC}{\ensuremath{900\pm340}}       
\newcommand{\hatcurPPtinfallorigxxxxxC}{\ensuremath{2970\pm870}}    
\newcommand{\hatcurXdistorigxxxxxC}{\ensuremath{192_{-11}^{+20}}}   
\newcommand{\hatcurXAvorigxxxxxC}{\ensuremath{0.028_{-0.028}^{+0.046}}} 
\newcommand{\hatcurXdistredorigxxxxxC}{\ensuremath{191_{-11}^{+20}}} 
\newcommand{\hatcurXEBVorigxxxxxC}{\ensuremath{0.0090_{-0.0090}^{+0.0150}}} 
\newcommand{\hatcurXmvisoredorigxxxxxC}{\ensuremath{9.5230\pm0.0010}} 
\newcommand{\hatcurXmiisoredorigxxxxxC}{\ensuremath{9.0000_{-0.0130}^{+0.0100}}} 
\newcommand{\hatcurXmjisoredorigxxxxxC}{\ensuremath{8.681\pm0.016}} 
\newcommand{\hatcurXmhisoredorigxxxxxC}{\ensuremath{8.448\pm0.021}} 
\newcommand{\hatcurXmkisoredorigxxxxxC}{\ensuremath{8.405\pm0.022}} 
\newcommand{\hatcurXviisoredorigxxxxxC}{\ensuremath{0.523\pm0.012}} 
\newcommand{\hatcurXvkisoredorigxxxxxC}{\ensuremath{1.118\pm0.022}} 
\newcommand{\hatcurXjhisoredorigxxxxxC}{\ensuremath{0.2320\pm0.0074}} 
\newcommand{\hatcurXjkisoredorigxxxxxC}{\ensuremath{0.2760\pm0.0069}} 
\newcommand{\hatcurCCpmraorigxxxxxC}{\ensuremath{25.50\pm0.60}}     
\newcommand{\hatcurCCpmdecorigxxxxxC}{\ensuremath{5.50\pm0.70}}     
\newcommand{\hatcurCCpmorigxxxxxC}{\ensuremath{26.09\pm0.92}}       

%% file: HTR094-024.orig.tex
\newcommand{\hatcurhtrorigxxxxxD}{HTR094-024}                       
\newcommand{\hatcurfieldorigxxxxxD}{\ensuremath{string}}            
\newcommand{\hatcurCCraorigxxxxxD}{\ensuremath{05^{\mathrm h}01^{\mathrm m}55.27{\mathrm s}}}                     
\newcommand{\hatcurCCdecorigxxxxxD}{\ensuremath{+50{\arcdeg}07{\arcmin}52.6{\arcsec}}}                    
\newcommand{\hatcurCCmagorigxxxxxD}{13.188}                         
\newcommand{\hatcurCCtwomassorigxxxxxD}{2MASS~05015525+5007526}     
\newcommand{\hatcurCCgscorigxxxxxD}{GSC~3352-00595}                 
\newcommand{\hatcurCCtassmvorigxxxxxD}{\ensuremath{13.188\pm0.029}} 
\newcommand{\hatcurCCtassmvshortorigxxxxxD}{\ensuremath{13.2}}      
\newcommand{\hatcurCCtassmBorigxxxxxD}{\ensuremath{14.040\pm0.056}} 
\newcommand{\hatcurCCtassmBshortorigxxxxxD}{\ensuremath{14.0}}      
\newcommand{\hatcurCCtassmIorigxxxxxD}{\ensuremath{12.05\pm0.12}}   
\newcommand{\hatcurCCtassmIshortorigxxxxxD}{\ensuremath{12.1}}      
\newcommand{\hatcurCCtassmgorigxxxxxD}{\ensuremath{13.550\pm0.045}} 
\newcommand{\hatcurCCtassmgshortorigxxxxxD}{\ensuremath{13.6}}      
\newcommand{\hatcurCCtassmrorigxxxxxD}{\ensuremath{12.915\pm0.033}} 
\newcommand{\hatcurCCtassmrshortorigxxxxxD}{\ensuremath{12.9}}      
\newcommand{\hatcurCCtassmiorigxxxxxD}{\ensuremath{12.675\pm0.051}} 
\newcommand{\hatcurCCtassmishortorigxxxxxD}{\ensuremath{12.7}}      
\newcommand{\hatcurCCtwomassJmagorigxxxxxD}{\ensuremath{11.598\pm0.021}} 
\newcommand{\hatcurCCtwomassHmagorigxxxxxD}{\ensuremath{11.263\pm0.029}} 
\newcommand{\hatcurCCtwomassKmagorigxxxxxD}{\ensuremath{11.141\pm0.020}} 
\newcommand{\hatcurCCcitJmagorigxxxxxD}{\ensuremath{11.610\pm0.021}} 
\newcommand{\hatcurCCcitHmagorigxxxxxD}{\ensuremath{11.257\pm0.029}} 
\newcommand{\hatcurCCcitKmagorigxxxxxD}{\ensuremath{11.165\pm0.020}} 
\newcommand{\hatcurCCbbJmagorigxxxxxD}{\ensuremath{11.667\pm0.023}} 
\newcommand{\hatcurCCbbHmagorigxxxxxD}{\ensuremath{11.279\pm0.030}} 
\newcommand{\hatcurCCbbKmagorigxxxxxD}{\ensuremath{11.185\pm0.020}} 
\newcommand{\hatcurCCesoJmagorigxxxxxD}{\ensuremath{11.670\pm0.025}} 
\newcommand{\hatcurCCesoHmagorigxxxxxD}{\ensuremath{11.276\pm0.035}} 
\newcommand{\hatcurCCesoKmagorigxxxxxD}{\ensuremath{11.184\pm0.021}} 
\newcommand{\hatcurCCesoJHmagorigxxxxxD}{\ensuremath{0.393\pm0.041}} 
\newcommand{\hatcurCCesoJKmagorigxxxxxD}{\ensuremath{0.487\pm0.032}} 
\newcommand{\hatcurCCesoHKmagorigxxxxxD}{\ensuremath{0.093\pm0.041}} 
\newcommand{\hatcurLCdiporigxxxxxD}{\ensuremath{12.7}}              
\newcommand{\hatcurLCrprstarorigxxxxxD}{\ensuremath{0.1003\pm0.0025}} 
\newcommand{\hatcurLCbsqorigxxxxxD}{\ensuremath{0.630_{-0.031}^{+0.033}}} 
\newcommand{\hatcurLCimporigxxxxxD}{\ensuremath{0.794_{-0.020}^{+0.021}}} 
\newcommand{\hatcurLCzetaorigxxxxxD}{\ensuremath{35.86\pm0.77}}     
\newcommand{\hatcurLCdurorigxxxxxD}{\ensuremath{0.0699\pm0.0016}}   
\newcommand{\hatcurLCdurshortorigxxxxxD}{\ensuremath{0.0699}}       
\newcommand{\hatcurLCdurhrorigxxxxxD}{\ensuremath{1.677\pm0.038}}   
\newcommand{\hatcurLCdurhrshortorigxxxxxD}{\ensuremath{1.677}}      
\newcommand{\hatcurLCqorigxxxxxD}{\ensuremath{0.03670\pm0.00083}}   
\newcommand{\hatcurLCqshortorigxxxxxD}{\ensuremath{0.037}}          
\newcommand{\hatcurLCingdurorigxxxxxD}{\ensuremath{0.0155\pm0.0017}} 
\newcommand{\hatcurLCPorigxxxxxD}{\ensuremath{1.9023147\pm0.0000021}} 
\newcommand{\hatcurLCPprecorigxxxxxD}{\ensuremath{1.9023147}}       
\newcommand{\hatcurLCPshortorigxxxxxD}{\ensuremath{1.9023}}         
\newcommand{\hatcurLCTorigxxxxxD}{\ensuremath{2456816.35232\pm0.00052}} 
\newcommand{\hatcurLCTshortorigxxxxxD}{\ensuremath{6816.35232\pm0.00052}} 
\newcommand{\hatcurLCTAorigxxxxxD}{\ensuremath{2454392.8034\pm0.0027}} 
\newcommand{\hatcurLCTBorigxxxxxD}{\ensuremath{2456941.90507\pm0.00056}} 
\newcommand{\hatcurLChatnetmAorigxxxxxD}{\ensuremath{12.61786\pm0.00018}} 
\newcommand{\hatcurLCiblendAorigxxxxxD}{\ensuremath{0.85\pm0.10}}   
\newcommand{\hatcurLChatnetmBorigxxxxxD}{\ensuremath{12.80531\pm0.00011}} 
\newcommand{\hatcurLCiblendBorigxxxxxD}{\ensuremath{0.912\pm0.063}} 
\newcommand{\hatcurLCrhoorigxxxxxD}{\ensuremath{1.82_{-0.39}^{+0.72}}} 
\newcommand{\hatcurSMEitefforigxxxxxD}{\ensuremath{5546\pm50}}      
\newcommand{\hatcurSMEizfehorigxxxxxD}{\ensuremath{0.399\pm0.080}}  
\newcommand{\hatcurSMEizfehshortorigxxxxxD}{\ensuremath{0.40}}      
\newcommand{\hatcurSMEiloggorigxxxxxD}{\ensuremath{4.43\pm0.10}}    
\newcommand{\hatcurSMEivsinorigxxxxxD}{\ensuremath{3.72\pm0.50}}    
\newcommand{\hatcurSMEivmacorigxxxxxD}{\ensuremath{0.0}}            
\newcommand{\hatcurSMEivmicorigxxxxxD}{\ensuremath{0.0}}            
\newcommand{\hatcurSMEiitefforigxxxxxD}{\ensuremath{5551\pm50}}     
\newcommand{\hatcurSMEiizfehorigxxxxxD}{\ensuremath{0.396\pm0.080}} 
\newcommand{\hatcurSMEiizfehshortorigxxxxxD}{\ensuremath{0.396}}    
\newcommand{\hatcurSMEiiloggorigxxxxxD}{\ensuremath{4.468\pm0.035}} 
\newcommand{\hatcurSMEiivsinorigxxxxxD}{\ensuremath{3.69\pm0.50}}   
\newcommand{\hatcurLBizorigxxxxxD}{\ensuremath{0.2481}}             
\newcommand{\hatcurLBiizorigxxxxxD}{\ensuremath{0.3193}}            
\newcommand{\hatcurLBiiorigxxxxxD}{\ensuremath{0.3253}}             
\newcommand{\hatcurLBiiiorigxxxxxD}{\ensuremath{0.3085}}            
\newcommand{\hatcurLBiIorigxxxxxD}{\ensuremath{0.3000}}             
\newcommand{\hatcurLBiiIorigxxxxxD}{\ensuremath{0.3124}}            
\newcommand{\hatcurLBigorigxxxxxD}{\ensuremath{0.6560}}             
\newcommand{\hatcurLBiigorigxxxxxD}{\ensuremath{0.1596}}            
\newcommand{\hatcurLBirorigxxxxxD}{\ensuremath{0.4330}}             
\newcommand{\hatcurLBiirorigxxxxxD}{\ensuremath{0.2849}}            
\newcommand{\hatcurLBiRorigxxxxxD}{\ensuremath{0.4032}}             
\newcommand{\hatcurLBiiRorigxxxxxD}{\ensuremath{0.2924}}            
\newcommand{\hatcurLBikeporigxxxxxD}{\ensuremath{0.1000}}           
\newcommand{\hatcurLBiikeporigxxxxxD}{\ensuremath{0.1000}}          
\newcommand{\hatcurISOmorigxxxxxD}{\ensuremath{1.043\pm0.022}}      
\newcommand{\hatcurISOmshortorigxxxxxD}{\ensuremath{1.04}}          
\newcommand{\hatcurISOmlongorigxxxxxD}{\ensuremath{1.043\pm0.022}}  
\newcommand{\hatcurISOrorigxxxxxD}{\ensuremath{0.991_{-0.034}^{+0.046}}} 
\newcommand{\hatcurISOrshortorigxxxxxD}{\ensuremath{0.99}}          
\newcommand{\hatcurISOrlongorigxxxxxD}{\ensuremath{0.991_{-0.034}^{+0.046}}} 
\newcommand{\hatcurISOrhoorigxxxxxD}{\ensuremath{1.51\pm0.17}}      
\newcommand{\hatcurISOrholongorigxxxxxD}{\ensuremath{1.51\pm0.17}}  
\newcommand{\hatcurISOloggorigxxxxxD}{\ensuremath{4.464\pm0.035}}   
\newcommand{\hatcurISOlumorigxxxxxD}{\ensuremath{0.835\pm0.083}}    
\newcommand{\hatcurISOlumshortorigxxxxxD}{\ensuremath{0.83}}        
\newcommand{\hatcurISOmvorigxxxxxD}{\ensuremath{5.06\pm0.11}}       
\newcommand{\hatcurISOviorigxxxxxD}{\ensuremath{0.780\pm0.016}}     
\newcommand{\hatcurISOageorigxxxxxD}{\ensuremath{2.7_{-2.5}^{+1.8}}} 
\newcommand{\hatcurISOsigmaorigxxxxxD}{\ensuremath{0.00170\pm0.00038}} 
\newcommand{\hatcurISOMJorigxxxxxD}{\ensuremath{3.790\pm0.097}}     
\newcommand{\hatcurISOMHorigxxxxxD}{\ensuremath{3.418\pm0.091}}     
\newcommand{\hatcurISOMKorigxxxxxD}{\ensuremath{3.357\pm0.090}}     
\newcommand{\hatcurISOJKorigxxxxxD}{\ensuremath{0.440\pm0.010}}     
\newcommand{\hatcurISOspecorigxxxxxD}{G}                            
\newcommand{\hatcurRVKorigxxxxxD}{\ensuremath{175\pm10}}            
\newcommand{\hatcurRVrkorigxxxxxD}{\ensuremath{0\pm0}}              
\newcommand{\hatcurRVrhorigxxxxxD}{\ensuremath{0\pm0}}              
\newcommand{\hatcurRVkorigxxxxxD}{\ensuremath{0\pm0}}               
\newcommand{\hatcurRVhorigxxxxxD}{\ensuremath{0\pm0}}               
\newcommand{\hatcurRVtroneorigxxxxxD}{\ensuremath{0\pm0}}           
\newcommand{\hatcurRVtrtwoorigxxxxxD}{\ensuremath{0\pm0}}           
\newcommand{\hatcurRVgammaAorigxxxxxD}{\ensuremath{-56\pm31}}       
\newcommand{\hatcurRVjitterAorigxxxxxD}{\ensuremath{15\pm30}}       
\newcommand{\hatcurRVjittertwosiglimAorigxxxxxD}{\ensuremath{<53.6}} 
\newcommand{\hatcurRVfitrmsAorigxxxxxD}{\ensuremath{0.0}}           
\newcommand{\hatcurRVgammaBorigxxxxxD}{\ensuremath{212\pm12}}       
\newcommand{\hatcurRVjitterBorigxxxxxD}{\ensuremath{39\pm12}}       
\newcommand{\hatcurRVjittertwosiglimBorigxxxxxD}{\ensuremath{<63.1}} 
\newcommand{\hatcurRVfitrmsBorigxxxxxD}{\ensuremath{0.0}}           
\newcommand{\hatcurRVeccenorigxxxxxD}{\ensuremath{0\pm0}}           
\newcommand{\hatcurRVeccentwosiglimorigxxxxxD}{\ensuremath{<0.000}} 
\newcommand{\hatcurRVomegaorigxxxxxD}{\ensuremath{0\pm0}}           
\newcommand{\hatcurPPiorigxxxxxD}{\ensuremath{83.12_{-0.52}^{+0.37}}} 
\newcommand{\hatcurPPgorigxxxxxD}{\ensuremath{28.8\pm3.5}}          
\newcommand{\hatcurPPloggorigxxxxxD}{\ensuremath{3.460\pm0.053}}    
\newcommand{\hatcurPParorigxxxxxD}{\ensuremath{6.61\pm0.26}}        
\newcommand{\hatcurPParelorigxxxxxD}{\ensuremath{0.03048\pm0.00021}} 
\newcommand{\hatcurPPrhoorigxxxxxD}{\ensuremath{1.49\pm0.25}}       
\newcommand{\hatcurPPmorigxxxxxD}{\ensuremath{1.102_{-0.071}^{+0.052}}} 
\newcommand{\hatcurPPmshortorigxxxxxD}{\ensuremath{1.10}}           
\newcommand{\hatcurPPmlongorigxxxxxD}{\ensuremath{1.102_{-0.071}^{+0.052}}} 
\newcommand{\hatcurPPmeorigxxxxxD}{\ensuremath{350_{-22}^{+17}}}    
\newcommand{\hatcurPPmeshortorigxxxxxD}{\ensuremath{350.3}}         
\newcommand{\hatcurPPmelongorigxxxxxD}{\ensuremath{350_{-22}^{+17}}} 
\newcommand{\hatcurPProrigxxxxxD}{\ensuremath{0.968_{-0.047}^{+0.061}}} 
\newcommand{\hatcurPPrshortorigxxxxxD}{\ensuremath{0.97}}           
\newcommand{\hatcurPPrlongorigxxxxxD}{\ensuremath{0.968_{-0.047}^{+0.061}}} 
\newcommand{\hatcurPPreorigxxxxxD}{\ensuremath{10.84_{-0.53}^{+0.69}}} 
\newcommand{\hatcurPPreshortorigxxxxxD}{\ensuremath{10.8}}          
\newcommand{\hatcurPPrelongorigxxxxxD}{\ensuremath{10.84_{-0.53}^{+0.69}}} 
\newcommand{\hatcurPPmrcorrorigxxxxxD}{\ensuremath{0.06}}           
\newcommand{\hatcurPPtefforigxxxxxD}{\ensuremath{1526\pm36}}        
\newcommand{\hatcurPPthetaorigxxxxxD}{\ensuremath{0.0655\pm0.0051}} 
\newcommand{\hatcurPPfluxperiorigxxxxxD}{\ensuremath{1.23\pm0.12}}  
\newcommand{\hatcurPPfluxperidimorigxxxxxD}{\ensuremath{9}}         
\newcommand{\hatcurPPfluxaporigxxxxxD}{\ensuremath{1.23\pm0.12}}    
\newcommand{\hatcurPPfluxapdimorigxxxxxD}{\ensuremath{9}}           
\newcommand{\hatcurPPfluxavgorigxxxxxD}{\ensuremath{1.23\pm0.12}}   
\newcommand{\hatcurPPfluxavgdimorigxxxxxD}{\ensuremath{9}}          
\newcommand{\hatcurPPfluxavglogorigxxxxxD}{\ensuremath{9.088\pm0.040}} 
\newcommand{\hatcurXsecphaseorigxxxxxD}{\ensuremath{0\pm0}}         
\newcommand{\hatcurXsecondaryorigxxxxxD}{\ensuremath{2456817.30348\pm0.00052}} 
\newcommand{\hatcurXsecdurorigxxxxxD}{\ensuremath{0.0699\pm0.0016}} 
\newcommand{\hatcurXsecingdurorigxxxxxD}{\ensuremath{0.0155\pm0.0017}} 
\newcommand{\hatcurPPphiconjorigxxxxxD}{\ensuremath{0\pm0}}         
\newcommand{\hatcurPPperiorigxxxxxD}{\ensuremath{2456815.87674\pm0.00052}} 
\newcommand{\hatcurPPaequivorigxxxxxD}{\ensuremath{0.0334\pm0.0015}} 
\newcommand{\hatcurPPtcircorigxxxxxD}{\ensuremath{65\pm18}}         
\newcommand{\hatcurPPtinfallorigxxxxxD}{\ensuremath{218\pm44}}      
\newcommand{\hatcurXdistorigxxxxxD}{\ensuremath{368\pm16}}          
\newcommand{\hatcurXAvorigxxxxxD}{\ensuremath{0.331\pm0.061}}       
\newcommand{\hatcurXdistredorigxxxxxD}{\ensuremath{361\pm15}}       
\newcommand{\hatcurXEBVorigxxxxxD}{\ensuremath{0.107\pm0.020}}      
\newcommand{\hatcurXmvisoredorigxxxxxD}{\ensuremath{13.183\pm0.029}} 
\newcommand{\hatcurXmiisoredorigxxxxxD}{\ensuremath{12.230\pm0.016}} 
\newcommand{\hatcurXmjisoredorigxxxxxD}{\ensuremath{11.672\pm0.014}} 
\newcommand{\hatcurXmhisoredorigxxxxxD}{\ensuremath{11.269\pm0.015}} 
\newcommand{\hatcurXmkisoredorigxxxxxD}{\ensuremath{11.182\pm0.016}} 
\newcommand{\hatcurXviisoredorigxxxxxD}{\ensuremath{0.953\pm0.023}} 
\newcommand{\hatcurXvkisoredorigxxxxxD}{\ensuremath{2.001\pm0.035}} 
\newcommand{\hatcurXjhisoredorigxxxxxD}{\ensuremath{0.4030\pm0.0085}} 
\newcommand{\hatcurXjkisoredorigxxxxxD}{\ensuremath{0.4910\pm0.0094}} 
\newcommand{\hatcurCCpmraorigxxxxxD}{\ensuremath{-9.5\pm1.7}}       
\newcommand{\hatcurCCpmdecorigxxxxxD}{\ensuremath{-19.7\pm2.2}}     
\newcommand{\hatcurCCpmorigxxxxxD}{\ensuremath{21.9\pm2.8}}         

%% file: HTR130-001.orig.tex
\newcommand{\hatcurhtrorigxxxxxE}{HTR130-001}                              
\newcommand{\hatcurfieldorigxxxxxE}{\ensuremath{string}}                   
\newcommand{\hatcurCCraorigxxxxxE}{\ensuremath{04^{\mathrm h}58^{\mathrm m}01.02{\mathrm s}}}                            
\newcommand{\hatcurCCdecorigxxxxxE}{\ensuremath{+48{\arcdeg}18{\arcmin}03.9{\arcsec}}}                           
\newcommand{\hatcurCCmagorigxxxxxE}{12.558}                                
\newcommand{\hatcurCCtwomassorigxxxxxE}{2MASS~04580102+4818038}            
\newcommand{\hatcurCCgscorigxxxxxE}{GSC~3348-01101}                        
\newcommand{\hatcurCCtassmvorigxxxxxE}{\ensuremath{12.558\pm0.035}}        
\newcommand{\hatcurCCtassmvshortorigxxxxxE}{\ensuremath{12.6}}             
\newcommand{\hatcurCCtassmBorigxxxxxE}{\ensuremath{13.493\pm0.065}}        
\newcommand{\hatcurCCtassmBshortorigxxxxxE}{\ensuremath{13.5}}             
\newcommand{\hatcurCCtassmIorigxxxxxE}{\ensuremath{11.68\pm0.11}}          
\newcommand{\hatcurCCtassmIshortorigxxxxxE}{\ensuremath{11.7}}             
\newcommand{\hatcurCCtassmgorigxxxxxE}{\ensuremath{13.10\pm0.17}}          
\newcommand{\hatcurCCtassmgshortorigxxxxxE}{\ensuremath{13.1}}             
\newcommand{\hatcurCCtassmrorigxxxxxE}{\ensuremath{12.31\pm0.12}}          
\newcommand{\hatcurCCtassmrshortorigxxxxxE}{\ensuremath{12.3}}             
\newcommand{\hatcurCCtassmiorigxxxxxE}{\ensuremath{12.04\pm0.12}}          
\newcommand{\hatcurCCtassmishortorigxxxxxE}{\ensuremath{12.0}}             
\newcommand{\hatcurCCtwomassJmagorigxxxxxE}{\ensuremath{11.144\pm0.029}}   
\newcommand{\hatcurCCtwomassHmagorigxxxxxE}{\ensuremath{10.803\pm0.041}}   
\newcommand{\hatcurCCtwomassKmagorigxxxxxE}{\ensuremath{10.701\pm0.026}}   
\newcommand{\hatcurCCcitJmagorigxxxxxE}{\ensuremath{11.157\pm0.029}}       
\newcommand{\hatcurCCcitHmagorigxxxxxE}{\ensuremath{10.797\pm0.040}}       
\newcommand{\hatcurCCcitKmagorigxxxxxE}{\ensuremath{10.725\pm0.026}}       
\newcommand{\hatcurCCbbJmagorigxxxxxE}{\ensuremath{11.212\pm0.031}}        
\newcommand{\hatcurCCbbHmagorigxxxxxE}{\ensuremath{10.819\pm0.042}}        
\newcommand{\hatcurCCbbKmagorigxxxxxE}{\ensuremath{10.745\pm0.026}}        
\newcommand{\hatcurCCesoJmagorigxxxxxE}{\ensuremath{11.215\pm0.033}}       
\newcommand{\hatcurCCesoHmagorigxxxxxE}{\ensuremath{10.815\pm0.047}}       
\newcommand{\hatcurCCesoKmagorigxxxxxE}{\ensuremath{10.744\pm0.027}}       
\newcommand{\hatcurCCesoJHmagorigxxxxxE}{\ensuremath{0.399\pm0.055}}       
\newcommand{\hatcurCCesoJKmagorigxxxxxE}{\ensuremath{0.472\pm0.042}}       
\newcommand{\hatcurCCesoHKmagorigxxxxxE}{\ensuremath{0.073\pm0.054}}       
\newcommand{\hatcurLCdiporigxxxxxE}{\ensuremath{9.3}}                      
\newcommand{\hatcurLCrprstarorigxxxxxE}{\ensuremath{0.0947\pm0.0021}}      
\newcommand{\hatcurLCbsqorigxxxxxE}{\ensuremath{0.092_{-0.060}^{+0.075}}}  
\newcommand{\hatcurLCimporigxxxxxE}{\ensuremath{0.30_{-0.12}^{+0.11}}}     
\newcommand{\hatcurLCzetaorigxxxxxE}{\ensuremath{16.89\pm0.12}}            
\newcommand{\hatcurLCdurorigxxxxxE}{\ensuremath{0.1308\pm0.0014}}          
\newcommand{\hatcurLCdurshortorigxxxxxE}{\ensuremath{0.1308}}              
\newcommand{\hatcurLCdurhrorigxxxxxE}{\ensuremath{3.139\pm0.033}}          
\newcommand{\hatcurLCdurhrshortorigxxxxxE}{\ensuremath{3.139}}             
\newcommand{\hatcurLCqorigxxxxxE}{\ensuremath{0.04940\pm0.00052}}          
\newcommand{\hatcurLCqshortorigxxxxxE}{\ensuremath{0.049}}                 
\newcommand{\hatcurLCingdurorigxxxxxE}{\ensuremath{0.0123\pm0.0011}}       
\newcommand{\hatcurLCPorigxxxxxE}{\ensuremath{2.6453233\pm0.0000042}}      
\newcommand{\hatcurLCPprecorigxxxxxE}{\ensuremath{2.6453233}}              
\newcommand{\hatcurLCPshortorigxxxxxE}{\ensuremath{2.6453}}                
\newcommand{\hatcurLCTorigxxxxxE}{\ensuremath{2457110.45306\pm0.00044}}    
\newcommand{\hatcurLCTshortorigxxxxxE}{\ensuremath{7110.45306\pm0.00044}}    
\newcommand{\hatcurLCTAorigxxxxxE}{\ensuremath{2456131.6834\pm0.0014}}     
\newcommand{\hatcurLCTBorigxxxxxE}{\ensuremath{2457292.98036\pm0.00061}}   
\newcommand{\hatcurLChatnetmorigxxxxxE}{\ensuremath{12.379220\pm0.000094}} 
\newcommand{\hatcurLCiblendorigxxxxxE}{\ensuremath{0.824\pm0.054}}         
\newcommand{\hatcurLCrhoorigxxxxxE}{\ensuremath{0.842\pm0.086}}            
\newcommand{\hatcurSMEitefforigxxxxxE}{\ensuremath{5617\pm50}}             
\newcommand{\hatcurSMEizfehorigxxxxxE}{\ensuremath{0.461\pm0.080}}         
\newcommand{\hatcurSMEizfehshortorigxxxxxE}{\ensuremath{0.46}}             
\newcommand{\hatcurSMEiloggorigxxxxxE}{\ensuremath{4.31\pm0.10}}           
\newcommand{\hatcurSMEivsinorigxxxxxE}{\ensuremath{3.53\pm0.50}}           
\newcommand{\hatcurSMEivmacorigxxxxxE}{\ensuremath{0.0}}                   
\newcommand{\hatcurSMEivmicorigxxxxxE}{\ensuremath{0.0}}                   
\newcommand{\hatcurSMEiitefforigxxxxxE}{\ensuremath{5601\pm50}}            
\newcommand{\hatcurSMEiizfehorigxxxxxE}{\ensuremath{0.449\pm0.080}}        
\newcommand{\hatcurSMEiizfehshortorigxxxxxE}{\ensuremath{0.449}}           
\newcommand{\hatcurSMEiiloggorigxxxxxE}{\ensuremath{4.305\pm0.030}}        
\newcommand{\hatcurSMEiivsinorigxxxxxE}{\ensuremath{3.55\pm0.50}}          
\newcommand{\hatcurLBizorigxxxxxE}{\ensuremath{0.2402}}                    
\newcommand{\hatcurLBiizorigxxxxxE}{\ensuremath{0.3260}}                   
\newcommand{\hatcurLBiiorigxxxxxE}{\ensuremath{0.3168}}                    
\newcommand{\hatcurLBiiiorigxxxxxE}{\ensuremath{0.3162}}                   
\newcommand{\hatcurLBiIorigxxxxxE}{\ensuremath{0.2914}}                    
\newcommand{\hatcurLBiiIorigxxxxxE}{\ensuremath{0.3201}}                   
\newcommand{\hatcurLBigorigxxxxxE}{\ensuremath{0.6478}}                    
\newcommand{\hatcurLBiigorigxxxxxE}{\ensuremath{0.1670}}                   
\newcommand{\hatcurLBirorigxxxxxE}{\ensuremath{0.4241}}                    
\newcommand{\hatcurLBiirorigxxxxxE}{\ensuremath{0.2923}}                   
\newcommand{\hatcurLBiRorigxxxxxE}{\ensuremath{0.3943}}                    
\newcommand{\hatcurLBiiRorigxxxxxE}{\ensuremath{0.3000}}                   
\newcommand{\hatcurLBikeporigxxxxxE}{\ensuremath{0.1000}}                  
\newcommand{\hatcurLBiikeporigxxxxxE}{\ensuremath{0.1000}}                 
\newcommand{\hatcurISOmorigxxxxxE}{\ensuremath{1.103\pm0.028}}             
\newcommand{\hatcurISOmshortorigxxxxxE}{\ensuremath{1.10}}                 
\newcommand{\hatcurISOmlongorigxxxxxE}{\ensuremath{1.103\pm0.028}}         
\newcommand{\hatcurISOrorigxxxxxE}{\ensuremath{1.227_{-0.044}^{+0.063}}}   
\newcommand{\hatcurISOrshortorigxxxxxE}{\ensuremath{1.23}}                 
\newcommand{\hatcurISOrlongorigxxxxxE}{\ensuremath{1.227_{-0.044}^{+0.063}}} 
\newcommand{\hatcurISOrhoorigxxxxxE}{\ensuremath{0.840\pm0.087}}           
\newcommand{\hatcurISOrholongorigxxxxxE}{\ensuremath{0.840\pm0.087}}       
\newcommand{\hatcurISOloggorigxxxxxE}{\ensuremath{4.302\pm0.030}}          
\newcommand{\hatcurISOlumorigxxxxxE}{\ensuremath{1.33_{-0.11}^{+0.15}}}    
\newcommand{\hatcurISOlumshortorigxxxxxE}{\ensuremath{1.33}}               
\newcommand{\hatcurISOmvorigxxxxxE}{\ensuremath{4.54\pm0.11}}              
\newcommand{\hatcurISOviorigxxxxxE}{\ensuremath{0.766\pm0.017}}            
\newcommand{\hatcurISOageorigxxxxxE}{\ensuremath{5.40\pm0.89}}             
\newcommand{\hatcurISOsigmaorigxxxxxE}{\ensuremath{0.000700\pm0.000096}}   
\newcommand{\hatcurISOMJorigxxxxxE}{\ensuremath{3.300\pm0.097}}            
\newcommand{\hatcurISOMHorigxxxxxE}{\ensuremath{2.946\pm0.093}}            
\newcommand{\hatcurISOMKorigxxxxxE}{\ensuremath{2.886\pm0.092}}            
\newcommand{\hatcurISOJKorigxxxxxE}{\ensuremath{0.420\pm0.010}}            
\newcommand{\hatcurISOspecorigxxxxxE}{G}                                   
\newcommand{\hatcurRVKorigxxxxxE}{\ensuremath{110\pm12}}                   
\newcommand{\hatcurRVrkorigxxxxxE}{\ensuremath{0\pm0}}                     
\newcommand{\hatcurRVrhorigxxxxxE}{\ensuremath{0\pm0}}                     
\newcommand{\hatcurRVkorigxxxxxE}{\ensuremath{0\pm0}}                      
\newcommand{\hatcurRVhorigxxxxxE}{\ensuremath{0\pm0}}                      
\newcommand{\hatcurRVtroneorigxxxxxE}{\ensuremath{0\pm0}}                  
\newcommand{\hatcurRVtrtwoorigxxxxxE}{\ensuremath{0\pm0}}                  
\newcommand{\hatcurRVgammaorigxxxxxE}{\ensuremath{127\pm10}}               
\newcommand{\hatcurRVjitterorigxxxxxE}{\ensuremath{32.3\pm9.5}}            
\newcommand{\hatcurRVjittertwosiglimorigxxxxxE}{\ensuremath{<50.5}}        
\newcommand{\hatcurRVfitrmsorigxxxxxE}{\ensuremath{.1fym}}                 %
\newcommand{\hatcurRVeccenorigxxxxxE}{\ensuremath{0\pm0}}                  
\newcommand{\hatcurRVeccentwosiglimorigxxxxxE}{\ensuremath{<0.000}}        
\newcommand{\hatcurRVomegaorigxxxxxE}{\ensuremath{0\pm0}}                  
\newcommand{\hatcurPPiorigxxxxxE}{\ensuremath{87.4\pm1.0}}                 
\newcommand{\hatcurPPgorigxxxxxE}{\ensuremath{15.3\pm2.2}}                 
\newcommand{\hatcurPPloggorigxxxxxE}{\ensuremath{3.186\pm0.065}}           
\newcommand{\hatcurPParorigxxxxxE}{\ensuremath{6.77_{-0.29}^{+0.22}}}      
\newcommand{\hatcurPParelorigxxxxxE}{\ensuremath{0.03868\pm0.00033}}       
\newcommand{\hatcurPPrhoorigxxxxxE}{\ensuremath{0.68\pm0.13}}              
\newcommand{\hatcurPPmorigxxxxxE}{\ensuremath{0.801\pm0.088}}              
\newcommand{\hatcurPPmshortorigxxxxxE}{\ensuremath{0.80}}                  
\newcommand{\hatcurPPmlongorigxxxxxE}{\ensuremath{0.801\pm0.088}}          
\newcommand{\hatcurPPmeorigxxxxxE}{\ensuremath{255\pm28}}                  
\newcommand{\hatcurPPmeshortorigxxxxxE}{\ensuremath{254.7}}                
\newcommand{\hatcurPPmelongorigxxxxxE}{\ensuremath{255\pm28}}              
\newcommand{\hatcurPProrigxxxxxE}{\ensuremath{1.131_{-0.053}^{+0.074}}}    
\newcommand{\hatcurPPrshortorigxxxxxE}{\ensuremath{1.13}}                  
\newcommand{\hatcurPPrlongorigxxxxxE}{\ensuremath{1.131_{-0.053}^{+0.074}}} 
\newcommand{\hatcurPPreorigxxxxxE}{\ensuremath{12.68_{-0.59}^{+0.83}}}     
\newcommand{\hatcurPPreshortorigxxxxxE}{\ensuremath{12.7}}                 
\newcommand{\hatcurPPrelongorigxxxxxE}{\ensuremath{12.68_{-0.59}^{+0.83}}} 
\newcommand{\hatcurPPmrcorrorigxxxxxE}{\ensuremath{0.10}}                  
\newcommand{\hatcurPPtefforigxxxxxE}{\ensuremath{1523\pm31}}               
\newcommand{\hatcurPPthetaorigxxxxxE}{\ensuremath{0.0492\pm0.0060}}        
\newcommand{\hatcurPPfluxperiorigxxxxxE}{\ensuremath{1.214_{-0.085}^{+0.115}}} 
\newcommand{\hatcurPPfluxperidimorigxxxxxE}{\ensuremath{9}}                
\newcommand{\hatcurPPfluxaporigxxxxxE}{\ensuremath{1.214_{-0.085}^{+0.115}}} 
\newcommand{\hatcurPPfluxapdimorigxxxxxE}{\ensuremath{9}}                  
\newcommand{\hatcurPPfluxavgorigxxxxxE}{\ensuremath{1.214_{-0.085}^{+0.115}}} 
\newcommand{\hatcurPPfluxavgdimorigxxxxxE}{\ensuremath{9}}                 
\newcommand{\hatcurPPfluxavglogorigxxxxxE}{\ensuremath{9.084\pm0.035}}     
\newcommand{\hatcurXsecphaseorigxxxxxE}{\ensuremath{0\pm0}}                
\newcommand{\hatcurXsecondaryorigxxxxxE}{\ensuremath{2457111.77572\pm0.00044}} 
\newcommand{\hatcurXsecdurorigxxxxxE}{\ensuremath{0.1308\pm0.0014}}        
\newcommand{\hatcurXsecingdurorigxxxxxE}{\ensuremath{0.0123\pm0.0011}}     
\newcommand{\hatcurPPphiconjorigxxxxxE}{\ensuremath{0\pm0}}                
\newcommand{\hatcurPPperiorigxxxxxE}{\ensuremath{2457109.79172\pm0.00044}} 
\newcommand{\hatcurPPaequivorigxxxxxE}{\ensuremath{0.0335\pm0.0013}}       
\newcommand{\hatcurPPtcircorigxxxxxE}{\ensuremath{94\pm25}}                
\newcommand{\hatcurPPtinfallorigxxxxxE}{\ensuremath{490\pm100}}            
\newcommand{\hatcurXdistorigxxxxxE}{\ensuremath{373_{-14}^{+19}}}          
\newcommand{\hatcurXAvorigxxxxxE}{\ensuremath{0.172\pm0.066}}              
\newcommand{\hatcurXdistredorigxxxxxE}{\ensuremath{371_{-14}^{+19}}}       
\newcommand{\hatcurXEBVorigxxxxxE}{\ensuremath{0.055\pm0.021}}             
\newcommand{\hatcurXmvisoredorigxxxxxE}{\ensuremath{12.561\pm0.034}}       
\newcommand{\hatcurXmiisoredorigxxxxxE}{\ensuremath{11.705\pm0.019}}       
\newcommand{\hatcurXmjisoredorigxxxxxE}{\ensuremath{11.195\pm0.018}}       
\newcommand{\hatcurXmhisoredorigxxxxxE}{\ensuremath{10.824\pm0.019}}       
\newcommand{\hatcurXmkisoredorigxxxxxE}{\ensuremath{10.751\pm0.021}}       
\newcommand{\hatcurXviisoredorigxxxxxE}{\ensuremath{0.856\pm0.026}}        
\newcommand{\hatcurXvkisoredorigxxxxxE}{\ensuremath{1.810\pm0.042}}        
\newcommand{\hatcurXjhisoredorigxxxxxE}{\ensuremath{0.3710\pm0.0087}}      
\newcommand{\hatcurXjkisoredorigxxxxxE}{\ensuremath{0.444\pm0.010}}        
\newcommand{\hatcurCCpmraorigxxxxxE}{\ensuremath{0\pm50}}                  
\newcommand{\hatcurCCpmdecorigxxxxxE}{\ensuremath{0\pm50}}                 
\newcommand{\hatcurCCpmorigxxxxxE}{\ensuremath{0\pm71}}                    

%% file: HTR384-014.orig.tex
\newcommand{\hatcurhtrorigxxxxxF}{HTR384-014}                              
\newcommand{\hatcurfieldorigxxxxxF}{\ensuremath{string}}                   
\newcommand{\hatcurCCraorigxxxxxF}{\ensuremath{17^{\mathrm h}58^{\mathrm m}17.40{\mathrm s}}}                            
\newcommand{\hatcurCCdecorigxxxxxF}{\ensuremath{+05{\arcdeg}45{\arcmin}40.9{\arcsec}}}                           
\newcommand{\hatcurCCmagorigxxxxxF}{13.753}                                
\newcommand{\hatcurCCtwomassorigxxxxxF}{2MASS~17581730+0545409}            
\newcommand{\hatcurCCgscorigxxxxxF}{GSC~0429-01697}                        
\newcommand{\hatcurCCtassmvorigxxxxxF}{\ensuremath{13.753\pm0.065}}        
\newcommand{\hatcurCCtassmvshortorigxxxxxF}{\ensuremath{13.8}}             
\newcommand{\hatcurCCtassmBorigxxxxxF}{\ensuremath{14.729\pm0.066}}        
\newcommand{\hatcurCCtassmBshortorigxxxxxF}{\ensuremath{14.7}}             
\newcommand{\hatcurCCtassmIorigxxxxxF}{\ensuremath{nff\pmnff}}             
\newcommand{\hatcurCCtassmIshortorigxxxxxF}{\ensuremath{0.0}}              
\newcommand{\hatcurCCtassmgorigxxxxxF}{\ensuremath{14.258\pm0.026}}        
\newcommand{\hatcurCCtassmgshortorigxxxxxF}{\ensuremath{14.3}}             
\newcommand{\hatcurCCtassmrorigxxxxxF}{\ensuremath{13.418\pm0.093}}        
\newcommand{\hatcurCCtassmrshortorigxxxxxF}{\ensuremath{13.4}}             
\newcommand{\hatcurCCtassmiorigxxxxxF}{\ensuremath{13.136\pm0.086}}        
\newcommand{\hatcurCCtassmishortorigxxxxxF}{\ensuremath{13.1}}             
\newcommand{\hatcurCCtwomassJmagorigxxxxxF}{\ensuremath{12.021\pm0.021}}   
\newcommand{\hatcurCCtwomassHmagorigxxxxxF}{\ensuremath{11.630\pm0.028}}   
\newcommand{\hatcurCCtwomassKmagorigxxxxxF}{\ensuremath{11.512\pm0.023}}   
\newcommand{\hatcurCCcitJmagorigxxxxxF}{\ensuremath{12.030\pm0.022}}       
\newcommand{\hatcurCCcitHmagorigxxxxxF}{\ensuremath{11.623\pm0.028}}       
\newcommand{\hatcurCCcitKmagorigxxxxxF}{\ensuremath{11.536\pm0.023}}       
\newcommand{\hatcurCCbbJmagorigxxxxxF}{\ensuremath{12.091\pm0.024}}        
\newcommand{\hatcurCCbbHmagorigxxxxxF}{\ensuremath{11.646\pm0.029}}        
\newcommand{\hatcurCCbbKmagorigxxxxxF}{\ensuremath{11.556\pm0.023}}        
\newcommand{\hatcurCCesoJmagorigxxxxxF}{\ensuremath{12.095\pm0.026}}       
\newcommand{\hatcurCCesoHmagorigxxxxxF}{\ensuremath{11.643\pm0.034}}       
\newcommand{\hatcurCCesoKmagorigxxxxxF}{\ensuremath{11.554\pm0.024}}       
\newcommand{\hatcurCCesoJHmagorigxxxxxF}{\ensuremath{0.452\pm0.040}}       
\newcommand{\hatcurCCesoJKmagorigxxxxxF}{\ensuremath{0.541\pm0.034}}       
\newcommand{\hatcurCCesoHKmagorigxxxxxF}{\ensuremath{0.089\pm0.042}}       
\newcommand{\hatcurLCdiporigxxxxxF}{\ensuremath{17.6}}                     
\newcommand{\hatcurLCrprstarorigxxxxxF}{\ensuremath{0.1223\pm0.0048}}      
\newcommand{\hatcurLCbsqorigxxxxxF}{\ensuremath{0.170_{-0.080}^{+0.087}}}  
\newcommand{\hatcurLCimporigxxxxxF}{\ensuremath{0.413_{-0.113}^{+0.094}}}  
\newcommand{\hatcurLCzetaorigxxxxxF}{\ensuremath{18.93_{-0.13}^{+0.19}}}   
\newcommand{\hatcurLCdurorigxxxxxF}{\ensuremath{0.1210\pm0.0019}}          
\newcommand{\hatcurLCdurshortorigxxxxxF}{\ensuremath{0.1210}}              
\newcommand{\hatcurLCdurhrorigxxxxxF}{\ensuremath{2.903\pm0.046}}          
\newcommand{\hatcurLCdurhrshortorigxxxxxF}{\ensuremath{2.903}}             
\newcommand{\hatcurLCqorigxxxxxF}{\ensuremath{0.03580\pm0.00057}}          
\newcommand{\hatcurLCqshortorigxxxxxF}{\ensuremath{0.036}}                 
\newcommand{\hatcurLCingdurorigxxxxxF}{\ensuremath{0.0158\pm0.0020}}       
\newcommand{\hatcurLCPorigxxxxxF}{\ensuremath{3.377733\pm0.000012}}        
\newcommand{\hatcurLCPprecorigxxxxxF}{\ensuremath{3.3777329}}              
\newcommand{\hatcurLCPshortorigxxxxxF}{\ensuremath{3.3777}}                
\newcommand{\hatcurLCTorigxxxxxF}{\ensuremath{2456382.94199\pm0.00054}}    
\newcommand{\hatcurLCTshortorigxxxxxF}{\ensuremath{6382.94199\pm0.00054}}    
\newcommand{\hatcurLCTAorigxxxxxF}{\ensuremath{2454964.2943\pm0.0050}}     
\newcommand{\hatcurLCTBorigxxxxxF}{\ensuremath{2456399.83065\pm0.00055}}   
\newcommand{\hatcurLChatnetmorigxxxxxF}{\ensuremath{13.32553\pm0.00020}}   
\newcommand{\hatcurLCiblendorigxxxxxF}{\ensuremath{0.513\pm0.061}}         
\newcommand{\hatcurLCrhoorigxxxxxF}{\ensuremath{1.31\pm0.18}}              
\newcommand{\hatcurSMEitefforigxxxxxF}{\ensuremath{5365\pm50}}             
\newcommand{\hatcurSMEizfehorigxxxxxF}{\ensuremath{0.428\pm0.080}}         
\newcommand{\hatcurSMEizfehshortorigxxxxxF}{\ensuremath{0.43}}             
\newcommand{\hatcurSMEiloggorigxxxxxF}{\ensuremath{4.40\pm0.10}}           
\newcommand{\hatcurSMEivsinorigxxxxxF}{\ensuremath{3.22\pm0.50}}           
\newcommand{\hatcurSMEivmacorigxxxxxF}{\ensuremath{0.0}}                   
\newcommand{\hatcurSMEivmicorigxxxxxF}{\ensuremath{0.0}}                   
\newcommand{\hatcurLBizorigxxxxxF}{\ensuremath{0.2730}}                    
\newcommand{\hatcurLBiizorigxxxxxF}{\ensuremath{0.3077}}                   
\newcommand{\hatcurLBiiorigxxxxxF}{\ensuremath{0.3576}}                    
\newcommand{\hatcurLBiiiorigxxxxxF}{\ensuremath{0.2893}}                   
\newcommand{\hatcurLBiIorigxxxxxF}{\ensuremath{0.3300}}                    
\newcommand{\hatcurLBiiIorigxxxxxF}{\ensuremath{0.2953}}                   
\newcommand{\hatcurLBigorigxxxxxF}{\ensuremath{0.7099}}                    
\newcommand{\hatcurLBiigorigxxxxxF}{\ensuremath{0.1164}}                   
\newcommand{\hatcurLBirorigxxxxxF}{\ensuremath{0.4752}}                    
\newcommand{\hatcurLBiirorigxxxxxF}{\ensuremath{0.2572}}                   
\newcommand{\hatcurLBiRorigxxxxxF}{\ensuremath{0.4428}}                    
\newcommand{\hatcurLBiiRorigxxxxxF}{\ensuremath{0.2670}}                   
\newcommand{\hatcurLBikeporigxxxxxF}{\ensuremath{0.1000}}                  
\newcommand{\hatcurLBiikeporigxxxxxF}{\ensuremath{0.1000}}                 
\newcommand{\hatcurISOmorigxxxxxF}{\ensuremath{0.976\pm0.022}}             
\newcommand{\hatcurISOmshortorigxxxxxF}{\ensuremath{0.98}}                 
\newcommand{\hatcurISOmlongorigxxxxxF}{\ensuremath{0.976\pm0.022}}         
\newcommand{\hatcurISOrorigxxxxxF}{\ensuremath{1.012_{-0.043}^{+0.057}}}   
\newcommand{\hatcurISOrshortorigxxxxxF}{\ensuremath{1.01}}                 
\newcommand{\hatcurISOrlongorigxxxxxF}{\ensuremath{1.012_{-0.043}^{+0.057}}} 
\newcommand{\hatcurISOrhoorigxxxxxF}{\ensuremath{1.32\pm0.18}}             
\newcommand{\hatcurISOrholongorigxxxxxF}{\ensuremath{1.32\pm0.18}}         
\newcommand{\hatcurISOloggorigxxxxxF}{\ensuremath{4.415\pm0.042}}          
\newcommand{\hatcurISOlumorigxxxxxF}{\ensuremath{0.761_{-0.072}^{+0.097}}} 
\newcommand{\hatcurISOlumshortorigxxxxxF}{\ensuremath{0.76}}               
\newcommand{\hatcurISOmvorigxxxxxF}{\ensuremath{5.21\pm0.12}}              
\newcommand{\hatcurISOviorigxxxxxF}{\ensuremath{0.834\pm0.014}}            
\newcommand{\hatcurISOageorigxxxxxF}{\ensuremath{7.2\pm2.0}}               
\newcommand{\hatcurISOsigmaorigxxxxxF}{\ensuremath{0.000800\pm0.000096}}   
\newcommand{\hatcurISOMJorigxxxxxF}{\ensuremath{3.83\pm0.11}}              
\newcommand{\hatcurISOMHorigxxxxxF}{\ensuremath{3.42\pm0.11}}              
\newcommand{\hatcurISOMKorigxxxxxF}{\ensuremath{3.35\pm0.11}}              
\newcommand{\hatcurISOJKorigxxxxxF}{\ensuremath{0.480\pm0.010}}            
\newcommand{\hatcurISOspecorigxxxxxF}{G}                                   
\newcommand{\hatcurRVKorigxxxxxF}{\ensuremath{87.5\pm2.9}}                 
\newcommand{\hatcurRVrkorigxxxxxF}{\ensuremath{0\pm0}}                     
\newcommand{\hatcurRVrhorigxxxxxF}{\ensuremath{0\pm0}}                     
\newcommand{\hatcurRVkorigxxxxxF}{\ensuremath{0\pm0}}                      
\newcommand{\hatcurRVhorigxxxxxF}{\ensuremath{0\pm0}}                      
\newcommand{\hatcurRVtroneorigxxxxxF}{\ensuremath{0\pm0}}                  
\newcommand{\hatcurRVtrtwoorigxxxxxF}{\ensuremath{0\pm0}}                  
\newcommand{\hatcurRVgammaAorigxxxxxF}{\ensuremath{7.2\pm2.5}}             
\newcommand{\hatcurRVjitterAorigxxxxxF}{\ensuremath{0.01\pm0.40}}          
\newcommand{\hatcurRVjittertwosiglimAorigxxxxxF}{\ensuremath{<1.1}}        
\newcommand{\hatcurRVfitrmsAorigxxxxxF}{\ensuremath{0.0}}                  
\newcommand{\hatcurRVgammaBorigxxxxxF}{\ensuremath{-103\pm27}}             
\newcommand{\hatcurRVjitterBorigxxxxxF}{\ensuremath{0.0\pm1.3}}            
\newcommand{\hatcurRVjittertwosiglimBorigxxxxxF}{\ensuremath{<2.7}}        
\newcommand{\hatcurRVfitrmsBorigxxxxxF}{\ensuremath{0.0}}                  
\newcommand{\hatcurRVgammaCorigxxxxxF}{\ensuremath{-69596.3\pm7.8}}        
\newcommand{\hatcurRVjitterCorigxxxxxF}{\ensuremath{17\pm10}}              
\newcommand{\hatcurRVjittertwosiglimCorigxxxxxF}{\ensuremath{<32.5}}       
\newcommand{\hatcurRVfitrmsCorigxxxxxF}{\ensuremath{0.0}}                  
\newcommand{\hatcurRVeccenorigxxxxxF}{\ensuremath{0\pm0}}                  
\newcommand{\hatcurRVeccentwosiglimorigxxxxxF}{\ensuremath{<0.000}}        
\newcommand{\hatcurRVomegaorigxxxxxF}{\ensuremath{0\pm0}}                  
\newcommand{\hatcurPPiorigxxxxxF}{\ensuremath{87.45\pm0.76}}               
\newcommand{\hatcurPPgorigxxxxxF}{\ensuremath{10.8_{-1.6}^{+2.2}}}         
\newcommand{\hatcurPPloggorigxxxxxF}{\ensuremath{3.032\pm0.070}}           
\newcommand{\hatcurPParorigxxxxxF}{\ensuremath{9.26\pm0.44}}               
\newcommand{\hatcurPParelorigxxxxxF}{\ensuremath{0.04370\pm0.00032}}       
\newcommand{\hatcurPPrhoorigxxxxxF}{\ensuremath{0.45\pm0.11}}              
\newcommand{\hatcurPPmorigxxxxxF}{\ensuremath{0.638\pm0.023}}              
\newcommand{\hatcurPPmshortorigxxxxxF}{\ensuremath{0.64}}                  
\newcommand{\hatcurPPmlongorigxxxxxF}{\ensuremath{0.638\pm0.023}}          
\newcommand{\hatcurPPmeorigxxxxxF}{\ensuremath{202.7\pm7.3}}               
\newcommand{\hatcurPPmeshortorigxxxxxF}{\ensuremath{202.7}}                
\newcommand{\hatcurPPmelongorigxxxxxF}{\ensuremath{202.7\pm7.3}}           
\newcommand{\hatcurPProrigxxxxxF}{\ensuremath{1.213\pm0.094}}              
\newcommand{\hatcurPPrshortorigxxxxxF}{\ensuremath{1.21}}                  
\newcommand{\hatcurPPrlongorigxxxxxF}{\ensuremath{1.213\pm0.094}}          
\newcommand{\hatcurPPreorigxxxxxF}{\ensuremath{13.6\pm1.0}}                
\newcommand{\hatcurPPreshortorigxxxxxF}{\ensuremath{13.6}}                 
\newcommand{\hatcurPPrelongorigxxxxxF}{\ensuremath{13.6\pm1.0}}            
\newcommand{\hatcurPPmrcorrorigxxxxxF}{\ensuremath{-0.09}}                 
\newcommand{\hatcurPPtefforigxxxxxF}{\ensuremath{1246\pm32}}               
\newcommand{\hatcurPPthetaorigxxxxxF}{\ensuremath{0.0476\pm0.0041}}        
\newcommand{\hatcurPPfluxperiorigxxxxxF}{\ensuremath{5.45_{-0.50}^{+0.66}}} 
\newcommand{\hatcurPPfluxperidimorigxxxxxF}{\ensuremath{8}}                
\newcommand{\hatcurPPfluxaporigxxxxxF}{\ensuremath{5.45_{-0.50}^{+0.66}}}  
\newcommand{\hatcurPPfluxapdimorigxxxxxF}{\ensuremath{8}}                  
\newcommand{\hatcurPPfluxavgorigxxxxxF}{\ensuremath{5.45_{-0.50}^{+0.66}}} 
\newcommand{\hatcurPPfluxavgdimorigxxxxxF}{\ensuremath{8}}                 
\newcommand{\hatcurPPfluxavglogorigxxxxxF}{\ensuremath{8.736\pm0.045}}     
\newcommand{\hatcurXsecphaseorigxxxxxF}{\ensuremath{0\pm0}}                
\newcommand{\hatcurXsecondaryorigxxxxxF}{\ensuremath{2456384.63085\pm0.00054}} 
\newcommand{\hatcurXsecdurorigxxxxxF}{\ensuremath{0.1210\pm0.0019}}        
\newcommand{\hatcurXsecingdurorigxxxxxF}{\ensuremath{0.0158\pm0.0020}}     
\newcommand{\hatcurPPphiconjorigxxxxxF}{\ensuremath{0\pm0}}                
\newcommand{\hatcurPPperiorigxxxxxF}{\ensuremath{2456382.09756\pm0.00054}} 
\newcommand{\hatcurPPaequivorigxxxxxF}{\ensuremath{0.0501\pm0.0026}}       
\newcommand{\hatcurPPtcircorigxxxxxF}{\ensuremath{144_{-48}^{+71}}}        
\newcommand{\hatcurPPtinfallorigxxxxxF}{\ensuremath{3330\pm780}}           
\newcommand{\hatcurXdistorigxxxxxF}{\ensuremath{438\pm23}}                 
\newcommand{\hatcurXAvorigxxxxxF}{\ensuremath{0.382\pm0.092}}              
\newcommand{\hatcurXdistredorigxxxxxF}{\ensuremath{429_{-19}^{+25}}}       
\newcommand{\hatcurXEBVorigxxxxxF}{\ensuremath{0.123\pm0.030}}             
\newcommand{\hatcurXmvisoredorigxxxxxF}{\ensuremath{13.751\pm0.061}}       
\newcommand{\hatcurXmiisoredorigxxxxxF}{\ensuremath{12.718\pm0.026}}       
\newcommand{\hatcurXmjisoredorigxxxxxF}{\ensuremath{12.095\pm0.016}}       
\newcommand{\hatcurXmhisoredorigxxxxxF}{\ensuremath{11.650\pm0.016}}       
\newcommand{\hatcurXmkisoredorigxxxxxF}{\ensuremath{11.551\pm0.018}}       
\newcommand{\hatcurXviisoredorigxxxxxF}{\ensuremath{1.033\pm0.043}}        
\newcommand{\hatcurXvkisoredorigxxxxxF}{\ensuremath{2.200\pm0.068}}        
\newcommand{\hatcurXjhisoredorigxxxxxF}{\ensuremath{0.445\pm0.011}}        
\newcommand{\hatcurXjkisoredorigxxxxxF}{\ensuremath{0.543\pm0.014}}        
\newcommand{\hatcurCCpmraorigxxxxxF}{\ensuremath{-16.2\pm4.3}}             
\newcommand{\hatcurCCpmdecorigxxxxxF}{\ensuremath{0.5\pm3.5}}              
\newcommand{\hatcurCCpmorigxxxxxF}{\ensuremath{16.2\pm5.5}}                

%% file: HTR405-001.orig.tex
\newcommand{\hatcurhtrorigxxxxxG}{HTR405-001}                              
\newcommand{\hatcurfieldorigxxxxxG}{\ensuremath{string}}                   
\newcommand{\hatcurCCraorigxxxxxG}{\ensuremath{04^{\mathrm h}35^{\mathrm m}53.84{\mathrm s}}}                            
\newcommand{\hatcurCCdecorigxxxxxG}{\ensuremath{+02{\arcdeg}25{\arcmin}52.7{\arcsec}}}                           
\newcommand{\hatcurCCmagorigxxxxxG}{12.771}                                
\newcommand{\hatcurCCtwomassorigxxxxxG}{2MASS~04355384+0225526}            
\newcommand{\hatcurCCgscorigxxxxxG}{GSC~0086-00341}                        
\newcommand{\hatcurCCtassmvorigxxxxxG}{\ensuremath{12.771\pm0.010}}        
\newcommand{\hatcurCCtassmvshortorigxxxxxG}{\ensuremath{12.8}}             
\newcommand{\hatcurCCtassmBorigxxxxxG}{\ensuremath{13.446\pm0.011}}        
\newcommand{\hatcurCCtassmBshortorigxxxxxG}{\ensuremath{13.4}}             
\newcommand{\hatcurCCtassmIorigxxxxxG}{\ensuremath{12.105\pm0.070}}        
\newcommand{\hatcurCCtassmIshortorigxxxxxG}{\ensuremath{12.1}}             
\newcommand{\hatcurCCtassmgorigxxxxxG}{\ensuremath{13.062\pm0.013}}        
\newcommand{\hatcurCCtassmgshortorigxxxxxG}{\ensuremath{13.1}}             
\newcommand{\hatcurCCtassmrorigxxxxxG}{\ensuremath{12.553\pm0.075}}        
\newcommand{\hatcurCCtassmrshortorigxxxxxG}{\ensuremath{12.6}}             
\newcommand{\hatcurCCtassmiorigxxxxxG}{\ensuremath{12.440\pm0.037}}        
\newcommand{\hatcurCCtassmishortorigxxxxxG}{\ensuremath{12.4}}             
\newcommand{\hatcurCCtwomassJmagorigxxxxxG}{\ensuremath{11.485\pm0.026}}   
\newcommand{\hatcurCCtwomassHmagorigxxxxxG}{\ensuremath{11.225\pm0.025}}   
\newcommand{\hatcurCCtwomassKmagorigxxxxxG}{\ensuremath{11.123\pm0.021}}   
\newcommand{\hatcurCCcitJmagorigxxxxxG}{\ensuremath{11.502\pm0.026}}       
\newcommand{\hatcurCCcitHmagorigxxxxxG}{\ensuremath{11.219\pm0.025}}       
\newcommand{\hatcurCCcitKmagorigxxxxxG}{\ensuremath{11.147\pm0.021}}       
\newcommand{\hatcurCCbbJmagorigxxxxxG}{\ensuremath{11.551\pm0.028}}        
\newcommand{\hatcurCCbbHmagorigxxxxxG}{\ensuremath{11.241\pm0.026}}        
\newcommand{\hatcurCCbbKmagorigxxxxxG}{\ensuremath{11.167\pm0.021}}        
\newcommand{\hatcurCCesoJmagorigxxxxxG}{\ensuremath{11.553\pm0.029}}       
\newcommand{\hatcurCCesoHmagorigxxxxxG}{\ensuremath{11.238\pm0.030}}       
\newcommand{\hatcurCCesoKmagorigxxxxxG}{\ensuremath{11.166\pm0.022}}       
\newcommand{\hatcurCCesoJHmagorigxxxxxG}{\ensuremath{0.315\pm0.040}}       
\newcommand{\hatcurCCesoJKmagorigxxxxxG}{\ensuremath{0.387\pm0.036}}       
\newcommand{\hatcurCCesoHKmagorigxxxxxG}{\ensuremath{0.072\pm0.038}}       
\newcommand{\hatcurLCdiporigxxxxxG}{\ensuremath{11.2}}                     
\newcommand{\hatcurLCrprstarorigxxxxxG}{\ensuremath{0.1022\pm0.0043}}      
\newcommand{\hatcurLCbsqorigxxxxxG}{\ensuremath{0.095_{-0.073}^{+0.089}}}  
\newcommand{\hatcurLCimporigxxxxxG}{\ensuremath{0.31_{-0.16}^{+0.12}}}     
\newcommand{\hatcurLCzetaorigxxxxxG}{\ensuremath{10.82\pm0.13}}            
\newcommand{\hatcurLCdurorigxxxxxG}{\ensuremath{0.2059\pm0.0032}}          
\newcommand{\hatcurLCdurshortorigxxxxxG}{\ensuremath{0.2059}}              
\newcommand{\hatcurLCdurhrorigxxxxxG}{\ensuremath{4.942\pm0.077}}          
\newcommand{\hatcurLCdurhrshortorigxxxxxG}{\ensuremath{4.942}}             
\newcommand{\hatcurLCqorigxxxxxG}{\ensuremath{0.05140\pm0.00080}}          
\newcommand{\hatcurLCqshortorigxxxxxG}{\ensuremath{0.051}}                 
\newcommand{\hatcurLCingdurorigxxxxxG}{\ensuremath{0.0210\pm0.0024}}       
\newcommand{\hatcurLCPorigxxxxxG}{\ensuremath{4.007207\pm0.000019}}        
\newcommand{\hatcurLCPprecorigxxxxxG}{\ensuremath{4.0072065}}              
\newcommand{\hatcurLCPshortorigxxxxxG}{\ensuremath{4.0072}}                
\newcommand{\hatcurLCTorigxxxxxG}{\ensuremath{2455735.82441\pm0.00097}}    
\newcommand{\hatcurLCTshortorigxxxxxG}{\ensuremath{5735.82441\pm0.00097}}    
\newcommand{\hatcurLCTAorigxxxxxG}{\ensuremath{2455078.6427\pm0.0030}}     
\newcommand{\hatcurLCTBorigxxxxxG}{\ensuremath{2455848.0261\pm0.0012}}     
\newcommand{\hatcurLChatnetmorigxxxxxG}{\ensuremath{12.52212\pm0.00013}}   
\newcommand{\hatcurLCiblendorigxxxxxG}{\ensuremath{0.717\pm0.069}}         
\newcommand{\hatcurLCrhoorigxxxxxG}{\ensuremath{0.333\pm0.040}}            
\newcommand{\hatcurSMEitefforigxxxxxG}{\ensuremath{6302\pm50}}             
\newcommand{\hatcurSMEizfehorigxxxxxG}{\ensuremath{-0.010\pm0.080}}        
\newcommand{\hatcurSMEizfehshortorigxxxxxG}{\ensuremath{-0.01}}            
\newcommand{\hatcurSMEiloggorigxxxxxG}{\ensuremath{3.99\pm0.10}}           
\newcommand{\hatcurSMEivsinorigxxxxxG}{\ensuremath{12.70\pm0.50}}          
\newcommand{\hatcurSMEivmacorigxxxxxG}{\ensuremath{0.0}}                   
\newcommand{\hatcurSMEivmicorigxxxxxG}{\ensuremath{0.0}}                   
\newcommand{\hatcurLBizorigxxxxxG}{\ensuremath{0.1378}}                    
\newcommand{\hatcurLBiizorigxxxxxG}{\ensuremath{0.3609}}                   
\newcommand{\hatcurLBiiorigxxxxxG}{\ensuremath{0.1885}}                    
\newcommand{\hatcurLBiiiorigxxxxxG}{\ensuremath{0.3692}}                   
\newcommand{\hatcurLBiIorigxxxxxG}{\ensuremath{0.1698}}                    
\newcommand{\hatcurLBiiIorigxxxxxG}{\ensuremath{0.3681}}                   
\newcommand{\hatcurLBigorigxxxxxG}{\ensuremath{0.4343}}                    
\newcommand{\hatcurLBiigorigxxxxxG}{\ensuremath{0.3181}}                   
\newcommand{\hatcurLBirorigxxxxxG}{\ensuremath{0.2646}}                    
\newcommand{\hatcurLBiirorigxxxxxG}{\ensuremath{0.3759}}                   
\newcommand{\hatcurLBiRorigxxxxxG}{\ensuremath{0.2430}}                    
\newcommand{\hatcurLBiiRorigxxxxxG}{\ensuremath{0.3756}}                   
\newcommand{\hatcurLBikeporigxxxxxG}{\ensuremath{0.1000}}                  
\newcommand{\hatcurLBiikeporigxxxxxG}{\ensuremath{0.1000}}                 
\newcommand{\hatcurISOmorigxxxxxG}{\ensuremath{1.369\pm0.046}}             
\newcommand{\hatcurISOmshortorigxxxxxG}{\ensuremath{1.37}}                 
\newcommand{\hatcurISOmlongorigxxxxxG}{\ensuremath{1.369\pm0.046}}         
\newcommand{\hatcurISOrorigxxxxxG}{\ensuremath{1.797_{-0.079}^{+0.119}}}   
\newcommand{\hatcurISOrshortorigxxxxxG}{\ensuremath{1.80}}                 
\newcommand{\hatcurISOrlongorigxxxxxG}{\ensuremath{1.797_{-0.079}^{+0.119}}} 
\newcommand{\hatcurISOrhoorigxxxxxG}{\ensuremath{0.332_{-0.054}^{+0.040}}} 
\newcommand{\hatcurISOrholongorigxxxxxG}{\ensuremath{0.332_{-0.054}^{+0.040}}} 
\newcommand{\hatcurISOloggorigxxxxxG}{\ensuremath{4.064\pm0.035}}          
\newcommand{\hatcurISOlumorigxxxxxG}{\ensuremath{4.56_{-0.42}^{+0.62}}}    
\newcommand{\hatcurISOlumshortorigxxxxxG}{\ensuremath{4.56}}               
\newcommand{\hatcurISOmvorigxxxxxG}{\ensuremath{3.11\pm0.12}}              
\newcommand{\hatcurISOviorigxxxxxG}{\ensuremath{0.547\pm0.013}}            
\newcommand{\hatcurISOageorigxxxxxG}{\ensuremath{2.87\pm0.30}}             
\newcommand{\hatcurISOsigmaorigxxxxxG}{\ensuremath{0.000300\pm0.000075}}   
\newcommand{\hatcurISOMJorigxxxxxG}{\ensuremath{2.22\pm0.11}}              
\newcommand{\hatcurISOMHorigxxxxxG}{\ensuremath{1.97\pm0.11}}              
\newcommand{\hatcurISOMKorigxxxxxG}{\ensuremath{1.92\pm0.11}}              
\newcommand{\hatcurISOJKorigxxxxxG}{\ensuremath{0.300\pm0.010}}            
\newcommand{\hatcurISOspecorigxxxxxG}{F}                                   
\newcommand{\hatcurRVKorigxxxxxG}{\ensuremath{78\pm16}}                    
\newcommand{\hatcurRVrkorigxxxxxG}{\ensuremath{0\pm0}}                     
\newcommand{\hatcurRVrhorigxxxxxG}{\ensuremath{0\pm0}}                     
\newcommand{\hatcurRVkorigxxxxxG}{\ensuremath{0\pm0}}                      
\newcommand{\hatcurRVhorigxxxxxG}{\ensuremath{0\pm0}}                      
\newcommand{\hatcurRVtroneorigxxxxxG}{\ensuremath{0\pm0}}                  
\newcommand{\hatcurRVtrtwoorigxxxxxG}{\ensuremath{0\pm0}}                  
\newcommand{\hatcurRVgammaAorigxxxxxG}{\ensuremath{10\pm12}}               
\newcommand{\hatcurRVjitterAorigxxxxxG}{\ensuremath{25\pm13}}              
\newcommand{\hatcurRVjittertwosiglimAorigxxxxxG}{\ensuremath{<52.6}}       
\newcommand{\hatcurRVfitrmsAorigxxxxxG}{\ensuremath{0.0}}                  
\newcommand{\hatcurRVgammaBorigxxxxxG}{\ensuremath{24485\pm13}}            
\newcommand{\hatcurRVjitterBorigxxxxxG}{\ensuremath{0.0\pm9.3}}            
\newcommand{\hatcurRVjittertwosiglimBorigxxxxxG}{\ensuremath{<22.7}}       
\newcommand{\hatcurRVfitrmsBorigxxxxxG}{\ensuremath{0.0}}                  
\newcommand{\hatcurRVeccenorigxxxxxG}{\ensuremath{0\pm0}}                  
\newcommand{\hatcurRVeccentwosiglimorigxxxxxG}{\ensuremath{<0.000}}        
\newcommand{\hatcurRVomegaorigxxxxxG}{\ensuremath{0\pm0}}                  
\newcommand{\hatcurPPiorigxxxxxG}{\ensuremath{87.3\pm1.2}}                 
\newcommand{\hatcurPPgorigxxxxxG}{\ensuremath{5.9\pm1.4}}                  
\newcommand{\hatcurPPloggorigxxxxxG}{\ensuremath{2.771_{-0.125}^{+0.093}}} 
\newcommand{\hatcurPParorigxxxxxG}{\ensuremath{6.56_{-0.38}^{+0.25}}}      
\newcommand{\hatcurPParelorigxxxxxG}{\ensuremath{0.05483\pm0.00062}}       
\newcommand{\hatcurPPrhoorigxxxxxG}{\ensuremath{0.162\pm0.048}}            
\newcommand{\hatcurPPmorigxxxxxG}{\ensuremath{0.75\pm0.16}}                
\newcommand{\hatcurPPmshortorigxxxxxG}{\ensuremath{0.75}}                  
\newcommand{\hatcurPPmlongorigxxxxxG}{\ensuremath{0.75\pm0.16}}            
\newcommand{\hatcurPPmeorigxxxxxG}{\ensuremath{240\pm50}}                  
\newcommand{\hatcurPPmeshortorigxxxxxG}{\ensuremath{239.7}}                
\newcommand{\hatcurPPmelongorigxxxxxG}{\ensuremath{240\pm50}}              
\newcommand{\hatcurPProrigxxxxxG}{\ensuremath{1.79\pm0.14}}                
\newcommand{\hatcurPPrshortorigxxxxxG}{\ensuremath{1.79}}                  
\newcommand{\hatcurPPrlongorigxxxxxG}{\ensuremath{1.79\pm0.14}}            
\newcommand{\hatcurPPreorigxxxxxG}{\ensuremath{20.1\pm1.6}}                
\newcommand{\hatcurPPreshortorigxxxxxG}{\ensuremath{20.1}}                 
\newcommand{\hatcurPPrelongorigxxxxxG}{\ensuremath{20.1\pm1.6}}            
\newcommand{\hatcurPPmrcorrorigxxxxxG}{\ensuremath{0.06}}                  
\newcommand{\hatcurPPtefforigxxxxxG}{\ensuremath{1741_{-35}^{+48}}}        
\newcommand{\hatcurPPthetaorigxxxxxG}{\ensuremath{0.0340\pm0.0073}}        
\newcommand{\hatcurPPfluxperiorigxxxxxG}{\ensuremath{2.07_{-0.16}^{+0.24}}} 
\newcommand{\hatcurPPfluxperidimorigxxxxxG}{\ensuremath{9}}                
\newcommand{\hatcurPPfluxaporigxxxxxG}{\ensuremath{2.07_{-0.16}^{+0.24}}}  
\newcommand{\hatcurPPfluxapdimorigxxxxxG}{\ensuremath{9}}                  
\newcommand{\hatcurPPfluxavgorigxxxxxG}{\ensuremath{2.07_{-0.16}^{+0.24}}} 
\newcommand{\hatcurPPfluxavgdimorigxxxxxG}{\ensuremath{9}}                 
\newcommand{\hatcurPPfluxavglogorigxxxxxG}{\ensuremath{9.317_{-0.035}^{+0.047}}} 
\newcommand{\hatcurXsecphaseorigxxxxxG}{\ensuremath{0\pm0}}                
\newcommand{\hatcurXsecondaryorigxxxxxG}{\ensuremath{2455737.82802\pm0.00097}} 
\newcommand{\hatcurXsecdurorigxxxxxG}{\ensuremath{0.2059\pm0.0032}}        
\newcommand{\hatcurXsecingdurorigxxxxxG}{\ensuremath{0.0210\pm0.0024}}     
\newcommand{\hatcurPPphiconjorigxxxxxG}{\ensuremath{0\pm0}}                
\newcommand{\hatcurPPperiorigxxxxxG}{\ensuremath{2455734.82261\pm0.00097}} 
\newcommand{\hatcurPPaequivorigxxxxxG}{\ensuremath{0.0257_{-0.0014}^{+0.0010}}} 
\newcommand{\hatcurPPtcircorigxxxxxG}{\ensuremath{62_{-21}^{+34}}}         
\newcommand{\hatcurPPtinfallorigxxxxxG}{\ensuremath{830\pm250}}            
\newcommand{\hatcurXdistorigxxxxxG}{\ensuremath{707_{-31}^{+46}}}          
\newcommand{\hatcurXAvorigxxxxxG}{\ensuremath{0.462\pm0.047}}              
\newcommand{\hatcurXdistredorigxxxxxG}{\ensuremath{691_{-30}^{+45}}}       
\newcommand{\hatcurXEBVorigxxxxxG}{\ensuremath{0.149\pm0.015}}             
\newcommand{\hatcurXmvisoredorigxxxxxG}{\ensuremath{12.7720\pm0.0100}}     
\newcommand{\hatcurXmiisoredorigxxxxxG}{\ensuremath{11.984\pm0.013}}       
\newcommand{\hatcurXmjisoredorigxxxxxG}{\ensuremath{11.547\pm0.014}}       
\newcommand{\hatcurXmhisoredorigxxxxxG}{\ensuremath{11.250\pm0.015}}       
\newcommand{\hatcurXmkisoredorigxxxxxG}{\ensuremath{11.170\pm0.016}}       
\newcommand{\hatcurXviisoredorigxxxxxG}{\ensuremath{0.788\pm0.014}}        
\newcommand{\hatcurXvkisoredorigxxxxxG}{\ensuremath{1.602\pm0.019}}        
\newcommand{\hatcurXjhisoredorigxxxxxG}{\ensuremath{0.2960\pm0.0062}}      
\newcommand{\hatcurXjkisoredorigxxxxxG}{\ensuremath{0.3770\pm0.0054}}      
\newcommand{\hatcurCCpmraorigxxxxxG}{\ensuremath{10.8\pm1.3}}              
\newcommand{\hatcurCCpmdecorigxxxxxG}{\ensuremath{1.2\pm1.8}}              
\newcommand{\hatcurCCpmorigxxxxxG}{\ensuremath{10.9\pm2.2}}                

%% file: planetinfo_multi_eccen_orig.tex
\input{HTR062-011.eccen.orig.tex}
\input{HTR080-003.eccen.orig.tex}
\input{HTR089-018.eccen.orig.tex}
\input{HTR094-024.eccen.orig.tex}
\input{HTR130-001.eccen.orig.tex}
\input{HTR384-014.eccen.orig.tex}
\input{HTR405-001.eccen.orig.tex}
\newcommand{\hatcurCCbbHmageccenorig}[1]{\ifnum#1=58 %
\hatcurCCbbHmageccenorigxxxxxA
\else
\ifnum#1=59 %
\hatcurCCbbHmageccenorigxxxxxB
\else
\ifnum#1=60 %
\hatcurCCbbHmageccenorigxxxxxC
\else
\ifnum#1=61 %
\hatcurCCbbHmageccenorigxxxxxD
\else
\ifnum#1=62 %
\hatcurCCbbHmageccenorigxxxxxE
\else
\ifnum#1=63 %
\hatcurCCbbHmageccenorigxxxxxF
\else
\ifnum#1=64 %
\hatcurCCbbHmageccenorigxxxxxG
\else
??????\fi
\fi
\fi
\fi
\fi
\fi
\fi
}
\newcommand{\hatcurCCbbJmageccenorig}[1]{\ifnum#1=58 %
\hatcurCCbbJmageccenorigxxxxxA
\else
\ifnum#1=59 %
\hatcurCCbbJmageccenorigxxxxxB
\else
\ifnum#1=60 %
\hatcurCCbbJmageccenorigxxxxxC
\else
\ifnum#1=61 %
\hatcurCCbbJmageccenorigxxxxxD
\else
\ifnum#1=62 %
\hatcurCCbbJmageccenorigxxxxxE
\else
\ifnum#1=63 %
\hatcurCCbbJmageccenorigxxxxxF
\else
\ifnum#1=64 %
\hatcurCCbbJmageccenorigxxxxxG
\else
??????\fi
\fi
\fi
\fi
\fi
\fi
\fi
}
\newcommand{\hatcurCCbbKmageccenorig}[1]{\ifnum#1=58 %
\hatcurCCbbKmageccenorigxxxxxA
\else
\ifnum#1=59 %
\hatcurCCbbKmageccenorigxxxxxB
\else
\ifnum#1=60 %
\hatcurCCbbKmageccenorigxxxxxC
\else
\ifnum#1=61 %
\hatcurCCbbKmageccenorigxxxxxD
\else
\ifnum#1=62 %
\hatcurCCbbKmageccenorigxxxxxE
\else
\ifnum#1=63 %
\hatcurCCbbKmageccenorigxxxxxF
\else
\ifnum#1=64 %
\hatcurCCbbKmageccenorigxxxxxG
\else
??????\fi
\fi
\fi
\fi
\fi
\fi
\fi
}
\newcommand{\hatcurCCcitHmageccenorig}[1]{\ifnum#1=58 %
\hatcurCCcitHmageccenorigxxxxxA
\else
\ifnum#1=59 %
\hatcurCCcitHmageccenorigxxxxxB
\else
\ifnum#1=60 %
\hatcurCCcitHmageccenorigxxxxxC
\else
\ifnum#1=61 %
\hatcurCCcitHmageccenorigxxxxxD
\else
\ifnum#1=62 %
\hatcurCCcitHmageccenorigxxxxxE
\else
\ifnum#1=63 %
\hatcurCCcitHmageccenorigxxxxxF
\else
\ifnum#1=64 %
\hatcurCCcitHmageccenorigxxxxxG
\else
??????\fi
\fi
\fi
\fi
\fi
\fi
\fi
}
\newcommand{\hatcurCCcitJmageccenorig}[1]{\ifnum#1=58 %
\hatcurCCcitJmageccenorigxxxxxA
\else
\ifnum#1=59 %
\hatcurCCcitJmageccenorigxxxxxB
\else
\ifnum#1=60 %
\hatcurCCcitJmageccenorigxxxxxC
\else
\ifnum#1=61 %
\hatcurCCcitJmageccenorigxxxxxD
\else
\ifnum#1=62 %
\hatcurCCcitJmageccenorigxxxxxE
\else
\ifnum#1=63 %
\hatcurCCcitJmageccenorigxxxxxF
\else
\ifnum#1=64 %
\hatcurCCcitJmageccenorigxxxxxG
\else
??????\fi
\fi
\fi
\fi
\fi
\fi
\fi
}
\newcommand{\hatcurCCcitKmageccenorig}[1]{\ifnum#1=58 %
\hatcurCCcitKmageccenorigxxxxxA
\else
\ifnum#1=59 %
\hatcurCCcitKmageccenorigxxxxxB
\else
\ifnum#1=60 %
\hatcurCCcitKmageccenorigxxxxxC
\else
\ifnum#1=61 %
\hatcurCCcitKmageccenorigxxxxxD
\else
\ifnum#1=62 %
\hatcurCCcitKmageccenorigxxxxxE
\else
\ifnum#1=63 %
\hatcurCCcitKmageccenorigxxxxxF
\else
\ifnum#1=64 %
\hatcurCCcitKmageccenorigxxxxxG
\else
??????\fi
\fi
\fi
\fi
\fi
\fi
\fi
}
\newcommand{\hatcurCCdececcenorig}[1]{\ifnum#1=58 %
\hatcurCCdececcenorigxxxxxA
\else
\ifnum#1=59 %
\hatcurCCdececcenorigxxxxxB
\else
\ifnum#1=60 %
\hatcurCCdececcenorigxxxxxC
\else
\ifnum#1=61 %
\hatcurCCdececcenorigxxxxxD
\else
\ifnum#1=62 %
\hatcurCCdececcenorigxxxxxE
\else
\ifnum#1=63 %
\hatcurCCdececcenorigxxxxxF
\else
\ifnum#1=64 %
\hatcurCCdececcenorigxxxxxG
\else
??????\fi
\fi
\fi
\fi
\fi
\fi
\fi
}
\newcommand{\hatcurCCesoHKmageccenorig}[1]{\ifnum#1=58 %
\hatcurCCesoHKmageccenorigxxxxxA
\else
\ifnum#1=59 %
\hatcurCCesoHKmageccenorigxxxxxB
\else
\ifnum#1=60 %
\hatcurCCesoHKmageccenorigxxxxxC
\else
\ifnum#1=61 %
\hatcurCCesoHKmageccenorigxxxxxD
\else
\ifnum#1=62 %
\hatcurCCesoHKmageccenorigxxxxxE
\else
\ifnum#1=63 %
\hatcurCCesoHKmageccenorigxxxxxF
\else
\ifnum#1=64 %
\hatcurCCesoHKmageccenorigxxxxxG
\else
??????\fi
\fi
\fi
\fi
\fi
\fi
\fi
}
\newcommand{\hatcurCCesoHmageccenorig}[1]{\ifnum#1=58 %
\hatcurCCesoHmageccenorigxxxxxA
\else
\ifnum#1=59 %
\hatcurCCesoHmageccenorigxxxxxB
\else
\ifnum#1=60 %
\hatcurCCesoHmageccenorigxxxxxC
\else
\ifnum#1=61 %
\hatcurCCesoHmageccenorigxxxxxD
\else
\ifnum#1=62 %
\hatcurCCesoHmageccenorigxxxxxE
\else
\ifnum#1=63 %
\hatcurCCesoHmageccenorigxxxxxF
\else
\ifnum#1=64 %
\hatcurCCesoHmageccenorigxxxxxG
\else
??????\fi
\fi
\fi
\fi
\fi
\fi
\fi
}
\newcommand{\hatcurCCesoJHmageccenorig}[1]{\ifnum#1=58 %
\hatcurCCesoJHmageccenorigxxxxxA
\else
\ifnum#1=59 %
\hatcurCCesoJHmageccenorigxxxxxB
\else
\ifnum#1=60 %
\hatcurCCesoJHmageccenorigxxxxxC
\else
\ifnum#1=61 %
\hatcurCCesoJHmageccenorigxxxxxD
\else
\ifnum#1=62 %
\hatcurCCesoJHmageccenorigxxxxxE
\else
\ifnum#1=63 %
\hatcurCCesoJHmageccenorigxxxxxF
\else
\ifnum#1=64 %
\hatcurCCesoJHmageccenorigxxxxxG
\else
??????\fi
\fi
\fi
\fi
\fi
\fi
\fi
}
\newcommand{\hatcurCCesoJKmageccenorig}[1]{\ifnum#1=58 %
\hatcurCCesoJKmageccenorigxxxxxA
\else
\ifnum#1=59 %
\hatcurCCesoJKmageccenorigxxxxxB
\else
\ifnum#1=60 %
\hatcurCCesoJKmageccenorigxxxxxC
\else
\ifnum#1=61 %
\hatcurCCesoJKmageccenorigxxxxxD
\else
\ifnum#1=62 %
\hatcurCCesoJKmageccenorigxxxxxE
\else
\ifnum#1=63 %
\hatcurCCesoJKmageccenorigxxxxxF
\else
\ifnum#1=64 %
\hatcurCCesoJKmageccenorigxxxxxG
\else
??????\fi
\fi
\fi
\fi
\fi
\fi
\fi
}
\newcommand{\hatcurCCesoJmageccenorig}[1]{\ifnum#1=58 %
\hatcurCCesoJmageccenorigxxxxxA
\else
\ifnum#1=59 %
\hatcurCCesoJmageccenorigxxxxxB
\else
\ifnum#1=60 %
\hatcurCCesoJmageccenorigxxxxxC
\else
\ifnum#1=61 %
\hatcurCCesoJmageccenorigxxxxxD
\else
\ifnum#1=62 %
\hatcurCCesoJmageccenorigxxxxxE
\else
\ifnum#1=63 %
\hatcurCCesoJmageccenorigxxxxxF
\else
\ifnum#1=64 %
\hatcurCCesoJmageccenorigxxxxxG
\else
??????\fi
\fi
\fi
\fi
\fi
\fi
\fi
}
\newcommand{\hatcurCCesoKmageccenorig}[1]{\ifnum#1=58 %
\hatcurCCesoKmageccenorigxxxxxA
\else
\ifnum#1=59 %
\hatcurCCesoKmageccenorigxxxxxB
\else
\ifnum#1=60 %
\hatcurCCesoKmageccenorigxxxxxC
\else
\ifnum#1=61 %
\hatcurCCesoKmageccenorigxxxxxD
\else
\ifnum#1=62 %
\hatcurCCesoKmageccenorigxxxxxE
\else
\ifnum#1=63 %
\hatcurCCesoKmageccenorigxxxxxF
\else
\ifnum#1=64 %
\hatcurCCesoKmageccenorigxxxxxG
\else
??????\fi
\fi
\fi
\fi
\fi
\fi
\fi
}
\newcommand{\hatcurCCgsceccenorig}[1]{\ifnum#1=58 %
\hatcurCCgsceccenorigxxxxxA
\else
\ifnum#1=59 %
\hatcurCCgsceccenorigxxxxxB
\else
\ifnum#1=60 %
\hatcurCCgsceccenorigxxxxxC
\else
\ifnum#1=61 %
\hatcurCCgsceccenorigxxxxxD
\else
\ifnum#1=62 %
\hatcurCCgsceccenorigxxxxxE
\else
\ifnum#1=63 %
\hatcurCCgsceccenorigxxxxxF
\else
\ifnum#1=64 %
\hatcurCCgsceccenorigxxxxxG
\else
??????\fi
\fi
\fi
\fi
\fi
\fi
\fi
}
\newcommand{\hatcurCCmageccenorig}[1]{\ifnum#1=58 %
\hatcurCCmageccenorigxxxxxA
\else
\ifnum#1=59 %
\hatcurCCmageccenorigxxxxxB
\else
\ifnum#1=60 %
\hatcurCCmageccenorigxxxxxC
\else
\ifnum#1=61 %
\hatcurCCmageccenorigxxxxxD
\else
\ifnum#1=62 %
\hatcurCCmageccenorigxxxxxE
\else
\ifnum#1=63 %
\hatcurCCmageccenorigxxxxxF
\else
\ifnum#1=64 %
\hatcurCCmageccenorigxxxxxG
\else
??????\fi
\fi
\fi
\fi
\fi
\fi
\fi
}
\newcommand{\hatcurCCpmdececcenorig}[1]{\ifnum#1=58 %
\hatcurCCpmdececcenorigxxxxxA
\else
\ifnum#1=59 %
\hatcurCCpmdececcenorigxxxxxB
\else
\ifnum#1=60 %
\hatcurCCpmdececcenorigxxxxxC
\else
\ifnum#1=61 %
\hatcurCCpmdececcenorigxxxxxD
\else
\ifnum#1=62 %
\hatcurCCpmdececcenorigxxxxxE
\else
\ifnum#1=63 %
\hatcurCCpmdececcenorigxxxxxF
\else
\ifnum#1=64 %
\hatcurCCpmdececcenorigxxxxxG
\else
??????\fi
\fi
\fi
\fi
\fi
\fi
\fi
}
\newcommand{\hatcurCCpmeccenorig}[1]{\ifnum#1=58 %
\hatcurCCpmeccenorigxxxxxA
\else
\ifnum#1=59 %
\hatcurCCpmeccenorigxxxxxB
\else
\ifnum#1=60 %
\hatcurCCpmeccenorigxxxxxC
\else
\ifnum#1=61 %
\hatcurCCpmeccenorigxxxxxD
\else
\ifnum#1=62 %
\hatcurCCpmeccenorigxxxxxE
\else
\ifnum#1=63 %
\hatcurCCpmeccenorigxxxxxF
\else
\ifnum#1=64 %
\hatcurCCpmeccenorigxxxxxG
\else
??????\fi
\fi
\fi
\fi
\fi
\fi
\fi
}
\newcommand{\hatcurCCpmraeccenorig}[1]{\ifnum#1=58 %
\hatcurCCpmraeccenorigxxxxxA
\else
\ifnum#1=59 %
\hatcurCCpmraeccenorigxxxxxB
\else
\ifnum#1=60 %
\hatcurCCpmraeccenorigxxxxxC
\else
\ifnum#1=61 %
\hatcurCCpmraeccenorigxxxxxD
\else
\ifnum#1=62 %
\hatcurCCpmraeccenorigxxxxxE
\else
\ifnum#1=63 %
\hatcurCCpmraeccenorigxxxxxF
\else
\ifnum#1=64 %
\hatcurCCpmraeccenorigxxxxxG
\else
??????\fi
\fi
\fi
\fi
\fi
\fi
\fi
}
\newcommand{\hatcurCCraeccenorig}[1]{\ifnum#1=58 %
\hatcurCCraeccenorigxxxxxA
\else
\ifnum#1=59 %
\hatcurCCraeccenorigxxxxxB
\else
\ifnum#1=60 %
\hatcurCCraeccenorigxxxxxC
\else
\ifnum#1=61 %
\hatcurCCraeccenorigxxxxxD
\else
\ifnum#1=62 %
\hatcurCCraeccenorigxxxxxE
\else
\ifnum#1=63 %
\hatcurCCraeccenorigxxxxxF
\else
\ifnum#1=64 %
\hatcurCCraeccenorigxxxxxG
\else
??????\fi
\fi
\fi
\fi
\fi
\fi
\fi
}
\newcommand{\hatcurCCtassmBeccenorig}[1]{\ifnum#1=58 %
\hatcurCCtassmBeccenorigxxxxxA
\else
\ifnum#1=59 %
\hatcurCCtassmBeccenorigxxxxxB
\else
\ifnum#1=60 %
\hatcurCCtassmBeccenorigxxxxxC
\else
\ifnum#1=61 %
\hatcurCCtassmBeccenorigxxxxxD
\else
\ifnum#1=62 %
\hatcurCCtassmBeccenorigxxxxxE
\else
\ifnum#1=63 %
\hatcurCCtassmBeccenorigxxxxxF
\else
\ifnum#1=64 %
\hatcurCCtassmBeccenorigxxxxxG
\else
??????\fi
\fi
\fi
\fi
\fi
\fi
\fi
}
\newcommand{\hatcurCCtassmBshorteccenorig}[1]{\ifnum#1=58 %
\hatcurCCtassmBshorteccenorigxxxxxA
\else
\ifnum#1=59 %
\hatcurCCtassmBshorteccenorigxxxxxB
\else
\ifnum#1=60 %
\hatcurCCtassmBshorteccenorigxxxxxC
\else
\ifnum#1=61 %
\hatcurCCtassmBshorteccenorigxxxxxD
\else
\ifnum#1=62 %
\hatcurCCtassmBshorteccenorigxxxxxE
\else
\ifnum#1=63 %
\hatcurCCtassmBshorteccenorigxxxxxF
\else
\ifnum#1=64 %
\hatcurCCtassmBshorteccenorigxxxxxG
\else
??????\fi
\fi
\fi
\fi
\fi
\fi
\fi
}
\newcommand{\hatcurCCtassmgeccenorig}[1]{\ifnum#1=58 %
\hatcurCCtassmgeccenorigxxxxxA
\else
\ifnum#1=59 %
\hatcurCCtassmgeccenorigxxxxxB
\else
\ifnum#1=60 %
\hatcurCCtassmgeccenorigxxxxxC
\else
\ifnum#1=61 %
\hatcurCCtassmgeccenorigxxxxxD
\else
\ifnum#1=62 %
\hatcurCCtassmgeccenorigxxxxxE
\else
\ifnum#1=63 %
\hatcurCCtassmgeccenorigxxxxxF
\else
\ifnum#1=64 %
\hatcurCCtassmgeccenorigxxxxxG
\else
??????\fi
\fi
\fi
\fi
\fi
\fi
\fi
}
\newcommand{\hatcurCCtassmgshorteccenorig}[1]{\ifnum#1=58 %
\hatcurCCtassmgshorteccenorigxxxxxA
\else
\ifnum#1=59 %
\hatcurCCtassmgshorteccenorigxxxxxB
\else
\ifnum#1=60 %
\hatcurCCtassmgshorteccenorigxxxxxC
\else
\ifnum#1=61 %
\hatcurCCtassmgshorteccenorigxxxxxD
\else
\ifnum#1=62 %
\hatcurCCtassmgshorteccenorigxxxxxE
\else
\ifnum#1=63 %
\hatcurCCtassmgshorteccenorigxxxxxF
\else
\ifnum#1=64 %
\hatcurCCtassmgshorteccenorigxxxxxG
\else
??????\fi
\fi
\fi
\fi
\fi
\fi
\fi
}
\newcommand{\hatcurCCtassmieccenorig}[1]{\ifnum#1=58 %
\hatcurCCtassmieccenorigxxxxxA
\else
\ifnum#1=59 %
\hatcurCCtassmieccenorigxxxxxB
\else
\ifnum#1=60 %
\hatcurCCtassmieccenorigxxxxxC
\else
\ifnum#1=61 %
\hatcurCCtassmieccenorigxxxxxD
\else
\ifnum#1=62 %
\hatcurCCtassmieccenorigxxxxxE
\else
\ifnum#1=63 %
\hatcurCCtassmieccenorigxxxxxF
\else
\ifnum#1=64 %
\hatcurCCtassmieccenorigxxxxxG
\else
??????\fi
\fi
\fi
\fi
\fi
\fi
\fi
}
\newcommand{\hatcurCCtassmIeccenorig}[1]{\ifnum#1=58 %
\hatcurCCtassmIeccenorigxxxxxA
\else
\ifnum#1=59 %
\hatcurCCtassmIeccenorigxxxxxB
\else
\ifnum#1=60 %
\hatcurCCtassmIeccenorigxxxxxC
\else
\ifnum#1=61 %
\hatcurCCtassmIeccenorigxxxxxD
\else
\ifnum#1=62 %
\hatcurCCtassmIeccenorigxxxxxE
\else
\ifnum#1=63 %
\hatcurCCtassmIeccenorigxxxxxF
\else
\ifnum#1=64 %
\hatcurCCtassmIeccenorigxxxxxG
\else
??????\fi
\fi
\fi
\fi
\fi
\fi
\fi
}
\newcommand{\hatcurCCtassmishorteccenorig}[1]{\ifnum#1=58 %
\hatcurCCtassmishorteccenorigxxxxxA
\else
\ifnum#1=59 %
\hatcurCCtassmishorteccenorigxxxxxB
\else
\ifnum#1=60 %
\hatcurCCtassmishorteccenorigxxxxxC
\else
\ifnum#1=61 %
\hatcurCCtassmishorteccenorigxxxxxD
\else
\ifnum#1=62 %
\hatcurCCtassmishorteccenorigxxxxxE
\else
\ifnum#1=63 %
\hatcurCCtassmishorteccenorigxxxxxF
\else
\ifnum#1=64 %
\hatcurCCtassmishorteccenorigxxxxxG
\else
??????\fi
\fi
\fi
\fi
\fi
\fi
\fi
}
\newcommand{\hatcurCCtassmIshorteccenorig}[1]{\ifnum#1=58 %
\hatcurCCtassmIshorteccenorigxxxxxA
\else
\ifnum#1=59 %
\hatcurCCtassmIshorteccenorigxxxxxB
\else
\ifnum#1=60 %
\hatcurCCtassmIshorteccenorigxxxxxC
\else
\ifnum#1=61 %
\hatcurCCtassmIshorteccenorigxxxxxD
\else
\ifnum#1=62 %
\hatcurCCtassmIshorteccenorigxxxxxE
\else
\ifnum#1=63 %
\hatcurCCtassmIshorteccenorigxxxxxF
\else
\ifnum#1=64 %
\hatcurCCtassmIshorteccenorigxxxxxG
\else
??????\fi
\fi
\fi
\fi
\fi
\fi
\fi
}
\newcommand{\hatcurCCtassmreccenorig}[1]{\ifnum#1=58 %
\hatcurCCtassmreccenorigxxxxxA
\else
\ifnum#1=59 %
\hatcurCCtassmreccenorigxxxxxB
\else
\ifnum#1=60 %
\hatcurCCtassmreccenorigxxxxxC
\else
\ifnum#1=61 %
\hatcurCCtassmreccenorigxxxxxD
\else
\ifnum#1=62 %
\hatcurCCtassmreccenorigxxxxxE
\else
\ifnum#1=63 %
\hatcurCCtassmreccenorigxxxxxF
\else
\ifnum#1=64 %
\hatcurCCtassmreccenorigxxxxxG
\else
??????\fi
\fi
\fi
\fi
\fi
\fi
\fi
}
\newcommand{\hatcurCCtassmrshorteccenorig}[1]{\ifnum#1=58 %
\hatcurCCtassmrshorteccenorigxxxxxA
\else
\ifnum#1=59 %
\hatcurCCtassmrshorteccenorigxxxxxB
\else
\ifnum#1=60 %
\hatcurCCtassmrshorteccenorigxxxxxC
\else
\ifnum#1=61 %
\hatcurCCtassmrshorteccenorigxxxxxD
\else
\ifnum#1=62 %
\hatcurCCtassmrshorteccenorigxxxxxE
\else
\ifnum#1=63 %
\hatcurCCtassmrshorteccenorigxxxxxF
\else
\ifnum#1=64 %
\hatcurCCtassmrshorteccenorigxxxxxG
\else
??????\fi
\fi
\fi
\fi
\fi
\fi
\fi
}
\newcommand{\hatcurCCtassmveccenorig}[1]{\ifnum#1=58 %
\hatcurCCtassmveccenorigxxxxxA
\else
\ifnum#1=59 %
\hatcurCCtassmveccenorigxxxxxB
\else
\ifnum#1=60 %
\hatcurCCtassmveccenorigxxxxxC
\else
\ifnum#1=61 %
\hatcurCCtassmveccenorigxxxxxD
\else
\ifnum#1=62 %
\hatcurCCtassmveccenorigxxxxxE
\else
\ifnum#1=63 %
\hatcurCCtassmveccenorigxxxxxF
\else
\ifnum#1=64 %
\hatcurCCtassmveccenorigxxxxxG
\else
??????\fi
\fi
\fi
\fi
\fi
\fi
\fi
}
\newcommand{\hatcurCCtassmvshorteccenorig}[1]{\ifnum#1=58 %
\hatcurCCtassmvshorteccenorigxxxxxA
\else
\ifnum#1=59 %
\hatcurCCtassmvshorteccenorigxxxxxB
\else
\ifnum#1=60 %
\hatcurCCtassmvshorteccenorigxxxxxC
\else
\ifnum#1=61 %
\hatcurCCtassmvshorteccenorigxxxxxD
\else
\ifnum#1=62 %
\hatcurCCtassmvshorteccenorigxxxxxE
\else
\ifnum#1=63 %
\hatcurCCtassmvshorteccenorigxxxxxF
\else
\ifnum#1=64 %
\hatcurCCtassmvshorteccenorigxxxxxG
\else
??????\fi
\fi
\fi
\fi
\fi
\fi
\fi
}
\newcommand{\hatcurCCtwomasseccenorig}[1]{\ifnum#1=58 %
\hatcurCCtwomasseccenorigxxxxxA
\else
\ifnum#1=59 %
\hatcurCCtwomasseccenorigxxxxxB
\else
\ifnum#1=60 %
\hatcurCCtwomasseccenorigxxxxxC
\else
\ifnum#1=61 %
\hatcurCCtwomasseccenorigxxxxxD
\else
\ifnum#1=62 %
\hatcurCCtwomasseccenorigxxxxxE
\else
\ifnum#1=63 %
\hatcurCCtwomasseccenorigxxxxxF
\else
\ifnum#1=64 %
\hatcurCCtwomasseccenorigxxxxxG
\else
??????\fi
\fi
\fi
\fi
\fi
\fi
\fi
}
\newcommand{\hatcurCCtwomassHmageccenorig}[1]{\ifnum#1=58 %
\hatcurCCtwomassHmageccenorigxxxxxA
\else
\ifnum#1=59 %
\hatcurCCtwomassHmageccenorigxxxxxB
\else
\ifnum#1=60 %
\hatcurCCtwomassHmageccenorigxxxxxC
\else
\ifnum#1=61 %
\hatcurCCtwomassHmageccenorigxxxxxD
\else
\ifnum#1=62 %
\hatcurCCtwomassHmageccenorigxxxxxE
\else
\ifnum#1=63 %
\hatcurCCtwomassHmageccenorigxxxxxF
\else
\ifnum#1=64 %
\hatcurCCtwomassHmageccenorigxxxxxG
\else
??????\fi
\fi
\fi
\fi
\fi
\fi
\fi
}
\newcommand{\hatcurCCtwomassJmageccenorig}[1]{\ifnum#1=58 %
\hatcurCCtwomassJmageccenorigxxxxxA
\else
\ifnum#1=59 %
\hatcurCCtwomassJmageccenorigxxxxxB
\else
\ifnum#1=60 %
\hatcurCCtwomassJmageccenorigxxxxxC
\else
\ifnum#1=61 %
\hatcurCCtwomassJmageccenorigxxxxxD
\else
\ifnum#1=62 %
\hatcurCCtwomassJmageccenorigxxxxxE
\else
\ifnum#1=63 %
\hatcurCCtwomassJmageccenorigxxxxxF
\else
\ifnum#1=64 %
\hatcurCCtwomassJmageccenorigxxxxxG
\else
??????\fi
\fi
\fi
\fi
\fi
\fi
\fi
}
\newcommand{\hatcurCCtwomassKmageccenorig}[1]{\ifnum#1=58 %
\hatcurCCtwomassKmageccenorigxxxxxA
\else
\ifnum#1=59 %
\hatcurCCtwomassKmageccenorigxxxxxB
\else
\ifnum#1=60 %
\hatcurCCtwomassKmageccenorigxxxxxC
\else
\ifnum#1=61 %
\hatcurCCtwomassKmageccenorigxxxxxD
\else
\ifnum#1=62 %
\hatcurCCtwomassKmageccenorigxxxxxE
\else
\ifnum#1=63 %
\hatcurCCtwomassKmageccenorigxxxxxF
\else
\ifnum#1=64 %
\hatcurCCtwomassKmageccenorigxxxxxG
\else
??????\fi
\fi
\fi
\fi
\fi
\fi
\fi
}
\newcommand{\hatcurfieldeccenorig}[1]{\ifnum#1=58 %
\hatcurfieldeccenorigxxxxxA
\else
\ifnum#1=59 %
\hatcurfieldeccenorigxxxxxB
\else
\ifnum#1=60 %
\hatcurfieldeccenorigxxxxxC
\else
\ifnum#1=61 %
\hatcurfieldeccenorigxxxxxD
\else
\ifnum#1=62 %
\hatcurfieldeccenorigxxxxxE
\else
\ifnum#1=63 %
\hatcurfieldeccenorigxxxxxF
\else
\ifnum#1=64 %
\hatcurfieldeccenorigxxxxxG
\else
??????\fi
\fi
\fi
\fi
\fi
\fi
\fi
}
\newcommand{\hatcurhtreccenorig}[1]{\ifnum#1=58 %
\hatcurhtreccenorigxxxxxA
\else
\ifnum#1=59 %
\hatcurhtreccenorigxxxxxB
\else
\ifnum#1=60 %
\hatcurhtreccenorigxxxxxC
\else
\ifnum#1=61 %
\hatcurhtreccenorigxxxxxD
\else
\ifnum#1=62 %
\hatcurhtreccenorigxxxxxE
\else
\ifnum#1=63 %
\hatcurhtreccenorigxxxxxF
\else
\ifnum#1=64 %
\hatcurhtreccenorigxxxxxG
\else
??????\fi
\fi
\fi
\fi
\fi
\fi
\fi
}
\newcommand{\hatcurISOageeccenorig}[1]{\ifnum#1=58 %
\hatcurISOageeccenorigxxxxxA
\else
\ifnum#1=59 %
\hatcurISOageeccenorigxxxxxB
\else
\ifnum#1=60 %
\hatcurISOageeccenorigxxxxxC
\else
\ifnum#1=61 %
\hatcurISOageeccenorigxxxxxD
\else
\ifnum#1=62 %
\hatcurISOageeccenorigxxxxxE
\else
\ifnum#1=63 %
\hatcurISOageeccenorigxxxxxF
\else
\ifnum#1=64 %
\hatcurISOageeccenorigxxxxxG
\else
??????\fi
\fi
\fi
\fi
\fi
\fi
\fi
}
\newcommand{\hatcurISOJKeccenorig}[1]{\ifnum#1=58 %
\hatcurISOJKeccenorigxxxxxA
\else
\ifnum#1=59 %
\hatcurISOJKeccenorigxxxxxB
\else
\ifnum#1=60 %
\hatcurISOJKeccenorigxxxxxC
\else
\ifnum#1=61 %
\hatcurISOJKeccenorigxxxxxD
\else
\ifnum#1=62 %
\hatcurISOJKeccenorigxxxxxE
\else
\ifnum#1=63 %
\hatcurISOJKeccenorigxxxxxF
\else
\ifnum#1=64 %
\hatcurISOJKeccenorigxxxxxG
\else
??????\fi
\fi
\fi
\fi
\fi
\fi
\fi
}
\newcommand{\hatcurISOloggeccenorig}[1]{\ifnum#1=58 %
\hatcurISOloggeccenorigxxxxxA
\else
\ifnum#1=59 %
\hatcurISOloggeccenorigxxxxxB
\else
\ifnum#1=60 %
\hatcurISOloggeccenorigxxxxxC
\else
\ifnum#1=61 %
\hatcurISOloggeccenorigxxxxxD
\else
\ifnum#1=62 %
\hatcurISOloggeccenorigxxxxxE
\else
\ifnum#1=63 %
\hatcurISOloggeccenorigxxxxxF
\else
\ifnum#1=64 %
\hatcurISOloggeccenorigxxxxxG
\else
??????\fi
\fi
\fi
\fi
\fi
\fi
\fi
}
\newcommand{\hatcurISOlumeccenorig}[1]{\ifnum#1=58 %
\hatcurISOlumeccenorigxxxxxA
\else
\ifnum#1=59 %
\hatcurISOlumeccenorigxxxxxB
\else
\ifnum#1=60 %
\hatcurISOlumeccenorigxxxxxC
\else
\ifnum#1=61 %
\hatcurISOlumeccenorigxxxxxD
\else
\ifnum#1=62 %
\hatcurISOlumeccenorigxxxxxE
\else
\ifnum#1=63 %
\hatcurISOlumeccenorigxxxxxF
\else
\ifnum#1=64 %
\hatcurISOlumeccenorigxxxxxG
\else
??????\fi
\fi
\fi
\fi
\fi
\fi
\fi
}
\newcommand{\hatcurISOlumshorteccenorig}[1]{\ifnum#1=58 %
\hatcurISOlumshorteccenorigxxxxxA
\else
\ifnum#1=59 %
\hatcurISOlumshorteccenorigxxxxxB
\else
\ifnum#1=60 %
\hatcurISOlumshorteccenorigxxxxxC
\else
\ifnum#1=61 %
\hatcurISOlumshorteccenorigxxxxxD
\else
\ifnum#1=62 %
\hatcurISOlumshorteccenorigxxxxxE
\else
\ifnum#1=63 %
\hatcurISOlumshorteccenorigxxxxxF
\else
\ifnum#1=64 %
\hatcurISOlumshorteccenorigxxxxxG
\else
??????\fi
\fi
\fi
\fi
\fi
\fi
\fi
}
\newcommand{\hatcurISOmeccenorig}[1]{\ifnum#1=58 %
\hatcurISOmeccenorigxxxxxA
\else
\ifnum#1=59 %
\hatcurISOmeccenorigxxxxxB
\else
\ifnum#1=60 %
\hatcurISOmeccenorigxxxxxC
\else
\ifnum#1=61 %
\hatcurISOmeccenorigxxxxxD
\else
\ifnum#1=62 %
\hatcurISOmeccenorigxxxxxE
\else
\ifnum#1=63 %
\hatcurISOmeccenorigxxxxxF
\else
\ifnum#1=64 %
\hatcurISOmeccenorigxxxxxG
\else
??????\fi
\fi
\fi
\fi
\fi
\fi
\fi
}
\newcommand{\hatcurISOMHeccenorig}[1]{\ifnum#1=58 %
\hatcurISOMHeccenorigxxxxxA
\else
\ifnum#1=59 %
\hatcurISOMHeccenorigxxxxxB
\else
\ifnum#1=60 %
\hatcurISOMHeccenorigxxxxxC
\else
\ifnum#1=61 %
\hatcurISOMHeccenorigxxxxxD
\else
\ifnum#1=62 %
\hatcurISOMHeccenorigxxxxxE
\else
\ifnum#1=63 %
\hatcurISOMHeccenorigxxxxxF
\else
\ifnum#1=64 %
\hatcurISOMHeccenorigxxxxxG
\else
??????\fi
\fi
\fi
\fi
\fi
\fi
\fi
}
\newcommand{\hatcurISOMJeccenorig}[1]{\ifnum#1=58 %
\hatcurISOMJeccenorigxxxxxA
\else
\ifnum#1=59 %
\hatcurISOMJeccenorigxxxxxB
\else
\ifnum#1=60 %
\hatcurISOMJeccenorigxxxxxC
\else
\ifnum#1=61 %
\hatcurISOMJeccenorigxxxxxD
\else
\ifnum#1=62 %
\hatcurISOMJeccenorigxxxxxE
\else
\ifnum#1=63 %
\hatcurISOMJeccenorigxxxxxF
\else
\ifnum#1=64 %
\hatcurISOMJeccenorigxxxxxG
\else
??????\fi
\fi
\fi
\fi
\fi
\fi
\fi
}
\newcommand{\hatcurISOMKeccenorig}[1]{\ifnum#1=58 %
\hatcurISOMKeccenorigxxxxxA
\else
\ifnum#1=59 %
\hatcurISOMKeccenorigxxxxxB
\else
\ifnum#1=60 %
\hatcurISOMKeccenorigxxxxxC
\else
\ifnum#1=61 %
\hatcurISOMKeccenorigxxxxxD
\else
\ifnum#1=62 %
\hatcurISOMKeccenorigxxxxxE
\else
\ifnum#1=63 %
\hatcurISOMKeccenorigxxxxxF
\else
\ifnum#1=64 %
\hatcurISOMKeccenorigxxxxxG
\else
??????\fi
\fi
\fi
\fi
\fi
\fi
\fi
}
\newcommand{\hatcurISOmlongeccenorig}[1]{\ifnum#1=58 %
\hatcurISOmlongeccenorigxxxxxA
\else
\ifnum#1=59 %
\hatcurISOmlongeccenorigxxxxxB
\else
\ifnum#1=60 %
\hatcurISOmlongeccenorigxxxxxC
\else
\ifnum#1=61 %
\hatcurISOmlongeccenorigxxxxxD
\else
\ifnum#1=62 %
\hatcurISOmlongeccenorigxxxxxE
\else
\ifnum#1=63 %
\hatcurISOmlongeccenorigxxxxxF
\else
\ifnum#1=64 %
\hatcurISOmlongeccenorigxxxxxG
\else
??????\fi
\fi
\fi
\fi
\fi
\fi
\fi
}
\newcommand{\hatcurISOmshorteccenorig}[1]{\ifnum#1=58 %
\hatcurISOmshorteccenorigxxxxxA
\else
\ifnum#1=59 %
\hatcurISOmshorteccenorigxxxxxB
\else
\ifnum#1=60 %
\hatcurISOmshorteccenorigxxxxxC
\else
\ifnum#1=61 %
\hatcurISOmshorteccenorigxxxxxD
\else
\ifnum#1=62 %
\hatcurISOmshorteccenorigxxxxxE
\else
\ifnum#1=63 %
\hatcurISOmshorteccenorigxxxxxF
\else
\ifnum#1=64 %
\hatcurISOmshorteccenorigxxxxxG
\else
??????\fi
\fi
\fi
\fi
\fi
\fi
\fi
}
\newcommand{\hatcurISOmveccenorig}[1]{\ifnum#1=58 %
\hatcurISOmveccenorigxxxxxA
\else
\ifnum#1=59 %
\hatcurISOmveccenorigxxxxxB
\else
\ifnum#1=60 %
\hatcurISOmveccenorigxxxxxC
\else
\ifnum#1=61 %
\hatcurISOmveccenorigxxxxxD
\else
\ifnum#1=62 %
\hatcurISOmveccenorigxxxxxE
\else
\ifnum#1=63 %
\hatcurISOmveccenorigxxxxxF
\else
\ifnum#1=64 %
\hatcurISOmveccenorigxxxxxG
\else
??????\fi
\fi
\fi
\fi
\fi
\fi
\fi
}
\newcommand{\hatcurISOreccenorig}[1]{\ifnum#1=58 %
\hatcurISOreccenorigxxxxxA
\else
\ifnum#1=59 %
\hatcurISOreccenorigxxxxxB
\else
\ifnum#1=60 %
\hatcurISOreccenorigxxxxxC
\else
\ifnum#1=61 %
\hatcurISOreccenorigxxxxxD
\else
\ifnum#1=62 %
\hatcurISOreccenorigxxxxxE
\else
\ifnum#1=63 %
\hatcurISOreccenorigxxxxxF
\else
\ifnum#1=64 %
\hatcurISOreccenorigxxxxxG
\else
??????\fi
\fi
\fi
\fi
\fi
\fi
\fi
}
\newcommand{\hatcurISOrhoeccenorig}[1]{\ifnum#1=58 %
\hatcurISOrhoeccenorigxxxxxA
\else
\ifnum#1=59 %
\hatcurISOrhoeccenorigxxxxxB
\else
\ifnum#1=60 %
\hatcurISOrhoeccenorigxxxxxC
\else
\ifnum#1=61 %
\hatcurISOrhoeccenorigxxxxxD
\else
\ifnum#1=62 %
\hatcurISOrhoeccenorigxxxxxE
\else
\ifnum#1=63 %
\hatcurISOrhoeccenorigxxxxxF
\else
\ifnum#1=64 %
\hatcurISOrhoeccenorigxxxxxG
\else
??????\fi
\fi
\fi
\fi
\fi
\fi
\fi
}
\newcommand{\hatcurISOrholongeccenorig}[1]{\ifnum#1=58 %
\hatcurISOrholongeccenorigxxxxxA
\else
\ifnum#1=59 %
\hatcurISOrholongeccenorigxxxxxB
\else
\ifnum#1=60 %
\hatcurISOrholongeccenorigxxxxxC
\else
\ifnum#1=61 %
\hatcurISOrholongeccenorigxxxxxD
\else
\ifnum#1=62 %
\hatcurISOrholongeccenorigxxxxxE
\else
\ifnum#1=63 %
\hatcurISOrholongeccenorigxxxxxF
\else
\ifnum#1=64 %
\hatcurISOrholongeccenorigxxxxxG
\else
??????\fi
\fi
\fi
\fi
\fi
\fi
\fi
}
\newcommand{\hatcurISOrlongeccenorig}[1]{\ifnum#1=58 %
\hatcurISOrlongeccenorigxxxxxA
\else
\ifnum#1=59 %
\hatcurISOrlongeccenorigxxxxxB
\else
\ifnum#1=60 %
\hatcurISOrlongeccenorigxxxxxC
\else
\ifnum#1=61 %
\hatcurISOrlongeccenorigxxxxxD
\else
\ifnum#1=62 %
\hatcurISOrlongeccenorigxxxxxE
\else
\ifnum#1=63 %
\hatcurISOrlongeccenorigxxxxxF
\else
\ifnum#1=64 %
\hatcurISOrlongeccenorigxxxxxG
\else
??????\fi
\fi
\fi
\fi
\fi
\fi
\fi
}
\newcommand{\hatcurISOrshorteccenorig}[1]{\ifnum#1=58 %
\hatcurISOrshorteccenorigxxxxxA
\else
\ifnum#1=59 %
\hatcurISOrshorteccenorigxxxxxB
\else
\ifnum#1=60 %
\hatcurISOrshorteccenorigxxxxxC
\else
\ifnum#1=61 %
\hatcurISOrshorteccenorigxxxxxD
\else
\ifnum#1=62 %
\hatcurISOrshorteccenorigxxxxxE
\else
\ifnum#1=63 %
\hatcurISOrshorteccenorigxxxxxF
\else
\ifnum#1=64 %
\hatcurISOrshorteccenorigxxxxxG
\else
??????\fi
\fi
\fi
\fi
\fi
\fi
\fi
}
\newcommand{\hatcurISOsigmaeccenorig}[1]{\ifnum#1=58 %
\hatcurISOsigmaeccenorigxxxxxA
\else
\ifnum#1=59 %
\hatcurISOsigmaeccenorigxxxxxB
\else
\ifnum#1=60 %
\hatcurISOsigmaeccenorigxxxxxC
\else
\ifnum#1=61 %
\hatcurISOsigmaeccenorigxxxxxD
\else
\ifnum#1=62 %
\hatcurISOsigmaeccenorigxxxxxE
\else
\ifnum#1=63 %
\hatcurISOsigmaeccenorigxxxxxF
\else
\ifnum#1=64 %
\hatcurISOsigmaeccenorigxxxxxG
\else
??????\fi
\fi
\fi
\fi
\fi
\fi
\fi
}
\newcommand{\hatcurISOspececcenorig}[1]{\ifnum#1=58 %
\hatcurISOspececcenorigxxxxxA
\else
\ifnum#1=59 %
\hatcurISOspececcenorigxxxxxB
\else
\ifnum#1=60 %
\hatcurISOspececcenorigxxxxxC
\else
\ifnum#1=61 %
\hatcurISOspececcenorigxxxxxD
\else
\ifnum#1=62 %
\hatcurISOspececcenorigxxxxxE
\else
\ifnum#1=63 %
\hatcurISOspececcenorigxxxxxF
\else
\ifnum#1=64 %
\hatcurISOspececcenorigxxxxxG
\else
??????\fi
\fi
\fi
\fi
\fi
\fi
\fi
}
\newcommand{\hatcurISOvieccenorig}[1]{\ifnum#1=58 %
\hatcurISOvieccenorigxxxxxA
\else
\ifnum#1=59 %
\hatcurISOvieccenorigxxxxxB
\else
\ifnum#1=60 %
\hatcurISOvieccenorigxxxxxC
\else
\ifnum#1=61 %
\hatcurISOvieccenorigxxxxxD
\else
\ifnum#1=62 %
\hatcurISOvieccenorigxxxxxE
\else
\ifnum#1=63 %
\hatcurISOvieccenorigxxxxxF
\else
\ifnum#1=64 %
\hatcurISOvieccenorigxxxxxG
\else
??????\fi
\fi
\fi
\fi
\fi
\fi
\fi
}
\newcommand{\hatcurLBigeccenorig}[1]{\ifnum#1=58 %
\hatcurLBigeccenorigxxxxxA
\else
\ifnum#1=59 %
\hatcurLBigeccenorigxxxxxB
\else
\ifnum#1=60 %
\hatcurLBigeccenorigxxxxxC
\else
\ifnum#1=61 %
\hatcurLBigeccenorigxxxxxD
\else
\ifnum#1=62 %
\hatcurLBigeccenorigxxxxxE
\else
\ifnum#1=63 %
\hatcurLBigeccenorigxxxxxF
\else
\ifnum#1=64 %
\hatcurLBigeccenorigxxxxxG
\else
??????\fi
\fi
\fi
\fi
\fi
\fi
\fi
}
\newcommand{\hatcurLBiieccenorig}[1]{\ifnum#1=58 %
\hatcurLBiieccenorigxxxxxA
\else
\ifnum#1=59 %
\hatcurLBiieccenorigxxxxxB
\else
\ifnum#1=60 %
\hatcurLBiieccenorigxxxxxC
\else
\ifnum#1=61 %
\hatcurLBiieccenorigxxxxxD
\else
\ifnum#1=62 %
\hatcurLBiieccenorigxxxxxE
\else
\ifnum#1=63 %
\hatcurLBiieccenorigxxxxxF
\else
\ifnum#1=64 %
\hatcurLBiieccenorigxxxxxG
\else
??????\fi
\fi
\fi
\fi
\fi
\fi
\fi
}
\newcommand{\hatcurLBiIeccenorig}[1]{\ifnum#1=58 %
\hatcurLBiIeccenorigxxxxxA
\else
\ifnum#1=59 %
\hatcurLBiIeccenorigxxxxxB
\else
\ifnum#1=60 %
\hatcurLBiIeccenorigxxxxxC
\else
\ifnum#1=61 %
\hatcurLBiIeccenorigxxxxxD
\else
\ifnum#1=62 %
\hatcurLBiIeccenorigxxxxxE
\else
\ifnum#1=63 %
\hatcurLBiIeccenorigxxxxxF
\else
\ifnum#1=64 %
\hatcurLBiIeccenorigxxxxxG
\else
??????\fi
\fi
\fi
\fi
\fi
\fi
\fi
}
\newcommand{\hatcurLBiigeccenorig}[1]{\ifnum#1=58 %
\hatcurLBiigeccenorigxxxxxA
\else
\ifnum#1=59 %
\hatcurLBiigeccenorigxxxxxB
\else
\ifnum#1=60 %
\hatcurLBiigeccenorigxxxxxC
\else
\ifnum#1=61 %
\hatcurLBiigeccenorigxxxxxD
\else
\ifnum#1=62 %
\hatcurLBiigeccenorigxxxxxE
\else
\ifnum#1=63 %
\hatcurLBiigeccenorigxxxxxF
\else
\ifnum#1=64 %
\hatcurLBiigeccenorigxxxxxG
\else
??????\fi
\fi
\fi
\fi
\fi
\fi
\fi
}
\newcommand{\hatcurLBiiieccenorig}[1]{\ifnum#1=58 %
\hatcurLBiiieccenorigxxxxxA
\else
\ifnum#1=59 %
\hatcurLBiiieccenorigxxxxxB
\else
\ifnum#1=60 %
\hatcurLBiiieccenorigxxxxxC
\else
\ifnum#1=61 %
\hatcurLBiiieccenorigxxxxxD
\else
\ifnum#1=62 %
\hatcurLBiiieccenorigxxxxxE
\else
\ifnum#1=63 %
\hatcurLBiiieccenorigxxxxxF
\else
\ifnum#1=64 %
\hatcurLBiiieccenorigxxxxxG
\else
??????\fi
\fi
\fi
\fi
\fi
\fi
\fi
}
\newcommand{\hatcurLBiiIeccenorig}[1]{\ifnum#1=58 %
\hatcurLBiiIeccenorigxxxxxA
\else
\ifnum#1=59 %
\hatcurLBiiIeccenorigxxxxxB
\else
\ifnum#1=60 %
\hatcurLBiiIeccenorigxxxxxC
\else
\ifnum#1=61 %
\hatcurLBiiIeccenorigxxxxxD
\else
\ifnum#1=62 %
\hatcurLBiiIeccenorigxxxxxE
\else
\ifnum#1=63 %
\hatcurLBiiIeccenorigxxxxxF
\else
\ifnum#1=64 %
\hatcurLBiiIeccenorigxxxxxG
\else
??????\fi
\fi
\fi
\fi
\fi
\fi
\fi
}
\newcommand{\hatcurLBiikepeccenorig}[1]{\ifnum#1=58 %
\hatcurLBiikepeccenorigxxxxxA
\else
\ifnum#1=59 %
\hatcurLBiikepeccenorigxxxxxB
\else
\ifnum#1=60 %
\hatcurLBiikepeccenorigxxxxxC
\else
\ifnum#1=61 %
\hatcurLBiikepeccenorigxxxxxD
\else
\ifnum#1=62 %
\hatcurLBiikepeccenorigxxxxxE
\else
\ifnum#1=63 %
\hatcurLBiikepeccenorigxxxxxF
\else
\ifnum#1=64 %
\hatcurLBiikepeccenorigxxxxxG
\else
??????\fi
\fi
\fi
\fi
\fi
\fi
\fi
}
\newcommand{\hatcurLBiireccenorig}[1]{\ifnum#1=58 %
\hatcurLBiireccenorigxxxxxA
\else
\ifnum#1=59 %
\hatcurLBiireccenorigxxxxxB
\else
\ifnum#1=60 %
\hatcurLBiireccenorigxxxxxC
\else
\ifnum#1=61 %
\hatcurLBiireccenorigxxxxxD
\else
\ifnum#1=62 %
\hatcurLBiireccenorigxxxxxE
\else
\ifnum#1=63 %
\hatcurLBiireccenorigxxxxxF
\else
\ifnum#1=64 %
\hatcurLBiireccenorigxxxxxG
\else
??????\fi
\fi
\fi
\fi
\fi
\fi
\fi
}
\newcommand{\hatcurLBiiReccenorig}[1]{\ifnum#1=58 %
\hatcurLBiiReccenorigxxxxxA
\else
\ifnum#1=59 %
\hatcurLBiiReccenorigxxxxxB
\else
\ifnum#1=60 %
\hatcurLBiiReccenorigxxxxxC
\else
\ifnum#1=61 %
\hatcurLBiiReccenorigxxxxxD
\else
\ifnum#1=62 %
\hatcurLBiiReccenorigxxxxxE
\else
\ifnum#1=63 %
\hatcurLBiiReccenorigxxxxxF
\else
\ifnum#1=64 %
\hatcurLBiiReccenorigxxxxxG
\else
??????\fi
\fi
\fi
\fi
\fi
\fi
\fi
}
\newcommand{\hatcurLBiizeccenorig}[1]{\ifnum#1=58 %
\hatcurLBiizeccenorigxxxxxA
\else
\ifnum#1=59 %
\hatcurLBiizeccenorigxxxxxB
\else
\ifnum#1=60 %
\hatcurLBiizeccenorigxxxxxC
\else
\ifnum#1=61 %
\hatcurLBiizeccenorigxxxxxD
\else
\ifnum#1=62 %
\hatcurLBiizeccenorigxxxxxE
\else
\ifnum#1=63 %
\hatcurLBiizeccenorigxxxxxF
\else
\ifnum#1=64 %
\hatcurLBiizeccenorigxxxxxG
\else
??????\fi
\fi
\fi
\fi
\fi
\fi
\fi
}
\newcommand{\hatcurLBikepeccenorig}[1]{\ifnum#1=58 %
\hatcurLBikepeccenorigxxxxxA
\else
\ifnum#1=59 %
\hatcurLBikepeccenorigxxxxxB
\else
\ifnum#1=60 %
\hatcurLBikepeccenorigxxxxxC
\else
\ifnum#1=61 %
\hatcurLBikepeccenorigxxxxxD
\else
\ifnum#1=62 %
\hatcurLBikepeccenorigxxxxxE
\else
\ifnum#1=63 %
\hatcurLBikepeccenorigxxxxxF
\else
\ifnum#1=64 %
\hatcurLBikepeccenorigxxxxxG
\else
??????\fi
\fi
\fi
\fi
\fi
\fi
\fi
}
\newcommand{\hatcurLBireccenorig}[1]{\ifnum#1=58 %
\hatcurLBireccenorigxxxxxA
\else
\ifnum#1=59 %
\hatcurLBireccenorigxxxxxB
\else
\ifnum#1=60 %
\hatcurLBireccenorigxxxxxC
\else
\ifnum#1=61 %
\hatcurLBireccenorigxxxxxD
\else
\ifnum#1=62 %
\hatcurLBireccenorigxxxxxE
\else
\ifnum#1=63 %
\hatcurLBireccenorigxxxxxF
\else
\ifnum#1=64 %
\hatcurLBireccenorigxxxxxG
\else
??????\fi
\fi
\fi
\fi
\fi
\fi
\fi
}
\newcommand{\hatcurLBiReccenorig}[1]{\ifnum#1=58 %
\hatcurLBiReccenorigxxxxxA
\else
\ifnum#1=59 %
\hatcurLBiReccenorigxxxxxB
\else
\ifnum#1=60 %
\hatcurLBiReccenorigxxxxxC
\else
\ifnum#1=61 %
\hatcurLBiReccenorigxxxxxD
\else
\ifnum#1=62 %
\hatcurLBiReccenorigxxxxxE
\else
\ifnum#1=63 %
\hatcurLBiReccenorigxxxxxF
\else
\ifnum#1=64 %
\hatcurLBiReccenorigxxxxxG
\else
??????\fi
\fi
\fi
\fi
\fi
\fi
\fi
}
\newcommand{\hatcurLBizeccenorig}[1]{\ifnum#1=58 %
\hatcurLBizeccenorigxxxxxA
\else
\ifnum#1=59 %
\hatcurLBizeccenorigxxxxxB
\else
\ifnum#1=60 %
\hatcurLBizeccenorigxxxxxC
\else
\ifnum#1=61 %
\hatcurLBizeccenorigxxxxxD
\else
\ifnum#1=62 %
\hatcurLBizeccenorigxxxxxE
\else
\ifnum#1=63 %
\hatcurLBizeccenorigxxxxxF
\else
\ifnum#1=64 %
\hatcurLBizeccenorigxxxxxG
\else
??????\fi
\fi
\fi
\fi
\fi
\fi
\fi
}
\newcommand{\hatcurLCbsqeccenorig}[1]{\ifnum#1=58 %
\hatcurLCbsqeccenorigxxxxxA
\else
\ifnum#1=59 %
\hatcurLCbsqeccenorigxxxxxB
\else
\ifnum#1=60 %
\hatcurLCbsqeccenorigxxxxxC
\else
\ifnum#1=61 %
\hatcurLCbsqeccenorigxxxxxD
\else
\ifnum#1=62 %
\hatcurLCbsqeccenorigxxxxxE
\else
\ifnum#1=63 %
\hatcurLCbsqeccenorigxxxxxF
\else
\ifnum#1=64 %
\hatcurLCbsqeccenorigxxxxxG
\else
??????\fi
\fi
\fi
\fi
\fi
\fi
\fi
}
\newcommand{\hatcurLCdipeccenorig}[1]{\ifnum#1=58 %
\hatcurLCdipeccenorigxxxxxA
\else
\ifnum#1=59 %
\hatcurLCdipeccenorigxxxxxB
\else
\ifnum#1=60 %
\hatcurLCdipeccenorigxxxxxC
\else
\ifnum#1=61 %
\hatcurLCdipeccenorigxxxxxD
\else
\ifnum#1=62 %
\hatcurLCdipeccenorigxxxxxE
\else
\ifnum#1=63 %
\hatcurLCdipeccenorigxxxxxF
\else
\ifnum#1=64 %
\hatcurLCdipeccenorigxxxxxG
\else
??????\fi
\fi
\fi
\fi
\fi
\fi
\fi
}
\newcommand{\hatcurLCdureccenorig}[1]{\ifnum#1=58 %
\hatcurLCdureccenorigxxxxxA
\else
\ifnum#1=59 %
\hatcurLCdureccenorigxxxxxB
\else
\ifnum#1=60 %
\hatcurLCdureccenorigxxxxxC
\else
\ifnum#1=61 %
\hatcurLCdureccenorigxxxxxD
\else
\ifnum#1=62 %
\hatcurLCdureccenorigxxxxxE
\else
\ifnum#1=63 %
\hatcurLCdureccenorigxxxxxF
\else
\ifnum#1=64 %
\hatcurLCdureccenorigxxxxxG
\else
??????\fi
\fi
\fi
\fi
\fi
\fi
\fi
}
\newcommand{\hatcurLCdurhreccenorig}[1]{\ifnum#1=58 %
\hatcurLCdurhreccenorigxxxxxA
\else
\ifnum#1=59 %
\hatcurLCdurhreccenorigxxxxxB
\else
\ifnum#1=60 %
\hatcurLCdurhreccenorigxxxxxC
\else
\ifnum#1=61 %
\hatcurLCdurhreccenorigxxxxxD
\else
\ifnum#1=62 %
\hatcurLCdurhreccenorigxxxxxE
\else
\ifnum#1=63 %
\hatcurLCdurhreccenorigxxxxxF
\else
\ifnum#1=64 %
\hatcurLCdurhreccenorigxxxxxG
\else
??????\fi
\fi
\fi
\fi
\fi
\fi
\fi
}
\newcommand{\hatcurLCdurhrshorteccenorig}[1]{\ifnum#1=58 %
\hatcurLCdurhrshorteccenorigxxxxxA
\else
\ifnum#1=59 %
\hatcurLCdurhrshorteccenorigxxxxxB
\else
\ifnum#1=60 %
\hatcurLCdurhrshorteccenorigxxxxxC
\else
\ifnum#1=61 %
\hatcurLCdurhrshorteccenorigxxxxxD
\else
\ifnum#1=62 %
\hatcurLCdurhrshorteccenorigxxxxxE
\else
\ifnum#1=63 %
\hatcurLCdurhrshorteccenorigxxxxxF
\else
\ifnum#1=64 %
\hatcurLCdurhrshorteccenorigxxxxxG
\else
??????\fi
\fi
\fi
\fi
\fi
\fi
\fi
}
\newcommand{\hatcurLCdurshorteccenorig}[1]{\ifnum#1=58 %
\hatcurLCdurshorteccenorigxxxxxA
\else
\ifnum#1=59 %
\hatcurLCdurshorteccenorigxxxxxB
\else
\ifnum#1=60 %
\hatcurLCdurshorteccenorigxxxxxC
\else
\ifnum#1=61 %
\hatcurLCdurshorteccenorigxxxxxD
\else
\ifnum#1=62 %
\hatcurLCdurshorteccenorigxxxxxE
\else
\ifnum#1=63 %
\hatcurLCdurshorteccenorigxxxxxF
\else
\ifnum#1=64 %
\hatcurLCdurshorteccenorigxxxxxG
\else
??????\fi
\fi
\fi
\fi
\fi
\fi
\fi
}
\newcommand{\hatcurLChatnetmAeccenorig}[1]{\ifnum#1=61 %
\hatcurLChatnetmAeccenorigxxxxxD
\else
??????\fi
}
\newcommand{\hatcurLChatnetmBeccenorig}[1]{\ifnum#1=61 %
\hatcurLChatnetmBeccenorigxxxxxD
\else
??????\fi
}
\newcommand{\hatcurLChatnetmeccenorig}[1]{\ifnum#1=58 %
\hatcurLChatnetmeccenorigxxxxxA
\else
\ifnum#1=59 %
\hatcurLChatnetmeccenorigxxxxxB
\else
\ifnum#1=62 %
\hatcurLChatnetmeccenorigxxxxxE
\else
\ifnum#1=63 %
\hatcurLChatnetmeccenorigxxxxxF
\else
\ifnum#1=64 %
\hatcurLChatnetmeccenorigxxxxxG
\else
??????\fi
\fi
\fi
\fi
\fi
}
\newcommand{\hatcurLCiblendAeccenorig}[1]{\ifnum#1=61 %
\hatcurLCiblendAeccenorigxxxxxD
\else
??????\fi
}
\newcommand{\hatcurLCiblendBeccenorig}[1]{\ifnum#1=61 %
\hatcurLCiblendBeccenorigxxxxxD
\else
??????\fi
}
\newcommand{\hatcurLCiblendeccenorig}[1]{\ifnum#1=58 %
\hatcurLCiblendeccenorigxxxxxA
\else
\ifnum#1=59 %
\hatcurLCiblendeccenorigxxxxxB
\else
\ifnum#1=62 %
\hatcurLCiblendeccenorigxxxxxE
\else
\ifnum#1=63 %
\hatcurLCiblendeccenorigxxxxxF
\else
\ifnum#1=64 %
\hatcurLCiblendeccenorigxxxxxG
\else
??????\fi
\fi
\fi
\fi
\fi
}
\newcommand{\hatcurLCimpeccenorig}[1]{\ifnum#1=58 %
\hatcurLCimpeccenorigxxxxxA
\else
\ifnum#1=59 %
\hatcurLCimpeccenorigxxxxxB
\else
\ifnum#1=60 %
\hatcurLCimpeccenorigxxxxxC
\else
\ifnum#1=61 %
\hatcurLCimpeccenorigxxxxxD
\else
\ifnum#1=62 %
\hatcurLCimpeccenorigxxxxxE
\else
\ifnum#1=63 %
\hatcurLCimpeccenorigxxxxxF
\else
\ifnum#1=64 %
\hatcurLCimpeccenorigxxxxxG
\else
??????\fi
\fi
\fi
\fi
\fi
\fi
\fi
}
\newcommand{\hatcurLCingdureccenorig}[1]{\ifnum#1=58 %
\hatcurLCingdureccenorigxxxxxA
\else
\ifnum#1=59 %
\hatcurLCingdureccenorigxxxxxB
\else
\ifnum#1=60 %
\hatcurLCingdureccenorigxxxxxC
\else
\ifnum#1=61 %
\hatcurLCingdureccenorigxxxxxD
\else
\ifnum#1=62 %
\hatcurLCingdureccenorigxxxxxE
\else
\ifnum#1=63 %
\hatcurLCingdureccenorigxxxxxF
\else
\ifnum#1=64 %
\hatcurLCingdureccenorigxxxxxG
\else
??????\fi
\fi
\fi
\fi
\fi
\fi
\fi
}
\newcommand{\hatcurLCPeccenorig}[1]{\ifnum#1=58 %
\hatcurLCPeccenorigxxxxxA
\else
\ifnum#1=59 %
\hatcurLCPeccenorigxxxxxB
\else
\ifnum#1=60 %
\hatcurLCPeccenorigxxxxxC
\else
\ifnum#1=61 %
\hatcurLCPeccenorigxxxxxD
\else
\ifnum#1=62 %
\hatcurLCPeccenorigxxxxxE
\else
\ifnum#1=63 %
\hatcurLCPeccenorigxxxxxF
\else
\ifnum#1=64 %
\hatcurLCPeccenorigxxxxxG
\else
??????\fi
\fi
\fi
\fi
\fi
\fi
\fi
}
\newcommand{\hatcurLCPprececcenorig}[1]{\ifnum#1=58 %
\hatcurLCPprececcenorigxxxxxA
\else
\ifnum#1=59 %
\hatcurLCPprececcenorigxxxxxB
\else
\ifnum#1=60 %
\hatcurLCPprececcenorigxxxxxC
\else
\ifnum#1=61 %
\hatcurLCPprececcenorigxxxxxD
\else
\ifnum#1=62 %
\hatcurLCPprececcenorigxxxxxE
\else
\ifnum#1=63 %
\hatcurLCPprececcenorigxxxxxF
\else
\ifnum#1=64 %
\hatcurLCPprececcenorigxxxxxG
\else
??????\fi
\fi
\fi
\fi
\fi
\fi
\fi
}
\newcommand{\hatcurLCPshorteccenorig}[1]{\ifnum#1=58 %
\hatcurLCPshorteccenorigxxxxxA
\else
\ifnum#1=59 %
\hatcurLCPshorteccenorigxxxxxB
\else
\ifnum#1=60 %
\hatcurLCPshorteccenorigxxxxxC
\else
\ifnum#1=61 %
\hatcurLCPshorteccenorigxxxxxD
\else
\ifnum#1=62 %
\hatcurLCPshorteccenorigxxxxxE
\else
\ifnum#1=63 %
\hatcurLCPshorteccenorigxxxxxF
\else
\ifnum#1=64 %
\hatcurLCPshorteccenorigxxxxxG
\else
??????\fi
\fi
\fi
\fi
\fi
\fi
\fi
}
\newcommand{\hatcurLCqeccenorig}[1]{\ifnum#1=58 %
\hatcurLCqeccenorigxxxxxA
\else
\ifnum#1=59 %
\hatcurLCqeccenorigxxxxxB
\else
\ifnum#1=60 %
\hatcurLCqeccenorigxxxxxC
\else
\ifnum#1=61 %
\hatcurLCqeccenorigxxxxxD
\else
\ifnum#1=62 %
\hatcurLCqeccenorigxxxxxE
\else
\ifnum#1=63 %
\hatcurLCqeccenorigxxxxxF
\else
\ifnum#1=64 %
\hatcurLCqeccenorigxxxxxG
\else
??????\fi
\fi
\fi
\fi
\fi
\fi
\fi
}
\newcommand{\hatcurLCqshorteccenorig}[1]{\ifnum#1=58 %
\hatcurLCqshorteccenorigxxxxxA
\else
\ifnum#1=59 %
\hatcurLCqshorteccenorigxxxxxB
\else
\ifnum#1=60 %
\hatcurLCqshorteccenorigxxxxxC
\else
\ifnum#1=61 %
\hatcurLCqshorteccenorigxxxxxD
\else
\ifnum#1=62 %
\hatcurLCqshorteccenorigxxxxxE
\else
\ifnum#1=63 %
\hatcurLCqshorteccenorigxxxxxF
\else
\ifnum#1=64 %
\hatcurLCqshorteccenorigxxxxxG
\else
??????\fi
\fi
\fi
\fi
\fi
\fi
\fi
}
\newcommand{\hatcurLCrhoeccenorig}[1]{\ifnum#1=58 %
\hatcurLCrhoeccenorigxxxxxA
\else
\ifnum#1=59 %
\hatcurLCrhoeccenorigxxxxxB
\else
\ifnum#1=60 %
\hatcurLCrhoeccenorigxxxxxC
\else
\ifnum#1=61 %
\hatcurLCrhoeccenorigxxxxxD
\else
\ifnum#1=62 %
\hatcurLCrhoeccenorigxxxxxE
\else
\ifnum#1=63 %
\hatcurLCrhoeccenorigxxxxxF
\else
\ifnum#1=64 %
\hatcurLCrhoeccenorigxxxxxG
\else
??????\fi
\fi
\fi
\fi
\fi
\fi
\fi
}
\newcommand{\hatcurLCrprstareccenorig}[1]{\ifnum#1=58 %
\hatcurLCrprstareccenorigxxxxxA
\else
\ifnum#1=59 %
\hatcurLCrprstareccenorigxxxxxB
\else
\ifnum#1=60 %
\hatcurLCrprstareccenorigxxxxxC
\else
\ifnum#1=61 %
\hatcurLCrprstareccenorigxxxxxD
\else
\ifnum#1=62 %
\hatcurLCrprstareccenorigxxxxxE
\else
\ifnum#1=63 %
\hatcurLCrprstareccenorigxxxxxF
\else
\ifnum#1=64 %
\hatcurLCrprstareccenorigxxxxxG
\else
??????\fi
\fi
\fi
\fi
\fi
\fi
\fi
}
\newcommand{\hatcurLCTAeccenorig}[1]{\ifnum#1=58 %
\hatcurLCTAeccenorigxxxxxA
\else
\ifnum#1=59 %
\hatcurLCTAeccenorigxxxxxB
\else
\ifnum#1=60 %
\hatcurLCTAeccenorigxxxxxC
\else
\ifnum#1=61 %
\hatcurLCTAeccenorigxxxxxD
\else
\ifnum#1=62 %
\hatcurLCTAeccenorigxxxxxE
\else
\ifnum#1=63 %
\hatcurLCTAeccenorigxxxxxF
\else
\ifnum#1=64 %
\hatcurLCTAeccenorigxxxxxG
\else
??????\fi
\fi
\fi
\fi
\fi
\fi
\fi
}
\newcommand{\hatcurLCTBeccenorig}[1]{\ifnum#1=58 %
\hatcurLCTBeccenorigxxxxxA
\else
\ifnum#1=59 %
\hatcurLCTBeccenorigxxxxxB
\else
\ifnum#1=60 %
\hatcurLCTBeccenorigxxxxxC
\else
\ifnum#1=61 %
\hatcurLCTBeccenorigxxxxxD
\else
\ifnum#1=62 %
\hatcurLCTBeccenorigxxxxxE
\else
\ifnum#1=63 %
\hatcurLCTBeccenorigxxxxxF
\else
\ifnum#1=64 %
\hatcurLCTBeccenorigxxxxxG
\else
??????\fi
\fi
\fi
\fi
\fi
\fi
\fi
}
\newcommand{\hatcurLCTeccenorig}[1]{\ifnum#1=58 %
\hatcurLCTeccenorigxxxxxA
\else
\ifnum#1=59 %
\hatcurLCTeccenorigxxxxxB
\else
\ifnum#1=60 %
\hatcurLCTeccenorigxxxxxC
\else
\ifnum#1=61 %
\hatcurLCTeccenorigxxxxxD
\else
\ifnum#1=62 %
\hatcurLCTeccenorigxxxxxE
\else
\ifnum#1=63 %
\hatcurLCTeccenorigxxxxxF
\else
\ifnum#1=64 %
\hatcurLCTeccenorigxxxxxG
\else
??????\fi
\fi
\fi
\fi
\fi
\fi
\fi
}
\newcommand{\hatcurLCzetaeccenorig}[1]{\ifnum#1=58 %
\hatcurLCzetaeccenorigxxxxxA
\else
\ifnum#1=59 %
\hatcurLCzetaeccenorigxxxxxB
\else
\ifnum#1=60 %
\hatcurLCzetaeccenorigxxxxxC
\else
\ifnum#1=61 %
\hatcurLCzetaeccenorigxxxxxD
\else
\ifnum#1=62 %
\hatcurLCzetaeccenorigxxxxxE
\else
\ifnum#1=63 %
\hatcurLCzetaeccenorigxxxxxF
\else
\ifnum#1=64 %
\hatcurLCzetaeccenorigxxxxxG
\else
??????\fi
\fi
\fi
\fi
\fi
\fi
\fi
}
\newcommand{\hatcurPPaequiveccenorig}[1]{\ifnum#1=58 %
\hatcurPPaequiveccenorigxxxxxA
\else
\ifnum#1=59 %
\hatcurPPaequiveccenorigxxxxxB
\else
\ifnum#1=60 %
\hatcurPPaequiveccenorigxxxxxC
\else
\ifnum#1=61 %
\hatcurPPaequiveccenorigxxxxxD
\else
\ifnum#1=62 %
\hatcurPPaequiveccenorigxxxxxE
\else
\ifnum#1=63 %
\hatcurPPaequiveccenorigxxxxxF
\else
\ifnum#1=64 %
\hatcurPPaequiveccenorigxxxxxG
\else
??????\fi
\fi
\fi
\fi
\fi
\fi
\fi
}
\newcommand{\hatcurPPareccenorig}[1]{\ifnum#1=58 %
\hatcurPPareccenorigxxxxxA
\else
\ifnum#1=59 %
\hatcurPPareccenorigxxxxxB
\else
\ifnum#1=60 %
\hatcurPPareccenorigxxxxxC
\else
\ifnum#1=61 %
\hatcurPPareccenorigxxxxxD
\else
\ifnum#1=62 %
\hatcurPPareccenorigxxxxxE
\else
\ifnum#1=63 %
\hatcurPPareccenorigxxxxxF
\else
\ifnum#1=64 %
\hatcurPPareccenorigxxxxxG
\else
??????\fi
\fi
\fi
\fi
\fi
\fi
\fi
}
\newcommand{\hatcurPPareleccenorig}[1]{\ifnum#1=58 %
\hatcurPPareleccenorigxxxxxA
\else
\ifnum#1=59 %
\hatcurPPareleccenorigxxxxxB
\else
\ifnum#1=60 %
\hatcurPPareleccenorigxxxxxC
\else
\ifnum#1=61 %
\hatcurPPareleccenorigxxxxxD
\else
\ifnum#1=62 %
\hatcurPPareleccenorigxxxxxE
\else
\ifnum#1=63 %
\hatcurPPareleccenorigxxxxxF
\else
\ifnum#1=64 %
\hatcurPPareleccenorigxxxxxG
\else
??????\fi
\fi
\fi
\fi
\fi
\fi
\fi
}
\newcommand{\hatcurPPfluxapdimeccenorig}[1]{\ifnum#1=58 %
\hatcurPPfluxapdimeccenorigxxxxxA
\else
\ifnum#1=59 %
\hatcurPPfluxapdimeccenorigxxxxxB
\else
\ifnum#1=60 %
\hatcurPPfluxapdimeccenorigxxxxxC
\else
\ifnum#1=61 %
\hatcurPPfluxapdimeccenorigxxxxxD
\else
\ifnum#1=62 %
\hatcurPPfluxapdimeccenorigxxxxxE
\else
\ifnum#1=63 %
\hatcurPPfluxapdimeccenorigxxxxxF
\else
\ifnum#1=64 %
\hatcurPPfluxapdimeccenorigxxxxxG
\else
??????\fi
\fi
\fi
\fi
\fi
\fi
\fi
}
\newcommand{\hatcurPPfluxapeccenorig}[1]{\ifnum#1=58 %
\hatcurPPfluxapeccenorigxxxxxA
\else
\ifnum#1=59 %
\hatcurPPfluxapeccenorigxxxxxB
\else
\ifnum#1=60 %
\hatcurPPfluxapeccenorigxxxxxC
\else
\ifnum#1=61 %
\hatcurPPfluxapeccenorigxxxxxD
\else
\ifnum#1=62 %
\hatcurPPfluxapeccenorigxxxxxE
\else
\ifnum#1=63 %
\hatcurPPfluxapeccenorigxxxxxF
\else
\ifnum#1=64 %
\hatcurPPfluxapeccenorigxxxxxG
\else
??????\fi
\fi
\fi
\fi
\fi
\fi
\fi
}
\newcommand{\hatcurPPfluxavgdimeccenorig}[1]{\ifnum#1=58 %
\hatcurPPfluxavgdimeccenorigxxxxxA
\else
\ifnum#1=59 %
\hatcurPPfluxavgdimeccenorigxxxxxB
\else
\ifnum#1=60 %
\hatcurPPfluxavgdimeccenorigxxxxxC
\else
\ifnum#1=61 %
\hatcurPPfluxavgdimeccenorigxxxxxD
\else
\ifnum#1=62 %
\hatcurPPfluxavgdimeccenorigxxxxxE
\else
\ifnum#1=63 %
\hatcurPPfluxavgdimeccenorigxxxxxF
\else
\ifnum#1=64 %
\hatcurPPfluxavgdimeccenorigxxxxxG
\else
??????\fi
\fi
\fi
\fi
\fi
\fi
\fi
}
\newcommand{\hatcurPPfluxavgeccenorig}[1]{\ifnum#1=58 %
\hatcurPPfluxavgeccenorigxxxxxA
\else
\ifnum#1=59 %
\hatcurPPfluxavgeccenorigxxxxxB
\else
\ifnum#1=60 %
\hatcurPPfluxavgeccenorigxxxxxC
\else
\ifnum#1=61 %
\hatcurPPfluxavgeccenorigxxxxxD
\else
\ifnum#1=62 %
\hatcurPPfluxavgeccenorigxxxxxE
\else
\ifnum#1=63 %
\hatcurPPfluxavgeccenorigxxxxxF
\else
\ifnum#1=64 %
\hatcurPPfluxavgeccenorigxxxxxG
\else
??????\fi
\fi
\fi
\fi
\fi
\fi
\fi
}
\newcommand{\hatcurPPfluxavglogeccenorig}[1]{\ifnum#1=58 %
\hatcurPPfluxavglogeccenorigxxxxxA
\else
\ifnum#1=59 %
\hatcurPPfluxavglogeccenorigxxxxxB
\else
\ifnum#1=60 %
\hatcurPPfluxavglogeccenorigxxxxxC
\else
\ifnum#1=61 %
\hatcurPPfluxavglogeccenorigxxxxxD
\else
\ifnum#1=62 %
\hatcurPPfluxavglogeccenorigxxxxxE
\else
\ifnum#1=63 %
\hatcurPPfluxavglogeccenorigxxxxxF
\else
\ifnum#1=64 %
\hatcurPPfluxavglogeccenorigxxxxxG
\else
??????\fi
\fi
\fi
\fi
\fi
\fi
\fi
}
\newcommand{\hatcurPPfluxperidimeccenorig}[1]{\ifnum#1=58 %
\hatcurPPfluxperidimeccenorigxxxxxA
\else
\ifnum#1=59 %
\hatcurPPfluxperidimeccenorigxxxxxB
\else
\ifnum#1=60 %
\hatcurPPfluxperidimeccenorigxxxxxC
\else
\ifnum#1=61 %
\hatcurPPfluxperidimeccenorigxxxxxD
\else
\ifnum#1=62 %
\hatcurPPfluxperidimeccenorigxxxxxE
\else
\ifnum#1=63 %
\hatcurPPfluxperidimeccenorigxxxxxF
\else
\ifnum#1=64 %
\hatcurPPfluxperidimeccenorigxxxxxG
\else
??????\fi
\fi
\fi
\fi
\fi
\fi
\fi
}
\newcommand{\hatcurPPfluxperieccenorig}[1]{\ifnum#1=58 %
\hatcurPPfluxperieccenorigxxxxxA
\else
\ifnum#1=59 %
\hatcurPPfluxperieccenorigxxxxxB
\else
\ifnum#1=60 %
\hatcurPPfluxperieccenorigxxxxxC
\else
\ifnum#1=61 %
\hatcurPPfluxperieccenorigxxxxxD
\else
\ifnum#1=62 %
\hatcurPPfluxperieccenorigxxxxxE
\else
\ifnum#1=63 %
\hatcurPPfluxperieccenorigxxxxxF
\else
\ifnum#1=64 %
\hatcurPPfluxperieccenorigxxxxxG
\else
??????\fi
\fi
\fi
\fi
\fi
\fi
\fi
}
\newcommand{\hatcurPPgeccenorig}[1]{\ifnum#1=58 %
\hatcurPPgeccenorigxxxxxA
\else
\ifnum#1=59 %
\hatcurPPgeccenorigxxxxxB
\else
\ifnum#1=60 %
\hatcurPPgeccenorigxxxxxC
\else
\ifnum#1=61 %
\hatcurPPgeccenorigxxxxxD
\else
\ifnum#1=62 %
\hatcurPPgeccenorigxxxxxE
\else
\ifnum#1=63 %
\hatcurPPgeccenorigxxxxxF
\else
\ifnum#1=64 %
\hatcurPPgeccenorigxxxxxG
\else
??????\fi
\fi
\fi
\fi
\fi
\fi
\fi
}
\newcommand{\hatcurPPieccenorig}[1]{\ifnum#1=58 %
\hatcurPPieccenorigxxxxxA
\else
\ifnum#1=59 %
\hatcurPPieccenorigxxxxxB
\else
\ifnum#1=60 %
\hatcurPPieccenorigxxxxxC
\else
\ifnum#1=61 %
\hatcurPPieccenorigxxxxxD
\else
\ifnum#1=62 %
\hatcurPPieccenorigxxxxxE
\else
\ifnum#1=63 %
\hatcurPPieccenorigxxxxxF
\else
\ifnum#1=64 %
\hatcurPPieccenorigxxxxxG
\else
??????\fi
\fi
\fi
\fi
\fi
\fi
\fi
}
\newcommand{\hatcurPPloggeccenorig}[1]{\ifnum#1=58 %
\hatcurPPloggeccenorigxxxxxA
\else
\ifnum#1=59 %
\hatcurPPloggeccenorigxxxxxB
\else
\ifnum#1=60 %
\hatcurPPloggeccenorigxxxxxC
\else
\ifnum#1=61 %
\hatcurPPloggeccenorigxxxxxD
\else
\ifnum#1=62 %
\hatcurPPloggeccenorigxxxxxE
\else
\ifnum#1=63 %
\hatcurPPloggeccenorigxxxxxF
\else
\ifnum#1=64 %
\hatcurPPloggeccenorigxxxxxG
\else
??????\fi
\fi
\fi
\fi
\fi
\fi
\fi
}
\newcommand{\hatcurPPmeccenorig}[1]{\ifnum#1=58 %
\hatcurPPmeccenorigxxxxxA
\else
\ifnum#1=59 %
\hatcurPPmeccenorigxxxxxB
\else
\ifnum#1=60 %
\hatcurPPmeccenorigxxxxxC
\else
\ifnum#1=61 %
\hatcurPPmeccenorigxxxxxD
\else
\ifnum#1=62 %
\hatcurPPmeccenorigxxxxxE
\else
\ifnum#1=63 %
\hatcurPPmeccenorigxxxxxF
\else
\ifnum#1=64 %
\hatcurPPmeccenorigxxxxxG
\else
??????\fi
\fi
\fi
\fi
\fi
\fi
\fi
}
\newcommand{\hatcurPPmeeccenorig}[1]{\ifnum#1=58 %
\hatcurPPmeeccenorigxxxxxA
\else
\ifnum#1=59 %
\hatcurPPmeeccenorigxxxxxB
\else
\ifnum#1=60 %
\hatcurPPmeeccenorigxxxxxC
\else
\ifnum#1=61 %
\hatcurPPmeeccenorigxxxxxD
\else
\ifnum#1=62 %
\hatcurPPmeeccenorigxxxxxE
\else
\ifnum#1=63 %
\hatcurPPmeeccenorigxxxxxF
\else
\ifnum#1=64 %
\hatcurPPmeeccenorigxxxxxG
\else
??????\fi
\fi
\fi
\fi
\fi
\fi
\fi
}
\newcommand{\hatcurPPmelongeccenorig}[1]{\ifnum#1=58 %
\hatcurPPmelongeccenorigxxxxxA
\else
\ifnum#1=59 %
\hatcurPPmelongeccenorigxxxxxB
\else
\ifnum#1=60 %
\hatcurPPmelongeccenorigxxxxxC
\else
\ifnum#1=61 %
\hatcurPPmelongeccenorigxxxxxD
\else
\ifnum#1=62 %
\hatcurPPmelongeccenorigxxxxxE
\else
\ifnum#1=63 %
\hatcurPPmelongeccenorigxxxxxF
\else
\ifnum#1=64 %
\hatcurPPmelongeccenorigxxxxxG
\else
??????\fi
\fi
\fi
\fi
\fi
\fi
\fi
}
\newcommand{\hatcurPPmeshorteccenorig}[1]{\ifnum#1=58 %
\hatcurPPmeshorteccenorigxxxxxA
\else
\ifnum#1=59 %
\hatcurPPmeshorteccenorigxxxxxB
\else
\ifnum#1=60 %
\hatcurPPmeshorteccenorigxxxxxC
\else
\ifnum#1=61 %
\hatcurPPmeshorteccenorigxxxxxD
\else
\ifnum#1=62 %
\hatcurPPmeshorteccenorigxxxxxE
\else
\ifnum#1=63 %
\hatcurPPmeshorteccenorigxxxxxF
\else
\ifnum#1=64 %
\hatcurPPmeshorteccenorigxxxxxG
\else
??????\fi
\fi
\fi
\fi
\fi
\fi
\fi
}
\newcommand{\hatcurPPmlongeccenorig}[1]{\ifnum#1=58 %
\hatcurPPmlongeccenorigxxxxxA
\else
\ifnum#1=59 %
\hatcurPPmlongeccenorigxxxxxB
\else
\ifnum#1=60 %
\hatcurPPmlongeccenorigxxxxxC
\else
\ifnum#1=61 %
\hatcurPPmlongeccenorigxxxxxD
\else
\ifnum#1=62 %
\hatcurPPmlongeccenorigxxxxxE
\else
\ifnum#1=63 %
\hatcurPPmlongeccenorigxxxxxF
\else
\ifnum#1=64 %
\hatcurPPmlongeccenorigxxxxxG
\else
??????\fi
\fi
\fi
\fi
\fi
\fi
\fi
}
\newcommand{\hatcurPPmrcorreccenorig}[1]{\ifnum#1=58 %
\hatcurPPmrcorreccenorigxxxxxA
\else
\ifnum#1=59 %
\hatcurPPmrcorreccenorigxxxxxB
\else
\ifnum#1=60 %
\hatcurPPmrcorreccenorigxxxxxC
\else
\ifnum#1=61 %
\hatcurPPmrcorreccenorigxxxxxD
\else
\ifnum#1=62 %
\hatcurPPmrcorreccenorigxxxxxE
\else
\ifnum#1=63 %
\hatcurPPmrcorreccenorigxxxxxF
\else
\ifnum#1=64 %
\hatcurPPmrcorreccenorigxxxxxG
\else
??????\fi
\fi
\fi
\fi
\fi
\fi
\fi
}
\newcommand{\hatcurPPmshorteccenorig}[1]{\ifnum#1=58 %
\hatcurPPmshorteccenorigxxxxxA
\else
\ifnum#1=59 %
\hatcurPPmshorteccenorigxxxxxB
\else
\ifnum#1=60 %
\hatcurPPmshorteccenorigxxxxxC
\else
\ifnum#1=61 %
\hatcurPPmshorteccenorigxxxxxD
\else
\ifnum#1=62 %
\hatcurPPmshorteccenorigxxxxxE
\else
\ifnum#1=63 %
\hatcurPPmshorteccenorigxxxxxF
\else
\ifnum#1=64 %
\hatcurPPmshorteccenorigxxxxxG
\else
??????\fi
\fi
\fi
\fi
\fi
\fi
\fi
}
\newcommand{\hatcurPPperieccenorig}[1]{\ifnum#1=58 %
\hatcurPPperieccenorigxxxxxA
\else
\ifnum#1=59 %
\hatcurPPperieccenorigxxxxxB
\else
\ifnum#1=60 %
\hatcurPPperieccenorigxxxxxC
\else
\ifnum#1=61 %
\hatcurPPperieccenorigxxxxxD
\else
\ifnum#1=62 %
\hatcurPPperieccenorigxxxxxE
\else
\ifnum#1=63 %
\hatcurPPperieccenorigxxxxxF
\else
\ifnum#1=64 %
\hatcurPPperieccenorigxxxxxG
\else
??????\fi
\fi
\fi
\fi
\fi
\fi
\fi
}
\newcommand{\hatcurPPphiconjeccenorig}[1]{\ifnum#1=58 %
\hatcurPPphiconjeccenorigxxxxxA
\else
\ifnum#1=59 %
\hatcurPPphiconjeccenorigxxxxxB
\else
\ifnum#1=60 %
\hatcurPPphiconjeccenorigxxxxxC
\else
\ifnum#1=61 %
\hatcurPPphiconjeccenorigxxxxxD
\else
\ifnum#1=62 %
\hatcurPPphiconjeccenorigxxxxxE
\else
\ifnum#1=63 %
\hatcurPPphiconjeccenorigxxxxxF
\else
\ifnum#1=64 %
\hatcurPPphiconjeccenorigxxxxxG
\else
??????\fi
\fi
\fi
\fi
\fi
\fi
\fi
}
\newcommand{\hatcurPPreccenorig}[1]{\ifnum#1=58 %
\hatcurPPreccenorigxxxxxA
\else
\ifnum#1=59 %
\hatcurPPreccenorigxxxxxB
\else
\ifnum#1=60 %
\hatcurPPreccenorigxxxxxC
\else
\ifnum#1=61 %
\hatcurPPreccenorigxxxxxD
\else
\ifnum#1=62 %
\hatcurPPreccenorigxxxxxE
\else
\ifnum#1=63 %
\hatcurPPreccenorigxxxxxF
\else
\ifnum#1=64 %
\hatcurPPreccenorigxxxxxG
\else
??????\fi
\fi
\fi
\fi
\fi
\fi
\fi
}
\newcommand{\hatcurPPreeccenorig}[1]{\ifnum#1=58 %
\hatcurPPreeccenorigxxxxxA
\else
\ifnum#1=59 %
\hatcurPPreeccenorigxxxxxB
\else
\ifnum#1=60 %
\hatcurPPreeccenorigxxxxxC
\else
\ifnum#1=61 %
\hatcurPPreeccenorigxxxxxD
\else
\ifnum#1=62 %
\hatcurPPreeccenorigxxxxxE
\else
\ifnum#1=63 %
\hatcurPPreeccenorigxxxxxF
\else
\ifnum#1=64 %
\hatcurPPreeccenorigxxxxxG
\else
??????\fi
\fi
\fi
\fi
\fi
\fi
\fi
}
\newcommand{\hatcurPPrelongeccenorig}[1]{\ifnum#1=58 %
\hatcurPPrelongeccenorigxxxxxA
\else
\ifnum#1=59 %
\hatcurPPrelongeccenorigxxxxxB
\else
\ifnum#1=60 %
\hatcurPPrelongeccenorigxxxxxC
\else
\ifnum#1=61 %
\hatcurPPrelongeccenorigxxxxxD
\else
\ifnum#1=62 %
\hatcurPPrelongeccenorigxxxxxE
\else
\ifnum#1=63 %
\hatcurPPrelongeccenorigxxxxxF
\else
\ifnum#1=64 %
\hatcurPPrelongeccenorigxxxxxG
\else
??????\fi
\fi
\fi
\fi
\fi
\fi
\fi
}
\newcommand{\hatcurPPreshorteccenorig}[1]{\ifnum#1=58 %
\hatcurPPreshorteccenorigxxxxxA
\else
\ifnum#1=59 %
\hatcurPPreshorteccenorigxxxxxB
\else
\ifnum#1=60 %
\hatcurPPreshorteccenorigxxxxxC
\else
\ifnum#1=61 %
\hatcurPPreshorteccenorigxxxxxD
\else
\ifnum#1=62 %
\hatcurPPreshorteccenorigxxxxxE
\else
\ifnum#1=63 %
\hatcurPPreshorteccenorigxxxxxF
\else
\ifnum#1=64 %
\hatcurPPreshorteccenorigxxxxxG
\else
??????\fi
\fi
\fi
\fi
\fi
\fi
\fi
}
\newcommand{\hatcurPPrhoeccenorig}[1]{\ifnum#1=58 %
\hatcurPPrhoeccenorigxxxxxA
\else
\ifnum#1=59 %
\hatcurPPrhoeccenorigxxxxxB
\else
\ifnum#1=60 %
\hatcurPPrhoeccenorigxxxxxC
\else
\ifnum#1=61 %
\hatcurPPrhoeccenorigxxxxxD
\else
\ifnum#1=62 %
\hatcurPPrhoeccenorigxxxxxE
\else
\ifnum#1=63 %
\hatcurPPrhoeccenorigxxxxxF
\else
\ifnum#1=64 %
\hatcurPPrhoeccenorigxxxxxG
\else
??????\fi
\fi
\fi
\fi
\fi
\fi
\fi
}
\newcommand{\hatcurPPrlongeccenorig}[1]{\ifnum#1=58 %
\hatcurPPrlongeccenorigxxxxxA
\else
\ifnum#1=59 %
\hatcurPPrlongeccenorigxxxxxB
\else
\ifnum#1=60 %
\hatcurPPrlongeccenorigxxxxxC
\else
\ifnum#1=61 %
\hatcurPPrlongeccenorigxxxxxD
\else
\ifnum#1=62 %
\hatcurPPrlongeccenorigxxxxxE
\else
\ifnum#1=63 %
\hatcurPPrlongeccenorigxxxxxF
\else
\ifnum#1=64 %
\hatcurPPrlongeccenorigxxxxxG
\else
??????\fi
\fi
\fi
\fi
\fi
\fi
\fi
}
\newcommand{\hatcurPPrshorteccenorig}[1]{\ifnum#1=58 %
\hatcurPPrshorteccenorigxxxxxA
\else
\ifnum#1=59 %
\hatcurPPrshorteccenorigxxxxxB
\else
\ifnum#1=60 %
\hatcurPPrshorteccenorigxxxxxC
\else
\ifnum#1=61 %
\hatcurPPrshorteccenorigxxxxxD
\else
\ifnum#1=62 %
\hatcurPPrshorteccenorigxxxxxE
\else
\ifnum#1=63 %
\hatcurPPrshorteccenorigxxxxxF
\else
\ifnum#1=64 %
\hatcurPPrshorteccenorigxxxxxG
\else
??????\fi
\fi
\fi
\fi
\fi
\fi
\fi
}
\newcommand{\hatcurPPtcirceccenorig}[1]{\ifnum#1=58 %
\hatcurPPtcirceccenorigxxxxxA
\else
\ifnum#1=59 %
\hatcurPPtcirceccenorigxxxxxB
\else
\ifnum#1=60 %
\hatcurPPtcirceccenorigxxxxxC
\else
\ifnum#1=61 %
\hatcurPPtcirceccenorigxxxxxD
\else
\ifnum#1=62 %
\hatcurPPtcirceccenorigxxxxxE
\else
\ifnum#1=63 %
\hatcurPPtcirceccenorigxxxxxF
\else
\ifnum#1=64 %
\hatcurPPtcirceccenorigxxxxxG
\else
??????\fi
\fi
\fi
\fi
\fi
\fi
\fi
}
\newcommand{\hatcurPPteffeccenorig}[1]{\ifnum#1=58 %
\hatcurPPteffeccenorigxxxxxA
\else
\ifnum#1=59 %
\hatcurPPteffeccenorigxxxxxB
\else
\ifnum#1=60 %
\hatcurPPteffeccenorigxxxxxC
\else
\ifnum#1=61 %
\hatcurPPteffeccenorigxxxxxD
\else
\ifnum#1=62 %
\hatcurPPteffeccenorigxxxxxE
\else
\ifnum#1=63 %
\hatcurPPteffeccenorigxxxxxF
\else
\ifnum#1=64 %
\hatcurPPteffeccenorigxxxxxG
\else
??????\fi
\fi
\fi
\fi
\fi
\fi
\fi
}
\newcommand{\hatcurPPthetaeccenorig}[1]{\ifnum#1=58 %
\hatcurPPthetaeccenorigxxxxxA
\else
\ifnum#1=59 %
\hatcurPPthetaeccenorigxxxxxB
\else
\ifnum#1=60 %
\hatcurPPthetaeccenorigxxxxxC
\else
\ifnum#1=61 %
\hatcurPPthetaeccenorigxxxxxD
\else
\ifnum#1=62 %
\hatcurPPthetaeccenorigxxxxxE
\else
\ifnum#1=63 %
\hatcurPPthetaeccenorigxxxxxF
\else
\ifnum#1=64 %
\hatcurPPthetaeccenorigxxxxxG
\else
??????\fi
\fi
\fi
\fi
\fi
\fi
\fi
}
\newcommand{\hatcurPPtinfalleccenorig}[1]{\ifnum#1=58 %
\hatcurPPtinfalleccenorigxxxxxA
\else
\ifnum#1=59 %
\hatcurPPtinfalleccenorigxxxxxB
\else
\ifnum#1=60 %
\hatcurPPtinfalleccenorigxxxxxC
\else
\ifnum#1=61 %
\hatcurPPtinfalleccenorigxxxxxD
\else
\ifnum#1=62 %
\hatcurPPtinfalleccenorigxxxxxE
\else
\ifnum#1=63 %
\hatcurPPtinfalleccenorigxxxxxF
\else
\ifnum#1=64 %
\hatcurPPtinfalleccenorigxxxxxG
\else
??????\fi
\fi
\fi
\fi
\fi
\fi
\fi
}
\newcommand{\hatcurRVecceneccenorig}[1]{\ifnum#1=58 %
\hatcurRVecceneccenorigxxxxxA
\else
\ifnum#1=59 %
\hatcurRVecceneccenorigxxxxxB
\else
\ifnum#1=60 %
\hatcurRVecceneccenorigxxxxxC
\else
\ifnum#1=61 %
\hatcurRVecceneccenorigxxxxxD
\else
\ifnum#1=62 %
\hatcurRVecceneccenorigxxxxxE
\else
\ifnum#1=63 %
\hatcurRVecceneccenorigxxxxxF
\else
\ifnum#1=64 %
\hatcurRVecceneccenorigxxxxxG
\else
??????\fi
\fi
\fi
\fi
\fi
\fi
\fi
}
\newcommand{\hatcurRVeccentwosiglimeccenorig}[1]{\ifnum#1=58 %
\hatcurRVeccentwosiglimeccenorigxxxxxA
\else
\ifnum#1=59 %
\hatcurRVeccentwosiglimeccenorigxxxxxB
\else
\ifnum#1=60 %
\hatcurRVeccentwosiglimeccenorigxxxxxC
\else
\ifnum#1=61 %
\hatcurRVeccentwosiglimeccenorigxxxxxD
\else
\ifnum#1=62 %
\hatcurRVeccentwosiglimeccenorigxxxxxE
\else
\ifnum#1=63 %
\hatcurRVeccentwosiglimeccenorigxxxxxF
\else
\ifnum#1=64 %
\hatcurRVeccentwosiglimeccenorigxxxxxG
\else
??????\fi
\fi
\fi
\fi
\fi
\fi
\fi
}
\newcommand{\hatcurRVfitrmsAeccenorig}[1]{\ifnum#1=59 %
\hatcurRVfitrmsAeccenorigxxxxxB
\else
\ifnum#1=60 %
\hatcurRVfitrmsAeccenorigxxxxxC
\else
\ifnum#1=61 %
\hatcurRVfitrmsAeccenorigxxxxxD
\else
\ifnum#1=63 %
\hatcurRVfitrmsAeccenorigxxxxxF
\else
\ifnum#1=64 %
\hatcurRVfitrmsAeccenorigxxxxxG
\else
??????\fi
\fi
\fi
\fi
\fi
}
\newcommand{\hatcurRVfitrmsBeccenorig}[1]{\ifnum#1=59 %
\hatcurRVfitrmsBeccenorigxxxxxB
\else
\ifnum#1=60 %
\hatcurRVfitrmsBeccenorigxxxxxC
\else
\ifnum#1=61 %
\hatcurRVfitrmsBeccenorigxxxxxD
\else
\ifnum#1=63 %
\hatcurRVfitrmsBeccenorigxxxxxF
\else
\ifnum#1=64 %
\hatcurRVfitrmsBeccenorigxxxxxG
\else
??????\fi
\fi
\fi
\fi
\fi
}
\newcommand{\hatcurRVfitrmsCeccenorig}[1]{\ifnum#1=60 %
\hatcurRVfitrmsCeccenorigxxxxxC
\else
\ifnum#1=63 %
\hatcurRVfitrmsCeccenorigxxxxxF
\else
??????\fi
\fi
}
\newcommand{\hatcurRVfitrmseccenorig}[1]{\ifnum#1=58 %
\hatcurRVfitrmseccenorigxxxxxA
\else
\ifnum#1=62 %
\hatcurRVfitrmseccenorigxxxxxE
\else
??????\fi
\fi
}
\newcommand{\hatcurRVgammaAeccenorig}[1]{\ifnum#1=59 %
\hatcurRVgammaAeccenorigxxxxxB
\else
\ifnum#1=60 %
\hatcurRVgammaAeccenorigxxxxxC
\else
\ifnum#1=61 %
\hatcurRVgammaAeccenorigxxxxxD
\else
\ifnum#1=63 %
\hatcurRVgammaAeccenorigxxxxxF
\else
\ifnum#1=64 %
\hatcurRVgammaAeccenorigxxxxxG
\else
??????\fi
\fi
\fi
\fi
\fi
}
\newcommand{\hatcurRVgammaBeccenorig}[1]{\ifnum#1=59 %
\hatcurRVgammaBeccenorigxxxxxB
\else
\ifnum#1=60 %
\hatcurRVgammaBeccenorigxxxxxC
\else
\ifnum#1=61 %
\hatcurRVgammaBeccenorigxxxxxD
\else
\ifnum#1=63 %
\hatcurRVgammaBeccenorigxxxxxF
\else
\ifnum#1=64 %
\hatcurRVgammaBeccenorigxxxxxG
\else
??????\fi
\fi
\fi
\fi
\fi
}
\newcommand{\hatcurRVgammaCeccenorig}[1]{\ifnum#1=60 %
\hatcurRVgammaCeccenorigxxxxxC
\else
\ifnum#1=63 %
\hatcurRVgammaCeccenorigxxxxxF
\else
??????\fi
\fi
}
\newcommand{\hatcurRVgammaeccenorig}[1]{\ifnum#1=58 %
\hatcurRVgammaeccenorigxxxxxA
\else
\ifnum#1=62 %
\hatcurRVgammaeccenorigxxxxxE
\else
??????\fi
\fi
}
\newcommand{\hatcurRVheccenorig}[1]{\ifnum#1=58 %
\hatcurRVheccenorigxxxxxA
\else
\ifnum#1=59 %
\hatcurRVheccenorigxxxxxB
\else
\ifnum#1=60 %
\hatcurRVheccenorigxxxxxC
\else
\ifnum#1=61 %
\hatcurRVheccenorigxxxxxD
\else
\ifnum#1=62 %
\hatcurRVheccenorigxxxxxE
\else
\ifnum#1=63 %
\hatcurRVheccenorigxxxxxF
\else
\ifnum#1=64 %
\hatcurRVheccenorigxxxxxG
\else
??????\fi
\fi
\fi
\fi
\fi
\fi
\fi
}
\newcommand{\hatcurRVjitterAeccenorig}[1]{\ifnum#1=59 %
\hatcurRVjitterAeccenorigxxxxxB
\else
\ifnum#1=60 %
\hatcurRVjitterAeccenorigxxxxxC
\else
\ifnum#1=61 %
\hatcurRVjitterAeccenorigxxxxxD
\else
\ifnum#1=63 %
\hatcurRVjitterAeccenorigxxxxxF
\else
\ifnum#1=64 %
\hatcurRVjitterAeccenorigxxxxxG
\else
??????\fi
\fi
\fi
\fi
\fi
}
\newcommand{\hatcurRVjitterBeccenorig}[1]{\ifnum#1=59 %
\hatcurRVjitterBeccenorigxxxxxB
\else
\ifnum#1=60 %
\hatcurRVjitterBeccenorigxxxxxC
\else
\ifnum#1=61 %
\hatcurRVjitterBeccenorigxxxxxD
\else
\ifnum#1=63 %
\hatcurRVjitterBeccenorigxxxxxF
\else
\ifnum#1=64 %
\hatcurRVjitterBeccenorigxxxxxG
\else
??????\fi
\fi
\fi
\fi
\fi
}
\newcommand{\hatcurRVjitterCeccenorig}[1]{\ifnum#1=60 %
\hatcurRVjitterCeccenorigxxxxxC
\else
\ifnum#1=63 %
\hatcurRVjitterCeccenorigxxxxxF
\else
??????\fi
\fi
}
\newcommand{\hatcurRVjittereccenorig}[1]{\ifnum#1=58 %
\hatcurRVjittereccenorigxxxxxA
\else
\ifnum#1=62 %
\hatcurRVjittereccenorigxxxxxE
\else
??????\fi
\fi
}
\newcommand{\hatcurRVjittertwosiglimAeccenorig}[1]{\ifnum#1=59 %
\hatcurRVjittertwosiglimAeccenorigxxxxxB
\else
\ifnum#1=60 %
\hatcurRVjittertwosiglimAeccenorigxxxxxC
\else
\ifnum#1=61 %
\hatcurRVjittertwosiglimAeccenorigxxxxxD
\else
\ifnum#1=63 %
\hatcurRVjittertwosiglimAeccenorigxxxxxF
\else
\ifnum#1=64 %
\hatcurRVjittertwosiglimAeccenorigxxxxxG
\else
??????\fi
\fi
\fi
\fi
\fi
}
\newcommand{\hatcurRVjittertwosiglimBeccenorig}[1]{\ifnum#1=59 %
\hatcurRVjittertwosiglimBeccenorigxxxxxB
\else
\ifnum#1=60 %
\hatcurRVjittertwosiglimBeccenorigxxxxxC
\else
\ifnum#1=61 %
\hatcurRVjittertwosiglimBeccenorigxxxxxD
\else
\ifnum#1=63 %
\hatcurRVjittertwosiglimBeccenorigxxxxxF
\else
\ifnum#1=64 %
\hatcurRVjittertwosiglimBeccenorigxxxxxG
\else
??????\fi
\fi
\fi
\fi
\fi
}
\newcommand{\hatcurRVjittertwosiglimCeccenorig}[1]{\ifnum#1=60 %
\hatcurRVjittertwosiglimCeccenorigxxxxxC
\else
\ifnum#1=63 %
\hatcurRVjittertwosiglimCeccenorigxxxxxF
\else
??????\fi
\fi
}
\newcommand{\hatcurRVjittertwosiglimeccenorig}[1]{\ifnum#1=58 %
\hatcurRVjittertwosiglimeccenorigxxxxxA
\else
\ifnum#1=62 %
\hatcurRVjittertwosiglimeccenorigxxxxxE
\else
??????\fi
\fi
}
\newcommand{\hatcurRVkeccenorig}[1]{\ifnum#1=58 %
\hatcurRVkeccenorigxxxxxA
\else
\ifnum#1=59 %
\hatcurRVkeccenorigxxxxxB
\else
\ifnum#1=60 %
\hatcurRVkeccenorigxxxxxC
\else
\ifnum#1=61 %
\hatcurRVkeccenorigxxxxxD
\else
\ifnum#1=62 %
\hatcurRVkeccenorigxxxxxE
\else
\ifnum#1=63 %
\hatcurRVkeccenorigxxxxxF
\else
\ifnum#1=64 %
\hatcurRVkeccenorigxxxxxG
\else
??????\fi
\fi
\fi
\fi
\fi
\fi
\fi
}
\newcommand{\hatcurRVKeccenorig}[1]{\ifnum#1=58 %
\hatcurRVKeccenorigxxxxxA
\else
\ifnum#1=59 %
\hatcurRVKeccenorigxxxxxB
\else
\ifnum#1=60 %
\hatcurRVKeccenorigxxxxxC
\else
\ifnum#1=61 %
\hatcurRVKeccenorigxxxxxD
\else
\ifnum#1=62 %
\hatcurRVKeccenorigxxxxxE
\else
\ifnum#1=63 %
\hatcurRVKeccenorigxxxxxF
\else
\ifnum#1=64 %
\hatcurRVKeccenorigxxxxxG
\else
??????\fi
\fi
\fi
\fi
\fi
\fi
\fi
}
\newcommand{\hatcurRVomegaeccenorig}[1]{\ifnum#1=58 %
\hatcurRVomegaeccenorigxxxxxA
\else
\ifnum#1=59 %
\hatcurRVomegaeccenorigxxxxxB
\else
\ifnum#1=60 %
\hatcurRVomegaeccenorigxxxxxC
\else
\ifnum#1=61 %
\hatcurRVomegaeccenorigxxxxxD
\else
\ifnum#1=62 %
\hatcurRVomegaeccenorigxxxxxE
\else
\ifnum#1=63 %
\hatcurRVomegaeccenorigxxxxxF
\else
\ifnum#1=64 %
\hatcurRVomegaeccenorigxxxxxG
\else
??????\fi
\fi
\fi
\fi
\fi
\fi
\fi
}
\newcommand{\hatcurRVrheccenorig}[1]{\ifnum#1=58 %
\hatcurRVrheccenorigxxxxxA
\else
\ifnum#1=59 %
\hatcurRVrheccenorigxxxxxB
\else
\ifnum#1=60 %
\hatcurRVrheccenorigxxxxxC
\else
\ifnum#1=61 %
\hatcurRVrheccenorigxxxxxD
\else
\ifnum#1=62 %
\hatcurRVrheccenorigxxxxxE
\else
\ifnum#1=63 %
\hatcurRVrheccenorigxxxxxF
\else
\ifnum#1=64 %
\hatcurRVrheccenorigxxxxxG
\else
??????\fi
\fi
\fi
\fi
\fi
\fi
\fi
}
\newcommand{\hatcurRVrkeccenorig}[1]{\ifnum#1=58 %
\hatcurRVrkeccenorigxxxxxA
\else
\ifnum#1=59 %
\hatcurRVrkeccenorigxxxxxB
\else
\ifnum#1=60 %
\hatcurRVrkeccenorigxxxxxC
\else
\ifnum#1=61 %
\hatcurRVrkeccenorigxxxxxD
\else
\ifnum#1=62 %
\hatcurRVrkeccenorigxxxxxE
\else
\ifnum#1=63 %
\hatcurRVrkeccenorigxxxxxF
\else
\ifnum#1=64 %
\hatcurRVrkeccenorigxxxxxG
\else
??????\fi
\fi
\fi
\fi
\fi
\fi
\fi
}
\newcommand{\hatcurRVtroneeccenorig}[1]{\ifnum#1=58 %
\hatcurRVtroneeccenorigxxxxxA
\else
\ifnum#1=59 %
\hatcurRVtroneeccenorigxxxxxB
\else
\ifnum#1=60 %
\hatcurRVtroneeccenorigxxxxxC
\else
\ifnum#1=61 %
\hatcurRVtroneeccenorigxxxxxD
\else
\ifnum#1=62 %
\hatcurRVtroneeccenorigxxxxxE
\else
\ifnum#1=63 %
\hatcurRVtroneeccenorigxxxxxF
\else
\ifnum#1=64 %
\hatcurRVtroneeccenorigxxxxxG
\else
??????\fi
\fi
\fi
\fi
\fi
\fi
\fi
}
\newcommand{\hatcurRVtrtwoeccenorig}[1]{\ifnum#1=58 %
\hatcurRVtrtwoeccenorigxxxxxA
\else
\ifnum#1=59 %
\hatcurRVtrtwoeccenorigxxxxxB
\else
\ifnum#1=60 %
\hatcurRVtrtwoeccenorigxxxxxC
\else
\ifnum#1=61 %
\hatcurRVtrtwoeccenorigxxxxxD
\else
\ifnum#1=62 %
\hatcurRVtrtwoeccenorigxxxxxE
\else
\ifnum#1=63 %
\hatcurRVtrtwoeccenorigxxxxxF
\else
\ifnum#1=64 %
\hatcurRVtrtwoeccenorigxxxxxG
\else
??????\fi
\fi
\fi
\fi
\fi
\fi
\fi
}
\newcommand{\hatcurSMEiiloggeccenorig}[1]{\ifnum#1=58 %
\hatcurSMEiiloggeccenorigxxxxxA
\else
\ifnum#1=59 %
\hatcurSMEiiloggeccenorigxxxxxB
\else
\ifnum#1=60 %
\hatcurSMEiiloggeccenorigxxxxxC
\else
\ifnum#1=61 %
\hatcurSMEiiloggeccenorigxxxxxD
\else
\ifnum#1=62 %
\hatcurSMEiiloggeccenorigxxxxxE
\else
??????\fi
\fi
\fi
\fi
\fi
}
\newcommand{\hatcurSMEiiteffeccenorig}[1]{\ifnum#1=58 %
\hatcurSMEiiteffeccenorigxxxxxA
\else
\ifnum#1=59 %
\hatcurSMEiiteffeccenorigxxxxxB
\else
\ifnum#1=60 %
\hatcurSMEiiteffeccenorigxxxxxC
\else
\ifnum#1=61 %
\hatcurSMEiiteffeccenorigxxxxxD
\else
\ifnum#1=62 %
\hatcurSMEiiteffeccenorigxxxxxE
\else
??????\fi
\fi
\fi
\fi
\fi
}
\newcommand{\hatcurSMEiivsineccenorig}[1]{\ifnum#1=58 %
\hatcurSMEiivsineccenorigxxxxxA
\else
\ifnum#1=59 %
\hatcurSMEiivsineccenorigxxxxxB
\else
\ifnum#1=60 %
\hatcurSMEiivsineccenorigxxxxxC
\else
\ifnum#1=61 %
\hatcurSMEiivsineccenorigxxxxxD
\else
\ifnum#1=62 %
\hatcurSMEiivsineccenorigxxxxxE
\else
??????\fi
\fi
\fi
\fi
\fi
}
\newcommand{\hatcurSMEiizfeheccenorig}[1]{\ifnum#1=58 %
\hatcurSMEiizfeheccenorigxxxxxA
\else
\ifnum#1=59 %
\hatcurSMEiizfeheccenorigxxxxxB
\else
\ifnum#1=60 %
\hatcurSMEiizfeheccenorigxxxxxC
\else
\ifnum#1=61 %
\hatcurSMEiizfeheccenorigxxxxxD
\else
\ifnum#1=62 %
\hatcurSMEiizfeheccenorigxxxxxE
\else
??????\fi
\fi
\fi
\fi
\fi
}
\newcommand{\hatcurSMEiizfehshorteccenorig}[1]{\ifnum#1=58 %
\hatcurSMEiizfehshorteccenorigxxxxxA
\else
\ifnum#1=59 %
\hatcurSMEiizfehshorteccenorigxxxxxB
\else
\ifnum#1=60 %
\hatcurSMEiizfehshorteccenorigxxxxxC
\else
\ifnum#1=61 %
\hatcurSMEiizfehshorteccenorigxxxxxD
\else
\ifnum#1=62 %
\hatcurSMEiizfehshorteccenorigxxxxxE
\else
??????\fi
\fi
\fi
\fi
\fi
}
\newcommand{\hatcurSMEiloggeccenorig}[1]{\ifnum#1=58 %
\hatcurSMEiloggeccenorigxxxxxA
\else
\ifnum#1=59 %
\hatcurSMEiloggeccenorigxxxxxB
\else
\ifnum#1=60 %
\hatcurSMEiloggeccenorigxxxxxC
\else
\ifnum#1=61 %
\hatcurSMEiloggeccenorigxxxxxD
\else
\ifnum#1=62 %
\hatcurSMEiloggeccenorigxxxxxE
\else
\ifnum#1=63 %
\hatcurSMEiloggeccenorigxxxxxF
\else
\ifnum#1=64 %
\hatcurSMEiloggeccenorigxxxxxG
\else
??????\fi
\fi
\fi
\fi
\fi
\fi
\fi
}
\newcommand{\hatcurSMEiteffeccenorig}[1]{\ifnum#1=58 %
\hatcurSMEiteffeccenorigxxxxxA
\else
\ifnum#1=59 %
\hatcurSMEiteffeccenorigxxxxxB
\else
\ifnum#1=60 %
\hatcurSMEiteffeccenorigxxxxxC
\else
\ifnum#1=61 %
\hatcurSMEiteffeccenorigxxxxxD
\else
\ifnum#1=62 %
\hatcurSMEiteffeccenorigxxxxxE
\else
\ifnum#1=63 %
\hatcurSMEiteffeccenorigxxxxxF
\else
\ifnum#1=64 %
\hatcurSMEiteffeccenorigxxxxxG
\else
??????\fi
\fi
\fi
\fi
\fi
\fi
\fi
}
\newcommand{\hatcurSMEivmaceccenorig}[1]{\ifnum#1=58 %
\hatcurSMEivmaceccenorigxxxxxA
\else
\ifnum#1=59 %
\hatcurSMEivmaceccenorigxxxxxB
\else
\ifnum#1=60 %
\hatcurSMEivmaceccenorigxxxxxC
\else
\ifnum#1=61 %
\hatcurSMEivmaceccenorigxxxxxD
\else
\ifnum#1=62 %
\hatcurSMEivmaceccenorigxxxxxE
\else
\ifnum#1=63 %
\hatcurSMEivmaceccenorigxxxxxF
\else
\ifnum#1=64 %
\hatcurSMEivmaceccenorigxxxxxG
\else
??????\fi
\fi
\fi
\fi
\fi
\fi
\fi
}
\newcommand{\hatcurSMEivmiceccenorig}[1]{\ifnum#1=58 %
\hatcurSMEivmiceccenorigxxxxxA
\else
\ifnum#1=59 %
\hatcurSMEivmiceccenorigxxxxxB
\else
\ifnum#1=60 %
\hatcurSMEivmiceccenorigxxxxxC
\else
\ifnum#1=61 %
\hatcurSMEivmiceccenorigxxxxxD
\else
\ifnum#1=62 %
\hatcurSMEivmiceccenorigxxxxxE
\else
\ifnum#1=63 %
\hatcurSMEivmiceccenorigxxxxxF
\else
\ifnum#1=64 %
\hatcurSMEivmiceccenorigxxxxxG
\else
??????\fi
\fi
\fi
\fi
\fi
\fi
\fi
}
\newcommand{\hatcurSMEivsineccenorig}[1]{\ifnum#1=58 %
\hatcurSMEivsineccenorigxxxxxA
\else
\ifnum#1=59 %
\hatcurSMEivsineccenorigxxxxxB
\else
\ifnum#1=60 %
\hatcurSMEivsineccenorigxxxxxC
\else
\ifnum#1=61 %
\hatcurSMEivsineccenorigxxxxxD
\else
\ifnum#1=62 %
\hatcurSMEivsineccenorigxxxxxE
\else
\ifnum#1=63 %
\hatcurSMEivsineccenorigxxxxxF
\else
\ifnum#1=64 %
\hatcurSMEivsineccenorigxxxxxG
\else
??????\fi
\fi
\fi
\fi
\fi
\fi
\fi
}
\newcommand{\hatcurSMEizfeheccenorig}[1]{\ifnum#1=58 %
\hatcurSMEizfeheccenorigxxxxxA
\else
\ifnum#1=59 %
\hatcurSMEizfeheccenorigxxxxxB
\else
\ifnum#1=60 %
\hatcurSMEizfeheccenorigxxxxxC
\else
\ifnum#1=61 %
\hatcurSMEizfeheccenorigxxxxxD
\else
\ifnum#1=62 %
\hatcurSMEizfeheccenorigxxxxxE
\else
\ifnum#1=63 %
\hatcurSMEizfeheccenorigxxxxxF
\else
\ifnum#1=64 %
\hatcurSMEizfeheccenorigxxxxxG
\else
??????\fi
\fi
\fi
\fi
\fi
\fi
\fi
}
\newcommand{\hatcurSMEizfehshorteccenorig}[1]{\ifnum#1=58 %
\hatcurSMEizfehshorteccenorigxxxxxA
\else
\ifnum#1=59 %
\hatcurSMEizfehshorteccenorigxxxxxB
\else
\ifnum#1=60 %
\hatcurSMEizfehshorteccenorigxxxxxC
\else
\ifnum#1=61 %
\hatcurSMEizfehshorteccenorigxxxxxD
\else
\ifnum#1=62 %
\hatcurSMEizfehshorteccenorigxxxxxE
\else
\ifnum#1=63 %
\hatcurSMEizfehshorteccenorigxxxxxF
\else
\ifnum#1=64 %
\hatcurSMEizfehshorteccenorigxxxxxG
\else
??????\fi
\fi
\fi
\fi
\fi
\fi
\fi
}
\newcommand{\hatcurXAveccenorig}[1]{\ifnum#1=58 %
\hatcurXAveccenorigxxxxxA
\else
\ifnum#1=59 %
\hatcurXAveccenorigxxxxxB
\else
\ifnum#1=60 %
\hatcurXAveccenorigxxxxxC
\else
\ifnum#1=61 %
\hatcurXAveccenorigxxxxxD
\else
\ifnum#1=62 %
\hatcurXAveccenorigxxxxxE
\else
\ifnum#1=63 %
\hatcurXAveccenorigxxxxxF
\else
\ifnum#1=64 %
\hatcurXAveccenorigxxxxxG
\else
??????\fi
\fi
\fi
\fi
\fi
\fi
\fi
}
\newcommand{\hatcurXdisteccenorig}[1]{\ifnum#1=58 %
\hatcurXdisteccenorigxxxxxA
\else
\ifnum#1=59 %
\hatcurXdisteccenorigxxxxxB
\else
\ifnum#1=60 %
\hatcurXdisteccenorigxxxxxC
\else
\ifnum#1=61 %
\hatcurXdisteccenorigxxxxxD
\else
\ifnum#1=62 %
\hatcurXdisteccenorigxxxxxE
\else
\ifnum#1=63 %
\hatcurXdisteccenorigxxxxxF
\else
\ifnum#1=64 %
\hatcurXdisteccenorigxxxxxG
\else
??????\fi
\fi
\fi
\fi
\fi
\fi
\fi
}
\newcommand{\hatcurXdistredeccenorig}[1]{\ifnum#1=58 %
\hatcurXdistredeccenorigxxxxxA
\else
\ifnum#1=59 %
\hatcurXdistredeccenorigxxxxxB
\else
\ifnum#1=60 %
\hatcurXdistredeccenorigxxxxxC
\else
\ifnum#1=61 %
\hatcurXdistredeccenorigxxxxxD
\else
\ifnum#1=62 %
\hatcurXdistredeccenorigxxxxxE
\else
\ifnum#1=63 %
\hatcurXdistredeccenorigxxxxxF
\else
\ifnum#1=64 %
\hatcurXdistredeccenorigxxxxxG
\else
??????\fi
\fi
\fi
\fi
\fi
\fi
\fi
}
\newcommand{\hatcurXEBVeccenorig}[1]{\ifnum#1=58 %
\hatcurXEBVeccenorigxxxxxA
\else
\ifnum#1=59 %
\hatcurXEBVeccenorigxxxxxB
\else
\ifnum#1=60 %
\hatcurXEBVeccenorigxxxxxC
\else
\ifnum#1=61 %
\hatcurXEBVeccenorigxxxxxD
\else
\ifnum#1=62 %
\hatcurXEBVeccenorigxxxxxE
\else
\ifnum#1=63 %
\hatcurXEBVeccenorigxxxxxF
\else
\ifnum#1=64 %
\hatcurXEBVeccenorigxxxxxG
\else
??????\fi
\fi
\fi
\fi
\fi
\fi
\fi
}
\newcommand{\hatcurXjhisoredeccenorig}[1]{\ifnum#1=58 %
\hatcurXjhisoredeccenorigxxxxxA
\else
\ifnum#1=59 %
\hatcurXjhisoredeccenorigxxxxxB
\else
\ifnum#1=60 %
\hatcurXjhisoredeccenorigxxxxxC
\else
\ifnum#1=61 %
\hatcurXjhisoredeccenorigxxxxxD
\else
\ifnum#1=62 %
\hatcurXjhisoredeccenorigxxxxxE
\else
\ifnum#1=63 %
\hatcurXjhisoredeccenorigxxxxxF
\else
\ifnum#1=64 %
\hatcurXjhisoredeccenorigxxxxxG
\else
??????\fi
\fi
\fi
\fi
\fi
\fi
\fi
}
\newcommand{\hatcurXjkisoredeccenorig}[1]{\ifnum#1=58 %
\hatcurXjkisoredeccenorigxxxxxA
\else
\ifnum#1=59 %
\hatcurXjkisoredeccenorigxxxxxB
\else
\ifnum#1=60 %
\hatcurXjkisoredeccenorigxxxxxC
\else
\ifnum#1=61 %
\hatcurXjkisoredeccenorigxxxxxD
\else
\ifnum#1=62 %
\hatcurXjkisoredeccenorigxxxxxE
\else
\ifnum#1=63 %
\hatcurXjkisoredeccenorigxxxxxF
\else
\ifnum#1=64 %
\hatcurXjkisoredeccenorigxxxxxG
\else
??????\fi
\fi
\fi
\fi
\fi
\fi
\fi
}
\newcommand{\hatcurXmhisoredeccenorig}[1]{\ifnum#1=58 %
\hatcurXmhisoredeccenorigxxxxxA
\else
\ifnum#1=59 %
\hatcurXmhisoredeccenorigxxxxxB
\else
\ifnum#1=60 %
\hatcurXmhisoredeccenorigxxxxxC
\else
\ifnum#1=61 %
\hatcurXmhisoredeccenorigxxxxxD
\else
\ifnum#1=62 %
\hatcurXmhisoredeccenorigxxxxxE
\else
\ifnum#1=63 %
\hatcurXmhisoredeccenorigxxxxxF
\else
\ifnum#1=64 %
\hatcurXmhisoredeccenorigxxxxxG
\else
??????\fi
\fi
\fi
\fi
\fi
\fi
\fi
}
\newcommand{\hatcurXmiisoredeccenorig}[1]{\ifnum#1=58 %
\hatcurXmiisoredeccenorigxxxxxA
\else
\ifnum#1=59 %
\hatcurXmiisoredeccenorigxxxxxB
\else
\ifnum#1=60 %
\hatcurXmiisoredeccenorigxxxxxC
\else
\ifnum#1=61 %
\hatcurXmiisoredeccenorigxxxxxD
\else
\ifnum#1=62 %
\hatcurXmiisoredeccenorigxxxxxE
\else
\ifnum#1=63 %
\hatcurXmiisoredeccenorigxxxxxF
\else
\ifnum#1=64 %
\hatcurXmiisoredeccenorigxxxxxG
\else
??????\fi
\fi
\fi
\fi
\fi
\fi
\fi
}
\newcommand{\hatcurXmjisoredeccenorig}[1]{\ifnum#1=58 %
\hatcurXmjisoredeccenorigxxxxxA
\else
\ifnum#1=59 %
\hatcurXmjisoredeccenorigxxxxxB
\else
\ifnum#1=60 %
\hatcurXmjisoredeccenorigxxxxxC
\else
\ifnum#1=61 %
\hatcurXmjisoredeccenorigxxxxxD
\else
\ifnum#1=62 %
\hatcurXmjisoredeccenorigxxxxxE
\else
\ifnum#1=63 %
\hatcurXmjisoredeccenorigxxxxxF
\else
\ifnum#1=64 %
\hatcurXmjisoredeccenorigxxxxxG
\else
??????\fi
\fi
\fi
\fi
\fi
\fi
\fi
}
\newcommand{\hatcurXmkisoredeccenorig}[1]{\ifnum#1=58 %
\hatcurXmkisoredeccenorigxxxxxA
\else
\ifnum#1=59 %
\hatcurXmkisoredeccenorigxxxxxB
\else
\ifnum#1=60 %
\hatcurXmkisoredeccenorigxxxxxC
\else
\ifnum#1=61 %
\hatcurXmkisoredeccenorigxxxxxD
\else
\ifnum#1=62 %
\hatcurXmkisoredeccenorigxxxxxE
\else
\ifnum#1=63 %
\hatcurXmkisoredeccenorigxxxxxF
\else
\ifnum#1=64 %
\hatcurXmkisoredeccenorigxxxxxG
\else
??????\fi
\fi
\fi
\fi
\fi
\fi
\fi
}
\newcommand{\hatcurXmvisoredeccenorig}[1]{\ifnum#1=58 %
\hatcurXmvisoredeccenorigxxxxxA
\else
\ifnum#1=59 %
\hatcurXmvisoredeccenorigxxxxxB
\else
\ifnum#1=60 %
\hatcurXmvisoredeccenorigxxxxxC
\else
\ifnum#1=61 %
\hatcurXmvisoredeccenorigxxxxxD
\else
\ifnum#1=62 %
\hatcurXmvisoredeccenorigxxxxxE
\else
\ifnum#1=63 %
\hatcurXmvisoredeccenorigxxxxxF
\else
\ifnum#1=64 %
\hatcurXmvisoredeccenorigxxxxxG
\else
??????\fi
\fi
\fi
\fi
\fi
\fi
\fi
}
\newcommand{\hatcurXsecdureccenorig}[1]{\ifnum#1=58 %
\hatcurXsecdureccenorigxxxxxA
\else
\ifnum#1=59 %
\hatcurXsecdureccenorigxxxxxB
\else
\ifnum#1=60 %
\hatcurXsecdureccenorigxxxxxC
\else
\ifnum#1=61 %
\hatcurXsecdureccenorigxxxxxD
\else
\ifnum#1=62 %
\hatcurXsecdureccenorigxxxxxE
\else
\ifnum#1=63 %
\hatcurXsecdureccenorigxxxxxF
\else
\ifnum#1=64 %
\hatcurXsecdureccenorigxxxxxG
\else
??????\fi
\fi
\fi
\fi
\fi
\fi
\fi
}
\newcommand{\hatcurXsecingdureccenorig}[1]{\ifnum#1=58 %
\hatcurXsecingdureccenorigxxxxxA
\else
\ifnum#1=59 %
\hatcurXsecingdureccenorigxxxxxB
\else
\ifnum#1=60 %
\hatcurXsecingdureccenorigxxxxxC
\else
\ifnum#1=61 %
\hatcurXsecingdureccenorigxxxxxD
\else
\ifnum#1=62 %
\hatcurXsecingdureccenorigxxxxxE
\else
\ifnum#1=63 %
\hatcurXsecingdureccenorigxxxxxF
\else
\ifnum#1=64 %
\hatcurXsecingdureccenorigxxxxxG
\else
??????\fi
\fi
\fi
\fi
\fi
\fi
\fi
}
\newcommand{\hatcurXsecondaryeccenorig}[1]{\ifnum#1=58 %
\hatcurXsecondaryeccenorigxxxxxA
\else
\ifnum#1=59 %
\hatcurXsecondaryeccenorigxxxxxB
\else
\ifnum#1=60 %
\hatcurXsecondaryeccenorigxxxxxC
\else
\ifnum#1=61 %
\hatcurXsecondaryeccenorigxxxxxD
\else
\ifnum#1=62 %
\hatcurXsecondaryeccenorigxxxxxE
\else
\ifnum#1=63 %
\hatcurXsecondaryeccenorigxxxxxF
\else
\ifnum#1=64 %
\hatcurXsecondaryeccenorigxxxxxG
\else
??????\fi
\fi
\fi
\fi
\fi
\fi
\fi
}
\newcommand{\hatcurXsecphaseeccenorig}[1]{\ifnum#1=58 %
\hatcurXsecphaseeccenorigxxxxxA
\else
\ifnum#1=59 %
\hatcurXsecphaseeccenorigxxxxxB
\else
\ifnum#1=60 %
\hatcurXsecphaseeccenorigxxxxxC
\else
\ifnum#1=61 %
\hatcurXsecphaseeccenorigxxxxxD
\else
\ifnum#1=62 %
\hatcurXsecphaseeccenorigxxxxxE
\else
\ifnum#1=63 %
\hatcurXsecphaseeccenorigxxxxxF
\else
\ifnum#1=64 %
\hatcurXsecphaseeccenorigxxxxxG
\else
??????\fi
\fi
\fi
\fi
\fi
\fi
\fi
}
\newcommand{\hatcurXviisoredeccenorig}[1]{\ifnum#1=58 %
\hatcurXviisoredeccenorigxxxxxA
\else
\ifnum#1=59 %
\hatcurXviisoredeccenorigxxxxxB
\else
\ifnum#1=60 %
\hatcurXviisoredeccenorigxxxxxC
\else
\ifnum#1=61 %
\hatcurXviisoredeccenorigxxxxxD
\else
\ifnum#1=62 %
\hatcurXviisoredeccenorigxxxxxE
\else
\ifnum#1=63 %
\hatcurXviisoredeccenorigxxxxxF
\else
\ifnum#1=64 %
\hatcurXviisoredeccenorigxxxxxG
\else
??????\fi
\fi
\fi
\fi
\fi
\fi
\fi
}
\newcommand{\hatcurXvkisoredeccenorig}[1]{\ifnum#1=58 %
\hatcurXvkisoredeccenorigxxxxxA
\else
\ifnum#1=59 %
\hatcurXvkisoredeccenorigxxxxxB
\else
\ifnum#1=60 %
\hatcurXvkisoredeccenorigxxxxxC
\else
\ifnum#1=61 %
\hatcurXvkisoredeccenorigxxxxxD
\else
\ifnum#1=62 %
\hatcurXvkisoredeccenorigxxxxxE
\else
\ifnum#1=63 %
\hatcurXvkisoredeccenorigxxxxxF
\else
\ifnum#1=64 %
\hatcurXvkisoredeccenorigxxxxxG
\else
??????\fi
\fi
\fi
\fi
\fi
\fi
\fi
}

%% file: HTR062-011.eccen.orig.tex
\newcommand{\hatcurhtreccenorigxxxxxA}{HTR062-011}                         
\newcommand{\hatcurfieldeccenorigxxxxxA}{\ensuremath{string}}              
\newcommand{\hatcurCCraeccenorigxxxxxA}{\ensuremath{04^{\mathrm h}35^{\mathrm m}23.17{\mathrm s}}}                       
\newcommand{\hatcurCCdececcenorigxxxxxA}{\ensuremath{+56{\arcdeg}52{\arcmin}05.5{\arcsec}}}                      
\newcommand{\hatcurCCmageccenorigxxxxxA}{12.971}                           
\newcommand{\hatcurCCtwomasseccenorigxxxxxA}{2MASS~04352318+5652055}       
\newcommand{\hatcurCCgsceccenorigxxxxxA}{GSC~3740-01482}                   
\newcommand{\hatcurCCtassmveccenorigxxxxxA}{\ensuremath{12.971\pm0.073}}   
\newcommand{\hatcurCCtassmvshorteccenorigxxxxxA}{\ensuremath{13.0}}        
\newcommand{\hatcurCCtassmBeccenorigxxxxxA}{\ensuremath{13.690\pm0.089}}   
\newcommand{\hatcurCCtassmBshorteccenorigxxxxxA}{\ensuremath{13.7}}        
\newcommand{\hatcurCCtassmIeccenorigxxxxxA}{\ensuremath{nff\pmnff}}        
\newcommand{\hatcurCCtassmIshorteccenorigxxxxxA}{\ensuremath{0.0}}         
\newcommand{\hatcurCCtassmgeccenorigxxxxxA}{\ensuremath{13.28\pm0.12}}     
\newcommand{\hatcurCCtassmgshorteccenorigxxxxxA}{\ensuremath{13.3}}        
\newcommand{\hatcurCCtassmreccenorigxxxxxA}{\ensuremath{12.74\pm0.12}}     
\newcommand{\hatcurCCtassmrshorteccenorigxxxxxA}{\ensuremath{12.7}}        
\newcommand{\hatcurCCtassmieccenorigxxxxxA}{\ensuremath{12.50\pm0.12}}     
\newcommand{\hatcurCCtassmishorteccenorigxxxxxA}{\ensuremath{12.5}}        
\newcommand{\hatcurCCtwomassJmageccenorigxxxxxA}{\ensuremath{11.429\pm0.022}} 
\newcommand{\hatcurCCtwomassHmageccenorigxxxxxA}{\ensuremath{11.075\pm0.020}} 
\newcommand{\hatcurCCtwomassKmageccenorigxxxxxA}{\ensuremath{10.978\pm0.023}} 
\newcommand{\hatcurCCcitJmageccenorigxxxxxA}{\ensuremath{11.441\pm0.022}}  
\newcommand{\hatcurCCcitHmageccenorigxxxxxA}{\ensuremath{11.069\pm0.020}}  
\newcommand{\hatcurCCcitKmageccenorigxxxxxA}{\ensuremath{11.002\pm0.023}}  
\newcommand{\hatcurCCbbJmageccenorigxxxxxA}{\ensuremath{11.498\pm0.024}}   
\newcommand{\hatcurCCbbHmageccenorigxxxxxA}{\ensuremath{11.091\pm0.021}}   
\newcommand{\hatcurCCbbKmageccenorigxxxxxA}{\ensuremath{11.022\pm0.023}}   
\newcommand{\hatcurCCesoJmageccenorigxxxxxA}{\ensuremath{11.501\pm0.026}}  
\newcommand{\hatcurCCesoHmageccenorigxxxxxA}{\ensuremath{11.087\pm0.026}}  
\newcommand{\hatcurCCesoKmageccenorigxxxxxA}{\ensuremath{11.021\pm0.024}}  
\newcommand{\hatcurCCesoJHmageccenorigxxxxxA}{\ensuremath{0.413\pm0.034}}  
\newcommand{\hatcurCCesoJKmageccenorigxxxxxA}{\ensuremath{0.481\pm0.035}}  
\newcommand{\hatcurCCesoHKmageccenorigxxxxxA}{\ensuremath{0.067\pm0.035}}  
\newcommand{\hatcurLCdipeccenorigxxxxxA}{\ensuremath{7.4}}                 
\newcommand{\hatcurLCrprstareccenorigxxxxxA}{\ensuremath{0.0878\pm0.0021}} 
\newcommand{\hatcurLCbsqeccenorigxxxxxA}{\ensuremath{0.106_{-0.069}^{+0.081}}} 
\newcommand{\hatcurLCimpeccenorigxxxxxA}{\ensuremath{0.33_{-0.13}^{+0.11}}} 
\newcommand{\hatcurLCzetaeccenorigxxxxxA}{\ensuremath{12.89\pm0.10}}       
\newcommand{\hatcurLCdureccenorigxxxxxA}{\ensuremath{0.1705\pm0.0022}}     
\newcommand{\hatcurLCdurshorteccenorigxxxxxA}{\ensuremath{0.1705}}         
\newcommand{\hatcurLCdurhreccenorigxxxxxA}{\ensuremath{4.092\pm0.053}}     
\newcommand{\hatcurLCdurhrshorteccenorigxxxxxA}{\ensuremath{4.092}}        
\newcommand{\hatcurLCqeccenorigxxxxxA}{\ensuremath{0.04250\pm0.00055}}     
\newcommand{\hatcurLCqshorteccenorigxxxxxA}{\ensuremath{0.043}}            
\newcommand{\hatcurLCingdureccenorigxxxxxA}{\ensuremath{0.0152\pm0.0016}}  
\newcommand{\hatcurLCPeccenorigxxxxxA}{\ensuremath{4.013847\pm0.000013}}   
\newcommand{\hatcurLCPprececcenorigxxxxxA}{\ensuremath{4.0138475}}         
\newcommand{\hatcurLCPshorteccenorigxxxxxA}{\ensuremath{4.0138}}           
\newcommand{\hatcurLCTeccenorigxxxxxA}{\ensuremath{2456791.03848\pm0.00071}} 
\newcommand{\hatcurLCTAeccenorigxxxxxA}{\ensuremath{2456132.7673\pm0.0020}} 
\newcommand{\hatcurLCTBeccenorigxxxxxA}{\ensuremath{2456999.7585\pm0.0011}} 
\newcommand{\hatcurLChatnetmeccenorigxxxxxA}{\ensuremath{12.85295\pm0.00013}} 
\newcommand{\hatcurLCiblendeccenorigxxxxxA}{\ensuremath{1\pm0}}            
\newcommand{\hatcurLCrhoeccenorigxxxxxA}{\ensuremath{0.59_{-0.18}^{+0.51}}} 
\newcommand{\hatcurSMEiteffeccenorigxxxxxA}{\ensuremath{5947\pm50}}        
\newcommand{\hatcurSMEizfeheccenorigxxxxxA}{\ensuremath{0.032\pm0.080}}    
\newcommand{\hatcurSMEizfehshorteccenorigxxxxxA}{\ensuremath{0.03}}        
\newcommand{\hatcurSMEiloggeccenorigxxxxxA}{\ensuremath{4.24\pm0.10}}      
\newcommand{\hatcurSMEivsineccenorigxxxxxA}{\ensuremath{4.81\pm0.50}}      
\newcommand{\hatcurSMEivmaceccenorigxxxxxA}{\ensuremath{0.0}}              
\newcommand{\hatcurSMEivmiceccenorigxxxxxA}{\ensuremath{0.0}}              
\newcommand{\hatcurSMEiiteffeccenorigxxxxxA}{\ensuremath{5931\pm50}}       
\newcommand{\hatcurSMEiizfeheccenorigxxxxxA}{\ensuremath{0.012\pm0.080}}   
\newcommand{\hatcurSMEiizfehshorteccenorigxxxxxA}{\ensuremath{0.012}}      
\newcommand{\hatcurSMEiiloggeccenorigxxxxxA}{\ensuremath{4.183\pm0.036}}   
\newcommand{\hatcurSMEiivsineccenorigxxxxxA}{\ensuremath{4.91\pm0.50}}     
\newcommand{\hatcurLBizeccenorigxxxxxA}{\ensuremath{0.1839}}               
\newcommand{\hatcurLBiizeccenorigxxxxxA}{\ensuremath{0.3408}}              
\newcommand{\hatcurLBiieccenorigxxxxxA}{\ensuremath{0.2401}}               
\newcommand{\hatcurLBiiieccenorigxxxxxA}{\ensuremath{0.3451}}              
\newcommand{\hatcurLBiIeccenorigxxxxxA}{\ensuremath{0.2206}}               
\newcommand{\hatcurLBiiIeccenorigxxxxxA}{\ensuremath{0.3449}}              
\newcommand{\hatcurLBigeccenorigxxxxxA}{\ensuremath{0.5087}}               
\newcommand{\hatcurLBiigeccenorigxxxxxA}{\ensuremath{0.2683}}              
\newcommand{\hatcurLBireccenorigxxxxxA}{\ensuremath{0.3231}}               
\newcommand{\hatcurLBiireccenorigxxxxxA}{\ensuremath{0.3457}}              
\newcommand{\hatcurLBiReccenorigxxxxxA}{\ensuremath{0.2999}}               
\newcommand{\hatcurLBiiReccenorigxxxxxA}{\ensuremath{0.3467}}              
\newcommand{\hatcurLBikepeccenorigxxxxxA}{\ensuremath{0.1000}}             
\newcommand{\hatcurLBiikepeccenorigxxxxxA}{\ensuremath{0.1000}}            
\newcommand{\hatcurISOmeccenorigxxxxxA}{\ensuremath{1.115_{-0.052}^{+0.080}}} 
\newcommand{\hatcurISOmshorteccenorigxxxxxA}{\ensuremath{1.11}}            
\newcommand{\hatcurISOmlongeccenorigxxxxxA}{\ensuremath{1.115_{-0.052}^{+0.080}}} 
\newcommand{\hatcurISOreccenorigxxxxxA}{\ensuremath{1.41\pm0.24}}          
\newcommand{\hatcurISOrshorteccenorigxxxxxA}{\ensuremath{1.41}}            
\newcommand{\hatcurISOrlongeccenorigxxxxxA}{\ensuremath{1.41\pm0.24}}      
\newcommand{\hatcurISOrhoeccenorigxxxxxA}{\ensuremath{0.56_{-0.17}^{+0.31}}} 
\newcommand{\hatcurISOrholongeccenorigxxxxxA}{\ensuremath{0.56_{-0.17}^{+0.31}}} 
\newcommand{\hatcurISOloggeccenorigxxxxxA}{\ensuremath{4.19\pm0.12}}       
\newcommand{\hatcurISOlumeccenorigxxxxxA}{\ensuremath{2.19\pm0.84}}        
\newcommand{\hatcurISOlumshorteccenorigxxxxxA}{\ensuremath{2.19}}          
\newcommand{\hatcurISOmveccenorigxxxxxA}{\ensuremath{3.96\pm0.35}}         
\newcommand{\hatcurISOvieccenorigxxxxxA}{\ensuremath{0.650\pm0.015}}       
\newcommand{\hatcurISOageeccenorigxxxxxA}{\ensuremath{5.5\pm1.4}}          
\newcommand{\hatcurISOsigmaeccenorigxxxxxA}{\ensuremath{0.00030\pm0.00021}} 
\newcommand{\hatcurISOMJeccenorigxxxxxA}{\ensuremath{2.88\pm0.35}}         
\newcommand{\hatcurISOMHeccenorigxxxxxA}{\ensuremath{2.56\pm0.35}}         
\newcommand{\hatcurISOMKeccenorigxxxxxA}{\ensuremath{2.51\pm0.35}}         
\newcommand{\hatcurISOJKeccenorigxxxxxA}{\ensuremath{0.370\pm0.010}}       
\newcommand{\hatcurISOspececcenorigxxxxxA}{G}                              
\newcommand{\hatcurRVKeccenorigxxxxxA}{\ensuremath{47.0\pm7.4}}            
\newcommand{\hatcurRVrkeccenorigxxxxxA}{\ensuremath{-0.08_{-0.12}^{+0.22}}} 
\newcommand{\hatcurRVrheccenorigxxxxxA}{\ensuremath{-0.04\pm0.29}}         
\newcommand{\hatcurRVkeccenorigxxxxxA}{\ensuremath{-0.015\pm0.072}}        
\newcommand{\hatcurRVheccenorigxxxxxA}{\ensuremath{-0.004_{-0.152}^{+0.091}}} 
\newcommand{\hatcurRVtroneeccenorigxxxxxA}{\ensuremath{0\pm0}}             
\newcommand{\hatcurRVtrtwoeccenorigxxxxxA}{\ensuremath{0\pm0}}             
\newcommand{\hatcurRVgammaeccenorigxxxxxA}{\ensuremath{2.5\pm5.0}}         
\newcommand{\hatcurRVjittereccenorigxxxxxA}{\ensuremath{4.8\pm4.1}}        
\newcommand{\hatcurRVjittertwosiglimeccenorigxxxxxA}{\ensuremath{<12.3}}   
\newcommand{\hatcurRVfitrmseccenorigxxxxxA}{\ensuremath{.1fym}}            %
\newcommand{\hatcurRVecceneccenorigxxxxxA}{\ensuremath{0.08\pm0.11}}       
\newcommand{\hatcurRVeccentwosiglimeccenorigxxxxxA}{\ensuremath{<0.325}}   
\newcommand{\hatcurRVomegaeccenorigxxxxxA}{\ensuremath{203\pm94}}          
\newcommand{\hatcurPPieccenorigxxxxxA}{\ensuremath{87.7_{-1.4}^{+1.0}}}    
\newcommand{\hatcurPPgeccenorigxxxxxA}{\ensuremath{6.8_{-1.5}^{+2.2}}}     
\newcommand{\hatcurPPloggeccenorigxxxxxA}{\ensuremath{2.83\pm0.12}}        
\newcommand{\hatcurPPareccenorigxxxxxA}{\ensuremath{7.83_{-0.88}^{+1.24}}} 
\newcommand{\hatcurPPareleccenorigxxxxxA}{\ensuremath{0.05126_{-0.00081}^{+0.00120}}} 
\newcommand{\hatcurPPrhoeccenorigxxxxxA}{\ensuremath{0.282_{-0.094}^{+0.156}}} 
\newcommand{\hatcurPPmeccenorigxxxxxA}{\ensuremath{0.394_{-0.031}^{+0.051}}} 
\newcommand{\hatcurPPmshorteccenorigxxxxxA}{\ensuremath{0.39}}             
\newcommand{\hatcurPPmlongeccenorigxxxxxA}{\ensuremath{0.394_{-0.031}^{+0.051}}} 
\newcommand{\hatcurPPmeeccenorigxxxxxA}{\ensuremath{125.1_{-9.8}^{+16.1}}} 
\newcommand{\hatcurPPmeshorteccenorigxxxxxA}{\ensuremath{125.1}}           
\newcommand{\hatcurPPmelongeccenorigxxxxxA}{\ensuremath{125.1_{-9.8}^{+16.1}}} 
\newcommand{\hatcurPPreccenorigxxxxxA}{\ensuremath{1.20\pm0.21}}           
\newcommand{\hatcurPPrshorteccenorigxxxxxA}{\ensuremath{1.20}}             
\newcommand{\hatcurPPrlongeccenorigxxxxxA}{\ensuremath{1.20\pm0.21}}       
\newcommand{\hatcurPPreeccenorigxxxxxA}{\ensuremath{13.4\pm2.4}}           
\newcommand{\hatcurPPreshorteccenorigxxxxxA}{\ensuremath{13.4}}            
\newcommand{\hatcurPPrelongeccenorigxxxxxA}{\ensuremath{13.4\pm2.4}}       
\newcommand{\hatcurPPmrcorreccenorigxxxxxA}{\ensuremath{0.60}}             
\newcommand{\hatcurPPteffeccenorigxxxxxA}{\ensuremath{1500\pm110}}         
\newcommand{\hatcurPPthetaeccenorigxxxxxA}{\ensuremath{0.0303\pm0.0051}}   
\newcommand{\hatcurPPfluxperieccenorigxxxxxA}{\ensuremath{1.31_{-0.16}^{+0.47}}} 
\newcommand{\hatcurPPfluxperidimeccenorigxxxxxA}{\ensuremath{9}}           
\newcommand{\hatcurPPfluxapeccenorigxxxxxA}{\ensuremath{1.04_{-0.42}^{+0.18}}} 
\newcommand{\hatcurPPfluxapdimeccenorigxxxxxA}{\ensuremath{9}}             
\newcommand{\hatcurPPfluxavgeccenorigxxxxxA}{\ensuremath{1.14\pm0.41}}     
\newcommand{\hatcurPPfluxavgdimeccenorigxxxxxA}{\ensuremath{9}}            
\newcommand{\hatcurPPfluxavglogeccenorigxxxxxA}{\ensuremath{9.06\pm0.13}}  
\newcommand{\hatcurXsecphaseeccenorigxxxxxA}{\ensuremath{0.491\pm0.047}}   
\newcommand{\hatcurXsecondaryeccenorigxxxxxA}{\ensuremath{2456793.01\pm0.19}} 
\newcommand{\hatcurXsecdureccenorigxxxxxA}{\ensuremath{0.169\pm0.043}}     
\newcommand{\hatcurXsecingdureccenorigxxxxxA}{\ensuremath{0.0149\pm0.0082}} 
\newcommand{\hatcurPPphiconjeccenorigxxxxxA}{\ensuremath{-0.17\pm0.26}}    
\newcommand{\hatcurPPperieccenorigxxxxxA}{\ensuremath{2456791.7\pm1.0}}    
\newcommand{\hatcurPPaequiveccenorigxxxxxA}{\ensuremath{0.0347_{-0.0041}^{+0.0054}}} 
\newcommand{\hatcurPPtcirceccenorigxxxxxA}{\ensuremath{210_{-110}^{+150}}} 
\newcommand{\hatcurPPtinfalleccenorigxxxxxA}{\ensuremath{3200_{-1500}^{+3600}}} 
\newcommand{\hatcurXdisteccenorigxxxxxA}{\ensuremath{504\pm85}}            
\newcommand{\hatcurXAveccenorigxxxxxA}{\ensuremath{0.587\pm0.092}}         
\newcommand{\hatcurXdistredeccenorigxxxxxA}{\ensuremath{487\pm82}}         
\newcommand{\hatcurXEBVeccenorigxxxxxA}{\ensuremath{0.189\pm0.030}}        
\newcommand{\hatcurXmvisoredeccenorigxxxxxA}{\ensuremath{12.981\pm0.067}}  
\newcommand{\hatcurXmiisoredeccenorigxxxxxA}{\ensuremath{12.025\pm0.026}}  
\newcommand{\hatcurXmjisoredeccenorigxxxxxA}{\ensuremath{11.483\pm0.015}}  
\newcommand{\hatcurXmhisoredeccenorigxxxxxA}{\ensuremath{11.112\pm0.015}}  
\newcommand{\hatcurXmkisoredeccenorigxxxxxA}{\ensuremath{11.014\pm0.017}}  
\newcommand{\hatcurXviisoredeccenorigxxxxxA}{\ensuremath{0.956\pm0.044}}   
\newcommand{\hatcurXvkisoredeccenorigxxxxxA}{\ensuremath{1.967\pm0.075}}   
\newcommand{\hatcurXjhisoredeccenorigxxxxxA}{\ensuremath{0.3710\pm0.0091}} 
\newcommand{\hatcurXjkisoredeccenorigxxxxxA}{\ensuremath{0.469\pm0.015}}   
\newcommand{\hatcurCCpmraeccenorigxxxxxA}{\ensuremath{-10.3\pm1.3}}        
\newcommand{\hatcurCCpmdececcenorigxxxxxA}{\ensuremath{9.1\pm1.4}}         
\newcommand{\hatcurCCpmeccenorigxxxxxA}{\ensuremath{13.7\pm1.9}}           

%% file: HTR080-003.eccen.orig.tex
\newcommand{\hatcurhtreccenorigxxxxxB}{HTR080-003}                         
\newcommand{\hatcurfieldeccenorigxxxxxB}{\ensuremath{string}}              
\newcommand{\hatcurCCraeccenorigxxxxxB}{\ensuremath{19^{\mathrm h}29^{\mathrm m}49.92{\mathrm s}}}                       
\newcommand{\hatcurCCdececcenorigxxxxxB}{\ensuremath{+62{\arcdeg}31{\arcmin}45.2{\arcsec}}}                      
\newcommand{\hatcurCCmageccenorigxxxxxB}{11.883}                           
\newcommand{\hatcurCCtwomasseccenorigxxxxxB}{2MASS~19295008+6231452}       
\newcommand{\hatcurCCgsceccenorigxxxxxB}{GSC~4234-02195}                   
\newcommand{\hatcurCCtassmveccenorigxxxxxB}{\ensuremath{11.883\pm0.065}}   
\newcommand{\hatcurCCtassmvshorteccenorigxxxxxB}{\ensuremath{11.9}}        
\newcommand{\hatcurCCtassmBeccenorigxxxxxB}{\ensuremath{12.581\pm0.094}}   
\newcommand{\hatcurCCtassmBshorteccenorigxxxxxB}{\ensuremath{12.6}}        
\newcommand{\hatcurCCtassmIeccenorigxxxxxB}{\ensuremath{11.073\pm0.078}}   
\newcommand{\hatcurCCtassmIshorteccenorigxxxxxB}{\ensuremath{11.1}}        
\newcommand{\hatcurCCtassmgeccenorigxxxxxB}{\ensuremath{12.16\pm0.10}}     
\newcommand{\hatcurCCtassmgshorteccenorigxxxxxB}{\ensuremath{12.2}}        
\newcommand{\hatcurCCtassmreccenorigxxxxxB}{\ensuremath{11.650\pm0.050}}   
\newcommand{\hatcurCCtassmrshorteccenorigxxxxxB}{\ensuremath{11.7}}        
\newcommand{\hatcurCCtassmieccenorigxxxxxB}{\ensuremath{11.478\pm0.050}}   
\newcommand{\hatcurCCtassmishorteccenorigxxxxxB}{\ensuremath{11.5}}        
\newcommand{\hatcurCCtwomassJmageccenorigxxxxxB}{\ensuremath{10.581\pm0.020}} 
\newcommand{\hatcurCCtwomassHmageccenorigxxxxxB}{\ensuremath{10.268\pm0.018}} 
\newcommand{\hatcurCCtwomassKmageccenorigxxxxxB}{\ensuremath{10.208\pm0.022}} 
\newcommand{\hatcurCCcitJmageccenorigxxxxxB}{\ensuremath{10.598\pm0.020}}  
\newcommand{\hatcurCCcitHmageccenorigxxxxxB}{\ensuremath{10.263\pm0.019}}  
\newcommand{\hatcurCCcitKmageccenorigxxxxxB}{\ensuremath{10.232\pm0.022}}  
\newcommand{\hatcurCCbbJmageccenorigxxxxxB}{\ensuremath{10.647\pm0.022}}   
\newcommand{\hatcurCCbbHmageccenorigxxxxxB}{\ensuremath{10.284\pm0.019}}   
\newcommand{\hatcurCCbbKmageccenorigxxxxxB}{\ensuremath{10.252\pm0.022}}   
\newcommand{\hatcurCCesoJmageccenorigxxxxxB}{\ensuremath{10.649\pm0.024}}  
\newcommand{\hatcurCCesoHmageccenorigxxxxxB}{\ensuremath{10.278\pm0.022}}  
\newcommand{\hatcurCCesoKmageccenorigxxxxxB}{\ensuremath{10.251\pm0.023}}  
\newcommand{\hatcurCCesoJHmageccenorigxxxxxB}{\ensuremath{0.370\pm0.030}}  
\newcommand{\hatcurCCesoJKmageccenorigxxxxxB}{\ensuremath{0.399\pm0.032}}  
\newcommand{\hatcurCCesoHKmageccenorigxxxxxB}{\ensuremath{0.028\pm0.032}}  
\newcommand{\hatcurLCdipeccenorigxxxxxB}{\ensuremath{12.8}}                
\newcommand{\hatcurLCrprstareccenorigxxxxxB}{\ensuremath{0.1020\pm0.0022}} 
\newcommand{\hatcurLCbsqeccenorigxxxxxB}{\ensuremath{0.673_{-0.032}^{+0.021}}} 
\newcommand{\hatcurLCimpeccenorigxxxxxB}{\ensuremath{0.820_{-0.020}^{+0.013}}} 
\newcommand{\hatcurLCzetaeccenorigxxxxxB}{\ensuremath{26.27\pm0.38}}       
\newcommand{\hatcurLCdureccenorigxxxxxB}{\ensuremath{0.0980\pm0.0019}}     
\newcommand{\hatcurLCdurshorteccenorigxxxxxB}{\ensuremath{0.0980}}         
\newcommand{\hatcurLCdurhreccenorigxxxxxB}{\ensuremath{2.352\pm0.045}}     
\newcommand{\hatcurLCdurhrshorteccenorigxxxxxB}{\ensuremath{2.352}}        
\newcommand{\hatcurLCqeccenorigxxxxxB}{\ensuremath{0.02370\pm0.00046}}     
\newcommand{\hatcurLCqshorteccenorigxxxxxB}{\ensuremath{0.024}}            
\newcommand{\hatcurLCingdureccenorigxxxxxB}{\ensuremath{0.0247\pm0.0024}}  
\newcommand{\hatcurLCPeccenorigxxxxxB}{\ensuremath{4.141949\pm0.000015}}   
\newcommand{\hatcurLCPprececcenorigxxxxxB}{\ensuremath{4.1419491}}         
\newcommand{\hatcurLCPshorteccenorigxxxxxB}{\ensuremath{4.1419}}           
\newcommand{\hatcurLCTeccenorigxxxxxB}{\ensuremath{2456750.50806\pm0.00057}} 
\newcommand{\hatcurLCTAeccenorigxxxxxB}{\ensuremath{2456129.2157\pm0.0023}} 
\newcommand{\hatcurLCTBeccenorigxxxxxB}{\ensuremath{2456791.92754\pm0.00060}} 
\newcommand{\hatcurLChatnetmeccenorigxxxxxB}{\ensuremath{11.72706\pm0.00010}} 
\newcommand{\hatcurLCiblendeccenorigxxxxxB}{\ensuremath{1\pm0}}            
\newcommand{\hatcurLCrhoeccenorigxxxxxB}{\ensuremath{1.10_{-0.12}^{+0.17}}} 
\newcommand{\hatcurSMEiteffeccenorigxxxxxB}{\ensuremath{5678\pm50}}        
\newcommand{\hatcurSMEizfeheccenorigxxxxxB}{\ensuremath{0.420\pm0.080}}    
\newcommand{\hatcurSMEizfehshorteccenorigxxxxxB}{\ensuremath{0.42}}        
\newcommand{\hatcurSMEiloggeccenorigxxxxxB}{\ensuremath{4.40\pm0.10}}      
\newcommand{\hatcurSMEivsineccenorigxxxxxB}{\ensuremath{3.05\pm0.50}}      
\newcommand{\hatcurSMEivmaceccenorigxxxxxB}{\ensuremath{0.0}}              
\newcommand{\hatcurSMEivmiceccenorigxxxxxB}{\ensuremath{0.0}}              
\newcommand{\hatcurSMEiiteffeccenorigxxxxxB}{\ensuremath{5665\pm50}}       
\newcommand{\hatcurSMEiizfeheccenorigxxxxxB}{\ensuremath{0.409\pm0.080}}   
\newcommand{\hatcurSMEiizfehshorteccenorigxxxxxB}{\ensuremath{0.409}}      
\newcommand{\hatcurSMEiiloggeccenorigxxxxxB}{\ensuremath{4.368\pm0.033}}   
\newcommand{\hatcurSMEiivsineccenorigxxxxxB}{\ensuremath{3.04\pm0.50}}     
\newcommand{\hatcurLBizeccenorigxxxxxB}{\ensuremath{0.2318}}               
\newcommand{\hatcurLBiizeccenorigxxxxxB}{\ensuremath{0.3284}}              
\newcommand{\hatcurLBiieccenorigxxxxxB}{\ensuremath{0.3052}}               
\newcommand{\hatcurLBiiieccenorigxxxxxB}{\ensuremath{0.3214}}              
\newcommand{\hatcurLBiIeccenorigxxxxxB}{\ensuremath{0.2807}}               
\newcommand{\hatcurLBiiIeccenorigxxxxxB}{\ensuremath{0.3245}}              
\newcommand{\hatcurLBigeccenorigxxxxxB}{\ensuremath{0.6266}}               
\newcommand{\hatcurLBiigeccenorigxxxxxB}{\ensuremath{0.1833}}              
\newcommand{\hatcurLBireccenorigxxxxxB}{\ensuremath{0.4088}}               
\newcommand{\hatcurLBiireccenorigxxxxxB}{\ensuremath{0.3012}}              
\newcommand{\hatcurLBiReccenorigxxxxxB}{\ensuremath{0.3800}}               
\newcommand{\hatcurLBiiReccenorigxxxxxB}{\ensuremath{0.3079}}              
\newcommand{\hatcurLBikepeccenorigxxxxxB}{\ensuremath{0.1000}}             
\newcommand{\hatcurLBiikepeccenorigxxxxxB}{\ensuremath{0.1000}}            
\newcommand{\hatcurISOmeccenorigxxxxxB}{\ensuremath{1.088\pm0.023}}        
\newcommand{\hatcurISOmshorteccenorigxxxxxB}{\ensuremath{1.09}}            
\newcommand{\hatcurISOmlongeccenorigxxxxxB}{\ensuremath{1.088\pm0.023}}    
\newcommand{\hatcurISOreccenorigxxxxxB}{\ensuremath{1.113\pm0.051}}        
\newcommand{\hatcurISOrshorteccenorigxxxxxB}{\ensuremath{1.11}}            
\newcommand{\hatcurISOrlongeccenorigxxxxxB}{\ensuremath{1.113\pm0.051}}    
\newcommand{\hatcurISOrhoeccenorigxxxxxB}{\ensuremath{1.11_{-0.12}^{+0.17}}} 
\newcommand{\hatcurISOrholongeccenorigxxxxxB}{\ensuremath{1.11_{-0.12}^{+0.17}}} 
\newcommand{\hatcurISOloggeccenorigxxxxxB}{\ensuremath{4.380\pm0.037}}     
\newcommand{\hatcurISOlumeccenorigxxxxxB}{\ensuremath{1.14\pm0.12}}        
\newcommand{\hatcurISOlumshorteccenorigxxxxxB}{\ensuremath{1.14}}          
\newcommand{\hatcurISOmveccenorigxxxxxB}{\ensuremath{4.69\pm0.12}}         
\newcommand{\hatcurISOvieccenorigxxxxxB}{\ensuremath{0.744\pm0.016}}       
\newcommand{\hatcurISOageeccenorigxxxxxB}{\ensuremath{4.1\pm1.2}}          
\newcommand{\hatcurISOsigmaeccenorigxxxxxB}{\ensuremath{0.00180\pm0.00019}} 
\newcommand{\hatcurISOMJeccenorigxxxxxB}{\ensuremath{3.49\pm0.11}}         
\newcommand{\hatcurISOMHeccenorigxxxxxB}{\ensuremath{3.14\pm0.10}}         
\newcommand{\hatcurISOMKeccenorigxxxxxB}{\ensuremath{3.08\pm0.10}}         
\newcommand{\hatcurISOJKeccenorigxxxxxB}{\ensuremath{0.400\pm0.010}}       
\newcommand{\hatcurISOspececcenorigxxxxxB}{G}                              
\newcommand{\hatcurRVKeccenorigxxxxxB}{\ensuremath{194.0\pm7.5}}           
\newcommand{\hatcurRVrkeccenorigxxxxxB}{\ensuremath{0.027\pm0.069}}        
\newcommand{\hatcurRVrheccenorigxxxxxB}{\ensuremath{-0.05_{-0.11}^{+0.15}}} 
\newcommand{\hatcurRVkeccenorigxxxxxB}{\ensuremath{0.0028_{-0.0072}^{+0.0119}}} 
\newcommand{\hatcurRVheccenorigxxxxxB}{\ensuremath{-0.004_{-0.025}^{+0.019}}} 
\newcommand{\hatcurRVtroneeccenorigxxxxxB}{\ensuremath{0\pm0}}             
\newcommand{\hatcurRVtrtwoeccenorigxxxxxB}{\ensuremath{0\pm0}}             
\newcommand{\hatcurRVgammaAeccenorigxxxxxB}{\ensuremath{-150.5\pm8.6}}     
\newcommand{\hatcurRVjitterAeccenorigxxxxxB}{\ensuremath{12\pm13}}         
\newcommand{\hatcurRVjittertwosiglimAeccenorigxxxxxB}{\ensuremath{<37.3}}  
\newcommand{\hatcurRVfitrmsAeccenorigxxxxxB}{\ensuremath{0.0}}             
\newcommand{\hatcurRVgammaBeccenorigxxxxxB}{\ensuremath{-21158.6\pm4.6}}   
\newcommand{\hatcurRVjitterBeccenorigxxxxxB}{\ensuremath{0.0\pm2.5}}       
\newcommand{\hatcurRVjittertwosiglimBeccenorigxxxxxB}{\ensuremath{<5.4}}   
\newcommand{\hatcurRVfitrmsBeccenorigxxxxxB}{\ensuremath{0.0}}             
\newcommand{\hatcurRVecceneccenorigxxxxxB}{\ensuremath{0.018\pm0.022}}     
\newcommand{\hatcurRVeccentwosiglimeccenorigxxxxxB}{\ensuremath{<0.071}}   
\newcommand{\hatcurRVomegaeccenorigxxxxxB}{\ensuremath{250\pm120}}         
\newcommand{\hatcurPPieccenorigxxxxxB}{\ensuremath{85.35\pm0.35}}          
\newcommand{\hatcurPPgeccenorigxxxxxB}{\ensuremath{33.2\pm4.0}}            
\newcommand{\hatcurPPloggeccenorigxxxxxB}{\ensuremath{3.521\pm0.052}}      
\newcommand{\hatcurPPareccenorigxxxxxB}{\ensuremath{10.01\pm0.44}}         
\newcommand{\hatcurPPareleccenorigxxxxxB}{\ensuremath{0.05191\pm0.00036}}  
\newcommand{\hatcurPPrhoeccenorigxxxxxB}{\ensuremath{1.50\pm0.27}}         
\newcommand{\hatcurPPmeccenorigxxxxxB}{\ensuremath{1.628\pm0.067}}         
\newcommand{\hatcurPPmshorteccenorigxxxxxB}{\ensuremath{1.63}}             
\newcommand{\hatcurPPmlongeccenorigxxxxxB}{\ensuremath{1.628\pm0.067}}     
\newcommand{\hatcurPPmeeccenorigxxxxxB}{\ensuremath{517\pm21}}             
\newcommand{\hatcurPPmeshorteccenorigxxxxxB}{\ensuremath{517.3}}           
\newcommand{\hatcurPPmelongeccenorigxxxxxB}{\ensuremath{517\pm21}}         
\newcommand{\hatcurPPreccenorigxxxxxB}{\ensuremath{1.107\pm0.066}}         
\newcommand{\hatcurPPrshorteccenorigxxxxxB}{\ensuremath{1.11}}             
\newcommand{\hatcurPPrlongeccenorigxxxxxB}{\ensuremath{1.107\pm0.066}}     
\newcommand{\hatcurPPreeccenorigxxxxxB}{\ensuremath{12.40\pm0.74}}         
\newcommand{\hatcurPPreshorteccenorigxxxxxB}{\ensuremath{12.4}}            
\newcommand{\hatcurPPrelongeccenorigxxxxxB}{\ensuremath{12.40\pm0.74}}     
\newcommand{\hatcurPPmrcorreccenorigxxxxxB}{\ensuremath{0.14}}             
\newcommand{\hatcurPPteffeccenorigxxxxxB}{\ensuremath{1266\pm30}}          
\newcommand{\hatcurPPthetaeccenorigxxxxxB}{\ensuremath{0.1401\pm0.0099}}   
\newcommand{\hatcurPPfluxperieccenorigxxxxxB}{\ensuremath{6.06\pm0.59}}    
\newcommand{\hatcurPPfluxperidimeccenorigxxxxxB}{\ensuremath{8}}           
\newcommand{\hatcurPPfluxapeccenorigxxxxxB}{\ensuremath{5.59\pm0.60}}      
\newcommand{\hatcurPPfluxapdimeccenorigxxxxxB}{\ensuremath{8}}             
\newcommand{\hatcurPPfluxavgeccenorigxxxxxB}{\ensuremath{5.80\pm0.55}}     
\newcommand{\hatcurPPfluxavgdimeccenorigxxxxxB}{\ensuremath{8}}            
\newcommand{\hatcurPPfluxavglogeccenorigxxxxxB}{\ensuremath{8.764\pm0.041}} 
\newcommand{\hatcurXsecphaseeccenorigxxxxxB}{\ensuremath{0.5018\pm0.0070}} 
\newcommand{\hatcurXsecondaryeccenorigxxxxxB}{\ensuremath{2456752.587\pm0.029}} 
\newcommand{\hatcurXsecdureccenorigxxxxxB}{\ensuremath{0.0978\pm0.0023}}   
\newcommand{\hatcurXsecingdureccenorigxxxxxB}{\ensuremath{0.0236\pm0.0058}} 
\newcommand{\hatcurPPphiconjeccenorigxxxxxB}{\ensuremath{0.13_{-0.54}^{+0.31}}} 
\newcommand{\hatcurPPperieccenorigxxxxxB}{\ensuremath{2456750.0\pm1.4}}    
\newcommand{\hatcurPPaequiveccenorigxxxxxB}{\ensuremath{0.0485\pm0.0023}}  
\newcommand{\hatcurPPtcirceccenorigxxxxxB}{\ensuremath{1490_{-350}^{+470}}} 
\newcommand{\hatcurPPtinfalleccenorigxxxxxB}{\ensuremath{2650_{-480}^{+630}}} 
\newcommand{\hatcurXdisteccenorigxxxxxB}{\ensuremath{272\pm13}}            
\newcommand{\hatcurXAveccenorigxxxxxB}{\ensuremath{0.023_{-0.023}^{+0.088}}} 
\newcommand{\hatcurXdistredeccenorigxxxxxB}{\ensuremath{269\pm13}}         
\newcommand{\hatcurXEBVeccenorigxxxxxB}{\ensuremath{0.0070_{-0.0070}^{+0.0290}}} 
\newcommand{\hatcurXmvisoredeccenorigxxxxxB}{\ensuremath{11.882\pm0.043}}  
\newcommand{\hatcurXmiisoredeccenorigxxxxxB}{\ensuremath{11.116\pm0.020}}  
\newcommand{\hatcurXmjisoredeccenorigxxxxxB}{\ensuremath{10.645\pm0.014}}  
\newcommand{\hatcurXmhisoredeccenorigxxxxxB}{\ensuremath{10.294\pm0.013}}  
\newcommand{\hatcurXmkisoredeccenorigxxxxxB}{\ensuremath{10.233\pm0.014}}  
\newcommand{\hatcurXviisoredeccenorigxxxxxB}{\ensuremath{0.764_{-0.020}^{+0.032}}} 
\newcommand{\hatcurXvkisoredeccenorigxxxxxB}{\ensuremath{1.648\pm0.048}}   
\newcommand{\hatcurXjhisoredeccenorigxxxxxB}{\ensuremath{0.3510\pm0.0092}} 
\newcommand{\hatcurXjkisoredeccenorigxxxxxB}{\ensuremath{0.412\pm0.012}}   
\newcommand{\hatcurCCpmraeccenorigxxxxxB}{\ensuremath{20.7\pm1.1}}         
\newcommand{\hatcurCCpmdececcenorigxxxxxB}{\ensuremath{-5.00\pm0.90}}      
\newcommand{\hatcurCCpmeccenorigxxxxxB}{\ensuremath{21.3\pm1.4}}           

%% file: HTR089-018.eccen.orig.tex
\newcommand{\hatcurhtreccenorigxxxxxC}{HTR089-018}                       
\newcommand{\hatcurfieldeccenorigxxxxxC}{\ensuremath{string}}            
\newcommand{\hatcurCCraeccenorigxxxxxC}{\ensuremath{01^{\mathrm h}53^{\mathrm m}07.76{\mathrm s}}}                     
\newcommand{\hatcurCCdececcenorigxxxxxC}{\ensuremath{+52{\arcdeg}03{\arcmin}14.1{\arcsec}}}                    
\newcommand{\hatcurCCmageccenorigxxxxxC}{9.523}                          
\newcommand{\hatcurCCtwomasseccenorigxxxxxC}{2MASS~01530777+5203140}     
\newcommand{\hatcurCCgsceccenorigxxxxxC}{GSC~3292-01330}                 
\newcommand{\hatcurCCtassmveccenorigxxxxxC}{\ensuremath{9.5230\pm0.0010}} 
\newcommand{\hatcurCCtassmvshorteccenorigxxxxxC}{\ensuremath{9.5}}       
\newcommand{\hatcurCCtassmBeccenorigxxxxxC}{\ensuremath{10.18\pm0.11}}   
\newcommand{\hatcurCCtassmBshorteccenorigxxxxxC}{\ensuremath{10.2}}      
\newcommand{\hatcurCCtassmIeccenorigxxxxxC}{\ensuremath{9.077\pm0.042}}  
\newcommand{\hatcurCCtassmIshorteccenorigxxxxxC}{\ensuremath{9.1}}       
\newcommand{\hatcurCCtassmgeccenorigxxxxxC}{\ensuremath{9.7580\pm0.0010}} 
\newcommand{\hatcurCCtassmgshorteccenorigxxxxxC}{\ensuremath{9.8}}       
\newcommand{\hatcurCCtassmreccenorigxxxxxC}{\ensuremath{9.5000\pm0.0010}} 
\newcommand{\hatcurCCtassmrshorteccenorigxxxxxC}{\ensuremath{9.5}}       
\newcommand{\hatcurCCtassmieccenorigxxxxxC}{\ensuremath{9.3160\pm0.0010}} 
\newcommand{\hatcurCCtassmishorteccenorigxxxxxC}{\ensuremath{9.3}}       
\newcommand{\hatcurCCtwomassJmageccenorigxxxxxC}{\ensuremath{8.677\pm0.052}} 
\newcommand{\hatcurCCtwomassHmageccenorigxxxxxC}{\ensuremath{8.396\pm0.029}} 
\newcommand{\hatcurCCtwomassKmageccenorigxxxxxC}{\ensuremath{8.368\pm0.031}} 
\newcommand{\hatcurCCcitJmageccenorigxxxxxC}{\ensuremath{8.697\pm0.050}} 
\newcommand{\hatcurCCcitHmageccenorigxxxxxC}{\ensuremath{8.392\pm0.029}} 
\newcommand{\hatcurCCcitKmageccenorigxxxxxC}{\ensuremath{8.392\pm0.031}} 
\newcommand{\hatcurCCbbJmageccenorigxxxxxC}{\ensuremath{8.742\pm0.055}}  
\newcommand{\hatcurCCbbHmageccenorigxxxxxC}{\ensuremath{8.412\pm0.030}}  
\newcommand{\hatcurCCbbKmageccenorigxxxxxC}{\ensuremath{8.412\pm0.031}}  
\newcommand{\hatcurCCesoJmageccenorigxxxxxC}{\ensuremath{8.742\pm0.055}} 
\newcommand{\hatcurCCesoHmageccenorigxxxxxC}{\ensuremath{8.406\pm0.033}} 
\newcommand{\hatcurCCesoKmageccenorigxxxxxC}{\ensuremath{8.411\pm0.032}} 
\newcommand{\hatcurCCesoJHmageccenorigxxxxxC}{\ensuremath{0.337\pm0.063}} 
\newcommand{\hatcurCCesoJKmageccenorigxxxxxC}{\ensuremath{0.332\pm0.064}} 
\newcommand{\hatcurCCesoHKmageccenorigxxxxxC}{\ensuremath{-0.005\pm0.047}} 
\newcommand{\hatcurLCdipeccenorigxxxxxC}{\ensuremath{5.8}}               
\newcommand{\hatcurLCrprstareccenorigxxxxxC}{\ensuremath{0.0674\pm0.0020}} 
\newcommand{\hatcurLCbsqeccenorigxxxxxC}{\ensuremath{0.141_{-0.098}^{+0.131}}} 
\newcommand{\hatcurLCimpeccenorigxxxxxC}{\ensuremath{0.38_{-0.17}^{+0.15}}} 
\newcommand{\hatcurLCzetaeccenorigxxxxxC}{\ensuremath{10.608\pm0.098}}   
\newcommand{\hatcurLCdureccenorigxxxxxC}{\ensuremath{0.2035\pm0.0028}}   
\newcommand{\hatcurLCdurshorteccenorigxxxxxC}{\ensuremath{0.2035}}       
\newcommand{\hatcurLCdurhreccenorigxxxxxC}{\ensuremath{4.883\pm0.066}}   
\newcommand{\hatcurLCdurhrshorteccenorigxxxxxC}{\ensuremath{4.883}}      
\newcommand{\hatcurLCqeccenorigxxxxxC}{\ensuremath{0.04240\pm0.00058}}   
\newcommand{\hatcurLCqshorteccenorigxxxxxC}{\ensuremath{0.042}}          
\newcommand{\hatcurLCingdureccenorigxxxxxC}{\ensuremath{0.0148\pm0.0025}} 
\newcommand{\hatcurLCPeccenorigxxxxxC}{\ensuremath{4.7947700\pm0.0000078}} 
\newcommand{\hatcurLCPprececcenorigxxxxxC}{\ensuremath{4.7947700}}       
\newcommand{\hatcurLCPshorteccenorigxxxxxC}{\ensuremath{4.7948}}         
\newcommand{\hatcurLCTeccenorigxxxxxC}{\ensuremath{2456745.09705\pm0.00091}} 
\newcommand{\hatcurLCTAeccenorigxxxxxC}{\ensuremath{2454764.8569\pm0.0033}} 
\newcommand{\hatcurLCTBeccenorigxxxxxC}{\ensuremath{2456941.6826\pm0.0010}} 
\newcommand{\hatcurLCrhoeccenorigxxxxxC}{\ensuremath{0.34\pm0.11}}       
\newcommand{\hatcurSMEiteffeccenorigxxxxxC}{\ensuremath{6223\pm50}}      
\newcommand{\hatcurSMEizfeheccenorigxxxxxC}{\ensuremath{-0.404\pm0.080}} 
\newcommand{\hatcurSMEizfehshorteccenorigxxxxxC}{\ensuremath{-0.40}}     
\newcommand{\hatcurSMEiloggeccenorigxxxxxC}{\ensuremath{3.59\pm0.10}}    
\newcommand{\hatcurSMEivsineccenorigxxxxxC}{\ensuremath{10.71\pm0.50}}   
\newcommand{\hatcurSMEivmaceccenorigxxxxxC}{\ensuremath{0.0}}            
\newcommand{\hatcurSMEivmiceccenorigxxxxxC}{\ensuremath{0.0}}            
\newcommand{\hatcurSMEiiteffeccenorigxxxxxC}{\ensuremath{6462\pm50}}     
\newcommand{\hatcurSMEiizfeheccenorigxxxxxC}{\ensuremath{-0.237\pm0.080}} 
\newcommand{\hatcurSMEiizfehshorteccenorigxxxxxC}{\ensuremath{-0.237}}   
\newcommand{\hatcurSMEiiloggeccenorigxxxxxC}{\ensuremath{4.052\pm0.049}} 
\newcommand{\hatcurSMEiivsineccenorigxxxxxC}{\ensuremath{10.42\pm0.50}}  
\newcommand{\hatcurLBizeccenorigxxxxxC}{\ensuremath{0.1281}}             
\newcommand{\hatcurLBiizeccenorigxxxxxC}{\ensuremath{0.3535}}            
\newcommand{\hatcurLBiieccenorigxxxxxC}{\ensuremath{0.1761}}             
\newcommand{\hatcurLBiiieccenorigxxxxxC}{\ensuremath{0.3644}}            
\newcommand{\hatcurLBiIeccenorigxxxxxC}{\ensuremath{0.1586}}             
\newcommand{\hatcurLBiiIeccenorigxxxxxC}{\ensuremath{0.3618}}            
\newcommand{\hatcurLBigeccenorigxxxxxC}{\ensuremath{0.4024}}             
\newcommand{\hatcurLBiigeccenorigxxxxxC}{\ensuremath{0.3343}}            
\newcommand{\hatcurLBireccenorigxxxxxC}{\ensuremath{0.2448}}             
\newcommand{\hatcurLBiireccenorigxxxxxC}{\ensuremath{0.3779}}            
\newcommand{\hatcurLBiReccenorigxxxxxC}{\ensuremath{0.2249}}             
\newcommand{\hatcurLBiiReccenorigxxxxxC}{\ensuremath{0.3760}}            
\newcommand{\hatcurLBikepeccenorigxxxxxC}{\ensuremath{0.1000}}           
\newcommand{\hatcurLBiikepeccenorigxxxxxC}{\ensuremath{0.1000}}          
\newcommand{\hatcurISOmeccenorigxxxxxC}{\ensuremath{1.312\pm0.082}}      
\newcommand{\hatcurISOmshorteccenorigxxxxxC}{\ensuremath{1.31}}          
\newcommand{\hatcurISOmlongeccenorigxxxxxC}{\ensuremath{1.312\pm0.082}}  
\newcommand{\hatcurISOreccenorigxxxxxC}{\ensuremath{1.75_{-0.15}^{+0.22}}} 
\newcommand{\hatcurISOrshorteccenorigxxxxxC}{\ensuremath{1.75}}          
\newcommand{\hatcurISOrlongeccenorigxxxxxC}{\ensuremath{1.75_{-0.15}^{+0.22}}} 
\newcommand{\hatcurISOrhoeccenorigxxxxxC}{\ensuremath{0.34\pm0.11}}      
\newcommand{\hatcurISOrholongeccenorigxxxxxC}{\ensuremath{0.34\pm0.11}}  
\newcommand{\hatcurISOloggeccenorigxxxxxC}{\ensuremath{4.068\pm0.080}}   
\newcommand{\hatcurISOlumeccenorigxxxxxC}{\ensuremath{4.76_{-0.83}^{+1.33}}} 
\newcommand{\hatcurISOlumshorteccenorigxxxxxC}{\ensuremath{4.76}}        
\newcommand{\hatcurISOmveccenorigxxxxxC}{\ensuremath{3.06\pm0.26}}       
\newcommand{\hatcurISOvieccenorigxxxxxC}{\ensuremath{0.506\pm0.012}}     
\newcommand{\hatcurISOageeccenorigxxxxxC}{\ensuremath{2.85\pm0.56}}      
\newcommand{\hatcurISOsigmaeccenorigxxxxxC}{\ensuremath{0.000200\pm0.000058}} 
\newcommand{\hatcurISOMJeccenorigxxxxxC}{\ensuremath{2.25\pm0.25}}       
\newcommand{\hatcurISOMHeccenorigxxxxxC}{\ensuremath{2.02\pm0.25}}       
\newcommand{\hatcurISOMKeccenorigxxxxxC}{\ensuremath{1.98\pm0.25}}       
\newcommand{\hatcurISOJKeccenorigxxxxxC}{\ensuremath{0.270\pm0.010}}     
\newcommand{\hatcurISOspececcenorigxxxxxC}{F}                            
\newcommand{\hatcurRVKeccenorigxxxxxC}{\ensuremath{50.1\pm4.8}}          
\newcommand{\hatcurRVrkeccenorigxxxxxC}{\ensuremath{0.11_{-0.20}^{+0.15}}} 
\newcommand{\hatcurRVrheccenorigxxxxxC}{\ensuremath{0.00\pm0.21}}        
\newcommand{\hatcurRVkeccenorigxxxxxC}{\ensuremath{0.022_{-0.038}^{+0.060}}} 
\newcommand{\hatcurRVheccenorigxxxxxC}{\ensuremath{0.000\pm0.077}}       
\newcommand{\hatcurRVtroneeccenorigxxxxxC}{\ensuremath{0\pm0}}           
\newcommand{\hatcurRVtrtwoeccenorigxxxxxC}{\ensuremath{0\pm0}}           
\newcommand{\hatcurRVgammaAeccenorigxxxxxC}{\ensuremath{8.7\pm5.6}}      
\newcommand{\hatcurRVjitterAeccenorigxxxxxC}{\ensuremath{13.5\pm4.9}}    
\newcommand{\hatcurRVjittertwosiglimAeccenorigxxxxxC}{\ensuremath{<22.6}} 
\newcommand{\hatcurRVfitrmsAeccenorigxxxxxC}{\ensuremath{0.0}}           
\newcommand{\hatcurRVgammaBeccenorigxxxxxC}{\ensuremath{53.4\pm6.8}}     
\newcommand{\hatcurRVjitterBeccenorigxxxxxC}{\ensuremath{0.1\pm4.3}}     
\newcommand{\hatcurRVjittertwosiglimBeccenorigxxxxxC}{\ensuremath{<10.8}} 
\newcommand{\hatcurRVfitrmsBeccenorigxxxxxC}{\ensuremath{0.0}}           
\newcommand{\hatcurRVgammaCeccenorigxxxxxC}{\ensuremath{6026.1\pm7.7}}   
\newcommand{\hatcurRVjitterCeccenorigxxxxxC}{\ensuremath{0.1\pm3.0}}     
\newcommand{\hatcurRVjittertwosiglimCeccenorigxxxxxC}{\ensuremath{<7.3}} 
\newcommand{\hatcurRVfitrmsCeccenorigxxxxxC}{\ensuremath{0.0}}           
\newcommand{\hatcurRVecceneccenorigxxxxxC}{\ensuremath{0.065\pm0.060}}   
\newcommand{\hatcurRVeccentwosiglimeccenorigxxxxxC}{\ensuremath{<0.193}} 
\newcommand{\hatcurRVomegaeccenorigxxxxxC}{\ensuremath{180\pm120}}       
\newcommand{\hatcurPPieccenorigxxxxxC}{\ensuremath{87.2\pm1.4}}          
\newcommand{\hatcurPPgeccenorigxxxxxC}{\ensuremath{9.4\pm2.3}}           
\newcommand{\hatcurPPloggeccenorigxxxxxC}{\ensuremath{2.975_{-0.118}^{+0.078}}} 
\newcommand{\hatcurPPareccenorigxxxxxC}{\ensuremath{7.47\pm0.74}}        
\newcommand{\hatcurPPareleccenorigxxxxxC}{\ensuremath{0.0609\pm0.0013}}  
\newcommand{\hatcurPPrhoeccenorigxxxxxC}{\ensuremath{0.41\pm0.16}}       
\newcommand{\hatcurPPmeccenorigxxxxxC}{\ensuremath{0.494\pm0.051}}       
\newcommand{\hatcurPPmshorteccenorigxxxxxC}{\ensuremath{0.49}}           
\newcommand{\hatcurPPmlongeccenorigxxxxxC}{\ensuremath{0.494\pm0.051}}   
\newcommand{\hatcurPPmeeccenorigxxxxxC}{\ensuremath{157\pm16}}           
\newcommand{\hatcurPPmeshorteccenorigxxxxxC}{\ensuremath{157.1}}         
\newcommand{\hatcurPPmelongeccenorigxxxxxC}{\ensuremath{157\pm16}}       
\newcommand{\hatcurPPreccenorigxxxxxC}{\ensuremath{1.14_{-0.11}^{+0.17}}} 
\newcommand{\hatcurPPrshorteccenorigxxxxxC}{\ensuremath{1.14}}           
\newcommand{\hatcurPPrlongeccenorigxxxxxC}{\ensuremath{1.14_{-0.11}^{+0.17}}} 
\newcommand{\hatcurPPreeccenorigxxxxxC}{\ensuremath{12.8_{-1.3}^{+1.9}}} 
\newcommand{\hatcurPPreshorteccenorigxxxxxC}{\ensuremath{12.8}}          
\newcommand{\hatcurPPrelongeccenorigxxxxxC}{\ensuremath{12.8_{-1.3}^{+1.9}}} 
\newcommand{\hatcurPPmrcorreccenorigxxxxxC}{\ensuremath{0.28}}           
\newcommand{\hatcurPPteffeccenorigxxxxxC}{\ensuremath{1672_{-66}^{+91}}} 
\newcommand{\hatcurPPthetaeccenorigxxxxxC}{\ensuremath{0.0401\pm0.0064}} 
\newcommand{\hatcurPPfluxperieccenorigxxxxxC}{\ensuremath{2.02_{-0.33}^{+0.56}}} 
\newcommand{\hatcurPPfluxperidimeccenorigxxxxxC}{\ensuremath{9}}         
\newcommand{\hatcurPPfluxapeccenorigxxxxxC}{\ensuremath{1.58\pm0.35}}    
\newcommand{\hatcurPPfluxapdimeccenorigxxxxxC}{\ensuremath{9}}           
\newcommand{\hatcurPPfluxavgeccenorigxxxxxC}{\ensuremath{1.76_{-0.26}^{+0.42}}} 
\newcommand{\hatcurPPfluxavgdimeccenorigxxxxxC}{\ensuremath{9}}          
\newcommand{\hatcurPPfluxavglogeccenorigxxxxxC}{\ensuremath{9.246_{-0.070}^{+0.092}}} 
\newcommand{\hatcurXsecphaseeccenorigxxxxxC}{\ensuremath{0.514\pm0.032}} 
\newcommand{\hatcurXsecondaryeccenorigxxxxxC}{\ensuremath{2456747.56\pm0.15}} 
\newcommand{\hatcurXsecdureccenorigxxxxxC}{\ensuremath{0.203\pm0.025}}   
\newcommand{\hatcurXsecingdureccenorigxxxxxC}{\ensuremath{0.0146\pm0.0048}} 
\newcommand{\hatcurPPphiconjeccenorigxxxxxC}{\ensuremath{0.12_{-0.34}^{+0.23}}} 
\newcommand{\hatcurPPperieccenorigxxxxxC}{\ensuremath{2456744.5\pm1.2}}  
\newcommand{\hatcurPPaequiveccenorigxxxxxC}{\ensuremath{0.0279\pm0.0028}} 
\newcommand{\hatcurPPtcirceccenorigxxxxxC}{\ensuremath{800\pm450}}       
\newcommand{\hatcurPPtinfalleccenorigxxxxxC}{\ensuremath{2800_{-1100}^{+1500}}} 
\newcommand{\hatcurXdisteccenorigxxxxxC}{\ensuremath{194_{-17}^{+25}}}   
\newcommand{\hatcurXAveccenorigxxxxxC}{\ensuremath{0.028_{-0.028}^{+0.046}}} 
\newcommand{\hatcurXdistredeccenorigxxxxxC}{\ensuremath{193_{-17}^{+25}}} 
\newcommand{\hatcurXEBVeccenorigxxxxxC}{\ensuremath{0.0090_{-0.0090}^{+0.0150}}} 
\newcommand{\hatcurXmvisoredeccenorigxxxxxC}{\ensuremath{9.5230\pm0.0010}} 
\newcommand{\hatcurXmiisoredeccenorigxxxxxC}{\ensuremath{9.0000_{-0.0130}^{+0.0100}}} 
\newcommand{\hatcurXmjisoredeccenorigxxxxxC}{\ensuremath{8.681\pm0.016}} 
\newcommand{\hatcurXmhisoredeccenorigxxxxxC}{\ensuremath{8.448\pm0.021}} 
\newcommand{\hatcurXmkisoredeccenorigxxxxxC}{\ensuremath{8.405\pm0.022}} 
\newcommand{\hatcurXviisoredeccenorigxxxxxC}{\ensuremath{0.523\pm0.012}} 
\newcommand{\hatcurXvkisoredeccenorigxxxxxC}{\ensuremath{1.118\pm0.022}} 
\newcommand{\hatcurXjhisoredeccenorigxxxxxC}{\ensuremath{0.2320\pm0.0074}} 
\newcommand{\hatcurXjkisoredeccenorigxxxxxC}{\ensuremath{0.2760\pm0.0069}} 
\newcommand{\hatcurCCpmraeccenorigxxxxxC}{\ensuremath{25.50\pm0.60}}     
\newcommand{\hatcurCCpmdececcenorigxxxxxC}{\ensuremath{5.50\pm0.70}}     
\newcommand{\hatcurCCpmeccenorigxxxxxC}{\ensuremath{26.09\pm0.92}}       

%% file: HTR094-024.eccen.orig.tex
\newcommand{\hatcurhtreccenorigxxxxxD}{HTR094-024}                       
\newcommand{\hatcurfieldeccenorigxxxxxD}{\ensuremath{string}}            
\newcommand{\hatcurCCraeccenorigxxxxxD}{\ensuremath{05^{\mathrm h}01^{\mathrm m}55.27{\mathrm s}}}                     
\newcommand{\hatcurCCdececcenorigxxxxxD}{\ensuremath{+50{\arcdeg}07{\arcmin}52.6{\arcsec}}}                    
\newcommand{\hatcurCCmageccenorigxxxxxD}{13.188}                         
\newcommand{\hatcurCCtwomasseccenorigxxxxxD}{2MASS~05015525+5007526}     
\newcommand{\hatcurCCgsceccenorigxxxxxD}{GSC~3352-00595}                 
\newcommand{\hatcurCCtassmveccenorigxxxxxD}{\ensuremath{13.188\pm0.029}} 
\newcommand{\hatcurCCtassmvshorteccenorigxxxxxD}{\ensuremath{13.2}}      
\newcommand{\hatcurCCtassmBeccenorigxxxxxD}{\ensuremath{14.040\pm0.056}} 
\newcommand{\hatcurCCtassmBshorteccenorigxxxxxD}{\ensuremath{14.0}}      
\newcommand{\hatcurCCtassmIeccenorigxxxxxD}{\ensuremath{12.05\pm0.12}}   
\newcommand{\hatcurCCtassmIshorteccenorigxxxxxD}{\ensuremath{12.1}}      
\newcommand{\hatcurCCtassmgeccenorigxxxxxD}{\ensuremath{13.550\pm0.045}} 
\newcommand{\hatcurCCtassmgshorteccenorigxxxxxD}{\ensuremath{13.6}}      
\newcommand{\hatcurCCtassmreccenorigxxxxxD}{\ensuremath{12.915\pm0.033}} 
\newcommand{\hatcurCCtassmrshorteccenorigxxxxxD}{\ensuremath{12.9}}      
\newcommand{\hatcurCCtassmieccenorigxxxxxD}{\ensuremath{12.675\pm0.051}} 
\newcommand{\hatcurCCtassmishorteccenorigxxxxxD}{\ensuremath{12.7}}      
\newcommand{\hatcurCCtwomassJmageccenorigxxxxxD}{\ensuremath{11.598\pm0.021}} 
\newcommand{\hatcurCCtwomassHmageccenorigxxxxxD}{\ensuremath{11.263\pm0.029}} 
\newcommand{\hatcurCCtwomassKmageccenorigxxxxxD}{\ensuremath{11.141\pm0.020}} 
\newcommand{\hatcurCCcitJmageccenorigxxxxxD}{\ensuremath{11.610\pm0.021}} 
\newcommand{\hatcurCCcitHmageccenorigxxxxxD}{\ensuremath{11.257\pm0.029}} 
\newcommand{\hatcurCCcitKmageccenorigxxxxxD}{\ensuremath{11.165\pm0.020}} 
\newcommand{\hatcurCCbbJmageccenorigxxxxxD}{\ensuremath{11.667\pm0.023}} 
\newcommand{\hatcurCCbbHmageccenorigxxxxxD}{\ensuremath{11.279\pm0.030}} 
\newcommand{\hatcurCCbbKmageccenorigxxxxxD}{\ensuremath{11.185\pm0.020}} 
\newcommand{\hatcurCCesoJmageccenorigxxxxxD}{\ensuremath{11.670\pm0.025}} 
\newcommand{\hatcurCCesoHmageccenorigxxxxxD}{\ensuremath{11.276\pm0.035}} 
\newcommand{\hatcurCCesoKmageccenorigxxxxxD}{\ensuremath{11.184\pm0.021}} 
\newcommand{\hatcurCCesoJHmageccenorigxxxxxD}{\ensuremath{0.393\pm0.041}} 
\newcommand{\hatcurCCesoJKmageccenorigxxxxxD}{\ensuremath{0.487\pm0.032}} 
\newcommand{\hatcurCCesoHKmageccenorigxxxxxD}{\ensuremath{0.093\pm0.041}} 
\newcommand{\hatcurLCdipeccenorigxxxxxD}{\ensuremath{12.7}}              
\newcommand{\hatcurLCrprstareccenorigxxxxxD}{\ensuremath{0.1003\pm0.0027}} 
\newcommand{\hatcurLCbsqeccenorigxxxxxD}{\ensuremath{0.641_{-0.041}^{+0.038}}} 
\newcommand{\hatcurLCimpeccenorigxxxxxD}{\ensuremath{0.801_{-0.026}^{+0.023}}} 
\newcommand{\hatcurLCzetaeccenorigxxxxxD}{\ensuremath{36.16\pm0.85}}     
\newcommand{\hatcurLCdureccenorigxxxxxD}{\ensuremath{0.0699\pm0.0020}}   
\newcommand{\hatcurLCdurshorteccenorigxxxxxD}{\ensuremath{0.0699}}       
\newcommand{\hatcurLCdurhreccenorigxxxxxD}{\ensuremath{1.678\pm0.048}}   
\newcommand{\hatcurLCdurhrshorteccenorigxxxxxD}{\ensuremath{1.678}}      
\newcommand{\hatcurLCqeccenorigxxxxxD}{\ensuremath{0.0368\pm0.0011}}     
\newcommand{\hatcurLCqshorteccenorigxxxxxD}{\ensuremath{0.037}}          
\newcommand{\hatcurLCingdureccenorigxxxxxD}{\ensuremath{0.0160\pm0.0021}} 
\newcommand{\hatcurLCPeccenorigxxxxxD}{\ensuremath{1.9023143\pm0.0000023}} 
\newcommand{\hatcurLCPprececcenorigxxxxxD}{\ensuremath{1.9023143}}       
\newcommand{\hatcurLCPshorteccenorigxxxxxD}{\ensuremath{1.9023}}         
\newcommand{\hatcurLCTeccenorigxxxxxD}{\ensuremath{2456808.74284\pm0.00057}} 
\newcommand{\hatcurLCTAeccenorigxxxxxD}{\ensuremath{2454392.8037\pm0.0029}} 
\newcommand{\hatcurLCTBeccenorigxxxxxD}{\ensuremath{2456941.90484\pm0.00062}} 
\newcommand{\hatcurLChatnetmAeccenorigxxxxxD}{\ensuremath{12.61786\pm0.00022}} 
\newcommand{\hatcurLCiblendAeccenorigxxxxxD}{\ensuremath{0.88\pm0.11}}   
\newcommand{\hatcurLChatnetmBeccenorigxxxxxD}{\ensuremath{12.80529\pm0.00011}} 
\newcommand{\hatcurLCiblendBeccenorigxxxxxD}{\ensuremath{0.926\pm0.065}} 
\newcommand{\hatcurLCrhoeccenorigxxxxxD}{\ensuremath{2.39_{-0.63}^{+0.84}}} 
\newcommand{\hatcurSMEiteffeccenorigxxxxxD}{\ensuremath{5546\pm50}}      
\newcommand{\hatcurSMEizfeheccenorigxxxxxD}{\ensuremath{0.399\pm0.080}}  
\newcommand{\hatcurSMEizfehshorteccenorigxxxxxD}{\ensuremath{0.40}}      
\newcommand{\hatcurSMEiloggeccenorigxxxxxD}{\ensuremath{4.43\pm0.10}}    
\newcommand{\hatcurSMEivsineccenorigxxxxxD}{\ensuremath{3.72\pm0.50}}    
\newcommand{\hatcurSMEivmaceccenorigxxxxxD}{\ensuremath{0.0}}            
\newcommand{\hatcurSMEivmiceccenorigxxxxxD}{\ensuremath{0.0}}            
\newcommand{\hatcurSMEiiteffeccenorigxxxxxD}{\ensuremath{5551\pm50}}     
\newcommand{\hatcurSMEiizfeheccenorigxxxxxD}{\ensuremath{0.396\pm0.080}} 
\newcommand{\hatcurSMEiizfehshorteccenorigxxxxxD}{\ensuremath{0.396}}    
\newcommand{\hatcurSMEiiloggeccenorigxxxxxD}{\ensuremath{4.468\pm0.035}} 
\newcommand{\hatcurSMEiivsineccenorigxxxxxD}{\ensuremath{3.69\pm0.50}}   
\newcommand{\hatcurLBizeccenorigxxxxxD}{\ensuremath{0.2481}}             
\newcommand{\hatcurLBiizeccenorigxxxxxD}{\ensuremath{0.3193}}            
\newcommand{\hatcurLBiieccenorigxxxxxD}{\ensuremath{0.3253}}             
\newcommand{\hatcurLBiiieccenorigxxxxxD}{\ensuremath{0.3085}}            
\newcommand{\hatcurLBiIeccenorigxxxxxD}{\ensuremath{0.3000}}             
\newcommand{\hatcurLBiiIeccenorigxxxxxD}{\ensuremath{0.3124}}            
\newcommand{\hatcurLBigeccenorigxxxxxD}{\ensuremath{0.6560}}             
\newcommand{\hatcurLBiigeccenorigxxxxxD}{\ensuremath{0.1596}}            
\newcommand{\hatcurLBireccenorigxxxxxD}{\ensuremath{0.4330}}             
\newcommand{\hatcurLBiireccenorigxxxxxD}{\ensuremath{0.2849}}            
\newcommand{\hatcurLBiReccenorigxxxxxD}{\ensuremath{0.4032}}             
\newcommand{\hatcurLBiiReccenorigxxxxxD}{\ensuremath{0.2924}}            
\newcommand{\hatcurLBikepeccenorigxxxxxD}{\ensuremath{0.1000}}           
\newcommand{\hatcurLBiikepeccenorigxxxxxD}{\ensuremath{0.1000}}          
\newcommand{\hatcurISOmeccenorigxxxxxD}{\ensuremath{1.043\pm0.023}}      
\newcommand{\hatcurISOmshorteccenorigxxxxxD}{\ensuremath{1.04}}          
\newcommand{\hatcurISOmlongeccenorigxxxxxD}{\ensuremath{1.043\pm0.023}}  
\newcommand{\hatcurISOreccenorigxxxxxD}{\ensuremath{0.990_{-0.039}^{+0.069}}} 
\newcommand{\hatcurISOrshorteccenorigxxxxxD}{\ensuremath{0.99}}          
\newcommand{\hatcurISOrlongeccenorigxxxxxD}{\ensuremath{0.990_{-0.039}^{+0.069}}} 
\newcommand{\hatcurISOrhoeccenorigxxxxxD}{\ensuremath{1.52_{-0.29}^{+0.19}}} 
\newcommand{\hatcurISOrholongeccenorigxxxxxD}{\ensuremath{1.52_{-0.29}^{+0.19}}} 
\newcommand{\hatcurISOloggeccenorigxxxxxD}{\ensuremath{4.466\pm0.053}}   
\newcommand{\hatcurISOlumeccenorigxxxxxD}{\ensuremath{0.833_{-0.083}^{+0.128}}} 
\newcommand{\hatcurISOlumshorteccenorigxxxxxD}{\ensuremath{0.83}}        
\newcommand{\hatcurISOmveccenorigxxxxxD}{\ensuremath{5.06\pm0.16}}       
\newcommand{\hatcurISOvieccenorigxxxxxD}{\ensuremath{0.781\pm0.016}}     
\newcommand{\hatcurISOageeccenorigxxxxxD}{\ensuremath{2.7\pm2.0}}        
\newcommand{\hatcurISOsigmaeccenorigxxxxxD}{\ensuremath{0.00210\pm0.00049}} 
\newcommand{\hatcurISOMJeccenorigxxxxxD}{\ensuremath{3.79\pm0.14}}       
\newcommand{\hatcurISOMHeccenorigxxxxxD}{\ensuremath{3.42\pm0.14}}       
\newcommand{\hatcurISOMKeccenorigxxxxxD}{\ensuremath{3.36\pm0.14}}       
\newcommand{\hatcurISOJKeccenorigxxxxxD}{\ensuremath{0.440\pm0.010}}     
\newcommand{\hatcurISOspececcenorigxxxxxD}{G}                            
\newcommand{\hatcurRVKeccenorigxxxxxD}{\ensuremath{174.2\pm8.5}}         
\newcommand{\hatcurRVrkeccenorigxxxxxD}{\ensuremath{-0.12\pm0.16}}       
\newcommand{\hatcurRVrheccenorigxxxxxD}{\ensuremath{-0.01\pm0.20}}       
\newcommand{\hatcurRVkeccenorigxxxxxD}{\ensuremath{-0.028_{-0.076}^{+0.035}}} 
\newcommand{\hatcurRVheccenorigxxxxxD}{\ensuremath{-0.000_{-0.068}^{+0.051}}} 
\newcommand{\hatcurRVtroneeccenorigxxxxxD}{\ensuremath{0\pm0}}           
\newcommand{\hatcurRVtrtwoeccenorigxxxxxD}{\ensuremath{0\pm0}}           
\newcommand{\hatcurRVgammaAeccenorigxxxxxD}{\ensuremath{-49\pm14}}       
\newcommand{\hatcurRVjitterAeccenorigxxxxxD}{\ensuremath{9\pm15}}        
\newcommand{\hatcurRVjittertwosiglimAeccenorigxxxxxD}{\ensuremath{<44.3}} 
\newcommand{\hatcurRVfitrmsAeccenorigxxxxxD}{\ensuremath{0.0}}           
\newcommand{\hatcurRVgammaBeccenorigxxxxxD}{\ensuremath{209\pm13}}       
\newcommand{\hatcurRVjitterBeccenorigxxxxxD}{\ensuremath{42\pm15}}       
\newcommand{\hatcurRVjittertwosiglimBeccenorigxxxxxD}{\ensuremath{<72.6}} 
\newcommand{\hatcurRVfitrmsBeccenorigxxxxxD}{\ensuremath{0.0}}           
\newcommand{\hatcurRVecceneccenorigxxxxxD}{\ensuremath{0.080\pm0.057}}   
\newcommand{\hatcurRVeccentwosiglimeccenorigxxxxxD}{\ensuremath{<0.175}} 
\newcommand{\hatcurRVomegaeccenorigxxxxxD}{\ensuremath{182\pm75}}        
\newcommand{\hatcurPPieccenorigxxxxxD}{\ensuremath{83.12_{-0.97}^{+0.52}}} 
\newcommand{\hatcurPPgeccenorigxxxxxD}{\ensuremath{29.1_{-4.4}^{+3.3}}}  
\newcommand{\hatcurPPloggeccenorigxxxxxD}{\ensuremath{3.463_{-0.071}^{+0.046}}} 
\newcommand{\hatcurPPareccenorigxxxxxD}{\ensuremath{6.63_{-0.45}^{+0.26}}} 
\newcommand{\hatcurPPareleccenorigxxxxxD}{\ensuremath{0.03047\pm0.00023}} 
\newcommand{\hatcurPPrhoeccenorigxxxxxD}{\ensuremath{1.50\pm0.28}}       
\newcommand{\hatcurPPmeccenorigxxxxxD}{\ensuremath{1.099\pm0.055}}       
\newcommand{\hatcurPPmshorteccenorigxxxxxD}{\ensuremath{1.10}}           
\newcommand{\hatcurPPmlongeccenorigxxxxxD}{\ensuremath{1.099\pm0.055}}   
\newcommand{\hatcurPPmeeccenorigxxxxxD}{\ensuremath{349\pm17}}           
\newcommand{\hatcurPPmeshorteccenorigxxxxxD}{\ensuremath{349.3}}         
\newcommand{\hatcurPPmelongeccenorigxxxxxD}{\ensuremath{349\pm17}}       
\newcommand{\hatcurPPreccenorigxxxxxD}{\ensuremath{0.968_{-0.050}^{+0.073}}} 
\newcommand{\hatcurPPrshorteccenorigxxxxxD}{\ensuremath{0.97}}           
\newcommand{\hatcurPPrlongeccenorigxxxxxD}{\ensuremath{0.968_{-0.050}^{+0.073}}} 
\newcommand{\hatcurPPreeccenorigxxxxxD}{\ensuremath{10.85_{-0.56}^{+0.82}}} 
\newcommand{\hatcurPPreshorteccenorigxxxxxD}{\ensuremath{10.9}}          
\newcommand{\hatcurPPrelongeccenorigxxxxxD}{\ensuremath{10.85_{-0.56}^{+0.82}}} 
\newcommand{\hatcurPPmrcorreccenorigxxxxxD}{\ensuremath{0.06}}           
\newcommand{\hatcurPPteffeccenorigxxxxxD}{\ensuremath{1527_{-38}^{+56}}} 
\newcommand{\hatcurPPthetaeccenorigxxxxxD}{\ensuremath{0.0657_{-0.0059}^{+0.0045}}} 
\newcommand{\hatcurPPfluxperieccenorigxxxxxD}{\ensuremath{1.41_{-0.16}^{+0.33}}} 
\newcommand{\hatcurPPfluxperidimeccenorigxxxxxD}{\ensuremath{9}}         
\newcommand{\hatcurPPfluxapeccenorigxxxxxD}{\ensuremath{1.07\pm0.15}}    
\newcommand{\hatcurPPfluxapdimeccenorigxxxxxD}{\ensuremath{9}}           
\newcommand{\hatcurPPfluxavgeccenorigxxxxxD}{\ensuremath{1.23_{-0.12}^{+0.19}}} 
\newcommand{\hatcurPPfluxavgdimeccenorigxxxxxD}{\ensuremath{9}}          
\newcommand{\hatcurPPfluxavglogeccenorigxxxxxD}{\ensuremath{9.089_{-0.044}^{+0.063}}} 
\newcommand{\hatcurXsecphaseeccenorigxxxxxD}{\ensuremath{0.482\pm0.034}} 
\newcommand{\hatcurXsecondaryeccenorigxxxxxD}{\ensuremath{2456809.660\pm0.065}} 
\newcommand{\hatcurXsecdureccenorigxxxxxD}{\ensuremath{0.0696\pm0.0047}} 
\newcommand{\hatcurXsecingdureccenorigxxxxxD}{\ensuremath{0.0151\pm0.0086}} 
\newcommand{\hatcurPPphiconjeccenorigxxxxxD}{\ensuremath{-0.20_{-0.20}^{+0.42}}} 
\newcommand{\hatcurPPperieccenorigxxxxxD}{\ensuremath{2456809.12\pm0.55}} 
\newcommand{\hatcurPPaequiveccenorigxxxxxD}{\ensuremath{0.0334_{-0.0023}^{+0.0017}}} 
\newcommand{\hatcurPPtcirceccenorigxxxxxD}{\ensuremath{63\pm19}}         
\newcommand{\hatcurPPtinfalleccenorigxxxxxD}{\ensuremath{220_{-66}^{+47}}} 
\newcommand{\hatcurXdisteccenorigxxxxxD}{\ensuremath{368_{-16}^{+26}}}   
\newcommand{\hatcurXAveccenorigxxxxxD}{\ensuremath{0.328\pm0.061}}       
\newcommand{\hatcurXdistredeccenorigxxxxxD}{\ensuremath{361_{-15}^{+25}}} 
\newcommand{\hatcurXEBVeccenorigxxxxxD}{\ensuremath{0.106\pm0.020}}      
\newcommand{\hatcurXmvisoredeccenorigxxxxxD}{\ensuremath{13.183\pm0.028}} 
\newcommand{\hatcurXmiisoredeccenorigxxxxxD}{\ensuremath{12.231\pm0.017}} 
\newcommand{\hatcurXmjisoredeccenorigxxxxxD}{\ensuremath{11.672\pm0.014}} 
\newcommand{\hatcurXmhisoredeccenorigxxxxxD}{\ensuremath{11.269\pm0.015}} 
\newcommand{\hatcurXmkisoredeccenorigxxxxxD}{\ensuremath{11.181\pm0.016}} 
\newcommand{\hatcurXviisoredeccenorigxxxxxD}{\ensuremath{0.952\pm0.023}} 
\newcommand{\hatcurXvkisoredeccenorigxxxxxD}{\ensuremath{2.001\pm0.034}} 
\newcommand{\hatcurXjhisoredeccenorigxxxxxD}{\ensuremath{0.4040\pm0.0085}} 
\newcommand{\hatcurXjkisoredeccenorigxxxxxD}{\ensuremath{0.4910\pm0.0093}} 
\newcommand{\hatcurCCpmraeccenorigxxxxxD}{\ensuremath{-9.5\pm1.7}}       
\newcommand{\hatcurCCpmdececcenorigxxxxxD}{\ensuremath{-19.7\pm2.2}}     
\newcommand{\hatcurCCpmeccenorigxxxxxD}{\ensuremath{21.9\pm2.8}}         

%% file: HTR130-001.eccen.orig.tex
\newcommand{\hatcurhtreccenorigxxxxxE}{HTR130-001}                         
\newcommand{\hatcurfieldeccenorigxxxxxE}{\ensuremath{string}}              
\newcommand{\hatcurCCraeccenorigxxxxxE}{\ensuremath{04^{\mathrm h}58^{\mathrm m}01.02{\mathrm s}}}                       
\newcommand{\hatcurCCdececcenorigxxxxxE}{\ensuremath{+48{\arcdeg}18{\arcmin}03.9{\arcsec}}}                      
\newcommand{\hatcurCCmageccenorigxxxxxE}{12.558}                           
\newcommand{\hatcurCCtwomasseccenorigxxxxxE}{2MASS~04580102+4818038}       
\newcommand{\hatcurCCgsceccenorigxxxxxE}{GSC~3348-01101}                   
\newcommand{\hatcurCCtassmveccenorigxxxxxE}{\ensuremath{12.558\pm0.035}}   
\newcommand{\hatcurCCtassmvshorteccenorigxxxxxE}{\ensuremath{12.6}}        
\newcommand{\hatcurCCtassmBeccenorigxxxxxE}{\ensuremath{13.493\pm0.065}}   
\newcommand{\hatcurCCtassmBshorteccenorigxxxxxE}{\ensuremath{13.5}}        
\newcommand{\hatcurCCtassmIeccenorigxxxxxE}{\ensuremath{11.68\pm0.11}}     
\newcommand{\hatcurCCtassmIshorteccenorigxxxxxE}{\ensuremath{11.7}}        
\newcommand{\hatcurCCtassmgeccenorigxxxxxE}{\ensuremath{13.10\pm0.17}}     
\newcommand{\hatcurCCtassmgshorteccenorigxxxxxE}{\ensuremath{13.1}}        
\newcommand{\hatcurCCtassmreccenorigxxxxxE}{\ensuremath{12.31\pm0.12}}     
\newcommand{\hatcurCCtassmrshorteccenorigxxxxxE}{\ensuremath{12.3}}        
\newcommand{\hatcurCCtassmieccenorigxxxxxE}{\ensuremath{12.04\pm0.12}}     
\newcommand{\hatcurCCtassmishorteccenorigxxxxxE}{\ensuremath{12.0}}        
\newcommand{\hatcurCCtwomassJmageccenorigxxxxxE}{\ensuremath{11.144\pm0.029}} 
\newcommand{\hatcurCCtwomassHmageccenorigxxxxxE}{\ensuremath{10.803\pm0.041}} 
\newcommand{\hatcurCCtwomassKmageccenorigxxxxxE}{\ensuremath{10.701\pm0.026}} 
\newcommand{\hatcurCCcitJmageccenorigxxxxxE}{\ensuremath{11.157\pm0.029}}  
\newcommand{\hatcurCCcitHmageccenorigxxxxxE}{\ensuremath{10.797\pm0.040}}  
\newcommand{\hatcurCCcitKmageccenorigxxxxxE}{\ensuremath{10.725\pm0.026}}  
\newcommand{\hatcurCCbbJmageccenorigxxxxxE}{\ensuremath{11.212\pm0.031}}   
\newcommand{\hatcurCCbbHmageccenorigxxxxxE}{\ensuremath{10.819\pm0.042}}   
\newcommand{\hatcurCCbbKmageccenorigxxxxxE}{\ensuremath{10.745\pm0.026}}   
\newcommand{\hatcurCCesoJmageccenorigxxxxxE}{\ensuremath{11.215\pm0.033}}  
\newcommand{\hatcurCCesoHmageccenorigxxxxxE}{\ensuremath{10.815\pm0.047}}  
\newcommand{\hatcurCCesoKmageccenorigxxxxxE}{\ensuremath{10.744\pm0.027}}  
\newcommand{\hatcurCCesoJHmageccenorigxxxxxE}{\ensuremath{0.399\pm0.055}}  
\newcommand{\hatcurCCesoJKmageccenorigxxxxxE}{\ensuremath{0.472\pm0.042}}  
\newcommand{\hatcurCCesoHKmageccenorigxxxxxE}{\ensuremath{0.073\pm0.054}}  
\newcommand{\hatcurLCdipeccenorigxxxxxE}{\ensuremath{9.3}}                 
\newcommand{\hatcurLCrprstareccenorigxxxxxE}{\ensuremath{0.0946\pm0.0021}} 
\newcommand{\hatcurLCbsqeccenorigxxxxxE}{\ensuremath{0.092_{-0.061}^{+0.076}}} 
\newcommand{\hatcurLCimpeccenorigxxxxxE}{\ensuremath{0.30_{-0.13}^{+0.11}}} 
\newcommand{\hatcurLCzetaeccenorigxxxxxE}{\ensuremath{16.89\pm0.12}}       
\newcommand{\hatcurLCdureccenorigxxxxxE}{\ensuremath{0.1307\pm0.0014}}     
\newcommand{\hatcurLCdurshorteccenorigxxxxxE}{\ensuremath{0.1307}}         
\newcommand{\hatcurLCdurhreccenorigxxxxxE}{\ensuremath{3.138\pm0.034}}     
\newcommand{\hatcurLCdurhrshorteccenorigxxxxxE}{\ensuremath{3.138}}        
\newcommand{\hatcurLCqeccenorigxxxxxE}{\ensuremath{0.04940\pm0.00055}}     
\newcommand{\hatcurLCqshorteccenorigxxxxxE}{\ensuremath{0.049}}            
\newcommand{\hatcurLCingdureccenorigxxxxxE}{\ensuremath{0.0123\pm0.0011}}  
\newcommand{\hatcurLCPeccenorigxxxxxE}{\ensuremath{2.6453233\pm0.0000039}} 
\newcommand{\hatcurLCPprececcenorigxxxxxE}{\ensuremath{2.6453233}}         
\newcommand{\hatcurLCPshorteccenorigxxxxxE}{\ensuremath{2.6453}}           
\newcommand{\hatcurLCTeccenorigxxxxxE}{\ensuremath{2457126.32500\pm0.00044}} 
\newcommand{\hatcurLCTAeccenorigxxxxxE}{\ensuremath{2456131.6834\pm0.0014}} 
\newcommand{\hatcurLCTBeccenorigxxxxxE}{\ensuremath{2457292.98036\pm0.00058}} 
\newcommand{\hatcurLChatnetmeccenorigxxxxxE}{\ensuremath{12.379220\pm0.000094}} 
\newcommand{\hatcurLCiblendeccenorigxxxxxE}{\ensuremath{0.831\pm0.053}}    
\newcommand{\hatcurLCrhoeccenorigxxxxxE}{\ensuremath{1.05_{-0.26}^{+0.49}}} 
\newcommand{\hatcurSMEiteffeccenorigxxxxxE}{\ensuremath{5617\pm50}}        
\newcommand{\hatcurSMEizfeheccenorigxxxxxE}{\ensuremath{0.461\pm0.080}}    
\newcommand{\hatcurSMEizfehshorteccenorigxxxxxE}{\ensuremath{0.46}}        
\newcommand{\hatcurSMEiloggeccenorigxxxxxE}{\ensuremath{4.31\pm0.10}}      
\newcommand{\hatcurSMEivsineccenorigxxxxxE}{\ensuremath{3.53\pm0.50}}      
\newcommand{\hatcurSMEivmaceccenorigxxxxxE}{\ensuremath{0.0}}              
\newcommand{\hatcurSMEivmiceccenorigxxxxxE}{\ensuremath{0.0}}              
\newcommand{\hatcurSMEiiteffeccenorigxxxxxE}{\ensuremath{5601\pm50}}       
\newcommand{\hatcurSMEiizfeheccenorigxxxxxE}{\ensuremath{0.449\pm0.080}}   
\newcommand{\hatcurSMEiizfehshorteccenorigxxxxxE}{\ensuremath{0.449}}      
\newcommand{\hatcurSMEiiloggeccenorigxxxxxE}{\ensuremath{4.305\pm0.030}}   
\newcommand{\hatcurSMEiivsineccenorigxxxxxE}{\ensuremath{3.55\pm0.50}}     
\newcommand{\hatcurLBizeccenorigxxxxxE}{\ensuremath{0.2402}}               
\newcommand{\hatcurLBiizeccenorigxxxxxE}{\ensuremath{0.3260}}              
\newcommand{\hatcurLBiieccenorigxxxxxE}{\ensuremath{0.3168}}               
\newcommand{\hatcurLBiiieccenorigxxxxxE}{\ensuremath{0.3162}}              
\newcommand{\hatcurLBiIeccenorigxxxxxE}{\ensuremath{0.2914}}               
\newcommand{\hatcurLBiiIeccenorigxxxxxE}{\ensuremath{0.3201}}              
\newcommand{\hatcurLBigeccenorigxxxxxE}{\ensuremath{0.6478}}               
\newcommand{\hatcurLBiigeccenorigxxxxxE}{\ensuremath{0.1670}}              
\newcommand{\hatcurLBireccenorigxxxxxE}{\ensuremath{0.4241}}               
\newcommand{\hatcurLBiireccenorigxxxxxE}{\ensuremath{0.2923}}              
\newcommand{\hatcurLBiReccenorigxxxxxE}{\ensuremath{0.3943}}               
\newcommand{\hatcurLBiiReccenorigxxxxxE}{\ensuremath{0.3000}}              
\newcommand{\hatcurLBikepeccenorigxxxxxE}{\ensuremath{0.1000}}             
\newcommand{\hatcurLBiikepeccenorigxxxxxE}{\ensuremath{0.1000}}            
\newcommand{\hatcurISOmeccenorigxxxxxE}{\ensuremath{1.085_{-0.029}^{+0.040}}} 
\newcommand{\hatcurISOmshorteccenorigxxxxxE}{\ensuremath{1.08}}            
\newcommand{\hatcurISOmlongeccenorigxxxxxE}{\ensuremath{1.085_{-0.029}^{+0.040}}} 
\newcommand{\hatcurISOreccenorigxxxxxE}{\ensuremath{1.16\pm0.12}}          
\newcommand{\hatcurISOrshorteccenorigxxxxxE}{\ensuremath{1.16}}            
\newcommand{\hatcurISOrlongeccenorigxxxxxE}{\ensuremath{1.16\pm0.12}}      
\newcommand{\hatcurISOrhoeccenorigxxxxxE}{\ensuremath{0.97_{-0.22}^{+0.35}}} 
\newcommand{\hatcurISOrholongeccenorigxxxxxE}{\ensuremath{0.97_{-0.22}^{+0.35}}} 
\newcommand{\hatcurISOloggeccenorigxxxxxE}{\ensuremath{4.340\pm0.077}}     
\newcommand{\hatcurISOlumeccenorigxxxxxE}{\ensuremath{1.19\pm0.27}}        
\newcommand{\hatcurISOlumshorteccenorigxxxxxE}{\ensuremath{1.19}}          
\newcommand{\hatcurISOmveccenorigxxxxxE}{\ensuremath{4.66\pm0.23}}         
\newcommand{\hatcurISOvieccenorigxxxxxE}{\ensuremath{0.766\pm0.016}}       
\newcommand{\hatcurISOageeccenorigxxxxxE}{\ensuremath{4.96_{-1.66}^{+0.98}}} 
\newcommand{\hatcurISOsigmaeccenorigxxxxxE}{\ensuremath{0.00080\pm0.00027}} 
\newcommand{\hatcurISOMJeccenorigxxxxxE}{\ensuremath{3.42\pm0.22}}         
\newcommand{\hatcurISOMHeccenorigxxxxxE}{\ensuremath{3.06\pm0.22}}         
\newcommand{\hatcurISOMKeccenorigxxxxxE}{\ensuremath{3.00\pm0.22}}         
\newcommand{\hatcurISOJKeccenorigxxxxxE}{\ensuremath{0.420\pm0.010}}       
\newcommand{\hatcurISOspececcenorigxxxxxE}{G}                              
\newcommand{\hatcurRVKeccenorigxxxxxE}{\ensuremath{110\pm13}}              
\newcommand{\hatcurRVrkeccenorigxxxxxE}{\ensuremath{0.02\pm0.15}}          
\newcommand{\hatcurRVrheccenorigxxxxxE}{\ensuremath{-0.19_{-0.19}^{+0.27}}} 
\newcommand{\hatcurRVkeccenorigxxxxxE}{\ensuremath{0.003\pm0.050}}         
\newcommand{\hatcurRVheccenorigxxxxxE}{\ensuremath{-0.044_{-0.101}^{+0.058}}} 
\newcommand{\hatcurRVtroneeccenorigxxxxxE}{\ensuremath{0\pm0}}             
\newcommand{\hatcurRVtrtwoeccenorigxxxxxE}{\ensuremath{0\pm0}}             
\newcommand{\hatcurRVgammaeccenorigxxxxxE}{\ensuremath{126\pm11}}          
\newcommand{\hatcurRVjittereccenorigxxxxxE}{\ensuremath{33\pm10}}          
\newcommand{\hatcurRVjittertwosiglimeccenorigxxxxxE}{\ensuremath{<52.5}}   
\newcommand{\hatcurRVfitrmseccenorigxxxxxE}{\ensuremath{.1fym}}            %
\newcommand{\hatcurRVecceneccenorigxxxxxE}{\ensuremath{0.080\pm0.066}}     
\newcommand{\hatcurRVeccentwosiglimeccenorigxxxxxE}{\ensuremath{<0.211}}   
\newcommand{\hatcurRVomegaeccenorigxxxxxE}{\ensuremath{261\pm97}}          
\newcommand{\hatcurPPieccenorigxxxxxE}{\ensuremath{87.7\pm1.1}}            
\newcommand{\hatcurPPgeccenorigxxxxxE}{\ensuremath{17.0\pm4.0}}            
\newcommand{\hatcurPPloggeccenorigxxxxxE}{\ensuremath{3.23\pm0.11}}        
\newcommand{\hatcurPPareccenorigxxxxxE}{\ensuremath{7.09_{-0.57}^{+0.76}}} 
\newcommand{\hatcurPPareleccenorigxxxxxE}{\ensuremath{0.03846_{-0.00035}^{+0.00046}}} 
\newcommand{\hatcurPPrhoeccenorigxxxxxE}{\ensuremath{0.79_{-0.22}^{+0.31}}} 
\newcommand{\hatcurPPmeccenorigxxxxxE}{\ensuremath{0.787\pm0.097}}         
\newcommand{\hatcurPPmshorteccenorigxxxxxE}{\ensuremath{0.79}}             
\newcommand{\hatcurPPmlongeccenorigxxxxxE}{\ensuremath{0.787\pm0.097}}     
\newcommand{\hatcurPPmeeccenorigxxxxxE}{\ensuremath{250\pm31}}             
\newcommand{\hatcurPPmeshorteccenorigxxxxxE}{\ensuremath{250.3}}           
\newcommand{\hatcurPPmelongeccenorigxxxxxE}{\ensuremath{250\pm31}}         
\newcommand{\hatcurPPreccenorigxxxxxE}{\ensuremath{1.07\pm0.12}}           
\newcommand{\hatcurPPrshorteccenorigxxxxxE}{\ensuremath{1.07}}             
\newcommand{\hatcurPPrlongeccenorigxxxxxE}{\ensuremath{1.07\pm0.12}}       
\newcommand{\hatcurPPreeccenorigxxxxxE}{\ensuremath{12.0\pm1.3}}           
\newcommand{\hatcurPPreshorteccenorigxxxxxE}{\ensuremath{12.0}}            
\newcommand{\hatcurPPrelongeccenorigxxxxxE}{\ensuremath{12.0\pm1.3}}       
\newcommand{\hatcurPPmrcorreccenorigxxxxxE}{\ensuremath{0.09}}             
\newcommand{\hatcurPPteffeccenorigxxxxxE}{\ensuremath{1488\pm71}}          
\newcommand{\hatcurPPthetaeccenorigxxxxxE}{\ensuremath{0.0519\pm0.0085}}   
\newcommand{\hatcurPPfluxperieccenorigxxxxxE}{\ensuremath{1.29_{-0.12}^{+0.21}}} 
\newcommand{\hatcurPPfluxperidimeccenorigxxxxxE}{\ensuremath{9}}           
\newcommand{\hatcurPPfluxapeccenorigxxxxxE}{\ensuremath{9.8_{-3.0}^{+2.2}}} 
\newcommand{\hatcurPPfluxapdimeccenorigxxxxxE}{\ensuremath{8}}             
\newcommand{\hatcurPPfluxavgeccenorigxxxxxE}{\ensuremath{1.10\pm0.23}}     
\newcommand{\hatcurPPfluxavgdimeccenorigxxxxxE}{\ensuremath{9}}            
\newcommand{\hatcurPPfluxavglogeccenorigxxxxxE}{\ensuremath{9.043\pm0.081}} 
\newcommand{\hatcurXsecphaseeccenorigxxxxxE}{\ensuremath{0.502\pm0.032}}   
\newcommand{\hatcurXsecondaryeccenorigxxxxxE}{\ensuremath{2457127.653\pm0.085}} 
\newcommand{\hatcurXsecdureccenorigxxxxxE}{\ensuremath{0.120\pm0.019}}     
\newcommand{\hatcurXsecingdureccenorigxxxxxE}{\ensuremath{0.0111\pm0.0028}} 
\newcommand{\hatcurPPphiconjeccenorigxxxxxE}{\ensuremath{0.06_{-0.50}^{+0.38}}} 
\newcommand{\hatcurPPperieccenorigxxxxxE}{\ensuremath{2457126.16\pm0.97}}  
\newcommand{\hatcurPPaequiveccenorigxxxxxE}{\ensuremath{0.0352_{-0.0029}^{+0.0038}}} 
\newcommand{\hatcurPPtcirceccenorigxxxxxE}{\ensuremath{116_{-45}^{+61}}}   
\newcommand{\hatcurPPtinfalleccenorigxxxxxE}{\ensuremath{640_{-230}^{+400}}} 
\newcommand{\hatcurXdisteccenorigxxxxxE}{\ensuremath{354\pm37}}            
\newcommand{\hatcurXAveccenorigxxxxxE}{\ensuremath{0.169\pm0.066}}         
\newcommand{\hatcurXdistredeccenorigxxxxxE}{\ensuremath{352\pm37}}         
\newcommand{\hatcurXEBVeccenorigxxxxxE}{\ensuremath{0.055\pm0.021}}        
\newcommand{\hatcurXmvisoredeccenorigxxxxxE}{\ensuremath{12.560\pm0.034}}  
\newcommand{\hatcurXmiisoredeccenorigxxxxxE}{\ensuremath{11.706\pm0.019}}  
\newcommand{\hatcurXmjisoredeccenorigxxxxxE}{\ensuremath{11.196\pm0.018}}  
\newcommand{\hatcurXmhisoredeccenorigxxxxxE}{\ensuremath{10.824\pm0.019}}  
\newcommand{\hatcurXmkisoredeccenorigxxxxxE}{\ensuremath{10.751\pm0.021}}  
\newcommand{\hatcurXviisoredeccenorigxxxxxE}{\ensuremath{0.854\pm0.026}}   
\newcommand{\hatcurXvkisoredeccenorigxxxxxE}{\ensuremath{1.810\pm0.042}}   
\newcommand{\hatcurXjhisoredeccenorigxxxxxE}{\ensuremath{0.3720\pm0.0088}} 
\newcommand{\hatcurXjkisoredeccenorigxxxxxE}{\ensuremath{0.445\pm0.010}}   
\newcommand{\hatcurCCpmraeccenorigxxxxxE}{\ensuremath{0\pm50}}             
\newcommand{\hatcurCCpmdececcenorigxxxxxE}{\ensuremath{0\pm50}}            
\newcommand{\hatcurCCpmeccenorigxxxxxE}{\ensuremath{0\pm71}}               

%% file: HTR384-014.eccen.orig.tex
\newcommand{\hatcurhtreccenorigxxxxxF}{HTR384-014}                         
\newcommand{\hatcurfieldeccenorigxxxxxF}{\ensuremath{string}}              
\newcommand{\hatcurCCraeccenorigxxxxxF}{\ensuremath{17^{\mathrm h}58^{\mathrm m}17.40{\mathrm s}}}                       
\newcommand{\hatcurCCdececcenorigxxxxxF}{\ensuremath{+05{\arcdeg}45{\arcmin}40.9{\arcsec}}}                      
\newcommand{\hatcurCCmageccenorigxxxxxF}{13.753}                           
\newcommand{\hatcurCCtwomasseccenorigxxxxxF}{2MASS~17581730+0545409}       
\newcommand{\hatcurCCgsceccenorigxxxxxF}{GSC~0429-01697}                   
\newcommand{\hatcurCCtassmveccenorigxxxxxF}{\ensuremath{13.753\pm0.065}}   
\newcommand{\hatcurCCtassmvshorteccenorigxxxxxF}{\ensuremath{13.8}}        
\newcommand{\hatcurCCtassmBeccenorigxxxxxF}{\ensuremath{14.729\pm0.066}}   
\newcommand{\hatcurCCtassmBshorteccenorigxxxxxF}{\ensuremath{14.7}}        
\newcommand{\hatcurCCtassmIeccenorigxxxxxF}{\ensuremath{nff\pmnff}}        
\newcommand{\hatcurCCtassmIshorteccenorigxxxxxF}{\ensuremath{0.0}}         
\newcommand{\hatcurCCtassmgeccenorigxxxxxF}{\ensuremath{14.258\pm0.026}}   
\newcommand{\hatcurCCtassmgshorteccenorigxxxxxF}{\ensuremath{14.3}}        
\newcommand{\hatcurCCtassmreccenorigxxxxxF}{\ensuremath{13.418\pm0.093}}   
\newcommand{\hatcurCCtassmrshorteccenorigxxxxxF}{\ensuremath{13.4}}        
\newcommand{\hatcurCCtassmieccenorigxxxxxF}{\ensuremath{13.136\pm0.086}}   
\newcommand{\hatcurCCtassmishorteccenorigxxxxxF}{\ensuremath{13.1}}        
\newcommand{\hatcurCCtwomassJmageccenorigxxxxxF}{\ensuremath{12.021\pm0.021}} 
\newcommand{\hatcurCCtwomassHmageccenorigxxxxxF}{\ensuremath{11.630\pm0.028}} 
\newcommand{\hatcurCCtwomassKmageccenorigxxxxxF}{\ensuremath{11.512\pm0.023}} 
\newcommand{\hatcurCCcitJmageccenorigxxxxxF}{\ensuremath{12.030\pm0.022}}  
\newcommand{\hatcurCCcitHmageccenorigxxxxxF}{\ensuremath{11.623\pm0.028}}  
\newcommand{\hatcurCCcitKmageccenorigxxxxxF}{\ensuremath{11.536\pm0.023}}  
\newcommand{\hatcurCCbbJmageccenorigxxxxxF}{\ensuremath{12.091\pm0.024}}   
\newcommand{\hatcurCCbbHmageccenorigxxxxxF}{\ensuremath{11.646\pm0.029}}   
\newcommand{\hatcurCCbbKmageccenorigxxxxxF}{\ensuremath{11.556\pm0.023}}   
\newcommand{\hatcurCCesoJmageccenorigxxxxxF}{\ensuremath{12.095\pm0.026}}  
\newcommand{\hatcurCCesoHmageccenorigxxxxxF}{\ensuremath{11.643\pm0.034}}  
\newcommand{\hatcurCCesoKmageccenorigxxxxxF}{\ensuremath{11.554\pm0.024}}  
\newcommand{\hatcurCCesoJHmageccenorigxxxxxF}{\ensuremath{0.452\pm0.040}}  
\newcommand{\hatcurCCesoJKmageccenorigxxxxxF}{\ensuremath{0.541\pm0.034}}  
\newcommand{\hatcurCCesoHKmageccenorigxxxxxF}{\ensuremath{0.089\pm0.042}}  
\newcommand{\hatcurLCdipeccenorigxxxxxF}{\ensuremath{17.6}}                
\newcommand{\hatcurLCrprstareccenorigxxxxxF}{\ensuremath{0.1219\pm0.0056}} 
\newcommand{\hatcurLCbsqeccenorigxxxxxF}{\ensuremath{0.186_{-0.079}^{+0.085}}} 
\newcommand{\hatcurLCimpeccenorigxxxxxF}{\ensuremath{0.432_{-0.103}^{+0.089}}} 
\newcommand{\hatcurLCzetaeccenorigxxxxxF}{\ensuremath{18.94\pm0.18}}       
\newcommand{\hatcurLCdureccenorigxxxxxF}{\ensuremath{0.1212\pm0.0022}}     
\newcommand{\hatcurLCdurshorteccenorigxxxxxF}{\ensuremath{0.1212}}         
\newcommand{\hatcurLCdurhreccenorigxxxxxF}{\ensuremath{2.909\pm0.053}}     
\newcommand{\hatcurLCdurhrshorteccenorigxxxxxF}{\ensuremath{2.909}}        
\newcommand{\hatcurLCqeccenorigxxxxxF}{\ensuremath{0.03590\pm0.00065}}     
\newcommand{\hatcurLCqshorteccenorigxxxxxF}{\ensuremath{0.036}}            
\newcommand{\hatcurLCingdureccenorigxxxxxF}{\ensuremath{0.0160\pm0.0023}}  
\newcommand{\hatcurLCPeccenorigxxxxxF}{\ensuremath{3.377729\pm0.000013}}   
\newcommand{\hatcurLCPprececcenorigxxxxxF}{\ensuremath{3.3777294}}         
\newcommand{\hatcurLCPshorteccenorigxxxxxF}{\ensuremath{3.3777}}           
\newcommand{\hatcurLCTeccenorigxxxxxF}{\ensuremath{2456386.31955\pm0.00057}} 
\newcommand{\hatcurLCTAeccenorigxxxxxF}{\ensuremath{2454964.2954\pm0.0054}} 
\newcommand{\hatcurLCTBeccenorigxxxxxF}{\ensuremath{2456399.83049\pm0.00058}} 
\newcommand{\hatcurLChatnetmeccenorigxxxxxF}{\ensuremath{13.32554\pm0.00019}} 
\newcommand{\hatcurLCiblendeccenorigxxxxxF}{\ensuremath{0.517\pm0.058}}    
\newcommand{\hatcurLCrhoeccenorigxxxxxF}{\ensuremath{1.31_{-0.26}^{+0.40}}} 
\newcommand{\hatcurSMEiteffeccenorigxxxxxF}{\ensuremath{5365\pm50}}        
\newcommand{\hatcurSMEizfeheccenorigxxxxxF}{\ensuremath{0.428\pm0.080}}    
\newcommand{\hatcurSMEizfehshorteccenorigxxxxxF}{\ensuremath{0.43}}        
\newcommand{\hatcurSMEiloggeccenorigxxxxxF}{\ensuremath{4.40\pm0.10}}      
\newcommand{\hatcurSMEivsineccenorigxxxxxF}{\ensuremath{3.22\pm0.50}}      
\newcommand{\hatcurSMEivmaceccenorigxxxxxF}{\ensuremath{0.0}}              
\newcommand{\hatcurSMEivmiceccenorigxxxxxF}{\ensuremath{0.0}}              
\newcommand{\hatcurLBizeccenorigxxxxxF}{\ensuremath{0.2730}}               
\newcommand{\hatcurLBiizeccenorigxxxxxF}{\ensuremath{0.3077}}              
\newcommand{\hatcurLBiieccenorigxxxxxF}{\ensuremath{0.3576}}               
\newcommand{\hatcurLBiiieccenorigxxxxxF}{\ensuremath{0.2893}}              
\newcommand{\hatcurLBiIeccenorigxxxxxF}{\ensuremath{0.3300}}               
\newcommand{\hatcurLBiiIeccenorigxxxxxF}{\ensuremath{0.2953}}              
\newcommand{\hatcurLBigeccenorigxxxxxF}{\ensuremath{0.7099}}               
\newcommand{\hatcurLBiigeccenorigxxxxxF}{\ensuremath{0.1164}}              
\newcommand{\hatcurLBireccenorigxxxxxF}{\ensuremath{0.4752}}               
\newcommand{\hatcurLBiireccenorigxxxxxF}{\ensuremath{0.2572}}              
\newcommand{\hatcurLBiReccenorigxxxxxF}{\ensuremath{0.4428}}               
\newcommand{\hatcurLBiiReccenorigxxxxxF}{\ensuremath{0.2670}}              
\newcommand{\hatcurLBikepeccenorigxxxxxF}{\ensuremath{0.1000}}             
\newcommand{\hatcurLBiikepeccenorigxxxxxF}{\ensuremath{0.1000}}            
\newcommand{\hatcurISOmeccenorigxxxxxF}{\ensuremath{0.977\pm0.022}}        
\newcommand{\hatcurISOmshorteccenorigxxxxxF}{\ensuremath{0.98}}            
\newcommand{\hatcurISOmlongeccenorigxxxxxF}{\ensuremath{0.977\pm0.022}}    
\newcommand{\hatcurISOreccenorigxxxxxF}{\ensuremath{1.024\pm0.067}}        
\newcommand{\hatcurISOrshorteccenorigxxxxxF}{\ensuremath{1.02}}            
\newcommand{\hatcurISOrlongeccenorigxxxxxF}{\ensuremath{1.024\pm0.067}}    
\newcommand{\hatcurISOrhoeccenorigxxxxxF}{\ensuremath{1.27_{-0.23}^{+0.32}}} 
\newcommand{\hatcurISOrholongeccenorigxxxxxF}{\ensuremath{1.27_{-0.23}^{+0.32}}} 
\newcommand{\hatcurISOloggeccenorigxxxxxF}{\ensuremath{4.404\pm0.057}}     
\newcommand{\hatcurISOlumeccenorigxxxxxF}{\ensuremath{0.78\pm0.11}}        
\newcommand{\hatcurISOlumshorteccenorigxxxxxF}{\ensuremath{0.78}}          
\newcommand{\hatcurISOmveccenorigxxxxxF}{\ensuremath{5.18\pm0.16}}         
\newcommand{\hatcurISOvieccenorigxxxxxF}{\ensuremath{0.834\pm0.014}}       
\newcommand{\hatcurISOageeccenorigxxxxxF}{\ensuremath{7.4\pm2.6}}          
\newcommand{\hatcurISOsigmaeccenorigxxxxxF}{\ensuremath{0.00080\pm0.00020}} 
\newcommand{\hatcurISOMJeccenorigxxxxxF}{\ensuremath{3.80\pm0.15}}         
\newcommand{\hatcurISOMHeccenorigxxxxxF}{\ensuremath{3.39\pm0.14}}         
\newcommand{\hatcurISOMKeccenorigxxxxxF}{\ensuremath{3.32\pm0.14}}         
\newcommand{\hatcurISOJKeccenorigxxxxxF}{\ensuremath{0.480\pm0.010}}       
\newcommand{\hatcurISOspececcenorigxxxxxF}{G}                              
\newcommand{\hatcurRVKeccenorigxxxxxF}{\ensuremath{88.1\pm3.4}}            
\newcommand{\hatcurRVrkeccenorigxxxxxF}{\ensuremath{-0.01\pm0.12}}         
\newcommand{\hatcurRVrheccenorigxxxxxF}{\ensuremath{-0.04_{-0.16}^{+0.22}}} 
\newcommand{\hatcurRVkeccenorigxxxxxF}{\ensuremath{-0.001\pm0.032}}        
\newcommand{\hatcurRVheccenorigxxxxxF}{\ensuremath{-0.004\pm0.049}}        
\newcommand{\hatcurRVtroneeccenorigxxxxxF}{\ensuremath{0\pm0}}             
\newcommand{\hatcurRVtrtwoeccenorigxxxxxF}{\ensuremath{0\pm0}}             
\newcommand{\hatcurRVgammaAeccenorigxxxxxF}{\ensuremath{6.8\pm3.1}}        
\newcommand{\hatcurRVjitterAeccenorigxxxxxF}{\ensuremath{0.0\pm1.0}}       
\newcommand{\hatcurRVjittertwosiglimAeccenorigxxxxxF}{\ensuremath{<2.4}}   
\newcommand{\hatcurRVfitrmsAeccenorigxxxxxF}{\ensuremath{0.0}}             
\newcommand{\hatcurRVgammaBeccenorigxxxxxF}{\ensuremath{-108\pm28}}        
\newcommand{\hatcurRVjitterBeccenorigxxxxxF}{\ensuremath{0.1\pm4.5}}       
\newcommand{\hatcurRVjittertwosiglimBeccenorigxxxxxF}{\ensuremath{<10.2}}  
\newcommand{\hatcurRVfitrmsBeccenorigxxxxxF}{\ensuremath{0.0}}             
\newcommand{\hatcurRVgammaCeccenorigxxxxxF}{\ensuremath{-69596.4\pm9.7}}   
\newcommand{\hatcurRVjitterCeccenorigxxxxxF}{\ensuremath{17\pm12}}         
\newcommand{\hatcurRVjittertwosiglimCeccenorigxxxxxF}{\ensuremath{<39.5}}  
\newcommand{\hatcurRVfitrmsCeccenorigxxxxxF}{\ensuremath{0.0}}             
\newcommand{\hatcurRVecceneccenorigxxxxxF}{\ensuremath{0.031\pm0.041}}     
\newcommand{\hatcurRVeccentwosiglimeccenorigxxxxxF}{\ensuremath{<0.135}}   
\newcommand{\hatcurRVomegaeccenorigxxxxxF}{\ensuremath{220\pm110}}         
\newcommand{\hatcurPPieccenorigxxxxxF}{\ensuremath{87.30\pm0.78}}          
\newcommand{\hatcurPPgeccenorigxxxxxF}{\ensuremath{10.8\pm2.0}}            
\newcommand{\hatcurPPloggeccenorigxxxxxF}{\ensuremath{3.034\pm0.080}}      
\newcommand{\hatcurPPareccenorigxxxxxF}{\ensuremath{9.15\pm0.60}}          
\newcommand{\hatcurPPareleccenorigxxxxxF}{\ensuremath{0.04372\pm0.00033}}  
\newcommand{\hatcurPPrhoeccenorigxxxxxF}{\ensuremath{0.44\pm0.13}}         
\newcommand{\hatcurPPmeccenorigxxxxxF}{\ensuremath{0.641\pm0.026}}         
\newcommand{\hatcurPPmshorteccenorigxxxxxF}{\ensuremath{0.64}}             
\newcommand{\hatcurPPmlongeccenorigxxxxxF}{\ensuremath{0.641\pm0.026}}     
\newcommand{\hatcurPPmeeccenorigxxxxxF}{\ensuremath{203.6\pm8.3}}          
\newcommand{\hatcurPPmeshorteccenorigxxxxxF}{\ensuremath{203.6}}           
\newcommand{\hatcurPPmelongeccenorigxxxxxF}{\ensuremath{203.6\pm8.3}}      
\newcommand{\hatcurPPreccenorigxxxxxF}{\ensuremath{1.22\pm0.11}}           
\newcommand{\hatcurPPrshorteccenorigxxxxxF}{\ensuremath{1.22}}             
\newcommand{\hatcurPPrlongeccenorigxxxxxF}{\ensuremath{1.22\pm0.11}}       
\newcommand{\hatcurPPreeccenorigxxxxxF}{\ensuremath{13.7\pm1.3}}           
\newcommand{\hatcurPPreshorteccenorigxxxxxF}{\ensuremath{13.7}}            
\newcommand{\hatcurPPrelongeccenorigxxxxxF}{\ensuremath{13.7\pm1.3}}       
\newcommand{\hatcurPPmrcorreccenorigxxxxxF}{\ensuremath{0.16}}             
\newcommand{\hatcurPPteffeccenorigxxxxxF}{\ensuremath{1253\pm43}}          
\newcommand{\hatcurPPthetaeccenorigxxxxxF}{\ensuremath{0.0471\pm0.0045}}   
\newcommand{\hatcurPPfluxperieccenorigxxxxxF}{\ensuremath{5.98_{-0.77}^{+1.07}}} 
\newcommand{\hatcurPPfluxperidimeccenorigxxxxxF}{\ensuremath{8}}           
\newcommand{\hatcurPPfluxapeccenorigxxxxxF}{\ensuremath{5.18\pm0.77}}      
\newcommand{\hatcurPPfluxapdimeccenorigxxxxxF}{\ensuremath{8}}             
\newcommand{\hatcurPPfluxavgeccenorigxxxxxF}{\ensuremath{5.57\pm0.78}}     
\newcommand{\hatcurPPfluxavgdimeccenorigxxxxxF}{\ensuremath{8}}            
\newcommand{\hatcurPPfluxavglogeccenorigxxxxxF}{\ensuremath{8.746\pm0.060}} 
\newcommand{\hatcurXsecphaseeccenorigxxxxxF}{\ensuremath{0.500\pm0.020}}   
\newcommand{\hatcurXsecondaryeccenorigxxxxxF}{\ensuremath{2456388.007\pm0.069}} 
\newcommand{\hatcurXsecdureccenorigxxxxxF}{\ensuremath{0.1209\pm0.0095}}   
\newcommand{\hatcurXsecingdureccenorigxxxxxF}{\ensuremath{0.0159\pm0.0029}} 
\newcommand{\hatcurPPphiconjeccenorigxxxxxF}{\ensuremath{-0.02_{-0.40}^{+0.25}}} 
\newcommand{\hatcurPPperieccenorigxxxxxF}{\ensuremath{2456386.4\pm1.0}}    
\newcommand{\hatcurPPaequiveccenorigxxxxxF}{\ensuremath{0.0495\pm0.0034}}  
\newcommand{\hatcurPPtcirceccenorigxxxxxF}{\ensuremath{139_{-57}^{+83}}}   
\newcommand{\hatcurPPtinfalleccenorigxxxxxF}{\ensuremath{3140_{-900}^{+1500}}} 
\newcommand{\hatcurXdisteccenorigxxxxxF}{\ensuremath{443\pm30}}            
\newcommand{\hatcurXAveccenorigxxxxxF}{\ensuremath{0.382\pm0.092}}         
\newcommand{\hatcurXdistredeccenorigxxxxxF}{\ensuremath{434\pm29}}         
\newcommand{\hatcurXEBVeccenorigxxxxxF}{\ensuremath{0.123\pm0.030}}        
\newcommand{\hatcurXmvisoredeccenorigxxxxxF}{\ensuremath{13.751\pm0.061}}  
\newcommand{\hatcurXmiisoredeccenorigxxxxxF}{\ensuremath{12.718\pm0.026}}  
\newcommand{\hatcurXmjisoredeccenorigxxxxxF}{\ensuremath{12.095\pm0.016}}  
\newcommand{\hatcurXmhisoredeccenorigxxxxxF}{\ensuremath{11.650\pm0.016}}  
\newcommand{\hatcurXmkisoredeccenorigxxxxxF}{\ensuremath{11.551\pm0.018}}  
\newcommand{\hatcurXviisoredeccenorigxxxxxF}{\ensuremath{1.033\pm0.042}}   
\newcommand{\hatcurXvkisoredeccenorigxxxxxF}{\ensuremath{2.200\pm0.068}}   
\newcommand{\hatcurXjhisoredeccenorigxxxxxF}{\ensuremath{0.445\pm0.010}}   
\newcommand{\hatcurXjkisoredeccenorigxxxxxF}{\ensuremath{0.543\pm0.014}}   
\newcommand{\hatcurCCpmraeccenorigxxxxxF}{\ensuremath{-16.2\pm4.3}}        
\newcommand{\hatcurCCpmdececcenorigxxxxxF}{\ensuremath{0.5\pm3.5}}         
\newcommand{\hatcurCCpmeccenorigxxxxxF}{\ensuremath{16.2\pm5.5}}           

%% file: HTR405-001.eccen.orig.tex
\newcommand{\hatcurhtreccenorigxxxxxG}{HTR405-001}                         
\newcommand{\hatcurfieldeccenorigxxxxxG}{\ensuremath{string}}              
\newcommand{\hatcurCCraeccenorigxxxxxG}{\ensuremath{04^{\mathrm h}35^{\mathrm m}53.84{\mathrm s}}}                       
\newcommand{\hatcurCCdececcenorigxxxxxG}{\ensuremath{+02{\arcdeg}25{\arcmin}52.7{\arcsec}}}                      
\newcommand{\hatcurCCmageccenorigxxxxxG}{12.771}                           
\newcommand{\hatcurCCtwomasseccenorigxxxxxG}{2MASS~04355384+0225526}       
\newcommand{\hatcurCCgsceccenorigxxxxxG}{GSC~0086-00341}                   
\newcommand{\hatcurCCtassmveccenorigxxxxxG}{\ensuremath{12.771\pm0.010}}   
\newcommand{\hatcurCCtassmvshorteccenorigxxxxxG}{\ensuremath{12.8}}        
\newcommand{\hatcurCCtassmBeccenorigxxxxxG}{\ensuremath{13.446\pm0.011}}   
\newcommand{\hatcurCCtassmBshorteccenorigxxxxxG}{\ensuremath{13.4}}        
\newcommand{\hatcurCCtassmIeccenorigxxxxxG}{\ensuremath{12.105\pm0.070}}   
\newcommand{\hatcurCCtassmIshorteccenorigxxxxxG}{\ensuremath{12.1}}        
\newcommand{\hatcurCCtassmgeccenorigxxxxxG}{\ensuremath{13.062\pm0.013}}   
\newcommand{\hatcurCCtassmgshorteccenorigxxxxxG}{\ensuremath{13.1}}        
\newcommand{\hatcurCCtassmreccenorigxxxxxG}{\ensuremath{12.553\pm0.075}}   
\newcommand{\hatcurCCtassmrshorteccenorigxxxxxG}{\ensuremath{12.6}}        
\newcommand{\hatcurCCtassmieccenorigxxxxxG}{\ensuremath{12.440\pm0.037}}   
\newcommand{\hatcurCCtassmishorteccenorigxxxxxG}{\ensuremath{12.4}}        
\newcommand{\hatcurCCtwomassJmageccenorigxxxxxG}{\ensuremath{11.485\pm0.026}} 
\newcommand{\hatcurCCtwomassHmageccenorigxxxxxG}{\ensuremath{11.225\pm0.025}} 
\newcommand{\hatcurCCtwomassKmageccenorigxxxxxG}{\ensuremath{11.123\pm0.021}} 
\newcommand{\hatcurCCcitJmageccenorigxxxxxG}{\ensuremath{11.502\pm0.026}}  
\newcommand{\hatcurCCcitHmageccenorigxxxxxG}{\ensuremath{11.219\pm0.025}}  
\newcommand{\hatcurCCcitKmageccenorigxxxxxG}{\ensuremath{11.147\pm0.021}}  
\newcommand{\hatcurCCbbJmageccenorigxxxxxG}{\ensuremath{11.551\pm0.028}}   
\newcommand{\hatcurCCbbHmageccenorigxxxxxG}{\ensuremath{11.241\pm0.026}}   
\newcommand{\hatcurCCbbKmageccenorigxxxxxG}{\ensuremath{11.167\pm0.021}}   
\newcommand{\hatcurCCesoJmageccenorigxxxxxG}{\ensuremath{11.553\pm0.029}}  
\newcommand{\hatcurCCesoHmageccenorigxxxxxG}{\ensuremath{11.238\pm0.030}}  
\newcommand{\hatcurCCesoKmageccenorigxxxxxG}{\ensuremath{11.166\pm0.022}}  
\newcommand{\hatcurCCesoJHmageccenorigxxxxxG}{\ensuremath{0.315\pm0.040}}  
\newcommand{\hatcurCCesoJKmageccenorigxxxxxG}{\ensuremath{0.387\pm0.036}}  
\newcommand{\hatcurCCesoHKmageccenorigxxxxxG}{\ensuremath{0.072\pm0.038}}  
\newcommand{\hatcurLCdipeccenorigxxxxxG}{\ensuremath{11.2}}                
\newcommand{\hatcurLCrprstareccenorigxxxxxG}{\ensuremath{0.1028\pm0.0043}} 
\newcommand{\hatcurLCbsqeccenorigxxxxxG}{\ensuremath{0.095_{-0.064}^{+0.085}}} 
\newcommand{\hatcurLCimpeccenorigxxxxxG}{\ensuremath{0.31_{-0.13}^{+0.12}}} 
\newcommand{\hatcurLCzetaeccenorigxxxxxG}{\ensuremath{10.85\pm0.16}}       
\newcommand{\hatcurLCdureccenorigxxxxxG}{\ensuremath{0.2054\pm0.0037}}     
\newcommand{\hatcurLCdurshorteccenorigxxxxxG}{\ensuremath{0.2054}}         
\newcommand{\hatcurLCdurhreccenorigxxxxxG}{\ensuremath{4.929\pm0.088}}     
\newcommand{\hatcurLCdurhrshorteccenorigxxxxxG}{\ensuremath{4.929}}        
\newcommand{\hatcurLCqeccenorigxxxxxG}{\ensuremath{0.05130\pm0.00091}}     
\newcommand{\hatcurLCqshorteccenorigxxxxxG}{\ensuremath{0.051}}            
\newcommand{\hatcurLCingdureccenorigxxxxxG}{\ensuremath{0.0209\pm0.0024}}  
\newcommand{\hatcurLCPeccenorigxxxxxG}{\ensuremath{4.007205\pm0.000020}}   
\newcommand{\hatcurLCPprececcenorigxxxxxG}{\ensuremath{4.0072047}}         
\newcommand{\hatcurLCPshorteccenorigxxxxxG}{\ensuremath{4.0072}}           
\newcommand{\hatcurLCTeccenorigxxxxxG}{\ensuremath{2455719.7956\pm0.0010}} 
\newcommand{\hatcurLCTAeccenorigxxxxxG}{\ensuremath{2455078.6430\pm0.0030}} 
\newcommand{\hatcurLCTBeccenorigxxxxxG}{\ensuremath{2455848.0262\pm0.0014}} 
\newcommand{\hatcurLChatnetmeccenorigxxxxxG}{\ensuremath{12.52212\pm0.00013}} 
\newcommand{\hatcurLCiblendeccenorigxxxxxG}{\ensuremath{0.713\pm0.074}}    
\newcommand{\hatcurLCrhoeccenorigxxxxxG}{\ensuremath{0.27\pm0.12}}         
\newcommand{\hatcurSMEiteffeccenorigxxxxxG}{\ensuremath{6302\pm50}}        
\newcommand{\hatcurSMEizfeheccenorigxxxxxG}{\ensuremath{-0.010\pm0.080}}   
\newcommand{\hatcurSMEizfehshorteccenorigxxxxxG}{\ensuremath{-0.01}}       
\newcommand{\hatcurSMEiloggeccenorigxxxxxG}{\ensuremath{3.99\pm0.10}}      
\newcommand{\hatcurSMEivsineccenorigxxxxxG}{\ensuremath{12.70\pm0.50}}     
\newcommand{\hatcurSMEivmaceccenorigxxxxxG}{\ensuremath{0.0}}              
\newcommand{\hatcurSMEivmiceccenorigxxxxxG}{\ensuremath{0.0}}              
\newcommand{\hatcurLBizeccenorigxxxxxG}{\ensuremath{0.1378}}               
\newcommand{\hatcurLBiizeccenorigxxxxxG}{\ensuremath{0.3609}}              
\newcommand{\hatcurLBiieccenorigxxxxxG}{\ensuremath{0.1885}}               
\newcommand{\hatcurLBiiieccenorigxxxxxG}{\ensuremath{0.3692}}              
\newcommand{\hatcurLBiIeccenorigxxxxxG}{\ensuremath{0.1698}}               
\newcommand{\hatcurLBiiIeccenorigxxxxxG}{\ensuremath{0.3681}}              
\newcommand{\hatcurLBigeccenorigxxxxxG}{\ensuremath{0.4343}}               
\newcommand{\hatcurLBiigeccenorigxxxxxG}{\ensuremath{0.3181}}              
\newcommand{\hatcurLBireccenorigxxxxxG}{\ensuremath{0.2646}}               
\newcommand{\hatcurLBiireccenorigxxxxxG}{\ensuremath{0.3759}}              
\newcommand{\hatcurLBiReccenorigxxxxxG}{\ensuremath{0.2430}}               
\newcommand{\hatcurLBiiReccenorigxxxxxG}{\ensuremath{0.3756}}              
\newcommand{\hatcurLBikepeccenorigxxxxxG}{\ensuremath{0.1000}}             
\newcommand{\hatcurLBiikepeccenorigxxxxxG}{\ensuremath{0.1000}}            
\newcommand{\hatcurISOmeccenorigxxxxxG}{\ensuremath{1.410_{-0.080}^{+0.121}}} 
\newcommand{\hatcurISOmshorteccenorigxxxxxG}{\ensuremath{1.41}}            
\newcommand{\hatcurISOmlongeccenorigxxxxxG}{\ensuremath{1.410_{-0.080}^{+0.121}}} 
\newcommand{\hatcurISOreccenorigxxxxxG}{\ensuremath{1.93_{-0.22}^{+0.40}}} 
\newcommand{\hatcurISOrshorteccenorigxxxxxG}{\ensuremath{1.93}}            
\newcommand{\hatcurISOrlongeccenorigxxxxxG}{\ensuremath{1.93_{-0.22}^{+0.40}}} 
\newcommand{\hatcurISOrhoeccenorigxxxxxG}{\ensuremath{0.27\pm0.12}}        
\newcommand{\hatcurISOrholongeccenorigxxxxxG}{\ensuremath{0.27\pm0.12}}    
\newcommand{\hatcurISOloggeccenorigxxxxxG}{\ensuremath{4.01\pm0.12}}       
\newcommand{\hatcurISOlumeccenorigxxxxxG}{\ensuremath{5.3_{-1.1}^{+2.4}}}  
\newcommand{\hatcurISOlumshorteccenorigxxxxxG}{\ensuremath{5.28}}          
\newcommand{\hatcurISOmveccenorigxxxxxG}{\ensuremath{2.95\pm0.39}}         
\newcommand{\hatcurISOvieccenorigxxxxxG}{\ensuremath{0.546\pm0.013}}       
\newcommand{\hatcurISOageeccenorigxxxxxG}{\ensuremath{2.71\pm0.41}}        
\newcommand{\hatcurISOsigmaeccenorigxxxxxG}{\ensuremath{0.00020\pm0.00012}} 
\newcommand{\hatcurISOMJeccenorigxxxxxG}{\ensuremath{2.06\pm0.39}}         
\newcommand{\hatcurISOMHeccenorigxxxxxG}{\ensuremath{1.80\pm0.38}}         
\newcommand{\hatcurISOMKeccenorigxxxxxG}{\ensuremath{1.76\pm0.38}}         
\newcommand{\hatcurISOJKeccenorigxxxxxG}{\ensuremath{0.300\pm0.010}}       
\newcommand{\hatcurISOspececcenorigxxxxxG}{F}                              
\newcommand{\hatcurRVKeccenorigxxxxxG}{\ensuremath{70\pm17}}               
\newcommand{\hatcurRVrkeccenorigxxxxxG}{\ensuremath{0.06\pm0.17}}          
\newcommand{\hatcurRVrheccenorigxxxxxG}{\ensuremath{0.16\pm0.29}}          
\newcommand{\hatcurRVkeccenorigxxxxxG}{\ensuremath{0.010_{-0.046}^{+0.065}}} 
\newcommand{\hatcurRVheccenorigxxxxxG}{\ensuremath{0.039_{-0.072}^{+0.176}}} 
\newcommand{\hatcurRVtroneeccenorigxxxxxG}{\ensuremath{0\pm0}}             
\newcommand{\hatcurRVtrtwoeccenorigxxxxxG}{\ensuremath{0\pm0}}             
\newcommand{\hatcurRVgammaAeccenorigxxxxxG}{\ensuremath{10\pm14}}          
\newcommand{\hatcurRVjitterAeccenorigxxxxxG}{\ensuremath{25\pm17}}         
\newcommand{\hatcurRVjittertwosiglimAeccenorigxxxxxG}{\ensuremath{<59.8}}  
\newcommand{\hatcurRVfitrmsAeccenorigxxxxxG}{\ensuremath{0.0}}             
\newcommand{\hatcurRVgammaBeccenorigxxxxxG}{\ensuremath{24487\pm14}}       
\newcommand{\hatcurRVjitterBeccenorigxxxxxG}{\ensuremath{0.0\pm8.6}}       
\newcommand{\hatcurRVjittertwosiglimBeccenorigxxxxxG}{\ensuremath{<19.7}}  
\newcommand{\hatcurRVfitrmsBeccenorigxxxxxG}{\ensuremath{0.0}}             
\newcommand{\hatcurRVecceneccenorigxxxxxG}{\ensuremath{0.10\pm0.13}}       
\newcommand{\hatcurRVeccentwosiglimeccenorigxxxxxG}{\ensuremath{<0.389}}   
\newcommand{\hatcurRVomegaeccenorigxxxxxG}{\ensuremath{100\pm110}}         
\newcommand{\hatcurPPieccenorigxxxxxG}{\ensuremath{86.9\pm2.1}}            
\newcommand{\hatcurPPgeccenorigxxxxxG}{\ensuremath{4.5\pm2.1}}             
\newcommand{\hatcurPPloggeccenorigxxxxxG}{\ensuremath{2.65_{-0.21}^{+0.16}}} 
\newcommand{\hatcurPPareccenorigxxxxxG}{\ensuremath{6.15_{-0.90}^{+0.67}}} 
\newcommand{\hatcurPPareleccenorigxxxxxG}{\ensuremath{0.0554_{-0.0011}^{+0.0015}}} 
\newcommand{\hatcurPPrhoeccenorigxxxxxG}{\ensuremath{0.116\pm0.078}}       
\newcommand{\hatcurPPmeccenorigxxxxxG}{\ensuremath{0.68_{-0.13}^{+0.18}}}  
\newcommand{\hatcurPPmshorteccenorigxxxxxG}{\ensuremath{0.68}}             
\newcommand{\hatcurPPmlongeccenorigxxxxxG}{\ensuremath{0.68_{-0.13}^{+0.18}}} 
\newcommand{\hatcurPPmeeccenorigxxxxxG}{\ensuremath{217_{-41}^{+58}}}      
\newcommand{\hatcurPPmeshorteccenorigxxxxxG}{\ensuremath{217.2}}           
\newcommand{\hatcurPPmelongeccenorigxxxxxG}{\ensuremath{217_{-41}^{+58}}}  
\newcommand{\hatcurPPreccenorigxxxxxG}{\ensuremath{1.94_{-0.25}^{+0.42}}}  
\newcommand{\hatcurPPrshorteccenorigxxxxxG}{\ensuremath{1.94}}             
\newcommand{\hatcurPPrlongeccenorigxxxxxG}{\ensuremath{1.94_{-0.25}^{+0.42}}} 
\newcommand{\hatcurPPreeccenorigxxxxxG}{\ensuremath{21.7_{-2.8}^{+4.7}}}   
\newcommand{\hatcurPPreshorteccenorigxxxxxG}{\ensuremath{21.7}}            
\newcommand{\hatcurPPrelongeccenorigxxxxxG}{\ensuremath{21.7_{-2.8}^{+4.7}}} 
\newcommand{\hatcurPPmrcorreccenorigxxxxxG}{\ensuremath{-0.07}}            
\newcommand{\hatcurPPteffeccenorigxxxxxG}{\ensuremath{1799_{-92}^{+164}}}  
\newcommand{\hatcurPPthetaeccenorigxxxxxG}{\ensuremath{0.0271_{-0.0078}^{+0.0102}}} 
\newcommand{\hatcurPPfluxperieccenorigxxxxxG}{\ensuremath{2.71_{-0.59}^{+2.93}}} 
\newcommand{\hatcurPPfluxperidimeccenorigxxxxxG}{\ensuremath{9}}           
\newcommand{\hatcurPPfluxapeccenorigxxxxxG}{\ensuremath{2.01\pm0.35}}      
\newcommand{\hatcurPPfluxapdimeccenorigxxxxxG}{\ensuremath{9}}             
\newcommand{\hatcurPPfluxavgeccenorigxxxxxG}{\ensuremath{2.36_{-0.45}^{+0.99}}} 
\newcommand{\hatcurPPfluxavgdimeccenorigxxxxxG}{\ensuremath{9}}            
\newcommand{\hatcurPPfluxavglogeccenorigxxxxxG}{\ensuremath{9.373_{-0.092}^{+0.152}}} 
\newcommand{\hatcurXsecphaseeccenorigxxxxxG}{\ensuremath{0.507\pm0.045}}   
\newcommand{\hatcurXsecondaryeccenorigxxxxxG}{\ensuremath{2455721.83\pm0.18}} 
\newcommand{\hatcurXsecdureccenorigxxxxxG}{\ensuremath{0.221\pm0.070}}     
\newcommand{\hatcurXsecingdureccenorigxxxxxG}{\ensuremath{0.024\pm0.027}}  
\newcommand{\hatcurPPphiconjeccenorigxxxxxG}{\ensuremath{0.02_{-0.17}^{+0.29}}} 
\newcommand{\hatcurPPperieccenorigxxxxxG}{\ensuremath{2455719.71\pm0.92}}  
\newcommand{\hatcurPPaequiveccenorigxxxxxG}{\ensuremath{0.0241_{-0.0036}^{+0.0026}}} 
\newcommand{\hatcurPPtcirceccenorigxxxxxG}{\ensuremath{37_{-28}^{+38}}}    
\newcommand{\hatcurPPtinfalleccenorigxxxxxG}{\ensuremath{650_{-320}^{+480}}} 
\newcommand{\hatcurXdisteccenorigxxxxxG}{\ensuremath{761_{-86}^{+157}}}    
\newcommand{\hatcurXAveccenorigxxxxxG}{\ensuremath{0.464\pm0.047}}         
\newcommand{\hatcurXdistredeccenorigxxxxxG}{\ensuremath{744_{-84}^{+152}}} 
\newcommand{\hatcurXEBVeccenorigxxxxxG}{\ensuremath{0.150\pm0.015}}        
\newcommand{\hatcurXmvisoredeccenorigxxxxxG}{\ensuremath{12.7720\pm0.0100}} 
\newcommand{\hatcurXmiisoredeccenorigxxxxxG}{\ensuremath{11.984\pm0.013}}  
\newcommand{\hatcurXmjisoredeccenorigxxxxxG}{\ensuremath{11.546\pm0.014}}  
\newcommand{\hatcurXmhisoredeccenorigxxxxxG}{\ensuremath{11.251\pm0.015}}  
\newcommand{\hatcurXmkisoredeccenorigxxxxxG}{\ensuremath{11.170\pm0.016}}  
\newcommand{\hatcurXviisoredeccenorigxxxxxG}{\ensuremath{0.788\pm0.014}}   
\newcommand{\hatcurXvkisoredeccenorigxxxxxG}{\ensuremath{1.602\pm0.019}}   
\newcommand{\hatcurXjhisoredeccenorigxxxxxG}{\ensuremath{0.2950\pm0.0065}} 
\newcommand{\hatcurXjkisoredeccenorigxxxxxG}{\ensuremath{0.3760\pm0.0055}} 
\newcommand{\hatcurCCpmraeccenorigxxxxxG}{\ensuremath{10.8\pm1.3}}         
\newcommand{\hatcurCCpmdececcenorigxxxxxG}{\ensuremath{1.2\pm1.8}}         
\newcommand{\hatcurCCpmeccenorigxxxxxG}{\ensuremath{10.9\pm2.2}}           

%% file: newcommand_spec_HTR062-011.tex
\newcommand{\hatcurxxxxxA}{HAT-P-58}
\newcommand{\hatcurbxxxxxA}{HAT-P-58b}
\newcommand{\hatcurcxxxxxA}{HAT-P-58c}

\newcommand{\hatcurplanetnumxxxxxA}{58}

\newcommand{\hatcurtwomassshortxxxxxA}{04352318+5652055}

\newcommand{\hatcurSindexxxxxxA}{\ensuremath{0.150\pm0.010}}
\newcommand{\hatcurRHKindexxxxxxA}{\ensuremath{-5.057\pm0.072}}

\newcommand{\hatcurRVgammaabsxxxxxA}{\ensuremath{-35.97\pm0.10}}                           
\newcommand{\hatcurRVgammainstxxxxxA}{TRES}                           

\newcommand{\hatcurCCtassvixxxxxA}{NULL}                  

\newcommand{\hatcurSMEversionxxxxxA}{ii}                                       

\newcommand{\hatcurisoshortxxxxxA}{YY}
\newcommand{\hatcurisofullxxxxxA}{Yonsei-Yale (YY)}
\newcommand{\hatcurisocitexxxxxA}{yi:2001}

\newcommand{\hatcurlumindxxxxxA}{\arstar}

\newcommand{\hatcurjhkfilsetxxxxxA}{ESO}

\newcommand{\hatcurSMEteffxxxxxA}{\ifthenelse{\equal{\hatcurSMEversionxxxxxA}{i}}{\hatcurSMEiteff{\hatcurplanetnumxxxxxA}}{\hatcurSMEiiteff{\hatcurplanetnumxxxxxA}}}
\newcommand{\hatcurSMEzfehxxxxxA}{\ifthenelse{\equal{\hatcurSMEversionxxxxxA}{i}}{\hatcurSMEizfeh{\hatcurplanetnumxxxxxA}}{\hatcurSMEiizfeh{\hatcurplanetnumxxxxxA}}}
\newcommand{\hatcurSMEzfehshortxxxxxA}{\ifthenelse{\equal{\hatcurSMEversionxxxxxA}{i}}{\hatcurSMEizfehshort{\hatcurplanetnumxxxxxA}}{\hatcurSMEiizfehshort{\hatcurplanetnumxxxxxA}}}
\newcommand{\hatcurSMEloggxxxxxA}{\ifthenelse{\equal{\hatcurSMEversionxxxxxA}{i}}{\hatcurSMEilogg{\hatcurplanetnumxxxxxA}}{\hatcurSMEiilogg{\hatcurplanetnumxxxxxA}}}
\newcommand{\hatcurSMEvsinxxxxxA}{\ifthenelse{\equal{\hatcurSMEversionxxxxxA}{i}}{\hatcurSMEivsin{\hatcurplanetnumxxxxxA}}{\hatcurSMEiivsin{\hatcurplanetnumxxxxxA}}}
\newcommand{\hatcurSMEvmacxxxxxA}{\ifthenelse{\equal{\hatcurSMEversionxxxxxA}{i}}{\hatcurSMEivmac{\hatcurplanetnumxxxxxA}}{\hatcurSMEiivmac{\hatcurplanetnumxxxxxA}}}
\newcommand{\hatcurSMEvmicxxxxxA}{\ifthenelse{\equal{\hatcurSMEversionxxxxxA}{i}}{\hatcurSMEivmic{\hatcurplanetnumxxxxxA}}{\hatcurSMEiivmic{\hatcurplanetnumxxxxxA}}}

%% file: newcommand_spec_HTR080-003.tex
\newcommand{\hatcurxxxxxB}{HAT-P-59}
\newcommand{\hatcurbxxxxxB}{HAT-P-59b}
\newcommand{\hatcurcxxxxxB}{HAT-P-59c}

\newcommand{\hatcurplanetnumxxxxxB}{59}

\newcommand{\hatcurtwomassshortxxxxxB}{19295008+6231452}

\newcommand{\hatcurRVgammaabsxxxxxB}{\ensuremath{-20.477\pm0.027}}                           
\newcommand{\hatcurRVgammainstxxxxxB}{TRES}                           

\newcommand{\hatcurCCtassvixxxxxB}{NULL}                  

\newcommand{\hatcurSMEversionxxxxxB}{ii}                                       

\newcommand{\hatcurisoshortxxxxxB}{YY}
\newcommand{\hatcurisofullxxxxxB}{Yonsei-Yale (YY)}
\newcommand{\hatcurisocitexxxxxB}{yi:2001}

\newcommand{\hatcurlumindxxxxxB}{\arstar}

\newcommand{\hatcurjhkfilsetxxxxxB}{ESO}

\newcommand{\hatcurSMEteffxxxxxB}{\ifthenelse{\equal{\hatcurSMEversionxxxxxB}{i}}{\hatcurSMEiteff{\hatcurplanetnumxxxxxB}}{\hatcurSMEiiteff{\hatcurplanetnumxxxxxB}}}
\newcommand{\hatcurSMEzfehxxxxxB}{\ifthenelse{\equal{\hatcurSMEversionxxxxxB}{i}}{\hatcurSMEizfeh{\hatcurplanetnumxxxxxB}}{\hatcurSMEiizfeh{\hatcurplanetnumxxxxxB}}}
\newcommand{\hatcurSMEzfehshortxxxxxB}{\ifthenelse{\equal{\hatcurSMEversionxxxxxB}{i}}{\hatcurSMEizfehshort{\hatcurplanetnumxxxxxB}}{\hatcurSMEiizfehshort{\hatcurplanetnumxxxxxB}}}
\newcommand{\hatcurSMEloggxxxxxB}{\ifthenelse{\equal{\hatcurSMEversionxxxxxB}{i}}{\hatcurSMEilogg{\hatcurplanetnumxxxxxB}}{\hatcurSMEiilogg{\hatcurplanetnumxxxxxB}}}
\newcommand{\hatcurSMEvsinxxxxxB}{\ifthenelse{\equal{\hatcurSMEversionxxxxxB}{i}}{\hatcurSMEivsin{\hatcurplanetnumxxxxxB}}{\hatcurSMEiivsin{\hatcurplanetnumxxxxxB}}}
\newcommand{\hatcurSMEvmacxxxxxB}{\ifthenelse{\equal{\hatcurSMEversionxxxxxB}{i}}{\hatcurSMEivmac{\hatcurplanetnumxxxxxB}}{\hatcurSMEiivmac{\hatcurplanetnumxxxxxB}}}
\newcommand{\hatcurSMEvmicxxxxxB}{\ifthenelse{\equal{\hatcurSMEversionxxxxxB}{i}}{\hatcurSMEivmic{\hatcurplanetnumxxxxxB}}{\hatcurSMEiivmic{\hatcurplanetnumxxxxxB}}}

%% file: newcommand_spec_HTR089-018.tex
\newcommand{\hatcurxxxxxC}{HAT-P-60}
\newcommand{\hatcurbxxxxxC}{HAT-P-60b}
\newcommand{\hatcurcxxxxxC}{HAT-P-60c}

\newcommand{\hatcurplanetnumxxxxxC}{60}

\newcommand{\hatcurtwomassshortxxxxxC}{01530777+5203140}

\newcommand{\hatcurSindexxxxxxC}{\ensuremath{0.1236\pm0.0022}}
\newcommand{\hatcurRHKindexxxxxxC}{\ensuremath{-5.309\pm0.033}}

\newcommand{\hatcurRVgammaabsxxxxxC}{\ensuremath{6.582\pm0.027}}                           
\newcommand{\hatcurRVgammainstxxxxxC}{TRES}                           

\newcommand{\hatcurCCtassvixxxxxC}{NULL}                  

\newcommand{\hatcurSMEversionxxxxxC}{ii}                                       

\newcommand{\hatcurisoshortxxxxxC}{YY}
\newcommand{\hatcurisofullxxxxxC}{Yonsei-Yale (YY)}
\newcommand{\hatcurisocitexxxxxC}{yi:2001}

\newcommand{\hatcurlumindxxxxxC}{\arstar}

\newcommand{\hatcurjhkfilsetxxxxxC}{ESO}

\newcommand{\hatcurSMEteffxxxxxC}{\ifthenelse{\equal{\hatcurSMEversionxxxxxC}{i}}{\hatcurSMEiteff{\hatcurplanetnumxxxxxC}}{\hatcurSMEiiteff{\hatcurplanetnumxxxxxC}}}
\newcommand{\hatcurSMEzfehxxxxxC}{\ifthenelse{\equal{\hatcurSMEversionxxxxxC}{i}}{\hatcurSMEizfeh{\hatcurplanetnumxxxxxC}}{\hatcurSMEiizfeh{\hatcurplanetnumxxxxxC}}}
\newcommand{\hatcurSMEzfehshortxxxxxC}{\ifthenelse{\equal{\hatcurSMEversionxxxxxC}{i}}{\hatcurSMEizfehshort{\hatcurplanetnumxxxxxC}}{\hatcurSMEiizfehshort{\hatcurplanetnumxxxxxC}}}
\newcommand{\hatcurSMEloggxxxxxC}{\ifthenelse{\equal{\hatcurSMEversionxxxxxC}{i}}{\hatcurSMEilogg{\hatcurplanetnumxxxxxC}}{\hatcurSMEiilogg{\hatcurplanetnumxxxxxC}}}
\newcommand{\hatcurSMEvsinxxxxxC}{\ifthenelse{\equal{\hatcurSMEversionxxxxxC}{i}}{\hatcurSMEivsin{\hatcurplanetnumxxxxxC}}{\hatcurSMEiivsin{\hatcurplanetnumxxxxxC}}}
\newcommand{\hatcurSMEvmacxxxxxC}{\ifthenelse{\equal{\hatcurSMEversionxxxxxC}{i}}{\hatcurSMEivmac{\hatcurplanetnumxxxxxC}}{\hatcurSMEiivmac{\hatcurplanetnumxxxxxC}}}
\newcommand{\hatcurSMEvmicxxxxxC}{\ifthenelse{\equal{\hatcurSMEversionxxxxxC}{i}}{\hatcurSMEivmic{\hatcurplanetnumxxxxxC}}{\hatcurSMEiivmic{\hatcurplanetnumxxxxxC}}}

%% file: newcommand_spec_HTR094-024.tex
\newcommand{\hatcurxxxxxD}{HAT-P-61}
\newcommand{\hatcurbxxxxxD}{HAT-P-61b}
\newcommand{\hatcurcxxxxxD}{HAT-P-61c}

\newcommand{\hatcurplanetnumxxxxxD}{61}

\newcommand{\hatcurtwomassshortxxxxxD}{05015525+5007526}

\newcommand{\hatcurSindexxxxxxD}{\ensuremath{0.240\pm0.012}}
\newcommand{\hatcurRHKindexxxxxxD}{\ensuremath{-4.719\pm0.032}}

\newcommand{\hatcurRVgammaabsxxxxxD}{\ensuremath{4.810\pm0.022}}                           
\newcommand{\hatcurRVgammainstxxxxxD}{TRES}                           

\newcommand{\hatcurCCtassvixxxxxD}{NULL}                  

\newcommand{\hatcurSMEversionxxxxxD}{ii}                                       

\newcommand{\hatcurisoshortxxxxxD}{YY}
\newcommand{\hatcurisofullxxxxxD}{Yonsei-Yale (YY)}
\newcommand{\hatcurisocitexxxxxD}{yi:2001}

\newcommand{\hatcurlumindxxxxxD}{\arstar}

\newcommand{\hatcurjhkfilsetxxxxxD}{ESO}

\newcommand{\hatcurSMEteffxxxxxD}{\ifthenelse{\equal{\hatcurSMEversionxxxxxD}{i}}{\hatcurSMEiteff{\hatcurplanetnumxxxxxD}}{\hatcurSMEiiteff{\hatcurplanetnumxxxxxD}}}
\newcommand{\hatcurSMEzfehxxxxxD}{\ifthenelse{\equal{\hatcurSMEversionxxxxxD}{i}}{\hatcurSMEizfeh{\hatcurplanetnumxxxxxD}}{\hatcurSMEiizfeh{\hatcurplanetnumxxxxxD}}}
\newcommand{\hatcurSMEzfehshortxxxxxD}{\ifthenelse{\equal{\hatcurSMEversionxxxxxD}{i}}{\hatcurSMEizfehshort{\hatcurplanetnumxxxxxD}}{\hatcurSMEiizfehshort{\hatcurplanetnumxxxxxD}}}
\newcommand{\hatcurSMEloggxxxxxD}{\ifthenelse{\equal{\hatcurSMEversionxxxxxD}{i}}{\hatcurSMEilogg{\hatcurplanetnumxxxxxD}}{\hatcurSMEiilogg{\hatcurplanetnumxxxxxD}}}
\newcommand{\hatcurSMEvsinxxxxxD}{\ifthenelse{\equal{\hatcurSMEversionxxxxxD}{i}}{\hatcurSMEivsin{\hatcurplanetnumxxxxxD}}{\hatcurSMEiivsin{\hatcurplanetnumxxxxxD}}}
\newcommand{\hatcurSMEvmacxxxxxD}{\ifthenelse{\equal{\hatcurSMEversionxxxxxD}{i}}{\hatcurSMEivmac{\hatcurplanetnumxxxxxD}}{\hatcurSMEiivmac{\hatcurplanetnumxxxxxD}}}
\newcommand{\hatcurSMEvmicxxxxxD}{\ifthenelse{\equal{\hatcurSMEversionxxxxxD}{i}}{\hatcurSMEivmic{\hatcurplanetnumxxxxxD}}{\hatcurSMEiivmic{\hatcurplanetnumxxxxxD}}}

%% file: newcommand_spec_HTR130-001.tex
\newcommand{\hatcurxxxxxE}{HAT-P-62}
\newcommand{\hatcurbxxxxxE}{HAT-P-62b}
\newcommand{\hatcurcxxxxxE}{HAT-P-62c}

\newcommand{\hatcurplanetnumxxxxxE}{62}

\newcommand{\hatcurtwomassshortxxxxxE}{04580102+4818038}

\newcommand{\hatcurRVgammaabsxxxxxE}{\ensuremath{50.424\pm0.025}}                           
\newcommand{\hatcurRVgammainstxxxxxE}{TRES}                           

\newcommand{\hatcurCCtassvixxxxxE}{NULL}                  

\newcommand{\hatcurSMEversionxxxxxE}{ii}                                       

\newcommand{\hatcurisoshortxxxxxE}{YY}
\newcommand{\hatcurisofullxxxxxE}{Yonsei-Yale (YY)}
\newcommand{\hatcurisocitexxxxxE}{yi:2001}

\newcommand{\hatcurlumindxxxxxE}{\arstar}

\newcommand{\hatcurjhkfilsetxxxxxE}{ESO}

\newcommand{\hatcurSMEteffxxxxxE}{\ifthenelse{\equal{\hatcurSMEversionxxxxxE}{i}}{\hatcurSMEiteff{\hatcurplanetnumxxxxxE}}{\hatcurSMEiiteff{\hatcurplanetnumxxxxxE}}}
\newcommand{\hatcurSMEzfehxxxxxE}{\ifthenelse{\equal{\hatcurSMEversionxxxxxE}{i}}{\hatcurSMEizfeh{\hatcurplanetnumxxxxxE}}{\hatcurSMEiizfeh{\hatcurplanetnumxxxxxE}}}
\newcommand{\hatcurSMEzfehshortxxxxxE}{\ifthenelse{\equal{\hatcurSMEversionxxxxxE}{i}}{\hatcurSMEizfehshort{\hatcurplanetnumxxxxxE}}{\hatcurSMEiizfehshort{\hatcurplanetnumxxxxxE}}}
\newcommand{\hatcurSMEloggxxxxxE}{\ifthenelse{\equal{\hatcurSMEversionxxxxxE}{i}}{\hatcurSMEilogg{\hatcurplanetnumxxxxxE}}{\hatcurSMEiilogg{\hatcurplanetnumxxxxxE}}}
\newcommand{\hatcurSMEvsinxxxxxE}{\ifthenelse{\equal{\hatcurSMEversionxxxxxE}{i}}{\hatcurSMEivsin{\hatcurplanetnumxxxxxE}}{\hatcurSMEiivsin{\hatcurplanetnumxxxxxE}}}
\newcommand{\hatcurSMEvmacxxxxxE}{\ifthenelse{\equal{\hatcurSMEversionxxxxxE}{i}}{\hatcurSMEivmac{\hatcurplanetnumxxxxxE}}{\hatcurSMEiivmac{\hatcurplanetnumxxxxxE}}}
\newcommand{\hatcurSMEvmicxxxxxE}{\ifthenelse{\equal{\hatcurSMEversionxxxxxE}{i}}{\hatcurSMEivmic{\hatcurplanetnumxxxxxE}}{\hatcurSMEiivmic{\hatcurplanetnumxxxxxE}}}

%% file: newcommand_spec_HTR384-014.tex
\newcommand{\hatcurxxxxxF}{HAT-P-63}
\newcommand{\hatcurbxxxxxF}{HAT-P-63b}
\newcommand{\hatcurcxxxxxF}{HAT-P-63c}

\newcommand{\hatcurplanetnumxxxxxF}{63}

\newcommand{\hatcurtwomassshortxxxxxF}{17581730+0545409}

\newcommand{\hatcurRVgammaabsxxxxxF}{\ensuremath{-68.994\pm0.057}}                           
\newcommand{\hatcurRVgammainstxxxxxF}{TRES}                           

\newcommand{\hatcurCCtassvixxxxxF}{NULL}                  

\newcommand{\hatcurSMEversionxxxxxF}{i}                                       

\newcommand{\hatcurisoshortxxxxxF}{YY}
\newcommand{\hatcurisofullxxxxxF}{Yonsei-Yale (YY)}
\newcommand{\hatcurisocitexxxxxF}{yi:2001}

\newcommand{\hatcurlumindxxxxxF}{\arstar}

\newcommand{\hatcurjhkfilsetxxxxxF}{ESO}

\newcommand{\hatcurSMEteffxxxxxF}{\ifthenelse{\equal{\hatcurSMEversionxxxxxF}{i}}{\hatcurSMEiteff{\hatcurplanetnumxxxxxF}}{\hatcurSMEiiteff{\hatcurplanetnumxxxxxF}}}
\newcommand{\hatcurSMEzfehxxxxxF}{\ifthenelse{\equal{\hatcurSMEversionxxxxxF}{i}}{\hatcurSMEizfeh{\hatcurplanetnumxxxxxF}}{\hatcurSMEiizfeh{\hatcurplanetnumxxxxxF}}}
\newcommand{\hatcurSMEzfehshortxxxxxF}{\ifthenelse{\equal{\hatcurSMEversionxxxxxF}{i}}{\hatcurSMEizfehshort{\hatcurplanetnumxxxxxF}}{\hatcurSMEiizfehshort{\hatcurplanetnumxxxxxF}}}
\newcommand{\hatcurSMEloggxxxxxF}{\ifthenelse{\equal{\hatcurSMEversionxxxxxF}{i}}{\hatcurSMEilogg{\hatcurplanetnumxxxxxF}}{\hatcurSMEiilogg{\hatcurplanetnumxxxxxF}}}
\newcommand{\hatcurSMEvsinxxxxxF}{\ifthenelse{\equal{\hatcurSMEversionxxxxxF}{i}}{\hatcurSMEivsin{\hatcurplanetnumxxxxxF}}{\hatcurSMEiivsin{\hatcurplanetnumxxxxxF}}}
\newcommand{\hatcurSMEvmacxxxxxF}{\ifthenelse{\equal{\hatcurSMEversionxxxxxF}{i}}{\hatcurSMEivmac{\hatcurplanetnumxxxxxF}}{\hatcurSMEiivmac{\hatcurplanetnumxxxxxF}}}
\newcommand{\hatcurSMEvmicxxxxxF}{\ifthenelse{\equal{\hatcurSMEversionxxxxxF}{i}}{\hatcurSMEivmic{\hatcurplanetnumxxxxxF}}{\hatcurSMEiivmic{\hatcurplanetnumxxxxxF}}}

%% file: newcommand_spec_HTR405-001.tex
\newcommand{\hatcurxxxxxG}{HAT-P-64}
\newcommand{\hatcurbxxxxxG}{HAT-P-64b}
\newcommand{\hatcurcxxxxxG}{HAT-P-64c}

\newcommand{\hatcurplanetnumxxxxxG}{64}

\newcommand{\hatcurtwomassshortxxxxxG}{04355384+0225526}

\newcommand{\hatcurSindexxxxxxG}{\ensuremath{0.1453\pm0.0068}}
\newcommand{\hatcurRHKindexxxxxxG}{\ensuremath{-5.062\pm0.057}}

\newcommand{\hatcurRVgammaabsxxxxxG}{\ensuremath{25.22\pm0.10}}                           
\newcommand{\hatcurRVgammainstxxxxxG}{TRES}                           

\newcommand{\hatcurCCtassvixxxxxG}{NULL}                  

\newcommand{\hatcurSMEversionxxxxxG}{i}                                       

\newcommand{\hatcurisoshortxxxxxG}{YY}
\newcommand{\hatcurisofullxxxxxG}{Yonsei-Yale (YY)}
\newcommand{\hatcurisocitexxxxxG}{yi:2001}

\newcommand{\hatcurlumindxxxxxG}{\arstar}

\newcommand{\hatcurjhkfilsetxxxxxG}{ESO}

\newcommand{\hatcurSMEteffxxxxxG}{\ifthenelse{\equal{\hatcurSMEversionxxxxxG}{i}}{\hatcurSMEiteff{\hatcurplanetnumxxxxxG}}{\hatcurSMEiiteff{\hatcurplanetnumxxxxxG}}}
\newcommand{\hatcurSMEzfehxxxxxG}{\ifthenelse{\equal{\hatcurSMEversionxxxxxG}{i}}{\hatcurSMEizfeh{\hatcurplanetnumxxxxxG}}{\hatcurSMEiizfeh{\hatcurplanetnumxxxxxG}}}
\newcommand{\hatcurSMEzfehshortxxxxxG}{\ifthenelse{\equal{\hatcurSMEversionxxxxxG}{i}}{\hatcurSMEizfehshort{\hatcurplanetnumxxxxxG}}{\hatcurSMEiizfehshort{\hatcurplanetnumxxxxxG}}}
\newcommand{\hatcurSMEloggxxxxxG}{\ifthenelse{\equal{\hatcurSMEversionxxxxxG}{i}}{\hatcurSMEilogg{\hatcurplanetnumxxxxxG}}{\hatcurSMEiilogg{\hatcurplanetnumxxxxxG}}}
\newcommand{\hatcurSMEvsinxxxxxG}{\ifthenelse{\equal{\hatcurSMEversionxxxxxG}{i}}{\hatcurSMEivsin{\hatcurplanetnumxxxxxG}}{\hatcurSMEiivsin{\hatcurplanetnumxxxxxG}}}
\newcommand{\hatcurSMEvmacxxxxxG}{\ifthenelse{\equal{\hatcurSMEversionxxxxxG}{i}}{\hatcurSMEivmac{\hatcurplanetnumxxxxxG}}{\hatcurSMEiivmac{\hatcurplanetnumxxxxxG}}}
\newcommand{\hatcurSMEvmicxxxxxG}{\ifthenelse{\equal{\hatcurSMEversionxxxxxG}{i}}{\hatcurSMEivmic{\hatcurplanetnumxxxxxG}}{\hatcurSMEiivmic{\hatcurplanetnumxxxxxG}}}

%% file: phfu_tab_combined_short.tex
HAT-P-58 & $ 56239.13511 $ & $   0.00565 $ & $   0.01008 $ & $ \cdots $ & $ r$ &     HATNet\\
HAT-P-58 & $ 56235.12147 $ & $  -0.01055 $ & $   0.01209 $ & $ \cdots $ & $ r$ &     HATNet\\
HAT-P-58 & $ 56207.02456 $ & $  -0.00900 $ & $   0.01093 $ & $ \cdots $ & $ r$ &     HATNet\\
HAT-P-58 & $ 56243.14926 $ & $  -0.00973 $ & $   0.01036 $ & $ \cdots $ & $ r$ &     HATNet\\
HAT-P-58 & $ 56194.98348 $ & $   0.03111 $ & $   0.01042 $ & $ \cdots $ & $ r$ &     HATNet\\
HAT-P-58 & $ 56211.03898 $ & $  -0.00313 $ & $   0.01037 $ & $ \cdots $ & $ r$ &     HATNet\\
HAT-P-58 & $ 56194.98363 $ & $  -0.00382 $ & $   0.00963 $ & $ \cdots $ & $ r$ &     HATNet\\
HAT-P-58 & $ 56375.60748 $ & $  -0.02041 $ & $   0.01786 $ & $ \cdots $ & $ r$ &     HATNet\\
HAT-P-58 & $ 56198.99820 $ & $  -0.02555 $ & $   0.01558 $ & $ \cdots $ & $ r$ &     HATNet\\
HAT-P-58 & $ 56383.63531 $ & $   0.01115 $ & $   0.01428 $ & $ \cdots $ & $ r$ &     HATNet\\